\documentclass[11pt, double, phd]{osudiss-2} 

\usepackage{graphicx} 
\usepackage{lipsum}
\usepackage{amsmath,amsfonts,amsthm}
\usepackage{commath}
\usepackage{amssymb}
\usepackage{verbatimbox}
\usepackage[utf8]{inputenc}
\usepackage{bm} 
\usepackage{booktabs} 
\usepackage[titletoc]{appendix}

\usepackage{bookmark} 
\hypersetup{colorlinks=true,urlcolor=blue,linkcolor=blue, citecolor=blue, breaklinks} 
\usepackage[all]{hypcap}

\usepackage[numbers,square,sort&compress]{natbib}

\usepackage{subfig}
\usepackage[export]{adjustbox}
\PassOptionsToPackage{obeyspaces}{url}

\newcommand{\be}{\begin{equation}}
\newcommand{\ee}{\end{equation}}
\newcommand{\bea}{\begin{eqnarray}}
\newcommand{\eea}{\end{eqnarray}}
\newcommand{\beal}{\begin{align}}
\newcommand{\enal}{\end{align}}
\newcommand{\bs}{\begin{subequations}}
\newcommand{\es}{\end{subequations}}
\newcommand{\besp}{\begin{split}}
\newcommand{\eesp}{\end{split}}

\newcommand{\feq}{f_\text{eq}}
\newcommand{\pxp}{(- p \cdot \Xi \cdot p)}

\newcommand{\dft}{\delta \tilde f}

\newcommand{\up}{u \cdot p}

\newcommand{\mzp}{(- z \cdot p)}

\newcommand{\ene}{\mathcal{E}}

\newcommand{\Peq}{\mathcal{P}_{\eq}}
\newcommand{\Pavg}{\bar{\mathcal{P}}}
\newcommand{\pperp}{p^{\{\mu\}}}

\newcommand{\Wperp}{W^\mu_{\perp z}}
\newcommand{\piperp}{\pi^\munu_{\perp}}
\newcommand{\munu}{{\mu\nu}}

\title{Far-from-equilibrium hydrodynamic simulations of ultrarelativistic nuclear collisions}
\author{Michael J. McNelis}
\advisorname{Ulrich Heinz}
\degree{Doctor of Philosophy}
\member{Michael Lisa}
\member{Richard Furnstahl}
\member{Yuanming Lu}
\authordegrees{M.S.}
\graduationyear{2021}
\unit{Graduate Program in Physics}

\begin{document}

\frontmatter

\begin{abstract}

We develop a far-from-equilibrium hydrodynamic model to evolve ultrarelativistic heavy-ion collisions in event-by-event simulations. Anisotropic hydrodynamics is designed to better handle the strong and highly anisotropic expansion during the early stages of the collision. The large gradients cause conventional second-order viscous hydrodynamic approaches to break down at early times. Anisotropic hydrodynamics evolves the large pressure anisotropies present in the quark-gluon plasma non-perturbatively, which prevents negative longitudinal pressures from developing even under extreme conditions. This increased stability allows us to start anisotropic hydrodynamics already at a very early longitudinal proper time to evolve the pre-hydrodynamic stage. In current pre-hydrodynamic models, the equation of state is not consistent with the QCD equation of state used in the subsequent fluid dynamic stage. Since our approach avoids this inconsistency, we are able to achieve a smooth transition to non-conformal viscous hydrodynamics as the gradients decrease over time. For our first phenomenological application, we apply our new simulation to model fluctuating Pb+Pb collisions at LHC energies ($\sqrt{s_\text{NN}} = 2.76$ TeV) and find that our preliminary calculations for the hadronic observables are in excellent agreement with the experimental data.

\end{abstract}

\dedication{To aspiring students with special needs} 
\begin{acknowledgments}

First, I would like to thank my advisor Ulrich Heinz for being an outstanding mentor. I was one of the most unconventional students to his eyes (always have been proud of it), but I am glad he helped me realize my full potential. I learned a lot about being a scientist from him, including how to approach very difficult problems from new perspectives and with great persistence, master my technical writing skills and confidently defend my work. The group atmosphere was very laid-back, flexible and open-minded, which gave me the freedom I needed to take the initiative and independently develop my own research projects. I also enjoyed the numerous opportunities he provided to travel across the country and abroad for conferences, summer schools and collaborations.

I am grateful to Jia Liu and Jing Qian for assisting me in my first project. Jing was a very warm person and a great source of encouragement when I struggled early on in my research. I would also like to thank Lipei Du and Derek Everett for being exceptional office mates and great friends. Lipei was a fun biking buddy to have here in Columbus and a good companion during my trips away from campus. Derek was highly knowledgeable in computer programming and his technical advice helped me refine my skills as a software developer. I particularly enjoyed the time we spent collaborating on one of the software packages developed in this thesis. The postdocs in our group, including Chandrodoy Chattopadhyay and Gojko Vujanovic, gave me numerous suggestions that were useful in my later projects. 

Next, I would like to thank my family for their love and ongoing support. Ever since I started college at Penn State, it was strange for me to live as an individual and be so far away from home for long periods of time. After more than ten years, I have grown very much accustomed to the independent lifestyle, but I have always enjoyed touching base with them occasionally on summer beach vacations and holiday gatherings. My mom was especially generous in providing me with additional financial support, which was a godsend when my old car broke down. Sadly, my grandmother passed away during my Ph.D. program, but I still hold fond memories of her when I was growing up and the intelligent discussions we had.

And finally, I would like to thank the Latin dance community in Columbus for making my time here enjoyable. Every week I would always look forward to go out salsa dancing, either to practice my moves or just blow off steam and have fun. I probably would have never done it at all if it was not for my physics friend Brady Hood who first introduced me to it. I learned a lot from him, as well as Stacy Coil and Deyannira Tirado, when I started dancing. I am also grateful for the friends I made at the Salsa Club at OSU, including Rowan McLachlan, Jorge Torres, Carly Schwarm, Brian Hribar, Rachel Sette, Claire Whillans and Dhwani Parikh. Rowan was an exceptional dancer and a phenomenal instructor who inspired me to take on a teaching and administrative role in the club. I had a lot of fun teaching salsa and bachata classes to the college students on campus during my last few years here and also do some community outreach activities, especially for students with disabilities who have few opportunities to dance. Unfortunately, the pandemic ended my time at Ohio State with a sour note but I was lucky to have Claire and Dhwani as my quarantine buddies. I really miss dancing with you guys and hope to return to the dance floor soon.

\end{acknowledgments}
\newcommand{\x}{x^\mu}
\newcommand{\xx}{x^\mu}
\newcommand{\g}{g_{\mu\nu}}
\newcommand{\gup}{g^{\mu\nu}}
\newcommand{\Du}{\Delta^{\mu\nu}}
\newcommand{\Duhat}{\hat{\Delta}^{\mu\nu}}
\newcommand{\Dd}{\Delta_{\mu\nu}}
\newcommand{\eq}{\mathrm{eq}}
\newcommand{\pres}{\mathcal{P}}
\newcommand{\uu}{u}
\newcommand{\enehat}{\hat{\mathcal{E}}}
\newcommand{\um}{u^\mu}
\newcommand{\umhat}{\hat{u}^\mu}
\newcommand{\uum}{u_\mu}
\newcommand{\un}{u^\nu}
\newcommand{\unhat}{\hat{u}^\nu}
\newcommand{\unn}{u_\nu}
\newcommand{\tmn}{T^{\mu\nu}}
\newcommand{\piu}{\pi^{\mu\nu}}
\newcommand{\parho}{\partial_\rho}
\newcommand{\pit}{\pi}
\newcommand{\trr}{T^{\rho\rho}}
\newcommand{\ttt}{T^{\theta\theta}}
\newcommand{\tpp}{T^{\phi\phi}}
\newcommand{\tvv}{T^{\varsigma\varsigma}}
\newcommand{\pr}{\partial_\rho}
\newcommand{\gh}{\hat{\Gamma}}
\newcommand{\p}{p_i}
\newcommand{\pp}{\hat{p}}
\newcommand{\po}{\hat{p}_\Omega^2}
\newcommand{\pv}{p^\varsigma}
\newcommand{\pvv}{p_\eta}
\newcommand{\pvvhat}{\hat{p}_\eta}
\newcommand{\pth}{\hat{p}_\theta}
\newcommand{\pph}{\hat{p}_\phi}
\newcommand{\ph}{\hat{p}^\rho}
\newcommand{\phrs}{\hat{p}^\rho_{RS}}
\newcommand{\tem}{T}
\newcommand{\temhat}{\hat{T}}
\newcommand{\lam}{\hat{\Lambda}}
\newcommand{\hint}{\hat{\mathcal{H}}}
\newcommand{\Idf}{\mathcal{I}^{{\delta \tilde{f}}}}
\newcommand{\tdf}{\delta \tilde{f}}
\newcommand{\tin}{T(\rho_0)}
\newcommand{\intp}{\int_{\hat{p}}} 
\newcommand{\R}{\mathcal{R}}
\newcommand{\Rhat}{\hat{\mathcal{R}}}
\newcommand{\Ih}{\hat{I}}
\newcommand{\It}{\tilde{I}}
\newcommand{\I}{\mathcal{I}}
\newcommand{\calI}{{\hat{\mathcal{I}}}}
\newcommand{\calIeq}{{\mathcal{I}_\mathrm{eq}}}
\newcommand{\II}{{\mathcal{I}^{\mathrm{eq}}}}
\newcommand{\J}{\mathcal{J}}
\newcommand{\Jhat}{\hat{\mathcal{J}}}
\newcommand{\lu}{z^\mu}
\newcommand{\luhat}{\hat{z}^\mu}
\newcommand{\ld}{z_\mu}
\newcommand{\ldhat}{\hat{z}_\mu}
\newcommand{\lh}{l}
\newcommand{\xu}{\Xi^{\mu\nu}}
\newcommand{\xuhat}{\hat{\Xi}^{\mu\nu}}
\newcommand{\xd}{\Xi_{\mu\nu}}
\newcommand{\she}{\pi}
\newcommand{\shehat}{\hat{\pi}}
\newcommand{\Pt}{\mathcal{P}_\perp}
\newcommand{\Pl}{\mathcal{P}_L}
\newcommand{\Pthat}{\hat{\mathcal{P}}_\perp}
\newcommand{\Plhat}{\hat{\mathcal{P}}_L}
\newcommand{\order}{\mathcal{O}}
\newcommand{\E}{\hat{E}_{\bold{\pp}\,\hat{u}}}
\newcommand{\El}{\hat{E}_{\bold{\pp}\,\lh}}
\newcommand{\dr}{\partial_\rho}
\newcommand{\ids}{\int_{\pp}}
\newcommand{\fp}{f_{\pp}}
\newcommand{\piti}{{\hat{\tilde{\pi}}}}
\newcommand{\xh}{\hat{x}}
\newcommand{\so}{SO(3)_q}
\newcommand{\gub}{SO(3)_q\otimes SO(1,1)\otimes Z_2}
\newcommand{\etas}{\eta / \mathcal{S}}
\newcommand{\zetas}{\zeta / \mathcal{S}}
\newcommand{\G}{\text{G}}
\newcommand{\tp}{\tau^\prime}
\newcommand{\tpi}{\tau_\pi}
\newcommand{\trel}{\tau_r}
\newcommand{\Kn}{\text{Kn}}
\newcommand{\dt}{\partial_\tau}
\newcommand{\dtsq}{\partial^2_\tau}
\newcommand{\dtcube}{\partial^3_\tau}
\newcommand{\bp}{{\bm p}}
\newcommand{\PL}{\mathcal{P}_L}
\newcommand{\Pperp}{\mathcal{P}_\perp}
\newcommand{\LL}{\mathcal{L}}
\newcommand{\W}{\mathcal{W}}
\newcommand{\half}{\frac{1}{2}}
\newcommand{\ab}{{\alpha\beta}}
\newcommand{\A}{\mathcal{A}}
\newcommand{\N}{\mathcal{N}}
\newcommand{\M}{\mathcal{M}}
\newcommand{\B}{\mathcal{B}}
\newcommand{\cpuvah}{{\sc VAH}}
\newcommand{\gpuvh}{{\sc GPU VH}}
\newcommand{\vh}{{\sc VH}}
\newcommand{\beshydro}{{\sc BEShydro}}
\newcommand{\trento}{{\sc T}$_\text{\sc R}${\sc{ENTo}}}
\newcommand{\PLhat}{\hat{\mathcal{P}}_L}
\newcommand{\pOp}{(p \cdot \Omega \cdot p)}

\newcommand{\mike}{\textcolor{blue}}
\begin{vita}
\dateitem{May, 2014}{B.S., Pennsylvania State University, State College, PA}
\dateitem{Aug, 2016}{M.S., The Ohio State University, Columbus, OH}

\begin{publist}

\pubitem{M. McNelis and U. Heinz, \textit{Modified equilibrium distributions for Cooper--Frye particlization}, Phys. Rev. C (under review), \href{https://arxiv.org/pdf/2103.03401.pdf}{arXiv:2103.03401}}

\pubitem{M. McNelis, D. Bazow and U. Heinz, \textit{Anisotropic fluid dynamical simulations of heavy-ion collisions}, Comput. Phys. Commun. (under review), \href{https://arxiv.org/pdf/2101.02827.pdf}{arXiv:2101.02827}}

\pubitem{M. McNelis and U. Heinz, \textit{Hydrodynamic generators in relativistic kinetic theory}, \href{https://journals.aps.org/prc/abstract/10.1103/PhysRevC.101.054901}{Phys. Rev. C 101, 054901 (2020)}, \href{https://arxiv.org/pdf/2001.09125.pdf}{arXiv:2001.09125 }}

\pubitem{M. McNelis, D. Everett and U. Heinz, \textit{Particlization in fluid dynamical simulations of heavy-ion collisions: The {\sc iS3D} module}, \href{https://www.sciencedirect.com/science/article/abs/pii/S0010465520302836?via\%3Dihub}{Comput. Phys. Commun. 258 (2021) 107604},  \href{https://arxiv.org/pdf/1912.08271.pdf}{arXiv:1912.08271 }}

\pubitem{M. McNelis, D. Bazow and U. Heinz, \textit{Viscous hydrodynamics for nonconformal anisotropic fluids}, \href{https://www.sciencedirect.com/science/article/pii/S0375947418302422?via\%3Dihub}{Nucl. Phys. A 982 (2019) 915-918}, \href{https://arxiv.org/pdf/1807.04223.pdf}{arXiv:1807.04223 }}

\pubitem{M. McNelis, D. Bazow and U. Heinz, \textit{(3+1)--dimensional anisotropic fluid dynamics with a lattice QCD equation of state}, \href{https://journals.aps.org/prc/abstract/10.1103/PhysRevC.97.054912}{Phys. Rev. C 97, 054912 (2018)}, \href{https://arxiv.org/pdf/1803.01810.pdf}{arXiv:1803.01810 }}

\pubitem{M. Martinez, M. McNelis and U. Heinz, \textit{Viscous anisotropic hydrodynamics for the Gubser flow}, \href{https://www.sciencedirect.com/science/article/pii/S0375947417300751?via\%3Dihub}{Nucl. Phys. A 967 (2017) 413-416}, \href{https://arxiv.org/pdf/1704.04727.pdf}{arXiv:1704.04727 }}

\pubitem{M. Martinez, M. McNelis and U. Heinz, \textit{Anisotropic fluid dynamics for Gubser flow}, \href{https://journals.aps.org/prc/abstract/10.1103/PhysRevC.95.054907}{Phys. Rev. C 95, 054907 (2017)}, \href{https://arxiv.org/pdf/1703.10955.pdf}{arXiv:1704.10955 }}

\end{publist}

\begin{fieldsstudy}
\majorfield{Physics}
\end{fieldsstudy}

\end{vita}

\tableofcontents 

\clearpage 
\listoffigures 
\clearpage  
\listoftables 

\mainmatter
\chapter{Introduction}
\label{ch1label}
\section{Quark-gluon plasma}
The quark-gluon plasma is one of the most extreme phases of matter in our universe. The incredibly high temperatures required to create it, $T \sim 10^{12}$ K, are orders of magnitude beyond those of any substance typically encountered in nature. This makes it one of the hottest phases of matter known to humankind, second only to the hypothetical electroweak plasma that preceded it after the Big Bang. The quark-gluon plasma is also unique because its constituents, the quarks and gluons, are color-charged (quarks also have electric charge) and interact primarily through the strong force instead of the electric force. On the contrary, most plasmas are composed of electrically charged ion and electron gases.

Apart from some of its thermodynamic properties~\cite{Wang:2016opj}, the quark-gluon plasma is not very well understood. This is mainly due to the differences in behavior from the hadron gas phase of QCD matter. At temperatures below the deconfinement temperature $T_c = 154$ MeV, the hadrons' valence quarks are bound together by gluons; hence the thermal degrees of freedom are the colorless hadrons themselves rather than the quarks and gluons. With increasing temperatures, however, the strength of the strong interaction decreases due to asymptotic freedom. Furthermore, the thermal quark-antiquark and gluon-antigluon pairs that are created Deybe screen the potential that confines the valence quarks. As a result, the quarks and gluons become deconfined, allowing them to permeate more freely across a volume larger than that of individual hadrons. At arbitrarily high temperatures, the quark-gluon plasma is expected to behave as a weakly coupled gas since the strong coupling constant diminishes~\cite{Laine:2016hma}. In this limit, perturbative QCD techniques can be used to deduce some of its transport properties~\cite{Arnold:2000dr,Arnold:2002zm,Arnold:2003zc,Ghiglieri:2018dgf,Ghiglieri:2018dib}. For the intermediate temperatures $T \sim 0.15 - 0.5$ GeV that are presently accessible in the world's largest particle accelerators, RHIC and LHC, there is growing experimental evidence that the quark-gluon plasma is a strongly coupled liquid with the lowest shear viscosity to entropy density ratio $\etas$ of any known fluid. Current phenomenological constraints put $\etas \sim 0.1$ at temperatures around the deconfinement temperature $T_c$ \cite{Everett:2020yty,Everett:2020xug}, which is very close to the conjectured Kovtun--Son--Starinets bound $\etas \geq 1/4\pi$~\cite{Kovtun:2003wp}. These findings are in line with the current understanding that QCD at finite temperature becomes non-perturbative in the intermediate region $T \sim 0.15 - 1.0$ GeV, making it extraordinarily difficult to evaluate the transport properties of QCD matter using first-principles~\cite{Bazavov:2019lgz}.
 
Besides the shear viscosity $\eta$, there are other transport properties such as the bulk viscosity $\zeta$ and baryon diffusion coefficient $\kappa_B$ that influence the dynamics of a non-equilibrium quark-gluon plasma.\footnote{%
    In this thesis, we will not consider the effects of baryon chemical potential $\mu_B$ and baryon diffusion current $V^\mu_B$ (i.e. we set $\mu_B = 0 =V_B^\mu$).} 
In viscous hydrodynamics, these transport coefficients are of first order since they couple to the first-order gradient forces that drive the fluid away from local equilibrium. Today, precise theoretical calculations for their values remain far out of reach. Even state-of-the-art lattice QCD calculations on the most advanced supercomputers are only capable of extracting the quark-gluon plasma's equilibrium equation of state $\Peq(T,\mu_B)$ at zero baryon chemical potential $\mu_B = 0$ but not its transport coefficients~\cite{Bazavov:2014pvz}. Alternatively, we can develop experiments to constrain the transport properties of the quark-gluon plasma. 

It is widely believed that the quark-gluon plasma filled the entire universe for a brief period of time $t \sim 10^{-12} - 10^{-5}$\,s after the Big Bang~\cite{Yagi:2005yb}. After the quark epoch, the rapidly expanding plasma cooled down and transitioned into a hadron resonance gas, which went on to decay into more stable hadrons and form the atomic nuclei that are prevalent today. Thus, we have no experimental access to the quark-gluon plasma from natural events; one possible exception is a binary neutron star merger, which we are now able to probe using gravitational waves~\cite{TheLIGOScientific:2017qsa}. Instead, we can produce it artificially in particle accelerators. Recreating such extreme conditions is accomplished by colliding beams of heavy ions (e.g. Au+Au, Pb+Pb) at very high energies. However, these ultrarelativistic nuclear collisions can only form femtoscopically small droplets of plasma, which have incredibly short lifetimes ($\tau_f \sim 10^{-23}$\,s) due to the large longitudinal and transverse expansion rates in such collisions. This makes them difficult to probe directly. To reconstruct the medium's transport properties, one can use so-called hybrid models, which simulate the various stages of a heavy-ion collision with individual physical models that are interfaced with each other, and fit the combined framework to large sets of soft-momentum ($p_T < 3$\,GeV) hadronic observables \cite{Bass:2000ib, Shen:2014vra, Bernhard:2016tnd, Bernhard:2018hnz, Everett:2020yty, Everett:2020xug, Nijs:2020ors, Nijs:2020roc}.\footnote{%
    High-energy jets, heavy quarks and electromagnetic radiation can also serve as experimental probes but their full integration into hybrid models \cite{Hirano:2012kj, Paquet:2015lta, Shen:2016zpp, Okai:2017ofp, Vujanovic:2019yih, Yao:2020xzw} is more 
    involved than for soft-momentum hadrons.}
The hybrid model must be accurate enough that it can make full use of the considerable precision of the available experimental data.
\section{Hybrid model simulations of heavy-ion collisions}
\subsection{Early models}
Older hydrodynamic evolution models from the early 2000s were simplistic in their design~\cite{Kolb:2003dz}. They consisted of three main stages, each of which is simulated by an individual program:
\begin{enumerate}
    \item {\sl Initial conditions:} An initial-state model sets the initial energy deposition by the participant nucleons of a colliding nuclei pair with impact parameter $b$. One assumes the medium forms instantaneously and uniformly on a line of constant longitudinal proper time $\tau = \sqrt{t^2-z^2}$, which is appropriate for high-energy collisions with approximate longitudinal boost-invariance. This also implies that macroscopic quantities are independent of the spacetime rapidity $\eta_s = \tanh^{-1}(z/t)$. The popular optical Glauber model~\cite{Miller:2007ri} outputs a smooth, event-averaged energy density profile in the transverse plane at $\eta_s = 0$ and does not consider any event-by-event fluctuations in the initial state. The initial fluid velocity profile is taken to be static.
    \item {\sl Relativistic hydrodynamics:} The medium is evolved macroscopically with (2+1)--dimensional ideal (i.e. perfect fluid) hydrodynamics, starting at some longitudinal proper time $\tau_0 \sim 0.5 - 1.0$ fm/$c$; the preceding collision dynamics in the interval $0 < \tau < \tau_0$ are ignored. The hydrodynamic equations are given by the energy-momentum conservation laws
    \be 
    \partial_\mu T^\munu_\text{ideal} = 0\,,
    \ee
    where the ideal part of the energy-momentum tensor $T^\munu$ is decomposed as\footnote{Unless stated otherwise, the net baryon current $J^\mu_B$ is set to zero.}
    \be
    T^\munu_\text{ideal} = \ene u^\mu u^\nu - \Peq \Delta^\munu\,.
    \ee
    Here $\ene$ and $\Peq$ are the energy density and equilibrium pressure in the local-rest-frame (LRF) of the fluid, $u^\mu$ is the fluid velocity and $\Delta^\munu = g^\munu - u^\mu u^\nu$ is the spatial projection tensor. The dissipative components of $T^\munu$, the shear stress tensor $\pi^\munu$ and bulk viscous pressure $\Pi$, are set to zero. Here the conservation laws are not enough to evolve the ideal energy-momentum tensor since it has five independent components.\footnote{The fluid velocity $u^\mu$ has three independent components from the normalization condition $u_\mu u^\mu = 1$} To close the hydrodynamic equations, one supplements a microscopic equation of state for the equilibrium pressure $\Peq(\ene)$, which can be provided by lattice QCD.
    \item {\sl Particlization:} Once the quark-gluon plasma cools down below the pseudocritical temperature $T_c$, it undergoes a phase transition into a hadron resonance gas. A particle detector then measures the incoming hadrons produced by the collision. In order to compare the theoretical model to experimental data, the macroscopic fluid is converted into microscopic hadrons using the Cooper--Frye formula~\cite{Cooper:1974mv}
    \be
    \label{eqch5:CooperFrye}
    E_p \frac{dN_n}{d^3p} = \frac{1}{(2\pi\hbar)^3}\int_{\Sigma} p \cdot d^3\sigma \,f_n(x,p)\,,
    \ee
    where the isothermal hypersurface $\Sigma(x)$ defines the fireball boundary from which hadrons are emitted and $f_n(x,p)$ is the single-particle distribution function of hadron species $n$. For simplicity, the hadronic distribution is set to the local-equilibrium distribution $f_n = f_{\eq,n}$ (see Eq.~\eqref{eqch5:feq}) while viscous corrections $\delta f_n = f_n - f_{\eq,n}$ are set to zero. It is also assumed that the hadrons kinetically decouple from each other immediately after emission. Resonance decay feed-down routines can be used to partially correct the continuous particle spectra~\eqref{eqch5:CooperFrye}, but other microscopic scattering mechanisms that play a role after thermal emission are neglected.
\end{enumerate}
It is important that the hydrodynamic model accurately represents the evolution of the quark-gluon plasma. As the simplest and most versatile fluid dynamic model, ideal hydrodynamics is a popular first-choice for modeling fluids in many fields of science such as astrophysics. However, it is only suitable for fluids whose viscous effects can be neglected.\footnote{%
    Obviously, ideal hydrodynamics cannot be used to constrain the transport coefficients of QCD matter since they do not appear in the model. Instead, the primary motivation for its early use was to predict signatures of hydrodynamic flow, if any, in the final-state hadronic observables.}

Historically, the simulation of heavy-ion collisions with relativistic hydrodynamics was met with great skepticism. This was mainly attributed to the prior belief that the quark-gluon plasma is a gaseous phase at intermediate temperatures. Even if a hot, dense plasma can form in high-energy nuclear collisions, the medium would be too viscous to be described by perfect fluid dynamics, especially given the fact that the system undergoes rapid longitudinal expansion. For that reason, microscopic models such as {\sc{AMPT}} were often preferred over macroscopic hydrodynamics~\cite{Zhang:1999bd}. The situation changed when the RHIC experiments involving Au+Au collisions $\sqrt{s_\text{NN}}=200$ GeV came online~\cite{Adler:2003kt} and reported a strong anisotropic flow, a key prediction of hydrodynamic models. Surprisingly, ideal hydrodynamic simulations were successful at qualitatively (and even semi-quantitatively) describing the elliptic flow coefficients $v_{2,\text{id}}$ of identified hadrons (e.g. $\pi^\pm, K^\pm, p\bar{p}$) ~\cite{Adler:2003kt, Adams:2004bi, Adams:2005dq}, while {\sc{AMPT}} underpredicted these observables. These results led to the new belief that the quark-gluon plasma was a near-perfect liquid capable of reaching thermalization within a short time scale $\tau_\text{therm} \sim 1$ fm/$c$~\cite{Heinz:2001xi}.

The conjecture that the quark-gluon plasma has a near-zero viscosity still holds up today but the idea of early thermalization turned out to be premature~\cite{Schlichting:2019abc}. Many attempts were made to find a mechanism that explains rapid thermalization but all of them predicted a time scale $\tau_\text{therm}$ several times longer than the best-fit parameter values for $\tau_0$ extracted from early model-to-data comparison studies. At best, thermalization is achieved during the middle-to-end of the fireball's lifetime. Furthermore, the KSS bound proposed shortly after meant that very strongly coupled fluids may have a small but nonzero shear viscosity. Due to the geometric profile in heavy-ion collisions, even a small $\etas$ value cannot fully compensate for the large gradient forces driving the medium away from local equilibrium. Therefore, viscous corrections to the hydrodynamic equations become important, especially during the early stages of the collision.
%
\begin{figure}[t]
\centering
\includegraphics[width=\linewidth]{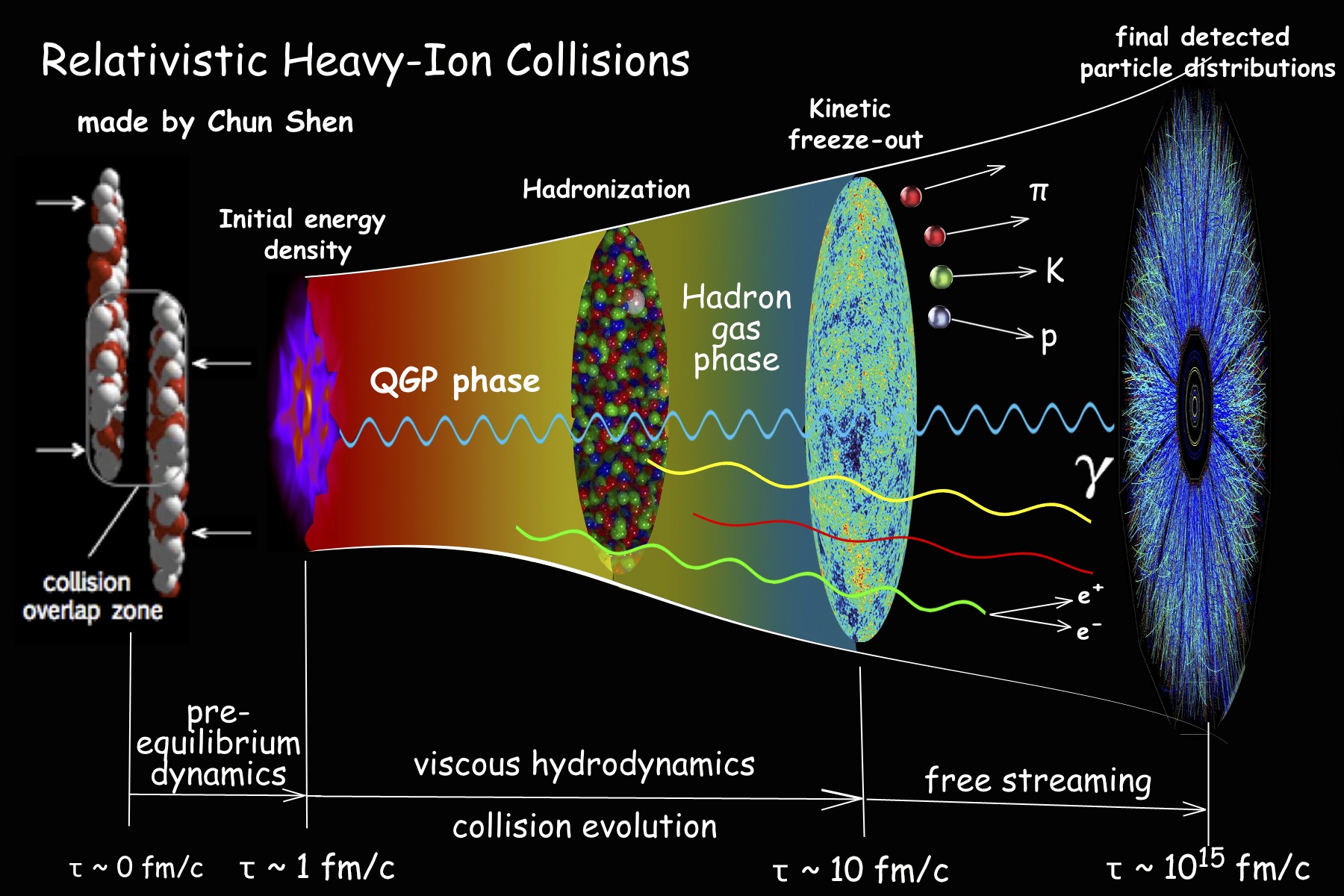}
\caption{A flow chart depicting the multiple stages of an ultrarelativistic heavy-ion collision. This figure was taken from Ref.~\cite{Shen:2015msa}.
\label{hybrid_model}
}
\end{figure}
\subsection{Current models}
The theoretical modeling of heavy-ion collisions has advanced considerably over the last two decades. An illustration of modern hybrid models is shown in Figure~\ref{hybrid_model}. Here we outline the five stages of the modular hybrid framework, which build on the models from the previous section: 
\begin{enumerate}
    \item {\sl Initial conditions:} Initial-state models now include quantum fluctuations in the initial energy density profile by sampling the colliding nucleons' position and the energy deposited by each inelastic collision. This is particularly important for producing the nonzero anisotropic flow observed experimentally in central collisions (i.e. $b \sim 0$ fm). Today, the most widely used initial-state models for high-energy Au+Au and Pb+Pb collisions at the RHIC and LHC, respectively, are Monte Carlo Glauber, \trento{} and IP--Glasma~\cite{Miller:2007ri,Moreland:2014oya, Schenke:2012wb}. For low-energy nuclear collisions, dynamical initialization models are more desirable but are still under development~\cite{Shen:2017bsr, Shen:2018pty, Akamatsu:2018olk, Du:2018mpf, Du:2019obx}.
    \item {\sl Pre-equilibrium dynamics:} A pre-equilibrium (or pre-hydrodynamic) model is used to evolve the initial profile before starting viscous hydrodynamics at $\tau_0 \sim 0.5 - 1$ fm/$c$. For simplicity, these models are typically conformal at leading order, i.e. the energy-momentum tensor is traceless ($T^\mu_\mu = 0$) and the bulk viscous pressure is $\Pi = 0$. The IP--Glasma profile is evolved with classical Yang--Mills dynamics~\cite{Schenke:2012wb, Gale:2012rq} while \trento{} is typically followed by the longitudinal free-streaming of massless partons~\cite{Liu:2015nwa, Bernhard:2016tnd, Bernhard:2018hnz, Everett:2020yty, Everett:2020xug}. Both of these pre-equilibrium models have zero longitudinal pressure $\PL$ prior to the hydrodynamic phase, even though the longitudinal expansion rate $\theta_L \sim 1/\tau$ is decreasing. Recently, effective kinetic theory models have been developed to enable a build-up of longitudinal pressure, making them more compatible with viscous hydrodynamics~\cite{Keegan:2016cpi, Kurkela:2018vqr, Kurkela:2018wud, Berges:2020fwq}.
    \item {\sl Relativistic hydrodynamics:}
    Second-order viscous hydrodynamics is used to evolve the full energy-momentum tensor of the quark-gluon plasma~\cite{Schenke:2010rr,Denicol:2012cn,Gale:2013da,Shen:2014vra,Ryu:2015vwa,Bazow:2016yra} 
    \be
        T^\munu = \ene u^\mu u^\nu - (\Peq{+}\Pi) \Delta^\munu + \pi^\munu\,.
    \ee
    In addition to the conservation laws $\partial_\mu T^\munu = 0$, microscopic evolution equations are needed for the shear stress $\pi^\munu$ and bulk viscous pressure $\Pi$. The Denicol--Niemi--Molnar--Rischke (DNMR) relaxation-type equations\footnote{%
        Hydrodynamic simulations of heavy-ion collisions often employ relaxation-type equations from relativistic kinetic theory, but they likely do not accurately capture the transient dynamics in strongly coupled fluids \cite{Florkowski:2017olj, vanderSchee:2013pia}.} 
    are commonly implemented in practice~\cite{Denicol:2012cn,Ryu:2015vwa} (see Chapters \ref{ch3label},~\ref{chapter4label} and Appendices \ref{appch3e},~\ref{app4e} for more details): 
    \bs 
    \label{eqch1:DNMR}
    \begin{align}
        \tau_\pi \Delta^\munu_\ab \dot{\pi}^\ab + \pi^\munu &= 2\eta \sigma^\munu + \J^\munu \\
        \tau_\Pi \dot{\Pi} + \Pi &= - \zeta\theta + \J \,,
    \end{align}
    \es
    where the dot denotes a comoving time derivative $\dot{a} = u^\mu\partial_\mu a$ and $\Delta^\munu_\ab = \frac{1}{2}(\Delta^\mu_\alpha \Delta^\nu_\beta + \Delta^\nu_\alpha \Delta^\mu_\beta - \frac{2}{3}\Delta^\munu \Delta_\ab)$ is the traceless double spatial projector, $\sigma_\munu = \Delta^\ab_\munu \partial_\alpha u_\beta$ is the velocity-shear tensor and $\theta = \partial_\mu u^\mu$ is the scalar expansion rate. Here ($\tau_{\pi}, \tau_\Pi$) are the shear and bulk relaxation times, $(- 2\eta \sigma^\munu, - \zeta \theta)$ are the first-order gradient forces, and $(\J^\munu, \J)$ are second-order driving terms. The finite relaxation times induce a delay in the system's response to the driving force terms on the r.h.s. of the relaxation equations~(\ref{eqch1:DNMR}a,b). As the system gradually thermalizes over time, the shear stress and bulk viscous pressure approach their respective Navier--Stokes solution $\pi^\munu_\text{NS} = 2\eta\sigma^\munu$ and $\Pi_\text{NS} = - \zeta \theta$ (the second-order contributions diminish in the near-equilibrium limit). \\
    \hspace*{6mm}The energy-momentum tensor components are initialized at $\tau_0$ using the $T^\munu$ output of the pre-equilibrium model. However, the initial bulk viscous pressure is positive, in spite of the system's expansion. This is a consequence of using a conformal pre-equilibrium model, whose equation of state does not match the realistic QCD equation of state used for the quark-gluon plasma phase.
    \item {\sl Particlization:} When converting a dissipative fluid to particles, the Cooper--Frye formula~\eqref{eqch5:CooperFrye} includes viscous corrections to the hadronic distribution $f_n = f_{\eq,n} + \delta f_n$. Two common models for the $\delta f_n$ correction are the 14--moment approximation~\cite{Grad:1963,Monnai:2009ad} and RTA Chapman--Enskog expansion~\cite{chapman1990mathematical,Anderson_Witting_1974,Jaiswal:2014isa}, both of which are linearized in the shear stress and bulk viscous pressure:
    \be
    \delta f_n(x,p) = c_{\pi,n}(x,p) \,p_{\langle\mu} p_{\nu\rangle}\,\pi^{\mu\nu}(x) \,+\, c_{\Pi,n}(x,p)\,\Pi(x)\,.
    \ee
    The linearized $\delta f_n$ models have different momentum dependencies via the expansion coefficients $(c_{\pi,n},c_{\Pi,n})$ but both fully capture the input $T^\munu$ on the hypersurface. This is done by adjusting the expansion coefficients to match the energy-momentum tensor of the hadron resonance gas
    \be
    \label{eqch1:Tmunu_hadron}
    T^\munu(x) = \sum_n \int_p p^\mu p^\nu (f_{\eq,n}(x,p) + \delta f_n(x,p))\,,
    \ee
    where $\int_p \equiv \int{\dfrac{d^3p}{(2\pi\hbar)^3E_p}}$, $E_p = \sqrt{p^2 + m_n^2}$ is the particle energy and the sum goes over all hadron resonance species. The viscous corrections have a considerable impact on the particle spectra: the shear stress decreases the anisotropic flow coefficients $v_k$ while the bulk viscous pressure decreases the mean transverse momentum $\langle p_T \rangle$ and alters the chemical abundance ratios between hadron species.\\
    \hspace*{5mm}In order to include all the effects of the hadronic cascade on the final-state particle spectra, one also needs to sample a discrete number of hadrons from the hypersurface. This is accomplished by using the Cooper--Frye integrand as a probability distribution to Monte Carlo sample the positions and momenta of the hadrons emitted from the fluid cells.
    \item {\sl Hadronic cascade:} The sampled hadrons then decay and rescatter in a Boltzmann transport simulator~\cite{Bass:1998ca,Weil:2016zrk}. The hadronic afterburner assumes that the hadron resonance gas is cold and dilute enough that it can be described in terms of relativistic kinetic theory; the fluid-to-particle transition is usually imposed at temperatures $10-20$ MeV below the pseudocritical temperature $T_c$. As the particle ensemble continues to expand and grow more dilute in the afterburner phase, the hadrons ultimately decouple from each other and free-stream to the particle detector. 
\end{enumerate}
Again we focus our attention on the viscous hydrodynamic stage, which is the stage of the hybrid model where the shear and bulk viscosities directly enter. Since we cannot yet compute $\etas$ and $\zetas$ theoretically, we parameterize their temperature dependence~\cite{Bernhard:2016tnd,Bernhard:2018hnz,Everett:2020xug,Everett:2020yty}. The functional forms are guided by known qualitative features: for example, we know that $(\etas)(T)$ has a minimum near the quark-hadron phase transition and increases with higher temperatures $T$, and $(\zetas)(T)$ peaks at some temperature near the pseudocritical temperature $T_c$ due to critical fluctuations~\cite{Csernai:2006zz,Kharzeev:2007wb,Ghiglieri:2018dib}. Then we fit the hybrid model to experimental data to phenomenologically constrain the viscosity model parameters. It is much more difficult to parameterize the relaxation times and transport coefficients in the second-order terms $(\J^\munu, \J)$ since we have very little first-principles information about their behavior. Instead we approximate them using a different microscopic model, usually kinetic theory~\cite{Denicol:2014vaa}. 

\begin{figure}[t]
\centering
\includegraphics[width=0.75\linewidth]{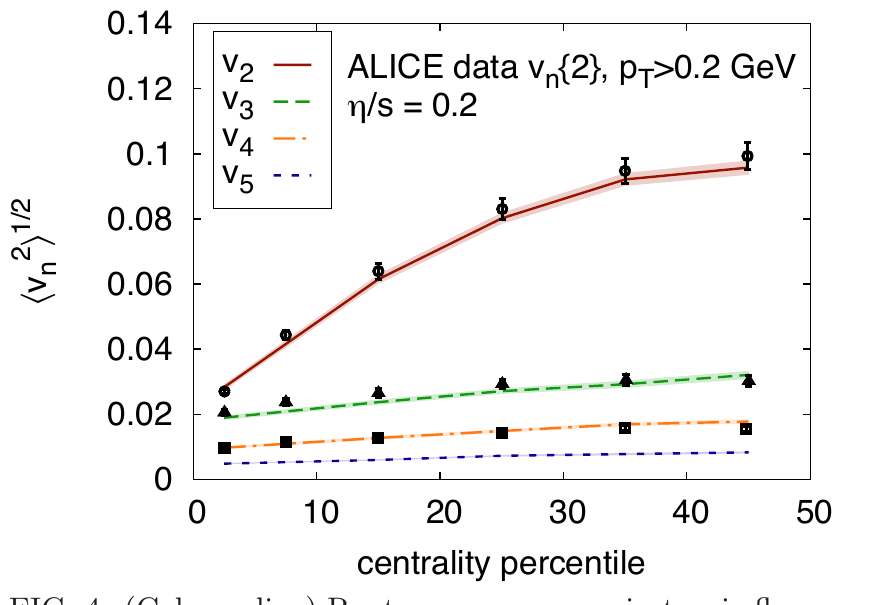}
\caption{The $p_T$--integrated anisotropic flow coefficients of charged hadrons in Pb+Pb collisions $\sqrt{s_\text{NN}} = 2.76$ TeV as a function of centrality. Calculations from a slightly older viscous hydrodynamic model (colored curves) are compared to the experimental measurements by the ALICE collaboration (black)~\cite{ALICE:2011ab}. This figure was taken from Ref.~\cite{Gale:2012rq}.
\label{vn_music}
}
\end{figure}

Second-order viscous hydrodynamics improves upon ideal hydrodynamics by adding viscous effects, but conventionally it only works for fluids that are both near local equilibrium (i.e. inverse Reynolds number $\text{Re}^{-1} \ll 1$) and have small spacetime gradients (i.e. Knudsen number Kn $\ll 1$) \cite{Denicol:2012cn, Rezzolla:2013rehy}. Unlike most fluids, however, the quark-gluon plasma is initially far from local equilibrium ($\text{Re}^{-1} \sim 1$) and sustains moderately large gradients (Kn $\sim 1$) for a good portion of its lifetime. This is primarily due to quantum fluctuations in the initial-state profile \cite{Kharzeev:2001yq, Kharzeev:2004if, Drescher:2006pi, Drescher:2006ca, Drescher:2007ax, Broniowski:2007nz, Schenke:2012wb, Loizides:2014vua, Moreland:2014oya, Welsh:2016siu, Bozek:2019wyr} and the rapid longitudinal expansion at early times. Under these extreme conditions, the validity of a fluid dynamical description for the quark-gluon plasma comes into question~\cite{Niemi:2014wta}. Nevertheless, viscous hydrodynamic simulations, in conjunction with the other multi-stage modules, have been widely successful at reproducing hadronic observables, such as the anisotropic flow coefficients $v_k$ shown in Figure~\ref{vn_music}~\cite{Luzum:2008cw, Holopainen:2010gz, Bozek:2011ua, Gale:2012rq, Shen:2014lye, Pang:2018zzo}. Hydrodynamics has also been found to be phenomenologically successful in collisions between light and heavy ions (e.g. He$^3$+Au, p+Pb) and in proton--proton collisions \cite{Bozek:2011if, Bozek:2012gr, Bozek:2013uha, Shen:2016zpp, Weller:2017tsr}, although the origin of collectivity in these small systems is still being debated \cite{Romatschke:2016hle, Zhao:2020pty, Plumberg:2020jod, Plumberg:2020jux}.

Recent studies suggest that the success of viscous hydrodynamics is attributed to an underlying resummed hydrodynamic theory that also extends to large gradients \cite{Heller:2015dha, Romatschke:2017ejr, Strickland:2017kux}. As the associated non-hydrodynamic modes decay on microscopic time scales $\tau_r {\,\ll\,} \tau_\text{hydro}$, the system approaches a non-equilibrium hydrodynamic attractor that is well approximated by viscous hydrodynamics. This happens within the time scale\footnote{%
    The ``hydrodynamization" time $\tau_\text{hydro}$ refers to the time when viscous hydrodynamics becomes applicable, but the hydrodynamic simulation starting time $\tau_0$ can be different from this.} 
$\tau_\text{hydro} \sim 1$\,fm/$c$, long before the system thermalizes \cite{Romatschke:2017vte, Romatschke:2017acs, Almaalol:2020rnu}. We note that \textit{far-from-equilibrium hydrodynamics} is a brand new field~\cite{Romatschke:2017ejr}, so it is not yet known whether the strongly coupled quark-gluon plasma possesses an attractor at intermediate temperatures $T \sim 0.15 - 0.5$ GeV. We will discuss this topic separately towards the end of this thesis in Chapter~\ref{chap8label}.

\section{Areas of ongoing improvement}
\begin{figure}[t]
\centering
\includegraphics[width=\linewidth]{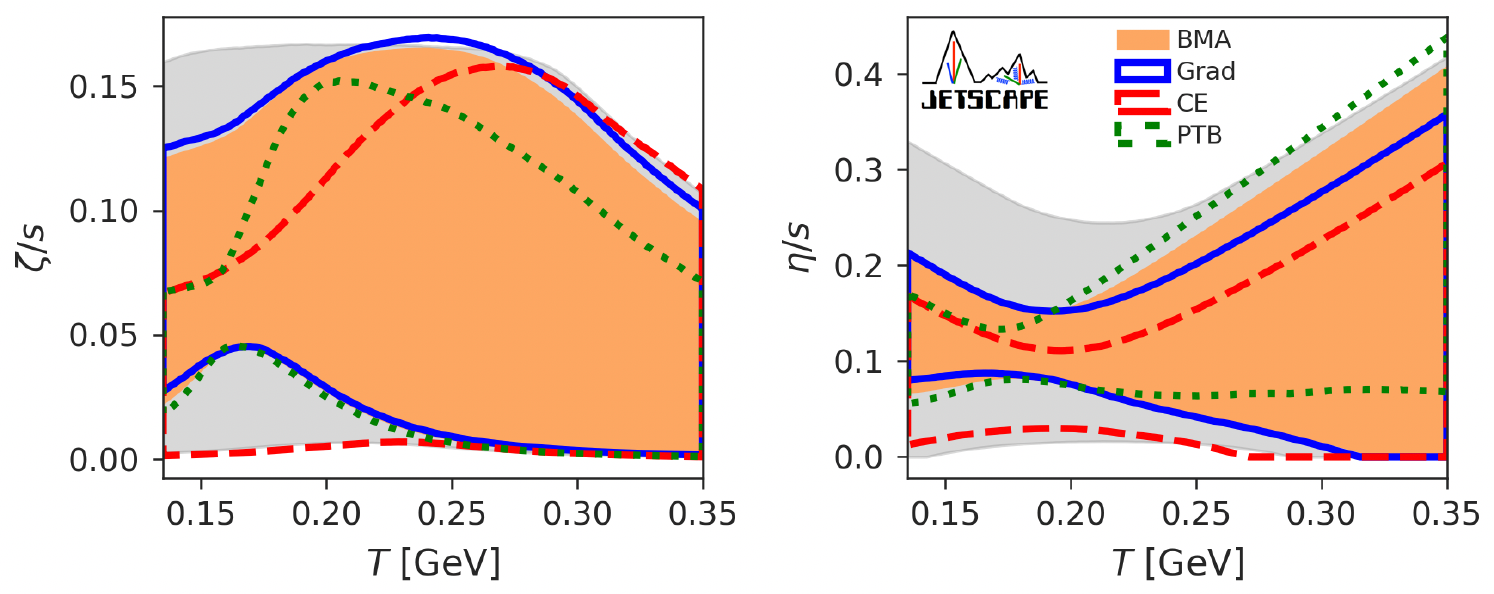}
\caption{The posterior of the shear and bulk viscosities in the 90\% confidence interval (colored) compared to the prior bands (shaded gray). The colored lines correspond to different particlization models for $\delta f_n$ (see Chapters~\ref{chap5label},~\ref{chap6label}) while the shaded orange region takes the Bayesian average of these models. This figure was taken from Ref.~\cite{Everett:2020yty}.
\label{SIMS_analysis}
}
\end{figure}

Within the past five years, hybrid models have obtained enough predictive power to start quantitatively constraining the shear and bulk viscosities of the quark-gluon plasma using heavy-ion experimental data and Bayesian inference \cite{Bernhard:2016tnd, Bernhard:2018hnz, Everett:2020yty, Everett:2020xug, Nijs:2020ors, Nijs:2020roc}. The most recent estimates for $\etas$ and $\zetas$ are shown in Figure~\ref{SIMS_analysis}. One sees that the shear viscosity is very well constrained except at high temperatures. However, precise information about the bulk viscosity remains elusive, even near the quark-hadron phase transition. The modeling of heavy-ion collisions clearly has plenty of room for additional improvements.\footnote{%
    As theoretical modeling continues to improve, it is     also important to keep track of and consider multiple model candidates. By dropping our attachment to one specific framework and comparing different model combinations, we can reduce model selection bias in the phenomenological constraints on $\etas$ and $\zetas$ (e.g. see Fig.~\ref{SIMS_analysis})~\cite{Everett:2020xug,Everett:2020yty}. The code packages developed in this thesis give the heavy-ion physics community new tools to perform such analyses, but it is not the primary objective here.}
Recently, most of the effort has gone into developing new models for the pre-hydrodynamic stage in high-energy collisions~\cite{Keegan:2016cpi, Kurkela:2018vqr, Kurkela:2018wud, Berges:2020fwq} and for dynamical initialization in low-energy collisions~\cite{Shen:2017bsr, Shen:2018pty, Akamatsu:2018olk, Du:2018mpf, Du:2019obx}. In this thesis we focus on improving the simulation of the fluid dynamic and particlization stages in high-energy nuclear collisions. Here we discuss the current issues with the viscous hydrodynamic and particlization models. 

\subsection{Fluid dynamic stage}
Despite the robustness of second-order viscous hydrodynamics, it is still susceptible to breaking down when gradients are very large (Kn $\gg 1$) \cite{Florkowski:2013lya, Martinez:2017ibh, Bazow:2016yra}. The largest gradients in heavy-ion collisions occur at very early times ($\tau{\,<\,}0.2$\,fm/$c$), especially around the edges of the fireball. If one starts the viscous hydrodynamics too early, one usually encounters negative longitudinal pressures $\PL$ due to huge pressure anisotropies~\cite{Bazow:2017ewq}; even the transverse pressure $\Pperp$ can turn negative if $\zetas$ peaks strongly near $T_c$. Not only does this cause excessive viscous heating\footnote{%
    Viscous heating refers to internal entropy production by dissipative processes, not due to external thermal sources.}
but large negative pressures also redirect the matter inward, potentially resulting in cavitation~\cite{Denicol:2015bpa}. These artifacts are a consequence of running the hydrodynamic code outside its region of applicability. Regulation schemes can be used to tamp down the viscous pressures and prevent the simulation from crashing, but they cloud the physical predictions of the original hydrodynamic theory \cite{Shen:2014vra,Bazow:2016yra}.\footnote{%
    Some regulation schemes are less extreme than others. For example, the hydrodynamic code {\sc MUSIC} only regulates the dissipative currents outside the fireball region \cite{Schenke:2010nt} while {\sc iEBE--VISHNU} regulates the entire grid \cite{Shen:2014vra}.}

Because of the technical issues involved, it is more suitable to use a pre-equilibrium model before transitioning to viscous hydrodynamics at $\tau_0 = \tau_\text{hydro}$. Still, the conformal approximation usually made in the former results in a mismatch to the latter's non-conformal equation of state, producing artificially {\it positive} bulk viscous pressures that can be as large as $\Pi \sim \Peq$ around the edges of the fireball \cite{NunesdaSilva:2020bfs}. In some situations, one cannot even switch between dynamical models instantaneously. For example, in low-energy collisions ($\sqrt{s_\text{NN}} \sim 10 - 50$ GeV) where the nuclear interpenetration time is comparable to the fireball lifetime, one needs to run viscous hydrodynamics in the background as soon as the participant nucleons start feeding thermal energy and net baryon number into the newly formed fireball \cite{Shen:2017bsr, Shen:2018pty, Akamatsu:2018olk, Du:2018mpf, Du:2019obx}.
\subsection{Particlization stage}
The linearized 14--moment approximation and RTA Chapman--Enskog expansion both suffer from the problem that, for moderately large values of shear stress and bulk viscous pressure, $f_{\eq,n} + \delta f_n$ can become negative at higher momentum, invalidating the truncation of the expansion at linear order and the interpretation
of $f_n(x,p)$ as a probability density from which particle positions and momenta can be sampled. These incidents occur either near the base of the hypersurface, where the pressure anisotropies are large at early times (although these freeze-out cells are often skipped by the pre-equilibrium module), or on a large fraction of the hypersurface if it happens to be in close proximity to the peak of $\zetas$, where the bulk viscous corrections tend to be substantial (e.g. see Fig.~\ref{FHydro}). The severity of both situations largely depends on the magnitude of the viscosities at particlization.

It is not necessarily a major problem as long as one evaluates the Cooper--Frye formula~\eqref{eqch5:CooperFrye} by numerical integration to obtain the continuous momentum spectra, because one can simply integrate blindly over the negative phase-space regions, hoping that they do not add much weight to the total result. However, when using the Cooper--Frye formula to sample particlization events, the regions of negative probability density $f_n(x,p)$ must be cut out by multiplying a step function inside the integral~\eqref{eqch5:CooperFrye} to enforce $|\delta f_n| \leq f_{\eq,n}$. The resulting modification to the Cooper--Frye formula violates energy-momentum conservation in the particlization process, i.e. the r.h.s. of Eq.~\eqref{eqch1:Tmunu_hadron} no longer reproduces the input energy-momentum tensor on the l.h.s. exactly. These violations manifest themselves in hadronic observables such as the mean transverse energy $\langle E_T \rangle$ and identified particle yields $dN_n/dy_p$. The effects of regulation are usually small in central collisions since most of the hypersurface is either partially or nearly thermalized. For peripheral collisions, whose short-lived fireballs experience significant viscous stresses, a large fraction of the hypersurface may need to be regulated. If the multiplicities at high centralities are significantly altered, then the linearized approaches should no longer be trusted.
\section{Anisotropic hydrodynamics}
The shortcomings of second-order viscous hydrodynamics have motivated the development of hydrodynamic models that can better handle far-from-equilibrium situations. The most promising candidate is anisotropic hydrodynamics \cite{Florkowski:2010cf, Ryblewski:2010bs, Martinez:2010sd, Martinez:2010sc, Martinez:2012tu, Ryblewski:2012rr, Bazow:2013ifa, Strickland:2014pga, Bazow:2015cha, Tinti:2015xwa, Molnar:2016vvu, Molnar:2016gwq, Alqahtani:2017mhy, McNelis:2018jho}, which treats the two largest dissipative effects arising in heavy-ion collisions (the pressure anisotropy $\PL - \Pperp$ and the bulk viscous pressure $\Pi$) non-perturbatively. In relativistic kinetic theory, this involves a dynamically parametrized anisotropic distribution function~\cite{Romatschke:2003ms}:
\be
\label{eqch1:fa}
f_a(x,p) = \exp\left[-\frac{1}{\Lambda(x)}\sqrt{m^2+\dfrac{p_{\perp,\text{LRF}}^2}{\alpha_\perp^2(x)} +\dfrac{p_{z,\text{LRF}}^2}{\alpha_L^2(x)}}\right]\,,
\ee
where $\Lambda(x)$ is the effective temperature and $\alpha_{L,\perp}(x)$ are the momentum anisotropy parameters. The longitudinal and transverse momentum scales are allowed to vary over a wide range to capture large dissipative effects in the longitudinal and transverse pressures $\PL$ and $\Pperp$. One then selects certain moments of the Boltzmann equation to evolve these anisotropy parameters.\footnote{The evolution of the effective temperature $\Lambda$ is controlled by the energy conservation law $u_\nu \partial_\mu T^\munu = 0$ after imposing the Landau matching condition for the energy density $\ene \equiv \int_p (u\cdot p)^2 f_a$.} Regardless of the choice for the kinetic equations, they are able to capture exact kinetic solutions more accurately than standard viscous hydrodynamics \cite{Florkowski:2013lya, Bazow:2013ifa, Bazow:2015cha, Tinti:2015xwa, Molnar:2016gwq, Nopoush:2014qba, Martinez:2017ibh}. 

Although anisotropic hydrodynamics is partially based on weakly coupled kinetic theory,\footnote{%
    Second-order viscous hydrodynamic simulations also rely on microscopic approaches such as relativistic kinetic theory to compute the relaxation times and other second-order transport coefficients \cite{Denicol:2012cn, Denicol:2014vaa, Ryu:2015vwa}.}
its ability to maintain positive longitudinal and transverse pressures makes it highly favourable for (3+1)--dimensional fluid dynamical simulations of heavy-ion collisions~\cite{Bazow:2017ewq}. Since they would be less prone to cavitation, they can run at early times ($\tau < 0.2$ fm/$c$) with little interference from viscous regulations and remain numerically stable. The leading order anisotropic distribution~\eqref{eqch1:fa} is also positive definite in spite of possibily large shear and bulk viscous effects, making it suitable for particlization.\footnote{%
    Additional deformations of the anisotropic distribution~\eqref{eqch1:fa} are needed to capture the smaller off-diagonal components of $T^\munu$ (see Chapter~\ref{chap5label})~\cite{Nopoush:2019vqc}.}
We can repurpose the hydrodynamic equations to some extent by replacing the equation of state with lattice QCD and the viscosities with parametric models. The anisotropic transport coefficients, which depend on the anisotropy parameters, are often computed with a parallel quasiparticle model for the quark-gluon plasma. However, before the work reported in this thesis, previous formulations of anisotropic hydrodynamics explicitly refer to this microsopic kinetic description by evolving the anisotropy parameters with non-hydrodynamic moments of the Boltzmann equation. Furthermore, the evolution of the viscous pressures is indirectly controlled by dynamical equations meant for higher-order moments of the distribution function~\cite{Alqahtani:2015qja}. Instead, we will here use macroscopic formulations~\cite{Tinti:2015xwa,Molnar:2016vvu,Molnar:2016gwq} to directly evolve the energy-momentum tensor components.

The application of anisotropic hydrodynamics to heavy-ion phenomenology is still relatively new; Au+Au collisions at RHIC ($\sqrt{s_\text{NN}} = 200$ GeV) and Pb+Pb collisions at the LHC ($\sqrt{s_\text{NN}} = 2.76$, 5.02 TeV) have been modeled reasonably well but so far only using smooth initial conditions and a few model parameters to adjust the fit to experimental data \cite{Alqahtani:2017jwl, Alqahtani:2017tnq, Almaalol:2018gjh}. Anisotropic hydrodynamics has yet to undergo the same extensive tests and trials as viscous hydrodynamics, but it can potentially serve as an additional discrete model for the fluid dynamical stage (as well as the pre-equilibrium stage) of heavy-ion collisions, complementing the existing arsenal.\footnote{%
    In phenomenological applications (see Chapter~\ref{chapter7label}), we would be comparing \cpuvah{} against a combination of pre-equilibrium dynamics and second-order viscous hydrodynamics. The differences between these models for the pre-hydrodynamic evolution are of particular interest since conformal pre-hydrodynamics (e.g. longitudinal free-streaming) and \cpuvah{} implement diametrically opposed assumptions about the strength of the microscopic interactions during this stage.}
This would further increase the flexibility of the Bayesian framework developed by the JETSCAPE collaboration \cite{Everett:2020yty, Everett:2020xug}, which has already incorporated a number of discrete models for the particlization stage (see Chapter~\ref{chap6label}) \cite{McNelis:2019auj}. The hope is that, by controlling large dissipative flows non-perturbatively, anisotropic hydrodynamics can more accurately constrain the transport coefficients of QCD matter.
\section{Thesis}
\subsection{Overview and goals}
In this thesis, we develop a full event-by-event simulation of heavy-ion collisions that evolves the far-from-equilibrium quark-gluon plasma with anisotropic fluid dynamics. The centerpiece of the new framework is the (3+1)--dimensional anisotropic hydrodynamic code \cpuvah{}. The hydrodynamic equations used in the code build on the formulations of anisotropic hydrodynamics developed over the past decade~\cite{Florkowski:2010cf, Martinez:2010sc, Martinez:2012tu, Bazow:2013ifa, Bazow:2015cha, Tinti:2015xwa, Alqahtani:2015qja, Molnar:2016vvu, McNelis:2018jho}. In particular, we extend the macroscopic approach~\cite{Molnar:2016vvu} to directly evolve the energy-momentum tensor of the quark-gluon plasma rather than the microscopic anisotropy parameters in Eq.~\eqref{eqch1:fa}. A purely macroscopic formulation is accomplished by applying the generalized Landau matching principle first proposed by Tinti~\cite{Tinti:2015xwa}. Specifically, we fix the momentum anisotropy parameters $\alpha_L$ and $\alpha_\perp$ to match the macroscopic longitudinal pressure $\PL$ and transverse pressure $\Pperp$, respectively. We also include in the transport equations the remaining residual shear stress components that are not captured by $\alpha_{L}$ and $\alpha_\perp$, but we treat them only as first-order perturbations. The resulting ($\PL,\Pperp$)--matching scheme~\cite{McNelis:2018jho} greatly simplifies the original relaxation equations from Ref.~\cite{Molnar:2016vvu} and parametrizes the anisotropic transport coefficients in terms of the macroscopic pressures. 

We also upgrade the particlization code {\sc iS3D} (originally called {\sc iSS}~\cite{Shen:2014vra}) to capture the far-off-equilibrium effects on the particle spectra from anisotropic hydrodynamics. The modified anisotropic distribution that we develop is positive definite, and it accurately reproduces all components of the energy-momentum tensor on the hypersurface. Unlike the linearized $\delta f_n$ corrections, it usually does not require viscous regulations; thus, the Monte Carlo sampler in {\sc iS3D} can better replicate the original continuous particle spectra. By integrating the particlization stage with anisotropic hydrodynamics, we can also construct the full hypersurface from the onset of the collision. This is especially useful for sampling particles from peripheral heavy-ion collisions or small collision systems: pre-equilibrium modules often skip a good portion of these hypersurfaces from times before the switching time $\tau_\text{hydro} \sim 1$ fm/$c$, which, in small collision systems, becomes comparable to the average fireball lifetime.

Both the \cpuvah{} and {\sc iS3D} modules are then integrated with the other components of the hybrid model (i.e. we use the initial-state model \trento{} and hadronic afterburner {\sc SMASH} \cite{Moreland:2014oya,Weil:2016zrk}) to simulate fluctuating heavy-ion collision events from start to finish. Our framework is primarily designed to improve the modeling of high-energy Au+Au collisions at RHIC ($\sqrt{s_\text{NN}} = 200$ GeV) and Pb+Pb collisions at the LHC ($\sqrt{s_\text{NN}} = 2.76$, 5.02 TeV), by replacing the pre-equilibrium and fluid dynamic stages with anisotropic hydrodynamics. The main advantage of this approach is that we can evolve the fluid continuously from the onset of the collision to particlization with a QCD equation of state. This ensures a smooth transition to non-conformal viscous hydrodynamics at the ``hydrodynamization" time $\tau_\text{hydro} \sim 1$ fm/$c$.

As a first phenomenological application in Chapter~\ref{chapter7label}, we run event-by-event simulations of Pb+Pb collisions ($\sqrt{s_\text{NN}} = 2.76$ TeV) with (2+1)--d anisotropic hydrodynamics, using the best-fit viscosity model parameters extracted by the JETSCAPE collaboration~\cite{Everett:2020xug,Everett:2020yty} (we also needed to adjust a few initial-state model parameters by hand). Our preliminary analysis demonstrates that our model can accurately reproduce the experimentally measured final-state hadronic observables at mid-rapidity ($y_p = 0$). In the near future, the ultimate goal would be to apply our framework in a full-fledged Bayesian analysis to obtain improved constraints for $\etas$ and $\zetas$ from high-energy Au+Au and Pb+Pb collisions. We also hope that our code can and will be utilized to improve the simulation of small collision systems (e.g. p+Pb), which is even more challenging since far-off-equilibrium effects persist throughout the fireball's short lifetime.
\subsection{Organization of the thesis}
This thesis is organized as follows: in Chapter 2, we test conformal anisotropic hydrodynamics (with $\PL$--matching only) against a known analytic solution for Gubser flow, where the fluid is driven particularly far from local equilibrium. In Chapter 3, we generalize the $\PL$--matching scheme to the non-conformal quark-gluon plasma by using the relativstic Boltzmann equation for medium-dependent masses as a starting point. In Chapter 4, we numerically implement our macroscopic set of anisotropic hydrodynamic equations in the (3+1)--dimensional simulation \cpuvah{} and compare it to standard second-order viscous hydrodynamics. In Chapter 5, we derive a positive definite, non-equilibrium hadronic distribution for Cooper-Frye particlization and compare the resulting particle spectra to those computed with a linearized $\delta f_n$ correction. In Chapter 6, we develop and test the Monte Carlo particle sampler {\sc iS3D} to sample hadrons emitted from the hypersurface and pass them to a hadronic afterburner. In Chapter 7, we combine our fluid dynamic and particlization modules in an event-by-event simulation to predict experimental observables in Pb+Pb collisions at the LHC ($\sqrt{s_\text{NN}} = 2.76$ TeV). In Chapter 8, we discuss separately new developments on resummed hydrodynamics and use a more systematic approach to construct models for far-from-equilibrium kinetic fluids. Finally, our conclusions and future outlook are summarized in Chapter 9.

\chapter{Anisotropic fluid dynamics in conformal Gubser flow}
\label{chap2label}

Gubser flow describes conformal systems that expand both along the longitudinal beam direction and azimuthally symmetrically in the transverse plane~\cite{Gubser:2010ze,Gubser:2010ui}. At early times, it has a very large longitudinal expansion rate, similar to that of Bjorken flow~\cite{Bjorken:1982qr}. But unlike Bjorken flow, it has a nonzero transverse velocity that grows rapidly over time, eventually overtaking the longitudinal expansion. This unique flow pattern ensures that the fluid almost always stays far from local equilibrium. Therefore, Gubser flow is an ideal testing ground for measuring the validity of hydrodynamic models subject to large gradients~\cite{Nopoush:2014qba}. In this chapter, we test the performance of a particular anisotropic hydrodynamic model called $\PL$--matching~\cite{Tinti:2015xwa,Molnar:2016gwq}, which is the foundation of the framework developed in this thesis. We find that it produces the most accurate macroscopic description of a conformal relativistic gas evolved by the Boltzmann equation in the relaxation time approximation (RTA) compared to standard second-order viscous hydrodynamics and other anisotropic hydrodynamic models~\cite{Marrochio:2013wla,Nopoush:2014qba,Denicol:2014tha}.

This chapter is based on material previously published in Ref.~\cite{Martinez:2017ibh}.
\section{The Gubser flow}
\label{sec:Gflow}
In the Milne coordinates $x^\mu = (\tau, x, y, \eta_s)$, the Gubser flow profile $u^\mu$ takes the following analytic form: 
\be
\label{eq:ch2gubser_velocity}
    u^\tau = \cosh\kappa \,,\quad
    u^x = \sinh\kappa \cos\phi \,,\quad 
    u^y = \sinh\kappa \sin\phi \,,\quad 
    u^\eta = 0\,,
\ee
where
\be
    \kappa(\tau, r) = {\tanh^{-1}}\bigg[\frac{2 q^2 \tau r}{1 + q^2(\tau^2 {+} r^2)}\bigg] \,,
\ee
$r{\,=\,}\sqrt{x^2{+}y^2}$ is the transverse radius, $\phi = {\tan^{-1}}(y/x)$ the azimuthal angle and $q$ is an energy scale that sets the transverse size of the system $R_\perp \sim 1/q$ at $\tau = 0^+$~\cite{Gubser:2010ze,Gubser:2010ui}.  This flow profile has the unique property of appearing static along the de Sitter line element $dS_3\otimes R$. To go from Milne coordinates to $dS_3\otimes R$ one first performs a Weyl rescaling of the metric,
\begin{equation}
d\hat{s}^2=\frac{ds^2}{\tau^2}
=\frac{-d\tau^2+dr^2+r^2 d\phi^2}{\tau^2}+d\eta^2\,,  
\label{metricdS3R}
\end{equation}
followed by the coordinate transformation $x^\mu=(\tau,r,\phi,\eta)\mapsto \xh^\mu=(\rho,\theta,\phi,\eta)$ where
\begin{subequations}
\allowdisplaybreaks
\label{eq:ch2rhotheta}
\begin{align}
\rho(\tilde\tau,\tilde r)& =-\mathrm{arcsinh}\left( \frac{1-\tilde\tau^2+\tilde r^2}
{2\tilde\tau }\right)\,,
\label{definerho} \\
\theta (\tilde\tau,\tilde r)& =\mathrm{arctan}\left(\frac{2\tilde r}{1+\tilde\tau^2-\tilde r^2}\right) \,,
\label{definetheta}
\end{align}
\end{subequations}
with $\tilde\tau = q\tau$ and $\tilde r = q r$. In these coordinates the Weyl transformed line element reads
\begin{equation}
d\hat{s}^{2}=-d\rho ^{2}+\cosh ^{2}\!\rho \left( d\theta ^{2}+\sin
^{2}\theta\, d\phi ^{2}\right) +d\eta ^{2}\,, 
\label{eq:ch2linedS3R}
\end{equation}
with the metric\footnote{In this chapter, we use the ``mostly plus'' convention $(-,+,+,+)$ for the metric tensor. For the remaining chapters, we use the ``mostly minus" convention $(+,-,-,-)$.} $\hat{g}_\munu=\mathrm{diag}\bigl(-1,\cosh^2\!\rho, \cosh^2\!\rho\sin^2\theta,1\bigr)$ and the square root of the metric determinant $\sqrt{-\hat{g}}=\cosh^2\!\rho\cos\theta$. The de Sitter coordinate $\rho\in (-\infty,\infty)$ acts as the time while $ \theta\in (0,2\pi)$ plays the role of an angle. The line element (\ref{eq:ch2linedS3R}) is invariant under rotations in the space spanned by $(\theta,\phi)$; the corresponding symmetry group is denoted as $SO(3)_q$~\cite{Gubser:2010ze}. Including the reflection symmetry $\eta\to-\eta$ and longitudinal boost invariance, the line element (\ref{eq:ch2linedS3R}) is invariant under the Gubser symmetry $\gub$ \cite{Gubser:2010ze}. The only normalized vector that is invariant under this symmetry is the static fluid velocity $\hat{u}^\mu=(1,0,0,0)$~\cite{Gubser:2010ze,Gubser:2010ui}. 

This symmetry also implies that macroscopic variables such as the energy density $\enehat(\xh)=\enehat(\rho)$ depend only on the de Sitter time \cite{Gubser:2010ze, Gubser:2010ui} while the on-shell phase-space distribution $f(\xh,\pp)=f(\rho,\po,\pp_\eta)$ depends only on $\rho$ \cite{Gubser:2010ze,Gubser:2010ui} and the momentum components $\po=\pp_\theta^2+\pp_\phi^2/\sin^2\theta$ and $\pp_\eta$ conjugate to the spatial coordinates $\theta, \phi$ and $\eta$ \cite{Denicol:2014xca,Denicol:2014tha}.\footnote{
    The temporal component of the de Sitter momentum ${\pp}^\mu = ({\pp}^\rho, {\pp}^\theta,{\pp}^\phi,{\pp}^\eta)$ is fixed by the on-shell condition:
    \be
    {\pp}^\rho = \sqrt{\frac{{\pp}_\Omega^2}{{\cosh^2}\rho} + {\pp}^2_\eta}
    \ee} We denote by variables with a hat all quantities that are expressed in Gubser coordinates $\xh^\mu$. 
\section{Fluid dynamics for Gubser flow}
\label{sec:fluid}

In this section we derive various fluid dynamical models (including the $\PL$--matching scheme) subject to Gubser flow from the relativistic Boltzmann equation for a system of massless particles ($\pp^\mu\pp_\mu=0$), using the relaxation time approximation (RTA) for the collision term. In Gubser coordinates, this RTA Boltzmann equation reads \cite{Denicol:2014xca,Denicol:2014tha}
\be
\label{eq:ch2boltzgubeq}
\partial_\rho f(\rho,\po,\pp_\eta)=\frac{\feq(\rho,\po,\pp_\eta) - f(\rho,\po,\pp_\eta)}{\hat{\tau}_r(\rho)}\,,
\ee
where $\hat{\tau}_r$ is the relaxation time,
\be
\feq(\rho,\po,\pp_\eta) = \exp\left[-\frac{(-\hat{u}\cdot\pp)}{\temhat(\rho)}\right]
\ee
is the local thermal equilibrium distribution,
\be
-\hat{u}\cdot\pp =\sqrt{\frac{\hat{p}^2_\Omega}{\cosh^2\!\rho} + \hat{p}^2_\eta}
\ee
is the particles' energy in the local rest frame (LRF) of the fluid, and $\temhat$ is the temperature. Conformal symmetry requires the relaxation time to take the form
\be
\hat{\tau}_r(\rho)= \frac{5\bar\eta}{\temhat(\rho)}\,,
\ee
where $\bar\eta\equiv\hat{\eta} / \hat{\mathcal{S}}$ is the specific shear viscosity. The exact solution of Eq.~(\ref{eq:ch2boltzgubeq}) \cite{Denicol:2014xca,Denicol:2014tha} will be given in Sec.~\ref{subsec:gubsol} below when we need it.

The macroscopic hydrodynamic variables that make up the energy-momentum tensor are obtained as momentum moments of the distribution function,
\be
\label{eq:ch2EMkin}
\temhat^{\mu\nu}(\hat{x})=\langle\,\pp^\mu\,\pp^\nu\,\rangle\,,
\ee
where 
\be
\langle\mathcal{O}(\hat{x})\rangle\equiv\int_{\pp}\,\mathcal{O}(\hat{x},\hat{p}) f(\hat{x},\hat{p}) = \int \frac{d\pp_\theta d\pp_\phi d\pp_\eta}{(2\pi)^3 \sqrt{-\hat{g}}\hat{E}_{\hat{p}}} \,\mathcal{O}(\hat{x},\hat{p}) f(\hat{x},\hat{p}),
\ee
denotes the momentum moment of the phase-space observable $\mathcal{O}(\hat{x},\hat{p})$.\footnote{%
	Note that we define the integration measure in terms of the covariant components $\pp_\mu$ \cite{Denicol:2014xca,Denicol:2014tha,Bazow:2016oky}.
	}
We evaluate these momentum integrals in the LRF where the on-shell energy $\hat{E}_{\hat{p}}=\pp^\rho=-\hat{u}\cdot\pp$. Momentum moments of the equilibrium distribution are denoted by $\langle\mathcal{O}(\hat{x},\hat{p})\rangle_{\mathrm{eq}}$.

Using the relation (\ref{eq:ch2EMkin}), any (approximate) dynamical solution for the distribution function can be used to derive a set of hydrodynamic and relaxation equations for the components of the energy-momentum tensor. Taking derivatives of the left hand side of Eq.~(\ref{eq:ch2EMkin}) leads to terms on the right hand side that involve momentum moments of derivatives of the distribution function. These must be evaluated using the Boltzmann equation (\ref{eq:ch2boltzgubeq}) and, in general, couple the hydrodynamic moments to higher-order, non-hydrodynamic moments of the distribution function. To close the set of equations one must truncate the momentum hierarchy using some approximation scheme. We now discuss the sets of hydrodynamic equations resulting from several such closing schemes for systems undergoing Gubser flow. 

\subsection{Viscous hydrodynamics}
\label{subsec:vischydro}

For the derivation of standard viscous hydrodynamics from kinetic theory one expands the distribution function around a local-equilibrium distribution:
\be
\label{eq:ch2isotropic}
f(\rho,\po,\pp_\eta)=f_\mathrm{eq}(\rho,\po,\pp_\eta) + \delta f (\rho,\po,\pp_\eta)\,,
\ee
where $\delta f$ encodes all deviations from local thermal equilibrium, in particular any local momentum anisotropies caused by global anisotropic expansion. 

In this situation it is convenient to decompose the metric tensor $\hat{g}^{\mu\nu}$ into the locally temporal and spatial projectors, $-\hat{u}^\mu \hat{u}^\nu$ and $\hat{\Delta}^{\mu\nu}=\hat{g}^{\mu\nu}+\hat{u}^\mu \hat{u}^\nu$, respectively, and use these to decompose the particle four-momentum as $\pp^\mu= (-\hat{u}\cdot\pp)\,\hat{u}^\mu
+\pp^{\langle\mu\rangle}$. Here the spatial projection $\hat{A}^{\langle\mu\rangle}=\hat{\Delta}^{\mu\nu}\hat{A}_\nu$ is a vector that is purely spatial in the LRF. For later use we also introduce $\hat{B}^{\langle\mu\nu\rangle}=\Duhat_{\alpha\beta}\hat{B}^{\alpha\beta}$ where the double projector $\Duhat_{\alpha\beta}=\half(\hat{\Delta}^{\mu}_{\alpha}\hat{\Delta}^{\nu}_{\beta}+\hat{\Delta}^{\mu}_{\beta}\hat{\Delta}^{\nu}_{\alpha}-\frac{2}{3}\Duhat\hat{\Delta}_{\alpha\beta})$ projects the tensor $\hat{B}^{\mu\nu}$ onto its symmetric, traceless and locally purely spatial (i.e. orthogonal to $\hat{u}^\mu$) part. 

In the Landau frame the most general form of the energy-momentum tensor is then
\be
\label{eq:ch2enemomvh}
\temhat^{\mu\nu}=\enehat\,\hat{u}^\mu\hat{u}^\mu+\hat{\mathcal{P}}\hat{\Delta}^{\mu\nu}+\hat{\pi}^{\mu\nu}\,,
\ee
where $\enehat$ is the LRF energy density, $\hat{\mathcal{P}}=\hat{\mathcal{P}}_\eq(\enehat)+\hat{\Pi}$ is the isotropic pressure, 
and $\shehat^{\mu\nu}$ is the shear stress tensor. For conformal systems such as the one studied here, the bulk viscous pressure $\hat{\Pi}$ vanishes, and the isotropic pressure $\hat{\mathcal{P}}$ is given by the thermal pressure $\hat{\mathcal{P}}_\eq(\enehat)=\enehat/3$ obtained from the conformal equation of state. These macroscopic quantities correspond to the following moments of the distribution function~\eqref{eq:ch2isotropic}: 
\begin{subequations}
\allowdisplaybreaks
\begin{align}
&\enehat=\hat{u}_\mu\hat{u}_\nu\temhat^{\mu\nu}=\langle (\hat{u}\cdot\pp)^2\rangle\,,\\
&\hat{\mathcal{P}}=\frac{1}{3}\hat{\Delta}_{\mu\nu}\temhat^{\mu\nu}=\frac{1}{3}\langle\hat{\Delta}_{\mu\nu}\pp^\mu\pp^\nu\rangle\,,\\
\label{eq:ch2shearvisc}
&\shehat^{\mu\nu}=\temhat^{\langle\mu\nu\rangle}=\langle\pp^{\langle\mu}\pp^{\nu\rangle}\rangle\,.
\end{align}
\end{subequations}
In addition to the choice of the LRF velocity as the timelike eigenvector of $\temhat^{\mu\nu}$, $\temhat^{\mu\nu} \hat{u}_\nu=\enehat\,\hat{u}^\mu$, uniqueness of the decomposition (\ref{eq:ch2isotropic}) requires fixing the temperature $\temhat$. This is done through the Landau matching condition \cite{Landaufluid} 
\be
\label{eq:ch2landaumatch}
\enehat := \langle (\hat{u}\cdot\pp)^2\rangle_\mathrm{eq} = \enehat_\mathrm{eq}(\temhat)
         = \frac{3\temhat^4}{\pi^2},
\ee 
which ensures that the parameter $\temhat$ in $f_\mathrm{eq}$ is adjusted such that $\delta f$ does not contribute to the LRF energy density. As a result, all deviations of the system from local equilibrium are encoded in the shear stress tensor: $\shehat^{\mu\nu}\equiv\langle\pp^{\langle\mu}\pp^{\nu\rangle}\rangle_{\delta}$ where $\langle\cdots\rangle_{\delta}$ indicates a momentum moment of $\delta f$.

The evolution equation for the energy density $\enehat$ is obtained from the time-like (i.e.\ $\hat{u}_\nu$) projection of the energy-momentum conservation law $\hat{D}_\mu \temhat^{\mu\nu}=0$ (where $\hat{D}_\mu$ is the covariant derivative). For systems with Gubser symmetry this yields \cite{Gubser:2010ze,Gubser:2010ui}
\be
\label{eq:ch2eqviscener}
\partial _{\rho} \enehat
+ \frac{8}{3}\,\enehat\tanh\rho
= \hat\pi^{\eta\eta}\tanh\rho\,.
\ee
Such systems have only one independent shear stress component $\hat\pi\equiv\shehat^{\eta\eta}$. To obtain an evolution equation for $\hat{\pi}$ one can start from the RTA Boltzmann equation and use the method of moments described in Ref.~\cite{Denicol:2012cn} (DNMR).\footnote{%
		For a complete discussion of different methods to derive hydrodynamics from relativistic kinetic theory 
		see Ref.~\cite{Denicol:2014loa}.}
Within the 14-moment approximation one obtains (see Refs.~\cite{Denicol:2014tha, Nopoush:2014qba})
\be
\label{eq:ch2eqviscshear}
  \partial _{\rho }\hat\pi+\frac{\hat\pi}{\hat{\tau}_r} + \frac{46}{21}\hat{\pi}\tanh\rho
  = \frac{16}{45}\enehat\tanh\rho\,.
\ee
We introduce the normalized shear stress $\hat{\bar\pi}\equiv3\hat\pi/(4\enehat)$ to rewrite the DNMR equations \eqref{eq:ch2eqviscener} --~\eqref{eq:ch2eqviscshear} as
\begin{subequations}
\allowdisplaybreaks
\label{eq:ch2vhydro}
\begin{align}
\label{eq:ch2vhydro1}
&\partial_\rho\ln\enehat = \frac{4}{3}(\hat{\bar\pi}-2)\tanh\rho\,,\\
\label{eq:ch2vhydro2}
&\partial_\rho\hat{\bar\pi} + \frac{\hat{\bar\pi}}{\hat{\tau}_r} =
   \frac{4}{3}\tanh\rho\left( \frac{1}{5} + \frac{5}{14}\hat{\bar\pi} - \hat{\bar\pi}^2 \right).
\end{align}
\end{subequations}
%

\subsection{Anisotropic hydrodynamics}
\label{subsec:ahydro}
Anisotropic hydrodynamics generalizes viscous hydrodynamics by allowing for a leading-order dissipative deformation of the distribution function due to anisotropic expansion of the system. If the expansion rate along a certain \textit{longitudinal} direction $\hat{z}^\mu$ is much larger or smaller than in the other directions, one can account for this by generalizing the decomposition (\ref{eq:ch2isotropic}) as \cite{Bazow:2013ifa,Molnar:2016vvu}
\be
\label{eq:ch2anisexp}
f(\rho,\po,\pp_\eta)=f_a(\rho,\po,\pp_\eta)+\tdf(\rho,\po,\pp_\eta)\,.
\ee
For the leading-order anisotropic distribution function we use the Romatschke-Strickland (RS) ansatz \cite{Romatschke:2003ms} 
\be
\allowdisplaybreaks
\label{eq:ch2RSansatz}
f_a(\rho,\po,\pp_\eta)= \exp\left[-\frac{E_\mathrm{RS}(\rho,\po,\pp_\eta;\xi(\rho))}{\hat{\Lambda}(\rho)}\right]\,,
\ee
which is a Boltzmann distribution but with an effective temperature $\hat\Lambda$ and momentum-anisotropic argument
\be
\label{eq:ch2ERS}
E_\mathrm{RS}(\rho,\po,\pp_\eta; \xi(\rho))
  \equiv \sqrt{(\hat{u}\cdot\pp)^2+\xi(\rho)(\hat{z}\cdot\pp)^2} 
  = \sqrt{\frac{\po}{\cosh^2\!\rho}+(1{+}\xi(\rho))\pp_\eta^2}.
\ee
The parameter $\xi$ encodes the strength of the leading-order local momentum anisotropy, and $\delta \tilde{f}$ takes into account residual dissipative corrections. Anisotropic hydrodynamics is expected to be an improvement over viscous hydrodynamics whenever the residual dissipative effects associated with $\delta \tilde{f}$ are smaller than the leading-order dissipative effects manifest in the local momentum anisotropy $\xi$. We demand that in the limit $\xi\to 0$, the anisotropic distribution function $f_a$ reduces to the local equilibrium distribution $f_\mathrm{eq}$ in Eq.~\eqref{eq:ch2isotropic}.

To account for the effects from the momentum anisotropy of the leading-order distribution function $f_a$ on the structure of the macroscopic energy-momentum tensor it is convenient to perform the tensor decomposition in terms of both the fluid velocity $\umhat$ and space-like longitudinal vector $\hat{z}^\mu$ which in the LRF is chosen to point in the $\eta_s$--direction: $\luhat=(0,0,0,1)$. The space orthogonal to these two vectors is spanned by the transverse spatial projector tensor $\xuhat=\Duhat-\luhat\hat{z}^\nu$ \cite{Huang:2009ue,Huang:2011dc,Gedalin1,Gedalin2,Molnar:2016vvu,Molnar:2016gwq}. The four-momentum can now be decomposed as $\pp^\mu=(-\hat{u}\cdot\pp)\unhat + (\hat{z}\cdot\pp) \luhat+\pp^{\{\mu\}}$, where $\pp^{\{\mu\}}\equiv\xuhat\pp_\nu$ are the transverse spatial momentum components and $\hat{z}\cdot\pp=\pp_\eta$ is the longitudinal momentum in the LRF.

This leads to the following decomposition of the energy-momentum tensor \eqref{eq:ch2EMkin} in the Landau frame \cite{Molnar:2016vvu}:
\be
\label{eq:ch2MNRemt}
\temhat^\munu=\enehat\,\umhat\hat{u}^\nu + \Plhat\luhat\hat{z}^{\nu} + \Pthat\xuhat + 2\hat{W}^{(\mu}_{\perp\hat{z}}\hat{z}^{\nu)} + \hat{\pi}_\perp^\munu,
\ee
with 
\bs
\allowdisplaybreaks
\label{eq:ch2anisodecomp}
\beal
\enehat&=\hat{u}_\mu \hat{u}_\nu \temhat^\munu\,\equiv \langle\,(-\,\hat{u}\cdot\pp)^2\,\rangle\,,\\
\Plhat&=\hat{z}_\mu\hat{z}_\nu\temhat^\munu\,\equiv\langle\,(\hat{z}\cdot\pp)^2\,\rangle\,,\\
\Pthat&=\frac{1}{2}\hat{\Xi}_{\mu\nu}\temhat^{\mu\nu}\,\equiv\frac{1}{2}\langle\,\hat{\Xi}_{\mu\nu}\pp^\mu\pp^\nu\,\rangle\,,\\
\hat{W}^{\mu}_{\perp\hat{z}}&=\hat\Xi^{\mu}_{\alpha}\temhat^{\alpha\beta}\hat{z}_\beta\,\equiv\langle\,(\hat{z}\cdot\pp)\,\pp^{\{\mu\}}\rangle\,, \\
\hat{\pi}_\perp^\munu&= \hat\Xi_{\alpha\beta}^{\mu\nu}\temhat^{\alpha\beta}\equiv\langle\,\pp^{\{\mu}\pp^{\nu\}}\,\rangle.
\end{align}
\es
In the last line we introduced the notation $\hat{B}^{\{ \mu\nu\}}\equiv\xuhat_{\alpha\beta}\hat{B}^{\alpha\beta}$ where the double projector $\xuhat_{\alpha\beta}\equiv\frac{1}{2}\big(\hat{\Xi}^\mu_\alpha\hat{\Xi}^\nu_\beta+\hat{\Xi}^\nu_\beta\hat{\Xi}^\mu_\alpha-\xuhat\hat{\Xi}_{\alpha\beta}\big)$ projects an arbitrary tensor $\hat{B}^{\mu\nu}$ on its symmetric, traceless and locally spatially transverse part. Comparison of Eqs.~\eqref{eq:ch2shearvisc} and \eqref{eq:ch2MNRemt} shows that the shear stress $\shehat^\munu$ can be further decomposed as \cite{Molnar:2016vvu,Molnar:2016gwq}
\be
\label{eq:ch2shearrelation}
\shehat^{\mu\nu}=\shehat^{\mu\nu}_{\perp}+2\hat{W}^{(\mu}_{\perp\hat{z}}\hat{z}^{\nu)}
+\frac{1}{3} \big(\Pthat{-}\Plhat\big) \big(\xuhat{-}2\,\luhat\hat{z}^\nu\big).
\ee  

The above decomposition~\eqref{eq:ch2MNRemt} is general. For systems with Gubser symmetry it simplifies considerably. Conformal symmetry requires $\enehat=2\Pthat+\Plhat$, corresponding to zero bulk viscous pressure $\Pi$ and a conformal equation of state $\enehat=3\hat{\mathcal{P}}_\eq(\enehat)=2\Pthat+\Plhat$. The $SO(3)_q\otimes Z(2)$ part of the Gubser symmetry implies
$\shehat^{\mu\nu}_{\perp}=0=\hat{W}^{\mu}_{\perp\hat{z}}$. This leaves 
\be
\label{eq:ch2anistmn}
\temhat^\munu=\enehat\,\umhat\hat{u}^\nu+\Plhat\luhat\hat{z}^{\nu}+\Pthat\xuhat
\ee
as the most general energy-momentum tensor for systems with Gubser symmetry. The single nonvanishing shear stress component $\hat\pi\equiv\shehat^{\eta\eta}$ defines the difference between the longitudinal and transverse pressures via  
\be
\label{eq:ch2pihat}
  \Plhat-\Pthat=\frac{3}{2}\shehat
\ee
and the shear stress tensor via
\be
\label{eq:ch2shearreduced}
\begin{split}
\hat{\pi}^{\mu\nu}&=\shehat\bigg(\luhat\hat{z}^\nu - \frac{1}{2}\xuhat\bigg)\,.
\end{split}
\ee  
Following \cite{Molnar:2016gwq} we define the scalar integrals
\be
\label{eq:ch2Inlq}
I_{nlq}\equiv\bigl\langle(-\hat{u}\cdot\pp)^{n-l-2q} (\hat{z}\cdot\pp)^l (\hat{\Xi}_{\mu\nu}\pp^\mu\pp^\nu)^q\bigr\rangle= \Ih_{nlq}(\hat{\Lambda},\xi)+\It_{nlq} \,,
\ee
%
%
%
where the first term on the r.h.s. denotes the leading order contribution from $f_a$ (which depends on the parameters $\hat\Lambda$ and $\xi$) and the second term the subleading contribution from $\delta\tilde{f}$ in Eq.~\eqref{eq:ch2anisexp}. For massless particles the dependences of the leading order $\Ih$ integrals on $\hat\Lambda$ and $\xi$ factorize (see Appendix \ref{appch2:anisint}, Eq.~\eqref{eq:ch2anisint2}). 

With these definitions the leading order RS distribution function \eqref{eq:ch2RSansatz} contributes to the energy-momentum tensor as follows: 
\be
\label{eq:ch2RStmn}
\temhat^\munu_\mathrm{RS}=\enehat_\mathrm{RS}\,\umhat\,\unhat\,+\,\Plhat^\mathrm{RS}\,\luhat\hat{z}^{\nu}\,+\,\Pthat^\mathrm{RS}\,\xuhat,
\ee
where
\bs
\allowdisplaybreaks
\label{eq:ch2RSmacquant}
\begin{align}
\label{eq:ch2RSenergydens}
\enehat_\mathrm{RS}&=\left\langle(-\,\hat{u}\cdot\pp)^2\right\rangle_a=\Ih_{200}\big(\hat{\Lambda},\xi\big)\,,\\
\Plhat^\mathrm{RS}&=\bigl\langle(\hat{z}\cdot\pp)^2\bigr\rangle_a=\Ih_{220}\big(\hat{\Lambda},\xi\big)\,,\\
\Pthat^\mathrm{RS}&=\frac{1}{2}\bigl\langle\hat{\Xi}_{\mu\nu}\pp^\mu\pp^\nu\bigr\rangle_a
   =\frac{1}{2}\Ih_{201}\big(\hat{\Lambda},\xi\big).
\end{align}
\es
Using Eq.~\eqref{eq:ch2pihat} and the mass-shell condition in the form $\hat{\Xi}_{\mu\nu}\pp^\mu\pp^\nu=(-\,\hat{u}\cdot\pp)^2-(\hat{z}\cdot\pp)^2$ these relations further imply
\be
\label{eq:ch2RSpi}
  \shehat_\mathrm{RS}=\bigl\langle(\hat{z}\cdot\pp)^2-\textstyle{\frac{1}{3}}(-\hat{u}\cdot\pp)^2\big\rangle_a
         =\Ih_{220}\big(\hat{\Lambda},\xi\big)-\textstyle{\frac{1}{3}}\Ih_{200}\big(\hat{\Lambda},\xi\big).
\ee
Conformal symmetry implies that the trace of the energy-momentum tensor to vanish exactly, hence
\be
\label{eq:ch2anisotrace}
\enehat_\mathrm{RS}= 2\Pthat^\mathrm{RS}+\Plhat^\mathrm{RS} \,.
\ee

As in the viscous hydrodynamic case, to make the decomposition \eqref{eq:ch2anisexp} of the distribution function unique we need a prescription for the kinetic evolution of the parameters $\hat{\Lambda}$ and $\xi$. We use the Landau matching condition \cite{Martinez:2010sc,Martinez:2010sd}
\be
\label{eq:ch2LMcond}
\enehat_\mathrm{RS}(\hat\Lambda,\xi) := \enehat_\mathrm{eq}(\temhat)
\ee
to relate the evolution between $\hat\Lambda$ and $\temhat$:
\be
\hat{\Lambda} := \frac{\temhat}{\big(\Rhat_{200}(\xi)\big)^{1/4}}\,,
\ee
where $\Rhat_{200}(\xi)$ is the $\xi$-dependent part of $\hat{I}_{200}(\hat\Lambda,\xi)$ (see Eq.~\eqref{eq:ch2anisint2}). This condition also ensures that the energy density receives no contribution from the residual deviation $\delta\tilde{f}$ in Eq.~\eqref{eq:ch2anisexp} (i.e. $\enehat = \enehat_\mathrm{RS}$). 

Various prescriptions for the evolution of the second parameter $\xi$ have been proposed \cite{Martinez:2010sc, Martinez:2010sd, Florkowski:2010cf, Ryblewski:2010bs, Florkowski:2011jg, Ryblewski:2011aq, Ryblewski:2012rr, Florkowski:2012as, Florkowski:2013uqa, Bazow:2013ifa, Tinti:2013vba, Florkowski:2014bba,Florkowski:2014txa,Nopoush:2014pfa,Denicol:2014mca,Bazow:2015cha,Nopoush:2015yga,Florkowski:2015cba,Alqahtani:2015qja,Bazow:2015zca,Tinti:2015xra,Alqahtani:2016rth,Florkowski:2016kjj,Nopoush:2016qas,Florkowski:2016zsi}. They correspond to different relaxation equations, derived from the Boltzmann equation, that evolve the leading order distribution $f_a$ (and the residual correction $\delta\tilde{f}$ if included). Since the energy-momentum tensor components depend on the decomposition~\eqref{eq:ch2anisexp}, different anisotropic hydrodynamic models can be distinguished by the way their anisotropy parameter $\xi(\rho)$ evolves. In the following subsections we discuss three such anisotropic hydrodynamic models. 
\subsection{$\mathcal{P}_L$ matching}
\label{subsec:plmatching}
We start with the simplest and, as it turns out, most effective prescription for $\xi$, the $\Plhat$--matching scheme first proposed in \cite{Bazow:2013ifa} and recently successfully implemented for systems undergoing Bjorken flow 
by Molnar {\it et\ al.}\ \cite{Molnar:2016vvu,Molnar:2016gwq}.\footnote{%
	We note that the $\Plhat$--matching condition \eqref{eq:ch2PLmatch} below is a special case of a more general 
	prescription proposed earlier by Tinti \cite{Tinti:2015xwa} for (3+1)-dimensional anisotropic hydrodynamics 
	which uses a generalized Romatschke-Strickland form for the leading order distribution function $f_a$
	that is flexible enough to capture {\em all} components of the energy-momentum tensor (by appropriately 
	matching its parameters), i.e. the residual deviation $\delta\tilde{f}$ of the distribution function contributes
	{\em nothing} to $\temhat^{\mu\nu}$. By adapting the derivations in \cite{Tinti:2015xwa} to Bjorken- and 
	Gubser-symmetric situations we checked that they lead to the same results as those reported in 
	Ref.~\cite{Molnar:2016gwq} and in this subsection, respectively, for the $\Plhat$--matching scheme.}
This prescription considers $\xi$ as a parameter that represents on the microscopic level the macroscopic longitudinal pressure $\Plhat$ (or, more precisely, the shear stress component $\shehat$ that is responsible for the longitudinal-transverse pressure difference $\Plhat{-}\Pthat$), in very much the same way as the temperature $\temhat$ represents the energy density $\enehat$. It should therefore be fixed by a corresponding ``Landau matching condition'' that adjusts the value of $\xi$ in $f_a$ such that $f_a$ fully captures all contributions to $\Plhat$ (or, equivalently, to $\shehat$), i.e. $\Plhat$ and $\shehat$ receive no contribution from the residual deviation $\delta\tilde{f}$ of the distribution function:
\be
\label{eq:ch2PLmatch}
\hat{\mathcal{P}}_L\equiv\hat{\mathcal{P}}^\mathrm{RS}_L(\Lambda,\xi) 
\quad \Longleftrightarrow\quad \shehat=\shehat_\mathrm{RS},\ \hat{\tilde{\pi}}=0 .
\ee
With this additional matching condition the leading order energy-momentum tensor \eqref{eq:ch2RStmn} and the full energy-momentum tensor \eqref{eq:ch2anistmn} become identical, i.e. the residual deviation $\delta\tilde{f}$ does not contribute at all to $\temhat^{\mu\nu}$.

Using the identities $\Plhat=\hat{\mathcal{P}}_\eq(\enehat) +\shehat=\enehat/3+\shehat$ and $\Pthat=\hat{\mathcal{P}}_\eq(\enehat) -\shehat/2 =\enehat/3-\shehat/2$ in \eqref{eq:ch2anistmn} the energy conservation law takes exactly the same form as in viscous hydrodynamics, Eqs.~\eqref{eq:ch2eqviscener} or \eqref{eq:ch2vhydro1}. The equation of motion for $\shehat$ is most easily obtained by using the RTA Boltzmann equation to derive an equation of motion for $\Plhat$ and then use $\shehat=\Plhat-\hat{\mathcal{P}}_\eq=\Plhat-\enehat/3$. A straightforward calculation yields
\be
\label{eq:ch2MNRlongpress}
 \parho\hat{\mathcal{P}}_L 
 =\parho\int_{\pp} \pvvhat^2 \, f 
 =\frac{\enehat{-}3\Plhat}{3\hat{\tau}_{r}}\,-\big(\Plhat{+}I_{240}\big)\tanh\rho\,,
\ee
where we used the mass-shell condition $\pp\cdot\pp=-(\ph)^2+\pp_\eta^2 + \po/\cosh^2\rho=0$ to eliminate $\po$ from the integration measure as well as the Landau matching conditions for $\enehat$ and $\Plhat$. Equation \eqref{eq:ch2MNRlongpress} is not closed because $I_{240}$ still involves an integral over the full distribution function; to close the equation we can approximate it by dropping the $\delta\tilde{f}$ contribution to $I_{240}$ \cite{Molnar:2016gwq} by replacing $I_{240} \to \Ih_{240}(\hat\Lambda,\xi)$. Substituting this approximation together with $\Plhat=\enehat/3+\shehat$ into Eq.~\eqref{eq:ch2MNRlongpress} yields the following equation for $\shehat$:
\be
\label{eq:ch2PLshear}
  \partial_\rho\shehat+\frac{\shehat}{\hat{\tau}_r} + \tanh\rho\Bigl(\frac{4}{3}\shehat + \Ih_{240}(\hat\Lambda,\xi)\Bigr)
  = \frac{5}{9}\enehat\tanh\rho\,,
\ee
which should be compared with Eq.~\eqref{eq:ch2eqviscshear} in viscous hydrodynamics. This equation can also be rewritten for the normalized shear stress $\hat{\bar{\pi}}=3\shehat/(4\enehat)$,
\be
\label{eq:ch2ahydroPl}
   \partial_\rho\hat{\bar\pi} + \frac{\hat{\bar\pi}}{\hat{\tau}_r} =
   \frac{4}{3}\tanh\rho\left(\frac{5}{16} + \hat{\bar\pi} - \hat{\bar\pi}^2 - \frac{9}{16}\mathcal{F}(\hat{\bar\pi})\right) ,
\ee
which should be compared with Eq.~\eqref{eq:ch2vhydro2}. Here
\be
\label{eq:ch2F}
   \mathcal{F}(\hat{\bar\pi}) \equiv \frac{\Rhat_{240}\bigl(\xi(\hat{\bar\pi})\bigr)}{\Rhat_{200}\bigl(\xi(\hat{\bar\pi})\bigr)},
\ee
where $\xi(\hat{\bar\pi})$ is the inverse of the function
\bea
\label{eq:ch2pibarxi}
   \hat{\bar\pi}(\xi) &=& \frac{3\shehat}{4\enehat}=\frac{3\Ih_{220}{-}\Ih_{200}}{4\Ih_{200}}
   = \frac{1}{4}\left(\frac{3\Rhat_{220}(\xi)}{\Rhat_{200}(\xi)}-1\right)\qquad
\eea
(see Eqs.~\eqref{eq:ch2RSmacquant} and \eqref{eq:ch2RSpi}), with the $\Rhat$ functions given in \eqref{eq:ch2Rfunctions}. Equation \eqref{eq:ch2pibarxi} can be used to compute (by numerical inversion) the de Sitter time evolution of $\xi(\rho)$ from the solution $\hat{\bar\pi}(\rho)$ of the anisotropic hydrodynamic equations.

Note that the coupled anisotropic hydrodynamic equations \eqref{eq:ch2vhydro1} and \eqref{eq:ch2ahydroPl} are formulated entirely in terms of the macroscopic hydrodynamic variables $\enehat$ and $\hat{\bar\pi}$, without taking recourse to the microscopic parameters $\hat\Lambda$ and $\xi$. The only differences between the DNMR equations \eqref{eq:ch2vhydro} and anisotropic hydrodynamics are somewhat different factors multiplying the constant and linear terms in $\hat{\bar\pi}$ and the appearance of the function $\mathcal{F}(\hat{\bar\pi})$ on the r.h.s. of Eq.~\eqref{eq:ch2ahydroPl}. This function is an additional driving force for the shear stress that depends on the system's dynamical state (reflected in its dependence on $\hat{\bar\pi}$ or the momentum anisotropy parameter $\xi$). While $\mathcal{F}$ is analytically known for the massless Boltzmann gas with Gubser symmetry studied here, it is not obvious how to calculate it in QCD from first principles for a system undergoing arbitrary anisotropic expansion. We regard $\mathcal{F}$ as an anisotropic driving force which must be suitably parametrized until a way of computing it from first principles has been found.
 
For comparison with the following subsections we also present the evolution equations in terms of the microscopic parameters $\hat{\Lambda}$ and $\xi$. Using the energy and $\Plhat$ matching conditions together with Eqs.~\eqref{eq:ch2RSmacquant} --~\eqref{eq:ch2RSpi}, we rewrite the energy conservation law \eqref{eq:ch2eqviscener} in terms of the scalar integrals \eqref{eq:ch2Inlq}:
\be
\label{eq:ch2lambdaPL}
  \partial_\rho\Ih_{200}(\hat{\Lambda},\xi) +
  \tanh\rho\,\left(3\,\Ih_{200}(\hat{\Lambda},\xi) - \Ih_{220}(\hat{\Lambda},\xi)\right)=0.
\ee
Chain rule differentiation $\partial_\rho\Ih_{200}(\hat{\Lambda},\xi)=(\partial_{\hat\Lambda}\Ih_{200})\,\partial_\rho\hat\Lambda$ $+\, (\partial_{\xi}\Ih_{200})\,\partial_\rho\xi$ turns this into an equation that couples the $\rho$ derivatives of $\hat\Lambda$ and $\xi$. They can be uncoupled by using Eq.~(\ref{eq:ch2ahydroPl}), rewritten (with $\hat{\bar{\pi}}(\xi)$ from \eqref{eq:ch2pibarxi}) as
\be
\label{eq:ch2drhoxi}
  \partial_\rho\xi + \frac{1}{\hat{\tau}_r}\,\frac{\hat{\bar{\pi}}(\xi)} {\partial_\xi\hat{\bar{\pi}}(\xi)} = - 2\tanh\rho\, (1{+}\xi).
\ee
The r.h.s. of this equation (which, due to subtle cancellations, turns out to be surprisingly simple when compared with that of Eq.~(\ref{eq:ch2ahydroPl})!) controls the free-streaming ($\hat{\tau}_r{\,\to\,}\infty$) evolution of the anisotropy parameter $\xi$, and thereby the late-time behavior of the hydrodynamic quantities.\footnote{%
	The factor $2\tanh\rho$ on the r.h.s. represents the scalar expansion rate  $\hat\theta=\hat{D}{\,\cdot\,}\hat{u}$ 
	of the Gubser flow \cite{Gubser:2010ze,Gubser:2010ui,Denicol:2014tha}.}

\subsection{The NRS prescription}
\label{subsec:LONRS}

Anisotropic hydrodynamics for Gubser symmetric systems was discussed previously by Nopoush, Ryblewski and Strickland (NRS)~\cite{Nopoush:2014qba}. Instead of using $\Plhat$--matching, they considered a linear combination of third order moments $\mathcal{I}^{\mu\nu\lambda}\equiv \int_{\pp} {\pp}^\mu {\pp}^\nu {\pp}^\lambda f$ of the distribution function, specifically the combination 
\be
\label{eq:ch2NRS1}
  \mathcal{I} \equiv (1{+}\xi)I_{320} - \frac{1}{2} I_{301} \equiv \calI(\hat\Lambda,\xi) +\tilde{\mathcal{I}},
\ee
where on the right hand side we split $\mathcal{I}$ into its leading order contribution $\calI(\hat\Lambda,\xi)$ (from $f_a$ in Eqs.~\eqref{eq:ch2RSansatz} --~\eqref{eq:ch2ERS} and the residual $\tilde{\mathcal{I}}$ (from $\delta\tilde{f}$), in analogy to Eq.~(\ref{eq:ch2Inlq}). Using the RTA Boltzmann equation, NRS derived the following equation of motion for $\calI(\hat\Lambda,\xi)$:
\be
\label{eq:ch2NRSsecondmom}
   \partial_\rho \calI - \Ih_{320}\partial_\rho \xi + \frac{\calI{-}\calIeq}{\hat{\tau}_r} 
   = - 2\tanh\rho \left(\calI - \frac{1}{2} \Ih_{301}\right) ,
\ee
where $\calIeq$ is the corresponding combination of third order moments of the local-equilibrium distribution $\feq$. In this derivation (which is accurate to leading order in the expansion of the distribution function around $f_a$) all contributions to $\mathcal{I}$ from $\delta\tilde{f}$ are neglected.

By dimensional analysis and thanks to the factorization of the $\hat\Lambda$ and $\xi$ dependencies for massless particles, $\calI$ is proportional to $\enehat^{5/4}$. Normalizing Eq.~\eqref{eq:ch2NRSsecondmom} by $\enehat^{5/4}$ one obtains after some algebra \cite{Nopoush:2014qba} the following equation of motion for the anisotropy parameter $\xi$:
\be
\label{eq:ch2NRSxidot}
  \partial_\rho\xi  + \frac{\xi(1{+}\xi)^{3/2}\,\Rhat^{5/4}_{200}(\xi)}{\hat{\tau}_r}
  = - 2\tanh\rho\, (1{+}\xi).
\ee
Eqs.~\eqref{eq:ch2lambdaPL} and~\eqref{eq:ch2NRSxidot} constitute the NRS scheme. This should be compared with Eq.~\eqref{eq:ch2lambdaPL} --~\eqref{eq:ch2drhoxi} in the $\Plhat$--matching scheme. Clearly the anisotropy parameter $\xi$ evolves differently in the NRS and $\Plhat$--matching schemes.

\subsection{Residual dissipative corrections to the NRS prescription}
\label{subsec:NLONRS}

Since with the NRS prescription $\xi$ evolves differently than in the $\Plhat$--matching scheme, the moment $\int_{\pp} \pp^2_\eta\, f_a$ no longer fully matches the macroscopic longitudinal pressure $\Plhat$. Instead, the latter receives an additional contribution $\piti$ from $\delta\tilde{f}$ which has to make up for the missing piece:
\be
\label{eq:ch2PLNRS} 
  \Plhat=\Plhat^\mathrm{RS} + \piti=\Ih_{220}(\hat\Lambda,\xi)+\int_{\pp} \pp_\eta^2\, \delta\tilde{f}.
\ee
This residual shear stress correction on the longitudinal pressure was not taken into account in \cite{Nopoush:2014qba}, i.e. NRS continued to solve for the energy conservation law with Eq.~\eqref{eq:ch2lambdaPL} which only accounts for the leading order contribution to $\Plhat$ from an incorrectly matched anisotropic distribution function $f_a$. Inclusion of the residual shear stress from $\delta\tilde{f}$ modifies the energy conservation law as follows:
\be
\label{eq:ch2NLONRSener2}
\partial_\rho\Ih_{200}(\hat{\Lambda},\xi)
+ \tanh\rho\Big(3\Ih_{200}(\hat{\Lambda},\xi){-}\Ih_{220}(\hat{\Lambda},\xi)\Big)
= \piti\tanh\rho.
\ee
Obviously, an additional equation of motion for $\piti$ is now needed. It is derived from the RTA Boltzmann equation using the standard procedure (see Ref.~\cite{Denicol:2012cn}): 
Writing $f=f_a+\delta\tilde{f}$, Eq.~\eqref{eq:ch2boltzgubeq} gives the following evolution equation for $\delta\tilde{f}$:
\be
\label{eq:ch2deltafeq}
   \partial_\rho \tdf = \frac{\feq-f_a-\tdf}{\hat{\tau}_{r}} - \partial_\rho f_a
\ee
where from Eqs.~\eqref{eq:ch2RSansatz} --~\eqref{eq:ch2ERS}
\be
\label{eq:ch2der-frs}
\besp
   \partial_\rho f_a = \biggl(\frac{\partial_\rho{\hat{\Lambda}}}{\hat{\Lambda}^2} E_\mathrm{RS} 
   + \frac{1}{\hat{\Lambda}\,E_\mathrm{RS}} \Big(\!\tanh\rho \frac{\hat{p}^2_\Omega}
   {{\cosh^2}\rho} - \frac{\partial_\rho{\xi}}{2}\,\hat{p}^2_\eta \Big)\! \biggr) f_a.
\end{split}
\ee
With this the residual shear stress $\piti\equiv\piti^{\eta\eta}$ evolves as
%
%
\be
\allowdisplaybreaks
\label{eq:ch2ressheareq}
\besp
  \parho\piti =\parho\intp\pp^{\langle\eta}\pp^{\eta\rangle}\tdf
  =&-\frac{\hat{\pi}_\mathrm{RS}{+}\piti}{\hat{\tau}_{r}}-\tanh\rho\left(\frac{4}{3}\piti{+}\It_{240}\right)
    -\frac{\partial_\rho{\hat{\Lambda}}}{\hat{\Lambda}^2}\,\left(\hint_{221}{-}\frac{1}{3}\hint_{201}\right)	
\\
    &-\frac{\tanh\rho}{\hat{\Lambda}}\left(\frac{4}{3}\hint_{42-1}{-}\hint_{44-1}-\frac{1}{3}\hint_{40-1}\right) \\
    &+\frac{\partial_\rho{\xi} }{2\hat{\Lambda}}\,\left(\hint_{44-1}{-}\frac{1}{3}\hint_{42-1}\right).
\end{split}
\ee
The anisotropic integrals $\hint_{nlr}$ are defined in Appendix~\ref{appch2:anisint}, Eq.~\eqref{eq:ch2Hfunc}, and $\It_{240}$ is the $\tdf$ contribution to $I_{240}$.
%
Equation~\eqref{eq:ch2ressheareq} is exact but not closed without an approximation for the residual deviation $\tdf$ in the integral $\It_{240}$.
We here use the 14-moment approximation for $\tdf$. Its specific form for our present situation is derived in Appendix~\ref{app:14mom}:
%
\be
\label{eq:ch214gradgub}
\tdf_{14} =\left[ \hat{\alpha} + \hat{\beta}\,\hat{u}\cdot\pp
                  + \frac{4}{3}\hat{\omega}(\hat{u}\cdot\pp)^2
        + \frac{1}{2} \hat{\omega}_{\langle\eta\eta\rangle} \Big(3(\hat{z}\cdot\pp)^2{-}(\hat{u}\cdot\pp)^2 \Big) \right] 
         f_a\,,
\ee
where (with $\alpha_{\tilde{\pi}}(\xi)$, $\beta_{\tilde{\pi}}(\xi)$, $\gamma_{\tilde{\pi}}(\xi)$ and $\kappa_{\tilde{\pi}}(\xi)$ given in Eqs.~\eqref{eq:ch2alphapi} -- \eqref{eq:ch2kappapi})
\bs
\allowdisplaybreaks
\label{eq:ch214gradcoefgub}
\begin{align}
  \hat{\alpha} &= \frac{\hat{\tilde{\pi}} \, \alpha_{\tilde{\pi}}(\xi)}{\J_{2}(\hat{\Lambda})}\,,
  \qquad
  \hat{\beta} = \frac{\hat{\tilde{\pi}} \, \beta_{\tilde{\pi}}(\xi)}{\J_{3}(\hat{\Lambda})} \,,
\qquad
  \hat{\omega} = \frac{\hat{\tilde{\pi}} \, \gamma_{\tilde{\pi}}(\xi)}{\J_{4}(\hat{\Lambda})} \,,
  \qquad
  \hat{\omega}_{\langle\eta\eta\rangle} 
  = \frac{\hat{\tilde{\pi}} \, \kappa_{\tilde{\pi}}(\xi)}{\J_{4}(\hat{\Lambda})}.
\end{align}
\es
Using this approximation for $\tdf$ the calculation of the term $\It_{240}$ in Eq.~\eqref{eq:ch2ressheareq} is straightforward. After some algebra, we arrive at the following closed evolution equation for the residual shear stress $\piti$: 
\be
\allowdisplaybreaks
\label{eq:ch2ressheareq2}
\besp
\parho\piti =
    & - \frac{\hat{\pi}_\mathrm{RS}{+}\piti}{\hat{\tau}_{r}}
         - \tanh\rho\left[\frac{4}{3}\piti + \hat{\alpha}\,\Ih_{240} - \hat{\beta}\,\Ih_{340} 
                              + \frac{4}{3}\hat{\omega}\,\Ih_{440} 
                              + \frac{1}{2}\hat{\omega}_{\langle\eta\eta\rangle}\left(3\,\Ih_{460}{-}\Ih_{440}\right)\,\right]
\\
   & - \frac{\partial_\rho{\hat{\Lambda}}}{\hat{\Lambda}^2}\,\left(\hint_{221}{-}\frac{1}{3}\hint_{201}\right)	
      - \frac{\tanh\rho}{\hat{\Lambda}}\left(\frac{4}{3}\hint_{42-1}{-}\hint_{44-1}-\frac{1}{3}\hint_{40-1}\right) \\
      &+ \frac{\partial_\rho{\xi} }{2\hat{\Lambda}}\,\left(\hint_{44-1}{-}\frac{1}{3}\hint_{42-1}\right).
\end{split}
\ee
The evolution equations \eqref{eq:ch2NLONRSener2}, \eqref{eq:ch2NRSxidot} and \eqref{eq:ch2ressheareq2} define the NLO NRS prescription for anisotropic hydrodynamics subject to Gubser flow. Note that the NLO residual dissipative corrections do not affect the evolution of the momentum anisotropy parameter $\xi$ which remains the same as in the leading-order NRS treatment of Ref.~\cite{Nopoush:2014qba}. The NLO corrections only modify the evolution of the energy density $\enehat$ and pressure anisotropy $\Plhat-\Pthat\sim\hat{\pi}=\hat{\pi}_\mathrm{RS}+\piti$.

We conclude this subsection by noting that, due to the appearance of non-hydrodynamic higher order moments $\Ih_{nlq}$ in the NRS and NLO NRS prescriptions, the latter lead to evolution equations that explicitly refer to the evolution of the microscopic momentum anisotropy parameter $\xi$ and cannot be formulated purely macroscopically. This sets them apart from the viscous hydrodynamic and $\Plhat$--matching anisotropic hydrodynamic formulations. It is therefore not clear how to generalize the NRS and NLO NRS prescriptions to strongly coupled situations where a microscopic kinetic description in terms of quasiparticle distribution functions is not possible.

\subsection{Exact solution of the RTA Boltzmann equation}
\label{subsec:gubsol}
For Gubser flow, the RTA Boltzmann equation \eqref{eq:ch2boltzgubeq} is solved exactly by \cite{Denicol:2014xca,Denicol:2014tha} 
\be
\label{eq:ch2exsol}
f(\rho;\po,\pvvhat) =D(\rho,\rho_0) f_0(\rho_0;\po,\pvvhat)\,+\,\frac{1}{5\bar{\eta}}\int_{\rho_0}^\rho d\rho'\,D(\rho,\rho')\,\hat{T}(\rho')\, \feq(\rho^\prime;\po,\pvvhat)\,,
\ee
where $D(\rho_2,\rho_1){\,=\,}\exp\!\left[-\frac{1}{5\bar{\eta}}\int_{\rho_1}^{\rho_2} d\rho^{\prime\prime} \,\hat{T}(\rho^{\prime\prime})\right]$ is the damping function and $f_0$ is the initial distribution at de Sitter time $\rho_0$, for which we take an RS distribution \eqref{eq:ch2RSansatz} with initial effective temperature $\hat{\Lambda}_0$ and momentum anisotropy $\xi_0$:
\be
\label{eq:ch2RS}
\!\!\!\!\!
f_{0}(\rho_0;\po,\pvvhat)= 
\exp\left[-\frac{1}{\hat{\Lambda}_0}\sqrt{\frac{\po}{\cosh^2\!\rho_0}+(1{+}\xi_0)\pp_\eta^2}\,\right]\,.
\ee
From the exact solution~\eqref{eq:ch2exsol} one can compute the exact evolution of the energy density $\enehat=\Ih_{200}$ and shear stress $\hat{\pi}=\Ih_{220}-\frac{1}{3}\Ih_{200}$ \cite{Denicol:2014xca,Denicol:2014tha}: 
\be
\allowdisplaybreaks
\label{eq:ch2exactene}
\besp
\enehat(\rho) 
        =\,&D(\rho,\rho_0)\Big(\frac{\cosh\rho_0}{\cosh\rho}\Big)^{\!4}\,
        \enehat_\mathrm{RS}\big(\lam_0, \xi_\mathrm{FS}(\rho;\rho_0,\xi_0)\big)\\
        &+ \frac{1}{5\bar{\eta}}\int_{\rho_0}^\rho d\rho'\,D(\rho,\rho')\,\hat{T}(\rho')\Big(\frac{\cosh\rho^\prime}{\cosh\rho}
           \Big)^{\!4}\,
         \enehat_\mathrm{RS}\big(\temhat(\rho^\prime), \xi_\mathrm{FS}(\rho;\rho^\prime,0)\big), 
\end{split}
\ee
\be
\allowdisplaybreaks
\label{eq:ch2exactpi}
\besp
\shehat(\rho)
        =\,&D(\rho,\rho_0)\Big(\frac{\cosh\rho_0}{\cosh\rho}\Big)^{\!4}\,
        \shehat_\mathrm{RS} \big(\lam_0, \xi_\mathrm{FS}(\rho;\rho_0,\xi_0)\big)\\
        &+ \frac{1}{5\bar{\eta}}\int_{\rho_0}^\rho d\rho'\,D(\rho,\rho')\,\hat{T}(\rho')\Big(\frac{\cosh\rho^\prime}{\cosh\rho}
            \Big)^{\!4}\,\shehat_\mathrm{RS}\big(\temhat(\rho^\prime), \xi_\mathrm{FS}(\rho;\rho^\prime,0)\big).
\end{split}
\ee
Here we defined $\xi_\mathrm{FS}(\rho;\rho_\alpha,\xi_\alpha) = -1 + (1+\xi_\alpha)\big(\frac{\cosh\rho_\alpha}{\cosh\rho}\big)^2$.  We solve the above integral equations for $\enehat$ and $\shehat$ numerically using the method described in Refs.~\cite{Banerjee:1989by,Florkowski:2013lza,Florkowski:2013lya,Denicol:2014tha}. The temperature is then obtained from the energy density by Landau matching: $\temhat = \big(\frac{1}{3}\pi^2 \enehat\big)^{1/4}$. When discussing the results in the following section we plot the evolution of the temperature rather than the energy density.
\section{Results}
\label{sec:results}
In this section we compare the numerical results obtained from the five different approaches discussed in the preceding section (second-order viscous hydrodynamics (DNMR), anisotropic hydrodynamics based on the $\Plhat$--matching, NRS and NLO NRS prescriptions, and the exact solution of the Boltzmann equation) for the evolution of the temperature and normalized shear stress.

For the benefit of the reader we summarize the set of equations that are being solved in each of the five cases:
\begin{enumerate}
\item
{\sl Viscous hydrodynamics (DNMR):} qquations~(\ref{eq:ch2vhydro}) for $\enehat$ and $\hat{\bar\pi}$, together with $\enehat=3\,\temhat^4/\pi^2$.
\item
{\sl Anisotropic hydrodynamics ($\Plhat$--matching):} equations \eqref{eq:ch2vhydro1} and~\eqref{eq:ch2ahydroPl} together with $\enehat{\,=\,}3\,\temhat^4/\pi^2$ and Eq.~\eqref{eq:ch2pibarxi} to obtain $\temhat$ and $\xi$ from $\enehat$ and $\hat{\bar\pi}$.
\item
{\sl Anisotropic hydrodynamics (NRS):} equations \eqref{eq:ch2lambdaPL} and \eqref{eq:ch2NRSxidot} for $\lam$ and $\xi$, from which we get $\temhat$ and $\hat{\bar\pi}$ using Eqs.~\eqref{eq:ch2LMcond} and \eqref{eq:ch2pibarxi}.
\item
{\sl Anisotropic hydrodynamics (NLO NRS):} equations \eqref{eq:ch2NRSxidot} and \eqref{eq:ch2NLONRSener2} for $\lam$ and $\xi$, together with Eq.~\eqref{eq:ch2ressheareq2} for the residual shear stress $\piti$ and Eqs.~\eqref{eq:ch2LMcond} and \eqref{eq:ch2pibarxi} to get $\temhat$ and the leading order $\hat{\bar\pi}(\xi)$ from $\lam$ and $\xi$. 
\item
{\sl Exact solution of the RTA Boltzmann equation:} equations \eqref{eq:ch2exactene} (rewritten in terms of $\temhat$) and \eqref{eq:ch2exactpi}.
\end{enumerate}

\begin{figure}[t]
\centering
\includegraphics[width=\linewidth]{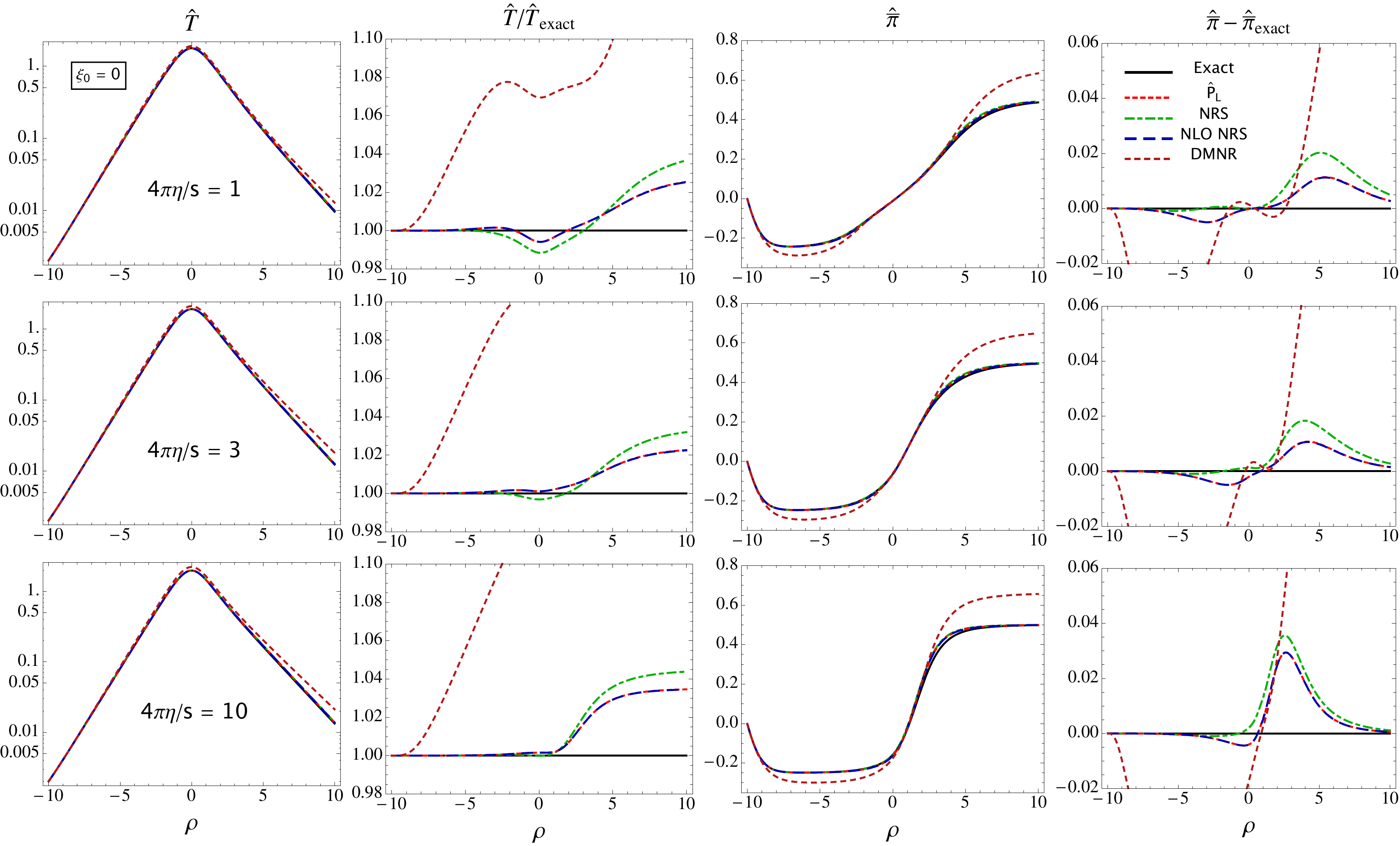}
\caption{The de Sitter time evolution of the temperature $\temhat$ and the normalized shear stress $\hat{\bar{\pi}}$ for the exact solution of the RTA Boltzmann equation (black solid lines) and four different hydrodynamic approximations: second-order viscous hydrodynamics (DNMR theory, short-dashed magenta lines), anisotropic hydrodynamics with $\Plhat$--matching (dotted red lines), leading order anisotropic hydrodynamics in the NRS scheme (dash-dotted green lines), and next-to-leading order anisotropic hydrodynamics in the NRS scheme amended by residual viscous corrections (long-dashed blue lines). For the initial momentum distribution we here assumed isotropy, i.e. $\xi_0=0$. The top, middle and bottom rows of panels correspond to specific shear viscosity $4\pi\hat\eta/\hat{\mathcal{S}} = 1$, 3, and 10, respectively. The four columns of plots show, from left to right, the $\rho$ evolution of the temperature $\temhat$, of the ratio between its hydrodynamic and exact kinetic evolution $\temhat/\temhat_\mathrm{exact}$, of the normalized shear stress $\hat{\bar\pi}$, and of the difference between its hydrodynamic and exact kinetic evolution $\hat{\bar\pi}-\hat{\bar\pi}_\mathrm{exact}$.
\label{F1chap2}
}
\end{figure}

All five models are started at an initial de Sitter time $\rho_0{\,=\,}-10$ with initial temperature $\temhat_0 = 0.002$ and run for three choices of the specific shear viscosity, $4\pi\hat\eta/\hat{\mathcal{S}}=1,\,3,$ and 10 (top, middle and bottom row of panels in Figs.~\ref{F1chap2} -- \ref{F3chap2}), and three choices of the initial momentum anisotropy, $\xi_0=0,\,100,$ and -0.9 (shown in Figs.~\ref{F1chap2}, \ref{F2chap2}, and \ref{F3chap2}, respectively). We now proceed to discuss these results in detail.
\subsection{Evolution of temperature and shear stress}
\label{sec:tempi}
\begin{figure}[t]
\centering
\includegraphics[width=\linewidth]{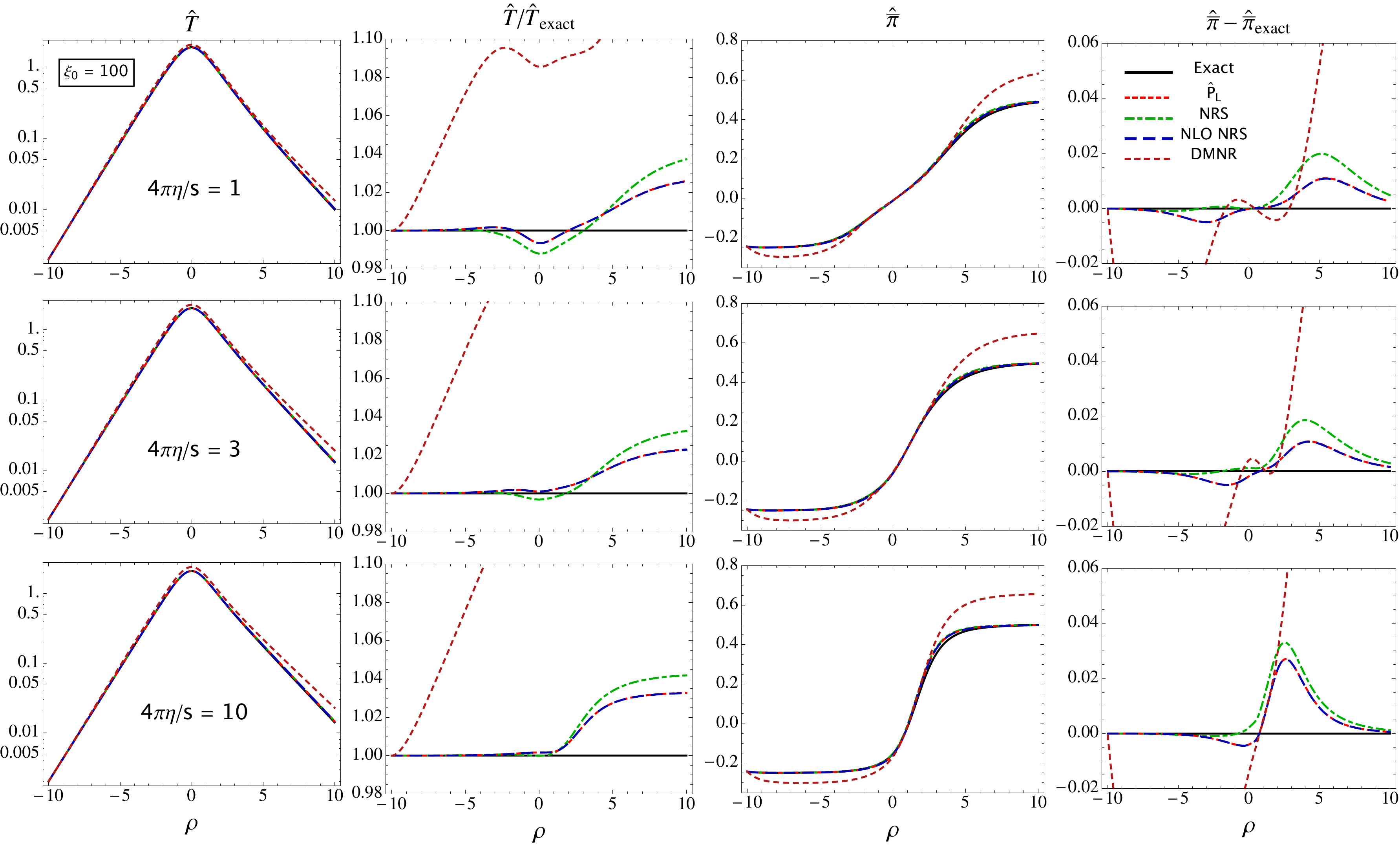}
\caption{Same as Fig.~\ref{F1chap2}, but for an initially highly oblate momentum distribution, $\xi_0=100$. 
\label{F2chap2}
}
\end{figure}
\begin{figure}[t]
\centering
\includegraphics[width=\linewidth]{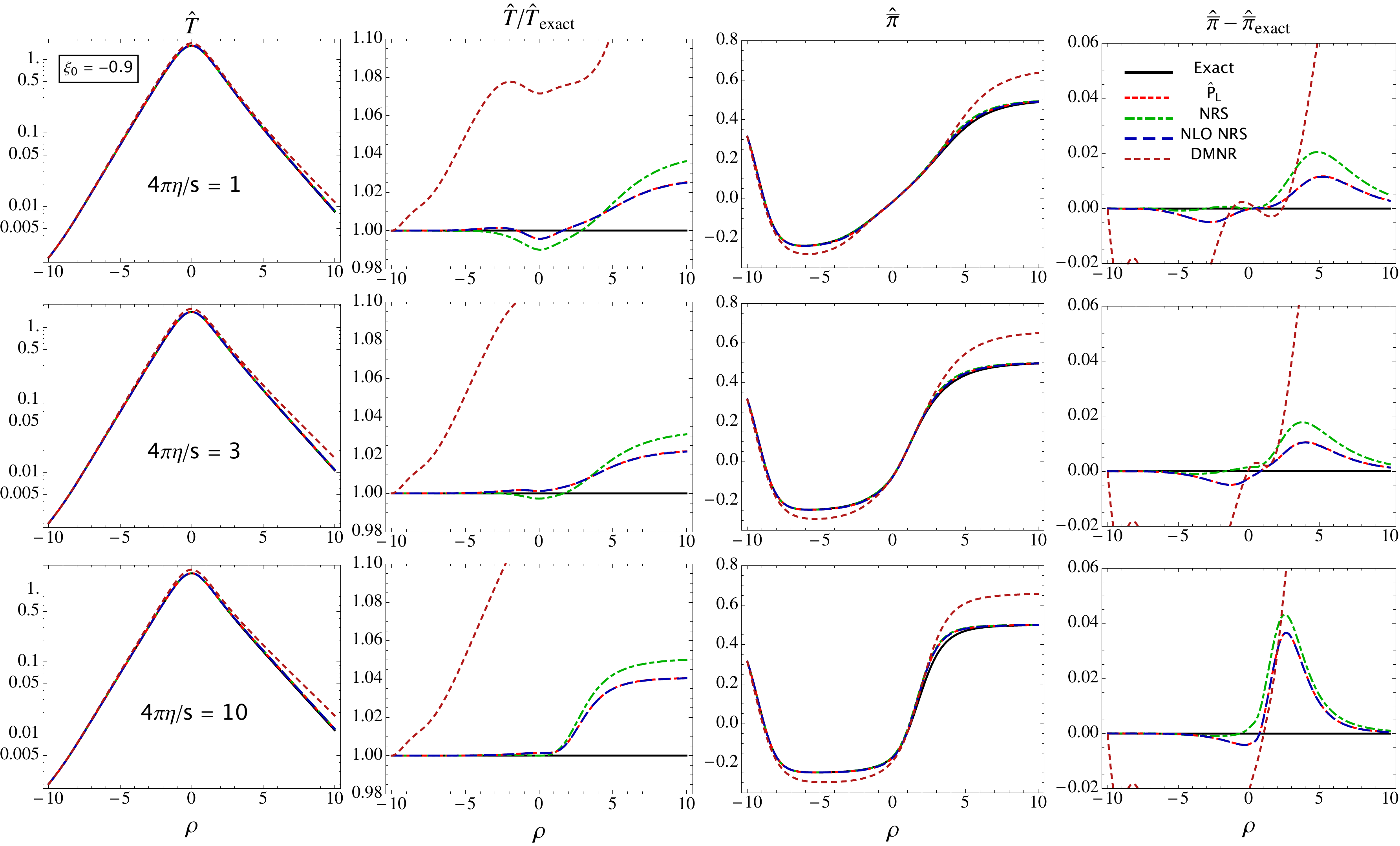}
\caption{Same as Fig.~\ref{F1chap2}, but for an initially highly prolate momentum distribution, $\xi_0=-0.9$. 
\label{F3chap2}
}
\end{figure}
Figures~\ref{F1chap2} -- \ref{F3chap2} show the de Sitter time evolution of the temperature $\temhat$ and of the normalized shear stress $\hat{\bar\pi}$, in the first and third column in absolute values, and in the second and fourth column relative to the exact solution of the RTA Boltzmann equation. It is obvious that all three anisotropic hydrodynamic schemes studied here vastly outperform standard second-order viscous hydrodynamics. Over the range of de Sitter times studied here, anisotropic hydrodynamics with $\Plhat$--matching never deviates from the exact solution by more than a few percent. The results from the leading-order NRS scheme (for which the momentum anisotropy parameter is not matched to the macroscopic pressure anisotropy) performs slightly worse, but not dramatically so. Once the residual shear stress caused by the non-optimal $\xi$-evolution in this approach is added to the formalism at next-to-leading order, the evolution of both $\temhat$ and $\hat{\bar\pi}$ agrees almost perfectly with that in the $\Plhat$--matching scheme.  

For all of the anisotropic hydrodynamic schemes the normalized shear stress $\hat{\bar\pi}$ correctly approaches the asymptotic free-streaming value $\frac{1}{2}$ predicted by the RTA Boltzmann equation \cite{Denicol:2014tha}, in contrast to standard second-order viscous fluid dynamics; the asymptotic temperature lies a few percent above the exact value. The asymptotic behavior at large de Sitter times is almost independent of the initial momentum anisotropy $\xi_0$: due to the rapid expansion the system quickly loses its memory of the initial state. As the specific shear viscosity increases (corresponding to increasing values of the microscopic relaxation time), the deviations between the hydrodynamic evolution of the macroscopic observables and that extracted from the exact solution of the Boltzmann equation grow a bit. However, even for $\hat\eta/\hat{\mathcal{S}}$ values 10 times larger than the ``minimal'' value $1/(4\pi)$ \cite{Policastro:2001yc,Kovtun:2004de} the deviations from the exact solution stay below 4\% for both the temperature ratio $\temhat/\temhat_\text{exact}$ and the difference $\hat{\bar\pi} - \hat{\bar\pi}_\text{exact}$ as long as the $\Plhat$--matching scheme is employed.

\section{Summary}
\label{sec:concl}
In this chapter we studied conformal systems undergoing Gubser flow that admit simultaneously a kinetic description via the RTA Boltzmann equation and a macroscopic hydrodynamic description. We reviewed the standard second-order viscous hydrodynamic formulation and compared it with three variants of anisotropic hydrodynamics, each of them implementing a different closing scheme. In the comparison with the evolution of the energy-momentum tensor obtained from the exact kinetic solution of the RTA Boltzmann equation, we found anisotropic hydrodynamics with the $\Plhat$--matching scheme to be the most accurate macroscopic realization of the underlying kinetic theory, confirming observations made previously in Ref.~\cite{Molnar:2016gwq} for Bjorken flow. In this approach the evolution of the microscopic momentum anisotropy parameter $\xi$ is matched to the expansion-driven macroscopic pressure anisotropy $\Plhat-\Pthat$. We also showed that an alternative prescription proposed in Ref.~\cite{Nopoush:2014qba} where the $\xi$ evolution does not match the pressure anisotropy can be made equally accurate, albeit with more work by solving a larger set of equations, by accounting for the residual shear stress associated with the non-optimal $\xi$ evolution (NLO NRS scheme). Both procedures lead to descriptions that deviate from the exact results by at most a few percent.

We emphasize that, in contrast to the NLO NRS scheme, anisotropic hydrodynamics with $\Pl$--matching can be formulated entirely macroscopically, without explicitly referring to the underlying microscopic kinetic description and its parameters. Here we showed that the momentum anisotropy parameter $\xi$ can be equivalently replaced by the pressure anisotropy $\Pl - \Pt$. This puts anisotropic hydrodynamics on the same footing as second-order viscous hydrodynamics in that it can be generalized from conformal systems with Gubser symmetry to non-conformal systems undergoing arbitrary expansion. To get there, we will now apply the techniques from the $\Pl$--matching scheme to derive new hydrodynamic equations for the anisotropically expanding quark-gluon plasma.

\chapter{Anisotropic fluid dynamics for the quark-gluon plasma}
\label{ch3label}
Here we generalize anisotropic hydrodynamics with $\PL$--matching to model the far-from-equilibrium dynamics of the non-conformal quark-gluon plasma in (3+1)--dimensions for various initial conditions. In addition to the large pressure anisotropies produced in the early stages of heavy-ion collisions, there may also be large bulk viscous effects present when the quark-gluon plasma converts to hadrons. Near the pseudocritical temperature $T_c$, the bulk viscosity $\zetas$ is expected to peak strongly due to large critical fluctuations.\footnote{%
    The overall magnitude of the bulk viscous pressure ultimately depends on critical slowing-down, where the response to global expansion is delayed by the same critical fluctuations~\cite{Berdnikov:1999ph,Song:2009rh,Rajagopal:2009yw}. The strength of this hydrodynamic response for the non-equilibrium quark-gluon plasma, however, is not well understood.} 
In our model, we capture these large dissipative flows non-perturbatively by extending the Landau matching to both the longitudinal and transverse pressures $\PL$ and $\Pperp$ (or equivalently the diagonal shear component $\pi^{zz} \propto \PL-\Pperp$ and bulk viscous pressure $\Pi$). We then treat the residual shear stresses perturbatively, similar to what is done for $\pi^\munu$ and $\Pi$ in second-order viscous hydrodynamics. To construct the framework, we derive the anisotropic hydrodynamic equations from the relativistic Boltzmann equation for a system of quasiparticles with a medium-dependent mass $m(T)$. With the generalized Landau matching conditions, we can parametrize the anisotropic transport coefficients of this kinetic theory model in terms of the QCD energy density $\ene$ and deviations of $\PL$ and $\Pperp$ from the QCD equilibrium pressure $\Peq(\ene)$ (the shear and bulk viscosities are modeled phenomenologically). This allows us to evolve the relaxation equations macroscopically in the same manner as standard viscous hydrodynamics. 

This chapter is based on material previously published in Ref.~\cite{McNelis:2018jho}.
\section{Anisotropic hydrodynamics}
\label{ch3sec2}

\subsection{Energy-momentum tensor decomposition}
\label{ch3sec2c}

Similar to the Gubser flow, the shear stress $\pi^\munu$ in ultrarelativistic heavy-ion collisions is initially highly anisotropic due to strongly different longitudinal and transverse expansion rates. This leads to large differences $\PL-\Pperp$ between the longitudinal and transverse pressures, especially at the onset of the collision. Therefore, the hydrodynamic equations for the non-equilibrium quark-gluon plasma are best formulated in the basis $\{u^\mu$, $z^\mu$, $\Xi^\munu\}$ used in the previous chapter. In ordinary Minkowski spacetime $x^\mu = (t,x,y,z)$, the energy-momentum tensor is decomposed as (suppressing the spacetime dependence)
\be
\label{eqch3:3}
  \tmn=\ene\um\un + \PL z^\mu z^\nu - \Pperp\xu  + 2W^{(\mu}_{\perp z} z^{\nu)} + \piu_\perp.
\ee
The round parentheses around pairs of Lorentz indices indicate symmetrization: $W^{(\mu}_{\perp z} z^{\nu)} \equiv\frac{1}{2}\bigl(W^{\mu}_{\perp z} z^{\nu}+W^{\nu}_{\perp z} z^{\mu}\bigr)$. The anisotropic decomposition (\ref{eqch3:3}) clearly separates the pressures $\PL$ and $\Pperp$ from the other dissipative components. Given an arbitrary energy-momentum tensor $T^\munu$, the anisotropic hydrodynamic quantities appearing in this decomposition can be obtained by the following projections:
\bs
\allowdisplaybreaks
\beal
\label{eqch3:4}
  \ene &= \uum\unn\tmn\,, \\
  \PL &=  z_\mu z_\nu\tmn\,,\\
  \Pperp &= - \frac{1}{2}\Xi_{\mu\nu}\tem^{\mu\nu}\,, \\
  W^{\mu}_{\perp z}\, &= - \Xi^{\mu}_{\alpha}\tem^{\alpha\nu}z_\nu\,, \\
  \piu_\perp &= \Xi_{\alpha\beta}^{\mu\nu}\tem^{\alpha\beta} \,.
\end{align}
\es
In the last line we introduced the symmetric traceless transverse projection tensor $\xu_{\alpha\beta} = \frac{1}{2}\big(\Xi^\mu_\alpha\Xi^\nu_\beta+\Xi^\nu_\beta\Xi^\mu_\alpha-\xu\Xi_{\alpha\beta}\big)$. The corresponding transverse shear stress tensor $\piperp$ describes two shear stress degrees of freedom that account for momentum diffusion currents along the transverse directions. It is traceless and orthogonal to both the fluid velocity $u^\mu$ and the direction of the pressure anisotropy $z^\mu$:
\be
\label{eqch3:5}
\begin{aligned}
   \pi^\mu_{\perp,\mu} = u_\mu \piperp = z_\mu \piperp = 0.
\end{aligned}
\ee
Another two shear stress degrees of freedom are encoded in the longitudinal-momentum diffusion current $\Wperp$  which is orthogonal to both $u^\mu$ and $z^\mu$:
\be
\label{eqch3:6}
\begin{aligned}
  u_\mu \Wperp = z_\mu \Wperp = 0.
\end{aligned}
\ee
The remaining fifth (and largest) shear stress component is given by $\Pl-\Pt$. Altogether, the five independent components of the standard shear stress tensor $\pi^\munu$ are related to those in the anisotropic decomposition (\ref{eqch3:3}) by
\be
\label{eqch3:7}
  \pi^{\mu\nu}=\she^{\mu\nu}_{\perp} + 2\,W^{(\mu}_{\perp z} z^{\nu)}
  +\frac{1}{3} \big(\Pl {-} \Pt\big) \big(2z^\mu z^\nu {-} \xu\big)
\ee
while the bulk viscous pressure $\Pi$ is related to the longitudinal and transverse pressures by
\be
\label{eqch3:8}
  \Pi = \frac{2\Pperp {+} \PL}{3} - \Peq.
\ee
Here the equilibrium pressure $\Peq$ is not an independent degree of freedom but related to the energy density $\ene$ through the equation of state of the fluid, $\Peq(\ene)$. 

\subsection{Hydrodynamic evolution equations}
\label{ch3sec2d}

Four of the ten evolution equations that control the dynamics of the energy-momentum tensor are obtained from  
the conservation laws for energy and momentum
\be
\label{eqch3:9}
  \partial_\mu T^\munu = 0. 
\ee
Projecting with $u_\nu$ on the temporal direction in the LRF provides an evolution equation for the LRF energy density:
\be
\label{eqch3:10}
  \dot\ene + (\ene{+}\Pperp) \theta_\perp + (\ene{+}\PL) \theta_L  + W^{\mu}_{\perp z}\bigl(D_z u_\mu{-}z_\nu \nabla_{\perp,\mu}u^\nu\bigr) - \piu_\perp \sigma_{\perp,\munu} = 0.
\ee
Here and below a dot over or a $D$ in front of a quantity denotes the co-moving time derivative, e.g. $D\ene\equiv \dot\ene \equiv  u^\mu \partial_\mu \ene$. $\theta_L = z_\mu D_z u^\mu$ is the scalar longitudinal expansion rate and $\theta_\perp = \nabla_{\perp,\mu}u^\mu$ is the scalar transverse expansion rate. The LRF longitudinal derivative and transverse gradient are written as $D_z = - z^\mu \partial_\mu$ and $\nabla_{\perp,\mu} = \Xi_\mu^\nu \partial_\nu$, respectively. $\sigma_{\perp,\munu} = \Xi^{\alpha\beta}_\munu \partial_\alpha u_\beta$ is the transverse velocity-shear tensor. 

The longitudinal projection $z_\nu \partial_\mu T^\munu = 0$ yields an equation for the longitudinal acceleration of the fluid in the LRF:
\be
\label{eqch3:11}
(\ene{+}\PL) z_\mu \dot{u}^\mu  =  - D_z \PL + (\PL{-}\Pperp) \tilde\theta_{\perp}
  - \Wperp \dot{u}_\mu - 2W^\mu_{\perp z} D_z z_\mu 
     + \nabla_{\perp,\mu} \Wperp + \piperp \tilde{\sigma}_{\perp,\mu\nu}.
\ee
Here $\tilde\theta_\perp\equiv \nabla_{\perp,\mu}z^\mu$ and $\tilde \sigma_{\perp,\munu} = \Xi^{\alpha\beta}_\munu \partial_\alpha z_\beta$.\footnote{%
	 Generically we use tildes to indicate quantities involving derivatives of $z^\mu$ instead of $u^\mu$.
	 }

An equation for the transverse acceleration is obtained from the transverse projection $\Xi^\alpha_\nu \partial_\mu T^\munu = 0$:
\be
\label{eqch3:12}
\begin{split}
  (\ene{+}\Pperp) \Xi^\alpha_\nu \dot{u}^\nu =\,& \nabla_\perp^\alpha \Pperp
    + (\PL{-}\Pperp) \Xi^\alpha_\nu D_z z^\nu - W^\alpha_{\perp z} \Big(\frac{3}{2} \tilde\theta_\perp{-}z_\nu \dot{u}^\nu\Big)  \\
      &- W_{\perp z, \nu}(\tilde\sigma^{\alpha\nu}_\perp{-}\tilde\omega^{\alpha\nu}_\perp) 
   + \Xi^\alpha_\nu D_z W^\nu_{\perp z} + \pi^{\alpha\nu}_\perp(\dot{u}_\nu{+}D_z z_\nu)  - \Xi^\alpha_\nu \nabla_{\perp,\mu} \piperp .
\end{split}
\ee
Here $\tilde \omega_\perp^{\alpha\nu} \equiv \Xi^{\alpha\mu} \Xi^{\nu\beta} \partial_{[\beta} z_{\mu]}$ where the square brackets indicate antisymmetrization: $\partial_{[\beta} z_{\mu]} \equiv \frac{1}{2}\bigl(\partial_{\beta} z_{\mu}-\partial_{\mu} z_{\beta}\bigr)$. Equations~(\ref{eqch3:10}) -- (\ref{eqch3:12}) agree (after adjustment of notation) with Eqs.\,(146) -- (148) in Ref.~\cite{Molnar:2016vvu}.

\section{Dissipative relaxation equations}
\label{ch3sec3}

To close the system of equations, we need six additional relaxation equations for $\PL$, $\Pperp$, $\Wperp$ and $\piperp$. Their dynamics is not controlled by macroscopic conservation laws but by microscopic interactions among the fluid's constituents. We will here derive them by assuming a weakly-coupled dilute fluid whose microscopic physics can be described by the relativistic Boltzmann-Vlasov equation for a single particle species with a medium-dependent mass:
\be
\label{eqch3:13}
p^\mu \partial_\mu f + m \, \partial^\mu m \, \partial_\mu^{(p)} f = C[f].
\ee
Here $f(x,p)$ is the single particle distribution function, $C[f]$ is the collision kernel, $m(x)$ is the medium-dependent effective mass, and $\partial_\mu^{(p)}$ is the momentum derivative.

\subsection{Anisotropic distribution function}
\label{ch3sec3a}

Similar to the decomposition~\eqref{eq:ch2anisexp} introduced in Chapter~\ref{chap2label}, we expand the distribution function around a leading-order anisotropic distribution:
\be
\label{eqch3:14}
  f(x,p) = f_a(x,p) + \dft(x,p)\,.
\ee
In this case, however, the Romatschke--Strickland distribution is not conformal~\cite{Romatschke:2003ms, Tinti:2013vba, Tinti:2015xwa}:
\be
\label{eqch3:15}
  f_a(x,p) = \left(\exp\left[\frac{\sqrt{\Omega_\munu(x) \, p^\mu p^\nu}}{\Lambda(x)} \,\right] + \Theta\right)^{-1}\,,
\ee
where $\Lambda(x)$ is the effective temperature and $\Theta \in (0, \pm 1)$ accounts for Boltzmann, Fermi--Dirac, and Bose--Einstein statistics, respectively. The leading-order momentum anisotropy is encoded in the ellipsoidal tensor 
\be
\label{eqch3:16}
   \Omega_\munu(x) = u_\mu(x) u_\nu(x) - \xi_\perp(x) \Xi_\munu(x)
    + \xi_L(x) z_\mu (x) z_\nu (x).
\ee
For non-conformal plasmas\footnote{In the conformal limit $m\to0$, the momentum anisotropy parameters reduce to $(\xi_\perp, \xi_L) \to (0,\xi)$.}, there are two spacetime dependent anisotropy parameters $\xi_L$ and $\xi_\perp$ that deform the longitudinal and transverse momentum space, respectively. The tensor contraction $p \cdot \Omega \cdot p$ can be rewritten as
\be
\label{eqch3:17}
    p \cdot \Omega \cdot p = m^2 + (1{+}\xi_\perp)p^2_{\perp,\mathrm{LRF}} 
                                                             + (1{+}\xi_L)p^2_{z,\mathrm{LRF}}\,,
\ee
where $p_{z,\mathrm{LRF}} = - z \cdot p$ is the LRF longitudinal momentum and $p^2_{\perp,\mathrm{LRF}} = - p \cdot \Xi \cdot p$ is the square of the LRF transverse momentum. The difference $\xi_L-\xi_\perp$ can be attributed to a manifestation of shear stress (resulting in a difference between the longitudinal and transverse pressures) while the sum  $\xi_L+\xi_\perp$ encodes a bulk viscous pressure \cite{Nopoush:2014pfa,Bazow:2015cha}. Introducing the notation $\alpha_{L,\perp}(x) = \bigl(1+\xi_{L,\perp}(x)\bigr)^{-1/2}$, Eq.~(\ref{eqch3:15}) can be written in LRF momentum components more conveniently as
\be
\label{eqch3:18}
  f_a = \left(\exp\left[\frac{1}{\Lambda} \sqrt{m^2 + \frac{p^2_{\perp,\mathrm{LRF}}}{\alpha_\perp^2} + \frac{p^2_{z,\mathrm{LRF}}}{\alpha_L^2}}\,\right]+\Theta\right)^{-1}.
\ee

To make the decomposition (\ref{eqch3:14}) unique one must specify the three parameters $\bm{\alpha}(x)\equiv\bigl(\alpha_\perp(x),\alpha_L(x)\bigr)$ and $\Lambda(x)$. We proceed as follows: the physical temperature $T$ of the system is defined by the LRF energy density via the thermodynamic relation $\ene(x) \equiv \ene\bigl(T(x)\bigr)$. To relate the effective temperature $\Lambda$ to $T$ we impose the generalized Landau matching condition 
\be
\label{eqch3:20}
  \delta \tilde\ene \equiv \int_p (\up)^2 \dft  = 0\,.
\ee
We define the Lorentz-invariant momentum space integral
\be
\label{eqch3:26}
  \int_p = \frac{g}{(2\pi)^3} \int d^4p \, 2\Theta(p^0) \delta(p^2{-}m^2) = \frac{g}{(2\pi)^3} \int \frac{d^3p}{E_p}\,,
\ee
where $g$ is a degeneracy factor counting the number of quantum states allowed for a particle with on-shell momentum $p^\mu$, and $\Theta(p^0)$ denotes the Heaviside step function. Equation~(\ref{eqch3:20}) states that $\Lambda(\bm{\alpha})$ must be chosen such that the residual deviation $\dft$ does not contribute to the energy density. This fixes $\Lambda(\bm{\alpha})$ as a function of $T$; the two agree in the limit $\bm{\alpha}\to(1,1)$ when the anisotropic distribution $f_a$ reduces to the local-equilibrium distribution $f_\eq = \left(\exp\left[u\cdot p / T\right]+\Theta\right)^{-1}$.

The momentum deformation parameters $\alpha_{L,\perp}(x)$ are fixed by similar generalized Landau matching conditions for the longitudinal and transverse pressures
\bs
\allowdisplaybreaks
\label{eqch3:21}
\beal
\PL &= \Peq + \Pi + \pi^{zz}_\mathrm{LRF}, \\
\Pperp &= \Peq + \Pi - \frac{1}{2} \pi^{zz}_\mathrm{LRF}.
\end{align}
\es 
Here $\pi^{zz}_\mathrm{LRF}$ is the LRF value of the longitudinal diagonal element of the shear stress tensor $\pi^\munu$ in the decomposition (\ref{eqch3:7}). Note that both the bulk viscous pressure $\Pi$ and shear stress component $\pi^{zz}_\mathrm{LRF}$ are here assumed to be ``large'' such that they must be accounted for already at leading order, by adjusting the parameters $\alpha_{L,\perp}(x)$ accordingly. This is done by demanding
\bs
\allowdisplaybreaks
\label{eqch3:22}
\beal
  \delta \tilde{\mathcal{P}}_L &\equiv \int_p \mzp^2 \dft  = 0, \\
  \delta \tilde{\mathcal{P}}_\perp  &\equiv  \frac{1}{2} \int_p \pxp \dft = 0.
\end{align}
\es
By imposing these conditions, $\alpha_{L,\perp}(x)$ are adjusted such that the longitudinal and transverse pressures $\PL$ and $\Pperp$ are everywhere fully accounted for by the leading-order distribution $f_a$, with zero residual contributions from $\tdf$. This is an application of the anisotropic matching scheme proposed by Tinti in \cite{Tinti:2015xwa} and generalizes the $P_L$--matching scheme used in Chapter~\ref{chap2label} \cite{Molnar:2016gwq,Martinez:2017ibh} to both $\PL$ and $\Pperp$ (or, equivalently, to the pressure anisotropy $\PL-\Pperp\sim\pi^{zz}_\mathrm{LRF}$ and bulk viscous pressure $\Pi$). In this matching scheme, the $\dft$ correction generates only the residual dissipative flows described by $\Wperp$ and $\piperp$, which break the cylindrical symmetry of the distribution function in the LRF and account for the remaining four smaller components of the shear stress tensor $\pi^\munu$ in Eq.~(\ref{eqch3:7}).

With the matching conditions (\ref{eqch3:20}) and (\ref{eqch3:22}a,b) we have the following kinetic theory expressions for the energy density and the longitudinal and transverse pressures:\footnote{%
   	The superscript $(k)$ on macroscopic quantities indicates their kinetic theory definition for a gas 
	of weakly-interacting quasiparticles. The purpose of this notation will become clear later when we introduce 
	a more realistic equation of state.
	}
\bs
\allowdisplaybreaks
\label{eqch3:23}
\beal
  \label{eqch3:23b} \ene^{(k)} &=\int_p (\up)^2 f_a = \I_{2000}, \\
  \label{eqch3:23c} \PL^{(k)} &= \int_p \mzp^2 f_a = \I_{2200}, \\
  \label{eqch3:23d} \Pperp^{(k)} &= \frac{1}{2} \int_p \pxp f_a = \I_{2010}.
\end{align}
\es
The anisotropic integrals $\I_{nrqs}$ over the leading-order distribution function $f_a$ that appear in these equations are defined in Eq.~(\ref{eqapp3:A1}).

\subsection{Relaxation equations I}
\label{ch3sec3b}

The relaxation equations for the dissipative flows are obtained by expressing the latter as moments of the distribution function and using the Boltzmann equation to describe its evolution, using the decomposition (\ref{eqch3:14}) and treating $\dft$ as a small perturbation. We start from   
\bs
\allowdisplaybreaks
\label{eqch3:24}
\beal
\dot{\mathcal{P}}^{(k)}_L &= D \int_p \mzp^2 f_a \,, \\
\dot{\mathcal{P}}^{(k)}_\perp &= \frac{1}{2} D \int_p \pxp f_a \,, \\
\dot{W}^{\{\mu\}}_{\perp z} &= \Xi^\mu_\nu D \int_p \mzp \, p^{\{\nu\}} \dft \,, \\
\dot{\pi}^{\{\munu\}}_{\perp} &= \Xi^\munu_{\alpha\beta} D \int_p p^{\{\alpha} p^{\beta\}} \dft \,,
\end{align}
\es
where we defined the compact notations \cite{Molnar:2016vvu}
\begin{eqnarray}
\allowdisplaybreaks
\label{eqch3:25}
  && a^{\{\mu\}} \equiv \Xi^\munu a_\nu,\quad
        b^{\{\munu\}} \equiv \Xi^\munu_{\alpha\beta}b^{\alpha\beta},
  \nonumber\\
  && \dot{a}^{\{\mu\}} \equiv \Xi^\munu \dot{a}_\nu,\quad
  \dot{b}^{\{\munu\}} \equiv \Xi^\munu_{\alpha\beta}\dot{b}^{\alpha\beta}
\end{eqnarray}
for the spatially transverse (in the LRF) components of a vector $a^\mu$ or its LRF time derivative $\dot{a}^\mu$ and the spatially transverse and traceless part of a tensor $b^\munu$ or its LRF time derivative $\dot{b}^\munu$. After moving the time derivative $D$ on the r.h.s. under the integral until it hits the distribution function $f_a$ or $\dft$, we use the decomposition $f=f_a+\dft$ together with 
\be
\label{eqch3:27}
  \partial_\mu = u_\mu D + z_\mu D_z + \nabla_{\perp,\mu}
\ee
to rewrite the Boltzmann-Vlasov equation \eqref{eqch3:13} in the form
\be
\allowdisplaybreaks
\label{eqch3:28}
\begin{split}
  \dot{f}_a + \delta\dot{\tilde f} =\,& \frac{C[f]- m \, \partial^\mu m \, \partial_\mu^{(p)} f}{\up} +  \frac{\mzp D_z f_a - \pperp \nabla_{\perp,\mu} f_a}{\up} \\
                                                &+ \frac{\mzp D_z \dft - \pperp \nabla_{\perp,\mu} \dft}{\up} \,.
\end{split}
\ee
Closing this equation requires an approximation for $\dft$. We here use the 14--moment approximation. 
 
\subsection{The 14--moment approximation}
\label{ch3sec3c}

The 14--moment approximation derives its name from approximating $\dft$ in terms of its 14 momentum moments with $p^\mu$ and $p^\mu p^\nu$~\cite{CPA:CPA3160020403, Israel:1979wp}. In our case the choice of the Landau frame, together with the generalized matching conditions (\ref{eqch3:20}) and (\ref{eqch3:22}a,b) and the absence of currents related to conserved charges, eliminate ten of these moments, leaving only four independent moments to construct $\dft$. These need to be matched to the residual dissipative flows $\Wperp$ and $\piperp$, which each have two degrees of freedom. The 14--moment approximation for $\dft$ can thus be written as \cite{Molnar:2016vvu}
\be
\label{eqch3:29}
   \frac{\dft}{f_a\bar{f}_a} = c_{\perp}^{\{\mu\}}\mzp p_{\{\mu\}} 
                                                              + c_{\perp}^{\{\munu\}} p_{\{\mu} \, p_{\nu\}}\,,
\ee
where $\bar{f}_a = 1 - \Theta f_a$. The coefficients $c_{\perp}^{\{\mu\}}$ and $c_{\perp}^{\{\munu\}}$ are computed by substituting Eq.~\eqref{eqch3:29} into the kinetic theory definitions of $\Wperp$ and $\piperp$,
\bs
\allowdisplaybreaks
\label{eqch3:30}
\beal
\Wperp &= \int_p  \mzp p^{\{\mu\}} \dft \,, \\
\piperp &= \int_p p^{\{\mu}p^{\nu\}} \dft \,,
\end{align}
\es
and decoupling the resulting set of linear equations:
\be   
\label{eqch3:31}
\begin{aligned}
 c_{\perp}^{\{\mu\}} = - \frac{\Wperp}{\J_{4210}}\,,\qquad c_{\perp}^{\{\munu\}} = \frac{\piperp}{2 \J_{4020}}\,. 
\end{aligned}
\ee
The anisotropic integrals $\J_{nrqs}$ appearing in these expressions
are defined in Eq.~(\ref{eqapp3:A2}). As expected, the coefficients are directly proportional to the residual dissipative flows: 
\be
\label{eqch3:32}
  \dft = \left( -\mzp \frac{p_{\{\mu\}}\Wperp}{\J_{4210}} 
        + \frac{p_{\{\mu}\,p_{\nu\}}\piperp}{2\,\J_{4020}}\right) f_a \bar{f}_a \,.
\ee
\subsection{Relaxation equations II}
\label{ch3sec3d}

Substituting the 14--moment approximation (\ref{eqch3:32}) for $\dft$ into Eqs.~(\ref{eqch3:24}) and (\ref{eqch3:28}), simplifying some of the~resulting terms by integrating by parts, and enforcing the generalized matching conditions, some algebra yields the following dissipative relaxation equations:\footnote{%
	\label{footchain}In deriving these equations one encounters terms involving the co-moving time derivative 
	$\dot{m}$ of the temperature-dependent quasiparticle mass that arise from the second term 
	on the l.h.s. of the Boltzmann equation (\ref{eqch3:13}). We eliminate them by using the 
	chain rule $\dot{m} = (dm/d\ene)\,\dot{\ene}$ where we take $dm/d\ene$ as external input 
	from the quasiparticle model discussed below and use Eq.\,(\ref{eqch3:10}) for $\dot{\ene}$.
	Equations~(\ref{eqchap3:relax_2}a-d) are found after combining the terms on the r.h.s. of Eq.~(\ref{eqch3:10}) with
	other terms involving the same dissipative forces. In doing so we also neglect contributions of 
	$\mathcal{O}(\text{Kn}\,\tilde{\text{R}}^{-1}_i \tilde{\text{R}}^{-1}_j)$, where $\tilde{\text{R}}^{-1}_i$ 
	is the residual inverse Reynolds number associated with the residual components 
	$\Wperp$ and $\piperp$.
	\label{fn5}
	}
\bs
\allowdisplaybreaks
\label{eqchap3:relax_2}
\beal
    \dot{\mathcal{P}}^{(k)}_L =&\,\frac{\Peq^{(k)}{-}\bar{\mathcal{P}}^{(k)}}{\tau_\Pi} - \frac{\PL^{(k)}{-}\Pperp^{(k)}}{3\tau_\pi / 2} + \bar{\zeta}^{L(k)}_z \theta_L + \bar{\zeta}^{L(k)}_\perp \theta_\perp - 2\Wperp \dot{z}_\mu + \bar{\lambda}^{L(k)}_{Wu} \Wperp D_z u_\mu
\\\nonumber
    & + \bar{\lambda}^{L(k)}_{W\perp} \Wperp z_\nu \nabla_{\perp,\mu} u^\nu - \bar{\lambda}^{L(k)}_{\pi} \piperp \sigma_{\perp,\munu} \,,
\\\nonumber
\\
    \dot{\mathcal{P}}^{(k)}_\perp 
    =&\, \frac{\Peq^{(k)}{-}\bar{\mathcal{P}}^{(k)}}{\tau_\Pi} + \frac{\PL^{(k)}{-}\Pperp^{(k)}}{3\tau_\pi} + \bar{\zeta}^{\perp(k)}_z \theta_L + \bar{\zeta}^{\perp(k)}_\perp \theta_\perp + \Wperp \dot{z}_\mu + \bar{\lambda}^{\perp(k)}_{Wu} \Wperp D_z u_\mu
\\\nonumber
      & - \bar{\lambda}^{\perp(k)}_{W\perp} \Wperp z_\nu \nabla_{\perp,\mu} u^\nu + \bar{\lambda}^{\perp(k)}_{\pi} \piperp \sigma_{\perp,\munu}\,,
\\\nonumber
\\ 
    \dot{W}^{\{\mu\}}_{\perp z} 
    =&\, - \frac{\Wperp}{\tau_\pi} + 2\bar{\eta}^W_u \Xi^\munu D_z u_\nu - 2\bar{\eta}^W_\perp z_\nu \nabla_\perp^\mu u^\nu - \big(\bar{\tau}^W_z \Xi^\munu {+} \piperp\big) \dot{z}_\nu - \bar{\lambda}^W_{W u} \Wperp  \theta_L
\\\nonumber
    & + \bar{\delta}^W_W \Wperp \theta_\perp + \bar{\lambda}^W_{W \perp} \sigma_\perp^\munu  W_{\perp z, \nu} + \omega_\perp^\munu W_{\perp z, \nu} + \bar{\lambda}^W_{\pi u} \piperp D_z u_\nu
\\\nonumber
    & - \bar{\lambda}^W_{\pi \perp} \piperp z_\lambda \nabla_{\perp,\nu} u^\lambda\,,
\\\nonumber
\\
    \dot{\pi}^{\{\munu\}}_{\perp} 
    =&\, - \frac{\piperp}{\tau_\pi} + 2 \bar{\eta}_\perp \sigma_\perp^\munu - 2 W_{\perp z}^{\{\mu} \dot{z}^{\nu\}} + \bar{\lambda}^\pi_\pi \piperp \theta_L - \bar{\delta}^\pi_\pi \piperp \theta_\perp - \bar{\tau}^\pi_\pi \pi_\perp^{\lambda \{\mu} \sigma^{\nu\}}_{\perp,\lambda}
\\\nonumber
    & + 2 \pi_\perp^{\lambda \{\mu} \omega^{\nu\}}_{\perp,\lambda}  - \bar{\lambda}^\pi_{W u} W_{\perp z}^{\{\mu} D_z u^{\nu\}} + \bar{\lambda}^\pi_{W \perp} W_{\perp z}^{\{\mu} z_\lambda \nabla_\perp^{\nu\}} u^\lambda\,. 
\end{align}
\es
Here $\Pavg^{(k)} = \frac{1}{3}(\PL^{(k)}{+}2\Pperp^{(k)})$ is the average pressure as given by kinetic theory, and $\omega_\perp^\munu\equiv\Xi^\mu_\alpha \Xi^\nu_\beta\, \partial^{[\beta} u^{\alpha]}$ is the transverse vorticity tensor. 

The structure of Eqs.\,(\ref{eqchap3:relax_2}a-d) is simpler than that of the corresponding equations derived in Ref.~\cite{Molnar:2016vvu}, not only by the absence of terms coupling to the conserved charge and diffusion currents (which only reflects the simplifying assumptions made here), but also as a result of imposing the generalized Landau matching conditions (\ref{eqch3:20}) and (\ref{eqch3:22}a,b), which optimize the evolution of the anisotropy parameters in $f_a$ and thus remove additional terms needed in Ref.~\cite{Molnar:2016vvu} to correct their evolution if not chosen optimally in the first place. 

The transport coefficients appearing on the right hand sides of Eqs.~(\ref{eqchap3:relax_2}a-d) are labeled following as much as possible the convention established in Ref.~\cite{Molnar:2016vvu}. Except for the relaxation times they are given in Sec.~\ref{chap4S2.6.3}.\footnote{%
    Please note the superscripts $(k)$ on the transport coefficients appearing on the right hand sides of 
    Eqs.~(\ref{eqchap3:relax_2}a-b) (also listed in Appendix~\ref{appch3c}). They reflect the fact that these control the evolution of the kinetic
    part of the longitudinal and transverse pressures. For the quasiparticle model introduced in the next section
    an additional mean-field enters which modifies these pressures and transport coefficients. The modified
    expressions will be denoted without the superscript in Eqs.~\eqref{eqch3:54} --~\eqref{eqch3:55} and in Chapter~\ref{chapter4label}.
    }
Generically they involve the ``anisotropic thermodynamic integrals'' over the anisotropic distribution function $f_a$ given in Appendix~\ref{appch3a}. Their validity, as well as specific relations between some of the transport coefficients listed in Chapter~\ref{chapter4label}, depends on the applicability of relativistic kinetic theory of a gas of weakly-interacting quasiparticles as the underlying microscopic theory, which is not guaranteed for quark-gluon plasma. Their generalization to a more realistic microscopic theory of QCD medium dynamics requires much additional work. For now, we will use the expressions given in Chapter~\ref{chapter4label} as order-of-magnitude estimates and placeholders for future sets of transport coefficients.

Equations~(\ref{eqchap3:relax_2}a-d) also involve two microscopic time scales: the bulk relaxation time $\tau_\Pi$ controls the relaxation of the bulk viscous pressure $\Pi$ whereas the shear relaxation time $\tau_\pi$ drives the relaxation of both the large shear stress component $\PL-\Pperp$ and the smaller ones described by $\Wperp$ and $\piperp$. That all shear stress components have the same relaxation time even if some of them become large is a model assumption that may be corrected in future improved calculations of the transport coefficients for strongly anisotropically expanding quark-gluon plasma. 

Formally, the relaxation times arise from a linearization of the collision term around the local equilibrium distribution $f_\eq$ (with temperature computed from the energy density):
\be
\label{eqch3:37}
  C[f]= \frac{p\cdot u(x)}{\tau_r(x)}\left(f_\eq(x,p)-f(x,p)\right).
\ee
Literal use of the relaxation time approximation (RTA) \cite{Anderson_Witting_1974} gives $\tau_\pi = \tau_\Pi = \tau_r$. However, strong coupling in the quark-gluon plasma in the temperature regime just above  the quark-hadron phase transition, as well as critical behavior near that phase transition, lead to very different temperature dependences of the bulk and shear viscosities and their associated relaxation times in QCD, especially around $T_c$ \cite{Csernai:2006zz, Paech:2006st, Arnold:2006fz, Kharzeev:2007wb, Karsch:2007jc}. In particular, the bulk relaxation time $\tau_\Pi$ is expected to be affected by ``critical slowing down'' \cite{Arnold:2006fz, Berdnikov:1999ph, Song:2009rh}, i.e. it should exhibit a strong peak near $T_c$. Since large bulk viscous effects near $T_c$ are one of the main motivations for this thesis, we feel compelled to account for them by introducing two different relaxation times $\tau_\pi$ and $\tau_\Pi$, and tying them to phenomenologically parametrized shear and bulk viscosities $\eta$ and $\zeta$ by taking the standard kinetic theory relations \cite{Denicol:2012cn}
\be
\label{eqch3:38}
   \tau_\pi = \eta / \beta_\pi \,, \qquad
   \tau_\Pi = \zeta / \beta_\Pi.
\ee
The (temperature dependent) isotropic thermodynamic integrals $\beta_\pi$ and $\beta_\Pi$ appearing in these relations are given further below in Eq.~(\ref{eqch3:beta}). The shear and bulk viscosities are first-order transport coefficients that appear in standard viscous hydroynamics -- they are also present here through the relaxation times $\tau_\pi$ and $\tau_\Pi$. When comparing anisotropic with viscous hydrodynamics further below in Sec.~\ref{ch3sec5} we will do so by using the same viscosities in both approaches (the coefficients $\beta_\pi$ and $\beta_\Pi$, however, can vary). 

\section{Anisotropic equation of state}
\label{ch3sec4}

While the relaxation equations~(\ref{eqchap3:relax_2}a-d) were derived from the Boltzmann equation, the equations remain structurally unchanged for strongly coupled fluids. They are purely macroscopic, i.e. all terms on the r.h.s.  have the form of some macroscopic driving force (proportional to the Knudsen or inverse Reynolds numbers or products thereof) multiplied by some transport coefficient. The kinetic origin of these equations is hidden in these transport coefficients. Applying the equations to strongly coupled fluids requires only that these transport coefficients, along with the equation of state relating the energy density and equilibrium pressure, are swapped out accordingly.

For the time being most of the transport coefficients of hot and dense QCD matter are still essentially unknown. While the shear and bulk viscosities will be taken as parameters whose functional forms are modeled phenomenologically and fitted to experimental observables, the remaining transport coefficients will be approximated using kinetic theory, for reasons of consistency with our derivation of the relaxation equations. Their evaluation requires microscopic kinetic inputs, namely the parameters $(\Lambda, \alpha_\perp, \alpha_L)$ and quasiparticle mass $m$ characterizing the anisotropic distribution function $f_a$. However, for the quark-gluon plasma's equation of state (EOS), which is very precisely known from lattice QCD calculations \cite{Borsanyi:2010cj, Bazavov:2014pvz}, we want to use first-principles theoretical input. 

In this section we discuss how to consistently incorporate such direct information from QCD into a hydrodynamic framework that was originally derived from kinetic theory. We introduce a parametric model for an anisotropic equation of state that allows the anisotropic hydrodynamic equations, including the dissipative relaxation equations for the longitudinal pressure, transverse pressure and residual shear stresses, to be solved on a purely macroscopic level. This differs from earlier implementations of the framework which relied on the solution of evolution equations for the microscopic kinetic parameters $(\Lambda, \alpha_\perp, \alpha_L)$ \cite{Martinez:2010sc,Martinez:2012tu,Molnar:2016gwq, Alqahtani:2016rth, Alqahtani:2017jwl, Alqahtani:2017tnq} (which, for the case of QCD, are not really well-defined). However, since we will need these microscopic parameters for the calculation of those transport coefficients computed from kinetic theory (a temporary necessity that will disappear as soon as we find ways of calculating these transport coefficients directly from QCD), we determine them from the macroscopic hydrodynamic quantities, using our parametric model for the anisotropic EOS.\footnote{%
	Note that the parametric model is not used for the QCD equilibrium pressure $\Peq(\ene)$ but only to parametrize the dissipative deviations of the
	longitudinal and transverse pressures from $\Peq(\ene)$, as well as for the calculation of the remaining 
	transport coefficients.
	}   

To construct this parametric model we follow Refs.~\cite{Alqahtani:2015qja, Alqahtani:2016rth, Tinti:2016bav} and parametrize the response of the pressure anisotropy and the bulk viscous pressure to anisotropic expansion using a quasiparticle EOS. The quasiparticles have a temperature-dependent mass that is chosen such that a weakly-interacting gas of these particles accurately mimics the QCD EOS. The transport coefficients are then worked out in this kinetic theory.\footnote{%
	Note that an accurate description of the QCD equation of state does not imply by any means that our
	kinetic theory model also predicts the correct transport properties of the quark-gluon plasma. 
	}
It is well known \cite{Biro:1990vj, Gorenstein:1995vm, Peshier:1995ty} that for thermodynamic consistency such an approach requires the introduction of a mean-field $B$ whose temperature dependence in equilibrium determines the temperature dependence of the quasiparticle's effective mass $m(T)$. The mean-field also receives additional dissipative corrections out of equilibrium \cite{Tinti:2016bav}.

\subsection{Integrating the lattice QCD EOS with a quasiparticle EOS}
\label{ch3sec4a}

The key question that needs to be addressed in anisotropic hydrodynamics is how much pressure anisotropy and bulk viscous pressure is generated by a given hydrodynamic expansion rate and its anisotropy. These are the two  largest and most important dissipative effects in our approach. The answer to this question depends on the microscopic properties of the medium. For QCD matter this response is presently not known. It is, however, a key ingredient in the hydrodynamic evolution model. In this subsection we model this response by that of a weakly interacting gas of quasiparticles with a medium-dependent mass $m(T)$. Within this model we can associate (within certain limits) any given deviations of the longitudinal and transverse pressures $\PL$ and $\Pperp$ from the equilibrium pressure $\Peq(\ene)$ with specific values for the microscopic parameters $\bigl(\Lambda, \bm{\alpha}, m\bigl(T)\bigr)\bigr)$ describing the anisotropic quasiparticle distribution function $f_a$. These values can then be used to compute the kinetic theory values for the transport coefficients. So while the equilibrium pressure is described by the lattice QCD EOS, the dissipative deviations of $\PL$ and $\Pperp$ from the equilibrium pressure are interpreted microscopically within a weakly interacting gas of massive Boltzmann particles. As we solve the hydrodynamic equations (\ref{eqch3:10}) -- (\ref{eqch3:12}) together with the dissipative relaxation equations (\ref{eqchap3:relax_2}a-d), we interpret the resulting deviations from local equilibrium within the quasiparticle model by writing
\bs
\allowdisplaybreaks
\label{eqch3:39}
\beal
  0 &= \ene^{(q)} - \ene_{\eq}^{(q)}\bigl(\ene\bigr), \\
  \PL - \Peq(\ene) &= \PL^{(q)} - \mathcal{P}_{\eq}^{(q)}\bigl(\ene\bigr), \\
  \Pperp - \Peq(\ene) &= \Pperp^{(q)} - \mathcal{P}_{\eq}^{(q)}\bigl(\ene\bigr).
\end{align}
\es
Here the superscript $(q)$ stands for ``quasiparticle model''. The zero on the l.h.s. of the first of these equations reflects the Landau matching condition $\ene=\ene(T)$ to the QCD energy density, which also provides us with the temperature $T$ at which the quasiparticle mass $m(T)$ and equilibrium mean-field $B_\eq(T)$ (see below) are evaluated. In the quasiparticle model the hydrodynamic quantities on the r.h.s. of~\eqref{eqch3:39} consist of kinetic and mean-field contributions  \cite{Alqahtani:2015qja}
\bs
\allowdisplaybreaks
\label{eqch3:40}
\beal
  \ene^{(q)} &= \ene^{(k)} + B,\\
  \PL^{(q)} &= \PL^{(k)} - B, \\
   \Pperp^{(q)} &= \Pperp^{(k)} - B. 
\end{align}
\es
The kinetic contributions are obtained from Eqs.~(\ref{eqch3:23}b,c,d):
\bs
\allowdisplaybreaks
\label{eqch3:41}
\beal
&\ene^{(k)}\bigl(\Lambda,\bm{\alpha}; m(T)\bigr)  = \I_{2000}\bigl(\Lambda,\bm{\alpha}; m(T)\bigr), \\
&\PL^{(k)}\bigl(\Lambda,\bm{\alpha}; m(T)\bigr)  = \I_{2200}\bigl(\Lambda,\bm{\alpha}; m(T)\bigr), \\
&\Pperp^{(k)}\bigl(\Lambda,\bm{\alpha}; m(T)\bigr)  = \I_{2010}\bigl(\Lambda,\bm{\alpha}; m(T)\bigr),
\end{align}
\es
where $T = T(\ene)$. The mean-field $B$ consists of an equilibrium part $B_\eq$ and a dissipative correction $\delta B$:
\be
\label{eqch3:42}
 B = B_\eq(T) + \delta B.
\ee
By Landau matching, the total quasiparticle energy density $\ene^{(q)}$ is fixed to its equilibrium value:
\be
\label{eqch3:43}
  \ene_\eq^{(q)}(\ene) = \ene^{(k)}_\eq(T) + B_{\eq}(T),
\ee
where $\ene^{(k)}_\eq(T) = \I_{2000}\bigl(T,\bm{1};m(T)\bigr)$. The Landau matching condition (\ref{eqch3:39}a) can then be rewritten as
\be
\label{eqch3:44}
  \I_{2000}\bigl(\Lambda,\bm{\alpha}; m(T)\bigr) = \I_{2000}\bigl(T,\bm{1};m(T)\bigr) - \delta B.
\ee
This establishes a relation between the temperature and the kinetic theory parameters, provided that $\delta B$ is determined. In the equilibrium limit, the quasiparticle pressure is
\be
\label{eqch3:45}
   \Peq^{(q)}(\ene) = \Peq^{(k)}(T) - B_\eq(T)\,,
\ee
where $\Peq^{(k)}(T) = \I_{2200}\bigl(T,\bm{1};m(T)\bigr)$. The equilibrium terms in (\ref{eqch3:43}) and (\ref{eqch3:45}) are all functions of temperature. 

For simplicity we assume that the quasiparticles have Boltzmann statistics $(\Theta = 0)$. To ensure that at asymptotically high temperatures the equilibrium pressure and energy density of this Boltzmann gas approach the corresponding values of a quark-gluon gas with $2(N_c^2{-}1)$ bosonic and $4N_cN_f$ fermionic degrees of freedom, we normalize them by applying to the quasiparticle distribution function a degeneracy factor
\be
\label{eqch3:46}
  g = \frac{\pi^4}{90}\Big[2(N_c^2{-}1) + \frac{7}{2} N_c N_f \Big] \,,
\ee
with $N_c = 3$ colors and  $N_f = 3$ flavors, counting $u$, $d$, and $s$ quarks only (heavier flavors are exponentially suppressed in the phenomenologically interesting temperature range and are therefore neglected). This degeneracy factor is part of the momentum integration measure $\int_p$ in the definition (\ref{eqch3:26}).
\begin{figure}[!t]
\centering
\includegraphics[width=0.45\linewidth]{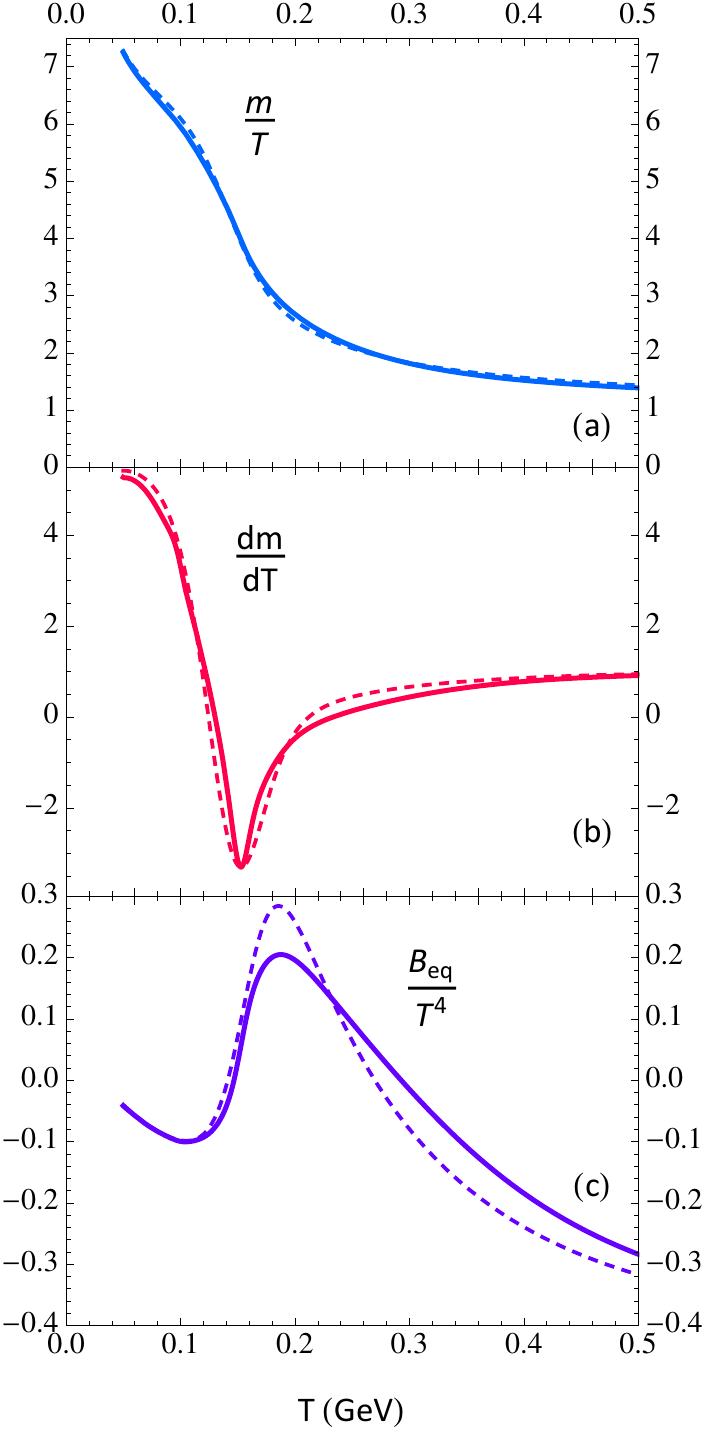}
\caption{The normalized quasiparticle mass $z=m/T$ (a), the derivative $dm/dT$ (b) and the equilibrium mean-field $B_\eq/T^4$ (c) obtained from Eqs.~(\ref{eqch3:47}) -- (\ref{eqch3:48}) using the lattice QCD EOS compiled by either the BEST Collaboration \cite{Parotto:2018pwk} (solid lines) or the Wuppertal-Budapest Collaboration~\cite{Borsanyi:2010cj, Alqahtani:2015qja} (dashed lines).
\label{F1chap3}}
\end{figure}
The thermal quasiparticle mass $m(T)$ is chosen such that the equilibrium pressure $\Peq^{(q)}(\ene)$ and energy density $\ene_\eq^{(q)}(\ene)$ of the quasiparticle model agree with their lattice QCD counterparts. Technically this is done by expressing the lattice QCD entropy density $\mathcal{S}$ in terms of the corresponding kinetic theory expression $\mathcal{S}^{(q)}$ for a gas of quasiparticles with mass $m(T)$ and Boltzmann statistics \cite{Alqahtani:2015qja}: 
\be
\label{eqch3:47}
   \mathcal{S} = \frac{\ene+\Peq}{T}
   = \frac{\ene_\eq^{(q)}+\Peq^{(q)}}{T}  = \frac{\ene_\eq^{(k)}+\Peq^{(k)}}{T}  =  \frac{g T^3 z^3}{2 \pi^2}  K_3(z),
\ee
where $K_n(z)$ is the modified Bessel function with $z = m(T)/T$. For thermodynamic consistency the r.h.s. must satisfy $\mathcal{S}^{(q)} = d\Peq^{(q)}/dT$, which is ensured by setting 
\be
\label{eqch3:48}
   B_\eq(T) = - \frac{g}{2\pi^2}\int^T_0 dT^{\,\prime}\, (T^{\,\prime})^3 z^2(T^{\,\prime}) K_1(z(T^{\,\prime}))\, \frac{dm(T^\prime)}{dT^\prime}\,.
\ee
Plots of the quasiparticle mass-to-temperature ratio $z$ and equilibrium mean-field $B_\eq$ as functions of $T$, using a 2010 lattice QCD EOS obtained by the Wuppertal-Budapest Collaboration \cite{Borsanyi:2010cj}, can be found in Ref.~\cite{Alqahtani:2015qja}. In this chapter we use the QCD EOS compiled by the Beam Energy Scan Theory (BEST) Collaboration \cite{Parotto:2018pwk}. The resulting slightly modified temperature dependences of $z(T)$, $dm(T)/dT$ and $B_\eq(T)$ are shown in Figure~\ref{F1chap3} as solid lines (together with the earlier results from Ref.~\cite{Alqahtani:2015qja} shown as dashed lines).

Equation~(\ref{eqch3:48}) determines the mean-field in equilibrium. Out of equilibrium it receives a non-equilibrium correction $\delta B$ \cite{Tinti:2016bav}.  As shown in Ref.~\cite{Tinti:2016bav}, thermodynamic consistency and energy-momentum conservation can be used to derive from the Boltzmann equation the following general evolution equation for $B$: 
\be
\label{eqch3:49}
  \dot{B} + m \dot m \int_p f + \int_p (\up)\, C[f] = 0. 
\ee
By Landau matching, the non-equilibrium correction to the quasiparticle energy density $\ene^{(q)} = \ene^{(k)}+B$ must vanish, hence
\be
\label{eqch3:50}
  B = B_\eq - \int_p (\up)^2 \delta f \,,
\ee
where $\delta f=f-f_\eq$. Substituting $f = f_a+\dft$ and using the relaxation time approximation
\be
\label{eqch3:51}
   C[f] \approx - \frac{(\up)\,\delta f}{\tau_r}\,,
\ee
Eq.~\eqref{eqch3:49} takes the form
\be
\label{eqch3:52}
  \dot{B} = \frac{B_\eq{-}B}{\tau_\Pi} - \frac{\dot{m}}{m} \Big(\ene^{(k)}{-}2\Pperp^{(k)}{-}\PL^{(k)} \Big).
\ee
Note that the expression in the parentheses is the trace of the kinetic contribution to the energy-momentum tensor $T^\munu$. Since the non-equilibrium component of the mean-field $\delta B = B - B_\eq$ contributes to the bulk viscous pressure, we have replaced in Eq.~(\ref{eqch3:52}) the relaxation time $\tau_r$ by the bulk relaxation time $\tau_\Pi$. The time derivative of the thermal mass can be expressed in terms of the energy conservation law~\eqref{eqch3:10} using the chain rule (see footnote~\ref{footchain}). 

Although Eq.~(\ref{eqch3:50}) shows that the mean-field $B$ is not an independent quantity, we find it most straightforward to use  Eq.~\eqref{eqch3:52} to evolve it dynamically. It does not directly enter the evolution equations for the components of the energy-momentum tensor as an independent variable, but is only needed for the model interpretation of the pressure anisotropy and bulk viscous pressure (which are hydrodynamic outputs) in terms of the microscopic parameters $(\Lambda,\bm{\alpha},m)$ needed for computing the kinetic transport coefficients. We use Eqs.~(\ref{eqch3:10}), (\ref{eqchap3:relax_2}a), (\ref{eqchap3:relax_2}b) and (\ref{eqch3:52}) to evolve $\ene$, $\PL^{(k)}$, $\Pperp^{(k)}$ and $B$. The physical pressures are obtained from $\PL=\PL^{(k)}-B$ and $\Pperp=\Pperp^{(k)}-B$. From $\ene$ we determine $T$ using the lattice EOS, and thus we know $m(T)$. Then we rewrite our anisotropic equation of state model (\ref{eqch3:39}) as%
\bs
\allowdisplaybreaks
\label{eqch3:53}
\beal
  \ene - B &= 
  \I_{2000}\bigl(\Lambda, \bm{\alpha}, m(T)\bigr) , \\
  \PL + B &= 
  \I_{2200}\bigl(\Lambda, \bm{\alpha}, m(T)\bigr), \\
  \Pperp + B &= 
  \I_{2010}\bigl(\Lambda, \bm{\alpha}, m(T)\bigr),
\end{align}
\es
solve these equations for the anisotropy parameters $(\Lambda, \alpha_\perp,\alpha_L)$, and compute the transport coefficients. The numerical method which we use to solve the nonlinear equations~\eqref{eqch3:53} will be presented in the next chapter. 

Of course, the values $\bigl(\Lambda, \bm{\alpha}, m\bigr)$ associated in this way with $\PL$, $\Pperp$ and $\ene$ are model dependent. A different parametrization of the lattice QCD EOS in terms of quasiparticles (for example, as a mixture of different types of quasiparticles with different quantum statistical properties, degeneracy factors and masses) would yield different results. 

With the anisotropic EOS model (\ref{eqch3:53}) we can finally write down the equations of motion for the total pressures $\PL$ and $\Pperp$, by combining Eqs.~(\ref{eqchap3:relax_2}a-b) with Eq.~(\ref{eqch3:52}):
\begin{eqnarray} 
\allowdisplaybreaks
\label{eqch3:54}
   \dot{\mathcal{P}}_L &=& \frac{\Peq{-}\bar{\mathcal{P}}}{\tau_\Pi}
   - \frac{\PL{-}\Pperp}{3\tau_\pi / 2} + \bar{\zeta}^L_z \theta_L + \bar{\zeta}^L_\perp \theta_\perp
   - 2\Wperp \dot{z}_\mu + \bar{\lambda}^L_{Wu} \Wperp D_z u_\mu
   \\\nonumber
  &&+\, \bar{\lambda}^L_{W\perp} \Wperp z_\nu \nabla_{\perp,\mu} u^\nu  
   - \bar{\lambda}^L_{\pi} \piperp \sigma_{\perp,\munu} ,\qquad
\\
\label{eqch3:55}
  \dot{\mathcal{P}}_\perp &=& \frac{\Peq{-}\bar{\mathcal{P}}}{\tau_\Pi} 
  + \frac{\PL{-}\Pperp}{3\tau_\pi} + \bar{\zeta}^\perp_z \theta_L + \bar{\zeta}^\perp_\perp \theta_\perp 
  + \Wperp \dot{z}_\mu + \bar{\lambda}^\perp_{Wu} \Wperp D_z u_\mu 
  \\\nonumber
   &&\,- \bar{\lambda}^\perp_{W\perp} \Wperp z_\nu \nabla_{\perp,\mu} u^\nu 
  + \bar{\lambda}^\perp_{\pi} \piperp \sigma_{\perp,\munu}.
\end{eqnarray}
Here we redefined the transport coefficients for the longitudinal and transverse pressures as detailed in Chapter~\ref{chapter4label}. This completes our formalism for non-conformal anisotropic hydrodynamics, where the equations of motion~\eqref{eqch3:10} --~\eqref{eqch3:12}, (\ref{eqchap3:relax_2}c), (\ref{eqchap3:relax_2}d), (\ref{eqch3:54}) and (\ref{eqch3:55}) are purely macroscopic and structurally independent of the underlying microscopic physics, while the transport coefficients (other than $\etas$ and $\zetas$) are evaluated with our specific quasiparticle kinetic model for the anisotropic equation of state.

\subsection{Initializing the mean-field and anisotropic parameters}
\label{ch3sec4d}

In the anisotropic equation of state model (\ref{eqch3:53}) there is an ambiguity between the initialization of the mean-field $B$ and of the kinetic terms $\ene^{(k)}$, $\PL^{(k)}$, and $\Pperp^{(k)}$. Standard hydrodynamic initial conditions for the energy-momentum tensor only provide $\ene$, $\PL$, and $\Pperp$ on the initialization hypersurface. The initial energy density profile $\ene$ also yields the initial temperature profile and thus the initial profile for the equilibrium part $B_\eq(T)$ of the mean-field. Its co-moving time derivative can be obtained by taking the equilibrium limit of Eq.~(\ref{eqch3:52}): 
\be
\label{eqch3:72}
   \dot{B}_\eq = -\frac{\dot m}{m}\left( \ene_\eq^{(k)}-3\Peq^{(k)}\right). 
\ee
To obtain a guess for the initial non-equilibrium deviation $\delta B$ we assume that $\delta B$ evolves on a time scale larger than the bulk viscous relaxation time $\tau_\Pi$. We can then ignore the time derivative of $\delta B$ on the left hand side of Eq.~(\ref{eqch3:52}) and obtain from the difference between Eqs.~(\ref{eqch3:52}) and (\ref{eqch3:72}) the ``asymptotic'' initial condition
\be
\label{eqch3:73}
   \delta B^{\mathrm{(asy)}} = \frac{3 \tau_\Pi \dot m}{m - 4 \tau_\Pi \dot m}\Pi.
\ee
As before (see footnote \ref{fn5}) $m$ and $\dot{m}$ can be expressed in terms of the energy density $\ene$ and its co-moving time derivative. Having thus specified the initial profile for the mean-field $B = B_\eq + \delta B$ we can proceed to solve for the initial anisotropic parameters from Eq.~\eqref{eqch3:53} numerically and compute the initial values for the transport coefficients. 

For far-from-equilibrium initial conditions, such as those provided by the IP-Glasma model \cite{Schenke:2012wb} where $\PL$ starts out with a very large negative value $\PL=-\ene$ and, after classical Yang-Mills evolution for a time of the order of the inverse saturation momentum, settles to around zero \cite{Gelis:2013rba,Wang:2034451}, the implied deviation $\PL-\Peq$ can become so large that, with this initial choice of $B$, the quasiparticle model cannot accommodate it within the allowed ranges for $(\Lambda,\alpha_\perp,\alpha_L)$. Since typically $B < 0$ at high temperatures (see Fig.~\ref{F1chap3}), the kinetic longitudinal pressure, $\PL^{(k)} = \PL+B$, may in this situation be negative. Specifically, the anisotropic parameter initialization for temperatures $T \gtrsim 0.5$ GeV is found to fail when $\PL/\Pperp \lesssim 0.08$. To overcome this problem, in the case of such extreme initial conditions for $\PL$ one can simply adjust the initial guess for $\delta B$ and increase the initial value for $B$ by hand until a solution for $(\Lambda,\alpha_\perp,\alpha_L)$ can be found. This is feasible for (0+1)--dimensional Bjorken flow (see next subsection) but much more difficult to do in (3+1)--dimensional anisotropic hydrodynamics. An alternative way of initializing the mean-field and anisotropic parameters for more realistic simulations of heavy-ion collision will be discussed in the next chapter.

\section{Non-conformal Bjorken flow}
\label{ch3sec5}
In this section we test our anisotropic hydrodynamic formalism by comparing it to second-order viscous hydrodynamics for the case of (0+1)--dimensional Bjorken expansion in Milne spacetime $x^\mu = (\tau,x,y,\eta_s)$, using the lattice QCD equation of state referenced in Fig.~\ref{F1chap3}. We begin by simplifying the anisotropic evolution equations (\ref{eqch3:10}) -- (\ref{eqch3:12}), (\ref{eqchap3:relax_2}c-d), and (\ref{eqch3:54}) -- (\ref{eqch3:55}) for systems with Bjorken symmetry (labeled ``vahydro"). The fluid velocity is $u^\mu = (1,0,0,0)$, the longitudinal and transverse expansion rates are $\theta_L = 1/\tau$ and $\theta_\perp = 0$, and the residual shear stress components $\Wperp$ and $\piperp$ vanish by symmetry. As a result, the anisotropic hydrodynamic equations simplify to 
\bs
\allowdisplaybreaks
\label{eqch3:ahydroeqs}
\beal
   &\dot\ene =  - \frac{\ene+\PL}{\tau},
   \\
   &\dot{\mathcal{P}}_L =  \frac{\Peq{-}\bar{\mathcal{P}}}{\tau_\Pi} 
      - \frac{\PL{-}\Pperp}{3\tau_\pi / 2} + \frac{\bar{\zeta}^L_z}{\tau},
   \\
   &\dot{\mathcal{P}}_\perp =  \frac{\Peq{-}\bar{\mathcal{P}}}{\tau_\Pi}
     + \frac{\PL{-}\Pperp}{3\tau_\pi} + \frac{\bar{\zeta}^\perp_z}{\tau},
   \\
   &\dot{B} = -\frac{B_\eq{-}B}{\tau_\Pi} 
   + \frac{\ene{+}\PL}{\tau m}\frac{dm}{d\ene}\big(\ene{-}2\Pperp{-}\PL{-}4B \big),
\end{align}
\es
where in Eq.~(\ref{eqch3:ahydroeqs}d) we used
\be
  \label{eqch3:75}
   \frac{\dot{m}}{m} = -\frac{\ene{+}\PL}{\tau m}\frac{dm}{d\ene}\,,
\ee
as well as $\ene^{(k)}-2\Pperp^{(k)}-\PL^{(k)} = \ene-2\Pperp-\PL-4B$. 

In this chapter, we use a temperature dependent parametrization for the specific shear viscosity $\eta/\mathcal{S}$ \cite{Bernhard:2016tnd}:
\be
\label{eqch3:78}
\eta/\mathcal{S} = \left\{
\begin{array}{ll}
      (\eta/\mathcal{S})_{\text{min}} + (\eta/\mathcal{S})_{\text{slope}}(T{-}T_c) & \text{for} \ T > T_c\,,
  \\
      (\eta/\mathcal{S})_{\text{min}} & \text{for} \ T\leq T_c \,,
\end{array} \right.
\ee
where $\mathcal S = \mathcal S(\ene)$ is the lattice QCD entropy density and $T_c = 154$\,MeV is the pseudo-critical temperature. The model parameters $(\eta/\mathcal S)_{\text{min}} = 0.08$ and $(\eta/\mathcal S)_{\text{slope}} = 0.85$\,GeV$^{-1}$ were extracted from an earlier global Bayesian analysis of LHC heavy-ion collision data \cite{Bernhard:2016tnd}.
Similarly, we use for the specific bulk viscosity $\zeta/\mathcal S$ the parameterization from Ref.~\cite{Denicol:2009am}: 
\be
\label{eqch3:79}
   \zeta/\mathcal{S} = (\zeta/\mathcal{S})_{\text{norm}} \,  w(T/T_p)\,,
\ee
where the function $w(x)$ is  given by
\be
\label{eqch3:80}
w(x) = \left\{
\begin{array}{ll}
      C_1 + \lambda_1 \exp\big[\frac{x-1}{\sigma_1}\big] + \lambda_2\exp\big[\frac{x-1}{\sigma_2}\big] 
      & (x{\,<\,}0.995), \\
      A_0 + A_1 x + A_2 x^2 &  (0.995{\,\leq\,}x{\,\leq\,}1.05), \\
      C_2 + \lambda_3 \exp\big[\frac{1-x}{\sigma_3}\big] + \lambda_4\exp\big[\frac{1-x}{\sigma_4}\big]  & (x{\,>\,}1.05),
\end{array} \right.
\ee
\begin{figure}[t]
\includegraphics[width=
\linewidth]{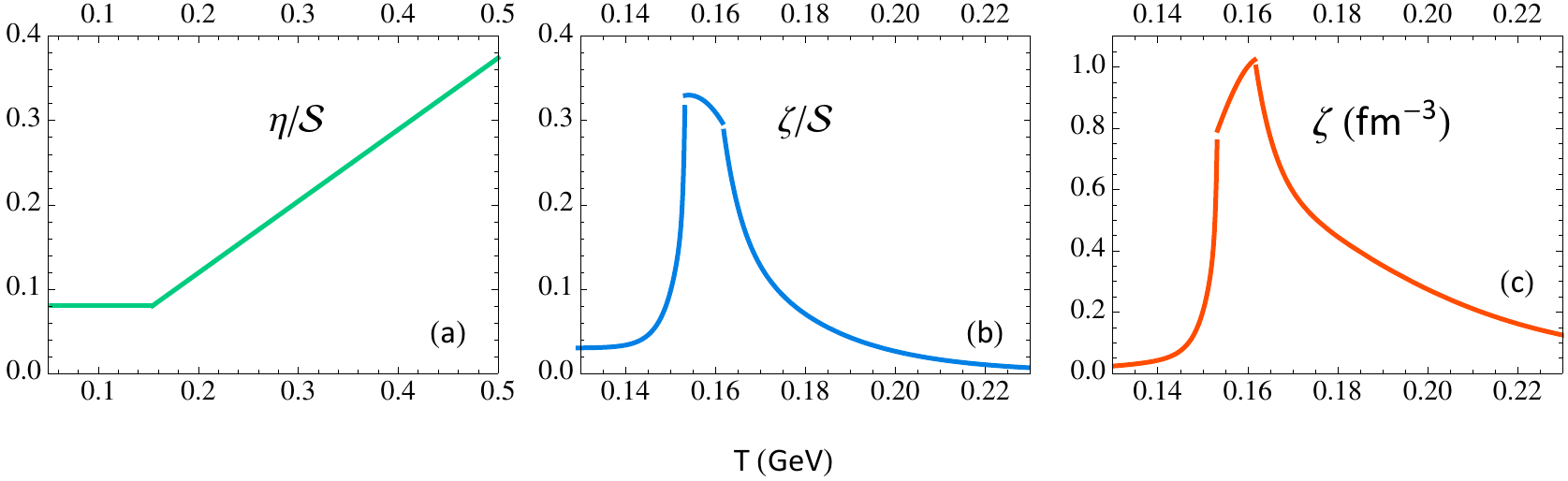}
\centering
\caption{The temperature dependence of (a) the specific shear viscosity $\eta / \mathcal{S}$, (b) the specific bulk viscosity $\zeta / \mathcal{S}$, and (c) the bulk viscosity $\zeta$ itself (fm$^{-3}$), given by the parameterizations (\ref{eqch3:78}) and (\ref{eqch3:79}) \cite{Bernhard:2016tnd}. We note that the peak of $\zeta / \mathcal{S}$ in (b) occurs at the temperature $T = 0.995 \, T_c \approx T_c$ whereas the bulk viscosity $\zeta = (\zeta / \mathcal{S}) \times \mathcal{S}$ in panel (c) peaks at the higher temperature $T = 1.05 \, T_c$. 
\label{FV}
}
\end{figure}
%
with $A_0 = -13.45$, $A_1 = 27.55$, $A_2 = -13.77$, $C_1 = 0.03$, $C_2 = 0.001$, $\lambda_1 = 0.9$, $\lambda_2 = 0.22$, $\lambda_3 = 0.9$, $\lambda_4 = 0.25$, $\sigma_1 = 0.0025$, $\sigma_2 = 0.022$, $\sigma_3 = 0.025$ and $\sigma_4 = 0.13$. For the normalization factor we choose $(\zeta / \mathcal S)_{\text{norm}} = 1.25$ \cite{Bernhard:2016tnd}, and we fix the location of the peak of the specific bulk viscosity by taking $T_p = T_c$. Figure~\ref{FV} shows the behavior of the specific shear and bulk viscosities as a function of temperature (a different parametrization is used in the next chapter). 

The relaxation times are then obtained from the kinetic theory relations (\ref{eqch3:38}), rewritten in the form
\be
\label{eqch3:81}
   \tau_\pi = \frac{\eta}{\mathcal{S}} \, \frac{\mathcal{S}}{\beta_\pi} , \qquad
   \tau_\Pi = \frac{\zeta}{\mathcal{S}} \, \frac{\mathcal{S}}{\beta_\Pi},
\ee
using the following quasiparticle versions of the $\beta$--coefficients \cite{Tinti:2016bav}:
\be
\label{eqch3:beta}
    \beta_\pi = \frac{\I_{32}}{T} \,, \qquad \qquad
    \beta_\Pi = \frac{5\beta_\pi}{3} - c_s^2(\ene{+}\Peq) 
       + c_s^2 m \frac{dm}{dT} \I_{11}\,,
\ee
where the thermodynamic integrals are defined in Eq.~\eqref{eqapp3:intI}. Here $c_s^2(\ene)$ is the squared speed of sound from lattice QCD. The system of ordinary differential equations \eqref{eqch3:ahydroeqs} is solved using Heun's method. After each intermediate and full time step the anisotropic parameters are updated by numerically solving Eq.~\eqref{eqch3:53}.

These anisotropic hydrodynamic results will be compared with those from second-order quasiparticle viscous hydrodynamics in the 14--moment approximation (labeled ``vhydro"). The corresponding evolution equations and transport coefficients are derived in Appendix~\ref{appch3e}. For Bjorken flow, the set of independent dynamical variables reduces to the energy density $\ene$, the shear stress $\pi = -\tau^2 \pi^{\eta\eta} = \frac{2}{3}(\Pperp{-}\PL)$ and the bulk viscous pressure $\Pi = \frac{1}{3}(2\Pperp{+}\PL) - \Peq$. Their evolution equations simplify to
\bs
\allowdisplaybreaks
\label{eqch3:vhydroeqs}
\beal
   & \dot\ene=  - \frac{\ene+\Peq + \Pi - \pi}{\tau} ,
   \\
   & \dot\pi = - \frac{\pi}{\tau_\pi} - \frac{4 \beta_\pi}{3\tau} - \frac{\left(\tau_{\pi\pi}
       {+}3\delta_{\pi\pi}\right)\pi - 2\lambda_{\pi\Pi}\Pi}{3\tau_\pi \tau},
   \\
   & \dot\Pi =  - \frac{\Pi}{\tau_\Pi} - \frac{\beta_\Pi}{\tau} {-} \frac{\delta_{\Pi\Pi}\Pi - \lambda_{\Pi\pi}\pi}{\tau_\Pi \tau}.
\end{align}
\es
For the non-equilibrium mean-field contribution $\delta B$ we use the second-order expression \cite{Tinti:2016bav}
\be
\label{eqch3:dB2nd}
  \delta B^{(2)} = - \frac{3 \tau_\Pi}{m} \frac{dm}{d\ene} (\ene{+}\Peq) \, \Pi \, \theta \,,
\ee
where $\theta = \partial_\mu u^\mu = 1/\tau$ is the scalar expansion rate. In Eq.~\eqref{eqch3:dB2nd}, we replaced the relaxation time $\tau_r$ by $\tau_\Pi$. The relaxation times $\tau_\pi$ and $\tau_\Pi$ are obtained from Eqs.~(\ref{eqch3:81}) --~\eqref{eqch3:beta} while the second-order transport coefficients $\tau_{\pi\pi}, \, \delta_{\pi\pi}, \, \lambda_{\pi\Pi}, \, \delta_{\Pi\Pi}$ and $\lambda_{\Pi\pi}$ are computed from the quasiparticle model in the 14--moment approximation (after expansion around a local-equilibrium distribution, see Appendix~\ref{appch3e}). We will also look at how the transport coefficients, including the relaxation times, affect the viscous hydrodynamic results when computed in the small fixed mass approximation $z \ll 1$, $dm/dT \approx 0$ and $B \approx 0$, which is commonly referred to as standard viscous hydrodynamics (labeled ``vhydro 2")~\cite{Denicol:2014vaa,Ryu:2015vwa,Ryu:2017qzn}.

\subsection{Equilibrium initial conditions}
\label{ch3sec5a}
\begin{figure}[t]
\centering
\includegraphics[width=\textwidth]{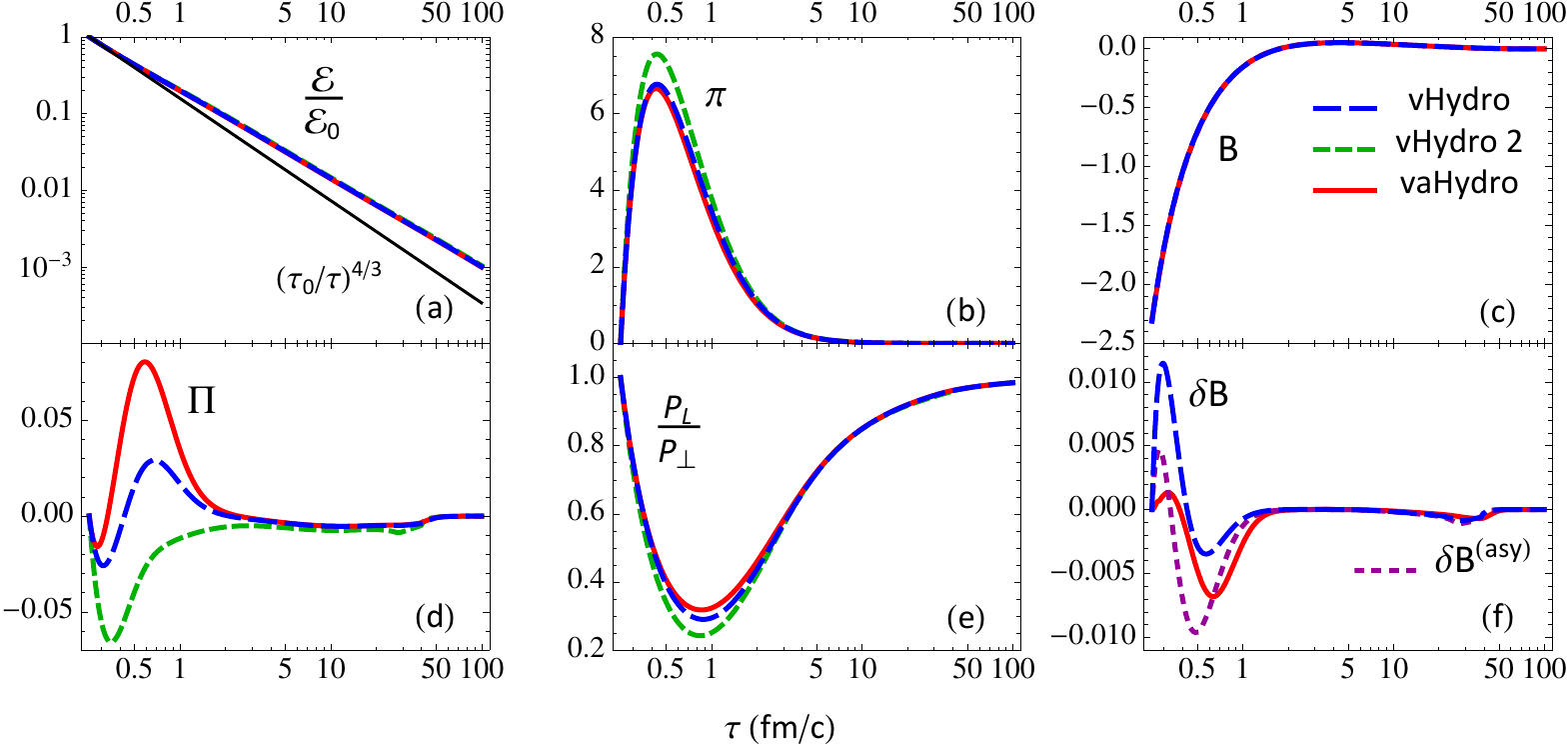}
\caption{The Bjorken evolution of the normalized energy density $\ene/\ene_0$, bulk viscous pressure $\Pi$, longitudinal shear stress component $\pi$, pressure anisotropy ratio $\PL/\Pperp$, total mean-field $B$, and non-equilibrium component of the mean-field $\delta B$, calculated in anisotropic hydrodynamics (solid red, vaHydro) as well as in second-order viscous hydrodynamics (dashed blue and green, vHydro and vHydro\,2). The solid red and dashed blue lines (vaHydro and vHydro) use transport coefficients derived from kinetic theory for medium-dependent quasiparticles while the dashed green lines (vHydro 2) use kinetic theory transport coefficients derived in the small fixed mass limit. The fluid starts out in thermal equilibrium at longitudinal proper time $\tau_0 = 0.25$\,fm/$c$ with initial temperature $T_0 = 0.5$\, GeV. In panel (f), the short-dashed purple line shows the ``asymptotic approximation'' \eqref{eqch3:73} for $\delta B$, computed using data from the anisotropic hydrodynamic evolution whereas the dashed blue line uses Eq.~(\ref{eqch3:dB2nd}) and data from the viscous hydrodynamic evolution. The macroscopic quantities $\Pi$, $\pi$, $B$, and $\delta B$ are plotted in units of GeV/fm$^3$.
\label{F2chap3}}
\end{figure}

\begin{figure}[t]
\centering
\includegraphics[width=\textwidth]{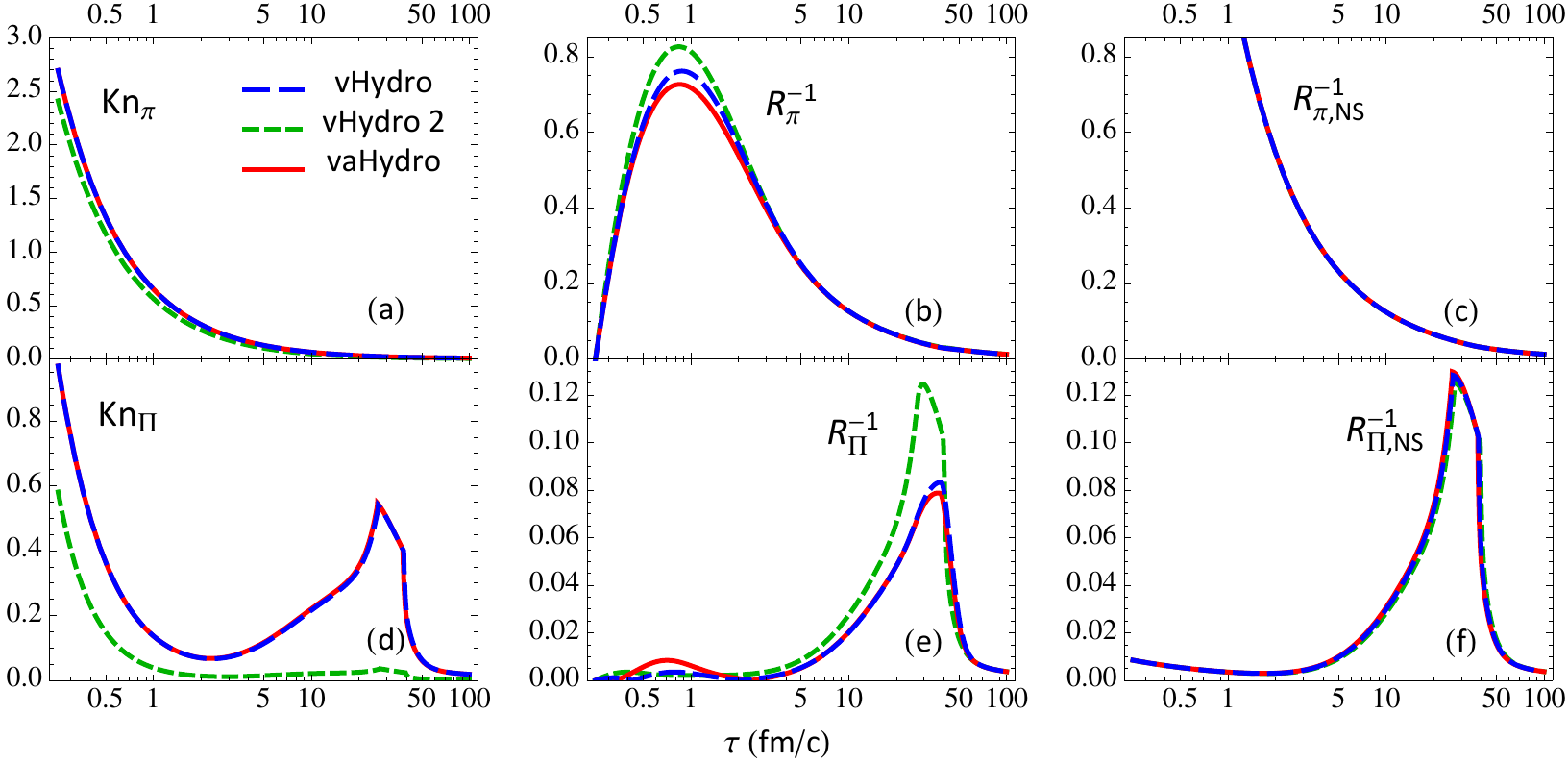}
\caption{The shear (top row) and bulk (bottom row) Knudsen and inverse Reynolds numbers numbers associated with Fig.~\ref{F2chap3}. For Bjorken flow, the formulas for the Knudsen and inverse Reynolds numbers reduce to $\text{Kn}_\pi = \tau_\pi \sqrt{\sigma_\munu \sigma^\munu} = \sqrt{2/3} \, \tau_\pi / \tau$, $\text{Kn}_\Pi = \tau_\Pi \theta =  \tau_\Pi / \tau$, $R^{-1}_\pi = \sqrt{\pi_\munu \pi^\munu} /\Peq =  \sqrt{3/2} \, \pi / \Peq$, and $R^{-1}_\Pi = |\Pi| / \Peq$. The last column (panels (c,f)) shows the Navier-Stokes limits of the shear and bulk inverse Reynolds numbers, $R^{-1}_{\pi,\mathrm{NS}} = \sqrt{8/3}\,\eta/(\tau\Peq)$ and $R^{-1}_{\Pi,\mathrm{NS}} = \zeta/(\tau\Peq)$.
\label{F3chap3}
}
\end{figure}

In Figure~\ref{F2chap3} we show the Bjorken evolution of the hydrodynamic variables in anisotropic hydrodynamics, including the total mean-field and its non-equilibrium component in the quasiparticle (QP) model used to compute the transport coefficients, and compare it with that in the second-order viscous hydrodynamic models. Figure~\ref{F3chap3} shows the same for the associated Knudsen and inverse Reynolds numbers. In this subsection we impose equilibrium initial conditions with initial temperature $T_0=0.5$\,GeV at longitudinal proper time $\tau_0=0.25$\,fm/$c$, i.e. all non-equilibrium effects are initially zero. Figure~\ref{F2chap3}a shows that all three models (anisotropic hydrodynamics with QP transport coefficients in solid red lines, second-order viscous hydrodynamics with QP transport coefficients in dashed blue lines and transport coefficients from a Boltzmann gas of small fixed masses in dashed green lines) produce almost identical evolutions for the energy density. The energy density decreases more slowly than for a conformal ideal fluid, indicated by the thin black line ${\sim\,}\tau^{-4/3}$. This is due to the smaller pressure of our EOS (which thus performs less longitudinal work) and to viscous heating. For reference we note that the system passes through the pseudo-critical temperature $T_c{\,=\,}154$\,MeV at $\tau_c{\,\sim\,}37$\,fm/$c$, with a small spread of less than 2\,fm/$c$ between the three models. 

Panel (c) shows that, if a QP model is used for the transport coefficients, the mean-field $B$ also evolves almost identically in anisotropic and viscous hydrodynamics. Small differences between anisotropic and viscous hydrodynamics with QP transport coefficients are observed in the evolution of the shear stress $\pi$ ($\order{(2\%)}$) and the pressure ratio $\PL/\Pperp$ ($\order{(10\%)}$): the effective resummation of shear viscous effects in anisotropic hydrodynamics leads to a slight reduction of the shear stress, resulting in a slightly reduced pressure anisotropy. Standard viscous hydrodynamics, with transport coefficients calculated in the small fixed mass expansion (dashed green lines), produces somewhat ($\order{(15\%)}$) larger shear stresses and stronger pressure anisotropies.

Given that the pressure anisotropy gets quite large, with $\PL/\Pperp$ decreasing to about 30\% at $\tau{\,\sim\,}1$\,fm/$c$, the excellent agreement between viscous and anisotropic hydrodynamics is somewhat unexpected. It suggests that the widely used standard viscous hydrodynamic approach is quite robust and quantitatively reliable even for large shear stresses. Similar observations were made before in Ref.~\cite{Bazow:phdthesis} as well as in studies of the Bjorken dynamics of strongly coupled theories where second-order viscous hydrodynamics could be directly compared with an exact numerical solution of the underlying strong-coupling dynamics \cite{Chesler:2015lsa, Chesler:2015bba}.   

The largest differences between anisotropic and viscous hydrodynamics are seen in the evolution of the bulk viscous pressure $\Pi$ (Fig.~\ref{F2chap3}d) and the non-equilibrium part of the mean-field $\delta B$ (Fig.~\ref{F2chap3}f). The two panels expose strong correlations between the evolutions of these two quantities. Both are small: (i) The bulk viscous pressure at early times is about 100 times smaller than the shear stress. While the evolution of $\Pi$ is qualitatively similar (although quantitatively different by more than a factor 2 at early times) for anisotropic and viscous hydrodynamics with QP transport coefficients, it exhibits {\em qualitatively different} dynamics in standard viscous hydrodynamics with transport coefficients computed from the small fixed mass expansion. (ii) Compared to the equilibrium mean-field, the non-equilibrium part $\delta B$ is about two orders of magnitude smaller (see panels (c) and (f) of Fig.~\ref{F2chap3}). Here one observes very different trajectories for $\delta B$ between the evolutions from anisotropic and quasiparticle viscous hydrodynamics, although their shapes are qualitatively similar. In addition, panel (f) shows for comparison the ``asymptotic approximation'' (\ref{eqch3:73}) (short-dashed purple curve) which should be compared to the exact numerical solution (red solid line). Obviously, the large expansion rate at early times makes the asymptotic trajectory, which is based on the assumption that $\delta B$ evolves more slowly than the bulk relaxation rate, a rather crude approximation.

Figure~\ref{F3chap3} shows the Knudsen and inverse Reynolds numbers associated with the shear and bulk viscous stresses. While the Knudsen and inverse Reynolds numbers associated with shear stress dominate the non-equilibrium dynamics at early times, those associated with bulk viscosity are the most relevant at late times when the system passes through the QCD phase transition.\footnote{%
	Note that the Knudsen numbers for vaHydro (solid red) and vHydro (dashed blue) are almost 
	identical due to the very similar energy density and temperature evolution, see Eq.~(\ref{eqch3:beta}) and 
	Fig.~\ref{F2chap3}a.
	}

In spite of the shear Knudsen number (Fig.~\ref{F3chap3}a) starting out large with a value of around 2.5, the shear inverse Reynolds number (Fig.~\ref{F3chap3}b) never exceeds a value of about 75\,--\,85\%. This results from the delay caused by the microscopic shear relaxation time which controls the approach of the shear stress $\pi$ from its zero starting point to its Navier-Stokes value and has at $\tau_0=0.25$ the value $\tau_\pi\approx 0.8$\,fm/$c$. By the time $R_\pi^{-1}$ reaches its Navier-Stokes limit (shown in Fig.~\ref{F3chap3}c), the shear Knudsen number has already dropped to values well below 1. We reiterate that at the peak value $\sim 3/4$ of the shear inverse Reynolds number the differences between anisotropic and standard viscous hydrodynamic evolution are less than 6\% as long as both are evaluated with transport coefficients computed from the same underlying kinetic theory.

Fig.\,\ref{F3chap3}e shows the evolution of the bulk inverse Reynolds number $R_\Pi^{-1}$, which peaks due to critical dynamics near the QCD phase transition temperature $T_c$; the corresponding Navier-Stokes value is shown in Fig.~\ref{F3chap3}f. Because $\tau_\Pi\propto\zeta$ (see Eq.~(\ref{eqch3:81})), the bulk relaxation rate slows down when the bulk viscosity peaks. This leads to ``critical slowing down'' of the evolution of the bulk viscous pressure $\Pi$, limiting its growth as the system cools down to $T_c$ \cite{Berdnikov:1999ph, Song:2009rh}. Comparing the solid red and dashed blue curves in Figs.~\ref{F3chap3}e,f we see that $\Pi$ and thus $R_\Pi^{-1}$ never reaches much more than about half of its peak Navier-Stokes value, and it also peaks later (around $\tau\sim 38$\,fm/$c \approx \tau_c$, corresponding to $T\approx0.995\,T_c$) than the Navier-Stokes limit which reaches its maximum already at $\tau{\,\sim\,}27$\,fm/$c$ (corresponding to $T\approx 1.05\,T_c$). One observes that even near its peak at $\tau \sim38$\,fm/$c$, $R_\Pi^{-1}$ evolves almost identically in anisotropic and quasiparticle viscous hydrodynamics. A marked difference is observed, however, when the system is evolved with standard hydrodynamics using transport coefficients from a massless Boltzmann gas without a mean-field (green dashed lines in Figs.~\ref{F3chap3}d,e). It turns out that the thermodynamic integral $\beta_\Pi$ in Eq.~(\ref{eqch3:beta}) is remarkably sensitive to the degree of nonconformality of the Boltzmann gas, giving rise to a much longer bulk viscous relaxation time in the QP model than for the light Boltzmann gas without a mean-field, especially in the neighborhood of $T_c$. This is reflected in the large difference between the dashed green line and the other two curves for the bulk Knudsen number shown in Fig.~\ref{F3chap3}d which causes the corresponding large difference in the evolution of the bulk inverse Reynolds number shown in panel (e): The much shorter relaxation time for the light Boltzmann gas allows the bulk viscous pressure to follow its Navier-Stokes limit (shown in Fig.~\ref{F3chap3}f) much more closely, causing $R_\Pi^{-1}$ to rise much more steeply and to a larger peak value as the system cools towards $T_c$ than in the other two approaches where $\beta_\Pi$ is calculated from the QP model.    

We have studied thermal equilibrium initial conditions with several other combinations of initial temperature $T_0$ and $\tau_0$, resulting in significantly different evolutions of the energy density and viscous pressure components (not shown here). Two features appear to be universal, however: (i) As long as we use transport coefficients computed from the same microscopic QP kinetic theory, the evolution of all components of the energy-momentum tensor, as well as of the mean-field $B$, shows only very small differences (of the same order as shown in Fig.~\ref{F2chap3}) between anisotropic and quasiparticle viscous hydrodynamics. (ii) Using instead transport coefficients for a Boltzmann gas of light fixed-mass particles, standard viscous hydrodynamics leads to significantly different evolutions for the bulk viscous pressure $\Pi$. For a meaningful comparison between anisotropic and second-order viscous hydrodynamics it is therefore important that a consistent set of transport coefficients is being employed. Also, for a medium with broken conformal invariance (such as the quark-gluon plasma and other forms of QCD matter) non-conformal effects on the transport coefficients can have a large effect on the evolution of the bulk viscous pressure which may not be properly captured when using transport coefficients derived from a theory with weakly interacting degrees of freedom that have small masses $m/T\ll1$. 
\subsection{Glasma-like initial conditions}
\label{ch3sec5bchap3}

\begin{figure}[t]
\includegraphics[width=\linewidth]{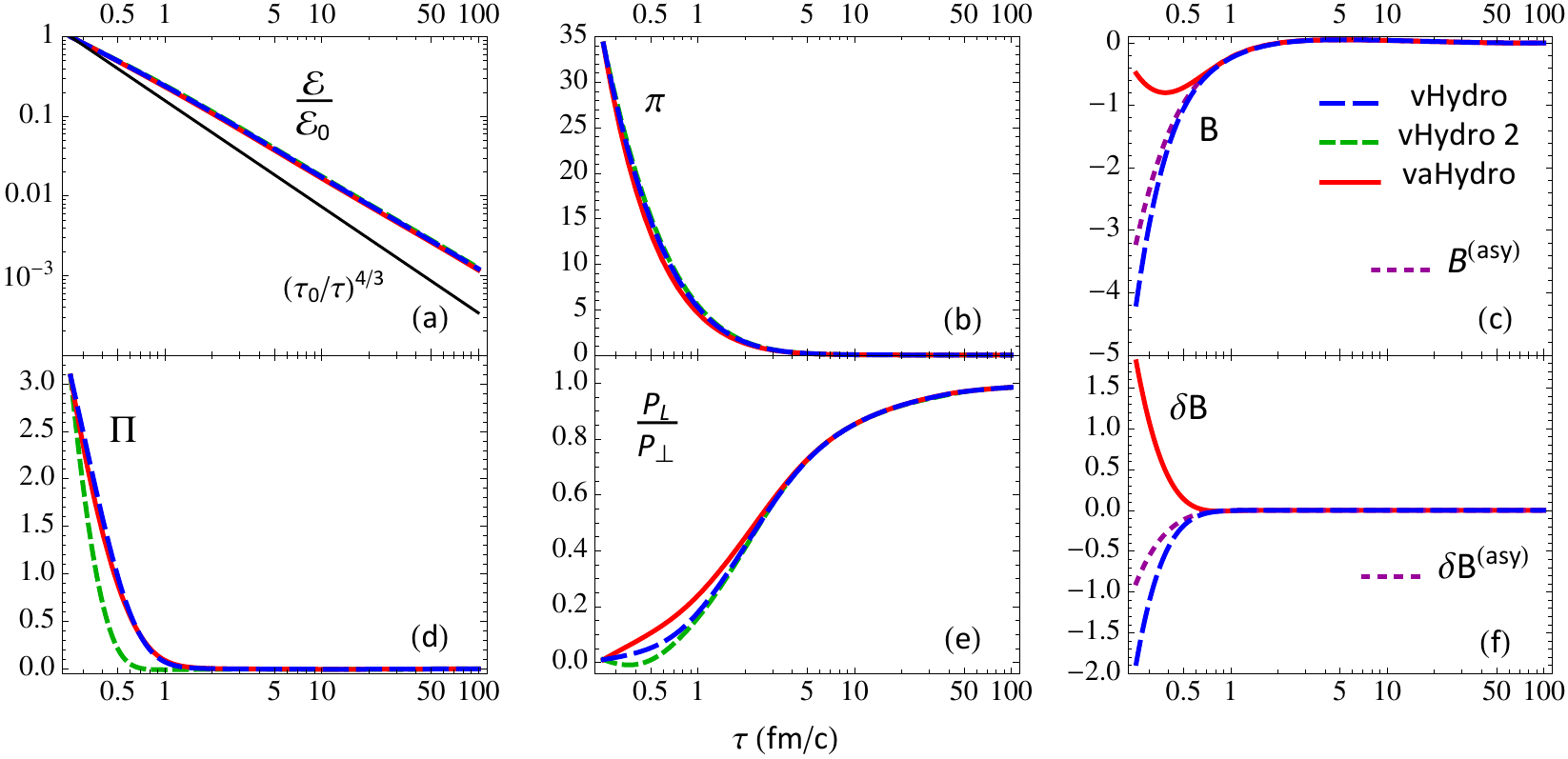}
\centering
\caption{Same as Fig.~\ref{F2chap3} but for Glasma-like initial conditions, with the same initial temperature at the 
same initial time. The initial pressures are set to $\mathcal{P}_{L0} = 4.975 \times 10^{-3} \, \ene_0$ 
and $\mathcal{P}_{\perp0} = 0.4975 \, \ene_0$. For the anisotropic hydrodynamic evolution (solid red
line) the magnitude of the initial mean-field $B_0$ is reduced to $15.3\,\%$ of the default value 
(purple short-dashed line) $B^{\mathrm{(asy)}} = B_\eq + \delta B^{\mathrm{(asy)}}$. For the quasiparticle 
viscous hydrodynamics (dashed blue lines), $B$
and $\delta B$ are determined as described in the text. $\Pi$, $\pi$, $B$, and $\delta B$ are plotted 
in units of GeV/fm$^3$. 
\label{F4chap3}
}
\end{figure}

\begin{figure}[t]
\centering
\includegraphics[width=\linewidth]{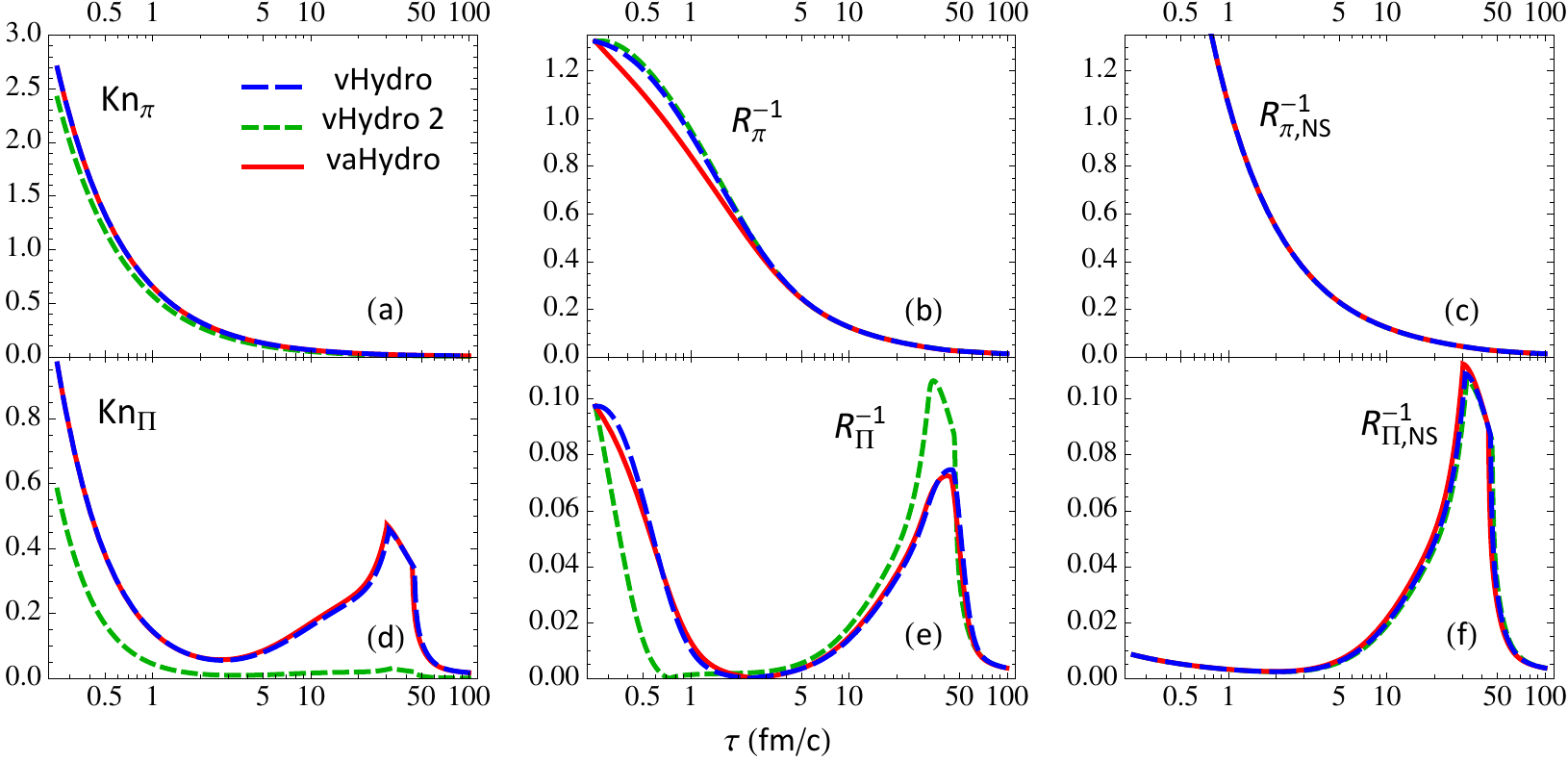}
\caption{The shear and bulk Knudsen numbers and inverse Reynolds numbers associated with Fig.~\ref{F4chap3}.
\label{F5chap3}
}
\end{figure}

In this subsection we repeat the exercise of the previous one for a different set of initial conditions, resembling those that one would get from matching the hydrodynamic evolution to a pre-equilibrium stage described by the IP-Glasma model \cite{Schenke:2012wb}. As already described, this model predicts approximately vanishing initial longitudinal pressure $\PL\approx 0$ and $\Pperp\approx\ene/2$ \cite{Gelis:2013rba}. (In practice, we set $\PL/\Pperp = 0.01$ initially). We use the same initial longitudinal proper time and temperature as before. The corresponding results are plotted in Figures~\ref{F4chap3} and \ref{F5chap3}.

%

For this extreme initial condition, the default magnitude of $B$ must be reduced by about $85\%$ in order to be able to successfully initialize the anisotropic microscopic parameters; for $B$ this is shown in Fig.~\ref{F4chap3}c while the implications for the anisotropic microscopic parameters will be discussed in the following subsection. The highly non-equilibrium initial conditions manifest themselves in large starting values for the shear and bulk stresses and the non-equilibrium mean-field. The initial shear stress (Fig.~\ref{F4chap3}b) is about five times larger than its peak value for equilibrium initial conditions. The bulk viscous pressure (Fig.~\ref{F4chap3}d) and non-equilibrium part of the mean-field (Fig.~\ref{F4chap3}f) are for the first fm/$c$ one to two orders of magnitude larger than for equilibrium initial conditions. In spite of this, anisotropic and viscous hydrodynamics still lead to almost identical evolution trajectories for the energy density (Fig.~\ref{F4chap3}a) and viscous pressures (Figs.~\ref{F4chap3}b,d) if QP transport coefficients are used, and if the latter are swapped out for those from a light Boltzmann gas, a significant change in the standard viscous hydrodynamic evolution is only seen for the bulk viscous pressure (dashed green curve in Fig.~\ref{F4chap3}d). The shear stress $\pi$ and the pressure ratio $\PL/\Pperp$ emphasize the differences in the hydrodynamic models somewhat at early times (Figs.~\ref{F4chap3}b,e), pushing the pressure ratio towards isotropy somewhat faster in anisotropic than in second-order viscous hydrodynamics, but all three models converge to a common late-time behavior for $\pi$ and $\PL/\Pperp$ after about 2\,fm/$c$ (i.e. after about 3 times the initial shear relaxation time of about 0.8\,fm/$c$). It is, however, not the case that equilibrium and Glasma-like initial conditions lead to the same temperature evolution of the system. A careful comparison of Figs.~\ref{F2chap3}a and \ref{F4chap3}a shows that for the non-equilibrium initial conditions viscous heating by the large initial bulk and shear stresses causes the energy density (and therefore temperature) to drop somewhat more slowly than for equilibrium initial conditions, especially at early times. 

Figure~\ref{F4chap3}f shows again that the ``asymptotic approximation'' (\ref{eqch3:73}) for $\delta B^{(\mathrm{asy})}$ (dashed purple line) is not a good approximation for the full numerical evolution of $\delta B$ shown by the solid red line. Since for the Glasma-like initial conditions the non-equilibrium mean-field contribution $\delta B$ is initially of the same order of magnitude as the equilibrium contribution $B_\eq$, the breakdown of this approximation is visible even in the evolution of the total mean-field $B$ (solid red line) which is not at all described by $B^{(\mathrm{asy})}\equiv B_\eq+\delta B^{(\mathrm{asy})}$. 

Looking at the Knudsen and inverse Reynolds numbers in Fig.~\ref{F5chap3} the only striking (although obvious) difference are the large starting values for both shear and bulk inverse Reynolds numbers when using Glasma-like initial conditions. Similar to the shear stress $\pi$ and pressure ratio $\PL/\Pperp$ in Figs.~\ref{F4chap3}b,e, these two observables exhibit noticeable differences at early times between anisotropic and viscous hydrodynamics.

\subsection{Evolution of the microscopic kinetic parameters}

Although the parameters $(\Lambda,\alpha_\perp,\alpha_L)$ describing the slope and anisotropy of the momentum distribution of the microscopic degrees of freedom are vestiges from an underlying kinetic theory whose traces we have tried to erase as much as possible in our formulation of anisotropic hydrodynamics (hoping that eventually we can obtain the transport coefficients of QCD matter from a more fundamental approach), it is interesting to ``look under the hood'' and see how our parametrized anisotropic EOS works, i.e. how the QP model adjusts its microscopic parameters to accommodate the macroscopic hydrodynamic initial conditions provided, and how it evolves them in response to the anisotropic hydrodynamic evolution of the energy-momentum tensor.

\begin{figure}[t]
\includegraphics[]{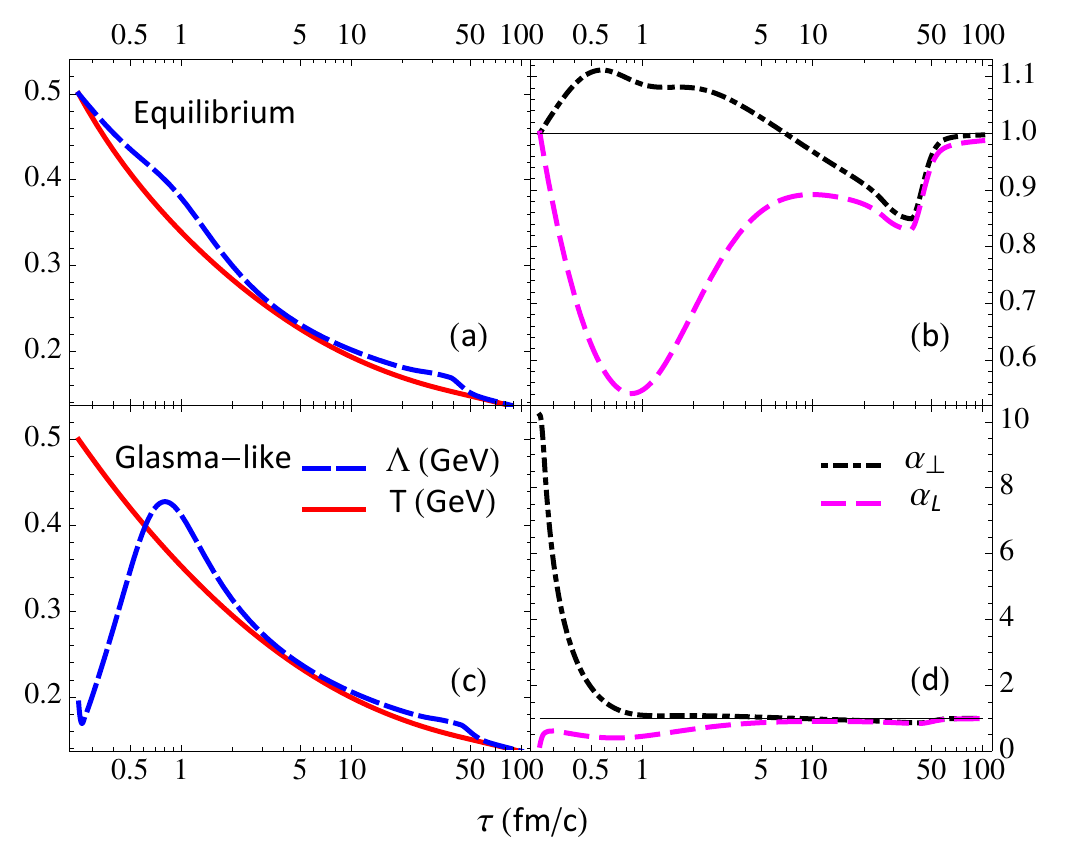}
\centering
\caption{The anisotropic hydrodynamic evolution of the effective temperature $\Lambda$ (dashed blue) compared to the temperature $T$ (solid red) is shown in the left panels for equilibrium (a) and Glasma-like initial conditions (c). The right panels show the associated evolution of the momentum anisotropy parameters $\alpha_L$ (dashed magenta) and $\alpha_\perp$ (dash-dotted black). The temperatures $\Lambda$ and $T$ are plotted in units of GeV. The deformation parameters $\alpha_L$ and $\alpha_\perp$ are unitless. 
\label{F6chap3}
}
\end{figure}

Figures~\ref{F6chap3}a,b compare, for equilibrium initial conditions, the evolution of the effective temperature parameter $\Lambda$ with that of the equilibrium temperature $T$ extracted from the energy density, and of the momentum anisotropy parameter $\alpha_L$ with that of $\alpha_\perp$, respectively. A comparison of Figs.~\ref{F6chap3}a,b with Figs.~\ref{F3chap3}b,e shows that large inverse Reynolds numbers in both the shear and bulk sectors correlate with effective temperatures $\Lambda{\,>\,}T$ and longitudinal momentum deformations $\alpha_L{\,<\,}1$. Large shear inverse Reynolds numbers correlate with $\alpha_\perp$ deviating from unity in the opposite direction (i.e. with $\alpha_\perp{\,>\,}1$), leading to narrower longitudinal and wider transverse momentum distributions than in the equilibrium distribution $f_\eq$, consistent with $\PL/\Pperp<1$. The presence of a large bulk viscosity peaking at $\tau \sim 40$ fm/$c$ push down both $\alpha_L$ and $\alpha_\perp$, corresponding to negative bulk viscous pressures. At very late times, when both the shear and bulk inverse Reynolds numbers approach zero, the momentum distribution approaches local equilibrium, $\alpha_{L,\perp}\to1$ and $\Lambda\to T$.

For Glasma-like initial conditions, shown in Figs.~\ref{F6chap3}c,d, these generic statements for the deformation parameters $\bm{\alpha}$ remain true but at early times the relationship between $T$ and $\Lambda$ is completely changed: the effective temperature $\Lambda$ starts out much smaller than $T$. A low effective temperature $\Lambda$ would narrow the microscopic momentum distribution in the transverse plane if it were not compensated by a very large ($\order(10)$) initial value of $\alpha_\perp$, which upholds the kinetic energy density and transverse pressure. On the other hand, $\alpha_L$ starts out almost at zero, reinforcing the narrowing of the longitudinal momentum distribution generated already by the small $\Lambda$ value and thereby causing a very small ratio of the kinetic contributions to $\PL$ and $\Pperp$. This is, of course, forced upon the system by the very anisotropic initial condition $\PL/\Pperp=0.01$. 
    
While the microscopic kinetic parameters $(\Lambda,\bm{\alpha})$ control only the kinetic contributions to energy density and pressures, the qualitative agreement of their tendencies extracted from this analysis of Figs.~\ref{F6chap3}c,d with those of the total energy density and pressures shown in Fig.~\ref{F4chap3} demonstrates that the mean-field $B$, even where large, cannot alter the sign of the pressure anisotropy (shear stress). Its value shifts the average kinetic pressure relative to the kinetic energy density and thereby has a large influence on the bulk viscous pressure.    

\section{Summary}
\label{ch3sec6}
 
In this chapter we derived a purely macroscopic formulation of anisotropic hydrodynamics in (3+1)--dimensions. To obtain the anisotropic hydrodynamic equations, we started from a microscopic description in terms of a relativistic Boltzmann-Vlasov equation for medium-dependent quasiparticles. The mean-field is constructed such that the equilibrium pressure of this kinetic theory agrees with the equation of state for strongly interacting QCD matter. The macroscopic equations of motion are then derived from an anisotropic 14--moment expansion, where the distribution function is split into a leading-order term $f_a$ and a residual correction $\dft$. The anisotropic distribution is constructed such that it non-perturbatively accounts for the two largest dissipative effects encountered in relativistic heavy-ion collisions: the large pressure anisotropy $\PL - \Pperp$ at early times and the large bulk viscous pressure $\Pi$ during the quark-hadron phase transition. These viscous effects are encoded in two momentum-anisotropy parameters $\alpha_L$ and $\alpha_\perp$ after imposing generalized Landau matching conditions for the longitudinal and transverse pressures. This matching scheme allows us to eliminate the microscopic parameters appearing in the kinetic theory, and to write down, for the first time, a set of macroscopic anisotropic hydrodynamic equations which make no explicit reference to their microscopic kinetic origin.

As a first application of this new approach we have here studied the Bjorken evolution of a longitudinally boost-invariant, transversely homogeneous system, evolving it both with anisotropic and second-order viscous hydrodynamics for comparison. We found remarkable agreement between the two approaches if both used mutually consistent quasiparticle transport coefficients, but noticed significant deviations from a standard viscous hydrodynamic model using transport coefficients for a weakly interacting Boltzmann gas with small masses. This suggests a remarkable robustness of second-order viscous hydrodynamics even in the presence of large shear and bulk viscous effects. A more concrete assessment of the relative strengths and weaknesses of anisotropic vs. viscous hydrodynamics for realistic heavy-ion collisions will be presented in the next chapter.

A key motivation for anisotropic hydrodynamics is that it maintains positive longitudinal and transverse pressures at the edges of the fireball, which prevents artificial cavitation in these regions. Furthermore, the residual shear stresses $\Wperp$ and $\piperp$ should be smaller than $\PL$ and $\Pperp$, which contain the largest viscous corrections. Bjorken does not allow us to test these expectations since it approximately describes only the central region of a smooth fireball. There, the temperatures are much higher than the pseudocritical temperature $T_c$, the transverse gradients are zero, and the residual dissipative flows thus vanish exactly. However, using (3+1)--dimensional anisotropic fluid dynamical simulations with fluctuating initial conditions will allow us to explore regions of the fireball that provide answers to these questions.
\chapter{Numerical implementation of anisotropic fluid dynamics}
\label{chapter4label}
 We now implement our formulation of non-conformal anisotropic hydrodynamics into a (3+1)--dimensional simulation for heavy-ion collisions called \cpuvah.\footnote{%
    The code package can be downloaded from the GitHub repository \url{https://github.com/mjmcnelis/cpu_vah}.}
The C++ module is based on the GPU--accelerated viscous hydrodynamic code \gpuvh{} \cite{Bazow:2016yra, Bazow:2017ewq} (although it itself has not yet been ported to GPUs). We solve the fluid dynamical equations on an Eulerian grid using the Kurganov--Tadmor algorithm \cite{Kurganov:2000}, a popular method also used in other hydrodynamic codes \cite{Schenke:2010nt, Schenke:2010rr, Bazow:2016yra, Pang:2018zzo}. The main improvement in our simulation is its ability to automatically adjust the time step $\Delta \tau_n$ after each iteration, as opposed to using a fixed value for the entire evolution. In an environment with a rapidly changing expansion rate, this offers a definitive advantage. The new adaptive Runge--Kutta method initially uses a fine time step to resolve the rapid longitudinal expansion at early times while speeding up the evolution at later times with a coarser time step given by the Courant--Friedrichs--Lewy (CFL) condition. Our numerical scheme allows us to start anisotropic hydrodynamics soon after the nuclear collision already at $\tau_0 = 0.05$ fm/$c$, so we can both model the far-off-equilibrium dynamics stage\footnote{\label{free_stream}%
    We assume the system is longitudinally free-streaming (i.e. $\PL/\Peq \approx 0$ and $\ene \propto 1/\tau$) in the time interval $0 < \tau \leq \tau_0$ before starting anisotropic hydrodynamics. This closely mirrors the situation found in other pre-hydrodynamic models \cite{Schenke:2012wb, Liu:2015nwa, Kurkela:2018wud}.}
at very early times with a non-conformal QCD equation of state and smoothly transition to second-order viscous hydrodynamics later.\footnote{%
    Anisotropic hydrodynamics reduces to second-order viscous hydrodynamics as the Knudsen and inverse Reynolds numbers decrease over time. Therefore, we do not have to switch to a separate viscous hydrodynamics model.} 
Thus, we can largely avoid the need to integrate \cpuvah{} with a separate pre-equilibrium dynamics module (e.g. longitudinal free-streaming of massless partons~\cite{Liu:2015nwa, Bernhard:2018hnz,Everett:2020xug,Everett:2020yty}).  

The code uses Milne coordinates $x^\mu = (\tau, x, y, \eta_s)$ and solves the hydrodynamic equations in natural units $\hbar = c = k_B = 1$ (i.e. the temperature $T$ has units of fm$^{-1}$ and the energy density units of fm$^{-4}$, etc). These units are converted back to physical units (e.g. $[\ene] =$ GeV/fm$^3$) when outputting and plotting the results.

This chapter is based on material submitted for publication in Ref.~\cite{McNelis:2021zji}.
\section{Hydrodynamic equations}
\label{chap4S2}
\subsection{Dynamical variables}
\label{chap4S2.2}
The dynamical variables that we propagate in the code are
\be
    \boldsymbol{q} = (T^{\tau\mu}, \PL, \Pperp, \Wperp, \piperp)\,,
\ee
where
\be
\label{eqchap4:Ttaumu}
    T^{\tau\mu} = \ene u^\tau u^\mu + \PL z^\tau z^\mu - \Pperp \Xi^{\tau\mu} + 2 W^{(\tau}_{\perp z} z^{\mu)} + \pi_\perp^{\tau\mu} 
\ee
are the time-like components of the energy-momentum tensor~\eqref{eqch3:3}; their evolution equations will be discussed below. Although $\Wperp$ and $\piperp$ each have only two independent components, we evolve all 14 of their components independently to simplify the workflow of the algorithm.\footnote{%
    When propagating these extraneous components numerically, slight violations of the orthogonality and tracelessness conditions \eqref{eqchap4:enforce} can occur. We correct for these errors at each step of the simulation with the regulation scheme described in Sec.~\ref{chap4S3.5}.}
In addition, we propagate the energy density $\ene$ and the fluid velocity's spatial components\footnote{The fluid velocity's temporal component is $u^\tau = \sqrt{1 + (u^x)^2 + (u^y)^2 + (\tau u^\eta)^2}$.} $\boldsymbol{u} =$ ($u^x$, $u^y$, $u^\eta$) since they appear in the hydrodynamic equations. They are inferred from the components $T^{\tau\mu}$; to reconstruct them, one also needs to know $z^\mu$. From the orthogonality conditions $z_\mu z^\mu = -1$ and $z_\mu u^\mu = 0$, there are only two nonzero components that depend on the fluid velocity:
\be
\label{eq:z_param}
    z^\mu = \frac{1}{\sqrt{1{+}u_\perp^2}}\left(\tau u^\eta, 0, 0, \frac{u^\tau}{\tau}\right)\,,
\ee
where $u_\perp = \sqrt{(u^x)^2 + (u^y)^2}$ is the transverse velocity. The solution of the inferred variables ($\ene$, $\boldsymbol{u}$) from the algebraic equations~\eqref{eqchap4:Ttaumu} will be discussed in Sec.~\ref{chap4S3.3}.

For (3+1)--dimensionally expanding fluids, we have a total of $20$ dynamical variables and four inferred variables to evolve on an Eulerian grid.\footnote{%
    Dynamical variables are evolved directly using the Kurganov--Tadmor algorithm (see Sec.~\ref{chap4S3.1}). Inferred variables are determined from the dynamical variables algebraically.}$^,$\footnote{%
    For anisotropic fluid dynamics with a QCD equation of state~\cite{McNelis:2018jho}, we further evolve the mean-field $B$ as a dynamical variable and the anisotropic variables ($\Lambda$, $\alpha_L$, $\alpha_\perp$) as additional inferred variables (see Secs.~\ref{chap4S2.4} and~\ref{chap4S2.6.3}).}
If the system is longitudinally boost-invariant, we do not need to propagate the components $T^{\tau\eta}$, $\Wperp$, $\pi_\perp^{\mu\eta}$ and $u^\eta$ since they vanish by symmetry.
\subsection{Conservation laws}
\label{chap4S2.3}
The evolution of the components $T^{\tau\mu}$ is given by the energy-momentum conservation laws
\be
    D_\mu T^\munu = 0 \,,
\ee
where the covariant derivative $D_\mu$ accounts for the curvilinear nature of the Milne coordinates. The conservation equations can be expanded as
\be
\label{eqchap4:conservation}
    \partial_\mu T^\munu + \Gamma^\mu_{\mu\lambda} T^{\lambda\nu} + \Gamma^\nu_{\mu\lambda} T^{\lambda\mu} = 0 \,,
\ee
where $\partial_\mu$ is the partial derivative and $\Gamma^{\mu}_{\nu\lambda}$ are the Christoffel symbols. In Milne spacetime, the only nonzero Christoffel symbols are 
\be
\begin{aligned}
    \Gamma^\tau_{\eta\eta} = \tau, \indent & \indent \Gamma^\eta_{\tau\eta} = \Gamma^\eta_{\eta\tau} = \frac{1}{\tau}\,.
\end{aligned}
\ee
Thus, the set of equations~\eqref{eqchap4:conservation} can be rewritten as 
\bs
\allowdisplaybreaks
\label{eqchap4:cons_eqs}
\beal
    \partial_\tau T^{\tau\tau} + \partial_i T^{\tau i} &= - \frac{T^{\tau\tau} {+} \tau^2 T^{\eta\eta}}{\tau} \,,
\\ 
    \partial_\tau T^{\tau x} + \partial_j T^{xj} &= - \frac{T^{\tau x}}{\tau} \,,
\\
    \partial_\tau T^{\tau y} + \partial_j T^{yj} &= - \frac{T^{\tau y}}{\tau} \,,
\\ 
    \partial_\tau T^{\tau\eta} + \partial_j T^{\eta j} &= - \frac{3T^{\tau\eta}}{\tau}\,,
\end{align}
\es
where the Latin indices $(i,j) \in (x,y,\eta)$ are summed over spatial components. We can eliminate the components $T^{\tau i}$ in (\ref{eqchap4:cons_eqs}a) by using the identity:
\be
\begin{split}
    T^{\tau i} &=  (\ene {+} \Pperp) u^\tau u^i + \LL^{\tau i} + \W^{\tau i} + \pi_\perp^{\tau i}
\\
    &= v^i T^{\tau\tau} + v^i (\Pperp {-} \LL^{\tau\tau} {-} \W^{\tau\tau} {-} \pi_\perp^{\tau\tau}) + \LL^{\tau i} + \W^{\tau i} + \pi_\perp^{\tau i} \,.
\end{split}
\ee
Here we introduced the three-velocity $v^i = u^i / u^\tau$ as well as the tensors $\LL^\munu = (\PL{-}\Pperp) z^\mu z^\nu$ and $\W^\munu = 2W^{(\mu}_{\perp z} z^{\nu)}$. Likewise, we can express the components $T^{ij}$ in (\ref{eqchap4:cons_eqs}b-d) in terms of $T^{\tau i}$:
\be
\begin{split}
    T^{ij} =& \,(\ene {+} \Pperp) u^i u^j - \Pperp g^{ij} + \LL^{ij} + \W^{ij} + \pi_\perp^{ij} 
\\ 
    =& \, v^j T^{\tau i} - v^j (\LL^{\tau i} {+} \W^{\tau i} {+} \pi_\perp^{\tau i}) - \Pperp g^{ij} + \LL^{ij} + \W^{ij} + \pi_\perp^{ij} \,.
\end{split}
\ee
After some algebra one obtains~\cite{Bazow:2017ewq}
\bs
\allowdisplaybreaks
\label{eqchap4:cons_flux}
\beal
    \partial_\tau T^{\tau\tau} + \partial_i (v^i T^{\tau\tau}) =& - \frac{T^{\tau\tau} {+} \tau^2 T^{\eta\eta}}{\tau} + (\LL^{\tau\tau} {+} \W^{\tau\tau} {+} \pi_\perp^{\tau\tau} {-} \Pperp) \partial_i v^i 
\\\nonumber 
    & + v^i \partial_i (\LL^{\tau\tau} {+} \W^{\tau\tau} {+} \pi_\perp^{\tau\tau} {-} \Pperp) - \partial_\eta \LL^{\tau\eta} - \partial_i \W^{\tau i} - \partial_i \pi_\perp^{\tau i} \,, \quad
\\\nonumber
\\
    \partial_\tau T^{\tau x} + \partial_i (v^i T^{\tau x}) =& - \frac{T^{\tau x}}{\tau} - \partial_x \Pperp + (\W^{\tau x} {+} \pi_\perp^{\tau x}) \partial_i v^i 
\\\nonumber
    &+ v^i \partial_i (\W^{\tau x} {+} \pi_\perp^{\tau x}) - \partial_\eta \W^{x\eta} - \partial_i \pi_\perp^{xi} \,,
\\\nonumber
\\
    \partial_\tau T^{\tau y} + \partial_i (v^i T^{\tau y}) =& - \frac{T^{\tau y}}{\tau} - \partial_y \Pperp + (\W^{\tau y} {+} \pi_\perp^{\tau y}) \partial_i v^i 
\\\nonumber
    &+ v^i \partial_i (\W^{\tau y} {+} \pi_\perp^{\tau y}) - \partial_\eta \W^{y\eta} - \partial_i \pi_\perp^{yi} \,,
\\\nonumber
\\
    \partial_\tau T^{\tau\eta} + \partial_i (v^i T^{\tau\eta}) =& - \frac{3T^{\tau\eta}}{\tau} - \frac{\partial_\eta \Pperp}{\tau^2} + (\LL^{\tau\eta} {+} \W^{\tau\eta} {+} \pi_\perp^{\tau\eta}) \partial_i v^i 
\\\nonumber
    &+ v^i \partial_i (\LL^{\tau\eta} {+} \W^{\tau\eta} {+} \pi_\perp^{\tau\eta}) - \partial_\eta \LL^{\eta\eta} - \partial_i \W^{\eta i} - \partial_i \pi_\perp^{\eta i}\,.
\end{align}
\es

For the numerical algorithm the hydrodynamic equations must be written in conservative flux form \cite{Kurganov:2000}:\footnote{%
    For a conformal fluid, the energy density's spatial derivatives are needed to evaluate the source term $\partial_i \Pperp = \frac{1}{2}(\partial_i \ene - \partial_i \PL)$.}
\be
\label{eqchap4:KT_eqs}
    \partial_\tau \boldsymbol{q}(x) + \partial_i\boldsymbol{F}^i(x) = \boldsymbol{S}(\tau,\boldsymbol{q}(x),\boldsymbol{u}(x),\ene(x),\partial_m\boldsymbol{q}(x), \partial_\mu\boldsymbol{u}(x))\,.
\ee
Here $\boldsymbol{F}^i = v^i \boldsymbol{q}$ are the currents, $\boldsymbol{S}$ are the source terms and $m \in (x,y,\eta)$ is a spatial index. Naturally, the evolution equations for $T^{\tau\mu}$ already assume this form. In the next subsection we will see that the relaxation equations for the dissipative flows require additional manipulations.
\subsection{Relaxation equations}
\label{chap4S2.4}
The relaxation equations for the dissipative flows $\PL$, $\Pperp$, $\Wperp$ and $\piperp$ were derived in the previous chapter but we restate them here for convenience:
\bs
\allowdisplaybreaks
\label{eqchap4:relax_1}
\beal
    \dot{\mathcal{P}}_L =&\,\frac{\Peq{-}\bar{\mathcal{P}}}{\tau_\Pi} - \frac{\PL{-}\Pperp}{3\tau_\pi / 2} + \bar{\zeta}^L_z \theta_L + \bar{\zeta}^L_\perp \theta_\perp - 2\Wperp \dot{z}_\mu + \bar{\lambda}^L_{Wu} \Wperp D_z u_\mu
\\\nonumber
    & + \bar{\lambda}^L_{W\perp} \Wperp z_\nu \nabla_{\perp,\mu} u^\nu - \bar{\lambda}^L_{\pi} \piperp \sigma_{\perp,\munu} \,,
\\\nonumber
\\
    \dot{\mathcal{P}}_\perp 
    =&\, \frac{\Peq{-}\bar{\mathcal{P}}}{\tau_\Pi} + \frac{\PL{-}\Pperp}{3\tau_\pi} + \bar{\zeta}^\perp_z \theta_L + \bar{\zeta}^\perp_\perp \theta_\perp + \Wperp \dot{z}_\mu + \bar{\lambda}^\perp_{Wu} \Wperp D_z u_\mu
\\\nonumber
      & - \bar{\lambda}^\perp_{W\perp} \Wperp z_\nu \nabla_{\perp,\mu} u^\nu + \bar{\lambda}^\perp_{\pi} \piperp \sigma_{\perp,\munu}\,,
\\\nonumber
\\ 
    \dot{W}^{\{\mu\}}_{\perp z} 
    =&\, - \frac{\Wperp}{\tau_\pi} + 2\bar{\eta}^W_u \Xi^\munu D_z u_\nu - 2\bar{\eta}^W_\perp z_\nu \nabla_\perp^\mu u^\nu - \big(\bar{\tau}^W_z \Xi^\munu {+} \piperp\big) \dot{z}_\nu 
\\\nonumber
    &- \bar{\lambda}^W_{W u} \Wperp  \theta_L + \bar{\delta}^W_W \Wperp \theta_\perp + \bar{\lambda}^W_{W \perp} \sigma_\perp^\munu  W_{\perp z, \nu} + \omega_\perp^\munu W_{\perp z, \nu} 
\\\nonumber
    &+ \bar{\lambda}^W_{\pi u} \piperp D_z u_\nu -  \bar{\lambda}^W_{\pi \perp} \piperp z_\lambda \nabla_{\perp,\nu} u^\lambda\,,
\\\nonumber
\\
    \dot{\pi}^{\{\munu\}}_{\perp} 
    =&\, - \frac{\piperp}{\tau_\pi} + 2 \bar{\eta}_\perp \sigma_\perp^\munu - 2 W_{\perp z}^{\{\mu} \dot{z}^{\nu\}} + \bar{\lambda}^\pi_\pi \piperp \theta_L - \bar{\delta}^\pi_\pi \piperp \theta_\perp - \bar{\tau}^\pi_\pi \pi_\perp^{\lambda \{\mu} \sigma^{\nu\}}_{\perp,\lambda}
\\\nonumber
    & + 2 \pi_\perp^{\lambda \{\mu} \omega^{\nu\}}_{\perp,\lambda}  - \bar{\lambda}^\pi_{W u} W_{\perp z}^{\{\mu} D_z u^{\nu\}} + \bar{\lambda}^\pi_{W \perp} W_{\perp z}^{\{\mu} z_\lambda \nabla_\perp^{\nu\}} u^\lambda \,. 
\end{align}
\es
Here a dot above any quantity denotes the co-moving time derivative $u^\gamma D_\gamma$. The transport coefficients that we use in the code will be discussed in Sec.~\ref{chap4S2.6}.

We recast the relaxation equations in conservative flux form by using the product rule identities
\bs
\allowdisplaybreaks
\beal
    \dot{W}_{\perp z}^{\{\mu\}} &= \Xi^\mu_\alpha u^\gamma D_\gamma W_{\perp z}^\alpha = u^\gamma D_\gamma \Wperp -  W_{\perp z}^\alpha u^\gamma D_\gamma \Xi^\mu_\alpha \,,
\\
    \dot{\pi}_{\perp}^{\{\munu\}} &= \Xi^\munu_{\alpha\beta} u^\gamma D_\gamma \pi_\perp^{\alpha\beta} = u^\gamma D_\gamma \piperp -  \pi_\perp^{\alpha\beta} u^\gamma D_\gamma \Xi^\munu_{\alpha\beta} 
\end{align}
\es
to rewrite the l.h.s. of Eqs.~(\ref{eqchap4:relax_1}a-d) as
\bs
\allowdisplaybreaks
\label{eqchap4:relax_3}
\beal
    &\dot{\mathcal{P}}_{L} = u^\gamma \partial_\gamma \mathcal{P}_{L}  \,,
\\
    &\dot{\mathcal{P}}_{\perp} = u^\gamma \partial_\gamma \mathcal{P}_{\perp}  \,,
\\
    &\dot{W}_{\perp z}^{\{\mu\}} = u^\gamma\partial_\gamma W_{\perp z}^\mu + u^\gamma \Gamma^\mu_{\gamma\lambda} W_{\perp z}^\lambda + W_{\perp z}^\alpha (u^\mu a_\alpha {-} z^\mu \dot{z}_\alpha)  \,,
\\
    &\dot{\pi}_\perp^{\{\munu\}} = u^\gamma\partial_\gamma \pi_\perp^{\munu} + u^\gamma \Gamma^\mu_{\gamma\lambda} \pi_\perp^{\nu\lambda} + u^\gamma \Gamma^\nu_{\gamma\lambda} \pi_\perp^{\mu\lambda} + \pi_\perp^{\mu\alpha}(u^\nu a_\alpha {-} z^\nu \dot{z}_\alpha) 
\\\nonumber
    &\qquad\quad + \pi_\perp^{\nu\alpha}(u^\mu a_\alpha {-} z^\mu \dot{z}_\alpha)\,,
\end{align}
\es
where $a^\mu = \dot{u}^\mu$ is the fluid acceleration. Finally, we use the product rule identity
\be
    u^\gamma \partial_\gamma \mathcal{P}_{L,\perp} = u^\tau \left[\partial_\tau \mathcal{P}_{L,\perp} + \partial_i (v^i \mathcal{P}_{L,\perp}) - \mathcal{P}_{L,\perp} \partial_i v^i \right]
\ee
to rewrite Eqs.~(\ref{eqchap4:relax_3}a-b) as
\bs
\allowdisplaybreaks
\label{eqchap4:relax_PLPT}
\beal
    \partial_\tau\PL + \partial_i(v^i \PL) &= \PL \partial_i v^i + \frac{1}{u^\tau} \left[ \frac{\Peq{-}\bar{\mathcal{P}}}{\tau_\Pi} - \frac{\PL{-}\Pperp}{3\tau_\pi / 2} + \mathcal{I}_L \right]\,,
\\
    \partial_\tau\Pperp + \partial_i(v^i \Pperp) &= \Pperp \partial_i v^i + \frac{1}{u^\tau} \left[\frac{\Peq{-}\bar{\mathcal{P}}}{\tau_\Pi} + \frac{\PL{-}\Pperp}{3\tau_\pi} + \mathcal{I}_\perp \right] \,;
\end{align}
\es
here
\bs
\allowdisplaybreaks
\beal
    \mathcal{I}_L =&\, \bar{\zeta}^L_z \theta_L + \bar{\zeta}^L_\perp \theta_\perp - 2\Wperp \dot{z}_\mu + \bar{\lambda}^L_{Wu} \Wperp D_z u_\mu + \bar{\lambda}^L_{W\perp} \Wperp z_\nu \nabla_{\perp,\mu} u^\nu 
\\\nonumber
  &- \bar{\lambda}^L_{\pi} \piperp \sigma_{\perp,\munu}  \,,
\\\nonumber
\\
    \mathcal{I}_\perp =&\, \bar{\zeta}^\perp_z \theta_L + \bar{\zeta}^\perp_\perp \theta_\perp + \Wperp \dot{z}_\mu + \bar{\lambda}^\perp_{Wu} \Wperp D_z u_\mu - \bar{\lambda}^\perp_{W\perp} \Wperp z_\nu \nabla_{\perp,\mu} u^\nu 
\\\nonumber
    &+ \bar{\lambda}^\perp_{\pi} \piperp \sigma_{\perp,\munu} 
\end{align}
\es
are the gradient source terms for $\PL$ and $\Pperp$. Similarly Eqs.~(\ref{eqchap4:relax_3}c-d) can be rewritten as
\bs
\allowdisplaybreaks
\label{eqchap4:relax_residual}
\beal
    \partial_\tau\Wperp + \partial_i(v^i \Wperp) &= \Wperp \partial_i v^i + \frac{1}{u^\tau} \left[-\frac{\Wperp}{\tau_\pi} + \mathcal{I}^\mu_W - \mathcal{P}^\mu_W - \mathcal{G}_W^\mu\right]\,,
\\
    \partial_\tau\piperp + \partial_i(v^i \piperp) &= \piperp \partial_i v^i + \frac{1}{u^\tau}\left[-\frac{\piperp}{\tau_\pi} + \mathcal{I}^\munu_\pi - \mathcal{P}^\munu_\pi - \mathcal{G}_\pi^\munu  \right] \,;
\end{align}
\es
here
\bs
\allowdisplaybreaks
\beal
    \mathcal{I}^\mu_W =&\, \Xi^\munu \big(2\bar{\eta}^W_u  D_z u_\nu {-} \bar{\tau}^W_z \dot{z}_\nu\big) - 2\bar{\eta}^W_\perp z_\nu \nabla_\perp^\mu u^\nu - \piperp\dot{z}_\nu  - \bar{\lambda}^W_{W u} \Wperp \theta_L
\\\nonumber
    & + \bar{\delta}^W_W \Wperp \theta_\perp + \bar{\lambda}^W_{W \perp} \sigma_\perp^\munu W_{\perp z, \nu} + \omega_\perp^\munu W_{\perp z, \nu} + \bar{\lambda}^W_{\pi u} \piperp D_z u_\nu 
\\\nonumber
    &- \bar{\lambda}^W_{\pi \perp} \piperp z_\lambda \nabla_{\perp,\nu} u^\lambda\,,
\\\nonumber
\\
    \mathcal{I}^\munu_\pi =&\, \Xi^\munu_{\alpha\beta}\Bigl[2 \pi_\perp^{\lambda (\alpha} \omega^{\beta)}_{\perp,\lambda} - \bar{\tau}^\pi_\pi \pi_\perp^{\lambda(\alpha} \sigma^{\beta)}_{\perp,\lambda} - 2 W_{\perp z}^{(\alpha} \dot{z}^{\beta)} - \bar{\lambda}^\pi_{W u} W_{\perp z}^{(\alpha} D_z u^{\beta)}
\\\nonumber
    & + \bar{\lambda}^\pi_{W \perp} W_{\perp z}^{(\alpha} z_\lambda \nabla_\perp^{\beta)} u^\lambda\Bigr] + 2 \bar{\eta}_\perp \sigma_\perp^\munu  + \bar{\lambda}^\pi_\pi \piperp \theta_L - \bar{\delta}^\pi_\pi \piperp \theta_\perp
\end{align}
\es
are the gradient source terms,
\bs
\allowdisplaybreaks
\beal
    \mathcal{P}^\mu_W =&\, W_{\perp z}^\alpha (u^\mu a_\alpha {-} z^\mu \dot{z}_\alpha) \,,
\\
    \mathcal{P}^\munu_\pi =&\, \pi_\perp^{\mu\alpha}(u^\nu a_\alpha {-} z^\nu \dot{z}_\alpha) + \pi_\perp^{\nu\alpha}(u^\mu a_\alpha {-} z^\mu \dot{z}_\alpha)
\end{align}
\es
are the transverse projection source terms, and
\bs
\allowdisplaybreaks
\label{eqchap4:geometric_source}
\beal
    \mathcal{G}^\mu_W =&\, u^\gamma \Gamma^\mu_{\gamma\lambda} W_{\perp z}^\lambda \,,
\\
    \mathcal{G}^\munu_\pi =&\, u^\gamma \Gamma^\mu_{\gamma\lambda} \pi_\perp^{\nu\lambda} + u^\gamma \Gamma^\nu_{\gamma\lambda} \pi_\perp^{\mu\lambda}
\end{align}
\es
are the geometric source terms for $\Wperp$ and $\piperp$. The individual components that make up the source terms in the relaxation equations (\ref{eqchap4:relax_PLPT}a,b) and (\ref{eqchap4:relax_residual}a,b) are listed in Appendices \ref{app4a} and \ref{app4b}.

If we evolve anisotropic fluid dynamics with a QCD equation of state, we also propagate the mean-field $B$ with the relaxation equation
\be
\label{eqchap4:relax_B_0}
    \dot{B} = \frac{B_\text{eq} {-} B}{\tau_\Pi} - \frac{\dot{m}}{m}(\ene {-} 2\Pperp {-} \PL {-} 4B)
\ee
or, in conservative flux form,
\be
\label{eqchap4:relax_B}
    \partial_\tau B + \partial_i(v^i B) = B \, \partial_i v^i + \frac{1}{u^\tau} \left[\frac{B_\text{eq} {-} B}{\tau_\Pi} - \frac{\dot{m}}{m}(\ene {-} 2\Pperp {-} \PL {-} 4B)\right]\,.
\ee
\subsection{Equation of state}
\label{chap4S2.5}
\begin{figure}[t]
\centering
\includegraphics[width=0.94\linewidth]{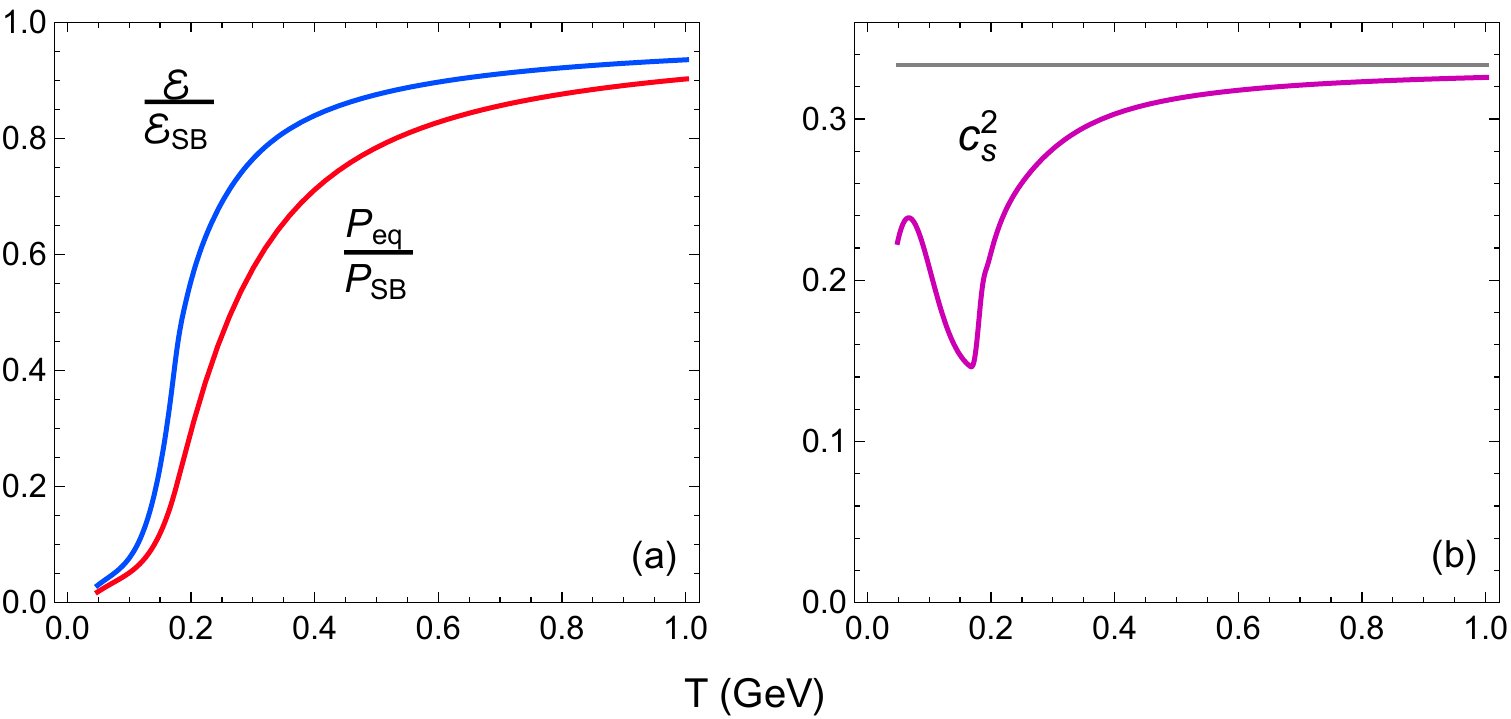}
\caption{(Color online)
\label{eos}
    {\sl Left:} The QCD energy density (blue) and equilibrium pressure  (red), normalized by their Stefan--Boltzmann limits \eqref{eqchap4:stefan}. 
    {\sl Right:} The squared speed of sound (purple) as a function of temperature; the gray line indicates the conformal limit $c_s^2 = \frac{1}{3}$.
}
\end{figure}
In the code there are two options for the quark-gluon plasma's equation of state $\Peq(\ene)$: conformal and QCD. The former assumes a non-interacting gas of massless quarks and gluons:
\be
\label{eqchap4:stefan}
    \Peq = \frac{\ene}{3} = \frac{g T^4}{\pi^2} \,,
\ee
where $T$ is the temperature and $g$ is the degeneracy factor given by Eq.~\eqref{eqch3:46}. This equation of state is primarily used to test the fluid dynamical simulation subject to either conformal Bjorken expansion or conformal Gubser expansion (see Secs.~\ref{chap4S4.1} and \ref{chap4S4.2}).

The QCD equation of state that we employ for more realistic simulations interpolates between the lattice QCD calculations provided by the HotQCD collaboration~\cite{Bazavov:2014pvz} and a hadron resonance gas composed of the hadrons that can be propagated in the hadronic afterburner code {\sc SMASH} \cite{Weil:2016zrk}.\footnote{%
    In hybrid model simulations of heavy-ion collisions, the equilibrium equation of state used in the hydrodynamic module must be consistent with that in the hadronic afterburner. Otherwise, serious violations of energy and momentum conservation can occur at the hadronization phase.}
Figure~\ref{eos} shows the energy density, equilibrium pressure and speed of sound as a function of temperature. The data table in the code covers the temperature range $T \in [0.05, 1.0]$\,GeV.\footnote{
    Thermodynamic functions that depends on the QCD equation of state (e.g. $\Peq(T)$) are evaluated using rational polynomial fits, which are computed by the directory \texttt{eos}. The user can update these fits when a new QCD equation of state becomes available.}
\subsection{Transport coefficients}
\label{chap4S2.6}
\begin{figure}[t]
\centering
\includegraphics[width=0.94\linewidth]{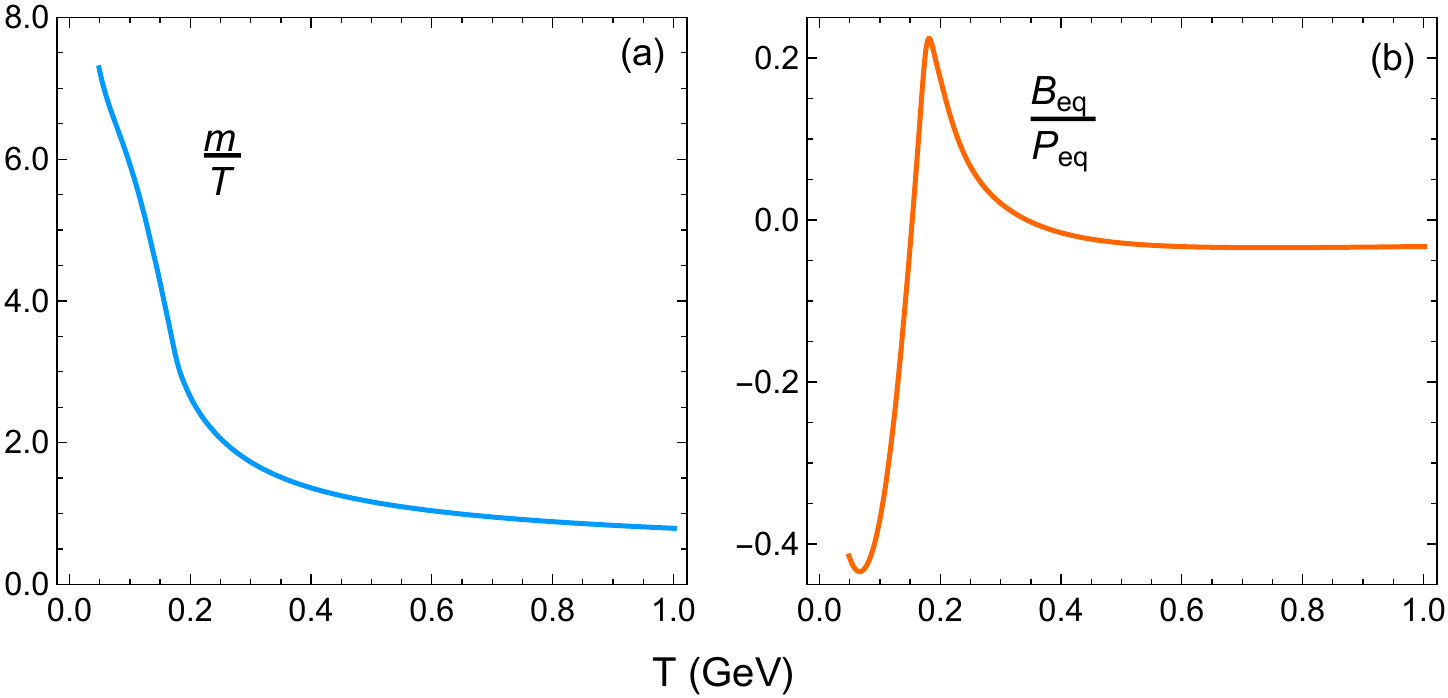}
\caption{(Color online)
\label{quasi}
    The quasiparticle mass to temperature ratio (blue, left) and equilibrium mean-field normalized to the QCD equilibrium pressure (orange, right) as functions of temperature (for conformal systems, $m = 0$ and $B_\text{eq} = 0$).
}
\end{figure}
Here we summarize the transport coefficients used in the relaxation equations in Sec.~\ref{chap4S2.4}. We parametrize the shear and bulk viscosities $(\etas)(T)$ and $(\zetas)(T)$ as a function of temperature, using the best-fit parametrization models from the JETSCAPE collaboration~\cite{Everett:2020yty,Everett:2020xug}. The relaxation times and anisotropic transport coefficients are computed with the quasiparticle kinetic theory model discussed in the previous chapter; the temperature-dependent mass $m(T)$ and equilibrium mean-field $B_\text{eq}(T)$ for the HotQCD equation of state are shown in Figure~\ref{quasi}.
\subsubsection{Shear and bulk viscosities}
\label{chap4S2.6.1}
\begin{figure}[t]
\centering
\includegraphics[width=0.94\linewidth]{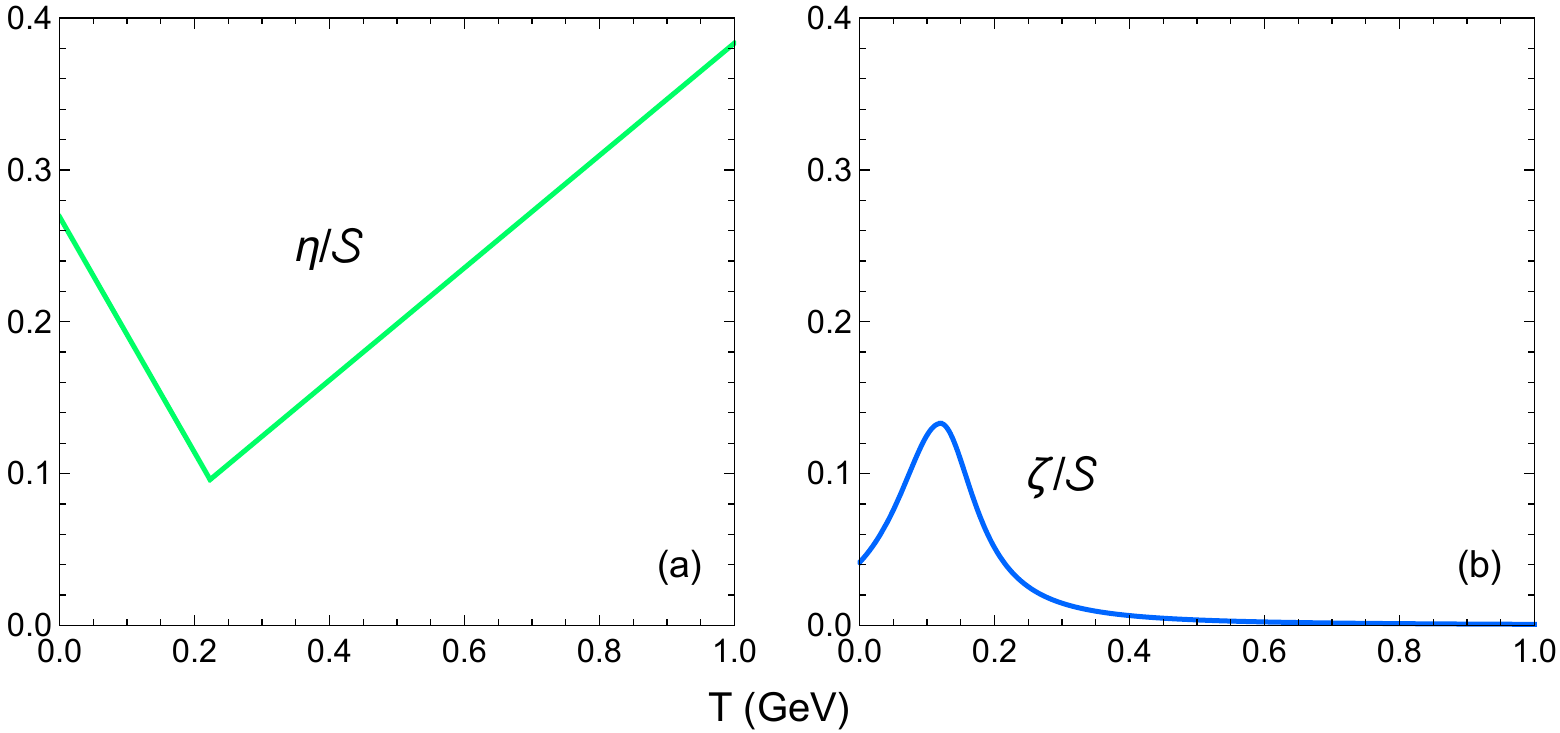}
\caption{(Color online)
\label{viscosities}
The temperature parametrization of $(\eta / \mathcal{S})(T)$ and $(\zeta / \mathcal{S})(T)$ used in this chapter (for conformal systems, we set $\eta / \mathcal{S} = 0.2$ and $\zeta / \mathcal{S} = 0$).
}
\end{figure}
The shear viscosity is modeled as a linear piecewise function with a kink at temperature $T_\eta$:
\be
\label{eqchap4:etas}
    (\eta/\mathcal{S})(T) = (\etas)_\text{kink} + (T {-} T_\eta)\left(a_\text{low} \Theta(T_\eta {-} T) + a_\text{high} \Theta(T {-} T_\eta)\right)\,,
\ee
where $(\etas)_\text{kink}$ is the value of $\eta/\mathcal{S}$ at $T_\eta$, $a_\text{low}$ and $a_\text{high}$ are the left and right slopes, respectively, and $\Theta$ is the Heaviside step function. The bulk viscosity is parametrized as a skewed Cauchy distribution:
\be
\label{eqchap4:zetas}
    (\zeta/\mathcal{S})(T) = \frac{(\zeta/\mathcal{S})_\text{max} \,\Lambda_\zeta(T)^2}{\Lambda_\zeta(T)^2 + (T {-} T_\zeta)^2}\,,
\ee
where $(\zeta/\mathcal{S})_\text{max}$ is the normalization factor, $T_\zeta$ is the peak temperature and 
\be
    \Lambda_\zeta(T) = w_\zeta \left(1 + \lambda_\zeta \,\mathrm{sgn}(T{-}T_\zeta)\right)\,,
\ee
with $w_\zeta$ and $\lambda_\zeta$ being the width and skewness parameters, respectively. 

The best-fit values for the viscosity parameters\footnote{%
    They correspond to a hybrid model whose particlization phase uses the 14-moment approximation for the $\delta f$ correction in the Cooper-Frye formula \cite{McNelis:2019auj, Everett:2020yty, Everett:2020xug} (the viscosity parameters used for the code test runs in Sec.~\ref{chap4S4} are slightly different because they were taken from an early draft of Ref.~\cite{Everett:2020xug}).}
are $(\etas)_\text{kink} = 0.096$, $T_\eta = 0.223$ GeV, $a_\text{low} = -0.776$ GeV$^{-1}$, $a_\text{high} = 0.37$ GeV$^{-1}$, $(\zetas)_\text{max} = 0.133$, $T_\zeta = 0.12$ GeV, $w_\zeta = 0.072$ GeV and $\lambda_\zeta = -0.122$ (see Table II in Ref.~\cite{Everett:2020xug}). The resulting temperature dependence of $\eta/\mathcal{S}$ and $\zeta/\mathcal{S}$ is shown in Figure~\ref{viscosities}. 

For conformal systems, we fix the shear viscosity to $\eta / \mathcal{S} = 0.2$ and the bulk viscosity to $\zeta / \mathcal{S} = 0$. 
\subsubsection{Shear and bulk relaxation times}
\label{chap4S2.6.2}
\begin{figure}[t]
\centering
\includegraphics[width=0.94\linewidth]{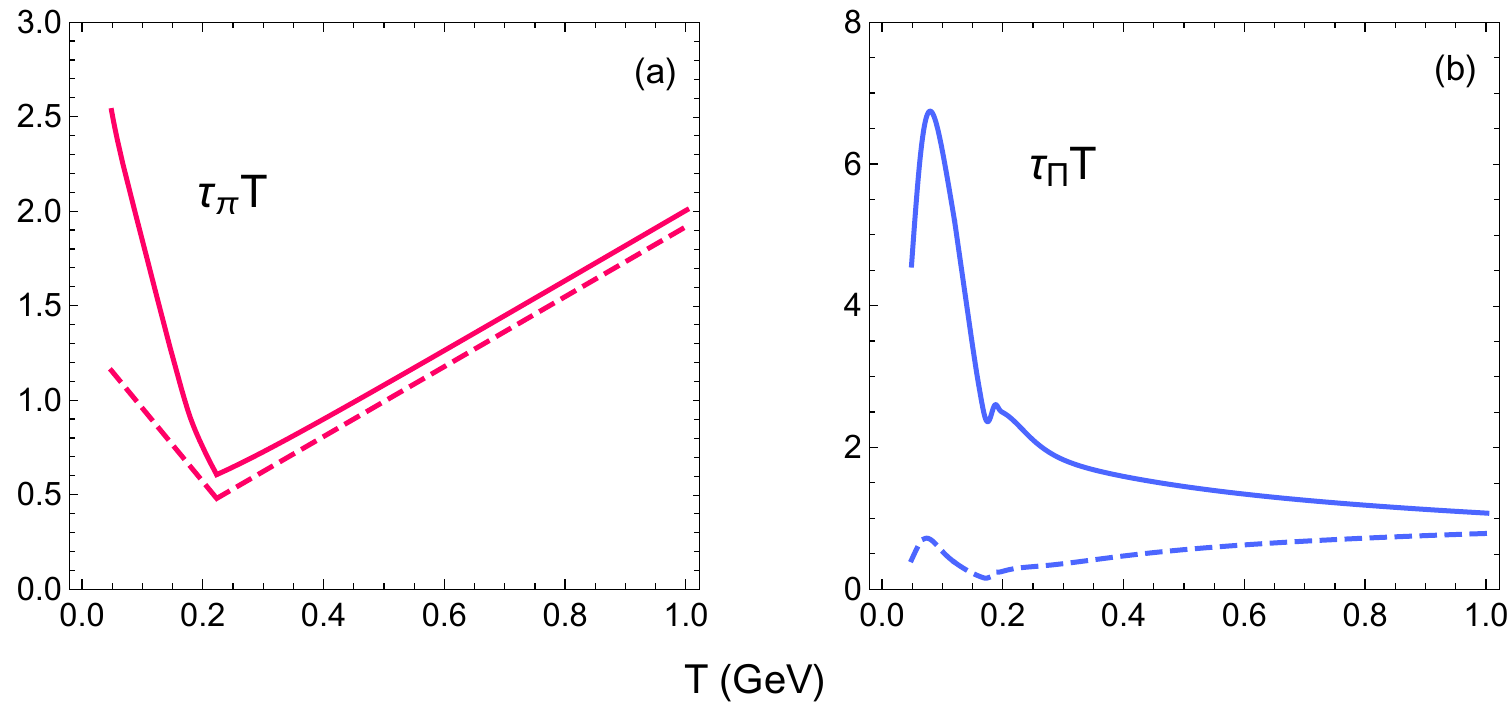}
\caption{(Color online)
\label{relaxation}
The dimensionless shear and bulk relaxation times computed with the quasiparticle kinetic model (solid color) and standard small-mass approximation (dashed color) (for conformal systems, we set $\tau_\pi T = 5\etas$ and $\tau_\Pi T = 0$).
}
\end{figure}
The normalized quasiparticle relaxation times $\tau_\pi T$ and $\tau_\Pi T$ are shown in Figure~\ref{relaxation}. These are compared to the relaxation times used in standard second-order viscous hydrodynamic model (\vh{}2)~\cite{Denicol:2014vaa,Bazow:2016yra}:
\be
\label{eqchap4:tau_r_small}
    (\tau_\pi T)_{|\text\vh{}2} = 5 \etas \,,\qquad
    (\tau_\Pi T)_{|\text\vh{}2} = \frac{\zeta \,T}{15(\ene {+} \Peq)\big(\frac{1}{3} {-} c_s^2\big)^2}\,.
\ee
One sees that the shear relaxation times are very similar to each other, except at low temperatures $T < 0.2$ GeV. On the other hand, the bulk relaxation times differ by about an order of magnitude for $T < 0.2$ GeV; this is due to the breakdown of the small-mass approximation, even at high temperatures $T \sim 1$ GeV. As a result, the evolution of the bulk viscous pressure $\Pi$ will be more greatly affected by critical slowing down in our anisotropic hydrodynamics model compared to standard viscous hydrodynamics.

For conformal kinetic plasmas ($m = 0$, $dm/dT = 0$), the shear relaxation time is $\tau_\pi = 5\eta / (\mathcal{S}T)$ and the bulk relaxation time is $\tau_\Pi = 0$.
\subsubsection{Anisotropic transport coefficients}
\label{chap4S2.6.3}
Finally, we list the anisotropic transport coefficients~\cite{McNelis:2018jho, Molnar:2016vvu} that are coupled to the gradient source terms in the relaxation equations for the longitudinal pressure $\PL$,
\bs
\allowdisplaybreaks
\label{eqchap4:pl_coeff}
\beal
    \bar{\zeta}^{L}_z & = \I_{2400} - 3 (\PL{+}B) + m\frac{dm}{d\ene}(\ene{+}\PL)\I_{0200} \,,
\\
    \bar{\zeta}^{L}_\perp & = \I_{2210} - \PL - B + m\frac{dm}{d\ene}(\ene{+}\Pperp)\I_{0200} \,,
\\
    \bar{\lambda}^{L}_{Wu} & = \frac{\I_{4410}}{\I_{4210}} + m\frac{dm}{d\ene}\I_{0200} \,,
\\
    \bar{\lambda}^{L}_{W\perp} & =  1 - \bar{\lambda}^{L}_{Wu}\,,
\\
    \bar{\lambda}^{L}_\pi & = \frac{\I_{4220}}{\I_{4020}} + m\frac{dm}{d\ene}\I_{0200} \,,
\end{align}
\es
the transverse pressure $\Pperp$,
\bs
\allowdisplaybreaks
\label{eqchap4:pt_coeff}
\beal
    \bar{\zeta}^{\perp}_z & = \I_{2210} - \Pperp - B + m\frac{dm}{d\ene}(\ene{+}\PL)\I_{0010}\,,
\\
    \bar{\zeta}^{\perp}_\perp & = 2(\I_{2020} {-} \Pperp {-} B) + m\frac{dm}{d\ene}(\ene{+}\Pperp)\I_{0010}\,,
\\
    \bar{\lambda}^{\perp}_{W\perp} & = \frac{2 \,\I_{4220}}{\I_{4210}} + m\frac{dm}{d\ene}\I_{0010}\,,
\\
    \bar{\lambda}^{\perp}_{Wu} & = \bar{\lambda}^{\perp}_{W\perp} - 1 \,,
\\
    \bar{\lambda}^{\perp}_\pi & = 1 - \frac{3\,\I_{4030}}{\I_{4020}} - m\frac{dm}{d\ene}\I_{0010}\,,
\end{align}
\es
the longitudinal momentum diffusion current $\Wperp$,
\bs
\allowdisplaybreaks
\label{eqB3}
\beal
    \bar{\eta}^W_u & = \frac{1}{2}\big(\PL + B  - \I_{2210})\,,
\\
    \bar{\eta}^W_\perp & = \frac{1}{2}(\Pperp + B  - \I_{2210})\,, 
\\
    \bar{\tau}^W_z & = \PL  - \Pperp \,,
\\
    \bar{\delta}^W_{W} & = \bar{\lambda}^W_{W\perp} - \frac{1}{2} + m\frac{dm}{d\ene} (\ene{+}\Pperp) \left(\frac{\I_{2210}}{\I_{4210}}\right) \,,
\\
    \bar{\lambda}^W_{Wu} & = 2 - \frac{\I_{4410}}{\I_{4210}} - m\frac{dm}{d\ene} (\ene{+}\PL) \left(\frac{\I_{2210}}{\I_{4210}}\right) \,,
\\
    \bar{\lambda}^W_{W\perp} & = \frac{2 \,\I_{4220}}{\I_{4210}} - 1,
\\
    \bar{\lambda}^W_{\pi u} & = \frac{\I_{4220}}{\I_{4020}} \,,
\\
    \bar{\lambda}^W_{\pi\perp} & = \bar{\lambda}^W_{\pi u} - 1 \,,
\end{align}
\es
and the transverse shear stress tensor $\piperp$:
\bs
\allowdisplaybreaks
\label{eqB4}
\beal
    \bar{\eta}_\perp & = \Pperp^{(k)}  - \I_{2020}, 
\\
    \bar{\delta}^\pi_\pi & = \frac{3}{4} \bar{\tau}^\pi_\pi + \frac{1}{2} - m\frac{dm}{d\ene} (\ene{+}\Pperp) \left(\frac{\I_{2020}}{\I_{4020}}\right), 
\\
    \bar{\tau}^\pi_\pi & =  2 - \frac{4\,\I_{4030}}{\I_{4020}},
\\
    \bar{\lambda}^\pi_{\pi} & = \bar{\lambda}^W_{\pi u} - 1 + m\frac{dm}{d\ene} (\ene{+}\PL) \left(\frac{\I_{2020}}{\I_{4020}}\right), 
\\
    \bar{\lambda}^\pi_{Wu} & =  \bar{\lambda}^W_{W\perp} -  1,
\\
    \bar{\lambda}^\pi_{W\perp} & = \bar{\lambda}^\pi_{Wu} + 2\,,
\end{align}
\es
where the anisotropic integrals $\I_{nrqs}$ are defined in Appendix~\ref{appch3a}. Since the transport coefficients depend on the leading-order anisotropic distribution~\eqref{eqch3:15}, we propagate the anisotropic variables $(\Lambda, \alpha_\perp, \alpha_L)$ as inferred variables in the code. We solve for them numerically using the method in Sec.~\ref{chap4S3.4}.

The anisotropic transport coefficients in the conformal limit are listed in Appendix~\ref{app4c}.
\section{Numerical scheme}
\label{chap4S3}
In this section we discuss the numerical implementation of the hydrodynamic equations in the code. The dynamical and inferred variables are evolved on an $(N_x{+}4) \times (N_y{+}4) \times (N_\eta{+}4)$ Eulerian grid, where $N_x$, $N_y$ and $N_\eta$ are the number of physical grid points along each spatial direction.\footnote{For longitudinally boost-invariant systems, the number of spacetime rapidity points is set to $N_\eta = 1$.} A grid point with cell index $(i,j,k)$ that corresponds to the lower left front corner of a fluid cell, has a spatial position
\bs
\allowdisplaybreaks
\beal
x_i &= \big[i - 2 - \frac{1}{2}(N_x {-} 1)\big] \Delta x \,,\\
y_j &= \big[j - 2 - \frac{1}{2}(N_y {-} 1)\big] \Delta y \,,\\
\eta_{s,k} &= \big[k - 2 - \frac{1}{2}(N_\eta {-} 1)\big] \Delta \eta_s \,,
\end{align}
\es
where $\Delta x$, $\Delta y$ and $\Delta \eta_s$ are the lattice spacings. Physical fluid cells have indices $i\in[2,N_x{+}1]$, $j\in[2,N_y{+}1]$ and $k\in[2,N_\eta{+}1]$. The numerical algorithm also requires six sets of ghost cells with depth two, which neighbor the physical grid's faces (for an illustration, see Fig.~3 in Ref.~\cite{Bazow:2016yra}). The boundary conditions that we impose on the ghost cells neighboring the two $(y,\eta_s)$ faces are\footnote{%
    In conformal anisotropic hydrodynamics, the energy density $\ene$ also requires ghost cell boundary conditions.}
\bs
\allowdisplaybreaks
\beal
\boldsymbol{q}_{0,j,k} &= \boldsymbol{q}_{1,j,k} = \boldsymbol{q}_{2,j,k} \,,\\
\boldsymbol{u}_{0,j,k} &= \boldsymbol{u}_{1,j,k} = \boldsymbol{u}_{2,j,k} \,,\\
\boldsymbol{q}_{N_x{+}2,j,k} &= \boldsymbol{q}_{N_x{+}3,j,k} = \boldsymbol{q}_{N_x{+}1 ,j,k} \,,\\
\boldsymbol{u}_{N_x{+}2,j,k} &= \boldsymbol{u}_{N_x{+}3,j,k} = \boldsymbol{u}_{N_x{+}1 ,j,k}\,,
\end{align}
\es
and similarly for $(x,\eta_s)$ and $(x,y)$ faces after permuting the grid indices and replacing $N_x \to N_y$ or $N_\eta$.

The dynamical variables $\boldsymbol{q}$ are updated using a two-stage Runge--Kutta (RK2) scheme, where the time derivatives are evaluated with the Kurganov--Tadmor (KT) algorithm~\cite{Schenke:2010nt, Bazow:2016yra,  Kurganov:2000}. After each intermediate Euler step in the RK2 scheme, we reconstruct the inferred variables from the dynamical variables. In addition, we regulate the residual shear stresses $\Wperp$ and $\piperp$ and the mean-field $B$. The code also provides the user with the option of using an adaptive time step to capture the fluid's longitudinal dynamics at very early times; this will be discussed at the end of the section.

\subsection{Kurganov-Tadmor algorithm}
\label{chap4S3.1}
The time derivative of the dynamical variables $\partial_\tau\boldsymbol{q}$ in the partial differential equations~\eqref{eqchap4:KT_eqs} can be computed on the Eulerian grid at any time $\tau$ using the KT algorithm~\cite{Kurganov:2000}:
\be
\label{eqchap4:KT_algorithm}
\begin{split}
(\partial_\tau \boldsymbol{q})_{ijk} = & -\frac{\boldsymbol{H}^x_{i{+}\half,j,k} {-} \boldsymbol{H}^x_{i{-}\half,j,k}}{\Delta x} - \frac{\boldsymbol{H}^y_{i,j{+}\half,k} {-} \boldsymbol{H}^y_{i,j{-}\half,k}}{\Delta y} - \frac{\boldsymbol{H}^\eta_{i,j,k{+}\half} {-} \boldsymbol{H}^\eta_{i,j,k{-}\half}}{\Delta \eta_s}\\&
+ \boldsymbol{S}_{ijk}\big(\tau, \boldsymbol{q}_{ijk}, \boldsymbol{u}_{ijk},\ene_{ijk}, (\partial_m \boldsymbol{q})_{ijk}, (\partial_\mu \boldsymbol{u})_{ijk}\big)\,,
\end{split}
\ee
where the numerical fluxes evaluated at the left and right faces of a staggered cell centered around the grid point $(i,j,k)$ are
\be
\label{eqchap4:flux}
\boldsymbol{H}^x_{i\pm\half,j,k} = \frac{1}{2}\Big[\boldsymbol{F}^{x+}_{i\pm\half,j,k} + \boldsymbol{F}^{x-}_{i\pm\half,j,k} - s^x_{i\pm\half,j,k}\big(\boldsymbol{q}^+_{i\pm\half,j,k} {-\,} \boldsymbol{q}^-_{i\pm\half,j,k}\big)\Big],
\ee
and similarly for $\boldsymbol{H}^y_{i,j\pm\frac{1}{2},k}$ and $\boldsymbol{H}^\eta_{i,j,k\pm\frac{1}{2}}$ after permuting the $\pm\half$ in the grid indices and the corresponding spatial component (i.e. $x \to y$ or $\eta$).

The first two terms in Eq.~\eqref{eqchap4:flux} take the average of the currents extrapolated to the staggered cell face $(i{+}\half,j,k)$ (or $(i{-}\half,j,k)$)  from the left ($-$) and right ($+$) sides. A first-order expression for the extrapolated currents can be computed using the chain rule:
\bs
\allowdisplaybreaks
\label{eqchap4:Fx_ex}
\beal
\boldsymbol{F}^{x-}_{i+\frac{1}{2},j,k} &= \boldsymbol{F}^x_{ijk} + \frac{\Delta x}{2}\Big[(\partial_x v^x)_{ijk} \boldsymbol{q}_{ijk} + v^x_{ijk} (\partial_x \boldsymbol{q})_{ijk} \Big] \,,\\
\boldsymbol{F}^{x+}_{i+\frac{1}{2},j,k} &= \boldsymbol{F}^x_{i+1,j,k} - \frac{\Delta x}{2}\Big[(\partial_x v^x)_{i+1,j,k} \boldsymbol{q}_{i+1,j,k} + v^x_{i+1,j,k}(\partial_x \boldsymbol{q})_{i+1,j,k}\Big] \,,\\
\boldsymbol{F}^{x-}_{i-\frac{1}{2},j,k} &= \boldsymbol{F}^x_{i-1,j,k} + \frac{\Delta x}{2}\Big[(\partial_x v^x)_{i-1,j,k}\boldsymbol{q}_{i-1,j,k} + v^x_{i-1,j,k}(\partial_x \boldsymbol{q})_{i-1,j,k}\Big]\,,\\
\boldsymbol{F}^{x+}_{i-\frac{1}{2},j,k} &= \boldsymbol{F}^x_{ijk} - \frac{\Delta x}{2}\Big[(\partial_x v^x)_{ijk}\boldsymbol{q}_{ijk} + v^x_{ijk}(\partial_x \boldsymbol{q})_{ijk}\Big] \,,
\end{align}
\es
where $\boldsymbol{F}^x_{ijk} = v^x_{ijk} \boldsymbol{q}_{ijk}$. Similarly, the extrapolated currents $\boldsymbol{F}^{y+}_{i,j\pm\frac{1}{2},k}$, $\boldsymbol{F}^{y-}_{i,j\pm\frac{1}{2},k}$, $\boldsymbol{F}^{\eta+}_{i,j,k\pm\frac{1}{2}}$ and $\boldsymbol{F}^{\eta-}_{i,j,k\pm\frac{1}{2}}$ at the remaining faces of the staggered cell are obtained by permuting the $\pm\half$ (or $\pm 1$) in the grid indices, the spatial components and derivatives, and the lattice spacing. 

The final term in Eq.~\eqref{eqchap4:flux} takes into account the wave propagation of the discontinuities $\boldsymbol{q}^+_{i\pm\half,j,k} - \boldsymbol{q}^-_{i\pm\half,j,k}$ at a finite speed~\cite{Kurganov:2000}. We define the local propagation speed component $s^x_{i\pm\half,j,k}$ at the staggered cell faces $(i{\pm}\half,j,k)$ as
\be
s^x_{i\pm\half,j,k} = \max\big(|v^{x-}_{i\pm\half,j,k}|, |v^{x+}_{i\pm\half,j,k}|\big)\,,
\ee
where the extrapolated velocities are
\bs
\allowdisplaybreaks
\label{eqchap4:vx_ex}
\beal
v^{x-}_{i+\frac{1}{2},j,k} &= v^x_{ijk} + \frac{\Delta x}{2}(\partial_x v^x)_{ijk} \,,\\
v^{x+}_{i+\frac{1}{2},j,k} &= v^x_{i+1,j,k} - \frac{\Delta x}{2}(\partial_x v^x)_{i+1,j,k} \,,\\
v^{x-}_{i-\frac{1}{2},j,k} &= v^x_{i-1,j,k} + \frac{\Delta x}{2}(\partial_x v^x)_{i-1,j,k}\,,\\
v^{x+}_{i-\frac{1}{2},j,k} &= v^x_{ijk} - \frac{\Delta x}{2}(\partial_x v^x)_{ijk} \,.
\end{align}
\es
The discontinuities $\boldsymbol{q}^+_{i\pm\half,j,k} - \boldsymbol{q}^-_{i\pm\half,j,k}$ propagating from the staggered cell faces $(i{\pm}\half,j,k)$  depend on the extrapolated dynamical variables
\bs
\allowdisplaybreaks
\label{eqchap4:q_ex}
\beal
\boldsymbol{q}^{-}_{i+\frac{1}{2},j,k} &= \boldsymbol{q}_{ijk} + \frac{\Delta x}{2}(\partial_x\boldsymbol{q})_{ijk} \,,\\
\boldsymbol{q}^{+}_{i+\frac{1}{2},j,k} &= \boldsymbol{q}_{i+1,j,k} - \frac{\Delta x}{2}(\partial_x \boldsymbol{q})_{i+1,j,k} \,,\\
\boldsymbol{q}^{-}_{i-\frac{1}{2},j,k} &= \boldsymbol{q}_{i-1,j,k} + \frac{\Delta x}{2}(\partial_x \boldsymbol{q})_{i-1,j,k}\,,\\
\boldsymbol{q}^{+}_{i-\frac{1}{2},j,k} &= \boldsymbol{q}_{ijk} - \frac{\Delta x}{2}(\partial_x \boldsymbol{q})_{ijk} \,.
\end{align}
\es
The formulae for the local propagation speed components $s^y_{i,j\pm\half,k}$ and $s^\eta_{i,j,k\pm\half}$, extrapolated velocities $v^{y+}_{i,j\pm\half,k}$, $v^{y-}_{i,j\pm\half,k}$, $v^{\eta+}_{i,j,k\pm\half}$ and $v^{\eta-}_{i,j,k\pm\half}$, and extrapolated dynamical variables $\boldsymbol{q}^+_{i,j\pm\half,k}$, $\boldsymbol{q}^-_{i,j\pm\half,k}$, $\boldsymbol{q}^+_{i,j,k\pm\half}$ and $\boldsymbol{q}^-_{i,j,k\pm\half}$ are analogous. 

The numerical spatial derivatives appearing in the extrapolated quantities~(\ref{eqchap4:Fx_ex}a-d),~(\ref{eqchap4:vx_ex}a-d) and~(\ref{eqchap4:q_ex}a-d)  are computed with a minmod flux limiter~\cite{Kurganov:2000}:
\bs
\allowdisplaybreaks
\beal
(\partial_x \boldsymbol{q})_{ijk} &= \mathcal{M}\Big(\Theta \, \frac{\boldsymbol{q}_{ijk} - \boldsymbol{q}_{i-1,j,k}}{\Delta x}, \frac{\boldsymbol{q}_{i+1,j,k} - \boldsymbol{q}_{i-1,j,k}}{2\Delta x},\Theta \, \frac{\boldsymbol{q}_{i+1,j,k} - \boldsymbol{q}_{ijk}}{\Delta x}\Big) \,,\\
(\partial_x v^x)_{ijk} &= \mathcal{M}\Big(\Theta \, \frac{v^x_{ijk} - v^x_{i-1,j,k}}{\Delta x}, \frac{v^x_{i+1,j,k} - v^x_{i-1,j,k}}{2\Delta x},\Theta \, \frac{v^x_{i+1,j,k} - v^x_{ijk}}{\Delta x}\Big) \,,
\end{align}
\es
where 
\be
\mathcal{M}(a,b,c) = \mathrm{minmod}(a, \mathrm{minmod}(b,c)) \,,
\ee
with
\be
\mathrm{minmod}(a,b) = \frac{\mathrm{sgn}(a) + \mathrm{sgn}(b)}{2} \times \min(|a|, |b|)\,;
\ee
the flux limiter parameter is set to $\Theta = 1.8$~\cite{Marrochio:2013wla}. The flux limiter derivatives $(\partial_y \boldsymbol{q})_{ijk}$, $(\partial_y v^y)_{ijk}$, $(\partial_\eta \boldsymbol{q})_{ijk}$ and $(\partial_\eta v^\eta)_{ijk}$ are analogous.

In contrast, the numerical spatial derivatives appearing in the source terms $\boldsymbol{S}_{ijk}$ are approximated with second-order central differences~\cite{Bazow:2016yra,Pang:2018zzo}:
\bs
\allowdisplaybreaks
\beal
(\partial_x \boldsymbol{q})_{ijk} &= \frac{\boldsymbol{q}_{i+1,j,k} - \boldsymbol{q}_{i-1,j,k}}{2\Delta x} \,,\\
(\partial_x \boldsymbol{u})_{ijk} &= \frac{\boldsymbol{u}_{i+1,j,k} - \boldsymbol{u}_{i-1,j,k}}{2\Delta x} \,,
\end{align}
\es
and similarly for $(\partial_y \boldsymbol{q})_{ijk}$, $(\partial_y \boldsymbol{u})_{ijk}$, $(\partial_\eta \boldsymbol{q})_{ijk}$ and $(\partial_\eta \boldsymbol{u})_{ijk}$. The fluid velocity's time derivative $(\partial_\tau \boldsymbol{u})_{ijk}$ also appears in the source terms; its evaluation will be discussed in the next subsection.
\subsection{Two-stage Runge--Kutta scheme}
Because the KT algorithm~\eqref{eqchap4:KT_algorithm} admits a semi-discrete form~\cite{Kurganov:2000} (i.e. the time derivative is continuous while the spatial derivatives are discrete), we can combine it with an RK2 ODE solver to evolve the system in time~\cite{Schenke:2010nt,Bazow:2016yra}. Given the dynamical variables $\boldsymbol{q}_{n,ijk} \equiv \boldsymbol{q}_{ijk}(\tau_n)$ at time $\tau = \tau_n$ (discrete times are labeled with index $n$), we evolve the system one time step $\Delta\tau_n$ with an intermediate Euler step (omitting the spatial indices):
\be
\label{eqchap4:intermediate_step}
\boldsymbol{q}_{\,\text{I},n+1} = \boldsymbol{q}_{n} + \Delta \tau _n \boldsymbol{E}(\tau_n, \boldsymbol{q}_n, \boldsymbol{u}_n,\ene_n; \boldsymbol{u}_{n-1}, \Delta\tau_{n-1}) \,,
\ee
where $\boldsymbol{E} = \partial_\tau \boldsymbol{q}$ is evaluated with r.h.s of Eq.~\eqref{eqchap4:KT_algorithm}. The time derivative $\partial_\tau \boldsymbol{u}$ in the source terms is approximated with a first-order backward difference~\cite{Bazow:2016yra,Schenke:2010rr}:
\be
(\partial_\tau \boldsymbol{u})_n = \frac{ \boldsymbol{u}_n - \boldsymbol{u}_{n-1}}{\Delta\tau_{n-1}}\,,
\ee
where $\boldsymbol{u}_{n-1}$ is the previous fluid velocity and $\Delta\tau_{n-1}$ is the previous time step.\footnote{%
    At the start of the hydrodynamic simulation, $n=0$ or $\tau = \tau_0$, the previous time step $\Delta \tau_{n-1}$ is set to the current time step $\Delta \tau_n$. Unless stated otherwise, we also initialize the previous fluid velocity as $\boldsymbol{u}_{n-1} = \boldsymbol{u}_n$.} 
From the intermediate variables \eqref{eqchap4:intermediate_step}, we reconstruct the inferred variables $(\ene_{\text{I},n+1}, \boldsymbol{u}_{\text{I},n+1})$ as well as the anisotropic variables ($\Lambda_{\text{I},n+1}$ $\alpha_{\perp,\text{I},n+1}$, $\alpha_{L,\text{I},n+1}$) (see Secs.~\ref{chap4S3.3} and~\ref{chap4S3.4}). Afterwards, we regulate the mean-field $B$ and residual shear stresses $\Wperp$ and $\piperp$ in $\boldsymbol{q}_{\,\text{I},n+1}$ (see Sec.~\ref{chap4S3.5}) and set the ghost cell boundary conditions for $\boldsymbol{q}_{\,\text{I},n+1}$ and $\boldsymbol{u}_{\text{I},n+1}$.

Next, we evolve the system with a second intermediate Euler step
\be
\label{eqchap4:2nd_intermediate_step}
\boldsymbol{Q}_{n+2} = \boldsymbol{q}_{\,\text{I},n+1} + \Delta \tau _n \boldsymbol{E}(\tau_n {+} \Delta\tau_n, \boldsymbol{q}_{\,\text{I},n+1}, \boldsymbol{u}_{\text{I},n+1},\ene_{\text{I},n+1}; \boldsymbol{u}_n, \Delta\tau_n) \,,
\ee
where the fluid velocity's time derivative is now evaluated as
\be
(\partial_\tau \boldsymbol{u})_{\text{I},n+1} = \frac{ \boldsymbol{u}_{\text{I},n+1} - \boldsymbol{u}_n}{\Delta\tau_n}\,.
\ee
In the RK2 scheme, we average the two intermediate Euler steps in $\boldsymbol{Q}_{n+2}$ to update the dynamical variables at $\tau = \tau_n + \Delta\tau_n$:
\be
\label{eqchap4:RK2}
\boldsymbol{q}_{n+1} = \frac{\boldsymbol{q}_n + \boldsymbol{Q}_{n+2}}{2}\,.
\ee
From this, we update the inferred variables ($\ene_{n+1}$, $\boldsymbol{u}_{n+1}, \Lambda_{n+1}$, $\alpha_{\perp,n+1}$, $\alpha_{L,n+1}$) and regulate the residual shear stresses and mean-field. Finally, we set the ghost cell boundary conditions for $\boldsymbol{q}_{n+1}$ and $\boldsymbol{u}_{n+1}$ and proceed with the next RK2 iteration.

\subsection{Reconstructing the energy density and fluid velocity}
\label{chap4S3.3}
Given the hydrodynamic variables $T^{\tau\mu}$, along with $\PL$, $\Pperp$, $\Wperp$ and $\pi_\perp^{\tau\mu}$, we can reconstruct the energy density and fluid velocity from Eq.~\eqref{eqchap4:Ttaumu}, rewritten as
\be
\label{eqchap4:Ttaumu_rewrite}
  T^{\tau\mu} = (\ene{+}\Pperp)u^\tau\um - \Pperp g^{\tau\mu} + \Delta\mathcal{P} z^\tau z^\mu
                      + 2W^{(\tau}_{\perp z} z^{\mu)} + \pi_\perp^{\tau\mu}\,,
\ee
where $\Delta \mathcal{P} = \PL{-}\Pperp$. One sees that the term $2W^{(\tau}_{\perp z} z^{\mu)}$ depends (through the components of $z^\mu$) on $u^\tau$ and $u^\eta$. It cannot be subtracted from $T^{\tau\mu}$ until a relation between $u^\tau$ and $u^\eta$ is found. We define the vector $K^\mu = T^{\tau\mu} - \pi_\perp^{\tau\mu}$ whose components read \cite{Bazow:phdthesis}
\bs
\allowdisplaybreaks
\label{eqch3:57}
\beal
  &K^\tau = (\ene{+}\Pperp) (u^\tau)^2 - \Pperp + \frac{\Delta\mathcal{P}(\tau u^\eta)^2}{1{+}u_\perp^2}
                 + \frac{2 W_{\perp z}^\tau \tau u^\eta}{\sqrt{1{+}u_\perp^2}}\,, \\
  &K^i = (\ene{+}\Pperp) u^\tau u^i  +   \frac{W_{\perp z}^i \tau u^\eta}{\sqrt{1{+}u_\perp^2}} \,,
              \quad\qquad (i = x,y) \\
  &K^\eta = (\ene{+}\Pperp) u^\tau u^\eta + \frac{ \Delta \mathcal{P} u^\tau u^\eta}{1{+}u_\perp^2} 
                 + W_{\perp z}^\tau \frac{ (u^\tau)^2 {+} (\tau u^\eta)^2}{\tau u^\tau \sqrt{1{+}u_\perp^2}}\,,
\end{align}
\es
where we used Eq.~\eqref{eq:z_param} to express $z^\mu$ in terms of $u^\mu$. In the last equation we also eliminated $W_{\perp z}^\eta$ via the orthogonality relation $z_\mu \Wperp = 0$. After taking the combination $\bigl(u^\eta\big)^2 K^\tau - u^\tau u^\eta K^\eta$, one obtains
\be
\label{eqch3:58}
  \frac{u^\eta}{u^\tau} = F = \frac{A - B \sqrt{1 + \tau^2(B^2{-}A^2)}}{1+(\tau B)^2}\,,
\ee
where
\be
\label{eqch3:59}
  A = \frac{K^\eta}{K^\tau {+} \PL}\,,\qquad
  B = \frac{W^\tau_{\perp z}}{\tau\left(K^\tau {+} \PL\right)}
\ee
are both known quantities. From this, we combine Eq.~\eqref{eqch3:58} with the normalization condition $u_\mu u^\mu = 1$ to express the components $z^\tau$ and $z^\eta$ in terms of hydrodynamic quantities:\footnote{%
    Since $F = u^\eta/u^\tau < \tau^{-1}$, the argument $1-(\tau F)^2$ in Eq.~\eqref{eqchap4:z_reconstruct} is always positive.
}
\be
\label{eqchap4:z_reconstruct}
    z^\tau = \frac{\tau F}{\sqrt{1 - (\tau F)^2}} \,,\qquad
    z^\eta = \frac{1}{\tau\sqrt{1 - (\tau F)^2}}\,.
\ee
Next we define the known vector $M^\mu{\,=\,}K^\mu{-}2W^{(\tau}_{\perp z} z^{\mu)}$, with components
\bs
\allowdisplaybreaks
\label{eqch3:62}
\beal
  &\label{eqch3:62a} M^\tau = (\ene{+}\Pperp) (u^\tau)^2 - \Pperp + \mathcal{L}^{\tau\tau}\,, \\
  &\label{eqch3:62b} M^i = (\ene{+}\Pperp) u^\tau u^i\,, \quad\qquad (i = x,y) \\
  &\label{eqch3:62c} M^\eta = (\ene{+}\Pperp) u^\tau u^\eta  + \mathcal{L}^{\tau\eta}\,,
\end{align}
\es
where $\mathcal{L}^\munu = \Delta \mathcal{P} z^\mu z^\nu$ was defined earlier. From Eq.~\eqref{eqch3:62b} one immediately obtains the transverse fluid velocity components and magnitude:
\bs
\allowdisplaybreaks
\label{eqch3:63}
\beal
  &\label{eqch3:63a} u^i = \frac{M^i}{u^\tau (\ene {+} \Pperp)}, \\
  &\label{eqch3:63b} u_\perp = \frac{M_\perp}{u^\tau (\ene {+} \Pperp)} ,
\end{align}
\es
where $M_\perp = \sqrt{(M^x)^2{+}(M^y)^2}$. The two remaining unknown variables are $u^\tau$ and $\ene$. By taking the combination $ u_\perp^2 M^\tau - u^\tau (u^x M^x + u^y M^y)$ one finds a relation between $u^\tau$ and $\ene$:
\be
\label{eqch3:64}
  u^\tau = \sqrt{\frac{M^\tau + \Pperp - \mathcal{L}^{\tau\tau}}{\ene + \Pperp}}.
\ee
This can now be used to express $(u^x,u^y,u^\eta)$ in Eqs.~(\ref{eqch3:58}) and~(\ref{eqch3:63}) in terms of $\ene$ and the other known quantities. With a bit of algebra, and making use of the relation $F = M^\eta / (M^\tau{+}\PL)$, the normalization condition $(u^\tau)^2 - u_\perp^2 - (\tau u^\eta)^2 = 1$ then yields the following explicit reconstruction formula for the energy density:
\be
\label{eqchap4:energy_recon}
    \ene = M^{\tau} - \mathcal{L}^{\tau\tau} - \frac{M_\perp^2}{M^\tau {+} \Pperp {-} \mathcal{L}^{\tau\tau}} - \frac{(\tau M^\eta)^2 (M^\tau {+} \Pperp {-} \mathcal{L}^{\tau\tau})}{(M^\tau {+} \PL)^2}\,,
\ee
Note that, since $\PL$ and $\Pperp$ are evolved directly, the r.h.s. of Eq.~\eqref{eqchap4:energy_recon} is entirely known. This is in contrast with the reconstruction formula for $\ene$ in second-order viscous hydrodynamics (see App.~\ref{app4:energy}), where one must solve some nonlinear equation $f(\ene) = 0$ numerically \cite{Shen:2014vra,Bazow:phdthesis}. 

In the cold, dilute regions surrounding the fireball (and, occasionally, in cold spots within the fluctuating fireball), the energy density can become much smaller than the freezeout energy density $\ene_\text{sw}$.\footnote{%
    In this chapter, we construct a particlization hypersurface of constant energy density $\ene_\text{sw} = 0.116$ GeV/fm$^3$, which corresponds to the switching temperature $T_\text{sw} = 0.136$ GeV. The lowest value used for the switching temperature in the JETSCAPE SIMS analysis is $T_\text{sw} = 0.135$ GeV~\cite{Everett:2020yty,Everett:2020xug}.} 
While these regions are not phenomenologically important, hydrodynamic simulations are susceptible to crashing there without intervention \cite{Shen:2014vra, Bazow:2016yra, Denicol:2018wdp}. To prevent this, we regulate the energy density with the formula
\be
\label{eqchap4:energy_reg}
    \ene \leftarrow \ene_+ + \ene_\text{min}\, e^{-\ene_+ / \ene_\text{min}}\,,
\ee
where $\ene_\text{min}$ is the minimum energy density allowed in the Eulerian grid and $\ene_+ = \max(0, \ene)$. For $\ene \gg \ene_\text{min}$, the regulation has virtually no effect on the energy density. As $\ene \to 0$, however, the energy density is smoothly regulated to $\ene_\text{min}$. Ideally, $\ene_\text{min}$ should be the lower limit of our QCD equation of state table $\ene_\text{low} = 3.5 \times 10^{-4}$ GeV/fm$^3$ (or $T_\text{low} = 0.05$ GeV).\footnote{
    Since we impose the cutoff $\ene \geq \ene_\text{min}$ to increase the code's stability, we never need to extrapolate the QCD equation of state to temperatures below $T_\text{low} = 0.05$\,GeV where the data table ends.}
However, we find that the code evolving non-conformal anisotropic hydrodynamics with fluctuating initial conditions encounters fewer technical difficulties if we instead use the larger value $\ene_\text{min} = 0.02$ GeV/fm$^3$, which is still about six times smaller than our choice for $\ene_\text{sw}$.\footnote{%
    The constraint $\ene \geq \ene_\text{min}$ is not imposed for conformal systems.}
We checked that the regulation scheme~\eqref{eqchap4:energy_reg} has little to no impact on the fluid's dynamics in regions where $\ene \sim \ene_\text{sw}$ or larger.

After regulating the energy density, we reconstruct the fluid velocity's spatial components using Eqs.~\eqref{eqch3:64}, \eqref{eqch3:63} and (\ref{eqch3:58}), rewritten as 
\bs
\allowdisplaybreaks
\beal
u^x &= \frac{M^x}{\sqrt{(\ene+\Pperp)(M^\tau + \Pperp - \mathcal{L}^{\tau\tau})}} \,,\\
u^y &= \frac{M^y}{\sqrt{(\ene+\Pperp)(M^\tau + \Pperp - \mathcal{L}^{\tau\tau})}} \,,\\
u^\eta &= F \sqrt{\frac{M^\tau + \Pperp - \mathcal{L}^{\tau\tau}}{\ene+\Pperp}} \,.
\end{align}
\es
Because of the regulation~\eqref{eqchap4:energy_reg}, there will be inconsistencies between the inferred variables ($\ene$, $\boldsymbol{u}$) and the components $T^{\tau\mu}$ but only in the cold dilute regions. 
%
%
\subsection{Reconstructing the anisotropic variables}
\label{chap4S3.4}

To compute the non-conformal anisotropic transport coefficients in Sec.~\ref{chap4S2.6.3}, we need to reconstruct the anisotropic variables
\be
  \boldsymbol{X} = \left(
    \begin{array}{c}
  \Lambda \\ \alpha_\perp \\ \alpha_L 
  \end{array}
  \right)
\ee
from $\ene$, $\PL$, $\Pperp$, $B$ and $m(\ene)$, by solving the system of equations~\eqref{eqch3:53}
\be
\label{eqchap4:nonlinear_system}
    \boldsymbol{f}(\boldsymbol{X}) = \boldsymbol{0}\,,
\ee
where
\be
\label{eqchap4:aniso_root_equation}
\boldsymbol{f}(\boldsymbol{X}) = \left(
  \begin{array}{c}
  \I_{2000}(\boldsymbol{X}) - \ene + B \\ \I_{2200}(\boldsymbol{X}) - \PL - B \\ \I_{2010}(\boldsymbol{X})-\Pperp-B 
  \end{array}
  \right).
\ee
We solve these nonlinear equations numerically using Newton's method. Taking the anisotropic variables prior to the intermediate Euler step~\eqref{eqchap4:intermediate_step} as our initial guess\footnote{
    At $\tau = \tau_0$ we initialize the anisotropic variables ($\Lambda_0$, $\alpha_{\perp,0}$, $\alpha_{L,0}$) by iterating $\boldsymbol{X}_g = (T, 1, 1)^T$ (where $T$ inside the parentheses is the temperature).} 
$\boldsymbol{X}_g = (\Lambda_n,\alpha_{\perp,n}, \alpha_{L,n})^T$, we iterate $\boldsymbol{X}$ along the direction given by the Newton step
\be
    \Delta \boldsymbol{X} = - \boldsymbol{J}^{-1} \boldsymbol{f}\,,
\ee
where $\boldsymbol{J}^{-1}$ is the inverse of the Jacobian
\be
\label{eqchap4:jacobian}
\boldsymbol{J} = \frac{\partial \boldsymbol{f}}{\partial \boldsymbol{X}} =    \left(
  \begin{array}{c c c}
      \dfrac{\I_{2001}(\boldsymbol{X})}{\Lambda^2} \, & \,\dfrac{2\,\I_{401-1}(\boldsymbol{X})}{\Lambda \alpha_\perp^3} \, & \, \dfrac{\I_{420-1}(\boldsymbol{X})}{\Lambda \alpha_L^3} \\ \\
              \dfrac{\I_{2201}(\boldsymbol{X})}{\Lambda^2}\, & \, \dfrac{2\, \I_{421-1}(\boldsymbol{X})}{\Lambda \alpha_\perp^3} \, 
                                                                          & \, \dfrac{\I_{440-1}(\boldsymbol{X})}{\Lambda \alpha_L^3} \\ \\
        \dfrac{\I_{2011}(\boldsymbol{X})}{\Lambda^2} \, & \, \dfrac{4\,\I_{402-1}(\boldsymbol{X})}{\Lambda \alpha_\perp^3} \,                                                      & \, \dfrac{\I_{421-1}(\boldsymbol{X})}{\Lambda \alpha_L^3} \\
  \end{array}
  \right)\,.
\ee
We can simplify the computation of a few matrix elements in Eq.~\eqref{eqchap4:jacobian} by using the identities\footnote{
    The numerical evaluation of Eq.~(\ref{eqchap4:jacobian_ids}c) is problematic in the isotropic limit $\alpha_\perp = \alpha_L$. In this case, it is better to compute $\I_{421-1}$ directly using the method outlined in App.~\ref{appch3a}.}
(see App.~\ref{appch3b})
\bs
\label{eqchap4:jacobian_ids}
\beal
  &\I_{420-1} = \Lambda \alpha_L^2 (\I_{2000}+\I_{2200}), \\
  &\I_{401-1} = \Lambda \alpha_\perp^2 (\I_{2000}+\I_{2010}), \\
  &\I_{421-1} = \frac{\Lambda \alpha_\perp^2 \alpha_L^2}{\alpha_L^2{-}\alpha_\perp^2} (\I_{2200}-\I_{2010}).
\end{align}
\es
Specifically, we iterate the solution as
\be
\boldsymbol{X} \leftarrow \boldsymbol{X} + \lambda \,\Delta\boldsymbol{X}\,,
\ee
where $\lambda \in [0,1]$ is a partial step that is optimized with a line-backtracking algorithm to improve the global convergence of Newton's method~\cite{Press:1992:NRC:148286}. This procedure is repeated until $\boldsymbol{X}$ has converged to the solution of Eq.~\eqref{eqchap4:nonlinear_system}.\footnote{%
    In our simulation tests, we found several instances when Newton's method failed to converge near the edges of the Eulerian grid. Whenever that happens we simply set the anisotropic variables to the initial guess $\boldsymbol{X}_g$.} 

For conformal systems ($B = 0$, $m = 0$, $\alpha_\perp = 1$), the anisotropic variables $\Lambda$ and $\alpha_L$ are much easier to solve. The system of equations \eqref{eqchap4:aniso_root_equation} reduces to
\bs
\allowdisplaybreaks
\label{eqchap4:aniso_eqs_conformal}
\beal
    \ene &= \frac{3 g \Lambda^4 \mathcal{R}_{200}(\alpha_L)}{2\pi^2} \,,
\\
    \PL &= \frac{g \Lambda^4 \mathcal{R}_{220}(\alpha_L)}{2\pi^2} \,,
\end{align}
\es
where the functions $\mathcal{R}_{200}$ and $\mathcal{R}_{220}$ are listed in Appendix~\ref{app4c}. We numerically invert the longitudinal pressure to energy density ratio for $\alpha_L$:
\be
\label{eqchap4:aL_conformal}
    \frac{\mathcal{R}_{220}(\alpha_L)}{\mathcal{R}_{200}(\alpha_L)} = \frac{\PL}{\ene} \,.
\ee
From this, we can evaluate $\Lambda$ as
\be
\label{eqchap4:Lambda_conformal}
    \Lambda = \bigg(\frac{\pi^2\ene}{3g\mathcal{R}_{200}(\alpha_L)}\bigg)^{1/4}\,.
\ee
Because the solutions~\eqref{eqchap4:aL_conformal} and~\eqref{eqchap4:Lambda_conformal} do not require the previous values for $\Lambda$ and $\alpha_L$ as input, we do not need to store them in memory during runtime.  
\subsection{Regulating the residual shear stresses and mean-field}
\label{chap4S3.5}
After reconstructing the inferred variables, we regulate the residual shear stress components $\Wperp$ and $\piperp$ such that they satisfy the orthogonality and traceless conditions
\bs
\allowdisplaybreaks
\label{eqchap4:enforce}
\beal
    u_\mu \Wperp &= 0 \,,
\\
    z_\mu \Wperp &= 0 \,,
\\
    u_\mu \piperp &= 0 \,,
\\
    z_\mu\piperp &= 0 \,,
\\
    \pi^\mu_{\perp,\mu} &= 0\,,
\end{align}
\es
and that their overall magnitude is smaller than the longitudinal and transverse pressures:
\be
\label{eqchap4:regulate_residual}
\sqrt{\pi_{\perp,\munu} \pi_\perp^\munu - 2 W_{\perp z,\mu} \Wperp} \leq \sqrt{\mathcal{P}_L^2 + 2\mathcal{P}_\perp^2}\,.
\ee
The former ensures that $\Wperp$ and $\piperp$ maintain their orthogonal and tracelessness properties within numerical accuracy during the hydrodynamic simulation. The latter prevents the residual shear stresses from overwhelming the anisotropic part of the energy-momentum tensor. Although Eq.~\eqref{eqchap4:regulate_residual} is not a mathematically required condition, we implement it because  negative transverse pressures could develop in regions with sharp transverse fluctuations and cause the code to break down~\cite{Bazow:2017ewq}.

First, we update these components in the following order:
\bs
\allowdisplaybreaks
\beal
W_{\perp z}^\tau &\leftarrow \frac{u^\tau(W_{\perp z}^x u^x + W_{\perp z}^y u^y)}{1+u_\perp^2}\,,\\
W_{\perp z}^\eta &\leftarrow \frac{W_{\perp z}^\tau u^\eta}{u^\tau}\,,\\
\pi_\perp^{yy} &\leftarrow \frac{2 \pi_\perp^{xy} u^x u^y - \pi_\perp^{xx}\left(1 {+} (u^y)^2\right)}{1 + (u^x)^2} \,,\\
\pi_\perp^{\tau x} &\leftarrow \frac{u^\tau(\pi_\perp^{xx} u^x + \pi_\perp^{xy}u^y)}{1+u_\perp^2}\,,\\
\pi_\perp^{\tau y} &\leftarrow \frac{u^\tau(\pi_\perp^{xy} u^x + \pi_\perp^{yy}u^y)}{1+u_\perp^2}\,,\\
\pi_\perp^{x\eta} &\leftarrow \frac{\pi_\perp^{\tau x}u^\eta}{u^\tau} \,,\\
\pi_\perp^{y\eta} &\leftarrow \frac{\pi_\perp^{\tau y}u^\eta}{u^\tau} \,,\\
\pi_\perp^{\tau\eta} &\leftarrow \frac{u^\tau(\pi_\perp^{x\eta}u^x + \pi_\perp^{y\eta}u^y)}{1+u_\perp^2} \,,\\
\pi_\perp^{\eta\eta} &\leftarrow \frac{\pi_\perp^{\tau \eta}u^\eta}{u^\tau}\,,\\
\pi_\perp^{\tau\tau} &\leftarrow \frac{\pi_\perp^{\tau x} u^x + \pi_\perp^{\tau y}u^y + \tau^2 \pi_\perp^{\tau\eta} u^\eta}{u^\tau} \,,
\end{align}
\es
while leaving the components $W_{\perp z}^x$, $W_{\perp z}^y$, $\pi_{\perp}^{xx}$ and $\pi_{\perp}^{xy}$ unchanged. Next, we rescale all the components by the same factor $\rho_\text{reg} \in [0,1]$:
\bs
\allowdisplaybreaks
\beal
\Wperp &\leftarrow \rho_\text{reg} \Wperp \,,\\
\piperp &\leftarrow \rho_\text{reg} \piperp\,,
\end{align}
\es
where\footnote{%
    For longitudinally boost-invariant systems, we replace the second argument in Eq.~\eqref{eqchap4:rescale_residual} by $\sqrt{2 \mathcal{P}_\perp^2 / \pi_\perp {\cdot\,} \pi_\perp}$.}
\be
\label{eqchap4:rescale_residual}
\rho_\text{reg} = \min\Bigg(1, \sqrt{\frac{\mathcal{P}_L^2 + 2\mathcal{P}_\perp^2}{\pi_\perp {\cdot\,} \pi_\perp - 2 W_{\perp z}{\,\cdot\,} W_{\perp z}}}\Bigg) \,,
\ee
with $\pi_\perp {\cdot\,} \pi_\perp = \pi_{\perp,\munu} \pi_\perp^\munu$. In the comparison tests discussed in Sec.~\ref{chap4S4}, we did not encounter a situation where the residual shear stresses are suppressed by Eq.~\eqref{eqchap4:rescale_residual}. This indicates that the residual shear stresses are naturally smaller than the leading-order anisotropic pressures during the simulation. Nevertheless, we keep this procedure in place as a precaution.

For hydrodynamic simulations with two or three spatial dimensions, we find that the relaxation equation \eqref{eqchap4:relax_B_0} for the mean-field $B$ can become unstable in certain spacetime regions ($T \sim 0.15 - 0.16$ GeV), causing the bulk viscous pressure $\Pi$ to grow positive without bound. The origin of this issue is not fully understood, and we will address it in future work. For now, we remove this instability by regulating the non-equilibrium mean-field component $\delta B = B - B_\text{eq}$ as
\be
\delta B \leftarrow \kappa_\text{reg} \delta B,
\ee
where
\be
\label{eqchap4:dB_reg}
\kappa_\text{reg} = \min\Big(1, -\frac{|B_\text{eq}|}{\delta B}\Big) \indent \forall \,\, \delta B < 0\,.
\ee
In practice, we only find it necessary to regulate the mean-field when $\delta B < 0$ (i.e. in regions with $\Pi > 0$).
%
%
%
\subsection{Adaptive time step}
\label{chap4S3.6}
Most relativistic hydrodynamic codes that simulate heavy-ion collisions evolve the system with a fixed time step $\Delta \tau_n = \Delta \tau$~\cite{Schenke:2010nt,Shen:2014vra}. Any choice for $\Delta \tau$ must satisfy the CFL condition so that the hydrodynamic simulation is at least dynamically stable \cite{Kurganov:2000}. However, the time step must also be small enough to resolve the fluid's evolution rate (in particular, the large longitudinal expansion rate $\theta_L \sim 1/\tau$ at early times) but, on the other hand, large enough to finish the simulation within a reasonable runtime. This balancing act places a practical limit on how early the user can start the hydrodynamic simulation (in practice, the smallest value typically used is $\tau_0 \sim 0.2$ fm/$c$ \cite{Gale:2012rq}). This is not much of a concern for second-order viscous hydrodynamics since it is anyhow prone to breaking down for very early initialization times, as discussed in Chapter~\ref{ch1label}. Anisotropic hydrodynamics, on the other hand, can handle the large pressure anisotropies occurring at early times much better; to realize its full potential it should be initialized at earlier times, but for that we need to move away from a fixed time step. In this section, we introduce a new adaptive stepsize method, which automatically adjusts the successive time step $\Delta \tau_{n+1}$ to be larger or smaller than $\Delta\tau_n$ in such a way that we can push back our fluid dynamical simulation to very early times ($\tau_0 \sim 0.01 - 0.05 \,\text{fm}/c$) without sacrificing numerical accuracy nor computational efficiency.\footnote{%
    This is also useful for constructing the particlization hypersurface in peripheral heavy-ion collisions or small collision systems (e.g. p+p), whose fireball lifetimes are not that much longer than the hydrodynamization time $\tau_\text{hydro}$ (see Chapter~\ref{ch1label}).}

For a system with no source terms ($\boldsymbol{S} = \boldsymbol{0}$) the KT algorithm is dynamically stable as long as the time step satisfies the CFL condition \cite{Kurganov:2000}
\be
\label{eqchap4:CFL_bound}
    \Delta \tau_n \leq \Delta\tau_\text{CFL} = \frac{1}{8} \min\left(\frac{\Delta x}{s^x_\text{max}(\tau_n)}, \frac{\Delta y}{s^y_\text{max}(\tau_n)}, \frac{\Delta\eta_s}{s^\eta_\text{max}(\tau_n)} \right)\,,
\ee
where $s^i_\text{max}(\tau_n)$ ($i=x,y,\eta$) are the maximum local propagation speed components in the Eulerian grid at time $\tau_n$. In Milne spacetime, the quark-gluon plasma's flow profile is not ultrarelativistic throughout most of its evolution (as long as one uses a QCD equation of state). Thus, the criterium \eqref{eqchap4:CFL_bound} allows one to maintain dynamical stability with an adaptive time step that is generally larger than the fixed time step obtained by taking the limit $s^i_\text{max} \to 1$:
\be
\label{eqchap4:CFL_fixed}
    \Delta \tau_n = \frac{1}{8} \min\left(\Delta x, \Delta y, \Delta \eta_s\right)\,.
\ee
For $\boldsymbol{S} \neq \boldsymbol{0}$, however, the time step $\Delta\tau_\text{CFL}$ from (\ref{eqchap4:CFL_bound}) is too coarse to resolve the fluid's gradients and relaxation rates at early times $\tau < 0.5$ fm/$c$ when the flow profile is very nonrelativistic. Therefore, at early times when the source terms are strongest, the adaptive time step should primarily depend on those source terms. The following implementation ensures that, as the source terms relax and the fluid velocity grows more relativistic over time, the adaptive time step naturally approaches the CFL bound \eqref{eqchap4:CFL_bound}.

First, we consider a homogeneous fluid undergoing Bjorken expansion (i.e. $\ene = T^{\tau\tau}$ and $\boldsymbol{u} = \boldsymbol{0}$). The system of differential equations~\eqref{eqchap4:KT_eqs} for the dynamical variables $\boldsymbol{q}(\tau)$ reduces to
\be
    \partial_\tau \boldsymbol{q}(\tau) = \boldsymbol{S}(\tau, \boldsymbol{q}),
\ee
and we are given the initial values $\boldsymbol{q}_n$ and time step $\Delta \tau_n$ at time $\tau_n$. We evolve the system one time step $\Delta \tau_n$ using the RK2 scheme~\eqref{eqchap4:RK2}:
\be
    \boldsymbol{q}_{n+1} = \boldsymbol{q}_n + \frac{\Delta \tau_n}{2}\left(\boldsymbol{S}(\tau_n, \boldsymbol{q}_n) + \boldsymbol{S}(\tau_n {+} \Delta \tau_n, \boldsymbol{q}_n {+} \Delta \tau_n \boldsymbol{S}(\tau_n, \boldsymbol{q}_n) \right)\,.
\ee
For the next iteration, we determine how much we need to adjust the time step $\Delta \tau_{n+1}$. Adaptive stepsize methods do this by estimating the local truncation error of each time step~\cite{Press:1992:NRC:148286}. In our method, we approximate the local truncation error of the next intermediate Euler step
\be
\label{eqchap4:next_Euler}
    \boldsymbol{q}_{\,\text{I},n+2} = \boldsymbol{q}_{n+1} + \Delta \tau_{n+1} \boldsymbol{S}(\tau_n {+} \Delta \tau_n, \boldsymbol{q}_{n+1})\,,
\ee
which is
\be
    \epsilon_{n+2} = \frac{1}{2}\norm{(\partial_\tau^2\boldsymbol{q})_{n+1}} \Delta \tau_{n+1}^2 + O(\Delta \tau_{n+1}^3)\,,
\ee
where $\norm{...}$ denotes the $\ell^2$-norm. If the second time derivative $(\partial^2_\tau\boldsymbol{q})_{n+1}$ at $\tau = \tau_n + \Delta\tau_n$ is known, we can set the local truncation error to the desired error tolerance (after dropping higher-order terms)
\be
\label{eqchap4:tolerance}
    \frac{1}{2}\norm{(\partial_\tau^2\boldsymbol{q})_{n+1}} \Delta \tau_{n+1}^2 = \delta_0 \times \max(N_q^{1/2}, \,\norm{\boldsymbol{q}_{\,\text{I},n+2}})\,,
\ee
where $\delta_0$ is the error tolerance parameter and $N_q$ is the number of dynamical variables. It is reasonable to use absolute or relative errors when $\norm{\boldsymbol{q}_{\,\text{I},n+2}}$ is small or large, respectively~\cite{Press:1992:NRC:148286}. We set the error tolerance parameter to $\delta_0 = 0.004$. 

To obtain an expression for the second time derivative, we compute the next intermediate Euler step~\eqref{eqchap4:next_Euler} using the old time step (denoted by $\star$):
\be
\label{eqchap4:next_Euler_star}
    \boldsymbol{q}^\star_{\,\text{I},n+2} = \boldsymbol{q}_{n+1} + \Delta \tau_n \boldsymbol{S}(\tau_n {+} \Delta \tau_n, \boldsymbol{q}_{n+1})\,.
\ee
This allows us to approximate $(\partial_\tau^2\boldsymbol{q})_{n+1}$ with central differences:
\be
\label{eqchap4:approx_2nd_derivative}
    (\partial_\tau^2\boldsymbol{q})_{n+1} = \frac{2(\boldsymbol{q}^\star_{\,\text{I},n+2} {-} 2 \boldsymbol{q}_{n+1} {+} \boldsymbol{q}_n)}{\Delta \tau_n^2} + O(\Delta\tau_n)\,.
\ee
Compared to the usual central difference formula there is an additional factor of $2$ that accounts for the local truncation error present in $\boldsymbol{q}_{\,\text{I},n+2}^\star$. Furthermore, the expression \eqref{eqchap4:approx_2nd_derivative} is numerically accurate to $O(\Delta\tau_n)$ rather than $O(\Delta\tau_n^2)$. After substituting Eqs.~\eqref{eqchap4:next_Euler} and \eqref{eqchap4:approx_2nd_derivative} in Eq.~\eqref{eqchap4:tolerance}, one has for the next time step
\be
\label{eqchap4:next_step}
    \Delta \tau_{n+1} = \max\big(\Delta \tau_{n+1}^\text{(abs)}, \Delta \tau_{n+1}^\text{(rel)}\big)\,,
\ee
where
\be
    \Delta\tau_{n+1}^\text{(abs)} = \Delta\tau_n \sqrt{\dfrac{\delta_0\, N_q^{1/2}}{\norm{\boldsymbol{q}^\star_{\,\text{I},n+2} {-} 2 \boldsymbol{q}_{n+1} {+} \boldsymbol{q}_n}}}
\ee
and $\Delta\tau_{n+1}^\text{(rel)}$ is the numerical solution to the algebraic equation 
\be
    \frac{N_q^{1/2}\big(\Delta \tau_{n+1}^\text{(rel)}\big)^2}{\big(\Delta \tau_{n+1}^\text{(abs)}\big)^2} = \sqrt{\norm{\boldsymbol{q}_{n+1}}^2 + 2 \boldsymbol{q}_{n+1} {\cdot\,} \boldsymbol{S}_{n+1} \Delta \tau_{n+1}^\text{(rel)} + \norm{\boldsymbol{S}_{n+1}}^2 \big(\Delta \tau_{n+1}^\text{(rel)}\big)^2} \,,
\ee
with $\boldsymbol{S}_{n+1} = \boldsymbol{S}(\tau_n {+} \Delta\tau_n, \boldsymbol{q}_{n+1})$. 

For the general case without Bjorken symmetry, $\boldsymbol{q}(x)$ varies across the grid. We then perform the calculation \eqref{eqchap4:next_step} (replacing $\boldsymbol{S}$ by $\boldsymbol{E}$) for all spatial grid points and take the minimum value. Afterwards, we place safety bounds to prevent the time step from changing too rapidly:
\be
    \left(1 {-} \alpha\right) \Delta\tau_n \leq \Delta \tau_{n+1} \leq (1 {+} \alpha)\Delta\tau_n \,,
\ee
where we set the control parameter to $\alpha = 0.5$. Finally, we impose the CFL bound~\eqref{eqchap4:CFL_bound} on $\Delta \tau_{n+1}$. With the new time step at hand, we resume computing the next RK2 iteration. Notice that there are no additional numerical evaluations of the flux and source terms in the adaptive RK2 scheme since we can recompute the next intermediate Euler step $\boldsymbol{q}_{\,\text{I},n+2}$ simply by adjusting the time step in $\boldsymbol{q}_{\,\text{I},n+2}^\star$. This allows our adaptive time step algorithm to be readily integrated into our numerical scheme.

When we start the hydrodynamic simulation, we initialize the time step $\Delta \tau_0$ to be 20 times smaller than the initial time $\tau_0$.\footnote{\label{adaptive_floor}%
    We do not allow the adaptive time step to become any smaller than this (i.e. we impose $\Delta\tau_n \geq 0.05\tau_0$).} 
At first, the adaptive time step $\Delta \tau_n$ tends to increase with time since the fluid's longitudinal expansion rate decreases. As the transverse flow builds up, $\Delta \tau_n$ eventually becomes bounded by the CFL condition~\eqref{eqchap4:CFL_bound}.
\subsection{Program summary}
\label{chap4S3.7}
\begin{figure}[!t]
\begin{center}
\includegraphics[width=0.8\linewidth]{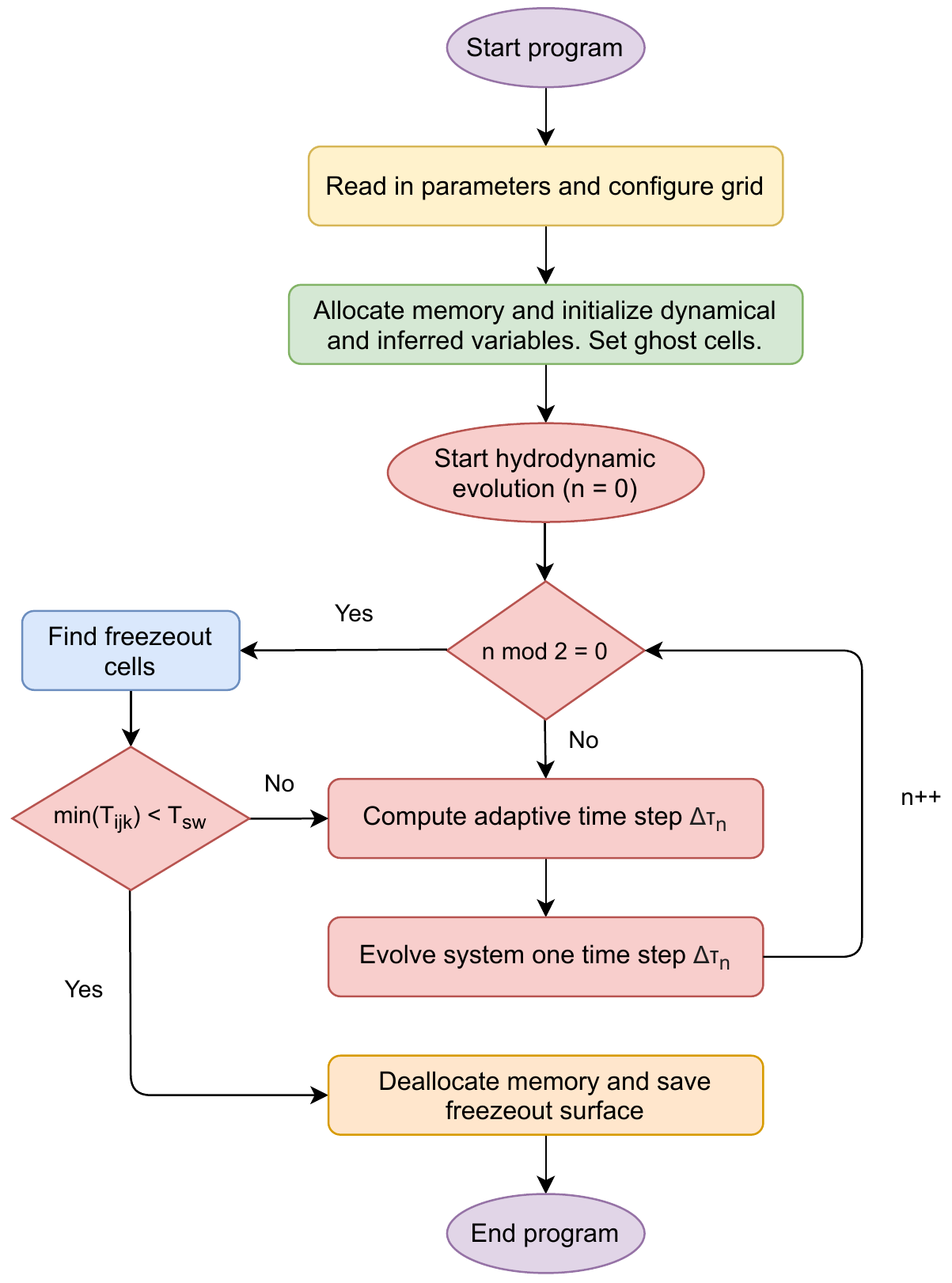}
\end{center}
\caption{(Color online)
\label{flowchart}
Program flowchart of \cpuvah.
}
\end{figure}

We close this section by summarizing the workflow of the anisotropic fluid dynamical simulation with a QCD equation of state. Figure~\ref{flowchart} shows the flowchart of the code's primary mode,\footnote{%
    The secondary (test) mode outputs the hydrodynamic evolution from the simulation and their corresponding semi-analytic solutions, if they exist.}
which constructs a hypersurface of constant energy density $\ene_\text{sw}$ for a particlization module:
\begin{enumerate}
    \item We read in the runtime parameters and configure the Eulerian grid.
    \item We allocate memory to store the dynamical and inferred variables in the Eulerian grid at a given time $\tau_n$. Altogether, there are three dynamical variables ($\texttt{q}$, $\texttt{qI}$, $\texttt{Q}$), two fluid velocity variables ($\texttt{u}$, $\texttt{up}$), one energy density variable $\texttt{e}$ and three anisotropic variables ($\texttt{lambda}$, $\texttt{aT}$, $\texttt{aL}$). They store one of the following variables during the RK2 iteration:
    \begin{enumerate}
    \allowdisplaybreaks
        \item \texttt{q} holds the current or updated dynamical variables $\boldsymbol{q}_n$ or $\boldsymbol{q}_{n+1}$.
        \item \texttt{qI} holds the intermediate dynamical variables $\boldsymbol{q}_{\,\text{I},n+1}$ (or $\boldsymbol{q}^\star_{\,\text{I},n+1}$).
        \item \texttt{Q} holds the previous, updated or current dynamical variables $\boldsymbol{q}_{n-1}$, $\boldsymbol{q}_{n+1}$ or $\boldsymbol{q}_n$.
        \item \texttt{u} holds the current, intermediate or updated fluid velocity $\boldsymbol{u}_n$, $\boldsymbol{u}_{\text{I},n+1}$ or $\boldsymbol{u}_{n+1}$.
        \item \texttt{up} holds the previous or current fluid velocity $\boldsymbol{u}_{n-1}$ or $\boldsymbol{u}_n$.
        \item \texttt{e} holds the current, intermediate or updated energy density $\ene_n$, $\ene_{\text{I},n+1}$ or $\ene_{n+1}$.
        \item \texttt{lambda}, \texttt{aT} and \texttt{aL} hold the current,  intermediate or updated anisotropic variables ($\Lambda_n$, $\alpha_{\perp,n}$, $\alpha_{L,n}$), ($\Lambda_{\text{I},n+1}$, $\alpha_{\perp,\text{I},n+1}$, $\alpha_{L,\text{I},n+1}$) or ($\Lambda_{n+1}$, $\alpha_{\perp,n+1}$, $\alpha_{L,n+1}$)
    \end{enumerate}
    
    \item We initialize the variables \texttt{q}, \texttt{u}, \texttt{up}, \texttt{e}, \texttt{lambda}, \texttt{aT} and \texttt{aL} as follows:\footnote{%
        If the code is run with Gubser initial conditions, the hydrodynamic variables are initialized differently (see Sec.~\ref{chap4S4.2}).}
    first, we compute (or read in) the energy density profile given by an initial-state model (e.g. \trento{} \cite{Moreland:2014oya}). Then we initialize the longitudinal and transverse pressures as
    \bs
    \allowdisplaybreaks
    \beal
    \PL &= \frac{3 R\,\Peq}{2+R} \,,\\
    \Pperp &= \frac{3\Peq}{2+R} \,,
    \end{align}
    \es
    where $R \in [0,1]$ is a pressure anisotropy ratio parameter set by the user (typically a small value $R \leq 0.3$). This model assumes that only the pressure anisotropy $\PL-\Pperp$ enters in the initial $\mathcal{P}_L$ and $\mathcal{P}_\perp$ while the initial bulk viscous pressure is $\Pi = 0$. 
    The initial fluid velocity is static (i.e. $\texttt{u} = \texttt{up} = \boldsymbol{0}$) and the residual shear stresses $\Wperp$ and $\piperp$ are set to zero. From this, we compute the components $T^{\tau\mu}$ with Eq.~\eqref{eqchap4:Ttaumu}. Finally, we set the mean-field to $B = B_\text{eq}$ and initialize the anisotropic variables by solving Eq.~\eqref{eqchap4:nonlinear_system}.\footnote{%
        This initialization scheme is similar to how conformal free-streaming modules are initialized at $\tau_0 \to 0$~\cite{Liu:2015nwa, Bernhard:2016tnd, Bernhard:2018hnz, Everett:2020yty, Everett:2020xug}.
        }
    
    \item We set the ghost cell boundary conditions for \texttt{q}, \texttt{e} and \texttt{u}.\footnote{%
        The ghost cells of \texttt{e} are only used in conformal anisotropic hydrodynamics to approximate the source terms $\propto \partial_i \ene$ on the grid's faces.}
        We also configure the freezeout finder and set the initial time step to $\Delta\tau_0 = 0.05\, \tau_0$.
    \item We start the hydrodynamic evolution at $\tau = \tau_0$ (or $n = 0$):
    \begin{enumerate}
        \item We call the freezeout finder every two time steps\footnote{The user can adjust the number of time steps between freezeout finder calls.} (i.e. $n \mod 2  = 0$) to search for freezeout cells between the Eulerian grids from the current and previous calls (if $n = 0$, we only load the initial grid to the freezeout finder). The hypersurface volume elements $d^3\sigma_\mu$ and their spacetime positions are constructed with the code {\sc CORNELIUS}~\cite{Huovinen:2012is}. The hydrodynamic variables at the hypersurface elements' centroid are approximated with a 4d linear interpolation.
        \item We compute the intermediate Euler step~\eqref{eqchap4:intermediate_step} and store the results $\boldsymbol{q}_{\,\text{I},n+1}$ in $\texttt{qI}$ (if $n>0$, we compute $\boldsymbol{q}^\star_{\,\text{I},n+1}$ and the adaptive time step $\Delta\tau_n$ using the algorithm in Sec.~\ref{chap4S3.6} before evaluating $\boldsymbol{q}_{\,\text{I},n+1}$). Afterwards, we swap the variables $\texttt{u} \leftrightarrow \texttt{up}$ so that $\texttt{up} \leftarrow \boldsymbol{u}_n$. 
        \item We reconstruct the intermediate inferred variables $\ene_{\text{I},n+1}$, $\boldsymbol{u}_{\text{I},n+1}$, $\Lambda_{\text{I},n+1}$, $\alpha_{\perp,\text{I},n+1}$ and $\alpha_{L,\text{I},n+1}$ from \texttt{qI} and store the results in \texttt{e}, \texttt{u}, \texttt{lambda}, \texttt{aT} and \texttt{aL}, respectively. We regulate the residual shear stresses and mean-field in \texttt{qI} and set the ghost cell boundary conditions for \texttt{qI}, \texttt{e} and \texttt{u}.
        \item We compute the second intermediate Euler step~\eqref{eqchap4:2nd_intermediate_step} and update the dynamical variables $\boldsymbol{q}_{n+1}$ via Eq.~\eqref{eqchap4:RK2}, which is stored in \texttt{Q}. Afterwards, we swap the variables $\texttt{q} \leftrightarrow \texttt{Q}$ so that $\texttt{q} \leftarrow \boldsymbol{q}_{n+1}$ and $\texttt{Q} \leftarrow \boldsymbol{q}_n$.
        \item We update the inferred variables $\ene_{n+1}$, $\boldsymbol{u}_{n+1}$, $\Lambda_{n+1}$, $\alpha_{\perp,n+1}$ and $\alpha_{L,n+1}$ from \texttt{q} and store the results in \texttt{e}, \texttt{u}, \texttt{lambda}, \texttt{aT} and \texttt{aL}, respectively. We regulate the residual shear stresses and mean-field in \texttt{q} and set the ghost cell boundary conditions for \texttt{q}, \texttt{e} and \texttt{u}.
        \item Steps (a) -- (e) are repeated until the temperature of all fluid cells in the grid are below the switching temperature (i.e. $\min(T_{ijk}) < T_\text{sw}$).
    \end{enumerate}
    \item After the hydrodynamic evolution, we deallocate the hydrodynamic variables and store the freezeout surface in memory, which can either be passed to another program or written to file. 
\end{enumerate}
\section{Validation tests and hydrodynamic model comparisons}
\label{chap4S4}

In this section we test the validity of our anisotropic fluid dynamical simulation for various initial-state configurations, using either the conformal or QCD equation of state. First, we run anisotropic hydrodynamics with conformal Bjorken and Gubser initial conditions \cite{Bjorken:1982qr, Gubser:2010ui, Gubser:2010ze}, whose semi-analytic solutions can be computed accurately using a fourth-order Runge--Kutta ODE solver \cite{Molnar:2016gwq, Martinez:2017ibh}. After these two validation tests, we compare (3+1)--dimensional conformal anisotropic hydrodynamics and second-order viscous hydrodynamics\footnote{\label{VH12}%
    The \cpuvah{} code can also run second-order viscous hydrodynamics, where the kinetic transport coefficients are computed with either quasiparticle masses $m(T)$ (see Fig.~\ref{quasi}a) \cite{Tinti:2016bav, McNelis:2018jho} or light masses (i.e. $m/T \ll 1$)~\cite{Denicol:2014vaa,Bazow:2016yra} (we use the same shear and bulk viscosities as in Fig.~\ref{viscosities}). We label the former model as \textit{quasiparticle viscous hydrodynamics} (or \vh{}) and the latter as \textit{standard viscous hydrodynamics} (or \vh{}2); for conformal systems, these two models are equivalent. For details on the numerical implementation of second-order viscous hydrodynamics, see Appendix \ref{app4e}.} 
for a central Pb+Pb collision with a smooth \trento{} initial condition \cite{Moreland:2014oya}.

Next, we run non-conformal anisotropic hydrodynamics and viscous hydrodynamics with Bjorken initial conditions and compare them to their semi-analytic solutions \cite{McNelis:2018jho}. Finally, we study the differences between (3+1)--dimensional non-conformal anisotropic hydrodynamics and viscous hydrodynamics in central Pb+Pb collisions with smooth or fluctuating \trento{} initial conditions, with the goal of identifying situations where \cpuvah{} offers definitive advantages in reliability over standard viscous hydrodynamics.

\subsection{Conformal Bjorken flow test}
\label{chap4S4.1}
For the first test, we evolve a conformal plasma undergoing Bjorken expansion~\cite{Bjorken:1982qr}. The only two degrees of freedom are the energy density $\ene = T^{\tau\tau}$ and longitudinal pressure $\PL$ (the transverse pressure is $\Pperp = \frac{1}{2}(\ene {-} \PL)$). The anisotropic hydrodynamic equations simplify to \cite{Molnar:2016gwq}
\bs
\allowdisplaybreaks
\label{eqchap4:semi_Bjorken}
\beal
    \partial_\tau \ene &= - \frac{\ene + \PL}{\tau} \,,
\\
    \partial_\tau \PL &= \frac{\ene {-} 3\PL}{3\tau_\pi} + \frac{\bar\zeta_z^L}{\tau}\,,
\end{align}
\es
where the conformal transport coefficient $\bar\zeta_z^L$ is given in Eq.~(\ref{eqchap4:pl_coeff}a).
%
\begin{figure}[t]
\includegraphics[width=\linewidth]{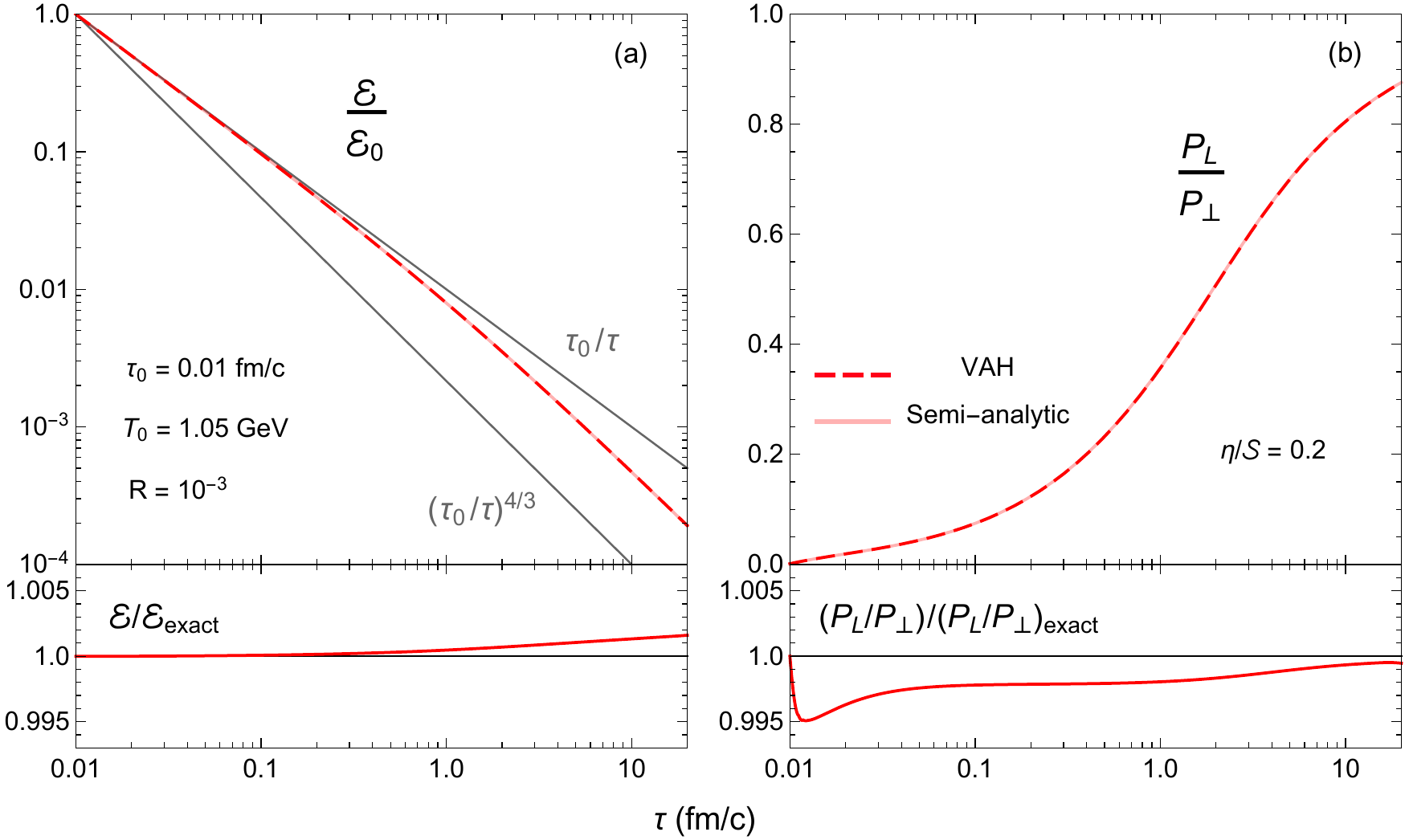}
\centering
\caption{(Color online)
\label{conformal_bjorken}
    Conformal Bjorken evolution of the normalized energy density $\ene/\ene_0$ (a) and pressure ratio $\PL/\Pperp$ (b) from the anisotropic hydrodynamic simulation (dashed red) and semi-analytic solution (transparent red). The gray lines in (a) show two different power laws for comparison. The subpanels at the bottom show the ratio between the numerical simulation and semi-analytic solution (solid red).
}
\end{figure}
%
%
\begin{figure}[t]
\includegraphics[width=0.6\linewidth]{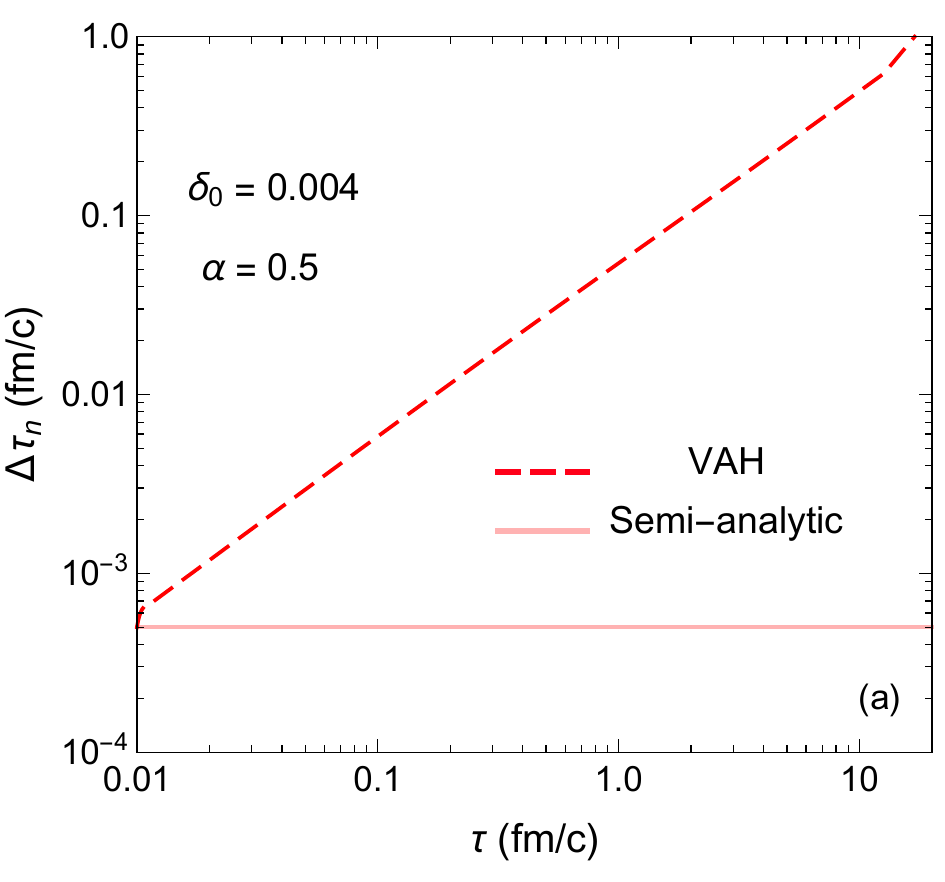}
\centering
\caption{(Color online)
\label{bjorken_adaptive}
The evolution of the adaptive time step $\Delta\tau_n$ (dashed red) in the conformal Bjorken simulation. The semi-analytic method (transparent red) uses a fixed time step of $\Delta\tau_n = 5 \times 10^{-4}$\,fm/$c$.
}
\end{figure}
%
We start the simulation\footnote{%
    Bjorken flow can be simulated on either one fluid cell (i.e. $N_x = N_y = N_\eta = 1)$ or a larger grid with homogeneous initial conditions.} 
at $\tau_0 = 0.01$\,fm/$c$ with an initial temperature $T_0 = 1.05$\,GeV, pressure ratio $R = 10^{-3}$ and shear viscosity $\etas = 0.2$. We evolve the system with an adaptive time step, initially set to $\Delta \tau_0 = 5 \times 10^{-4}$\,fm/$c$, until the temperature drops below $T_\text{sw} = 0.136$\,GeV at $\tau_f \sim 20$ fm/$c$. Figure~\ref{conformal_bjorken} shows the evolution of the energy density normalized to its initial value $\ene_0$ and the pressure ratio $\PL / \Pperp$. At very early times, the system is approximately free-streaming due to the rapid longitudinal expansion, resulting in a large Knudsen number and causing the energy density to decrease like $\ene \approx \ene_0 \tau_0/ \tau$. Over time, the longitudinal expansion rate $\theta_L = 1/\tau$ decreases and the system approaches local equilibrium, i.e. $\ene\propto\tau^{-4/3}$ and $\PL / \Pperp \to 1$. 

We compare the (0+1)--d \cpuvah{} simulation to the semi-analytic solution of \eqref{eqchap4:semi_Bjorken}, which uses a fixed time step $\Delta \tau = 5 \times 10^{-4}$\,fm/$c$. The simulation is in excellent agreement with the semi-analytic solution, with numerical errors staying below $0.5\%$. We also checked the convergence of the simulation curves as we decrease the error tolerance parameter $\delta_0$ in the adaptive time step algorithm. Figure~\ref{bjorken_adaptive} shows the evolution of the adaptive time step $\Delta \tau_n$ for the parameters $\delta_0 = 0.004$ and $\alpha = 0.5$. One sees that the time step increases linearly with time and is not bounded by the CFL condition \eqref{eqchap4:CFL_bound} since the flow is stationary (i.e. $\Delta\tau_\text{CFL} \to \infty)$. As a result, the adaptive RK2 scheme quickly reaches the switching temperature in 145 time steps, compared to about 40000 steps for the semi-analytic method. 


\subsection{Conformal Gubser flow test}
\label{chap4S4.2}
In the next test, we evolve a conformal fluid subject to Gubser flow \cite{Gubser:2010ze, Gubser:2010ui}, which was discussed in Chapter~\ref{chap2label}. The anisotropic hydrodynamic equations for $\hat\ene$ and $\PLhat$ in the de Sitter space $\hat{x}^\mu = (\rho,\theta,\phi,\eta)$ are given by \cite{Martinez:2017ibh}
\bs
\allowdisplaybreaks
\label{eqchap4:semi_Gubser}
\allowdisplaybreaks
\beal
    \partial_\rho \hat\ene &= (\PLhat - 3\hat\ene) \tanh\rho \,,
\\
    \partial_\rho \PLhat &= \frac{\hat\ene {-} 3\PLhat}{3\hat\tau_\pi} - \big(4 \PLhat {+} \hat{\bar\zeta}_z^L\big)\tanh\rho \,,
\end{align}
\es
while the residual shear stresses $\hat{W}_{\perp z}^\mu$ and  $\hat{\pi}_\perp^\munu$ vanish under Gubser symmetry \cite{Martinez:2017ibh}. In the semi-analytic solution \eqref{eqchap4:semi_Gubser}, we start the evolution at $\rho_0 = -9.2$, which corresponds to a corner in a $14$\,fm\,$\times 14$\,fm transverse grid ($r = 7 \sqrt{2}$\,fm) at the initial time $\tau_0 = 0.01$\,fm/$c$, with an initial temperature $\hat{T}_0 = 0.0017$ and pressure ratio $\hat{R} = 10^{-3}$. We set the inverse fireball size to $q = 1.0$\,fm$^{-1}$ and the shear viscosity to $\eta/\mathcal{S} = 0.2$. We evolve the system with a fixed stepsize $\Delta \rho = 10^{-4}$ until $\rho_f = 1.1$, which corresponds to the center of our Eulerian grid ($r = 0$\,fm) at $\tau_f = 3.01$\,fm/$c$. 

Next, we map the semi-analytic solution for the de Sitter energy density $\hat{\ene}(\rho)$ and longitudinal pressure $\PLhat(\rho)$ back to Milne coordinates \cite{Denicol:2014tha}:
\be
\label{eqchap4:semi_Gubser_map}
    \ene(\tau, r) = \frac{\hat\ene\big(\rho(\tau,r)\big)}{\tau^4} \,,\qquad
    \PL(\tau, r) = \frac{\PLhat\big(\rho(\tau,r)\big)}{\tau^4}\,.
\ee
\begin{figure}[thbp]
 \makebox[\textwidth][c]{\includegraphics[width=0.83\linewidth]{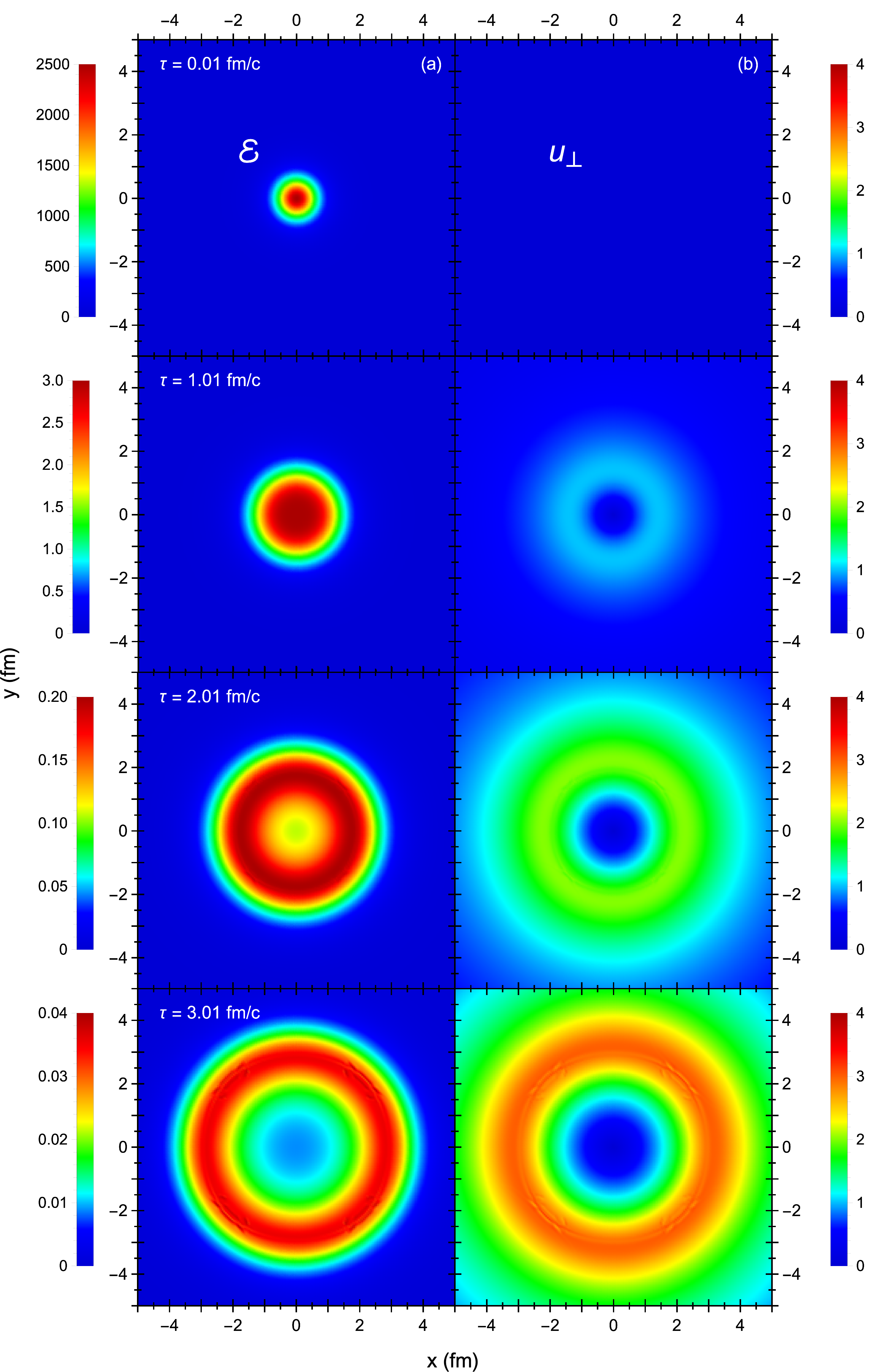}}
\caption{(Color online)
\label{gubser_2d}
    Conformal Gubser evolution of the energy density $\ene$ (GeV/fm$^3$) (left column) and transverse fluid velocity $u_\perp$ (right column) in the anisotropic hydrodynamic simulation.
}
\end{figure}
\begin{figure}[!t]
\includegraphics[width=\linewidth]{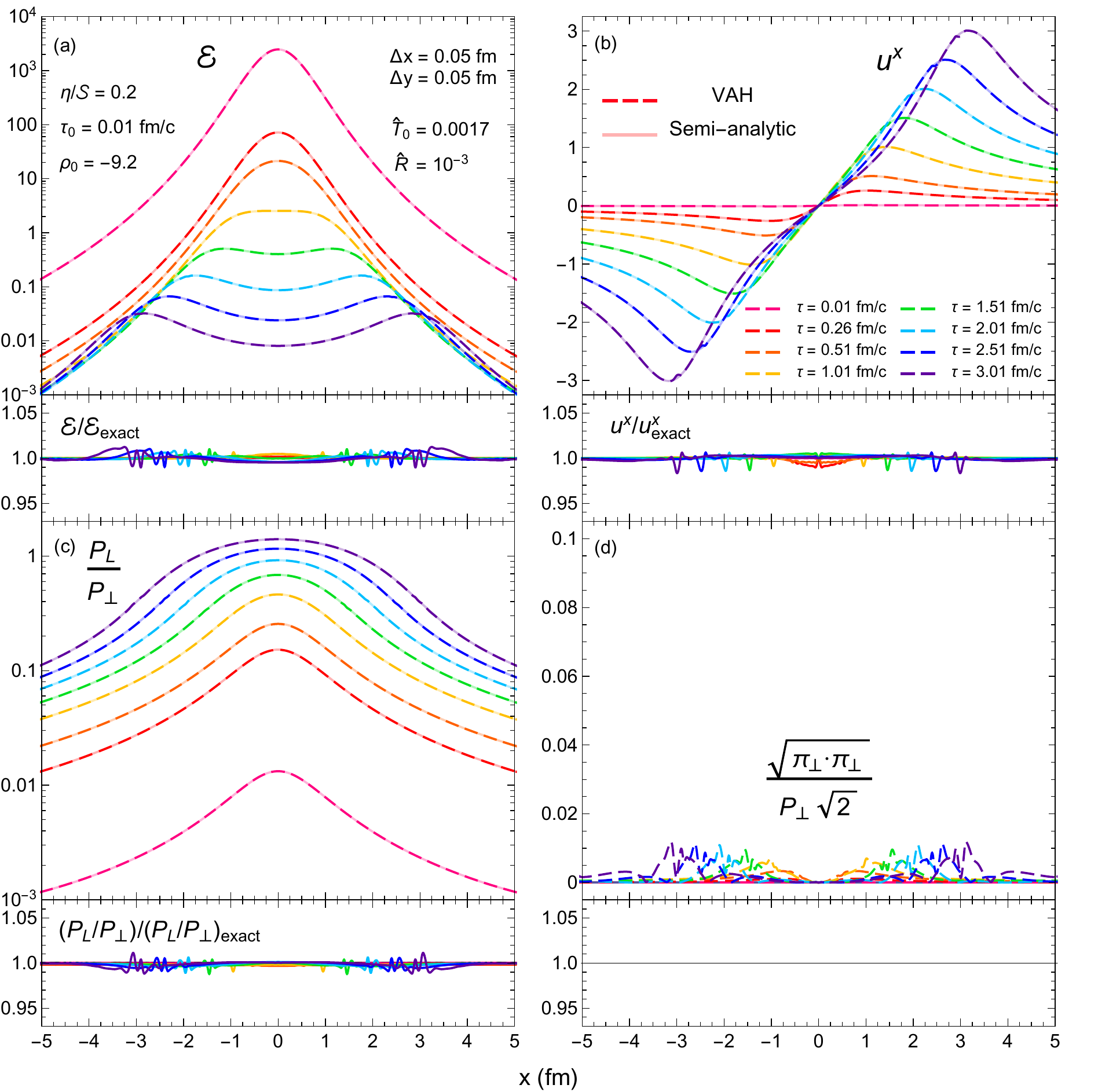}
\centering
\caption{(Color online)
\label{gubser_test}
    Conformal Gubser evolution of (a) $\ene$ (GeV/fm$^3$), (b) $u^x$, (c) $\PL/\Pperp$, and (d) $\sqrt{\pi_\perp {\cdot\,} \pi_\perp }/(\Pperp\sqrt{2})$ along the $x$--axis ($y=0$), given by the anisotropic hydrodynamic simulation (dashed color) and semi-analytic solution (transparent color). The subpanels below panels (a-c) show the ratio between numerical simulation and semi-analytic solution (solid color).
}
\end{figure}
We use this map to set the initial energy density and longitudinal pressure profiles $\ene(\tau_0, x, y)$ and $\PL(\tau_0, x, y)$; the initial temperature at the center of the fireball is $T_{0,\text{center}} = 1.05$\,GeV. We set the initial transverse shear stress to $\piperp = 0$. Finally, we use Eq.~\eqref{eq:ch2gubser_velocity} to initialize the current fluid velocity $\texttt{u} \leftarrow u^i(\tau_0,x,y)$ and previous fluid velocity $\texttt{up} \leftarrow u^i(\tau_0 - \Delta\tau_0,x,y)$, where the initial time step is $\Delta\tau_0 = 5 \times 10^{-4}$\,fm/$c$. We evolve the Gubser profile on a $14$\,fm\,$\times 14$\,fm transverse grid with a lattice spacing of $\Delta x = \Delta y = 0.05$\,fm. 

\begin{figure}[!t]
\includegraphics[width=0.6\linewidth]{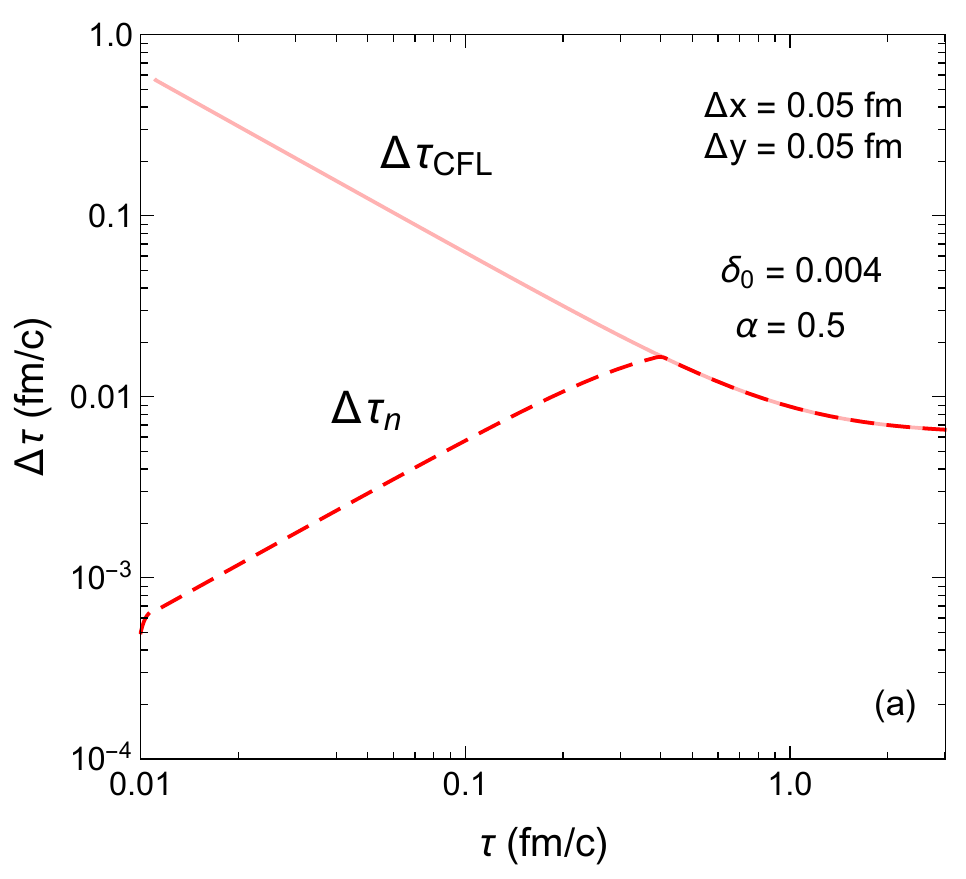}
\centering
\caption{(Color online)
\label{gubser_timestep}
The evolution of the adaptive time step $\Delta \tau_n$ (dashed red) and CFL bound $\Delta \tau_\text{CFL}$ (transparent red) in the conformal Gubser simulation.
}
\end{figure}

Figure~\ref{gubser_2d} shows the evolution of the energy density and transverse fluid velocity $u_\perp = \sqrt{(u^x)^2 + (u^y)^2}$ in the transverse plane at various time frames.\footnote{%
    In order to output the hydrodynamic quantities at specific times, we readjust the adaptive time step whenever we are close to these output times (this is not shown in Fig.~\ref{gubser_timestep}).} 
At early times, the very hot and compact fireball expands rapidly along the longitudinal direction and quickly cools down without much change to its transverse shape. Over time, the transverse expansion overtakes the longitudinal expansion, pushing the medium radially outward as a \textit{ring of fire}. One sees that the energy density and transverse fluid velocity maintain their azimuthal symmetry throughout the evolution. However, there are some numerical fluctuations around the lines $y = \pm \,x$ at later times, especially near the peak of $u_\perp$. This is a consequence of using a Cartesian grid with a finite spatial resolution.

Figure~\ref{gubser_test} shows the evolution of the energy density $\ene$, fluid velocity component $u^x$, pressure anisotropy ratio $\PL/\Pperp$ and transverse shear inverse Reynolds number $\sqrt{\pi_\perp {\cdot\,} \pi_\perp }$\,/\,$(\Pperp\sqrt{2})$ along the $x$--axis ($y=0$) at multiple time frames. Overall, there is very good agreement between the simulation and semi-analytic solution, except near the local extrema of $u^x$, where the numerical errors are about $1{-}2\%$; these errors can be reduced by using a finer lattice spacing. The transverse shear stress $\piperp$ should remain zero throughout the simulation. However, the transverse shear inverse Reynolds number from the \cpuvah{} simulation is nonzero (albeit small, $\lesssim 2\%$) since the Eulerian grid does not perfectly preserve Gubser symmetry.

Figure~\ref{gubser_timestep} shows the evolution of the adaptive time step in the Gubser simulation. In contrast to Fig.~\ref{bjorken_adaptive}, here $\Delta\tau_n$ becomes bounded by the CFL condition~\eqref{eqchap4:CFL_bound} at $\tau \sim 0.4$ fm/$c$ due to the increasing transverse expansion rate. As a result, the Gubser simulation finishes in 407 time steps, compared to 480 steps if we had used the fixed time step $\Delta \tau = \Delta x / 8$. Although we only gain a slight $1.15\times$ speedup, the adaptive time step is able to resolve the fluid's early-time dynamics more accurately than the fixed time step~\eqref{eqchap4:CFL_fixed} because it is initially independent of the lattice spacing. This property is especially useful when simulating more realistic nuclear collisions, where the lattice spacing required to resolve the participant nucleons' transverse energy deposition is several times coarser than the one used in this test (e.g. $\Delta x = \Delta y \sim 0.1 - 0.3$ fm).
\subsection{(3+1)--d conformal hydrodynamic models in central Pb+Pb collisions}
\label{chap4S4.3}
Here we compare (3+1)--dimensional conformal anisotropic hydrodynamics (\cpuvah{}) to conformal second-order viscous hydrodynamics (\vh{}) (see Appendix~\ref{app4e}), for a central Pb+Pb collision at zero impact parameter ($b = 0$ fm) with static (in Milne coordinates) and smooth initial conditions. To generate the latter we use an azimuthally symmetric \trento{} energy density profile averaged over 2000 fluctuating events \cite{Moreland:2014oya} and extended along the $\eta_s$--direction with a smooth rapidity plateau~\cite{Pang:2018zzo} (see Appendix~\ref{app4d} for more information on the \trento{} energy deposition model for high-energy nuclear collisions). The initial temperature at the center of the fireball is $T_{0,\text{center}} = 1.05$ GeV at a starting time of $\tau_0 = 0.01\,\text{fm}/c$ which is the same for both types of hydrodynamic simulations. The fluid is evolved from an initial pressure ratio $R = 10^{-3}$ with a constant specific shear viscosity $\etas = 0.2$; the runs stop once all the fluid cells are below $T_\text{sw} = 0.136$\,GeV, which happens after $\tau_f \sim 7-8$\,fm/$c$.

\begin{figure}[!t]
\includegraphics[width=\linewidth]{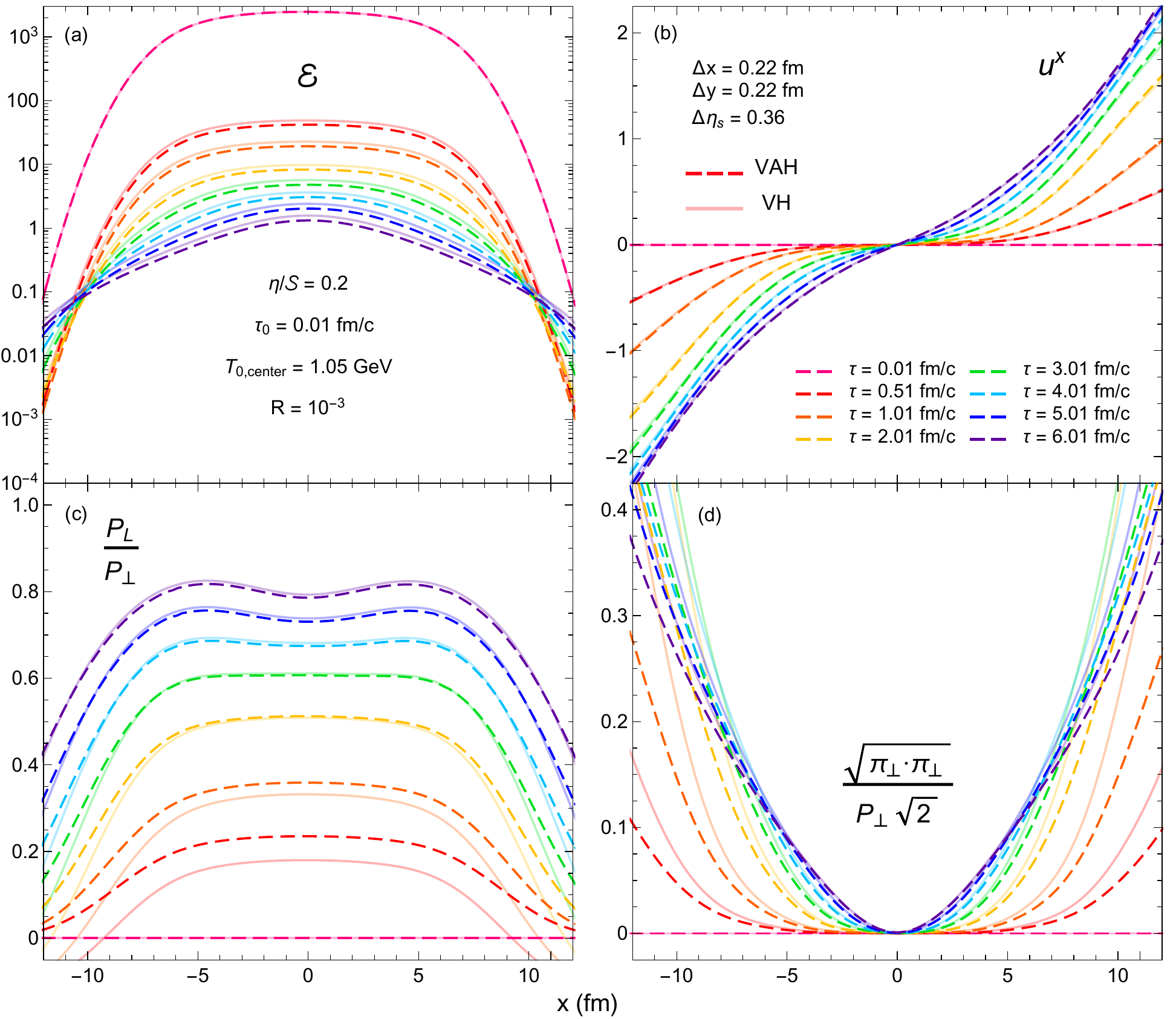}
\centering
\caption{(Color online)
\label{conformal_trento_x}
The evolution of (a) $\ene$ (GeV/fm$^3$), (b) $u^x$, (c) $\PL/\Pperp$ and (d) $\sqrt{\pi_\perp {\cdot\,} \pi_\perp }/(\Pperp\sqrt{2})$ along the $x$--axis ($y{\,=\,}\eta_s{\,=\,}0$), given by conformal anisotropic hydrodynamics (\cpuvah{}, dashed color) and second-order viscous hydrodynamics (\vh{}, transparent color), for the smooth (3+1)--dimensional \trento{} initial condition described in the text.
}
\end{figure}

Figure~\ref{conformal_trento_x} shows the evolution of $\ene$, $u^x$, $\PL/\Pperp$ and $\sqrt{\pi_\perp {\cdot\,} \pi_\perp}$\,/\, $(\Pperp\sqrt{2})$ along the $x$--axis ($y=\eta_s=0)$. Initially, the pressure anisotropy $\PL{-}\Pperp$ in second-order viscous hydrodynamics is so large that the longitudinal pressure turns negative, especially near the grid's edges. In comparison, the pressure ratio $\PL/\Pperp$ in anisotropic hydrodynamics is, at early times, only moderately larger in the central fireball region but quite dramatically different near its transverse edge, staying positive everywhere. The resulting differences in early-time viscous heating create a disparity between the two models for the normalization of the energy density which persists to late times even after the $\PL/\Pperp$ ratios have converged. On the other hand, the transverse flows $u^x$ predicted by the two models are nearly identical even at very early times since they are primarily driven by the transverse pressure gradients $\partial_x \Pperp$ which, in the smooth \trento{} profile, are smaller at early times than the longitudinal gradients $\sim 1/\tau$. The transverse shear stress $\piperp$, which tends to counteract the fluid's transverse acceleration, is larger in viscous hydrodynamics than in anisotropic hydrodynamics, especially along the edges of the fireball.\footnote{The transverse shear stress $\piperp$ is generally nonzero even for azimuthally symmetric flow profiles as long as they do not possess Gubser symmetry.} Overall, however, the hydrodynamic variables are not substantially different along the transverse directions.

\begin{figure}[!t]
\includegraphics[width=\linewidth]{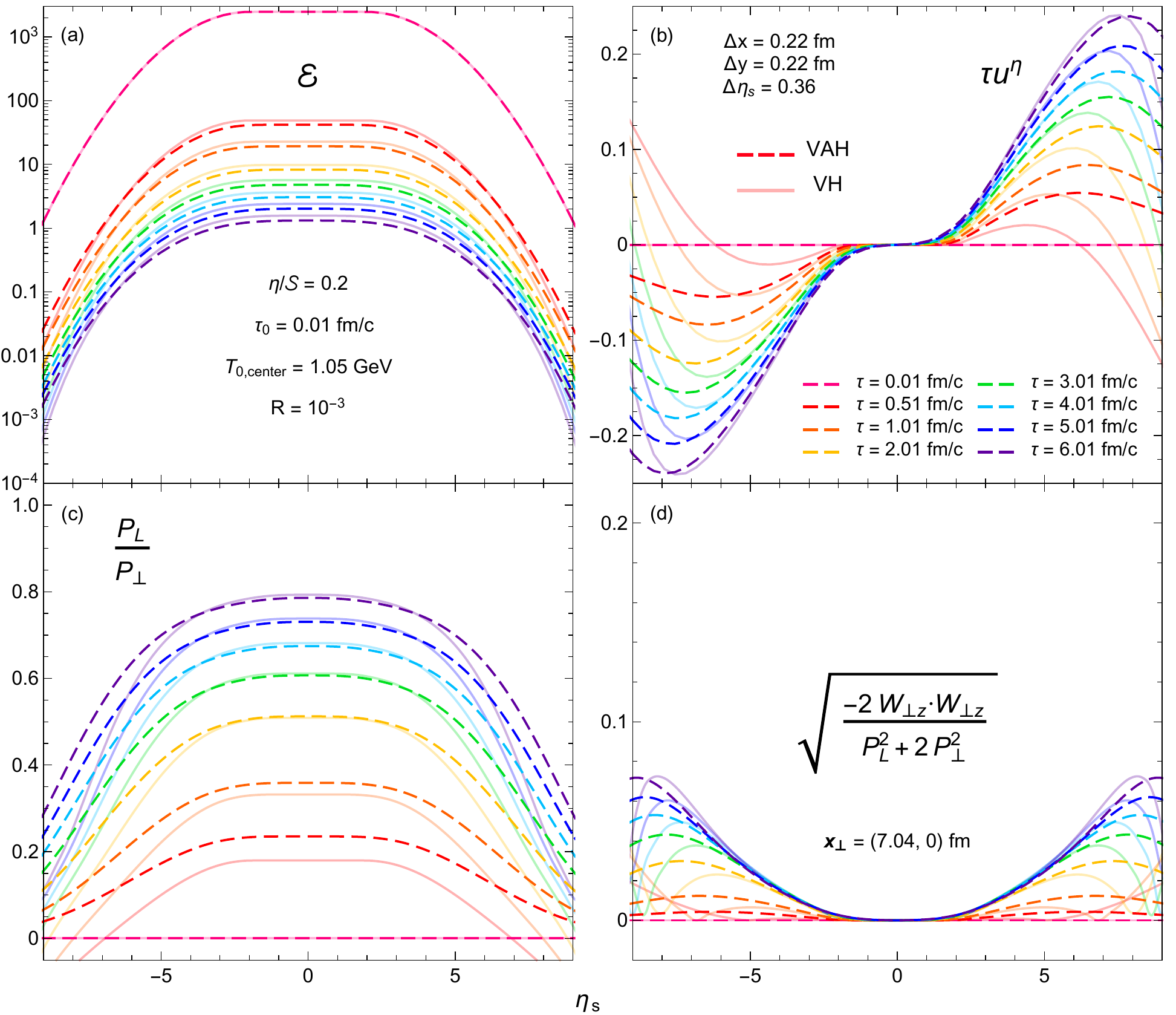}
\centering
\caption{(Color online)
\label{conformal_trento_z}
The evolution of (a) $\ene$ (GeV/fm$^3$), (b) $\tau u^\eta$, (c) $\PL/\Pperp$ and (d) $\sqrt{2 W_{\perp z} {\,\cdot\,} W_{\perp z}}$ /$\sqrt{\mathcal{P}_L^2 {+} 2\mathcal{P}_\perp^2}$ along the $\eta_s$--axis ($x{\,=\,}y{\,=\,}0$), computed with conformal anisotropic hydrodynamics (\cpuvah{}, dashed color) and second-order viscous hydrodynamics (\vh{}, transparent color), for the smooth (3+1)--d \trento{} initial condition described in the text. Note that in panel (d) the $\eta_s$--axis is shifted transversely to $(x,y)=(7.04,0)$\,fm.
}
\end{figure}

The longitudinal profile, however, evolves very differently in anisotropic hydrodynamics (\cpuvah{}) compared to second-order viscous hydrodynamics (\vh{}). This is shown in Figure~\ref{conformal_trento_z} where we plot the evolution of the dimensionless longitudinal velocity $\tau u^\eta$ (as well as $\ene$ and $\PL/\Pperp$) along the $\eta_s$--axis ($x=y=0$). Similar to Fig.~\ref{conformal_trento_x}, the shear stress in viscous hydrodynamics quickly overpowers the equilibrium pressure, making $\PL$ negative initially. Along the longitudinal direction this causes a strong reversal of the longitudinal flow $\tau u^\eta$ at around $|\eta_s| \sim 7$. This unphysical feature (primarily driven by the strong longitudinal expansion rate at early times) results in a longitudinal contraction (in $\eta_s$) of the fireball near its forward and backward edges at the beginning of the simulation. In anisotropic hydrodynamics, the longitudinal pressure remains always positive, allowing for a stronger longitudinal flow profile whose signs are consistent with the gradients of the rapidity distribution of the energy density \eqref{app4:plateau} (and thus of the thermal pressure). This gives anisotropic hydrodynamics a clear advantage over second-order viscous hydrodynamics when simulating the longitudinal dynamics of a heavy-ion collision. 

\begin{figure}[!t]
\includegraphics[width=0.6\linewidth]{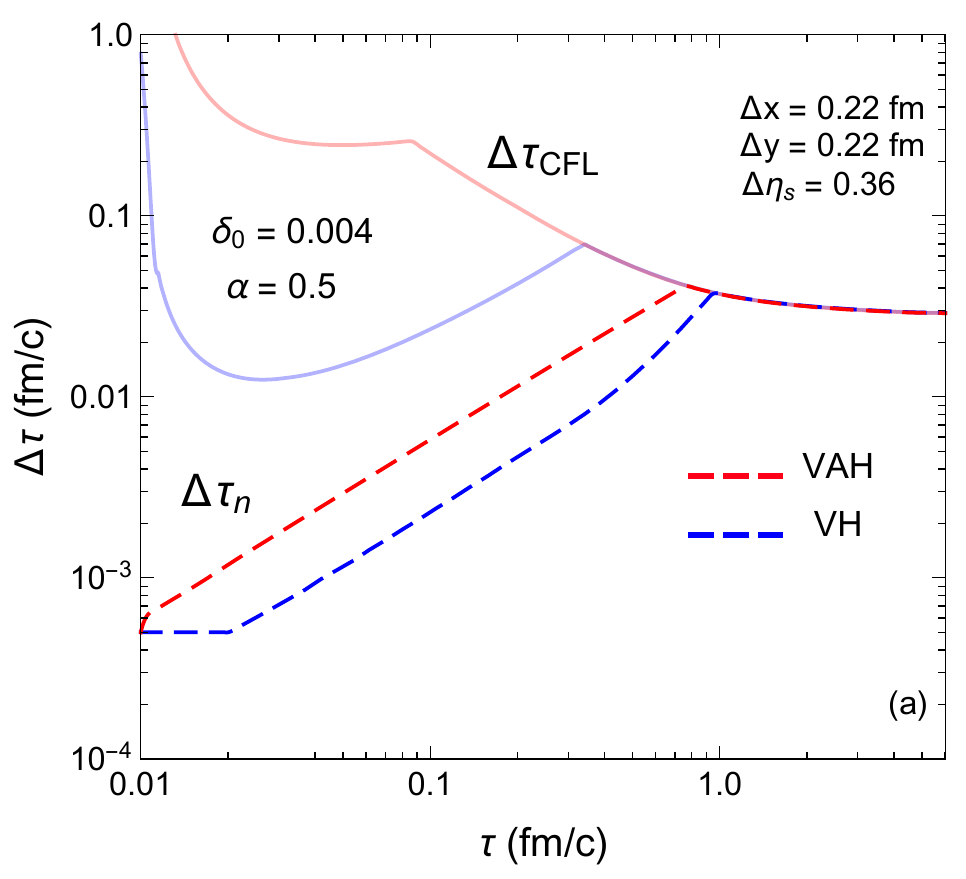}
\centering
\caption{(Color online)
\label{conformal_trento_timestep}
The evolution of the adaptive time step $\Delta \tau_n$ (dashed color) and CFL bound $\Delta \tau_\text{CFL}$ (transparent color) in conformal anisotropic hydrodynamics (red) and second-order viscous hydrodynamics (blue) for the smooth (3+1)--d \trento{} initial condition.
}
\end{figure}

In panel (d) of Fig.~\ref{conformal_trento_z} we also plot the spacetime rapidity dependence of the inverse Reynolds number $\sqrt{-2W_{\perp z}{\,\cdot\,} W_{\perp z} / (\mathcal{P}_L^2 + 2\mathcal{P}_\perp^2)}$. The longitudinal momentum diffusion current $\Wperp$ is nonzero only in regions that have both longitudinal and transverse gradients. For this reason we shift the $\eta_s$--axis transversely to $(x,y) = (7.04,0)$\,fm, which corresponds to the mid-right region of the grid. As expected, the momentum diffusion current is weakest around mid-rapidity, $\eta_s \sim 0$, where the fireball profile is approximately longitudinally boost-invariant, and strongest along the sloping edges of the rapidity plateau \eqref{app4:plateau}. However, this longitudinal edge region is also the place where its inverse Reynolds number differs most strongly between anisotropic and standard viscous hydrodynamics. Overall, $\Wperp$ only makes up a tiny fraction of the total shear stress \eqref{eqch3:7} in anisotropic hydrodynamics. The situation may change for initial conditions whose longitudinal and transverse gradients are larger than the ones used in this comparison study. 

Finally, we plot in Figure~\ref{conformal_trento_timestep} the adaptive time step for each of the two hydrodynamic simulations. One sees that in \vh{} $\Delta\tau_n$ stagnates until $\tau{\,\sim\,}0.02$\,fm/$c$ (see footnote \ref{adaptive_floor}) while the one in \cpuvah{} starts increasing immediately. This indicates that at early times viscous hydrodynamics generally has a faster evolution rate and therefore requires a smaller initial time step compared to anisotropic hydrodynamics. Nevertheless, both adaptive time steps remain below their CFL bound (initially dominated by the longitudinal velocity $u^\eta$) until $\tau{\,\sim\,}0.8{-}1$ fm/$c$. Notice that the CFL bounds for \vh{} and \cpuvah{} become virtually identical since they have almost the same transverse velocity profile (see Fig.~\ref{conformal_trento_x}b).
\subsection{Non-conformal Bjorken flow test}
\label{chap4S4.4}
Now we turn to testing our non-conformal hydrodynamic simulation with the QCD equation of state from Fig.~\ref{eos}. First, we run the \cpuvah{} simulation with Bjorken initial conditions and compare it to the semi-analytic solution of Eq.~\eqref{eqch3:ahydroeqs}. We use the shear and bulk viscosities given by Eqs.~\eqref{eqchap4:etas} --~\eqref{eqchap4:zetas} (see Fig.~\ref{viscosities}) and take the relaxation times from Fig.~\ref{relaxation}. We start the simulation at $\tau_0 = 0.05$ fm/$c$ with initial temperature $T_0 = 0.718$ GeV and initial pressure ratio $R = 0.3$, and evolve the system until $\tau_f \sim 80$ fm/$c$ when the temperature falls below $T_\text{sw} = 0.136$\,GeV (we plot results only up to $\tau = 20$ fm/$c$). We also repeat this for quasiparticle viscous hydrodynamics (\vh{}) and standard viscous hydrodynamics (\vh{}2) (see footnote \ref{VH12}).\footnote{%
    \vh{} and \vh{}2 use the same equation of state $\Peq(\ene)$ as \cpuvah{} but different transport coefficients; see Sec.~\ref{ch3sec5} and App.~\ref{appch3e}.}
Ideally, we would have preferred using smaller values for $\tau_0$ and $R$ to better match the longitudinally free-streaming initial condition used in the conformal hydrodynamic simulations (see Secs.~\ref{chap4S4.1} --~\ref{chap4S4.3} and footnote \ref{free_stream}). However, we found that at earlier times the \cpuvah{} simulation has greater difficulty reconstructing the anisotropic variables ($\Lambda$, $\alpha_\perp$, $\alpha_L$), especially directly at initialization. This indicates that our anisotropic hydrodynamic model has its limitations. 

\begin{figure}[!t]
\includegraphics[width=\linewidth]{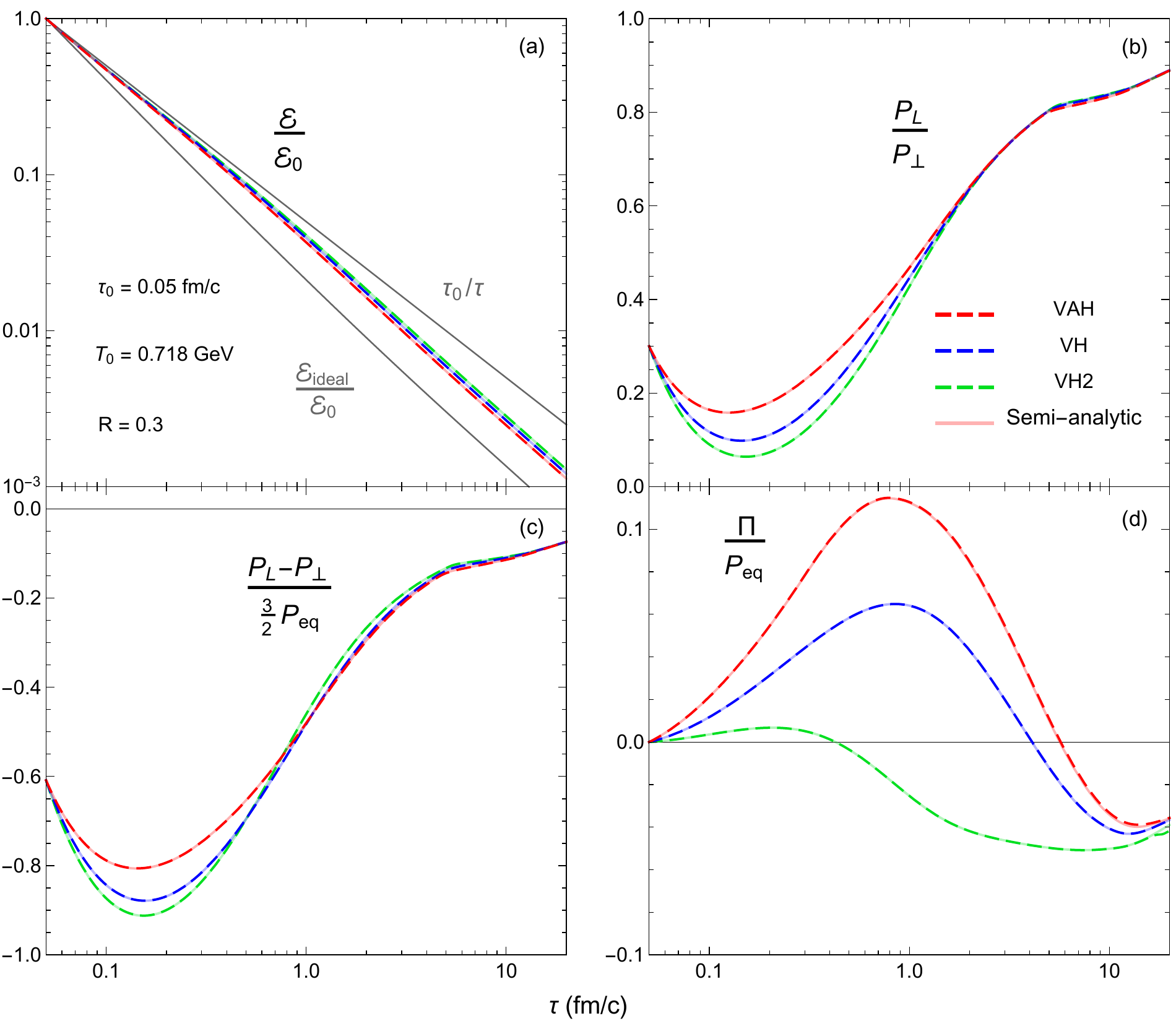}
\centering
\caption{(Color online)
\label{nonconformal_bjorken}
Non-conformal Bjorken evolution of (a) $\ene/\ene_0$, (b) the pressure ratio $\PL/\Pperp$, (c) the pressure anisotropy $\frac{2}{3}(\PL{-}\Pperp)$, and (d) the bulk viscous pressure $\Pi$ (the latter two normalized to the equilibrium pressure $\Peq$), given by anisotropic hydrodynamics (\cpuvah{}, dashed red), quasiparticle viscous hydrodynamics (\vh{}, dashed blue) and standard viscous hydrodynamics (\vh{}2, dashed green), along with their semi-analytic solutions (transparent color). The lower gray curve in panel (a) is from an ideal hydrodynamic calculation (i.e. $\PL = \Pperp = \Peq(\ene)$ and $\etas = \zetas = 0$). 
}
\end{figure}

Figures~\ref{nonconformal_bjorken}a,b show the evolution of the normalized energy density $\ene/\ene_0$ and pressure ratio $\PL/\Pperp$ from the three hydrodynamic simulations. We also disentangle the longitudinal and transverse pressures' viscous components $\PL {-} \Peq = \frac{2}{3}\Delta \mathcal{P} {+} \Pi$ and $\Pperp {-} \Peq = -\frac{1}{3}\Delta \mathcal{P} {+} \Pi$ into the pressure anisotropy $\Delta \mathcal{P} = \PL{-}\Pperp$ and bulk viscous pressure $\Pi = \frac{1}{3}(\PL{+}2\Pperp)-\Peq$; their evolutions relative to $\Peq$ are shown in Figs.~\ref{nonconformal_bjorken}c,d. Although our initial condition for the pressure ratio is somewhat ad hoc, it quickly reaches its minimum value at $\tau \sim 0.1$ fm/c, allowing for the energy density to closely follow its free-streaming limit at the beginning of the simulation. Similar to the comparison study from the previous chapter, the $\PL/\Pperp$ ratio is typically larger in anisotropic hydrodynamics compared to the two viscous hydrodynamic models (until $\tau \sim 1$ fm/$c$). Although in Bjorken flow the bulk viscous pressure $\Pi$ is much smaller than the pressure anisotropy $\Delta\mathcal{P}$, it evolves very differently in standard viscous hydrodynamics (\vh{}2) compared to quasiparticle viscous hydrodynamics (\vh{}) and anisotropic hydrodynamics (\cpuvah{}). Since the dimensionless bulk relaxation time $\tau_\Pi T < 1$ in \vh2 (see the dashed curve in Fig.~\ref{relaxation}b), the bulk viscous pressure quickly reaches its Navier-Stokes solution $\Pi_\text{NS} = -\zeta/\tau$ at $\tau \sim 2$\,fm/$c$. In contrast, the bulk relaxation time used in \cpuvah{} and \vh{} is significantly larger (solid curve in Fig.~\ref{relaxation}b). As a consequence, it takes a much longer time for $\Pi$ to relax to $\Pi_\text{NS}$ \cite{Tinti:2016bav, McNelis:2018jho}.

One also sees that the simulations agree very well with their corresponding semi-analytic solutions (continuous lines in transparent color). Although we do not display them explicitly here, the numerical errors are small enough that we can unambiguously distinguish the different dynamics of the three hydrodynamic models in comparison studies discussed in the next subsections.

\subsection{(3+1)--d non-conformal hydrodynamic models in central Pb+Pb collisions}
\label{chap4S4.5}
\subsubsection{Smooth \trento{} initial conditions}
\label{chap4S4.5.1}
Next we run non-conformal hydrodynamics for a central Pb+Pb collision with smooth, azimuthally symmetric \trento{} initial conditions; we set the initial time to $\tau_0 = 0.05$\,fm/$c$ and the initial pressure ratio to $R = 0.3$. The initial energy density profile is almost identical to the one in Sec.~\ref{chap4S4.3}, except the normalization $\propto 1/\tau_0$ is five times smaller, making the initial temperature at the center of the fireball $T_{0,\text{center}} = 0.718$\,GeV. We evolve the system until, at $\tau_f{\,\sim\,}15{\,-\,}16$\,fm/$c$, all fluid cells are below $T_\text{sw} = 0.136$\,GeV.
\begin{figure}[thbp]
\includegraphics[width=0.94\linewidth]{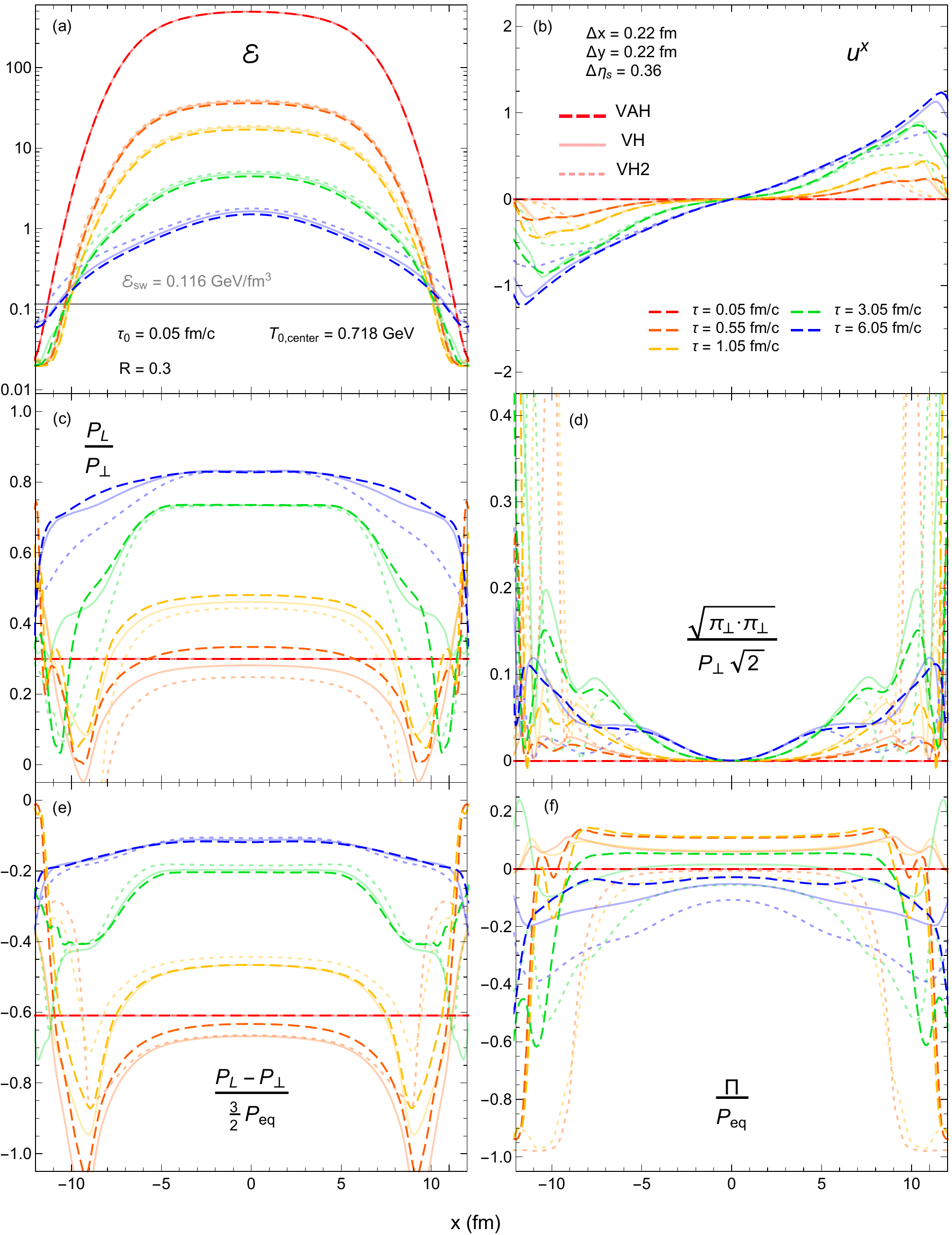}
\centering
\caption{(Color online)
\label{lattice_trento_x}
    The evolution of (a) $\ene$ (GeV/fm$^3$), (b) $u^x$, (c) $\PL/\Pperp$, (d) $\sqrt{\pi_\perp {\cdot\,} \pi_\perp }/(\Pperp\sqrt{2})$, (e) $\frac{2}{3}(\PL{-}\Pperp)/\Peq$ and (f) $\Pi/\Peq$ along the $x$--axis ($y{\,=\,}\eta_s{\,=\,}0$), given by non-conformal anisotropic hydrodynamics (\cpuvah{}, dashed color), quasiparticle viscous hydrodynamics (\vh{}, transparent color) and standard viscous hydrodynamics (\vh2, dotted transparent color), for the smooth (3+1)--d \trento{} initial condition. The minimum energy density parameter is set to $\ene_\text{min} = 0.02$\,GeV/fm$^3$.
}
\end{figure}

Figure~\ref{lattice_trento_x} shows, for the first $\Delta\tau = 6$ fm/$c$ of the simulation, the evolution of $\ene$, $u^x$, $\PL/\Pperp$, $\sqrt{\pi_\perp \cdot \pi_\perp}$/$(\Pperp\sqrt{2})$, $\frac{2}{3}(\PL-\Pperp)$/$\Peq$ and $\Pi/\Peq$ given by the three hydrodynamic models along the $x$--axis ($y=\eta_s = 0$). Here the fireball maintains higher temperatures for a longer duration than in conformal hydrodynamics since the QCD equilibrium pressure is much weaker than its Stefan--Boltzmann limit \eqref{eqchap4:stefan} (see Fig.~\ref{eos}a). But similar to Fig.~\ref{conformal_trento_x}a, the energy density in anisotropic hydrodynamics cools down at a slightly faster rate compared to viscous hydrodynamics. This is initially due to the moderately larger $\PL/\Pperp$ ratio in the central fireball region, which closely follows the Bjorken evolution in Fig.~\ref{nonconformal_bjorken}b. The viscous corrections to the longitudinal pressure increase as we move towards the edges of the fireball, but $\PL$ still remains positive in anisotropic hydrodynamics (\cpuvah{}). The longitudinal pressure in standard viscous hydrodynamics (\vh{}2), however, is strongly negative at early times (and also, to a lesser extent, for the quasiparticle case \vh{}), although it does not significantly alter the fluid's transverse dynamics. The $\PL/\Pperp$ ratios start to overlap at later times, mainly driven by the pressure anisotropies' convergence in Fig.~\ref{lattice_trento_x}e.

Compared to Fig.~\ref{conformal_trento_x}b, the transverse flow $u^x$ in non-conformal hydrodynamics is significantly weaker due to the softer QCD equation of state. The dip in the speed of sound at the quark-hadron phase transition (see Fig.~\ref{eos}b) also flattens the transverse velocity gradients near the edges of the fireball. This causes a significant decrease in the transverse shear stress relative to the conformal case in Fig.~\ref{conformal_trento_x}d. Among the hydrodynamic models, we see that \vh{}2 produces the smallest $u^x$ around the edges of the fireball since it has the largest negative bulk viscous pressure there. This is due to its ability to converge to the Navier-Stokes solution shortly after the simulation begins. In contrast, the bulk viscous pressure in \vh{} hovers slightly above zero across the grid before falling down toward negative values at $\tau \sim 3$ fm/$c$. Even then, it hardly catches up to the bulk viscous pressure curves in \vh{}2 since the differences between their bulk relaxation times grow as the fireball cools down. Overall, the bulk viscous pressure in \vh{} is much weaker compared to \vh{}2, which leads to a stronger transverse flow. Finally, the \cpuvah{} simulation has a bulk viscous pressure that is qualitatively similar to \vh{} (except at the edges of the grid) but slightly lags behind it. As a result, it generates the largest transverse flow out of the three simulations. The study here illustrates how the evolution of the bulk viscous pressure and transverse fluid velocity strongly depends on the choice of hydrodynamic model, specifically the temperature-dependent model for the bulk relaxation time $\tau_\Pi$. But if our hydrodynamic code were used as a stand-alone model to evolve both the pre-hydrodynamic and fluid dynamic stages, the phenomenological constraints for $\zetas$ may not deviate that strongly from existing hybrid models, which simulate the pre-hydrodynamic phase with conformal free-streaming.\footnote{
    The VAH framework describes the far-off-equilibrium stage  hydrodynamically from such early times that there is no need for a separate pre-hydrodynamic model, as typically used in other approaches to evolve the system from the initial energy deposition to the time at which standard viscous hydrodynamics (\vh{}2) becomes applicable. If the pre-hydrodynamic evolution is described by conformal free-streaming  \cite{Bernhard:2018hnz, Everett:2020yty, Everett:2020xug}, it leads to large positive bulk viscous pressures that partially negate the effects of the bulk viscosity in the beginning of the \vh{}2 simulation. Therefore, replacing the free-streaming and \vh{}2 modules by \cpuvah{} may not necessarily require much larger $\zetas$ values.}

\begin{figure}[thbp]
\includegraphics[width=0.94\linewidth]{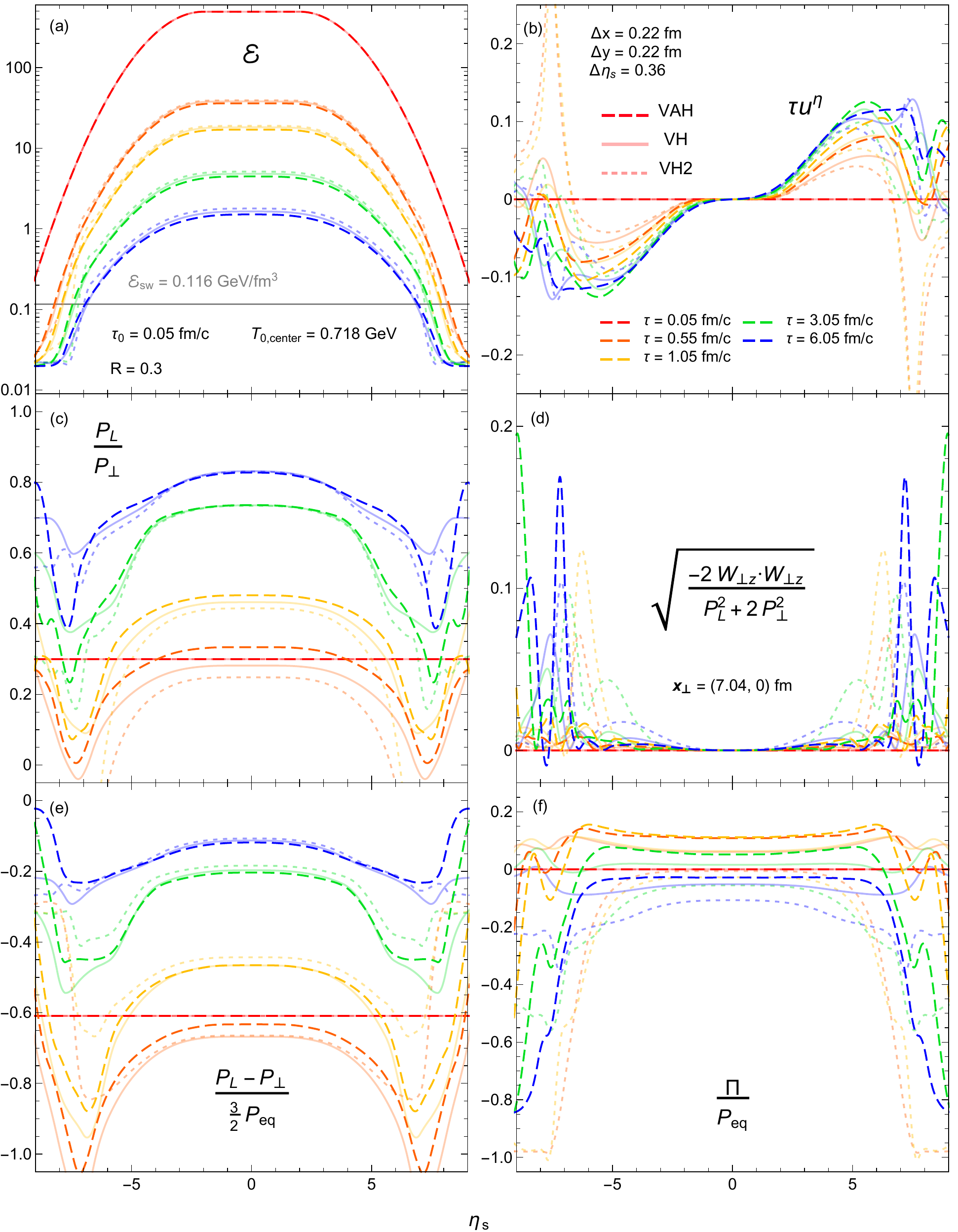}
\centering
\caption{(Color online)
\label{lattice_trento_z}
    The evolution of (a) $\ene$ (GeV/fm$^3$), (b) $\tau u^\eta$, (c) $\PL/\Pperp$, (d) $\sqrt{2 W_{\perp z} {\,\cdot\,} W_{\perp z}}$ /$\sqrt{\mathcal{P}_L^2 {+} 2\mathcal{P}_\perp^2}$, (e) $\frac{2}{3}(\PL{-}\Pperp) / \Peq$, and (f) $\Pi/\Peq$ along the $\eta_s$--axis ($x=y=0$), given by non-conformal anisotropic hydrodynamics (\cpuvah{}, dashed color), quasiparticle viscous hydrodynamics (\vh{}, transparent color) and standard viscous hydrodynamics (\vh2, dotted transparent color), for the smooth (3+1)--d \trento{} initial condition. Note that in panel (d) the $\eta_s$--axis is shifted transversely to $(x,y)=(7.04,0)$\,fm.
}
\end{figure}

Next, we compare the fireballs' longitudinal evolution along the $\eta_s$--axis ($x=y=0$) in Figure~\ref{lattice_trento_z}. When comparing the longitudinal evolution of the energy density and viscous pressures between the three hydrodynamic models, we note qualitatively similar features as already observed in Fig.~\ref{lattice_trento_x} for their transverse evolution. While the transverse velocities shown in Fig.~\ref{lattice_trento_x}b varied between the models as a result of differences in the transverse pressures (mainly the bulk viscous pressures), here differences in their longitudinal pressures result in very different longitudinal flow profiles (Figure~\ref{lattice_trento_z}b). Most notably, the large negative $\PL$ in standard viscous hydrodynamics (\vh{}2) initially causes $\tau u^\eta$ to reverse sign very sharply in the forward and backward rapidity regions. Only by imposing strong regulations on the \vh{}2 simulation is it able to recover the correct direction of longitudinal flow at later times as the gradients relax. The same breakdown at early times can also be seen in quasiparticle viscous hydrodynamics (\vh{}) although there the situation is not nearly as bad. With \cpuvah{} we can maintain positive longitudinal pressures so that the fireball can expand without imploding in regions where the longitudinal gradients are large. This greatly reduces the amount of regulation needed for the viscous pressures.

\begin{figure}[!t]
\includegraphics[width=0.6\linewidth]{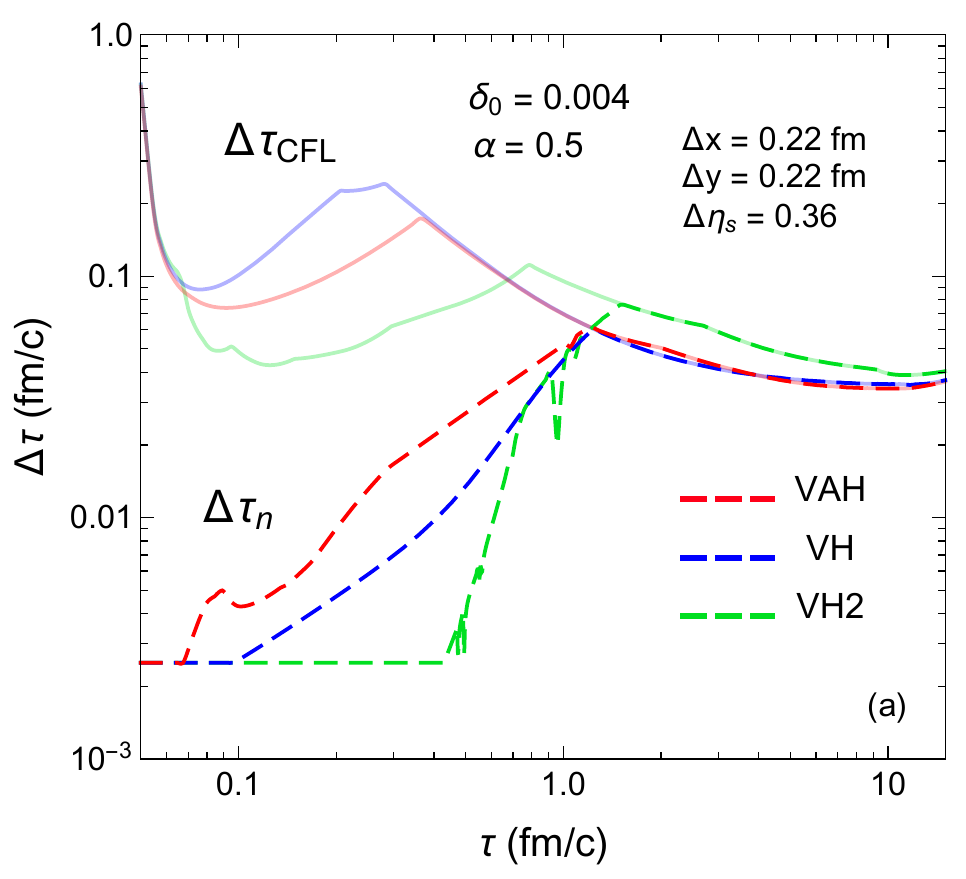}
\centering
\caption{(Color online)
\label{lattice_trento_timestep}
    The evolution of the adaptive time step $\Delta \tau_n$ (dashed color) and CFL bound $\Delta \tau_\text{CFL}$ (transparent color) in non-conformal anisotropic hydrodynamics (\cpuvah{}, red), quasiparticle viscous hydrodynamics (\vh{}, blue) and standard viscous hydrodynamics (\vh2, green), for the smooth (3+1)--d \trento{} initial condition.
}
\end{figure}

Finally, in Figure~\ref{lattice_trento_timestep} we plot the adaptive time step for each of the three hydrodynamic simulations. One sees that in anisotropic hydrodynamics (\cpuvah{}) $\Delta\tau_n$ increases steadily until hitting the CFL bound at $\tau\sim 1$ fm/$c$. Its behavior is similar in quasiparticle viscous hydrodynamics (\vh{}). The two CFL bounds are slightly separated due to differences between their transverse velocity profiles (see Fig.~\ref{lattice_trento_x}b). On the other hand, the adaptive time step in standard viscous hydrodynamics (\vh2) does not pick up until $\tau \sim 0.4$ fm/$c$ (see footnote~\ref{adaptive_floor}). Even then, it fluctuates up and down before reaching its CFL bound, which is higher than the other two bounds since its transverse flow is suppressed by a large bulk viscous pressure. This indicates that standard viscous hydrodynamics has a difficult time resolving the fireball evolution in the presence of large gradients. 

\subsubsection{Fluctuating \trento{} initial conditions}
\label{chap4S4.5.2}

Finally we study the differences among the three non-conformal hydrodynamic models for a central Pb+Pb collision with a fluctuating initial energy density profile, using the same model parameters as in Sec.~\ref{chap4S4.5.1}. In the \trento{} model used in this chapter, the energy density profile only has fluctuations in the transverse directions; the longitudinal profile is modeled with a finite rapidity plateau with smooth slopes at its longitudinal ends (see Appendix~\ref{app4d}).

\begin{figure}[!t]
 \makebox[\textwidth][c]{\includegraphics[width=0.99\linewidth]{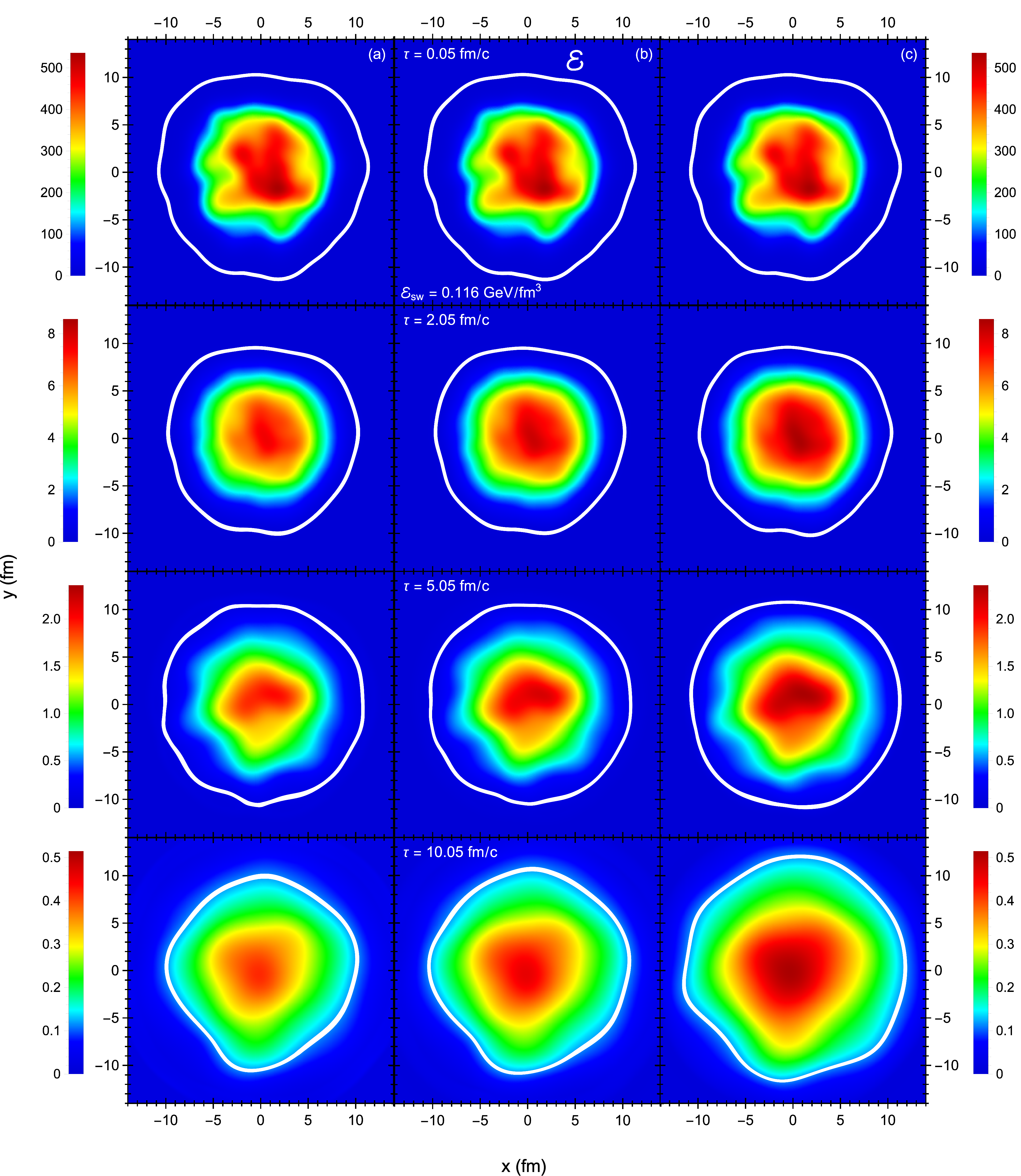}}
\centering
\caption{(Color online)
\label{fluctuating_profile}
    The evolution of the QCD energy density profile (GeV/fm$^3$) in the central transverse plane $(\eta_s{\,=\,}0)$, given by non-conformal anisotropic hydrodynamics (\cpuvah{}, left column), quasiparticle viscous hydrodynamics (\vh{}, middle column) and standard viscous hydrodynamics (\vh{}2, right column) for the fluctuating (3+1)--d \trento{} event described in the text. The white contour lines are $\eta_s{\,=\,}0$ slices at the shown time frames of a particlization hypersurface of constant energy density $\ene_\text{sw} = 0.116$\,GeV/fm$^3$.
}
\end{figure}

Figure~\ref{fluctuating_profile} shows the evolution of the fluctuating energy density profile in the transverse plane $(\eta_s = 0)$, along with the transverse slice of a particlization hypersurface of constant energy density $\ene_\text{sw} = 0.116$\,GeV/fm$^3$, for the three hydrodynamic simulations. Since anisotropic hydrodynamics generally has a smaller shear stress than viscous hydrodynamics (especially at early times), the transverse fluctuations across its fireball show the least dissipation or smearing. As a result, it can convert the initial-state eccentricities into anisotropic flow slightly more efficiently. The pressure anisotropy $\PL{-}\Pperp$ and bulk viscous pressure $\Pi$ mainly influence the overall size of the fireball (and particlization hypersurface), affecting the final-state particle yields. The initially strong longitudinal expansion rapidly cools down the system and temporarily shrinks the fireball size at early times; a more positive longitudinal pressure helps cool the fireball even further. Compared to the two viscous hydrodynamic models, anisotropic hydrodynamics has a larger longitudinal pressure, which means its hypersurface has a slightly narrower waist. At later times, the transverse expansion overtakes the longitudinal expansion, increasing the fireball size. Although anisotropic hydrodynamics has the largest transverse flow, it also has the smallest bulk viscous pressure, enabling the fireball to cool faster and evaporate more quickly. This ultimately results in a smaller maximum fireball size. In contrast, standard viscous hydrodynamics has a much larger bulk viscous pressure. This extends the fireball's lifetime by about $1$ fm$/c$ relative to the one in anisotropic hydrodynamics, allowing it to grow larger in size. 

\begin{figure}[thbp]
\includegraphics[width=0.98\linewidth]{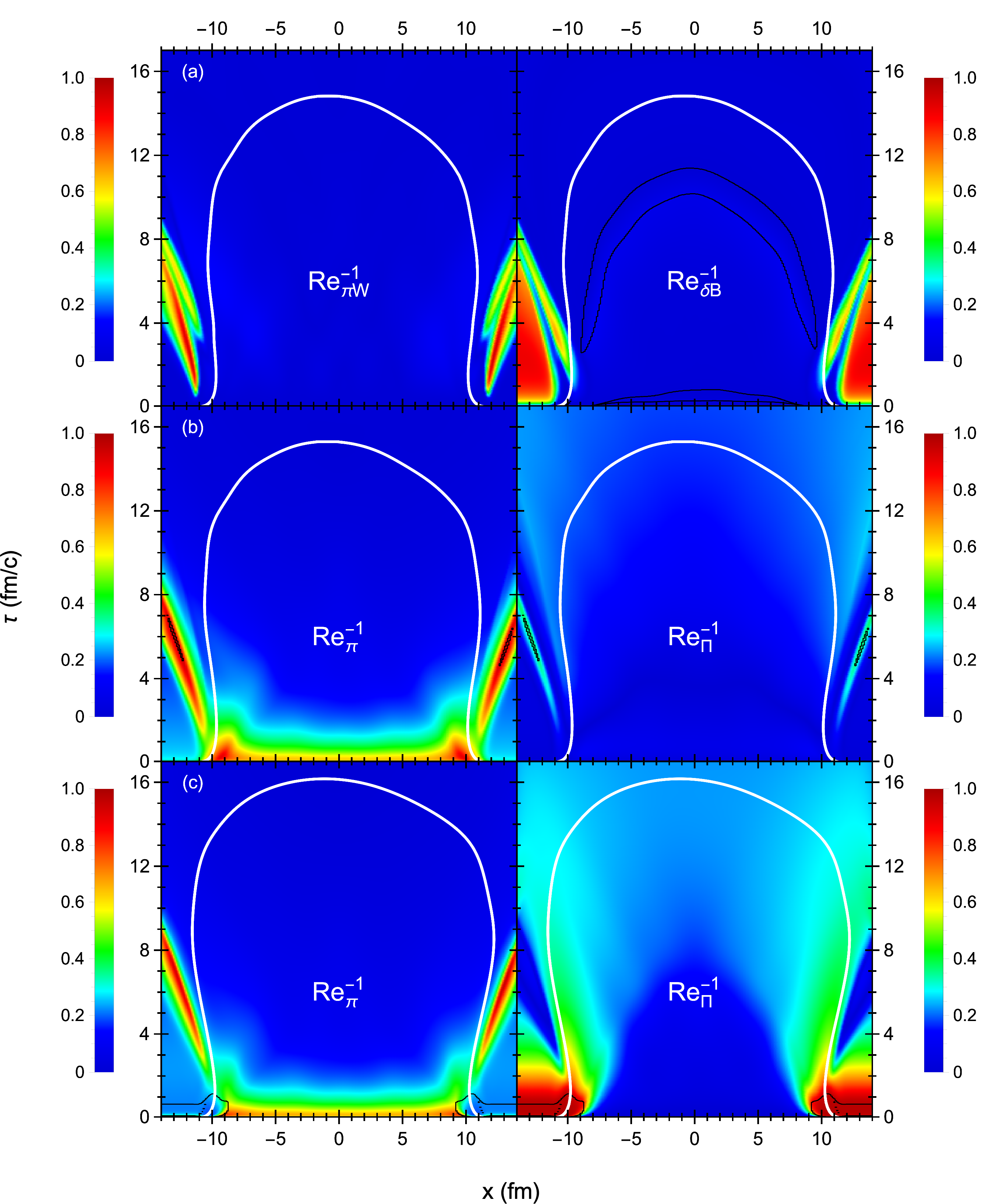}
\centering
\caption{(Color online)
\label{freezeout_x}
    The $\tau{-}x$ slice at $y{\,=\,}\eta_s{\,=\,}0$ of the residual shear and mean-field inverse Reynolds numbers in anisotropic hydrodynamics (\cpuvah{}, top row) and the shear and bulk inverse Reynolds numbers in quasiparticle viscous hydrodynamics (\vh{}, middle row) and standard viscous hydrodynamics (\vh{}2, bottom row), for the fluctuating (3+1)--d \trento{} event. The white contour lines are $\tau{-}x$ slices ($y=\eta_s=0$) of the same particlization hypersurface as Fig.~\ref{fluctuating_profile}. Spacetime regions that are regulated according to Sec.~\ref{chap4S3.5} and App.~\ref{app4e} are circled in black.
}
\end{figure}

\begin{figure}[thbp]
\includegraphics[width=0.98\linewidth]{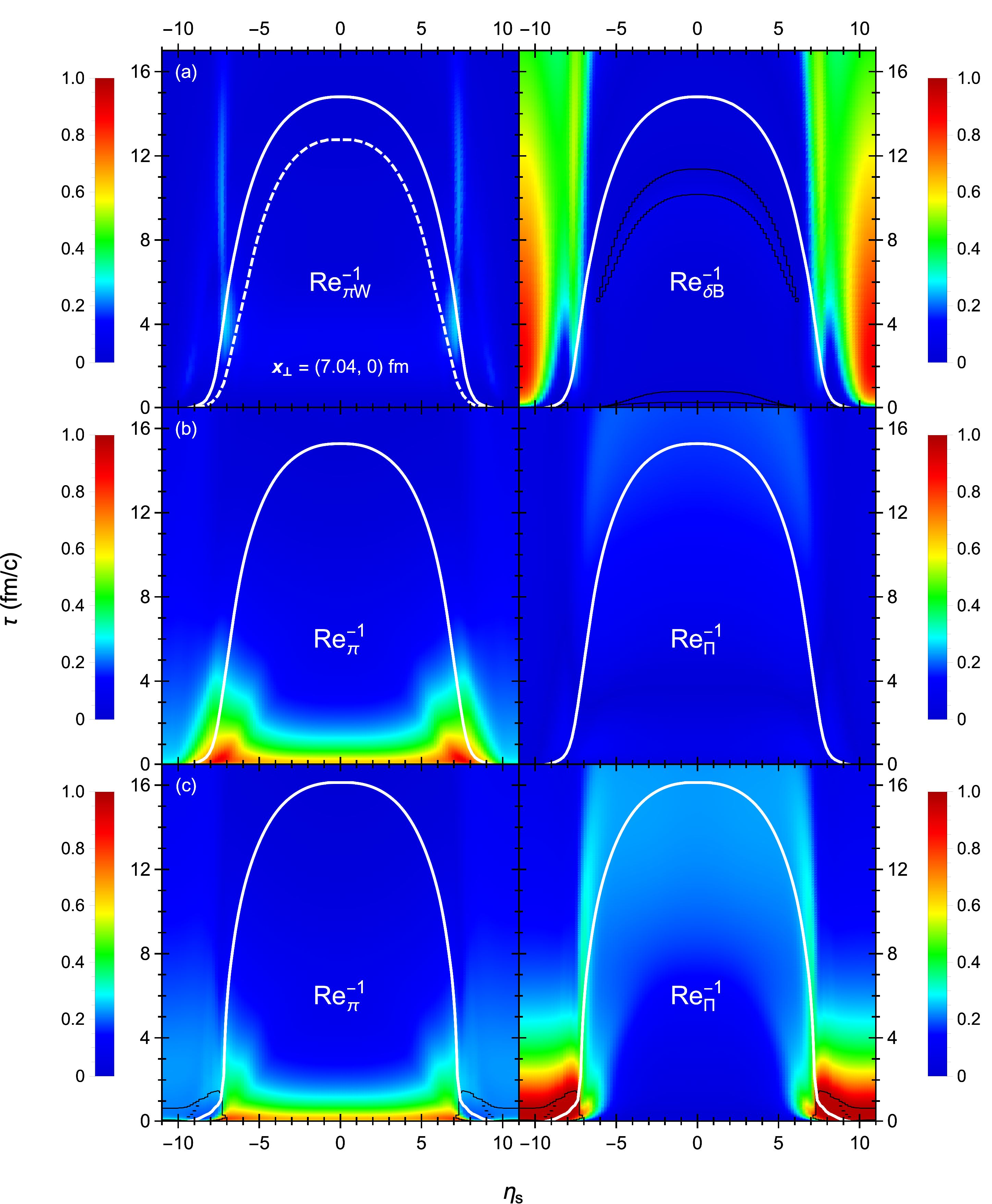}
\centering
\caption{(Color online)
\label{freezeout_z}
    The same as Fig.~\ref{freezeout_x} but for the $\tau{-}\eta_s$ plane at $x{\,=\,}y{\,=\,}0$. The $\eta_s$--axis in the upper left panel is shifted transversely to $(x,y) = (7.04, 0)$\,fm; the corresponding hypersurface slice (dashed white) is smaller than the one at $x{\,=\,}y{\,=\,}0$ (solid white).
}
\end{figure}

In Figures~\ref{freezeout_x} and~\ref{freezeout_z} we plot the inverse Reynolds numbers from the three hydrodynamic simulations in the $\tau{-}x$ plane at $y{\,=\,}\eta_s{\,=\,}0$ and in the $\tau{-}\eta_s$ plane at $x{\,=\,}y{\,=\,}0$, respectively. Traditionally, the inverse Reynolds number measures the validity of the hydrodynamic expansion around local equilibrium \cite{Bazow:2016yra, Shen:2014vra}. Since anisotropic hydrodynamics expands the energy-momentum tensor~\eqref{eqch3:3} around an anisotropic background (i.e. $T^\munu_a = \ene u^\nu u^\nu + \PL z^\mu z^\nu - \frac{1}{2}\Pperp \Xi^\munu$), its validity is measured by the residual shear inverse Reynolds number~\cite{Bazow:2017ewq}
\bs
\allowdisplaybreaks
\beal
    \text{Re}^{-1}_{\pi W} =&\, \sqrt{\frac{\pi_\perp {\cdot\,} \pi_\perp - 2 W_{\perp z} {\,\cdot\,} W_{\perp z}}{\mathcal{P}_L^2 + 2\mathcal{P}_\perp^2}}\,,
\end{align}
\es
as opposed to the shear and bulk inverse Reynolds numbers in second-order viscous hydrodynamics \cite{Bazow:2016yra, Shen:2014vra}: 
\be
    \text{Re}^{-1}_{\pi} =\, \frac{\sqrt{\pi \cdot \pi}}{\Peq\sqrt{3}}\,,\qquad
    \text{Re}^{-1}_{\Pi} =\, \frac{|\Pi|}{\Peq}\,,
\ee
with $\pi {\,\cdot\,} \pi = \pi_\munu \pi^\munu$. One sees that the residual shear inverse Reynolds numbers stay much smaller than unity inside the fireball, indicating that a first-order expansion in $O(\text{Re}^{-1}_{\pi W})$ is sufficient to capture the residual shear corrections to the anisotropic hydrodynamic equations. While there is no need to regulate their strength here, the regulation scheme might be needed for collision events with sharper initial-state fluctuations.


We also plot the contribution of the non-equilibrium mean-field component $\delta B$ to the bulk viscous pressure in anisotropic hydrodynamics:
\be
    \text{Re}_{\delta B}^{-1} = \frac{|\delta B|}{\Peq}\,.
\ee
We find that $\delta B$ is moderately large and positive near the edges of the fireball but small and negative inside the fireball. However, it is necessary to regulate the mean-field in the latter region via Eq.~\eqref{eqchap4:dB_reg} to prevent unstable growth. Specifically, the instability seems to originate near the quark-hadron phase transition, where the driving force proportional to the kinetic trace anomaly $\ene^{(k)} - 2\mathcal{P}_\perp^{(k)} - \mathcal{P}_L^{(k)} = \ene - 2\Pperp - \PL - 4 B$ is at its strongest while the bulk relaxation rate $\tau_\Pi^{-1}$ is near a minimum.  After applying the regulation~\eqref{eqchap4:dB_reg} for a brief period, $\delta B$ evolves freely from that point onward. While this initial instability is unfortunate, the overall impact of the non-equilibrium mean-field component on the longitudinal and transverse pressures inside the fireball region is limited.

In second-order viscous hydrodynamics, the shear inverse Reynolds number is large for a short period of time $\Delta\tau \sim 1$ fm/$c$ after the collision. The bulk inverse Reynolds number in standard viscous hydrodynamics is also very large near the peak of the bulk viscosity $\zetas$ during the same time frame. To prevent the code from crashing, our regulation scheme suppresses both the shear stress and bulk viscous pressure (see Appendix~\ref{app4e}). Although it is still technically possible to run the heavy-ion simulation with standard viscous hydrodynamics, we cannot escape the effects of regulation around the base of the particlization hypersurface. In contrast, the viscous pressures in quasiparticle viscous hydrodynamics often do not require regulation\footnote{Even in the absence of regulation, the longitudinal pressure in quasiparticle viscous hydrodynamics can still be negative in some regions at early times ($\tau < 0.6$ fm/$c$) (e.g. see Figs.~\ref{lattice_trento_x} and~\ref{lattice_trento_z}).} since the bulk inverse Reynolds number is quite small.

We close this section with a comparison of the particlization hypersurfaces in the $\tau{-}x$ and $\tau{-}\eta_s$ planes. In Fig.~\ref{freezeout_x} one sees that standard viscous hydrodynamics has the largest hypersurface along the $x$--direction since its large bulk viscous pressure generates the most viscous heating. It also has the longest lifetime $\tau_f \sim 17$\,fm/$c$, for the same reason. In contrast, it has a pretty narrow waist along the $\eta_s$--direction because its longitudinal flow profile $u^\eta$ initially contracts the medium in the rapidity direction. In anisotropic hydrodynamics, the longitudinal flow begins transporting the medium away from the collision zone, following the direction of the longitudinal pressure gradients. As a result, its hypersurface has a wider waist than the one in standard viscous hydrodynamics (about $\Delta\eta_s \sim 0.5$ larger for $\tau \sim 1-2$ fm/$c$) along the $\eta_s$--direction. 

\section{Summary}
\label{chap4S6}

In this chapter we developed a (3+1)--dimensional anisotropic fluid dynamical simulation that evolves both the far-off-equilibrium and viscous hydrodynamic stages of ultrarelativistic nuclear collisions. We validated the code's performance by reproducing the semi-analytic solutions of conformal and non-conformal Bjorken flow and conformal Gubser flow on a Cartesian grid. Thanks to the adaptive time step algorithm derived in Sec.~\ref{chap4S3.6}, we can accurately capture the very early-time dynamics of both Bjorken and Gubser flow while finishing the simulation within a reasonable number of iterations. We also compared anisotropic hydrodynamics to two variants of second-order viscous hydrodynamics in central Pb+Pb collisions. Apart from an apparent sensitivity of the bulk viscous pressure evolution to the bulk relaxation time, the three hydrodynamic models have similar transverse dynamics, at least in the mid-rapidity region $\eta_s \sim 0$. However, near the longitudinal edges of the fireball, the fluid's longitudinal evolution varies greatly between the models. We showed that the longitudinal pressure in (3+1)--dimensional anisotropic hydrodynamics stays positive even in the presence of large gradients at very early times, unlike in viscous hydrodynamics. This causes the fluid to initially expand outward along the spacetime rapidity direction, as expected from the outward-pointing longitudinal gradients of the thermal pressure, instead of collapsing inward as seen in standard viscous hydrodynamics. This reduces the risk of cavitation at the beginning of the simulation. With the new hydrodynamic code presented here, we can for the first time model even the very early far-from-equilibrium dynamical stage at $\tau_0{\,\ll\,}\tau_\text{hydro}$ with a realistic, non-conformal QCD equation of state, as opposed to the conformal equation of state implicit in other pre-hydrodynamic models \cite{Chesler:2010bi, Chesler:2013lia, Liu:2015nwa, Chesler:2015bba, Attems:2016tby, Heller:2016rtz, Keegan:2016cpi, Kurkela:2018wud}.

Later in Chapter~\ref{chapter7label}, we will run \cpuvah{} in the JETSCAPE framework with the Maximum a Posteriori (MAP) model parameters extracted in Refs.~\cite{Everett:2020yty, Everett:2020xug}. Since the JETSCAPE SIMS hybrid model \cite{Everett:2020yty, Everett:2020xug} uses conformal free-streaming and standard viscous hydrodynamics to evolve the pre-equilibrium and hydrodynamic stages, we will replace these two modules with our code and study the changes to hadronic observables such as the mean transverse momentum  $\langle p_T\rangle$ and $p_T$--integrated anisotropic flow coefficients $v_n$.\footnote{%
    Several model parameters associated with the conformal free-streaming module will not be used in \cpuvah{}.}

With this chapter, we temporarily close our current discussion about anisotropic fluid dynamics and its numerical implementation for heavy-ion collisions. In the next two chapters we turn to the particlization stage, where hadrons are emitted from the expanding quark-gluon plasma after it has cooled down below the pseudocritical temperature $T_c$. This step is necessary to turn our fluid dynamical output into real particles whose characteristics can be compared with experimental measurements. We will review several models for the viscous corrections to the hadronic distribution function $f_n(x,p)$, and also introduce two new models that are positive definite. Finally, we develop and test another software package called {\sc iS3D}, which samples a discrete number of hadrons from the Cooper--Frye formula~\eqref{eqch5:CooperFrye} using one of these particlization models. 
\chapter{Modified equilibrium distributions for Cooper--Frye particlization}
\label{chap5label}
In this chapter we introduce a positive definite hadronic distribution that is suitable for describing the transition from anisotropic hydrodynamics to a microscopic kinetic description during the late stages of heavy-ion collisions in the presence of moderately large viscous corrections. First we construct a modified equilibrium distribution using macroscopic input from standard viscous hydrodynamics, and test its hydrodynamic output for a stationary hadron resonance gas subject to either shear stress, bulk viscous pressure, or baryon diffusion current at a given freeze-out temperature and baryon chemical potential. While it does not exactly reproduce all components of the net baryon current and energy-momentum tensor, it improves upon the linearized non-equilibrium corrections $\delta f_n$ commonly used in the Cooper--Frye formula~\eqref{eqch5:CooperFrye} since the latter typically lead to unphysical negative distribution functions at large particle momenta. We further refine the modified equilibrium distribution with input from anisotropic hydrodynamics. This modified anisotropic distribution is compatible with \cpuvah{} by accurately reproducing all components in the energy-momentum tensor decomposition~\eqref{eqch3:3} while staying positive definite. Finally, we compare the particle spectra and $p_T$--differential elliptic flow coefficients from the Cooper--Frye formula computed with the modified distributions and with linearized $\delta f_n$ corrections for two different (2+1)--dimensional hypersurfaces corresponding to central and non-central Pb+Pb collisions at the LHC. 

This chapter is based on material submitted for publication in Ref.~\cite{McNelis:2021acu}.
\section{Linearized viscous corrections} 
\label{sec2}
First we review the 14--moment approximation and RTA Chapman--Enskog expansion, the two most popular methods for linearizing the non-equilibrium correction to the hadronic distribution function $f_n = f_{\eq,n} + \delta f_n$. Here the local-equilibrium distribution is given by
\be
\label{eqch5:feq}
 f_{\mathrm{eq},n}(x,p) = \frac{g_n}{\exp\biggl[\dfrac{p \cdot u(x)}{T(x)} - b_n \alpha_B(x)\biggr] + \Theta_n} \,,
\ee
where $g_n=2s_n + 1$ and $b_n$ are the spin degeneracy and baryon number of species $n$\footnote{%
    Different isospin (electric charge) and strangeness states within a hadronic multiplet (e.g.\ $K^+$ and $K^0$) are counted as separate hadron species; however, we here ignore chemical potentials associated with these additional conserved charges.}
, $u^\mu(x)$ is the fluid velocity, $T(x)$ is the temperature, $\alpha_B(x) = \mu_B(x)/T(x)$ is the baryon chemical potential-to-temperature ratio and $\Theta_n \in [1,-1]$ accounts for the quantum statistics of fermions and bosons, respectively. Unless stated otherwise, the baryon chemical potential $\alpha_B$ is nonzero.
\subsection{14--moment approximation}
\label{sec2a}

The 14--moment approximation is a moments expansion of the distribution function around $f_{\eq,n}$ that is truncated to the 14 lowest momentum moments ($p^\mu$, $p^\mu p^\nu$)~\cite{CPA:CPA3160020403}. This approximation assumes that the distribution function can be adequately characterized by just its hydrodynamic moments (i.e. $\int_p p^\mu f_n$, $\int_p p^\mu p^\nu f_n$) ~\cite{Denicol:2012cn}. For a multi-component gas with nonzero baryon chemical potential, the 14--moment approximation reads~\cite{Monnai:2009ad} 
\be
\label{eqch5:14_moment}
  \delta f^{14}_n = f_{\eq,n} \bar{f}_{\eq,n} \left( b_n c_\mu p^\mu + c_\munu p^\mu p^\nu \right)\,,
\ee
where $\bar{f}_{\eq,n} = 1 - g_n^{-1} \Theta_n f_{\eq,n}$. For simplicity, we follow common practice and take the expansion coefficients $c_\mu$ and $c_\munu$ to be species-independent. To solve for the coefficients, one rewrites Eq.~\eqref{eqch5:14_moment} in irreducible form:
\be
\label{eqch5:14_moment_irreducible}
  \frac{\delta f^{14}_n}{f_{\eq,n} \bar{f}_{\eq,n}} = c_T m_n^2 + b_n \big(c_B(\up) {+} c_V^{\langle\mu\rangle} p_{\langle\mu\rangle}\big) + c_E (u {\,\cdot\,}p)^2 + c_Q^{\langle\mu\rangle} (u {\,\cdot\,}p) p_{\langle\mu\rangle} + c_\pi^{\langle\munu\rangle} p_{\langle\mu} p_{\nu\rangle}\,,
\ee
where $p^{\langle\mu\rangle} = \Delta^\mu_\nu p^\nu$ and $p^{\langle\mu} p^{\nu\rangle} = \Delta^\munu_{\alpha\beta} p^\alpha p^\beta$, with $\Delta^\munu = g^\munu - u^\mu u^\nu$ and $\Delta^\munu_{\alpha\beta} =\frac{1}{2}\big(\Delta^\mu_\alpha \Delta^\nu_\beta{+}\Delta^\mu_\beta \Delta^\nu_\alpha \big) - \frac{1}{3}\Delta^\munu \Delta_{\alpha\beta}$, are purely spatial in the LRF and traceless. The irreducible coefficients in (\ref{eqch5:14_moment_irreducible}) are
\bs
\allowdisplaybreaks
\beal
  c_B &= u_\mu c^{\mu} \,,
\\
  c_V^{\langle\mu\rangle} &= \Delta^\mu_\nu c^{\nu} \,,
\\
  c_T &= g_\munu c^\munu \,,
\\
  c_E &= u_\mu u_\nu  c^{\munu} \,,
\\
  c_Q^{\langle\mu\rangle} &= u_\alpha \Delta^\mu_\beta c^{\alpha\beta} \,,
\\
  c_\pi^{\langle\munu\rangle} &= \Delta^\munu_{\alpha\beta} \, c^{\alpha\beta}\,.
\end{align}
\es
One solves for these coefficients by inserting Eq.~\eqref{eqch5:14_moment_irreducible} into the Landau matching conditions for the net baryon density $n_B(x)$ and energy density $\ene(x)$, the definition of the Landau frame $T^\munu u_\nu = \ene u^\mu$ (which implies a vanishing heat current $Q^\mu(x) = 0$), and the kinetic definitions for the bulk viscous pressure $\Pi(x)$, baryon diffusion current $V_B^\mu(x)$ and shear stress tensor $\pi^\munu(x)$:
\bs
\allowdisplaybreaks
\label{eqch5:matching}
\beal
  \delta n_B &= \sum_n \int_{p} b_n \, (\up) \delta f_n = 0 \,,
\\
  \delta \ene &= \sum_n \int_{p} (\up)^2 \delta f_n = 0 \,,
\\
  Q^\mu &= \sum_n \int_{p} (\up) p^{\langle\mu\rangle} \delta f_n = 0 \,,
\\
  \Pi &= \frac{1}{3} \sum_n \int_{p} (- p {\,\cdot\,} \Delta {\,\cdot\,} p) \delta f_n \,,
\\
  V_B^\mu &= \sum_n \int_{p} b_n \, p^{\langle\mu\rangle} \delta f_n \,,
\\
  \pi^\munu &= \sum_n \int_{p} p^{\langle\mu} p^{\nu\rangle} \delta f_n \,.
\end{align}
\es
Here the sum runs over the number of resonances $N_R$ and $\int_{p} \equiv \int \dfrac{d^3p}{(2\pi\hbar)^3 E}$. After some algebra one obtains
\bs
\allowdisplaybreaks
\label{eq7}
\beal
  c_T &= \frac{\Pi \, \mathcal{P}}{\A_{21}\mathcal{P} + \N_{31}\mathcal{Q} + \J_{41}\mathcal{R}} \,,
\\
  c_B &= \frac{\Pi \, \mathcal{Q}}{\A_{21}\mathcal{P} + \N_{31}\mathcal{Q} + \J_{41}\mathcal{R}} \,,
\\
  c_E &= \frac{\Pi \, \mathcal{R} }{\A_{21}\mathcal{P} + \N_{31}\mathcal{Q} + \J_{41}\mathcal{R}} \,,
\\
  c_V^{\langle\mu\rangle} &= \frac{V_B^\mu \J_{41}}{\N^2_{31} - \M_{21} \J_{41}} \,,
\\
  c_Q^{\langle\mu\rangle} &= -\frac{V_B^\mu \N_{31}} {\N^2_{31} - \M_{21} \J_{41}} \,,
\\
  c_\pi^{\langle\munu\rangle} &= \frac{\pi^\munu}{2(\ene{+}\Peq)T^2}\,,
\end{align}
\es
where $\Peq(\ene,n_B)$ is the equilibrium pressure and
\bs
\label{eq8}
\beal
  \mathcal{P} = &\, \N^2_{30} - \J_{40}\M_{20} \,, 
\\
  \mathcal{Q} = &\, \B_{10}\J_{40} - \A_{20}\N_{30} \,, 
\\
  \mathcal{R} = &\, \A_{20}\M_{20} - \B_{10}\N_{30}\,. 
\end{align}
\es
The thermal integrals $\J_{kq}$, $\mathcal{N}_{kq}$, $\mathcal{M}_{kq}$, $\A_{kq}$, and $\B_{kq}$ are defined in Appendix~\ref{app5:integrals}. It is well known that the coefficients of the 14--moment approximation are linearly proportional to the dissipative flows of the fluid~\cite{DeGroot:1980dk}. An obvious problem with truncating the $\delta f_n^{14}$ correction to first order is that it can overwhelm the local-equilibrium distribution at sufficiently high momentum, potentially turning the total distribution function negative. This problem grows worse with larger dissipative flows. In principle, one can systematically improve the moments expansion by including the non-hydrodynamic moments ($\int_p p^\mu p^\nu p^\lambda f_n$, etc.)~\cite{Denicol:2012cn}. However, using them as macroscopic input along with $J_B^\mu$ and $T^\munu$ would require solving a set of evolution equations for these higher-order moments together with the fluid dynamical simulation.

\subsection{First-order RTA Chapman--Enskog expansion}
\label{sec2b}

Another common approach to obtaining a linear $\delta f_n$ correction for the Cooper--Frye formula is the RTA Chapman--Enskog expansion, which is a perturbative series of the form~\cite{chapman1990mathematical}
\be
\label{eqch5:gradient_series}
  f_n =  f_{\eq,n} + \sum^\infty_{k=1} \delta f_n^{(k)} = f_{\eq,n} + \sum^\infty_{k=1} \epsilon^k h_n^{(k)}\,,
\ee 
where the expansion parameter and coefficients $\epsilon$ and $h_n^{(k)}$ are determined by solving the Boltzmann equation in the relaxation time approximation (RTA) perturbatively~\cite{Anderson_Witting_1974,Jaiswal:2014isa}:
\be
\label{eqch5:RTA}
  p^\mu \partial_\mu f_n = \frac{(\up)(f_{\eq,n} - f_n)}{\tau_r}\,.
\ee
Here we take the relaxation time $\tau_r(x)$ to be momentum and species independent. If the hydrodynamic gradients are small compared to the relaxation rate $\tau_r^{-1}$, the expansion parameter is the directional derivative
\be
\label{eq11}
  \epsilon = - s^\mu \partial_\mu
\ee
with
\be
\label{eqch5:s}
  s^\mu(x) = \dfrac{\tau_r(x)\, p^\mu}{u(x) {\,\cdot\,} p}\,.
\ee
Inserting the series \eqref{eqch5:gradient_series} into Eq.~\eqref{eqch5:RTA} one finds at first order in gradients
\be
\label{eqch5:df_RTA}
\delta f^{(1)}_n = - s^\mu \partial_\mu f_{\eq,n} \,.
\ee 
Using Eq. (\ref{eqch5:feq}) and expanding the derivative one obtains
\be
\label{eqch5:df1}
\begin{split}
  \frac{\delta f_n^{(1)}}{f_{\eq,n} \bar{f}_{\eq,n}} =& - \tau_r \bigg(b_n \dot{\alpha}_B + \frac{(\up)\,\dot{T}}{T^2} + \frac{(- p {\,\cdot\,} \Delta {\,\cdot\,} p)\theta}{3(\up)T} 
 + \frac{b_n p^{\langle\mu\rangle} \nabla_\mu \alpha_B}{\up}  \\
 & - \frac{p^{\langle\mu\rangle} (\dot{u}_\mu {-} \nabla_\mu {\ln} T)}{T} - \frac{\sigma_\munu p^{\langle\mu} p^{\nu\rangle}}{(\up)T}\bigg) \,,
\end{split}
\ee
where $\theta = \partial_\mu u^\mu$ is the scalar expansion rate and $\sigma_\munu = \partial_{\langle\mu} u_{\nu\rangle} \equiv \Delta^{\alpha\beta}_\munu \partial_\beta u_\alpha$ is the velocity shear tensor. We denote the LRF time derivative as $\dot{a} = u^\mu \partial_\mu a$ and the spatial gradient in the LRF as $\nabla^\mu a = \Delta^\munu \partial_\nu a$. To simplify Eq.~\eqref{eqch5:df1} in terms of the hydrodynamic quantities one makes use of the conservation equations for the net baryon number, energy and momentum,
\bs
\allowdisplaybreaks
\label{eqch5:conservation_laws}
\beal
  \partial_\mu J_B^\mu & = 0 \,,
\\
  u_\nu \partial_\mu T^\munu & = 0 \,,
\\
  \Delta^\mu_\nu \partial_\lambda T^{\lambda\nu} & = 0\,,
\end{align}
\es
to eliminate the time derivatives:
\bs
\allowdisplaybreaks
\label{eq15}
\beal
  \dot\alpha_B &\approx \mathcal{G} \theta\,,
\\
  \dot T &\approx \mathcal{F} \theta \,,
\\
  \dot u^\mu &\approx \nabla^\mu \ln{T} + \frac{n_B T}{\ene {+} \Peq} \nabla^\mu \alpha_B\,.
\end{align}
\es
Here we only keep the first-order terms; the coefficients $\mathcal{G}$ and $\mathcal{F}$ are listed in Appendix~\ref{app5:conservation}. The spatial gradients are substituted by dissipative flows using the Navier--Stokes relations:
\bs
\allowdisplaybreaks
\label{eq16}
\beal
  \Pi &\approx - \zeta \theta \,,
\\
  V_B^\mu &\approx \kappa_B \nabla^\mu\alpha_B \,,
\\
  \pi^\munu &\approx 2 \eta \sigma^\munu \,.
\end{align}
\es
The first-order RTA Chapman--Enskog expansion \eqref{eqch5:df1} then reduces to 
\be
\label{eqch5:Chapman_Enskog}
\begin{split}
  \frac{\delta f^{\text{CE}}_n}{f_{\eq,n} \bar{f}_{\eq,n}} =\,& \frac{\Pi}{\beta_\Pi}\bigg(b_n \mathcal{G} +\frac{(\up)\mathcal F}{T^2} + \frac{(-p \cdot \Delta \cdot p)}{3(\up)T}\bigg) 
+ \frac{V_B^\mu p_{\langle\mu\rangle}}{\beta_V}\bigg(\frac{n_B}{\ene {+} \Peq} - \frac{b_n}{\up}\bigg) \\
&+ \frac{\pi_\munu p^{\langle\mu} p^{\nu\rangle}}{2\beta_\pi (\up)T}\,,
\end{split}
\ee
where $\beta_\Pi$, $\beta_V$ and $\beta_\pi$ are the ratios of the bulk viscosity, baryon diffusion coefficient and shear viscosity, respectively, to the relaxation time. It is straightforward to check that $\delta f^{\text{CE}}_n$ satisfies the Landau matching and frame conditions (\ref{eqch5:matching}a-c). The coefficients $\beta_\Pi$, $\beta_V$ and $\beta_\pi$ can be extracted by inserting Eq.~\eqref{eqch5:Chapman_Enskog} into Eqs.~(\ref{eqch5:matching}d-f). The resulting expressions are 
\bs
\allowdisplaybreaks
\label{eq18}
\beal
  \beta_\Pi &= \mathcal{G} n_B T + \frac{\mathcal{F} (\ene{+}\Peq)}{T} + \frac{5 \J_{32}}{3T} \,,
\\
  \beta_V &= \mathcal{M}_{11} - \frac{n_B^2 T}{\ene{+}\Peq} \,,
\\
  \beta_\pi &= \frac{\J_{32}}{T} \,.
\end{align}
\es
Like the 14--moment approximation, the first-order Chapman--Enskog expansion reproduces $J_B^\mu$ and $T^\munu$ exactly, but the total distribution function can turn negative at high momentum. In this case, one could improve the expansion scheme by adding higher-order gradient corrections, but this approach is tedious and will not be pursued here. 

\section{Modified equilibrium distributions} 
\label{sec3}
Both the 14--moment approximation (\ref{eqch5:14_moment_irreducible}) and the first-order RTA Chapman--Enskog expansion (\ref{eqch5:Chapman_Enskog}) parametrize the momentum dependence of $\delta f_n(x,p)$ in terms of the available hydrodynamic information in such a way that the dissipative flows are exactly reproduced from the corresponding hydrodynamic moments of $\delta f_n$. However, their total distribution function $f_{\eq,n} + \delta f_n$ suffers from the negative probability problem at sufficiently high momenta due to the truncation of the expansion at first order in the dissipative flows. One way to circumvent this problem is to manipulate the argument of the exponential function in the local-equilibrium distribution \eqref{eqch5:feq} with viscous corrections, as was done in Ref.~\cite{Pratt:2010jt}. By construction, this avoids the negative probability problem but, as we will see, at the expense of not being able to match the dissipative flows exactly to the corresponding moments of this modified equilibrium distribution.
\subsection{Pratt--Torrieri distribution}
Over the past decade, work has been done on constructing a non-equilibrium distribution function suitable for Cooper--Frye particlization that does not rely on a linearized expansion scheme. In the original work by Pratt and Torrieri~\cite{Pratt:2010jt} (see also \cite{Dusling:2011fd} for a related ansatz), dissipative perturbations are added to the ``auxiliary fields'' ($T, u^\mu, \alpha_B$) in the Boltzmann factor, effectively transforming the local-equilibrium distribution \eqref{eqch5:feq} into a quasi-equilibrium distribution. The resulting modified equilibrium distribution is given by the formula
\be
\label{eqch5:feqmod_original}
  f_{\eq,n}^\text{PT} = 
  \frac{\mathcal{Z} g_n}
       {\exp\biggl[\dfrac{\sqrt{{\bm{p}'^2{+}m_n^2}}}{T {+} \delta T} - b_n(\alpha_B {+} \delta \alpha_B)\biggr] + \Theta_n}\,,
\ee
where $\delta T = \lambda_T \Pi$ and $\delta\alpha_B = \lambda_{\alpha_B} \Pi$ are bulk viscous corrections to the effective temperature and chemical potential, while $p^{\,\prime}_i = - X_i \cdot p^{\,\prime}$ are modified local-rest-frame (LRF) momentum components, with $X^\mu_i = (X^\mu, Y^\mu, Z^\mu)$ being the spatial basis vectors (i.e. the four-vectors that reduce in the LRF to the directional unit vectors along the $x$, $y$, $z$ axes). The fluid velocity perturbation $\delta u^\mu$ is encoded in the momentum space transformation
\be
\label{eqch5:rescale}
  p_i = A_{ij} p^{\,\prime}_j \,,
\ee
where $p_i = - X_i \cdot p$ are the usual LRF momentum components and
\be
\label{eqch5:Aij}
  A_{ij} = (1 {+} \lambda_\Pi \Pi)\delta_{ij} + \lambda_\pi \pi_{ij}
\ee
is a symmetric matrix that deforms the momentum space linearly with the bulk viscous pressure $\Pi$ and LRF shear stress tensor $\pi_{ij}= X^\mu_i X^\nu_j \pi_\munu$.\footnote{%
    The scalar coefficients ($\lambda_T$, $\lambda_{\alpha_B}$, $\lambda_\Pi$, $\lambda_\pi$) are functions of temperature and chemical potential ($T$, $\alpha_B$).}$^,$\footnote{%
    Ref.~\cite{Pratt:2010jt} did not consider a nonzero baryon diffusion current $V_B^\mu$.} 
The normalization factor 
\be
\label{eqch5:Z_PT}
 \mathcal{Z} = \frac{1}{{\det}A}
\ee
rescales the distribution function such that its particle density agrees with that of a local-equilibrium distribution with temperature $T+\delta T$ and chemical potential $\alpha_B+\delta\alpha_B$.

The modified equilibrium distribution (\ref{eqch5:feqmod_original}) is constructed such that its momentum moments reproduce all components of $J_B^\mu$ and $T^\munu$ at first order in the (small) dissipative flows. As the shear stress $\pi_{ij}$ and bulk viscous pressure $\Pi$ become larger, however, the mismatch between the hydrodynamic input and the output when re-computing them as moments of the modified distribution (\ref{eqch5:feqmod_original}) increases. This raises the question whether the modifications to the local-equilibrium distribution can minimize this mismatch even for moderately large viscous corrections. Tests conducted in Ref.~\cite{Pratt:2010jt} for the case of vanishing bulk viscous pressure ($\Pi=0$) showed that Eq.~\eqref{eqch5:feqmod_original} accurately reproduces the target shear stress as well as the energy and charge densities even for moderately large shear stress modifications. However, when repeated for moderately large bulk viscous pressures, this time setting the shear stress to $\pi_{ij} = 0$, the modified equilibrium distribution's hydrodynamic output strongly deviated from the input values. This suggests that the PT distribution \eqref{eqch5:feqmod_original} can be further improved.
\subsection{PTM distribution}
\label{sec3a}
In this chapter we derive the modified equilibrium distribution with a more systematic approach to search for additional corrections to the PT distribution~\eqref{eqch5:feqmod_original}. Our starting point is the RTA Chapman--Enskog expansion. Equation \eqref{eqch5:df_RTA} is the first-order term of the Taylor series
\be
\label{eq21}
  f_{\eq,n}(x{\,-\,}s, p) \approx f_{\eq,n}(x,p) - s^\mu \partial_\mu f_{\eq,n}(x,p)\,,
\ee
with $s^\mu$ given by Eq.~(\ref{eqch5:s}). Therefore we have as a first approximation for the modified equilibrium distribution
\be
\label{eqch5:feqmod_start}
  f^\text{(mod)}_{\eq,n}(x,p) = f_{\eq,n}(x{\,-\,}s, p)\,,
\ee
which retains the same analytic form as $f_{\eq,n}$, except for a shift in the position arguments of the auxiliary fields $T(x)$, $\alpha_B(x)$ and $u^\mu(x)$. Note that the shift depends on the particle momentum $p^\mu$ at which the distribution function is evaluated. Assuming small gradients we can evaluate these corrections to first-order:
\bs
\label{eqch5:shifts}
\beal
  T(x{\,-\,}s) &\approx T(x) - s^\nu\partial_\nu T(x) \,,
\\
  \alpha_B(x{\,-\,}s) &\approx \alpha_B(x) - s^\nu\partial_\nu \alpha_B(x) \,,
\\
  u^\mu(x{\,-\,}s) &\approx u^\mu(x) - s^\nu\partial_\nu u^\mu(x)\,.
\end{align}
\es
The perturbations can be rewritten as
\bs
\allowdisplaybreaks
\label{eqch5:perturbations}
\beal
  \delta T &= - \tau_r \dot T - \frac{\tau_r p^{\langle\mu\rangle}\nabla_\mu T}{\up} \,,
\\
  \delta \alpha_B &= - \tau_r \dot \alpha_B - \frac{\tau_r p^{\langle\mu\rangle}\nabla_\mu \alpha_B}{\up} \,,
\\\!\!\!\!
  \delta u {\,\cdot\,}p &= \frac{\tau_r \theta (- p {\,\cdot\,} \Delta {\,\cdot\,} p)}{3(\up)} - \tau_r \dot u_\mu  p^{\langle\mu\rangle} - \frac{\tau_r \sigma_\munu p^{\langle\mu} p^{\nu\rangle}}{\up}\,.
\end{align}
\es
As before, we can eliminate the time derivatives and spatial gradients by using the conservation laws and Navier--Stokes relations. To obtain more precise expressions for the perturbations, we first linearize the $\delta f_n$ correction in Eq.~\eqref{eqch5:feqmod_start} as
\be
\label{eqch5:feqmod_expand}
  \delta f_n \approx f_{\eq,n} \bar{f}_{\eq,n} \bigg(b_n \delta\alpha_B + \frac{(\up)\delta T}{T^2} - \frac{\delta u {\,\cdot\,}p}{T} \bigg)\,.
\ee
After substituting Eqs.~\eqref{eqch5:perturbations}, \eqref{eq15} and \eqref{eq16} in Eq.~\eqref{eqch5:feqmod_expand} we recover the RTA Chapman--Enskog expansion \eqref{eqch5:Chapman_Enskog}. We now compare Eqs.~\eqref{eqch5:Chapman_Enskog} and \eqref{eqch5:feqmod_expand} to read off the perturbations:\footnote{%
    We here drop the spatial temperature gradients arising from this procedure in Eqs.~(\ref{eqch5:perturbations}a,c) since they cancel in Eq.~\eqref{eqch5:feqmod_expand}.}
\bs
\allowdisplaybreaks
\label{eqch5:perturb}
\beal
  \delta T &= \frac{\Pi \mathcal{F}}{\beta_\Pi} \,,
\\
  \delta \alpha_B &= \frac{\Pi \mathcal{G}}{\beta_\Pi} - \frac{V_B^\mu p_{\langle\mu\rangle}}{\beta_V(\up)} \,,
\\\!\!\!\!\!\!\!\!
  \delta u {\,\cdot\,}p & ={-}\frac{\Pi({-}p{\cdot} \Delta{\cdot}p)}{3\beta_\Pi(\up)} - \frac{V_B^\mu p_{\langle\mu\rangle} n_B T}{\beta_V(\ene{+}\Peq)} - \frac{\pi_\munu p^{\langle\mu} p^{\nu\rangle}}{2\beta_\pi(\up)}.
\end{align}
\es
These corrections are now used to modify the exponent of Eq. (\ref{eqch5:feqmod_start}) to obtain\footnote{%
    Equation~\eqref{eqch5:feqmod_1} is closely related to the ``maximum entropy" (ME) distribution recently proposed in Ref.~\cite{Everett:2021ulz}, which modifies the exponent of the local-equilibrium distribution with additional Lagrange multipliers. At first order in the dissipative flows, it was shown \cite{Everett:2021ulz} that the non-equilibrium corrections due to these Lagrange multipliers reduce to the perturbative terms in Eq.~\eqref{eqch5:feqmod_1}.}
\be
\label{eqch5:feqmod_1}
  f_{\eq,n}^{\text{(mod)}} = \frac{g_n}{\exp\biggl[\dfrac{(u {+} \delta u) {\,\cdot\,} p}{T {+} \delta T} - b_n(\alpha_B {+} \delta\alpha_B)\biggr] + \Theta_n}\,.
\ee
As long as the viscous corrections are not too large (e.g. $|\Pi| < 3\beta_\Pi$), this expression is positive definite but does not reproduce the components of $J_B^\mu$ and $T^\munu$ exactly since it contains higher-order terms beyond the first-order RTA Chapman--Enskog expansion. These discrepancies are at least second order in gradients and should therefore be negligible if the viscous corrections are small:
\bs
\allowdisplaybreaks
\label{eqch5:small_viscous_limit}
\beal
  |\Pi| & \ll \beta_\Pi \,,
\\
  \sqrt{V_{B,\mu} V_B^\mu} & \ll \beta_V \,,
\\
  \sqrt{\pi_\munu \pi^\munu} & \ll 2 \beta_\pi.
\end{align}
\es
However, Eq.~\eqref{eqch5:feqmod_1} turns out to quickly lose its usefulness for even moderate viscous corrections which are found to result in serious violations of the matching to $J_B^\mu$ and $T^\munu$. As it stands, Eq.~\eqref{eqch5:feqmod_1} is not yet sufficient and needs additional improvements in order to mitigate the errors arising from the higher-order terms.\footnote{%
    One option is to replace the perturbations in Eq.~\eqref{eqch5:feqmod_1} by the Lagrange multipliers in Ref.~\cite{Everett:2021ulz} and adjust them so that the maximum entropy distribution exactly reproduces the hydrodynamic moments. However, an exact (numerical) calculation of these Lagrange multipliers for moderately large viscous corrections is still outstanding.}

The main source of errors in the modified equilibrium distribution \eqref{eqch5:feqmod_1} comes from the momentum dependent terms in the perturbations  (\ref{eqch5:perturb}b,c). In the limit of small gradients, the deformations and shifts to the momentum space are linearly proportional to the dissipative flows. For larger viscous corrections, however, these effects propagate into the hydrodynamic quantities non-linearly. In order to control these errors, one should recast the momentum-dependent perturbations (\ref{eqch5:perturb}b,c) in such a way that the viscous corrections to the dimensionless momentum scales $\langle |\bar{p}_i| \rangle = \langle |p_i| \rangle / (T {+} \delta T)$ are strictly linear in the dissipative flows.

The local momentum transformation \eqref{eqch5:rescale}, first introduced in Ref.~\cite{Pratt:2010jt}, provides an effective way of dealing with the momentum-dependent perturbations in Eq.~\eqref{eqch5:perturb}. Following their prescription, we rewrite Eq.~\eqref{eqch5:feqmod_1} as
\be
\label{eqch5:feqmod_2}
  f_{\eq,n}^{\text{PTM}} = \frac{\mathcal{Z}_n g_n}{\exp\biggl[\dfrac{\sqrt{\bm{p}'^2 {+} m_n^2}}{T {+} \beta_\Pi^{-1}\Pi \mathcal{F}} - b_n \Big(\alpha_B {+} \dfrac{\Pi \mathcal{G}}{\beta_\Pi}\Big)\biggr] + \Theta_n} \,.
\ee
Starting from a \textit{thermal} distribution of momenta $p'$ we construct a map $p = M(p^{\,\prime})$ to the LRF momenta $p$ such that the deformations and shifts are linearly proportional to the dissipative flows. Under the constraint of reducing to the momentum-dependent perturbations in the limit of small gradients, one finds the following transformation between $p$ and $p'$:
\be
\label{eqch5:general_rescaling}
  p_i = A_{ij} p'_j - q_i \sqrt{\bm{p}'^2 {+} m_n^2} + b_n T a_i \,,
\ee
where
\bs
\allowdisplaybreaks
\label{eqch5:rescale_matrix}
\beal
  A_{ij} &= \biggl(1{+}\frac{\Pi}{3\beta_\Pi}\biggr)\,\delta _{ij} + \frac{\pi_{ij}}{2\beta_\pi} \,,
\\
  q_{i} &= \frac{ V_{B,i} \, n_B T}{\beta_V(\ene{+}\Peq)} \,,
\\
a_{i} &= \frac{V_{B,i}}{\beta_V}\,,
\end{align}
\es
with $V_{B,i} =- X_i {\,\cdot\,} V_B$ being the LRF spatial components of the baryon diffusion current. The deformation matrix $A_{ij}$ is identical to the one in Eqs.~(\ref{eqch5:rescale}) --~(\ref{eqch5:Aij})~\cite{Pratt:2010jt}, i.e. $\lambda_\Pi=1/(3\beta_\Pi)$ and $\lambda_\pi=1/(2\beta_\pi)$. Compared to Eqs.~(\ref{eqch5:rescale}) --~(\ref{eqch5:Aij})~\cite{Pratt:2010jt}, Eqs.~(\ref{eqch5:general_rescaling}) --~(\ref{eqch5:rescale_matrix}) are further generalized to include baryon diffusion effects. 

For the normalization factor $\mathcal{Z}_n$ we fix the particle density of the modified equilibrium distribution to that given by the RTA Chapman--Enskog expansion:
\be
\label{eqch5:n_linear}
  n^{(1)}_n  = n_{\eq,n} + \frac{\Pi}{\beta_\Pi}\bigg(n_{\eq,n}+ \N_{10,n} \mathcal G + \frac{\J_{20,n} \mathcal F}{T^2} \bigg)\,,
\ee
which contains a bulk viscous correction to the equilibrium particle density $n_{\eq,n}(T,\alpha_B)$~\cite{Pratt:2010jt, Bernhard:2018hnz}. This suppresses the nonlinear shear and bulk viscous corrections to the modified particle density
\be
\label{eqch5:mod_density}
  n^{\text{PTM}}_{n} = \mathcal{Z}_n \, {\det}A \,\times n_{\eq,n}(T{+}\beta_\Pi^{-1}\Pi \, \mathcal{F},\alpha_B{+}\beta_\Pi^{-1}\Pi \, \mathcal{G}),
\ee
which (if $\mathcal{Z}_n = 1$) are another major source of errors in matching to the hydrodynamic output. Our choice for the normalization factor is therefore
\be
\label{eqch5:renorm}
  \mathcal{Z}_n = \frac{1}{{\det}A} \times \frac{n_n^{(1)}}{n_{\eq,n}(T{+}\beta_\Pi^{-1}\Pi \, \mathcal{F},\alpha_B{+}\beta_\Pi^{-1}\Pi \, \mathcal{G})}\,,
\ee
which is similar to Eq.~\eqref{eqch5:Z_PT} except it contains an additional factor. One should keep in mind that the linearized density $n^{(1)}_n$ of light hadrons, especially pions, may turn negative if the bulk viscous pressure is too negative.

The ``PTM distribution" (\ref{eqch5:feqmod_2}) cannot be applied in every situation. The requirements for using $f_{\eq,n}^\text{PTM}$ in the Cooper--Frye formula are that the determinant of the Jacobian
\be
\label{eqch5:Jacobian}
  \det\bigg(\frac{\partial p_i}{\partial p'_j}\bigg) 
  = {\det}A\,\bigg(1{\,-\,}\frac{q_i A^{-1}_{ij}p'_j}{\sqrt{\bm{p}'^2 {+} m_n^2}}\bigg)
\ee
is positive for any value of $p'$, the deformation matrix $A_{ij}$ is invertible, and the normalization factor $\mathcal{Z}_n$ is nonnegative. These conditions can be violated when the dissipative flows are too large. What to do in this case will be discussed in Sec.~\ref{sec5} where we study certain situations where the modified equilibrium distribution breaks down.
\subsection{Reproducing the hydrodynamic quantities}
\label{sec3c}

As already noted, the nonlinear dependence of the modified equilibrium distribution (\ref{eqch5:feqmod_2}) on the dissipative flows makes it impossible to achieve an exact matching of the hydrodynamic moments of this distribution with all components of $T^\munu$ and $J_B^\mu$. This is different from the linearized parametrizations described in Sec.~\ref{sec2}. In this subsection, we study the matching violations for hadrons emitted from a stationary time-like freeze-out cell with the modified equilibrium distribution \eqref{eqch5:feqmod_2}.

\begin{figure*}[t]
\centering
\includegraphics[width=\textwidth]{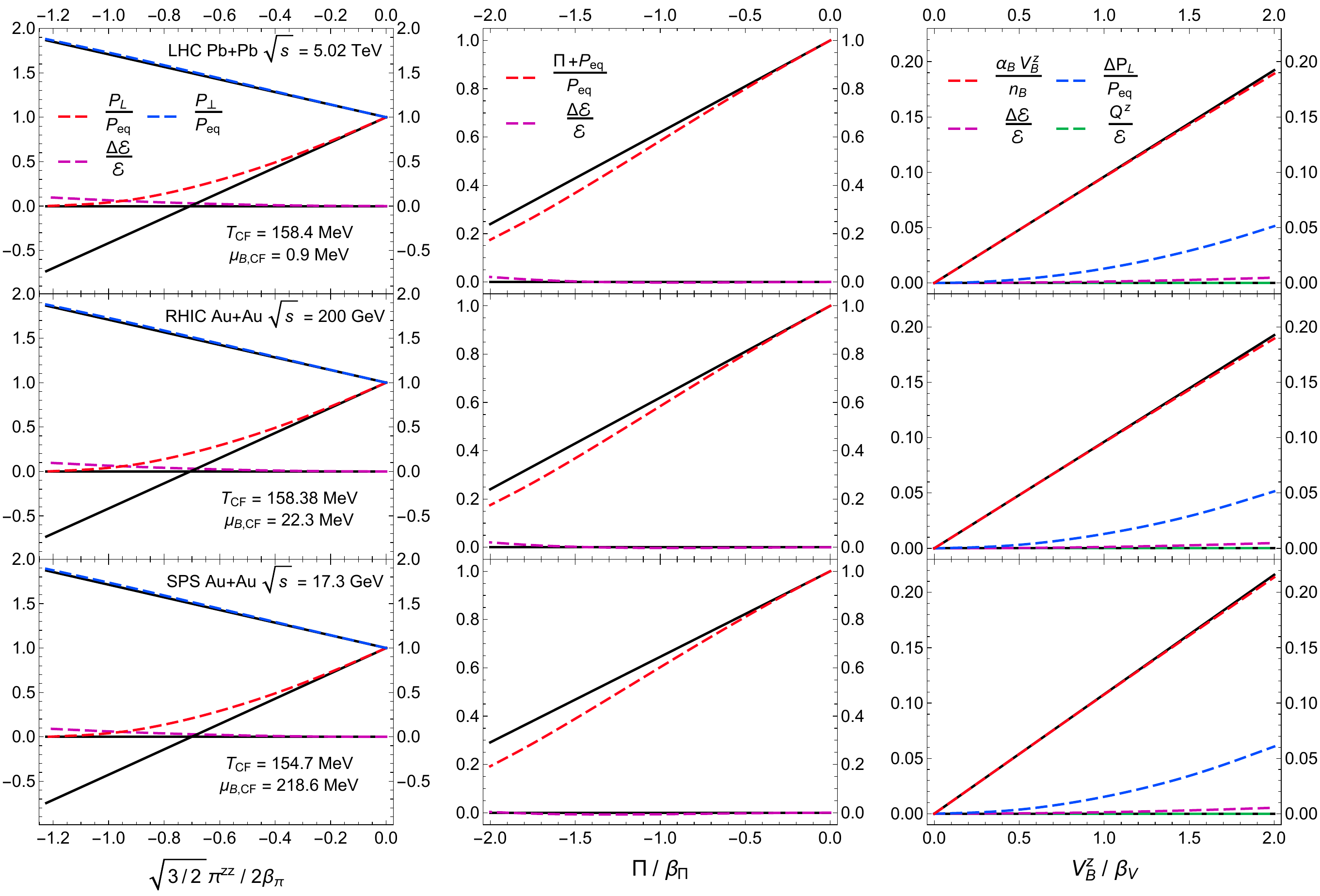}
\caption{
    The reproduction of selected components of the net baryon current $J_B^\mu$ and energy-momentum tensor $T^\munu$ using the Cooper--Frye prescription with the PTM distribution (\ref{eqch5:feqmod_2}), for a stationary hadron resonance gas subject to either shear stress (left column), bulk viscous pressure (middle column) or a baryon diffusion current (right column). The temperature and baryon chemical potential $(T_\text{CF},\mu_{B,\text{CF}})$ at chemical freeze-out are indicated in the legend (top to bottom rows). Solid black curves represent the hydrodynamic input (i.e. the target $J_B^\mu$ and $T^\munu$) while the dashed color lines are the output components of $J_B^\mu$ and $T^\munu$ obtained by computing the corresponding moments of $f_{\eq}^{\text{PTM}}$. 
\label{hydro_output}
}
\end{figure*}

The system is composed of the set of hadron species included in the hadronic afterburner code URQMD~\cite{Bleicher:1999xi} ($N_R \sim 320$). These hadrons are assumed to be produced from a fluid in chemical equilibrium at the chemical freeze-out (CF) temperature $T_\text{CF}$ and baryon chemical potential $\mu_{B,\text{CF}}$, which can vary across different collision systems. The values used here (shown in the legend of Figure~\ref{hydro_output}) are taken from a statistical model fit to the hadron abundance ratios measured in Pb+Pb collisions at LHC energies ($\sqrt{s_\mathrm{NN}}= 5.02$ TeV) and in Au+Au collisions at RHIC and SPS energies ($\sqrt{s_\text{NN}} = 200$ GeV and 17.3 GeV, respectively)~\cite{Andronic:2017pug}. For clarity, we vary either the input shear stress, bulk viscous pressure or baryon diffusion current while fixing the other dissipative flows to zero. 

%

In the first example (left column of Fig.~\ref{hydro_output}) we consider the case where in a given freeze-out cell the hydrodynamic $T^\munu$ features a negative pressure anisotropy $\frac{2}{3}(\PL{-}\Pperp) = \pi^{zz} = -2\pi^{xx} = -2\pi^{yy}$, combined with zero bulk viscous pressure and baryon diffusion current. Starting from equilibrium at the right edge of the plots, we decrease $\pi^{zz}$ going left, causing the longitudinal pressure $\PL = \Peq + \pi^{zz}$ to decrease and the transverse pressure $\Pperp = \Peq + \frac{1}{2}(\pi^{xx} {+} \pi^{yy})$ to increase. The energy density should not change due to the Landau matching condition (\ref{eqch5:matching}a). These are the hydrodynamic components of $T^\munu$ that we aim to reproduce, shown by the solid black curves. The colored dashed curves are the kinetic outputs, calculated as moments of the modified equilibrium distribution with these hydrodynamic inputs.\footnote{%
    Due to the normalization factor \eqref{eqch5:renorm}, the modified equilibrium distribution conserves the net baryon number exactly (i.e. $\delta n_B = 0$).}$^,$\footnote{%
    For only shear stress inputs, the outputs $Q^\mu$ and $V_B^\mu$ vanish by symmetry and are therefore not of interest here.} 
One sees that for small $\pi^{zz}$ the kinetic output closely follows the initial hydrodynamic input. This is because in the limit of small dissipative flows the modified equilibrium distribution reduces to the linear RTA Chapman--Enskog expansion for which the kinetic output reproduces the hydrodynamic input exactly. As $\pi^{zz}$ further decreases to larger negative values, the kinetic output $T^\munu$ begins to deviate from the hydrodynamic target. In particular, for a positive definite distribution function such as Eq.~(\ref{eqch5:feqmod_2}) the kinetic output $\PL$ stays always above zero even if the hydrodynamic input for $\pi^{zz}$ is so large and negative that the total longitudinal pressure in the fluid is negative. On the other hand, the kinetic outputs for $\ene$ and $\Pperp$ agree well with their hydrodynamic targets even for large pressure anisotropies. These trends are confirmed for all three combinations $(T_\text{CF},\mu_{B,\text{CF}})$ studied in the three rows of Fig.~\ref{hydro_output}. Technically, the modified equilibrium distribution breaks down completely for $\pi^{zz} \leq -2\beta_\pi$ (which causes ${\det}A \leq 0$), and one should not expect $f_{\eq,n}^{\text{PTM}}$ to work well for very large pressure anisotropies. However, for moderately large $|\pi^{zz}| \leq \frac{1}{2}\Peq$, $f_{\eq,n}^{\text{PTM}}$ reproduces the target $T^\munu$ components quite well, with $|\Delta\PL| / \PL  \leq 9.0\%$, $|\Delta\Pperp| / \Pperp \leq 0.5\%$ and $|\Delta\ene| / \ene \leq 0.8\%$.\footnote{%
    We also successfully replicated the shear stress test shown in Fig.~2 of Ref.~\cite{Pratt:2010jt}.}

In the middle column of Fig.~\ref{hydro_output} we vary the bulk viscous pressure $\Pi$ at vanishing shear stress and baryon diffusion. We checked that for this hydrodynamic input the kinetic outputs for $\pi^\munu$, $Q^\mu$ and $V_B^\mu$ vanish. The plots show that, for all three choices of freeze-out parameters, the energy matching condition $\Delta\ene = 0$ holds very well even for large (negative) values of $\Pi$. The kinetic output for the total isotropic pressure $\Peq+\Pi$ tends to somewhat underpredict the hydrodynamic target, with the error staying below 10\% for moderately large bulk viscous pressures $|\Pi| \leq \frac{1}{2}\Peq$. For more negative input values of $\Pi$, $\Pi \lesssim -2\beta_\Pi$, the particle density for pions turns negative and the modified equilibrium distribution becomes invalid.

In the right column of Fig.~\ref{hydro_output} we test the reproduction of the baryon diffusion current (taken arbitrarily to point in the $z$--direction) at zero shear and bulk viscous stress. One sees that the hydrodynamic input for the ratio $\alpha_B V_B^z /n_B$ (which is approximately the same for all three freeze-out parameter pairs) is well reproduced by the kinetic output from the  modified equilibrium distribution. The energy matching and Landau frame ($Q^z=0$) conditions are also reproduced with excellent precision. However, a nonzero input for $V_B^z \ne 0$ also generates nonzero kinetic outputs for the bulk and shear viscous stresses even when their hydrodynamic inputs are zero: For $V_B^z/\beta_V=2$ we find a 5\% positive difference $\Delta \PL/\Peq$ between the hydrodynamic input and the kinetic output for the longitudinal pressure of which, after decomposition into shear and bulk viscous contributions, 1.5\% can be attributed to an induced bulk viscous pressure $\Pi$ and 3.5\% to an induced shear stress $\pi^{zz}$, both with positive signs. For moderately large baryon diffusion currents $|V_B^z| \leq \beta_V$, the hydrodynamic output errors are  $|\Delta V_B^z / V_B^z| \leq 0.5\%$, $|\Delta\PL| / \Peq \leq 1.3\%$, $|\Delta\ene| / \ene \leq 0.2\%$ and $|Q^z| / \ene \approx 0$.


\subsection{Pratt--Torrieri--Bernhard distribution}
\label{sec3d}

Another variant of Pratt and Torrieri's idea~\cite{Pratt:2010jt} was implemented by Bernhard in Ref.~\cite{Bernhard:2018hnz}:
\be
\label{eqch5:Jonah}
  f_{\eq,n}^{\text{PTB}} = \frac{\mathcal{Z} g_n}{\exp\Bigl[\dfrac{\sqrt{\bm{p}'^2 {+} m_n^2}}{T}\Bigr] + \Theta_n}\,.
\ee
In this ``PTB distribution" the baryon chemical potential and diffusion current are neglected ($\alpha_B = V_B^\mu = 0$), and the effective temperature is not modified ($\delta T = 0$). The momentum transformation rule is $p_i = A_{ij} p^\prime_j$ with
\be
\label{eqch5:lambda}
  A_{ij} = (1 {+} \lambda_\Pi)\delta_{ij} + \frac{\pi_{ij}}{2\beta_\pi} \,.
\ee
The shear stress modification is the same as in Eq.~(\ref{eqch5:rescale_matrix}a) while the bulk pressure term is replaced by the isotropic scale parameter $\lambda_\Pi$. The normalization factor $\mathcal{Z}$ is taken as species-independent:
\be
\mathcal{Z} = \frac{z_\Pi}{{\det}A}\,,
\ee
leaving particle abundance ratios unchanged from their chemical equilibrium values. The parameters $\lambda_\Pi$ and $z_\Pi$ are fixed such that in the absence of shear stress ($\pi_{ij} = 0$) the energy density and bulk viscous pressure are exactly matched~\cite{Bernhard:2018hnz}:
\bs
\allowdisplaybreaks
\label{eqch5:bulkparameters}
\beal
  \ene'(\lambda_\Pi, z_\Pi,T) &= \ene\,,  
\\
  \mathcal{P}'(\lambda_\Pi, z_\Pi,T) &= \Peq + \Pi\,;
\end{align}
\es
here $\ene'$ and $\mathcal{P}'$ are the kinetic theory output for the energy density and isotropic pressure computed from the PTB distribution \eqref{eqch5:Jonah}. 

%
\begin{figure}[t]
\centering
\includegraphics[width=0.7\linewidth]{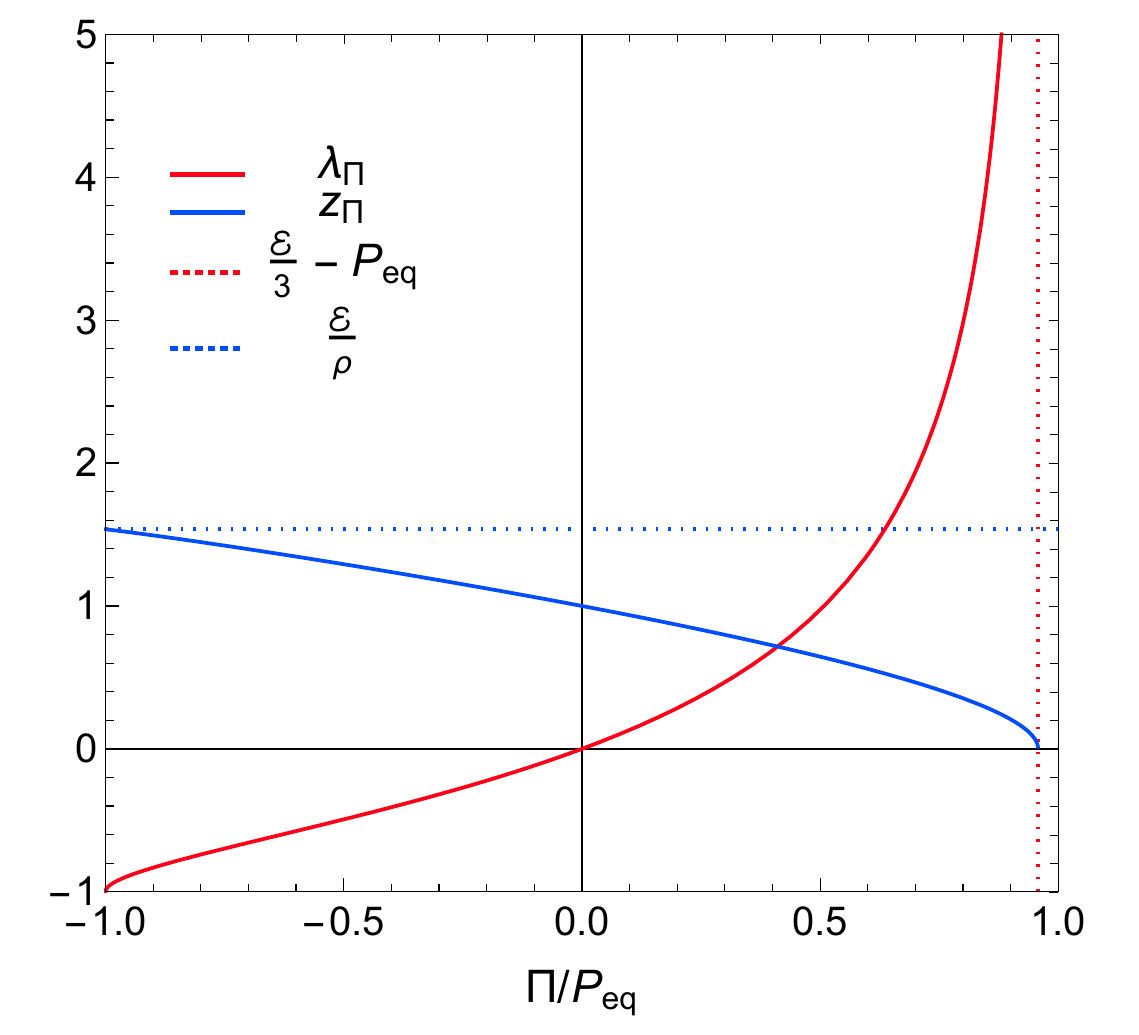}
\caption{
    The isotropic scale parameter $\lambda_\Pi$ (solid red) and normalization factor $z_\Pi$ (solid blue) as a function of $\Pi/\Peq$ are computed for a hadron resonance gas at a fixed temperature $T = 150$ MeV. The dotted blue and red lines are the upper bounds for $z_\Pi$ and $\Pi / \Peq$, respectively. This plot corresponds to Fig.~3.12 in Ref.~\cite{Bernhard:2018hnz}, where $\Delta \langle p \rangle / \langle p \rangle_0 = \lambda_\Pi$ and $\Delta n/n_0 = z_\Pi - 1$. 
\label{FJonah}
}
\end{figure}
%

The parametrization procedure is as follows: for the distribution function \eqref{eqch5:Jonah} to be well-defined the isotropic scale parameter must satisfy $\lambda_\Pi \in (-1,\infty)$. At a given temperature $T$ one sets up a grid in $\lambda_\Pi$ and computes the corresponding values for $z_\Pi$ and $\Pi$ numerically by rewriting Eqs.~(\ref{eqch5:bulkparameters}a,b) as
\bs
\allowdisplaybreaks
\label{eqch5:zbulk}
\beal
  z_\Pi &= \frac{\ene}{\mathcal{L}_{20}(\lambda_\Pi, T)} \,, 
\\
  \Pi &= z_\Pi \mathcal{L}_{21}(\lambda_\Pi, T) - \Peq \,.
\end{align}
\es
The functions $\mathcal{L}_{kq}$ are defined in Appendix~\ref{app5:integrals}. One can then interpolate the data with respect to $\Pi$ to construct the functions $\lambda_\Pi(\Pi; T)$ and $z_\Pi(\Pi; T)$; an example is shown in Figure~\ref{FJonah}. Generally, these are nonlinear functions of $\Pi$. In the limit of small bulk viscous pressure, they linearize to
\bs
\allowdisplaybreaks
\label{dz}
\beal
  z_\Pi &\approx  1 + \delta z_\Pi = 1 - \frac{3\Pi \Peq}{5 \beta_\pi \ene - 3\Peq(\ene {+} \Peq)} \,,
\\
  \lambda_\Pi &\approx \delta \lambda_\Pi = \frac{\Pi \ene}{5 \beta_\pi \ene - 3\Peq(\ene {+} \Peq)}\,.
\end{align}
\es
Equations~(\ref{eqch5:zbulk}a,b) imply that $z_\Pi$ and $\Pi$ are bounded by
\bs
\allowdisplaybreaks
\label{eqch5:bounds}
\beal
  0 < &\,z_\Pi < \frac{\ene}{\rho}\, ,
\\
  -\Peq < &\,\Pi < \frac{\ene}{3} - \Peq \,,
\end{align}
\es
where $\rho = \sum_n m_n n_{\eq,n}$ is the equilibrium mass density. When $\Pi$ lies outside the bound\footnote{%
    Violations of the upper bound in Eq.~(\ref{eqch5:bounds}b) can occur at the transition from a conformal pre-hydrodynamic model to non-conformal viscous hydrodynamics, where the mismatch between the conformal and QCD equations of state gives exactly $\Pi = \ene/3 - \Peq(\ene)$.}
(\ref{eqch5:bounds}b) or ${\det}A \leq 0$, we consider the modified equilibrium distribution \eqref{eqch5:Jonah} to have broken down. As will be discussed in Sec.~\ref{sec5b},  we then resort to linearizing Eq.~\eqref{eqch5:Jonah} around local equilibrium, i.e. we write $f_{\eq,n}^{\text{PTB}}\approx f_{\eq,n}+\delta f_n$ with
\be
\label{eqch5:linear_Jonah}
  \delta f_n = f_{\eq,n} \left(\delta z_\Pi - 3 \delta \lambda_\Pi\right) \,+\,
f_{\eq,n} \bar{f}_{\eq,n}\left(\frac{\delta \lambda_\Pi (-p \cdot \Delta \cdot p)}{(u\cdot p)T} + \frac{\pi_\munu p^{\langle\mu} p^{\nu\rangle}}{2\beta_\pi (u\cdot p)T}\right)\,.
\ee
The effectiveness of the PTB distribution (\ref{eqch5:Jonah}) in reproducing the components of a given energy-momentum tensor $T^\munu$ was studied in Ref.~\cite{Bernhard:2018hnz} to which we refer the reader for comparison with Fig.~\ref{hydro_output}.

\section{Modified anisotropic distribution}
\label{sec4}
We can also apply the methodology developed in Sec.~\ref{sec2} to modify an anisotropic distribution function that has already been deformed at leading order. Here we start with the same type of Romatschke--Strickland distribution used in Chapter~\ref{ch3label}:\footnote{%
    For simplicity, we set the effective baryon chemical potential to $\tilde{\mu}_B = 0$.}
\be
\label{eqch5:fa}
  f_{a,n}(x,p) = \frac{g_n}{\exp\biggl[\dfrac{\sqrt{p \cdot \Omega(x) \cdot p + m_n^2}}{\Lambda(x)}\biggr] + \Theta_n}\,,
\ee
where here the ellipsoidal tensor is purely spatial:
\be
\Omega_\munu = \frac{Z_\mu Z_\nu}{\alpha_L^2} - \frac{\Xi_\munu}{\alpha_\perp^2}\,,
\ee
with $Z^\mu{\,\equiv\,}X_3^\mu$ being the longitudinal basis vector (we used $z^\mu$ in the other chapters). We rewrite the anisotropic distribution \eqref{eqch5:fa} to match the notation used in the PTM distribution \eqref{eqch5:feqmod_2}:
\be
\label{eqch5:fa2}
  f_{a,n} = \frac{g_n}{\exp\biggl[\dfrac{\sqrt{\bm{p}'^2 {+} m_n^2}}{\Lambda}\biggr] + \Theta_n} \,,
\ee
where $p_i = A_{ij} p'_j$ with
\be
\label{eqch5:Aij_mod}
   A_{ij} = \alpha_\perp \delta_{ij} + (\alpha_L {-} \alpha_\perp) Z_i Z_j \,,
\ee
and $Z_i = - X_i {\,\cdot\,} Z = (0,0,1)$ being the LRF components of the longitudinal basis vector. 
\subsection{Anisotropic RTA Chapman--Enskog expansion}
\label{sec4a}
To modify the anisotropic distribution~\eqref{eqch5:fa}, we first need a linearized expression for the residual correction
\be
\label{eqch5:vah}
f_n(x,p) = f_{a,n}(x,p) + \delta\tilde{f}_n(x,p)\,.
\ee
Instead of the 14--moment approximation used in Chapter~\ref{ch3label}, we derive $\delta \tilde{f}$ from the RTA Chapman--Enskog expansion. After substituting Eq.~\eqref{eqch5:vah} in the RTA Boltzmann equation~\eqref{eqch5:RTA}, the first-order expression for $\delta \tilde{f}_n$ is
\be
\label{eqch5:ACE1}
  \delta\tilde{f}_n = - s^\mu \partial_\mu f_{a,n} + f_{\eq,n} - f_{a,n}\,,
\ee
where the second-order term $-s^\mu \partial_\mu \delta \tilde{f}_n$ was neglected. The first term in Eq.~\eqref{eqch5:ACE1} can be expanded as
\be
  - s^\mu \partial_\mu f_{a,n} = f_{a,n} \bar{f}_{a,n} \left(\dfrac{E_{a,n} \delta\Lambda}{\Lambda^2} - \frac{p \cdot\delta\Omega\cdot p}{2 E_{a,n} \Lambda}\right)\,,
\ee
where $\bar{f}_{a,n} = 1 - g_n^{-1} \Theta_n f_{a,n}$, $E_{a,n} = \sqrt{p {\,\cdot\,} \Omega {\,\cdot\,} p + m_n^2}$,
\be
\begin{split}
  \frac{p \cdot \delta \Omega \cdot p}{2} =\,\,& \frac{(\alpha_L^2 {-} \alpha_\perp^2)(-Z\cdot p)(\delta Z \cdot p)}{\alpha_\perp^2 \alpha_L^2} + \frac{(u {\,\cdot\,}p)(\delta u {\,\cdot\,}p)}{\alpha_\perp^2} - \frac{\delta \alpha_L (-Z \cdot p)^2}{\alpha_L^3} \\
&\, - \frac{\delta \alpha_\perp (-p \cdot \Xi \cdot p)}{\alpha_\perp^3} \,,
\end{split}
\ee
and the perturbations $\delta u^\mu = s^\nu\partial_\nu u^\mu$ and $\delta Z^\mu = s^\nu\partial_\nu Z^\mu$ are\footnote{%
    We find that the $\delta\tilde{f}_n$ terms proportional to $\delta \Lambda = s^\mu\partial_\mu \Lambda$, $\delta \alpha_L = s^\mu\partial_\mu \alpha_L$ and $\delta \alpha_\perp = s^\mu\partial_\mu \alpha_\perp$ do not contribute to $\Wperp$ and $\piperp$, so we can set these perturbations to zero.}
\bs
\allowdisplaybreaks
\label{eqch5:dUdZ_full}
\beal
  \delta u {\,\cdot\,}p =\,\,& \frac{-\tau_r}{\up}\Big[(u\cdot p)\big((-Z\cdot p)Z^\mu {+} p^{\{\mu\}}\big)\dot{u}_\mu - (-Z\cdot p)^2 \theta_L  - \frac{1}{2}(-p\cdot\Xi\cdot p)\theta_\perp
\\ \nonumber
  & + p^{\{\mu} p^{\nu\}} \sigma_{\perp,\munu} + (-Z\cdot p)p_{\{\mu\}}\big(Z_\nu \nabla_\perp^\mu u^\nu {-} D_z u^\mu\big) \Big] \,,
\\ 
  \delta Z \cdot p =\,\,& \frac{-\tau_r}{\up} \Big[{-}(u\cdot p)^2 Z_\mu \dot{u}^\mu + (u {\,\cdot\,}p)(-Z\cdot p)\theta_L + (u {\,\cdot\,}p)(-Z\cdot p)\theta_L + p^{\{\mu} p^{\nu\}} \sigma_{z,\munu}
\\ \nonumber
  &-(-Z\cdot p)p_{\{\mu\}}D_z Z^\mu - \frac{1}{2}(-p\cdot\Xi\cdot p)\nabla_{\perp\mu}Z^\mu + (u\cdot p)p_{\{\mu\}}\big(\dot{Z}^\mu {-} Z_\nu \nabla_\perp^\mu u^\nu\big) 
   \Big]\,,
\end{align}
\es
where we used the decomposition~\eqref{eqch3:27} for the partial derivative.

For the sake of simplicity, we only consider the terms with nonzero contributions to $\Wperp$ and $\piperp$, whose kinetic definitions are
\bs
\allowdisplaybreaks
\label{eqch5:Wpi_defs}
\beal
  \Wperp &= \sum_n \int_p (-Z\cdot p)p^{\{\mu\}} \delta\tilde{f}_n \,,
\\
  \piperp &= \sum_n \int_p p^{\{\mu} p^{\nu\}} \delta\tilde{f}_n\,.
\end{align}
\es
By symmetry, the term $f_{\eq,n} - f_{a,n}$ in Eq.~\eqref{eqch5:ACE1} and corrections $\propto (\delta\Lambda, \delta\alpha_L, \delta\alpha_\perp)$ have zero contributions, so we can effectively eliminate them. The only nonzero contributions from $\delta u {\,\cdot\,}p$ and $\delta Z {\,\cdot\,} p$ in Eqs.~(\ref{eqch5:dUdZ_full}a,b) are 
\bs
\allowdisplaybreaks
\label{eqch5:dUdZ}
\beal
  \delta u {\,\cdot\,}p \rightarrow&  - \frac{\tau_r(-Z \cdot p)p_{\{\mu\}}\left(Z_\nu \nabla_\perp^\mu u^\nu {-} D_z u^\mu \right)}{\up} - \frac{\tau_r p_{\{\mu} p_{\nu\}}\sigma_\perp^\munu}{\up} \,,
\\
  \delta Z \cdot p \rightarrow& - \tau_r p_{\{\mu\}}\big(\dot{Z}^\mu {-} Z_\nu \nabla_\perp^\mu u^\nu\big)\,.
\end{align}
\es
The $\delta \tilde{f}_n$ correction then reduces to
\be
\label{eqch5:ACEgrad}
\begin{split}
  \delta\tilde{f}_n &= \frac{\tau_r f_{a,n}\bar{f}_{a,n}}{E_{a,n}\Lambda}\left[\frac{p_{\{\mu} p_{\nu\}}\sigma_\perp^\munu}{\alpha_\perp^2} - \frac{({-}Z{\,\cdot\,} P)p_{\{\mu\}} \kappa_\perp^\mu}{\alpha_\perp \alpha_L}\right]\,,
\end{split}
\ee
where
\be
  \kappa_\perp^\mu = \frac{(\alpha_\perp^2 {-} \alpha_L^2)\Xi^\mu_\nu \dot{Z}^\nu - \alpha_\perp^2 Z_\nu \nabla_\perp^\mu u^\nu + \alpha_L^2 \Xi^\mu_\nu D_z u^\nu}{\alpha_\perp \alpha_L}\,.
\ee
The gradients $\sigma_\perp^\munu$ and $\kappa_\perp^\mu$ are proportional to the ``Navier--Stokes" values for $\piperp$ and $\Wperp$, respectively. After inserting Eq.~\eqref{eqch5:ACEgrad} in Eq.~\eqref{eqch5:Wpi_defs} one obtains
\be
  \piperp =  2 \tau_r \beta_\pi^\perp \sigma_\perp^\munu \,,\qquad
  \Wperp =  \tau_r \beta_W^\perp \kappa_\perp^\mu\,,
\ee
where
\be
  \beta_\pi^\perp = \frac{\mathcal{J}^{(R)}_{402-1}}{\alpha_\perp^2\Lambda} \,,
  \qquad
  \beta_W^\perp = \frac{\mathcal{J}^{(R)}_{421-1}}{\alpha_L \alpha_\perp \Lambda},
\ee
and the integrals $\mathcal{J}^{(R)}_{krqs}$ are defined in Appendix~\ref{app5:integrals} (compared to the integrals $\mathcal{J}_{nrqs}$ in Chapters \ref{ch3label} and \ref{chapter4label}, they sum over the hadron resonances). Thus, the first-order anisotropic RTA Chapman--Enskog expansion is
\be
\label{eqch5:ACE2}
  \!\!\!\!
  \delta\tilde{f}_n = f_{a,n} \bar{f}_{a,n}\left(\frac{p_{\{\mu} p_{\nu\}} \piperp}{2 \beta_\pi^\perp E_{a,n} \alpha_\perp^2 \Lambda} - \frac{(-Z \cdot p)p_{\{\mu\}} \Wperp}{\beta_W^\perp E_{a,n} \alpha_\perp \alpha_L \Lambda}\right)\!.\!\!
\ee
One can show that Eq.~\eqref{eqch5:ACE2} satisfies the anisotropic matching conditions for $\ene$, $\PL$ and $\Pperp$ and the Landau frame constraints:
\bs
\allowdisplaybreaks
\label{eqch5:aniso_matching}
\beal
  \delta\tilde\ene &= \sum_n \int_p (u\cdot p)^2 \delta\tilde{f}_n = 0 \,,
\\
  \delta\tilde{\mathcal{P}}_L &= \sum_n \int_p (-Z\cdot p)^2 \delta\tilde{f}_n = 0 \,,
\\
  \delta\tilde{\mathcal{P}}_\perp &= \frac{1}{2}\sum_n \int_p (-p \cdot \Xi \cdot p) \delta\tilde{f}_n = 0 \,,
\\
  Q_L &= \sum_n \int_p (u\cdot p)(-Z\cdot p) \delta\tilde{f}_n = 0 \,,
\\
  Q^\mu_\perp &= \sum_n \int_p (u\cdot p) p^{\{\mu\}} \delta\tilde{f}_n = 0 \,,
\end{align}
\es
where $Q_L = {\,-\,} Z {\,\cdot\,} Q$ is the LRF longitudinal heat flow and $Q_\perp^\mu = \Xi^\mu_\nu Q^\nu$ is the transverse heat current. 

\subsection{PTMA distribution}
\label{sec4b}
With the anisotropic RTA Chapman--Enskog expansion \eqref{eqch5:ACE2} at hand we can proceed to modify the leading order anisotropic distribution~\eqref{eqch5:fa2}. First, we insert a perturbation $\delta\Omega$ in $f_{a,n}$, keeping $\delta\Lambda = 0$:
\be
\label{eqch5:famod_approx}
  f_{a,n}^{\text{(mod)}} = \frac{g_n}{\exp\biggl[\dfrac{\sqrt{p {\,\cdot\,}(\Omega {+} \delta\Omega) {\,\cdot\,} p {\,+\,} m_n^2}}{\Lambda}\biggr] + \Theta_n}\,.
\ee
Linearizing this modified anisotropic distribution and comparing it to Eq.~\eqref{eqch5:ACE2} one finds
\be
\label{eqch5:famod_pert}
  p \cdot \delta \Omega \cdot p = -\frac{p_{\{\mu} p_{\nu\}} \piperp}{\beta_\pi^\perp \alpha_\perp^2} + \frac{2(-Z \cdot p)p_{\{\mu\}} \Wperp}{\beta_W^\perp \alpha_\perp \alpha_L}\,.
\ee
Next we rewrite Eq.~\eqref{eqch5:famod_approx} as\footnote{%
    The generalization of Eq.~\eqref{eqch5:famod1} for nonzero baryon chemical potential and diffusion is non-trivial and will be left to future work.}
\be
\label{eqch5:famod1}
  f^{\text{PTMA}}_{a,n} = \frac{\mathcal{Z} g_n}{\exp\biggl[\dfrac{\sqrt{\bm{p}''^2 {+} m_n^2}}{\Lambda}\biggr] + \Theta_n}\,,
\ee
where the normalization $\mathcal{Z}$ will be discussed further below and
\be
  p_i = B_{ij} p^{\prime\prime}_j = C_{im} A_{mj} p^{\prime\prime}_j \,.
\ee
By comparing with Eq.~\eqref{eqch5:Aij_mod} one sees that the residual shear transformation $C_{im}$ further deforms the anisotropic momentum space. The additional deformations, which are linearly proportional to $\piperp$ and $\Wperp$, are assumed to be smaller than the anisotropy parameters $\alpha_L$ and $\alpha_\perp$. The matrix $C_{im}$ is constructed such that the total deformation matrix $B_{ij} = C_{im} A_{mj}$ is symmetric and reproduces the perturbation $\delta\Omega$ in Eq.~\eqref{eqch5:famod_approx} for small residual shear stresses. After some algebra one finds
\be
\label{eqch5:Cmatrix}
  C_{im} = \delta_{im} + \frac{\pi_{\perp,im}}{2\beta_\pi^\perp} + \frac{\alpha_\perp W_{\perp z, i} Z_m + \alpha_L W_{\perp z, m} Z_i}{\beta_W^\perp(\alpha_\perp {+} \alpha_L)}\,.
\ee
Here $\pi_{\perp,im} = X_i {\,\cdot\,} \pi_\perp {\cdot\,} X_m$ and $W_{\perp z,i} = - X_i {\,\cdot\,} W_{\perp,z}$ are the LRF residual shear stress components. Although $C_{im}$ in Eq.~\eqref{eqch5:Cmatrix} is not symmetric, $B_{ij}$ is:
\be
\label{eqch5:Bij}
  B_{ij} = A_{ij} + \frac{\alpha_\perp \pi_{\perp,ij}}{2\beta_\pi^\perp} + \frac{\alpha_\perp \alpha_L (W_{\perp z, i} Z_j + W_{\perp z, j} Z_i)}{\beta_W^\perp(\alpha_\perp {+} \alpha_L)}\,.
\ee
Finally we renormalize the particle density to the one given by the leading order anisotropic distribution:
\be
\label{eqch5:aniso_density}
  n_{a,n} = {\det} A \times n_{\eq,n}(\Lambda,0)\,,
\ee
where $n_{a,n}$ is the anisotropic particle density and ${\det}A = \alpha_\perp^2\alpha_L$ (there is no contribution from the anisotropic RTA Chapman--Enskog correction $\delta\tilde{f}_n$). The modified particle density from Eq.~\eqref{eqch5:famod1} is
\be
 n_{a,n}^{\text{PTMA}} = \mathcal{Z} \, {\det} C \times n_{a,n} \,,
\ee
which leads us to the normalization factor 
\be
\label{eqch5:Z_mod}
  \mathcal{Z} = \frac{1}{{\det} C}\,.
\ee
%

\begin{figure*}[t]
\includegraphics[width=\linewidth]{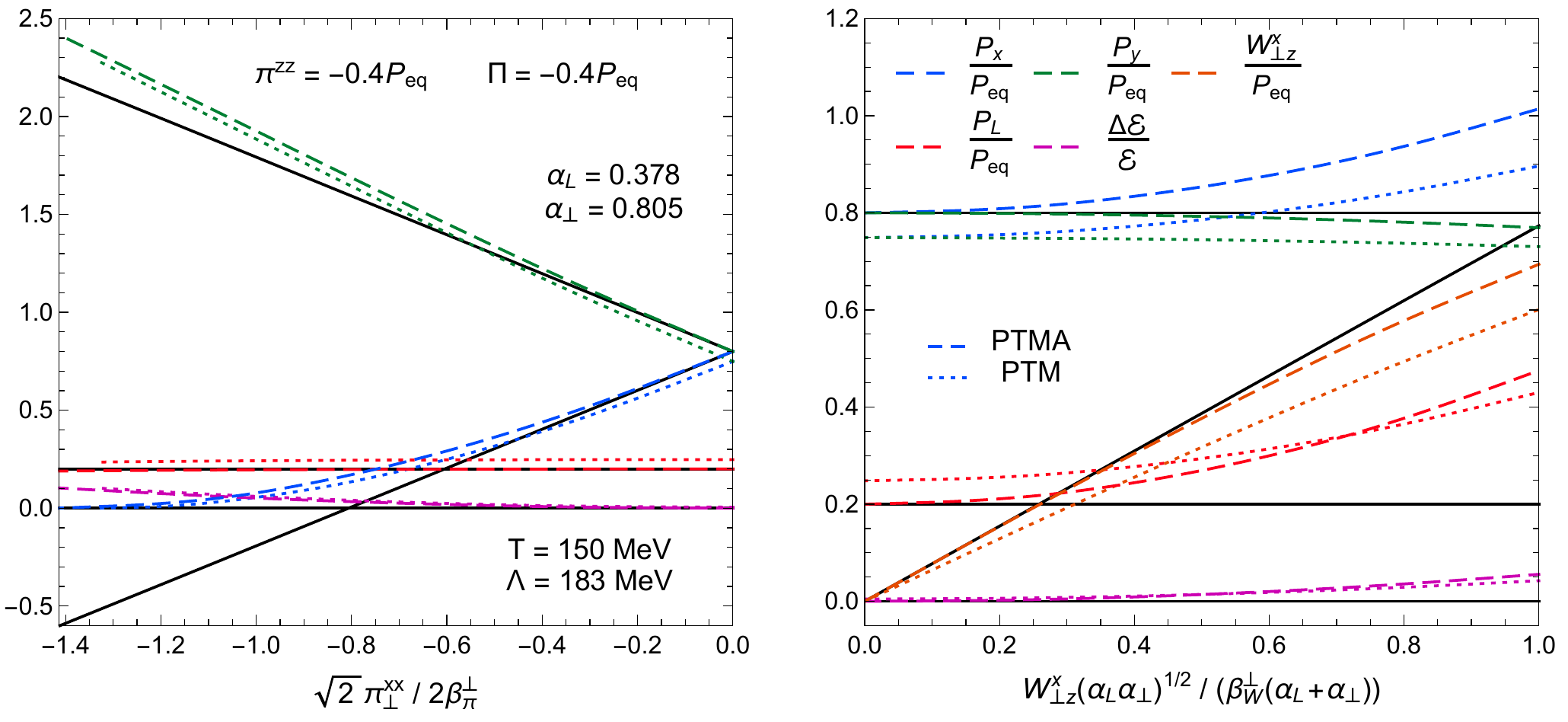}
\caption{
    The reproduction of selected components of the energy-momentum tensor $T^\munu$ using either the PTMA distribution (\ref{eqch5:famod1}) (dashed color) or the PTM distribution (\ref{eqch5:feqmod_2}) (dotted color) for a stationary hadron resonance gas at temperature $T = 150$ MeV with zero net baryon density ($\alpha_B = 0$). The system is assumed to have a fixed pressure anisotropy $\pi^{zz} = -\frac{2}{5}\Peq$ and bulk viscous pressure $\Pi = -\frac{2}{5}\Peq$ and is further subjected to either transverse shear stress (left panel) or longitudinal momentum diffusion (right panel). The solid black lines show the components of the hydrodynamic input $T^\munu$. 
\label{famod_output}
}
\end{figure*}

In Figure~\ref{famod_output}, we test the reproduction of the energy-momentum tensor by the ``PTMA distribution" \eqref{eqch5:famod1}, similar to the test done in Fig.~\ref{hydro_output}. We consider a hadron resonance gas at temperature $T = 150$ MeV and baryon chemical potential $\alpha_B = 0$ with an input $T^\munu$ featuring the two fixed dissipative flows $\pi^{zz} = \Pi = -\frac{2}{5}\Peq$ (or, equivalently, $\PL = \frac{1}{5}\Peq$ and $\Pperp = \frac{4}{5}\Peq$). These viscous pressures are captured by the leading order distribution $f_{a,n}$ by numerically adjusting (``Landau matching'') the anisotropy parameters $\Lambda$, $\alpha_L$ and $\alpha_\perp$ accordingly. 

In the left panel of Fig.~\ref{famod_output} we explore the reliability of the modified anisotropic distribution in reproducing a nonzero transverse shear stress $\pi_\perp^{xx} = -\pi_\perp^{yy}$. Moving from the right edge of the plot leftward we decrease $\pi_\perp^{xx}$, causing the pressures $\mathcal{P}_x = \Pperp + \pi_\perp^{xx}$ and $\mathcal{P}_y = \Pperp + \pi_\perp^{yy}$ to decrease and increase, respectively. According to the anisotropic matching conditions \eqref{eqch5:aniso_matching}, the energy density and longitudinal pressure should remain constant as we do so. The hydrodynamic input (solid black lines) and kinetic output $T^\munu$ components (colored dashed lines) from this test are very similar to those shown in the left panels of Fig.~\ref{hydro_output}, with the kinetic output for $\mathcal{P}_x$ (blue dashed) approaching zero for large negative $\pi_\perp^{xx}$ while the output for  $\mathcal{P}_y$ (green dashed) overestimates the hydrodynamic target value but to a lesser degree. The kinetic outputs for $\ene$ and $\PL$ are in very good agreement with their hydrodynamic target values. The modified anisotropic distribution breaks down for $\pi_\perp^{xx} < -2\beta_\pi^\perp$ (or $\det C < 0$) but for moderately large values $|\pi_\perp^{xx}| \leq \frac{1}{6}(\PL {+} 2\Pperp)$ the discrepancies between the input and output values are $|\Delta \mathcal{P}_x|/\mathcal{P}_x \leq 4.0 \%$, $|\Delta \mathcal{P}_y|/\mathcal{P}_y \leq 1.5 \%$, $|\Delta \PL|/\PL \leq 0.17 \%$ and $|\Delta \ene|/\ene \leq 0.47 \%$. From Chapter~\ref{chapter4label} we know that the transverse shear stress is typically much smaller than the longitudinal and transverse pressures in the mid-rapidity region $\eta_s = 0$. Thus, for (2+1)--d event-by-event simulations, the PTMA distribution should accurately capture all components of the energy-momentum tensor on a hypersurface provided by anisotropic hydrodynamics.

For the second test (right panel of Fig.~\ref{famod_output}) we start from the same system but add a longitudinal momentum diffusion component $W_{\perp z}^x$ of increasing magnitude. The energy density and pressure components $\mathcal{P}_x$, $\mathcal{P}_y$ and $\PL$ should not change as we do so. Overall, we find that the kinetic outputs for $\ene$, $\mathcal{P}_y$ and $W_{\perp z}^x$ are in good agreement with this expectation. However, the kinetic outputs for the pressure components $\mathcal{P}_x$ and $\PL$ are seen to exhibit stronger sensitivity to the errors caused by mismatched nonlinear terms in $W_{\perp z}^x$. Still, for moderate values of $|W_{\perp z}^x| \leq \frac{1}{6}(\PL {+} 2\Pperp)$, the output errors stay below $|\Delta W_{\perp,z}^x / W_{\perp,z}^x| \leq 1.6 \%$, $|\Delta \mathcal{P}_x|/\mathcal{P}_x \leq 4.0 \%$, $|\Delta \mathcal{P}_y|/\mathcal{P}_y \leq 0.55 \%$, $|\Delta \PL|/\PL \leq 21 \%$ and $|\Delta \ene|/\ene \leq 0.84 \%$

Finally, we repeat the previous tests with the PTM modified equilibrium distribution~\eqref{eqch5:feqmod_2} and plot 
the corresponding kinetic output $T^\munu$ as dotted lines in Fig.~\ref{famod_output} for comparison. One observes that the PTM distribution follows essentially the same trends as the PTMA distribution, but it does not fully capture the pressure anisotropy and bulk viscous pressure at zero residual shear stress, similar to what we saw in Fig.~\ref{hydro_output}. On the other hand, by imposing the generalized Landau matching conditions (\ref{eqch5:aniso_matching}b,c) the leading order anisotropic distribution precisely reproduces the longitudinal and transverse pressures in that limit.
\section{Continuous particle spectra}
\label{sec5}
In this section we compute the continuous momentum spectra of identified hadrons $(\pi^+, K^+, p)$ using the Cooper--Frye formula \eqref{eqch5:CooperFrye}. To generate the hypersurfaces, we run the {\sc VAH} code from Chapter~\ref{chapter4label} to evolve central and non-central Pb+Pb collisions using standard viscous hydrodynamics with smooth initial conditions. For simplicity, we only consider the central slice ($\eta_s=0$) in the transverse plane and assume longitudinal boost-invariance to extend the solution in the spacetime rapidity direction. For the different hadron phase-space distribution models discussed in Secs.~\ref{sec2} --~\ref{sec4}, we will compare the azimuthally-averaged transverse momentum spectra 
\be
\label{eqch5:pT_spectra}
  \frac{dN_n}{2\pi p_T dp_T dy_p} = \int_0^{2\pi}\, \frac{d\phi_p}{2\pi}\frac{dN_n}{p_T dp_T d\phi_p dy_p}
\ee
and the $p_T$--differential elliptic flow coefficient
\be
\label{eqch5:v2}
  v_{2,n}(p_T) = \frac{\int_0^{2\pi} d\phi_p \cos(2\phi_p)\,\dfrac{dN_n}{p_T dp_T d\phi_p dy_p}}{\int_0^{2\pi} d\phi_p\, \dfrac{dN_n}{p_T dp_T d\phi_p dy_p}}
\ee
at mid-rapidity ($y_p = 0$). Specifically, we compare results obtained with the 14--moment approximation \eqref{eqch5:14_moment_irreducible}, the RTA Chapman--Enskog expansion \eqref{eqch5:Chapman_Enskog}, the PTM and PTB modified equilibrium distributions \eqref{eqch5:feqmod_2} and \eqref{eqch5:Jonah}, and the PTMA modified anisotrpic distribution \eqref{eqch5:famod1}. 

\subsection{Setup}
\label{sec5a}

We evolve an azimuthally symmetric, event-averaged \trento{} transverse energy density profile with (2+1)--dimensional second-order viscous hydrodynamics~\cite{Moreland:2014oya, McNelis:2021zji}. We start the simulation at the longitudinal proper time $\tau_0 = 0.5$ fm/$c$ with an initial central temperature of $T_{0,\text{center}} = 400$ MeV. The spatial components of the fluid velocity $u^\mu$ and bulk viscous pressure $\Pi$ are initialized to zero. The shear stress tensor is initialized as
\be
\pi^\munu = \frac{1}{3}\big(\PL {-} \Pperp\big) \left(\Delta^\munu + 3Z^\mu Z^\nu\right)\,,
\ee
where the initial pressure anisotropy is set to $\PL - \Pperp = \frac{12}{11}\Peq$ and the initial longitudinal basis vector is $Z^\mu = (0,0,0,\tau_0^{-1})$. The baryon chemical potential $\alpha_B$ and baryon diffusion current $V_B^\mu$ are fixed to zero for the entire simulation.\footnote{%
    The current version of {\sc VAH}~\cite{McNelis:2021zji} does not propagate the net baryon density and baryon diffusion current.}
For the equilibrium pressure $\mathcal{P}_\eq(\ene)$, we use the lattice QCD equation of state from the HotQCD collaboration~\cite{Bazavov:2014pvz}. The shear and bulk viscosities are modeled using the temperature-dependent parameterizations~\eqref{eqchap4:etas} and~\eqref{eqchap4:zetas}. In this chapter we use the same shear viscosity parameters as Chapter~\ref{chapter4label} but change the bulk viscosity parameters to $(\zeta/\mathcal{S})_\text{max} = 0.1$, $w_\zeta = 0.05$ GeV and $\lambda_\zeta = 0$. For exploration purposes, we vary the temperature $T_\zeta$ at which $\zeta / \mathcal{S}$ peaks, setting it to either $220$ MeV or $160$ MeV. 

\begin{figure}[!t]
\centering
\includegraphics[width=\linewidth]{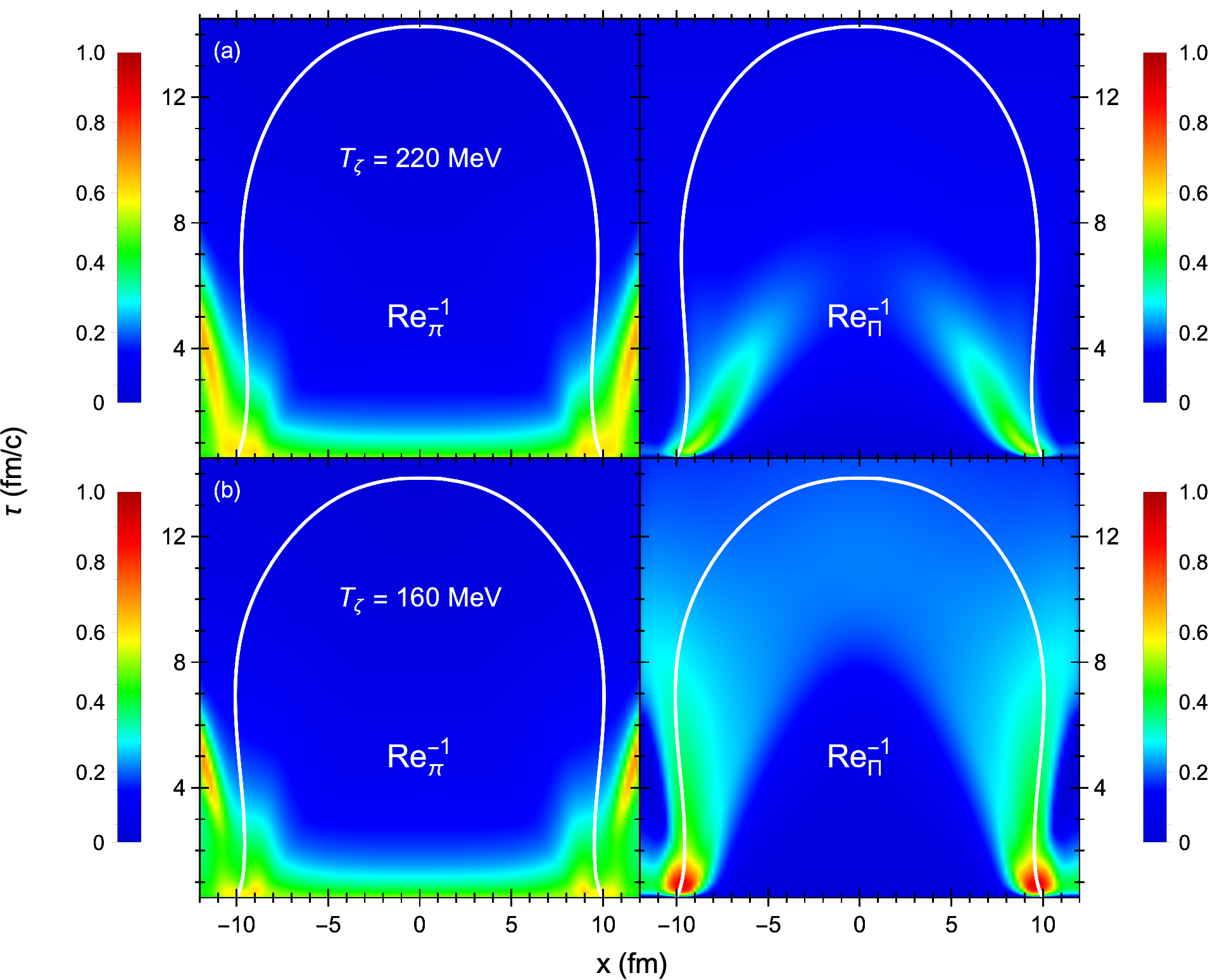}
\caption{
    The $\tau{-}x$ slice $(y = \eta_s = 0)$ of the shear inverse Reynolds number $\text{Re}^{-1}_\pi = \sqrt{\pi^\munu \pi_\munu} / (\Peq\sqrt{3})$ (left panels) and bulk inverse Reynolds number $\text{Re}^{-1}_\Pi = |\Pi| / \Peq$ (right panels) showing the strength of the $\delta f_n$ corrections on a particlization hypersurface of constant temperature $T_\text{sw} = 150$ MeV (white contour) from a (2+1)--d central Pb+Pb collision with smooth \trento{} initial conditions.  We vary the peak temperature of the specific bulk viscosity $\zeta / \mathcal{S}$ to either $T_\zeta = 220$ MeV (top row) or $T_\zeta = 160$ MeV (bottom row).
\label{feqmod_hydro_slice}
}
\end{figure}

From the hydrodynamic simulation we generate an isothermal particlization hypersurface of temperature $T_\text{sw} = 150$ MeV, using the freeze-out finder code {\sc CORNELIUS} \cite{Huovinen:2012is}. For the central Pb+Pb collision, the longitudinally boost-invariant hypersurface at spacetime rapidity $\eta_s = 0$ contains about $6.8 {\,\times\,} 10^4$ freeze-out cells. Figure~\ref{feqmod_hydro_slice} shows a $\tau{-}x$ slice of the particlization surface at $y = \eta_s = 0$, as well as the shear and bulk inverse Reynolds numbers to gauge the strength of the shear and bulk $\delta f_n$ corrections.\footnote{%
    We note the setup used in this Section is different from the one in an earlier study to test the {\sc iS3D} particle sampler (see Chapter~\ref{chap6label}). There we found that the viscous corrections were so large that the PTMA distribution was unusable for a significant part of the hypersurface. The hypersurface shown in Fig.~\ref{feqmod_hydro_slice} was computed with parameter settings that allow for a meaningful comparison between the PTM and PTMA distributions.
    \label{fn14}} 
Finally, we evaluate the Cooper--Frye formula with the code {\sc iS3D} (see Chapter~\ref{chap6label})~\cite{McNelis:2019auj}. To compute the longitudinally boost-invariant particle spectra, the hypersurface volume needs to be extended to the $\eta_s$ dimension and centered around the momentum rapidity $y_p$.  We perform the numerical integration of the Cooper--Frye formula along the $\eta_s$--direction using Gauss--Legendre integration on a 48-point grid $(y_p {\,-\,} \eta_s)_j$ with integration weights $\omega_j$ given by
\bs
\allowdisplaybreaks
\label{eqch5:integration}
\beal
  (y_p {\,-\,} \eta_s)_j &= \sinh^{-1}\left(\frac{x_j}{1{\,-\,}x_j^2}\right) \,,
\\
  \omega_j &= w_j \frac{1{\,+\,}x_j^2}{|1{\,-\,}x_j^2|\sqrt{1{\,-\,}x_j^2{\,+\,}x_j^4}}\,,
\end{align}
\es
where $x_j$ and $w_j$ are the Gauss--Legendre roots and weights, respectively.
\subsection{Breakdown of the modified distributions and technical issues}
\label{sec5b}
As stated in the previous sections, the modified equilibrium distributions~\eqref{eqch5:feqmod_2} and~\eqref{eqch5:Jonah} break down in freeze-out cells with large viscous corrections. For the first scenario where the specific bulk viscosity's peak temperature is $T_\zeta = 220$ MeV (top row in Fig.~\ref{feqmod_hydro_slice}), both the shear and bulk viscous corrections are small enough to use the modified equilibrium distribution for all freeze-out cells. For $T_\zeta = 160$ MeV (bottom row in Fig.~\ref{feqmod_hydro_slice}) the bulk viscosity peaks much closer to the particlization hypersurface. Not only are the bulk viscous corrections much larger in this case, but the shear corrections are also enhanced due to shear-bulk coupling effects in the hydrodynamic simulation \cite{Denicol:2014mca}. Together they cause the modified equilibrium distribution to break down for about 4400 freeze-out cells at $r \sim 10$ fm and $0.5$ fm/$c$ $< \tau < 1.3$ fm/$c$. There are several options to handle the particle spectra contribution from these freeze-out cells: (i) ignore such freeze-out cells entirely, (ii) use for them the local-equilibrium distribution $f_n = f_{\eq,n}$, which is positive definite but neglects the viscous components of $T^\munu$, (iii) linearize in such cells the modified equilibrium distribution $f_{\eq,n}^{(\text{mod})} \approx f_{\eq,n} + \delta f_n$, which may turn negative at high momenta but captures all components of $T^\munu$. Here we choose the third option, i.e. for the PTM distribution we switch to the Chapman--Enskog expansion~\eqref{eqch5:Chapman_Enskog} when ${\det} A < 10^{-5}$ or when the normalization factor $\mathcal{Z}_n$ of any hadron (usually the lightest pion $\pi^0$) turns negative. For the PTB distribution, we use $\delta f_n$ from Eq.~\eqref{eqch5:linear_Jonah} when ${\det} A < 10^{-5}$ or $-\Peq < \Pi < \ene/3 - \Peq$. In this study, the impact of these manipulations on the total particle spectra is negligible since the number of ``bad'' freeze-out cells is small. However, if a significant fraction of the hypersurface requires one to use something other than the modified equilibrium distribution, then Eqs.~\eqref{eqch5:feqmod_2} and \eqref{eqch5:Jonah} should probably not be used for particlization.

Even if the modified equilibrium distribution can be used, a very small ${\det}A$ value can cause its width along the rapidity direction $y_p {\,-\,} \eta_s$ to be extremely narrow. This leads to numerical errors in the particle yields if the spacetime rapidity grid $(y_p {\,-\,} \eta_s)_j$ is not fine enough. To resolve this technical issue, we rescale the spatial grid $(y_p {\,-\,} \eta_s)_j$ by the distribution's rapidity width $\delta y_p$. One can estimate $\delta y_p$ from a diagonal deformation matrix $A_{ij} {\,=\,} \mathrm{diag}(1{\,+\,}\bar\Pi{\,-\,}\frac{1}{2}\bar{\pi}_{zz},1{\,+\,}\bar\Pi{\,-\,}\frac{1}{2}\bar{\pi}_{zz},1{\,+\,}\bar\Pi{\,+\,}\bar{\pi}_{zz})$:
\be
\delta y_p \sim 1+\bar\Pi+\bar{\pi}_{zz}\,,
\ee
where $\bar\Pi = \Pi/(3\beta_\Pi)$ (or $\lambda_\Pi$) and $\bar{\pi}_{zz} = \pi_{zz}/(2\beta_\pi)$. Relative to the transverse momentum space, the rapidity distribution becomes very narrow when $\bar{\pi}_{zz} \approx - (1+\bar\Pi)$. In this limit, the rapidity width is proportional to
\be
\delta y_p \propto \frac{{\det}A}{({\det}A_\Pi)^{2/3}}\,,
\ee
where ${\det}A_\Pi = (1{\,+\,}\bar\Pi)^3$. Therefore, for each longitudinally boost-invariant freeze-out cell $d^2\sigma_{\mu,i}$ we rescale the spacetime rapidity grid points and weights~\eqref{eqch5:integration} as
\bs
\label{eqch5:integration_rescale}
\beal
(y_p {\,-\,} \eta_s)_{i,j} &= \frac{{\det}A_i}{({\det}A_{\Pi,i})^{2/3}} \times (y_p {\,-\,} \eta_s)_j \,,\\
\omega_{i,j} &= \frac{{\det}A_i}{({\det}A_{\Pi,i})^{2/3}} \times \omega_j\,.
\end{align}
\es
This rescaling trick is found to work well even for small values of ${\det}A \sim 10^{-5}$. For (3+1)--d hypersurfaces, however, this method cannot be used because the freeze-out finder fixes the freeze-out cells' spacetime rapidity. Instead, we switch to a linearized $\delta f_n$ correction if the rapidity width is too small (e.g. ${\det}A / ({\det}A_\Pi)^{2/3} < 0.01$).

The modified anisotropic distribution~\eqref{eqch5:famod1} can also break down, usually because the longitudinal pressure $\PL$ turns negative during the viscous hydrodynamic simulation. When this happens one cannot construct a solution for the momentum deformation parameter $\alpha_L$ in the leading order anisotropic distribution \eqref{eqch5:fa}. For the hypersurface with large bulk viscous corrections we find $\PL{\,<\,}0$ in about 7800 freeze-out cells during the period $0.5$\,fm/$c < \tau < 2.3$\,fm/$c$ (see footnote \ref{fn14}). For these freeze-out cells, we here simply replace Eq.~\eqref{eqch5:famod1} by the local-equilibrium distribution \eqref{eqch5:feq}. In practice it would be more appropriate to use the PTMA distribution on hypersurfaces constructed from anisotropic hydrodynamic simulations in which the occurrence of negative longitudinal pressures is largely avoided \cite{McNelis:2021zji}.
\begin{figure*}[t]
\centering
\includegraphics[width=\linewidth]{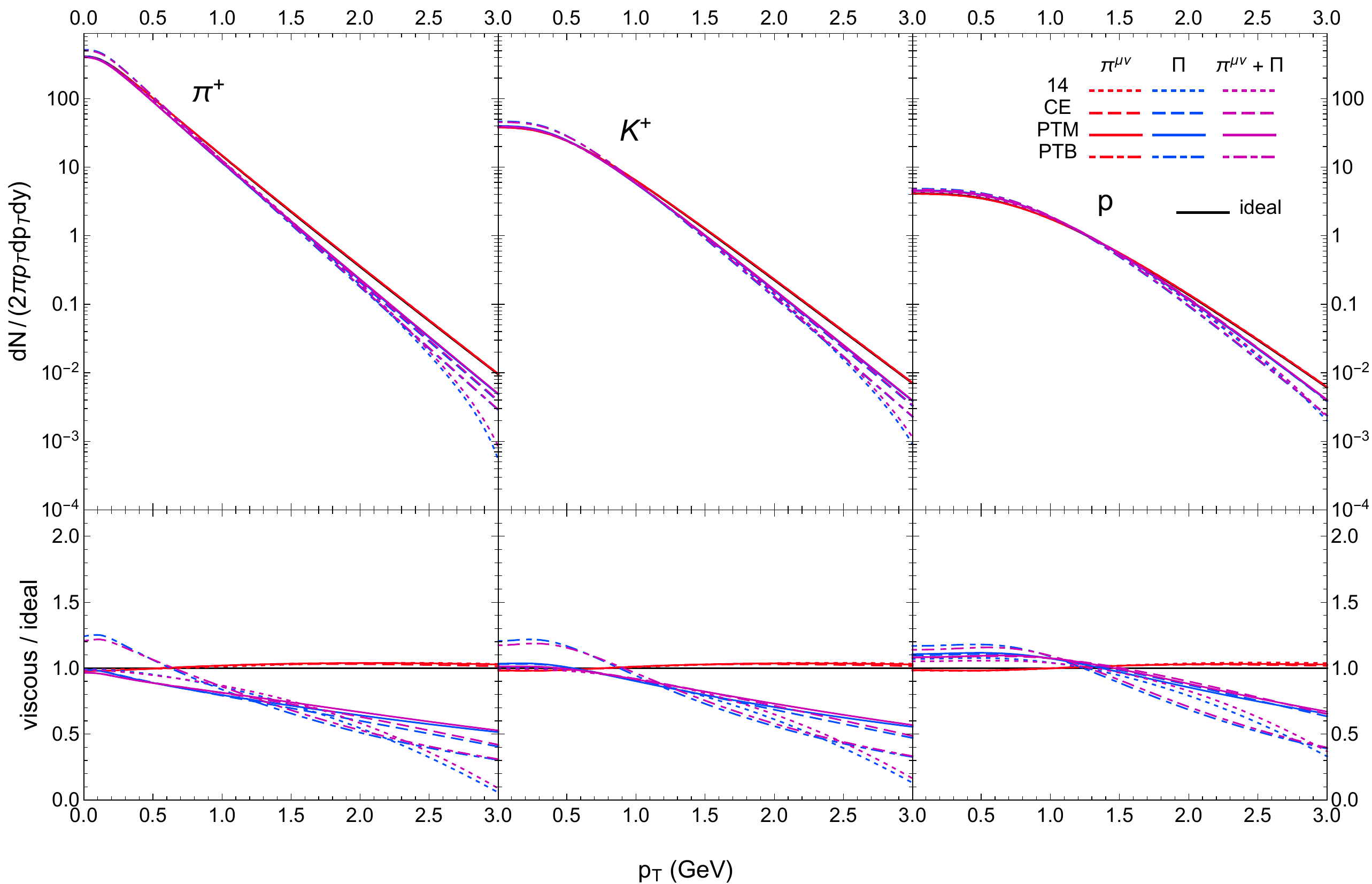}
\caption{
    The azimuthally-averaged transverse momentum spectra of $(\pi^+, K^+, p)$ (top panels) computed with the Cooper--Frye formula for the (2+1)--d central Pb+Pb collision described in Sec.~\ref{sec5a}. The peak temperature of $\zeta/\mathcal{S}$ is set to $T_\zeta = 220$ MeV. We compare the shear (red), bulk (blue) and combined shear + bulk (purple) $\delta f_n$ corrections of the 14--moment approximation (dotted color), RTA Chapman--Enskog expansion (dashed color), PTM distribution (solid color) and PTB distribution (dot-dashed color) relative to the \textit{ideal} spectra with $\delta f_n = 0$ (solid black). The bottom panels show the ratio of the particle spectra with $\delta f_n$ corrections to the \textit{ideal} spectra. 
\label{dNdpT_small_bulk}
}
\end{figure*}
\subsection{Central collisions}
\label{sec5c}
Figure~\ref{dNdpT_small_bulk} shows the continuous transverse momentum spectra~\eqref{eqch5:pT_spectra} of $(\pi^+, K^+, p)$, without resonance decay contributions or hadronic rescattering, for the central Pb+Pb collision with a specific bulk viscosity peak temperature of $T_\zeta = 220$ MeV. We will study the shear and bulk viscous corrections to the \textit{ideal} spectra (i.e. $\delta f_n = 0$) computed with the four models for $\delta f_n$ discussed in Sec.~\ref{sec2}.

As seen in the figure, the shear stress (although small) slightly flattens the $p_T$ spectra (red curves), without affecting the total yields, since it slows the longitudinal expansion and pushes the particles outward in the transverse direction. The bulk viscous pressure has the opposite effect by counteracting the scalar expansion rate and reducing the average pressure, thereby softening the slope of the $p_T$ spectra (blue curves). One observes the ``shoulder'' at low values of $p_T$ that is typical for thermal flow spectra \cite{Schnedermann:1993ws}; there the bulk corrections are more pronounced in protons than in pions and kaons. Furthermore, the bulk viscous correction also decreases the total pion and kaon yields while increasing the proton yield. The PTB distribution is the only exception to these trends: for non-zero bulk viscous pressure it both raises the shoulder and increases the yields of all particles, which sets it apart from the other $\delta f_n$ corrections. Finally, the purple curves show the combined effects of the shear and bulk viscous corrections to the spectra. Overall, the slope of the total $p_T$ spectra is steeper than the \textit{ideal} one since here the bulk viscous pressure is larger than the shear stress for most of the hypersurface.

We also compare the $\delta f_n$ corrections of the 14--moment approximation, RTA Chapman--Enskog expansion, PTM distribution and PTB distribution. We see that the momentum dependence of the viscous correction varies considerably between models: the ratio of the total particle spectra to the $\textit{ideal}$ one at high $p_T$ is approximately quadratic for the 14--moment approximation and linear for the RTA Chapman--Enskog expansion, as illustrated by the bottom panels in Fig.~\ref{dNdpT_small_bulk}. The PTM spectra is almost identical to the one computed with the Chapman--Enskog expansion because the shear stress and bulk viscous pressure are small; differences between the two approaches emerge only at large values of $p_T$. Overall, there are no significant differences among the first three $\delta f_n$ models across most of the $p_T$ spectrum in Fig.~\ref{dNdpT_small_bulk}. The PTB distribution, on the other hand, has a moderate excess of low $p_T$ pions and kaons relative to the other $\delta f_n$ corrections.

\begin{figure*}[t]
\centering
\includegraphics[width=\linewidth]{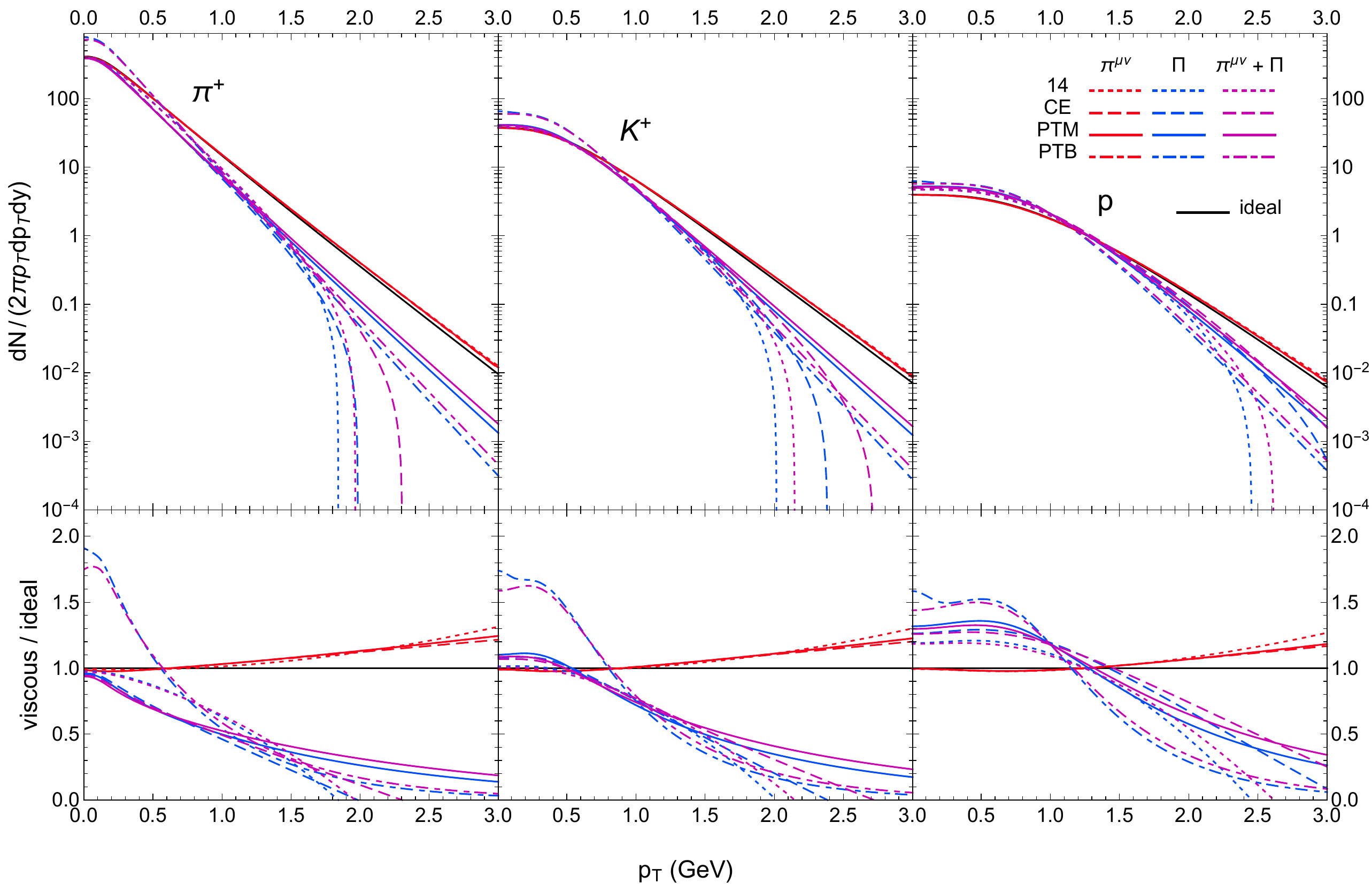}
\caption{
    Same as Fig.~\ref{dNdpT_small_bulk} but with a $\zeta / \mathcal{S}$ peak temperature of $T_\zeta = 160$ MeV. 
\label{dNdpT_large_bulk}
}
\end{figure*}
\begin{figure*}[t]
\centering
\includegraphics[width=\linewidth]{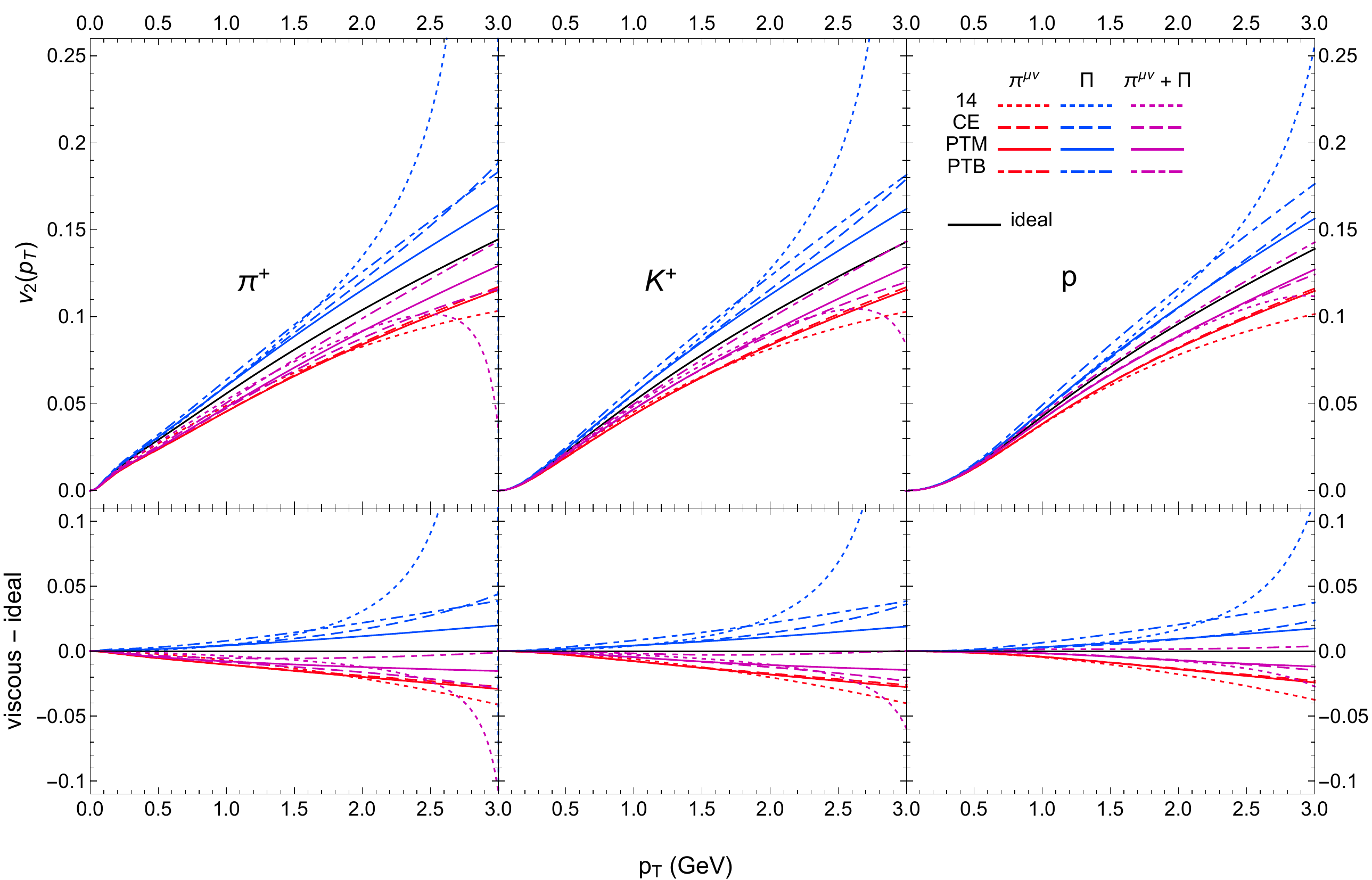}
\caption{(Color online)
The $p_T$--differential elliptic flow coefficient of $(\pi^+, K^+, p)$ for the (2+1)--d non-central Pb+Pb collision (top panels). The peak temperature of $\zeta / \mathcal{S}$ is set to $T_\zeta = 220$ MeV. Similar to Fig.~\ref{dNdpT_small_bulk}, we plot the shear and bulk viscous corrections of each $\delta f_n$ model to the \textit{ideal} $v_2(p_T)$ (solid black). The bottom panels show the differences between $v_2(p_T)$ with $\delta f_n$ corrections and the \textit{ideal} $v_2(p_T)$.
\label{v2_small_bulk}
}
\end{figure*}

In Figure~\ref{dNdpT_large_bulk}  we show how the central Pb+Pb collision spectra from Fig.~\ref{dNdpT_small_bulk} change when we move the peak of the specific bulk viscosity from $T_\zeta = 220$ MeV to 160 MeV. This obviously increases the magnitude of the bulk viscous pressure on the particlization hypersurface at $T_\mathrm{sw}=150$ MeV. The shear-bulk coupling effect in the hydrodynamic simulation then also increases the strength of the shear viscous corrections on the hypersurface, resulting in much flatter $p_T$ slopes than in the previous case. Surprisingly, the shear viscous corrections from the four $\delta f_n$ models are still very close to each other (for PTM and PTB this is expected because these distributions use the same shear stress modification). Instead, the largest differences are found in their bulk viscous corrections. Because the bulk viscous pressure on the hypersurface is quite large (lower right panel in Fig.~\ref{feqmod_hydro_slice}), the linearized bulk viscous correction in the 14--moment approximation causes the pion spectra to turn negative already at intermediate momentum $p_T \sim 1.8$ GeV, whereas the kaon and proton spectra are negative for $p_T > 2$ GeV and 2.4 GeV, respectively. The bulk viscous correction in the RTA Chapman--Enskog expansion is less severe than the 14--moment approximation, but the pion and kaon spectra still turn negative for $p_T \gtrsim 2$ GeV (the proton spectra remain positive up to $p_T = 3$ GeV). In contrast, the PTM spectra are positive definite by construction even for moderately large bulk viscous pressures, maintaining an exponential tail at high values of $p_T$.\footnote{%
    Without regulation, the linearized $\delta f_n$ corrections from the ``bad" freeze-out cells (see Sec.~\ref{sec5b}) eventually turn the PTM and PTB spectra negative but this occurs outside the experimental range of interest for soft hadron emission $p_T < 3$ GeV.} 
One also observes a slight excess of protons at low $p_T$ (this effect is much less visible for the kaon). The PTB spectra also remain positive but have significantly steeper slopes than the PTM spectra; the shoulder enhancements at low $p_T$ are also much larger, especially for pions.

After combining the shear and bulk viscous corrections, the $p_T$ spectra become positive except for the one computed with the 14--moment approximation, which has a strong quadratic momentum dependence from its bulk viscous correction. 

\begin{figure*}[t]
\centering
\includegraphics[width=\linewidth]{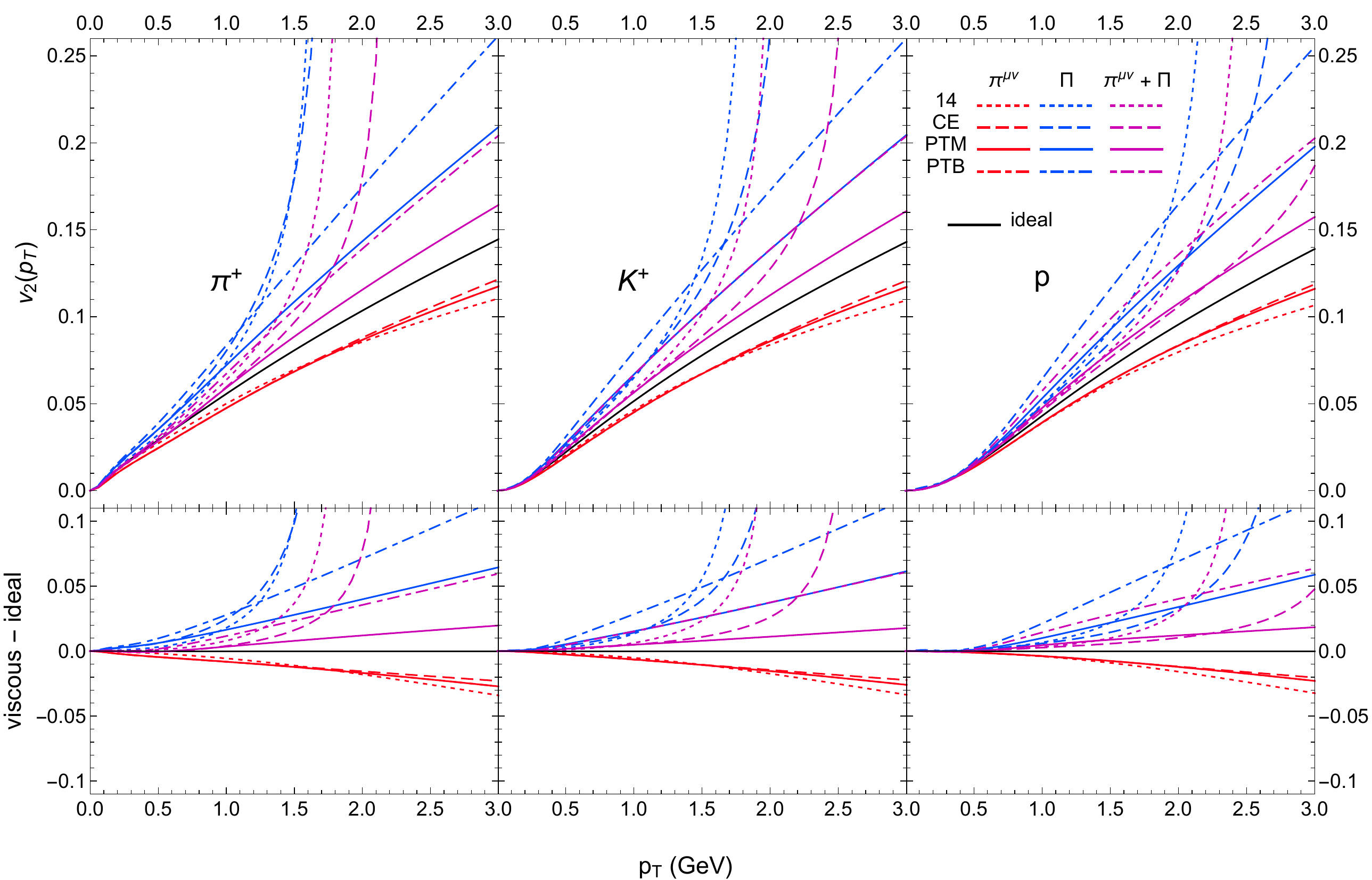}
\caption{
    Same as Fig.~\ref{v2_small_bulk} but with a $\zeta / \mathcal{S}$ peak temperature of $T_\zeta = 160$ MeV. 
\label{v2_large_bulk}
}
\end{figure*}

\subsection{Non-central collisions}
\label{sec5d}

Next we repeat the same hydrodynamic simulations for a nonzero impact parameter $b = 5$\,fm, to study the effects of the different viscous corrections $\delta f_n$ on the $p_T$--differential elliptic flow coefficient, shown in Fig.~\ref{v2_small_bulk} for $\pi^+$, $K^+$, and $p$. Here we start with $T_\zeta = 220$ MeV, resulting in relatively weak viscous stresses on the particlization hypersurface. The shear viscous corrections are seen to decrease the differential elliptic flow $v_2(p_T)$, counteracting the effects of anisotropic transverse flow. The bulk viscous pressure, on the other hand, tends to increase $v_2(p_T)$ by suppressing the radial flow and making the $p_T$ spectra \eqref{eqch5:pT_spectra} steeper. Individually, the small shear and bulk viscous corrections to the \textit{ideal} $v_2(p_T)$ (defined by setting $\delta f_n{\,=\,}0$) are roughly linear in $p_T$, except for the bulk $\delta f_n$ correction of the 14--moment approximation, which causes $v_2(p_T)$ to diverge at high $p_T$ when the spectrum (which enters the denominator of $v_n(p_T)$) passes through zero. Overall, there is a net suppression on the differential elliptic flow since it is more sensitive to the shear viscous corrections at the space-like edges of the hypersurface, whose fluid cells have undergone the strongest transverse acceleration. Interestingly, the only exception is the PTB elliptic flow, whose shear and bulk viscous corrections nearly cancel each other. Similar to the spectra discussed in the previous subsection, the four $\delta f_n$ models produce very similar viscous corrections to the elliptic flow when $T_\zeta = 220$ MeV. Clear distinctions between these models appear only for large transverse momenta $p_T > 2$ GeV.

When $T_\zeta$ is lowered to 160 MeV (see Fig.~\ref{v2_large_bulk}), the much larger bulk viscous pressure on the space-like part of the hypersurface results in the divergence of the linearized bulk viscous corrections to $v_2(p_T)$ at intermediate values of $p_T{\,\sim\,}1.5{-}2.5$ GeV, for all three particle species considered, reflecting the corresponding sign change of their azimuthally averaged $p_T$ spectra \eqref{eqch5:pT_spectra}. The shear viscous corrections help offset this effect by flattening the slope of the $p_T$ spectra, but the pion and kaon elliptic flows still diverge, albeit now at slightly higher $p_T$ (as does the proton $v_2(p_T)$ from the 14--moment approximation). For realistic event-by-event simulations where the hadrons are Monte Carlo sampled from the hypersurface prior to the afterburner phase, these divergences are removed by enforcing the regulation $f_{\eq,n} + \delta f_n \geq 0$ in the Cooper--Frye formula~\cite{Shen:2014vra,McNelis:2019auj}.

In the PTM distribution, on the other hand, the bulk viscous modifications prevent the $p_T$ spectra from turning negative, and the resulting elliptic flow curves are well behaved even at high $p_T$. After including the shear stress modification, the PTM differential elliptic flows decrease but remain above the \textit{ideal} curves since the bulk viscous pressure overwhelms the shear stress on the hypersurface. The PTB elliptic flow coefficients also stay finite at high $p_T$ but are significantly larger than those computed with the PTM distribution. 
\begin{figure*}[!htbp]
\centering
\includegraphics[width=\linewidth]{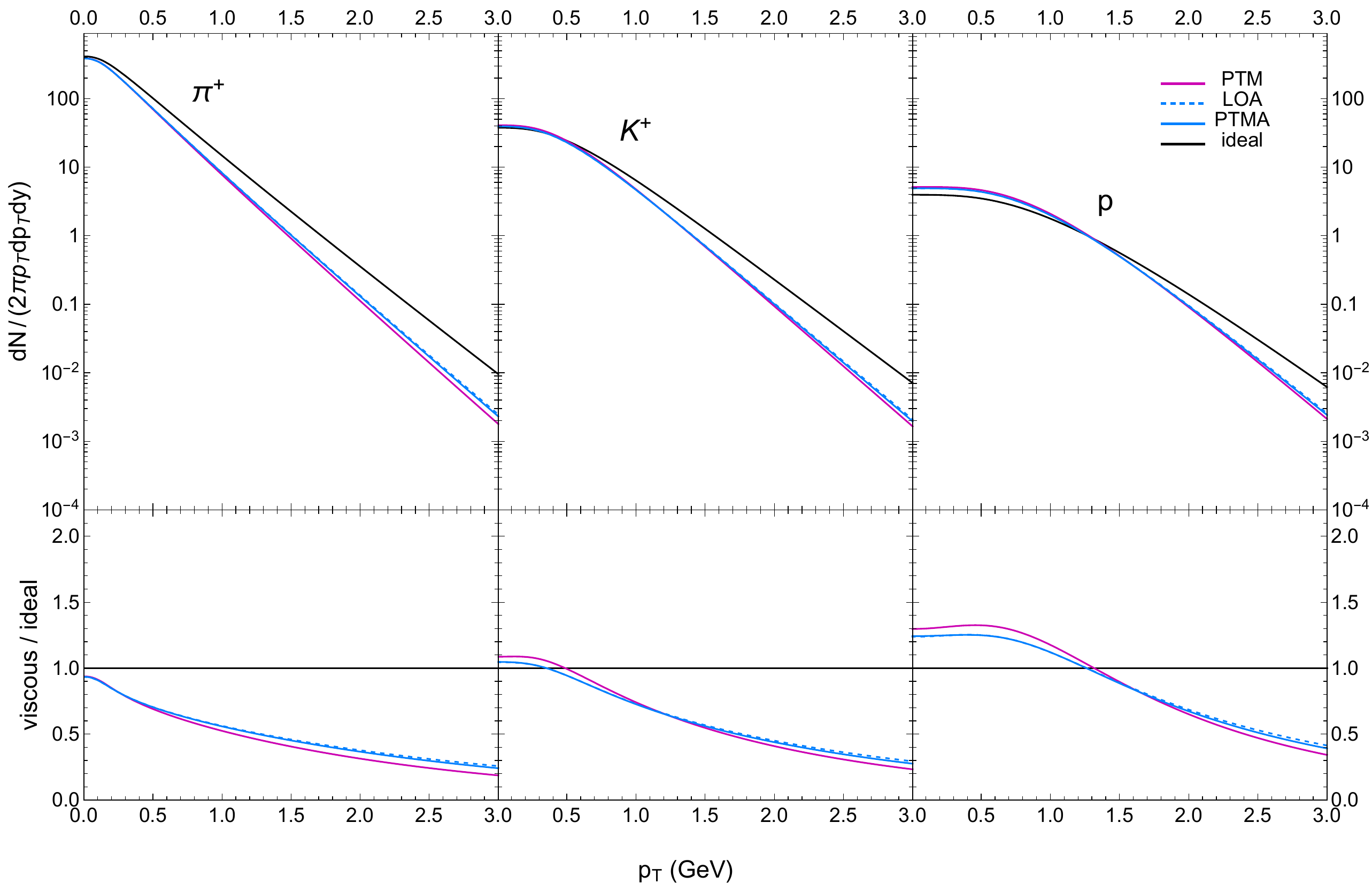}
\caption{
    The azimuthally-averaged transverse momentum spectra of ($\pi^+,K^+,p)$ (top panels) for the central Pb+Pb collision with $T_\zeta = 160$ MeV. We compare the full PTMA distribution (solid light blue) to the PTM distribution with shear and bulk modifications (solid purple) and the leading order anisotropic distribution~\eqref{eqch5:fa} without residual shear corrections (dotted light blue). The bottom panels show the ratio of the particle spectra with viscous corrections to the \textit{ideal} spectra.
\label{dN_dpT_famod}
}
\includegraphics[width=\linewidth]{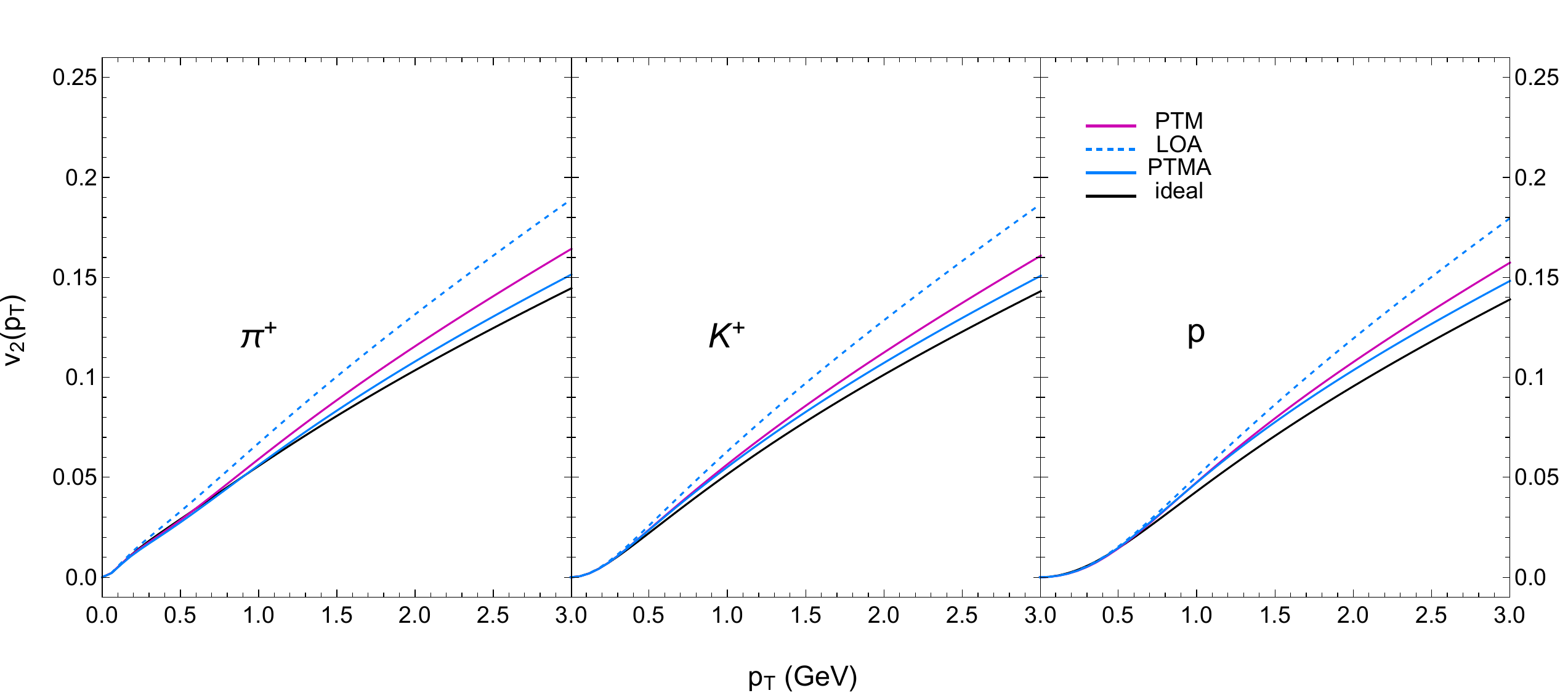}
\caption{
    The $p_T$--differential elliptic flow of ($\pi^+,K^+,p)$  for the non-central Pb+Pb collision with $T_\zeta = 160$ MeV. Similar to Fig.~\ref{dN_dpT_famod}, we compare the PTMA distribution (solid light blue) to the PTM distribution (solid purple) and leading order anisotropic distribution (dotted light blue).
\label{v2_famod}
}
\end{figure*}
\subsection{Modified anisotropic distribution}
We close this section by looking at the particle $p_T$ spectra and differential elliptic flows computed with the PTMA distribution \eqref{eqch5:famod1}. Figure~\ref{dN_dpT_famod} shows the resulting transverse momentum spectra for the same central Pb+Pb collision as in Fig.~\ref{dNdpT_large_bulk}. Compared to the PTM spectra, which here include both shear and bulk viscous corrections, the PTMA kaon and proton distributions have a slightly higher mean transverse momentum, indicated by corresponding shifts in the slope and shoulder, and the pion yield slightly increases. We know from Fig.~\ref{hydro_output} that the PTM distribution underpredicts the input isotropic pressure $\Peq + \Pi$ because the bulk viscous modifications to the effective temperature and isotropic momentum space in Eq.~\eqref{eqch5:feqmod_2} do not perfectly reproduce the input bulk viscous pressure. The anisotropic parameters ($\Lambda, \alpha_\perp,\alpha_L$) in the PTMA distribution are optimized to correct for these errors, outputting a slightly larger isotropic pressure than the PTM distribution. One also notices that the PTMA spectra are virtually identical to the ones computed with leading order anisotropic distribution~\eqref{eqch5:fa}, which excludes the residual shear corrections in Eq.~\eqref{eqch5:Bij}. Due to the approximate azimuthal symmetry of the hypersurface, we expect the transverse shear stress $\piperp$ ($\Wperp = 0$ by longitudinal boost-invariance) to have little to no impact on the azimuthally-averaged particle spectra. 

Figure~\ref{v2_famod} shows the elliptic flow coefficients from the PTMA distribution for the same non-central Pb+Pb collision as in Fig.~\ref{v2_large_bulk}. Although the PTM and PTMA models yield very similar results, the smaller bulk viscous modification in the PTMA distribution slightly brings down the $v_2(p_T)$ curves. Relative to the leading order anisotropic distribution, we see that the transverse shear corrections play a larger role in non-central collisions, damping the anisotropic flow.

\section{Summary}
\label{sec6}
In this chapter we have developed a positive definite hadronic distribution for Cooper--Frye particlization that improves upon the Pratt--Torrieri distribution in Ref.~\cite{Pratt:2010jt}. At an intermediate stage of the derivation, we used the RTA Boltzmann equation to identify the non-equilibrium corrections to the Boltzmann factor. This ensures that in the limit of small dissipative flows our ``modified equilibrium" distribution reduces exactly to the first-order RTA Chapman--Enskog expansion. Even with these dissipative modifications, the distribution function remains positive definite for arbitrarily large momenta, but there is a trade-off: it no longer matches the input energy-momentum tensor and net baryon current exactly. The mismatch becomes significant already for moderately large viscous corrections.

To minimize these errors, we slightly restructured the dissipative perturbations by linearizing the viscous corrections to the momentum scales of the local-equilibrium distribution as well as their contributions to the corresponding particle yields. The resulting PTM modified equilibrium distribution bears close resemblance to the Pratt--Torrieri distribution~\eqref{eqch5:feqmod_original} but it can better reproduce the input hydrodynamic quantities in freeze-out cells subject to moderately large bulk viscous pressures. Although the output of the energy-momentum tensor calculated with the modified equilibrium distribution does not perfectly match the input $T^{\mu\nu}$, one can further improve it by applying the same technique to modify the leading order anisotropic distribution (which already matches to some large dissipative effects). We will use this PTMA distribution for our event-by-event simulation in Chapter~\ref{chapter7label}.

We also compared the PTM distribution to the linearized 14--moment approximation and the first-order RTA Chapman--Enskog expansion, by using the Cooper--Frye formula to compute the momentum spectra and $p_T$--differential elliptic flow coefficients of hadrons emitted from a particlization hypersurface. For small viscous corrections the azimuthally averaged transverse momentum spectra and $p_T$--differential elliptic flows generated by all of these $\delta f_n$ models are very similar. When the dissipative flows on the hypersurface (in our case mostly the bulk viscous pressure) become moderately large, the transverse momentum spectra of the linearized $\delta f_n$ corrections turn negative at intermediate $p_T$ values while those of the PTM and PTMA distributions remain positive at all $p_T$, by construction. This simultaneously prevents the $p_T$--differential elliptic flow coefficients from developing singularities caused by zero crossings of the $p_T$ spectra, a problem that plagues the linearized viscous corrections whenever the hypersurface features large bulk viscous pressures.

In practical applications, the continuous particle spectra from the Cooper--Frye formula are not sufficient to make quantitative predictions for experimental observables. Instead, one usually samples hadronic emission events from the hypersurface, each of which contains a finite number of real particles; the subsequent particle interactions can then be simulated in a Boltzmann transport code to improve the final-state hadronic distributions. However, this sampling procedure requires the Cooper--Frye integrand to be regulated to remain positive definite in certain phase-space regions. If the hypersurface has large viscous corrections, the continuous particle spectra can be significantly altered by the regulations at high momenta, especially those computed with a linearized $\delta f_n$ correction. A Monte Carlo simulation should be able to replicate the regulated continuous spectra after event-averaging many particlization events. This ensures the user that the deviations of the sampled spectra from the original Cooper--Frye formula are due to only these regulations and not internal errors in the code.

\chapter{Monte Carlo simulation of the particlization phase}
\label{chap6label}
Finally, we simulate the particlization stage using the {\sc iS3D} code, which was developed from the C++ module {\sc iSS} in the {\sc iEBE--VISHNU} event-by-event simulation code package \cite{Shen:2014vra}.\footnote{%
    The code package can be downloaded from the GitHub repository \url{https://github.com/derekeverett/iS3D}.
    \label{fn7}}
The most important routine in the code is the particle sampler, which Monte Carlo samples particlized events from a hypersurface given by the preceding hydrodynamic simulation (e.g. \cpuvah{}).\footnote{%
    Some routines in the particle sampler algorithm were inspired by earlier work reported in Refs.~\cite{Bernhard:2018hnz, Pang:2018zzo}.} 
It is capable of sampling the hadronic distributions from the Cooper--Frye formula~\eqref{eqch5:CooperFrye} using one of five choices for the $\delta f_n$ correction: the 14--moment approximation, the RTA Chapman--Enskog expansion, the PTM and PTB modified equilibrium distributions and the PTMA modified anisotropic distribution. This gives the heavy-ion physics community new tools to explore the sensitivity of heavy-ion experimental data to different choices for $\delta f_n$, which is presently the main source of theoretical uncertainty in the particlization stage~\cite{Everett:2020xug,Everett:2020yty}. We validate the particle sampler by conducting high-precision tests comparing the event-averaged sampled momentum spectra and spacetime distributions with those from the numerically evaluated continuous Cooper--Frye formula. The {\sc iS3D} code is also compatible with the state-of-the-art hadronic afterburner code {\sc SMASH}~\cite{Weil:2016zrk}, to which we will feed our particlization events as input in Chapter~\ref{chapter7label}.

In the code we convert the hydrodynamic variables on the particlization hypersurface back to physical units (i.e. the temperature has units of GeV, etc). We express the Lorentz-invariant momentum spectra as
\be
E_p\frac{dN_n}{d^3p} = \frac{dN_n}{p_T dp_T d\phi_p dy_p}\,,
\ee
where $p_T = \sqrt{(p^x)^2 + (p^y)^2}$ is the transverse momentum, $\phi_p = \tan^{-1}(p^y/p^x)$ is the azimuthal angle and $y_p={\tanh^{-1}}\left(p^z/E_p\right)$ is the momentum rapidity.

This chapter is based on material published in Ref.~\cite{McNelis:2019auj} which, however, only considered the first four models for the $\delta f_n$ correction. The latest version of the code also implements the PTMA distribution, but we do not present any new sampler validation tests here.
\section{Setup}
\label{chap6S3}

In this section, we discuss the collision systems used for generating the particlization hypersurfaces and computing the hadronic observables needed in our tests. We note that this older setup is different from the one used in the previous chapter.

\subsection{Central Pb+Pb collision}
\label{chap6S3.1}

We use the code {\sc GPU VH} \cite{Bazow:2016yra} (which is similar to \cpuvah{}) to evolve a longitudinally boost-invariant central Pb+Pb collision with standard (2+1)--dimensional second-order viscous hydrodynamics.\footnote{%
    For the purpose of testing the {\sc iS3D} sampler, we assume longitudinal boost-invariance for simplicity. The code can also sample particlization hypersurfaces from (3+1)--dimensional hydrodynamic simulations.} 
For longitudinally boost-invariant systems, the hadronic distribution function $f_n(x,p)= f_n(\tau,x,y, p_T,\phi_p, y_p\,{-}\,\eta_s)$ can depend only on the difference between $y_p$ and $\eta_s$, and macroscopic fields (e.g. the temperature $T(x)$) must be independent of $\eta_s$. Furthermore, we will only be interested in mid-rapidity observables at $y_p = \eta_s = 0$. 

We use smooth optical Glauber initial conditions \cite{Miller:2007ri} with a central temperature of $T_0 = 600$ MeV. We start the hydrodynamic simulation at a longitudinal proper time $\tau_0 = 0.25$ fm/$c$, with the fluid velocity's spatial components $u^i$, shear stress tensor $\pi^\munu$ and bulk pressure $\Pi$ initialized to zero. We set the specific shear viscosity to $\eta / \mathcal{S} = 0.2$. The specific bulk viscosity is parameterized as $(\zeta / \mathcal{S})(T) = (\zeta/\mathcal{S})_\mathrm{norm} \, w(T/T_p)$, where $w(x)$ is the function in Eq.~\eqref{eqch3:80}.
%
\begin{figure}[t!]
\centering
\includegraphics[width=\textwidth]{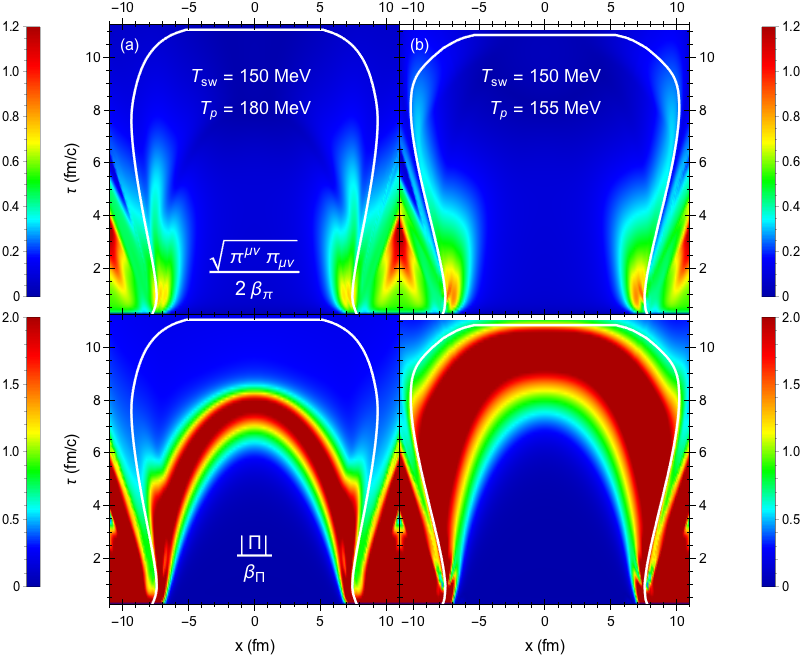}
\caption{(Color online)
    The $\tau{-}x$ slice at $y=\eta_s=0$ of the shear Knudsen number $\mathrm{Kn}_\pi = \sqrt{\pi^\munu \pi_\munu} / (2\beta_\pi)$ (top panels), bulk Knudsen number $\text{Kn}_{_\Pi} = |\Pi| / (\beta_\Pi)$ (bottom panels), and particlization hypersurface at temperature $T_\mathrm{sw} = 150$\,MeV (white contour) from the (2+1)--d hydrodynamic simulation of a central Pb+Pb collision. Results for two choices for the temperature $T_p$ at which the bulk viscosity $\zeta / \mathcal{S}$ peaks ($T_p=180$ MeV (a) and $T_p=155$ MeV (b)) are shown in the left and right columns. 
\label{FHydro}
}
\end{figure}
%
%
We set the normalization factor to ($\zeta/\mathcal{S})_\text{norm} = 1$ and the peak temperature to either $T_p = $ 180 MeV or 155 MeV, as specified in each case below. \footnote{%
    Note that the hydrodynamic code {\sc GPU VH} does not evolve the net baryon density and baryon diffusion current~\cite{Bazow:2016yra}. Therefore, we set $\alpha_B = 0 = V_B^\mu$ in this chapter. While the effects of baryon chemical potential are fully implemented in {\sc iS3D}, we have not yet tested them.}
Here $\zetas$ has a much larger peak value than the parameterization used in the previous chapter. 

A particlization hypersurface of constant temperature $T_\mathrm{sw}{\,=\,}150$ MeV is generated using the freezeout surface finder code CORNELIUS \cite{Huovinen:2012is}. For a central Pb+Pb collision with these parameters, the boost-invariant hypersurface contains about $N_\Sigma = 1.9 \times 10^5$ freezeout cells.\footnote{%
    These freezeout cells are located at the slice $\eta_s{\,=\,}y_{p}^\text{CM}$ where $y_{p}^\text{CM}$ is the center-of-momentum rapidity of the collision system (here we set $y_{p}^\text{CM}{\,=\,}0$). Since the Cooper--Frye formula involves an integral over $\eta_s$, we invoke longitudinal boost-invariance to replicate freezeout cells from the slice $\eta_s{\,=\,}y_{p}^\text{CM}$ at different spacetime rapidities $\eta_s{\,\ne\,}y_{p}^\text{CM}$.}
Figure~\ref{FHydro} shows the $\tau-x$ slice at $y = \eta_s = 0$ of the shear and bulk Knudsen numbers as well as the particlization surface generated from the simulation. Note that the shear and bulk viscous corrections are much larger than in the previous setup in Chapter~\ref{chap5label}.

For central collision systems we are interested in the azimuthally-averaged transverse momentum spectra
\be
\label{eqch6:central_pT_spectra}
    \frac{dN_n}{2\pi p_T dp_T dy_p} = \int_0^{2\pi} 
    \frac{d\phi_p}{2\pi} \frac{dN_n}{p_T dp_T d\phi_p dy_p}
    = \int_0^{2\pi} \frac{d\phi_p}{2\pi}
      \frac{1}{(2\pi\hbar)^3}\int_\Sigma p \cdot d^3\sigma \, f_n\,,
\ee
as well as the temporal and (azimuthally averaged) radial distributions 
\begin{eqnarray}
\allowdisplaybreaks
\label{eqch6:time_formula}
    \frac{dN_n}{\tau d\tau d\eta_s} &=& 
    \frac{\partial^2}{\tau \partial\tau \partial\eta_s}
    \int_p \int_\Sigma p \cdot d^3 \sigma \, f_n, 
\\
\label{eqch6:radial_formula}
    \frac{dN_n}{2\pi r dr d\eta_s} &=& 
    \frac{\partial^2}{2\pi r \partial r \partial\eta_s}
    \int_p \int_\Sigma p \cdot d^3 \sigma \, f_n \,,
\end{eqnarray}
where $r{\,=\,}\sqrt{x^2{+}y^2}$. Due to the azimuthal symmetry of the optical Glauber initial condition (which represents an ensemble average over fluctuating initial conditions with random orientations in the transverse plane) there are no interesting azimuthally sensitive observables to be computed from this particlization surface.\footnote{%
    Azimuthal fluctuations arising from finite-number statistical effects in the individually sampled events are of physical interest but without value for code verification.}

\subsection{Non-central Pb+Pb collision}
\label{chap6S3.2}

As an example for a non-central collision fireball we evolve, for the same Glauber model and viscosity parameters, a smooth hydrodynamic event with nonzero impact parameter $b = 5$ fm. The resulting particlization surface emits particles with anisotropic flow, which is encoded in the differential flow coefficients
\be
\label{eqch6:vn}
    v_{k,n}(p_T) = 
    \frac{\displaystyle{\int_0^{2\pi}} \!\! d\phi_p\,    
          e^{ik\phi_p}\,\frac{dN_n}{p_T\,dp_T\,d\phi_p dy_p}}
         {\displaystyle{\int_0^{2\pi}} \!\! d\phi_p\,  
         \frac{dN_n}{p_Tdp_Td\phi_p dy_p}}
    = \frac{\displaystyle{\int_0^{2\pi}} \!\! d\phi_p\,    
          e^{ik\phi_p}\,\! \int_\Sigma p \cdot d^3\sigma\,f_n}
         {\displaystyle{\int_0^{2\pi}} \!\! d\phi_p\,  
         \! \int_\Sigma p \cdot d^3\sigma\,f_n}.
\ee
In particular, we will be interested in computing the elliptic and quadrangular flow coefficients $v_2(p_T)$ and $v_4(p_T)$ for non-central collisions.\footnote{%
    Due to the $x\leftrightarrow-x$ reflection symmetry of the fireball in the optical Glauber limit, all odd flow coefficients vanish and the factor $e^{ik\phi_p}$ under the integral in Eq.~(\ref{eqch6:vn}) can be replaced by $\cos(k\phi_p)$.}
    
\section{Continuous Cooper--Frye distributions}
\label{chap6S4}

For each collision event, the Cooper--Frye formulae (\ref{eqch6:central_pT_spectra}) --~(\ref{eqch6:vn}) describe continuous distributions of the emitted hadrons. These can be regarded as the statistical ensemble average of fluctuating discrete distributions if one interprets the Cooper--Frye integrand $p\cdot d^3\sigma\,f_n$ as a probability distribution. We will use them to check the accuracy and precision of the {\sc iS3D} sampler. In this section we review the numerical computation of Eqs.~(\ref{eqch6:central_pT_spectra}) --~(\ref{eqch6:vn}) for both (2+1)--d and (3+1)--d hypersurfaces.

\subsection{Integration routine}
\label{chap6S4.1}
To compute the continuous transverse momentum spectra \eqref{eqch6:central_pT_spectra} and aniso\-tropic flow coefficients \eqref{eqch6:vn} we integrate the Cooper--Frye formula numerically:
\be
\label{eqch6:continuous_spectra}
    \frac{dN_n}{p_T dp_T d\phi_p dy_p} = 
    \sum^{N_\Sigma}_i p \cdot d^3\sigma_i \, 
    f_n(x^\mu_i, p_T,\phi_p, y_p)\,.
\ee
Here $d^3\sigma_i$ are the discrete hypersurface elements at positions $x^\mu_i = (\tau_i, x_i, y_i, \eta_{s,i})$. The algorithm for computing the momentum spectra \eqref{eqch6:continuous_spectra} is straightforward: one simply loops over the freezeout cells $i$ and adds their contribution to the spectra of each particle species at different momentum points. After integrating over the freezeout surface, we use Gaussian quadrature to compute the observables~\eqref{eqch6:central_pT_spectra} and~\eqref{eqch6:vn}.

An exact calculation of the spacetime distributions \eqref{eqch6:time_formula} and \eqref{eqch6:radial_formula} requires knowledge about the partial derivatives of $d^3 \sigma$ and $f_n$. Since the freezeout surface finder code does not provide this information, we use a zeroth-order approximation by computing the particle yield from each freezeout cell:
\be
\label{eqch6:yield_per_cell}
  \Delta N_{n,i} = \int p_T \, dp_T \, d\phi_p \, dy_p \, p \cdot d^3\sigma_i \, f_n(x^\mu_i, p_T,\phi_p, y_p)\,,
\ee
which is evaluated with Gaussian quadrature. We then construct the spacetime distributions by binning the weights \eqref{eqch6:yield_per_cell} in a uniform spacetime grid with $\Delta \tau = 0.1$ fm/$c$, $\Delta r = 0.2$ fm, and $\Delta \eta_s = 0.1$.

For the longitudinally boost-invariant hypersurfaces used in our validation tests, the integration routine is slightly modified. The momentum spectra are evaluated using the formula
\be
\label{eqch6:continuous_spectra_2D}
  \frac{dN_n}{p_T dp_T d\phi_p dy_p} = 
  \sum^{N_\Sigma}_i   p\cdot d^2\sigma_i 
  \sum^{N_{\eta_s}}_j \omega_j\, f_n
  \bigl(x^\mu_{\perp i}, p_T,\phi_p,(y_p {-} \eta_s)_j\bigr).
\ee
Here the first sum over $i$ contains only hydrodynamically generated freezeout cells in the transverse plane at spacetime rapidity $\eta_s=y_{p}^\text{CM}$, i.e. (2+1)--dimensional surface elements $d^2\sigma_i$ at positions $x^\mu_{\perp i} = (\tau_i, x_i, y_i, \eta_s=y_{p}^\text{CM})$. The second sum over $j$ represents the integration over $\eta_s$, where the grid points and integration weights are given by Eqs.~(\ref{eqch5:integration}a,b) (or Eqs.~(\ref{eqch5:integration_rescale}a-b) if $f_n$ is a modified equilibrium distribution). Table~\ref{tab:momentum_table} summarizes the formulas used for the remaining momentum grids $(p_T, \phi_p)$. 
\begin{table}[t]
\centering
\setlength{\tabcolsep}{0.875em} 
{\renewcommand{\arraystretch}{2.25}
 \begin{tabular}{|c|c|c|} 
 \hline
 & $\dfrac{dN}{2\pi p_T dp_T dy_p}$ \,\,\,\,\,\,\,\, $v_k(p_T)$ & \,\,\,$\dfrac{dN}{\tau d\tau d\eta_s}$ \,\,\,\,\,\,\,\, $\dfrac{dN}{2\pi rdr d\eta_s}$\,\,\,\\ [0.875ex]
 \hline 
 $p_{T,j} \, (\mathrm{GeV})$ & $0.03j - 0.015\,\,\,$ & $\left(\dfrac{1+x_j}{1-x_j}\right)^{1/2}$ \\ [0.875ex]\hline
 $\phi_{p,j}$ & $\pi(1+x_j)$ & $\pi(1+x_j)$ \\ [0.875ex]\hline 
 $(y_{p}^\text{CM} - \eta_s)_j$ & $\sinh^{-1}{\left(\dfrac{x_j}{1-x_j^2}\right)}$ & $\sinh^{-1}{\left(\dfrac{x_j}{1-x_j^2}\right)}$ \\ [0.875ex]
 \hline
 \end{tabular}}
 \caption{The momentum tables used in the continuous Cooper--Frye formula to compute the momentum spectra and spacetime distributions for a (2+1)--d hypersurface. We use a 100--point uniform $p_T$ grid for the momentum spectra. The remaining entries use a 48--point non-uniform grid, where $x_j$ are the roots of the Legendre polynomial $P_{48}(x)$.}
\label{tab:momentum_table}
\end{table}

The spacetime distributions for a longitudinally boost-invariant (2+1)--d hypersurface are $\eta_s$--independent. In this case, we evaluate the particle yield per unit spacetime rapidity of each freezeout cell:
\be
\label{eqch6:dNdeta_per_cell}
  \frac{\Delta N_{n,i}}{\Delta\eta_s} = \int p_T \, dp_T \, d\phi_p \, dy_p \, p \cdot d^2\sigma_i \, f_n(x^\mu_{\perp i}, p_T,\phi_p,y_p{-}\eta_s)\,,
\ee
and bin the weights in the ($\tau,r$)--grid.\footnote{%
    For the momentum rapidity integral, we use the same Gaussian quadrature~(\ref{eqch5:integration_rescale}a,b) except the $y_p$--grid is centered around the spacetime rapidity $\eta_s = y_{p}^\text{CM}$.}
\section{Sampling particles from the Cooper--Frye Formula} 
\label{chap6S5}
 
The Cooper--Frye formula~\eqref{eqch5:CooperFrye} converts hydrodynamic output into hadron momentum spectra, but this conversion must be done before hydrodynamics breaks down. The final kinetic stage, in which the hadrons and hadronic resonances continue to rescatter (albeit at ever-decreasing rates) until they ultimately decouple and decay or free-stream to the detector, must be handled microscopically. This is usually done with the help of Monte Carlo implementations of kinetic equations in which real hadrons propagate on classical trajectories and rescatter stochastically. 

To initiate such a hadronic rescattering cascade requires the conversion of the hydrodynamic output on the switching surface into particles with positions and momenta, by interpreting the Cooper--Frye integrand as a probability density in phase-space and sampling it stochastically. Both theoretical arguments and model-to-data comparisons suggest that hydrodynamics is the more precise dynamical description of the fireball for temperatures above the pseudocritical temperature $T_c \approx 155$ MeV \cite{Borsanyi:2010cj, Borsanyi:2013bia, Bazavov:2014pvz}, whereas hadronic transport is a more reliable model below $T_c$ where the quark-gluon plasma liquid has fragmented into a gas of hadronic resonances \cite{Bernhard:2016tnd,Shen:2014lye}. In this section we describe the sampling mode of {\sc iS3D} which provides such a \textit{particlization scheme}.  

One must keep in mind, however, that the Cooper--Frye integrand in Eq.~\eqref{eqch5:CooperFrye} is not always positive-definite, and this must be fixed before it can be used as a probability density. There are two possible sources of negative contributions: first, the particlization surface typically contains regions with space-like normal vectors such that for certain ranges of momenta $p\cdot d^3\sigma < 0$, corresponding to particles being reabsorbed by the fireball. Second, when using a linearized form for the viscous correction to the distribution function $f_n$, the latter can turn negative at high momentum. To sample the Cooper--Frye integrand probabilistically, it must be rendered positive-definite with the following modification:
\be
\label{eqch6:sampledCFF}
  E_p \frac{dN_n}{d^3p} = \frac{1}{(2\pi\hbar)^3}\int_\Sigma p \cdot d^3\sigma \, f_n \, \Theta(p \cdot d^3\sigma) \, \Theta(f_n) \,.
\ee
Here $\Theta$ is the Heaviside step function. As a consequence, the sampled particle spectra will deviate from the original Cooper--Frye formula (e.g. energy-momentum and charge conservation are slightly violated). As will be discussed below, the deviations from the original spectra are typically small, except for very soft or hard momentum particles.

In the following sections, we discuss the methodology for sampling particles from Eq.~\eqref{eqch6:sampledCFF}, which is carried out in two steps: (i) For each freezeout cell we sample the number of hadrons emitted and their type. (ii) For each hadron produced, we then sample its momentum ~\cite{Pratt:2010jt, Bernhard:2018hnz, Shen:2014vra, Pang:2018zzo}.

\subsection{Sampling the number of hadrons from each freezeout cell}
\label{chap6sec2b}

Interpreting the number of particles given by the hydrodynamic output as the mean of a Poisson distribution, one can sample the number of hadrons $N$ from
\be
\label{eqch6:poisson}
  P(N) = \frac{\exp(-\Delta N_\text{h})(\Delta N_\text{h})^N}
              {N!} \,,
\ee
where
\be
\label{eqch6:meanHadrons}
\Delta N_\text{h} = \sum_n \Delta N_n = \sum_n \int_p \, p \cdot d^3\sigma \, f_n \, \Theta(p \cdot d^3\sigma) \, \Theta(f_n)
\ee
is the mean number of hadrons emitted from the selected freezeout cell. After sampling the total number of hadrons, we sample their species from the discrete probability distribution 
\be
\label{eqch6:discrete}
D_n = \frac{\Delta N_n}{\Delta N_\text{h}} \,.
\ee
The C++ 11 library contains Poisson and discrete distribution classes which we use to sample Eqs.~\eqref{eqch6:poisson} and~\eqref{eqch6:discrete}, respectively. 

The prerequisite for sampling the numbers of hadrons is computing the mean number of each hadron species. However, enforcing the outflow of particles ($p \cdot d^3\sigma > 0$) in Eq.~\eqref{eqch6:meanHadrons} presents a complication in evaluating the momentum-space integral. If one ignores the effect of the function $\Theta(p \cdot d^3\sigma)$, then $\Delta N_n$ in Eq.~\eqref{eqch6:meanHadrons} reduces to (in the absence of diffusion current)
\be
\label{eqch6:meanHadronsNoOutflow}
\Delta N_n \approx (u \cdot d^3\sigma) \int_p (u \cdot p) \, f_n \,\Theta(f_n) = (u \cdot d^3\sigma) \, n_n \,,
\ee
which is simply the particle number density $n_n$ multiplied by the time-like hypersurface volume element $u \cdot d^3\sigma$ in the local rest frame. For time-like and light-like freezeout cells the dot product $p \cdot d^3\sigma$ is always positive and Eq.~\eqref{eqch6:meanHadrons} reduces to Eq.~\eqref{eqch6:meanHadronsNoOutflow}. For space-like cells, there are momentum space regions with $p \cdot d^3\sigma < 0$ that are cut out by the function $\Theta(p \cdot d^3\sigma)$, increasing the particle yield. If a considerable fraction of the emission occurs from space-like domains on the hypersurface one must include the outflow effect on the mean hadron number.

Enforcing positivity of the distribution function adds another dimension of complexity to the evaluation of Eq.~\eqref{eqch6:meanHadrons}. If $f_n$ contains linear viscous corrections, then the function $\Theta(f_n)$ effectively regulates $\delta f_n$ such that 
\be
\label{eqch6:df_reg}
  \delta f_n \leftarrow \delta f_{n,\mathrm{reg}} = 
  \max\big({-}f_{\eq,n},\,\min(\delta f_n, \,f_{\eq,n})
      \big)\,.
\ee
Here, we place an additional bound such that $|\delta f_{n,\text{reg}}| \leq f_{\eq,n}$ even if $\delta f_n$ is positive.\footnote{%
    Although the upper bound is not required, it facilitates the calculation of the maximum hadron number \eqref{eqch6:meanHadronsMax}.} 
The primary culprit for causing negative $f_n$ is the linearized bulk viscous correction, causing it to be regulated to zero at high momentum. 

An exact evaluation of the mean hadron number \eqref{eqch6:meanHadrons} would require identifying the boundary along which one or the other $\Theta$ function vanishes, which is hard. We circumvent this problem with a stochastic trick \cite{Pratt:2010jt,Bernhard:2018hnz}, by making use of the inequality
\be
  \int_p p \cdot d^3\sigma \,\Theta(p \cdot d^3\sigma) \, 
  \left(f_{\eq,n} {+} \delta f_{n,\mathrm{reg}}\right) 
  \,\leq\, 2 |d^3\sigma| \int_p (u\cdot p) f_{\eq, n} \,,
\ee
where
\be
\label{maxvol}
  |d^3\sigma| = (u \cdot d^3\sigma) + \sqrt{(u\cdot d^3\sigma)^2 
                - d^3\sigma \cdot d^3\sigma}\,.
\ee
Thus, we can establish an upper limit for $\Delta N_n$ in Eq.~\eqref{eqch6:meanHadrons}: 
\be
\label{eqch6:meanHadronsMax}
  \Delta N_n \leq \Delta N_{n,\text{max}} 
  = 2\,|d^3\sigma|\,n_{\eq,n}\,.
\ee
We then sample additional particles by replacing $\Delta N_n$ in Eqs.~\eqref{eqch6:poisson} -- \eqref{eqch6:discrete} with $\Delta N_{n,\text{max}}$ \cite{Pratt:2010jt,Bernhard:2018hnz}. After sampling their type and momentum (as described in the following subsection), we keep these particles with probability
\be
\label{eqch6:keep_weight}
w_\mathrm{keep}(p) = w_{d\sigma} \times w_{\delta f} = \frac{p\cdot d^3\sigma\ \Theta(p\cdot d^3\sigma)}{(u \cdot p)|d^3\sigma|} \times \frac{1}{2}\left(1 + \frac{\delta f_{n,\text{reg}}}{f_{\eq,n}}\right) \,,
\ee
where $w_{d\sigma}$ and $w_{\delta f}$ are called the flux and viscous weights whose product satisfies $0 \leq w_{\text{keep}}\leq 1$. In this way the right fraction of particles is discarded to recover, after sampling many events, the correct mean hadron number
\be
\label{eqch6:meanHadrons_linear}
\Delta N_\text{h} = \sum_n \int_p \, p \cdot d^3\sigma \, \left(f_{\eq,n} {+} \delta f_{n,\text{reg}}\right) \, \Theta(p \cdot d^3\sigma)\,.
\ee

If $f_n$ is one of the modified equilibrium distributions (\ref{eqch5:feqmod_2}) or (\ref{eqch5:Jonah}), no regulation \eqref{eqch6:df_reg} of $f_n$ is needed, i.e. the viscous weight $w_{\delta f}$ is set to 1, and the maximum hadron number is
\be
\label{eqch6:meanHadronsMax_mod}
\Delta N_n \leq \Delta N_{n,\text{max}} = |d^3\sigma|\,n_{R,n}\,,
\ee
where the renormalized particle density is $n_{R,n} = n_n^{(1)}$ for the PTM distribution~\eqref{eqch5:feqmod_2} (see also Eq.~\eqref{eqch5:n_linear}) and $n_{R,n} = z_\Pi\,n_{\eq,n}$ for the PTB distribution~\eqref{eqch5:Jonah}. After sampling many events this procedure recovers the correct mean hadron number
\be
\Delta N_\text{h} = \sum_n \int_p \, p \cdot d^3\sigma \, f^\mathrm{mod}_{\eq,n} \, \Theta(p \cdot d^3\sigma)\,,
\ee
where the superscript `mod' stands generically for either PTB or PTM. For the PTMA distribution, Eq.~\eqref{eqch6:meanHadronsMax_mod} is replaced by
\be
\label{eqch6:meanHadronsMax_famod}
\Delta N_n \leq \Delta N_{n,\text{max}} = |d^3\sigma|\,n_{a,n}\,,
\ee
where $n_{a,n}$ is given by Eq.~\eqref{eqch5:aniso_density}.
\subsection{Sampling the particle momentum}
\label{chap6sec2c}

After sampling a hadron and its type from a freezeout cell, we sample its local-rest-frame momentum $\bp_{_\mathrm{LRF}}$ from the probability density function $Q_n(\bp) \, d^3p$, where $Q_n$ is either (suppressing the $x^\mu$ dependence)
\be
\label{eqch6:PDF}
  Q_n(\bp) = 
  \frac{2 |d^3\sigma| \, f_{\eq,n}(\bp)}
       {\Delta N_{n,\text{max}}}
\ee
for the linearized $\delta f_n$ corrections or 
\be
\label{eqch6:PDF_mod}
Q_n(\bp) = \frac{|d^3\sigma| \, f^\mathrm{mod}_{\eq,n}(\bp')}{\Delta N_{n,\text{max}}}
\ee
for the modified equilibrium distributions. We sample the momentum from Eqs.~\eqref{eqch6:PDF}, \eqref{eqch6:PDF_mod} using the acceptance-rejection (AR) method. Conceptually, the method involves drawing a momentum sample from a proposal distribution $R_n(\bm{s})$,
\be
Q_n(\bp) \, d^3p = C \times R_n(\bm{s}) \, d^3s \times w_n(\bp) \,,
\ee
where $C$ is a normalization constant and $\bp = {\bm M}({\bm s})$ is some coordinate transformation, and accepting the sampled momentum with probability $w_n(\bp)$. If the sample is rejected, the procedure is repeated until a sampled momentum assignment is accepted. Once the assigned momentum for this particle has been accepted, the weight $w_\mathrm{keep}(p)$ in Eq.~\eqref{eqch6:keep_weight} (which enforces that the momentum points outward and the viscous correction remains within the regulated range) can be calculated and used to decide whether to keep the particle or discard it. 

The proposal distribution $R_n({\bm s})$ must be chosen judiciously such that the associated weight satisfies the condition $0 \leq w_n(\bp) \leq 1$. To increase the acceptance rate the weight should also be as close to unity as possible.\footnote{%
    In our sampling routine the average acceptance rate for the momentum sampling loop is about 60\% for the modified equilibrium distributions and about half of that for the linearized $\delta f_n$ corrections because for the latter about twice as many particles than ultimately desired must be sampled to account for the factor $\frac{1}{2}$ in the viscous weight $w_{_{\delta f}}$ in Eq.~\eqref{eqch6:keep_weight}.}
In the following subsections, we describe the choice of $R_n({\bm s})$ and the associated weight for sampling the momenta of pions and other (heavier) hadrons from either Eq.~\eqref{eqch6:PDF} or \eqref{eqch6:PDF_mod}. 

\subsubsection{Pions with linear viscous corrections}

For pions with linear viscous corrections, it is efficient to sample the momentum from a massless Boltzmann distribution~\cite{Pratt:2014vja}
\be
\label{eqch6:massless}
R_n({\bm s}) \, d^3s = \exp\left(-p/T \right) \, p^2 \, dp \, d\cos\theta \, d\phi \,,
\ee
where we use the spherical coordinates ${\bm s} = (p, \cos\theta, \phi)$:
\bs
\label{eqch6:sphericaliS3D}
\beal
& p_x = p \sin\theta \cos\phi, \\
& p_y = p \sin\theta \sin\phi, \\
& p_z = p \cos\theta \,.
\end{align}
\es
The distribution~\eqref{eqch6:massless} can be easily sampled using Scott Pratt's trick, which employs an additional coordinate transformation
\bs
\label{eqch6:PrattLight}
\beal
& p = - T \, \ln(r_1 \, r_2 \, r_3) \,,\\
& \cos\theta = \frac{\ln(r_1/r_2)}{\ln(r_1\,r_2)} \,,\\
& \phi = 2 \pi \left(\frac{\ln(r_1\,r_2)}{\ln(r_1\,r_2\,r_3)}\right)^2 \,,
\end{align}
\es
where the random variables $(r_1,r_2,r_3) \in (0,1]$. The massless Boltzmann distribution~\eqref{eqch6:massless} can then be rewritten as (ignoring constant factors)
\be
R_n({\bm s}) \, d^3s = dr_1 dr_2 dr_3 \,.
\ee
As a result, one can simply sample $(r_1,r_2,r_3)$ uniformly and apply the transformations~\eqref{eqch6:PrattLight} and~\eqref{eqch6:sphericaliS3D} for the momentum components $\bm p$.

The weight factor for pions is 
\be
\label{eqch6:pion_weight}
w_n(\bm p) = \frac{\exp\left(p/T\right)}{\exp\left(E/T\right) - 1}\,,
\ee
where $E = \sqrt{p^2 + m_n^2}$. One finds that the thermal weight $0 \leq w_n(p) \leq 1$ for all values of $p$ when $m / T > 0.8554$. For the lightest pion mass, this corresponds to $T < T_\text{max} = 157.8$ MeV. This is acceptable because particlization typically occurs several MeV below the pseudocritical temperature $T_c = 155$ MeV.\footnote{For situations where the switching temperature $T_\text{sw} \geq T_\text{max}$, one can renormalize the thermal weight by its maximum value, which we solve for numerically.} 

\subsubsection{Heavy hadrons with linear viscous corrections}

For heavy hadrons, we sample the momentum from the Boltzmann distribution~\cite{Pang:2018zzo,Pratt:2014vja}
\be
\label{eqch6:classical}
R_n({\bm s}) \, d^3s = \exp\left(b_n \alpha_B{-}E/T\right) \, p^2 \, dp \, d\cos\theta \, d\phi \,.
\ee
We use an intermediate variable transformation $k = E-m_n$, with $k$ being the kinetic energy, to rewrite Eq.~\eqref{eqch6:classical} as (omitting constant factors)
\be
R_n({\bm s}) \, d^3s = \,\frac{p}{E} \, S_n({\bm k}) \, d^3k\,,
\ee
where ${\bm k} \equiv (k,\cos\theta,\phi)$ and
\be
\label{eqch6:sampleK}
S_n({\bm k}) \, d^3k = \, \exp\left(-k/T \right) \, \left(k^2 + 2km_n + m_n^2\right) \, dk \, d\cos\theta \, d\phi \,.
\ee
Thus, we can sample the kinetic energy and angles from the distribution $S_n({\bm k})$ (moving the factor $p/E$ over to the weight $w_n({\bm p})$) to compute the energy $E$, radial momentum $p = \sqrt{E^2{-}m_n^2}$, and momentum components $p_i$ with Eq.~\eqref{eqch6:sphericaliS3D}. We write Eq.~(\ref{eqch6:sampleK}) as the sum of three distributions
\begin{equation}
\label{Si}    
  S_n({\bm k}) \, d^3k = 
  \Bigl(S_1({\bm k}) + 2 m_n S_2({\bm k}) 
        + m_n^2 S_3({\bm k})\Bigr)\, d^3k\,,
\end{equation}
with
\bs
\label{eqch6:heavy3}
\beal
& S_1({\bm k}) \, d^3k = \exp\left(-k/T \right) \, k^2 \, dk \, d\cos\theta \, d\phi\,, \\
& S_2({\bm k}) \, d^3k = \exp\left(-k/T \right) \, k \, dk \, d\cos\theta \, d\phi\,, \\
& S_3({\bm k}) \, d^3k = \exp\left(-k/T \right) \, dk \, d\cos\theta \, d\phi \,.
\end{align}
\es
The first distribution (\ref{eqch6:heavy3}a) is identical to the massless Boltzmann distribution~\eqref{eqch6:massless} after replacing the variable $p$ with $k$, so we can sample it with the same technique. For the second distribution (\ref{eqch6:heavy3}b), we use the coordinates
\beal
\label{eqch6:PrattSemi}
 k = - T \, \ln(r_1 \, r_2), \quad
 \phi = \frac{2 \pi \, \ln(r_1)}{\ln(r_1\,r_2)} \,,
\end{align}
where $(r_1, r_2) \in (0,1]$. One can check that $S_{2}({\bm k}) \, d^3k \sim dr_1 \, dr_2 \, d\cos\theta$, which means we can sample $(r_1, r_2, \cos\theta)$ uniformly. For the third distribution (\ref{eqch6:heavy3}c), we substitute 
\be
k = - T\,\ln(r_1) \,,
\ee
where $r_1 \in (0,1]$ and sample $(r_1, \cos\theta, \phi)$ uniformly since $S_{3}({\bm k}) \, d^3k \sim dr_1 \, \, d\cos\theta \, d\phi$.

Each time we draw a momentum sample, instead of drawing it from the total distribution \eqref{eqch6:sampleK} we draw it from one of these three simpler distributions, selected randomly with probabilities $(I_1,I_2,I_3)/I_\text{tot}$ where
\bs
\label{eqch6:integratedWeights}
\beal
& I_1 = \int_0^\infty dk \, k^2 \, \exp\left(-k/T \right) = 2 T^3 \,,\\
& I_2 = 2 m_n \int_0^\infty dk \, k \, \exp\left(-k/T \right) = 2 m_n T^2 \,,\\
& I_3 = m_n^2 \int_0^\infty dk \exp\left(-k/T \right) = m^2_n T
\end{align}
\es
and $I_\text{tot} = I_1 + I_2 + I_3$ \cite{Pang:2018zzo, Pratt:2014vja}. After many draws this ensures that, on average, the momenta have been drawn from the desired distribution \eqref{eqch6:sampleK}.

The weight for heavy hadrons is then chosen as
\be
w_n(\bm p) = \frac{p}{E} \, \frac{\exp\left(E/T - b_n \alpha_B\right)}{\exp\left(E/T - b_n \alpha_B\right) + \Theta_n}\,.
\ee
For baryons ($\Theta_n = 1$) this thermal weight satisfies $0 \leq w_n(p) \leq 1$ for all values of $p$. For mesons ($\Theta_n =-1$) the weight condition holds for $m/T > 1.008$.\footnote{%
    For pions and heavy mesons, the thermal weight $w_n(p)$ can exceed unity for situations where the electric and net-strangeness chemical potentials are nonzero. For this reason, we leave the generalization to sampling particles with nonzero $(\mu_Q, \mu_S)$ to future work.}
Since even the mass of the lightest of them, the kaon, is three times larger than the typical switching temperature, the weight condition $0 \leq w_n(p) \leq 1$ is satisfied for all heavy hadrons. 

\subsubsection{Modified equilibrium distribution}

To sample momenta from the PTM or PTB modified equilibrium distributions, we use the momentum transformation \eqref{eqch5:general_rescaling} or \eqref{eqch5:lambda} to rewrite Eq.~\eqref{eqch6:PDF_mod} as (ignoring constant factors)
\be
\label{eqch6:modified_PDF_2}
  Q_n(\bp) \, d^3p = 
  \left[ 1-\frac{q_i A^{-1}_{ij} p'_j}
                {\sqrt{{p'}^2 {+} m_n^2}}
  \right]\,
  f_{\eq,n}^{\text{mod}}(\bm{p'}) \, d^3p^\prime \,,
\ee
where $q_i$ is given by~(\ref{eqch5:rescale_matrix}b) ($q_i = 0$ for the PTB distribution). One can then sample the modified momentum components $\bm{p'}$ from $f_{\eq,n}^{\text{mod}}(\bm{p'})$ and apply the viscous transformation \eqref{eqch5:general_rescaling} or \eqref{eqch5:lambda} for $\bm{p}$ \cite{Pratt:2010jt,Bernhard:2018hnz}. For pions, we sample $\bm{p'}$ from the modified massless Boltzmann distribution
\be
\label{eqch6:massless_modified}
R_n(\bm s) \, d^3s = \exp\left(-p'/T' \right) \, (p^\prime)^2 \, dp' \, d\cos\theta' \, d\phi' \,,
\ee
where we use the \textit{modified} spherical coordinates ${\bm s} = (p^\prime, \cos\theta^\prime, \phi^\prime)$:
\bs
\label{eqch6:spherical_prime}
\beal
& p'_x = p' \sin\theta' \cos\phi' \,, \\
& p'_y = p' \sin\theta' \sin\phi' \,, \\
& p'_z = p' \cos\theta' \,.
\end{align}
\es
It is straightforward to sample ($p^\prime$, $\cos\theta^\prime$, $\phi^\prime$) with Scott Pratt's trick. The modified weight for pions is
\be
   w_n(\bm p) = w_{q} \times
   \frac{\exp(p'/T')}
        {\exp(E'/T') - 1}\,,
\ee
where
\be
\label{eqch6:diffusion_weight}
  w_{q} = \frac{1 - q_i A^{-1}_{ij} p'_j/E'}
               {1 + \sqrt{A^{-1}_{rs} A^{-1}_{rs}\, {\bm q}^2}}
\ee
is the baryon diffusion weight and $E' = \sqrt{{p'}^2 {+} m_n^2}$.\footnote{%
    Here we assume that the numerator of the diffusion weight \eqref{eqch6:diffusion_weight} is positive since we require a positive Jacobian determinant \eqref{eqch5:Jacobian}. Future work will address whether the PTM distribution may break down for large baryon diffusion currents.}
It is important to note that the modified temperature $T'$ in the PTM distribution increases with negative bulk pressure, which could make the pion effectively too light (i.e. lead to $m/T' < 0.8554$). If the bulk pressure is too large, the thermal weight can be renormalized by its maximum value (as long as it is finite) to ensure that it stays below unity.

For heavy hadrons, we sample $\bm{p'}$ from the modified Boltzmann distribution
\be
   R_n({\bm s}) \, d^3s = \exp
   \left( b_n \alpha'_B - E'/T'\right) \, 
   {p'}^2 dp' \, d\cos\theta'\, d\phi' \,,
\ee
where (in the PTM distribution) $\alpha'_B$ is the modified baryon chemical potential. After substituting $k' = E'-m_n$, we have
\be
\begin{split}
   R_n({\bm s}) \, d^3s 
   = & \, \frac{p'}{E'} \, S_n({\bm k'}) \, d^3k' \\
   = & \, \frac{p'}{E'} \, \exp(-k'/T') \, 
     \bigl({k'}^2 + 2k' m_n + m_n^2\bigr) \, dk^\prime \, d\cos\theta^\prime \, d\phi^\prime \,.
\end{split}
\ee
The procedure for sampling ${\bm k'} = (k',\, \cos\theta',\, \phi')$ is analogous to sampling $\bm{k}$ from Eq.~\eqref{eqch6:sampleK}. The modified weight for heavy hadrons is then
\be
  w_n(\bm p) = w_q \times \frac{p'}{E'} \, 
  \frac{\exp(E'/T' - b_n \alpha'_B) }
       {\exp(E'/T' - b_n \alpha'_B) + \Theta_n} \,.
\ee

The procedure for sampling the LRF momentum from the PTMA distribution is the identical to that for the PTM distribution after replacing the effective temperature $T^\prime$ by $\Lambda$ and the momentum transformation by $B_{ij}$ in Eq.~\eqref{eqch5:Bij} (the effective baryon chemical potential and diffusion vector are set to $\alpha_B^\prime = 0 = q_i$).
%
\begin{figure}[t!]
\begin{center}
  \includegraphics[width=0.8\linewidth]{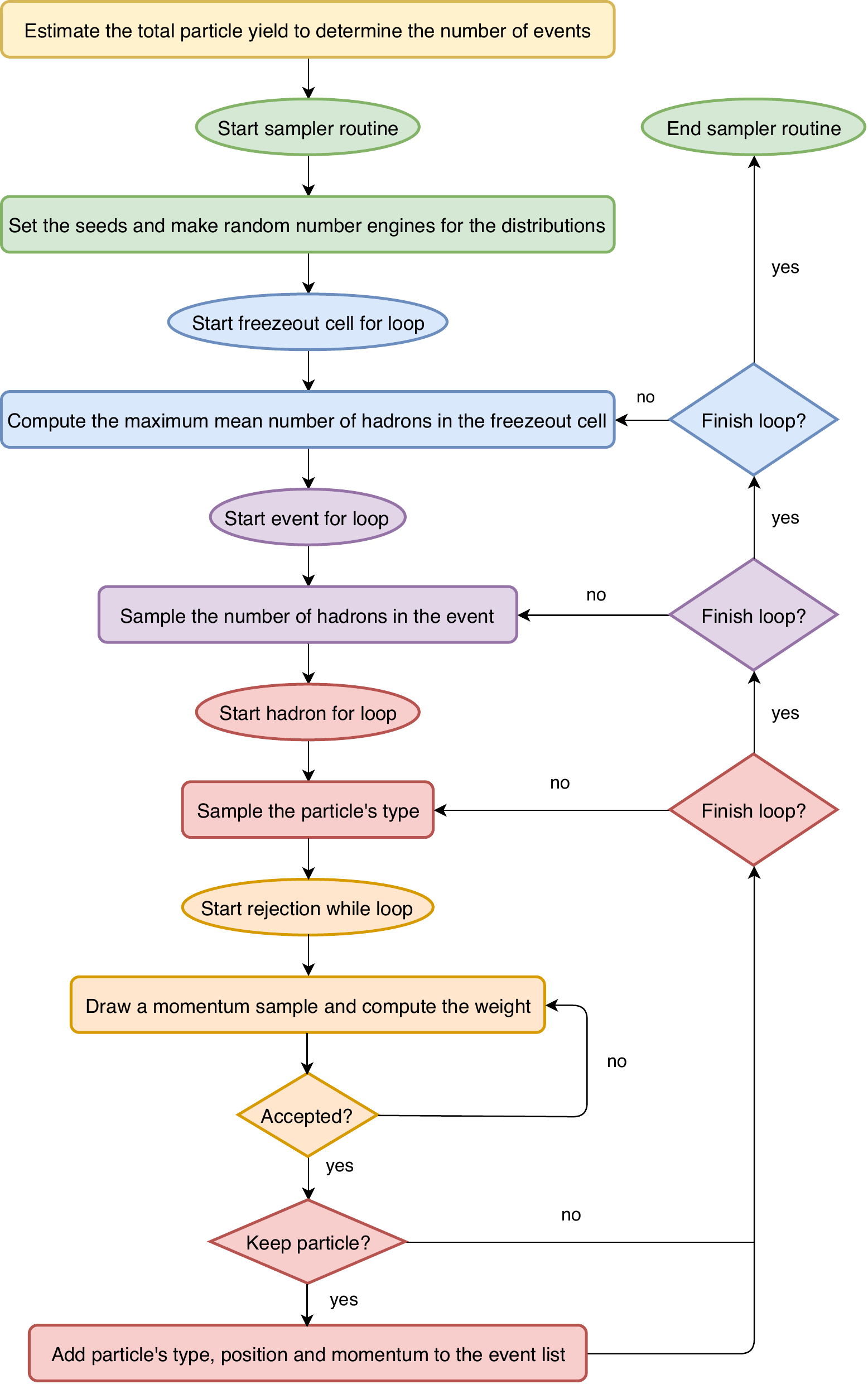}
\end{center}
  \caption{Program flow chart of the particle sampler routine in {\sc iS3D}.
  \label{sampler_flowchart}
}
\end{figure}

\subsection{Program flow chart}
\label{chap6S5.3}

Figure~\ref{sampler_flowchart} shows the program flow chart of the particle sampling routine in iS3D. Here we summarize the steps of the sampling procedure: 
\begin{enumerate}
    \item Before calling the sampler routine, we estimate the total particle yield per collision event (without including the effects of outflow or regulating $\delta f_n$)
    \be
    N_\text{yield} \approx \sum_n \int_\Sigma \int_p  (u \cdot d^3\sigma) (u\cdot p) \, \left(f_{\eq,n} {+} \delta f_n\right)
    \ee
    to determine the number of sampled events $N_\text{event} = N_\text{sampled} / N_\text{yield}$ needed to accumulate the desired statistics of approximately $N_\text{sampled}$ particles from the switching hypersurface ($N_\text{sampled}$ is a user input parameter).

    \item We call the sampler routine, initializing the seeds and random number engines for each of the three distributions $P(N)$, $D_n$ and $Q_n({\bm p})$ (see Secs.~\ref{chap6sec2b} --~\ref{chap6sec2c}).

    \item We loop over the freezeout cells. For each freezeout cell, we evaluate the position $x^\mu$, surface volume element $d^3 \sigma_\mu$, hydrodynamic quantities ($u^\mu$, $T$, $\mathcal{E}$, $\mathcal{P}_\text{eq}$, $\pi^{\mu\nu}$, $\Pi$), and $\delta f_n$ coefficients; we skip over freezeout cells with negative time-like volumes (i.e. $u \cdot d^3\sigma < 0$). Then, we construct the hadron number and hadron species distributions $P(N)$ and $D_n$ by computing the maximum number of each hadron species emitted from the freezeout cell \eqref{eqch6:meanHadronsMax} or \eqref{eqch6:meanHadronsMax_mod}.

    \item For a given freezeout cell, we loop over the events. For each event, we sample the number of hadrons from the distribution $P(N)$. It is efficient to nest the event for-loop in this way because we only need to access the freezeout cell information and compute the max hadron number once. 

    \item For a given event, we loop over the sampled hadrons. For each hadron, we sample its species from the distribution $D_n$. Next, we sample the particle's LRF momentum $\bp_{_\mathrm{LRF}}=(p_x,p_y,p_z)$ from the distribution $Q_n(\bp)$ using the AR method. We keep the particle with probability $w_\text{keep}$ and, if accepted,  we compute its lab frame momentum from
    \be
    \label{lab_mom}
      p^\mu = E u^\mu + p_x X^\mu + p_y Y^\mu + p_z Z^\mu 
    \ee
    and also set the particle's lab frame position to that of the freezeout cell. Finally, we append the particle to the list of particles sampled in the event.

    \item After the sampler routine is finished, we write the particle data list of each event to file. This file follows the OSCAR format \cite{OSCAR} to allow for integration with hadronic afterburners such as {\sc URQMD} and {\sc SMASH} ~\cite{Bass:1998ca,Weil:2016zrk}.
\end{enumerate}

For the special case of a (2+1)--dimensional hydrodynamic switching surface with longitudinal boost-invariance (as used in the performance tests presented here) the sampler routine is modified in two ways \cite{Bernhard:2018hnz}: first, we note that for boost-invariant switching surfaces, the surface finder CORNELIUS \cite{Huovinen:2012is} assumes by default a longitudinal extension of one unit of spacetime rapidity $\Delta\eta_s=1$. When computing the mean number of hadrons according to Eq.~(\ref{eqch6:meanHadrons}) we multiply the mid-rapidity surface volume element $d^2\sigma_i$ by a factor $2y_{p,\mathrm{max}}$ to obtain $\Delta\eta_s=2y_{p,\mathrm{max}}$ (the rapidity cutoff $y_{p,\mathrm{max}}$ is a parameter set by the user). Second, after sampling the LRF momentum and accepting the particle, we compute its lab frame momentum from Eq.~(\ref{lab_mom}). Expressing the Milne lab frame momentum components as $p^\tau=m_T\cosh(y_p{-}\eta_s)$ and $p^\eta = (m_T/\tau) \sinh(y_p{-}\eta_s)$, the particle's momentum rapidity $y_p$ relative to its spacetime rapidity $\eta_s$ is $y_p-\eta_s =  \tanh^{-1}\!\left(\tau p^\eta / p^\tau\right)$. We then generate a boost-invariant (i.e. constant) distribution $dN_n / dy_p$ over the range $y_p\in[-y_{p,\mathrm{max}}, y_{p,\mathrm{max}}]$  by sampling $y_p$ uniformly within the interval $[-y_{p,\text{max}}, y_{p,\text{max}}]$. This also yields a spacetime rapidity distribution $dN_n / d\eta_s$ after assigning the particle the spacetime rapidity
\be
\label{kinematic}
  \eta_s =  y_p - \tanh^{-1}\!\left(\frac{\tau p^\eta}{p^\tau}\right)\,.
\ee
The resulting pair $(y_p,\eta_s)$ is attached to the accepted particle before it is written to the sampled particle list for the event. For sufficiently large $y_{p,\text{max}}$, the uniform sampling of $y_p$, followed by the kinematic constraint (\ref{kinematic}), ensures after event-averaging a constant (i.e. perfectly boost-invariant) and correctly normalized mean yield $dN_n / dy_p$ in the range $y_p\in[-y_{p,\text{max}}, y_{p,\text{max}}]$, combined with a spacetime rapidity distribution $dN_n / d\eta_s$ that is approximately constant except for edge effects localized near $\eta_s=\pm y_{p,\mathrm{max}}$. 

\section{Particle sampler performance}
\label{chap6S6}

In this section we test the performance of our particle sampler by comparing the event-averaged particle spacetime distributions and momentum spectra to the positive-definite Cooper--Frye formula \eqref{eqch6:sampledCFF} for the central and non-central Pb+Pb collision systems described in Sec.~\ref{chap6S3}. Our hadron resonance gas consists of the $N_R = 444$ hadron species that can be propagated in {\sc SMASH} \cite{Weil:2016zrk}. For each hypersurface from the (2+1)--d hydrodynamic model we multiply the volume by a factor 10 (by setting $y_{p,\text{max}}{\,=\,}5$) and sample a total of approximately $N_\text{sampled} \approx 10^{11}$ particles. This corresponds to sampling between five to ten million events, depending on the kind of hypersurface and choice for $\delta f_n$. When testing the sampler, we bin the particles in position and momentum grids during runtime to construct the sampled distributions, rather than accumulating all of the sampled particle data for later processing in a separate analysis. This avoids running into RAM limitations and file I/O bottlenecks that result from generating so many particles.
\begin{figure}[t]
\includegraphics[width=\textwidth]{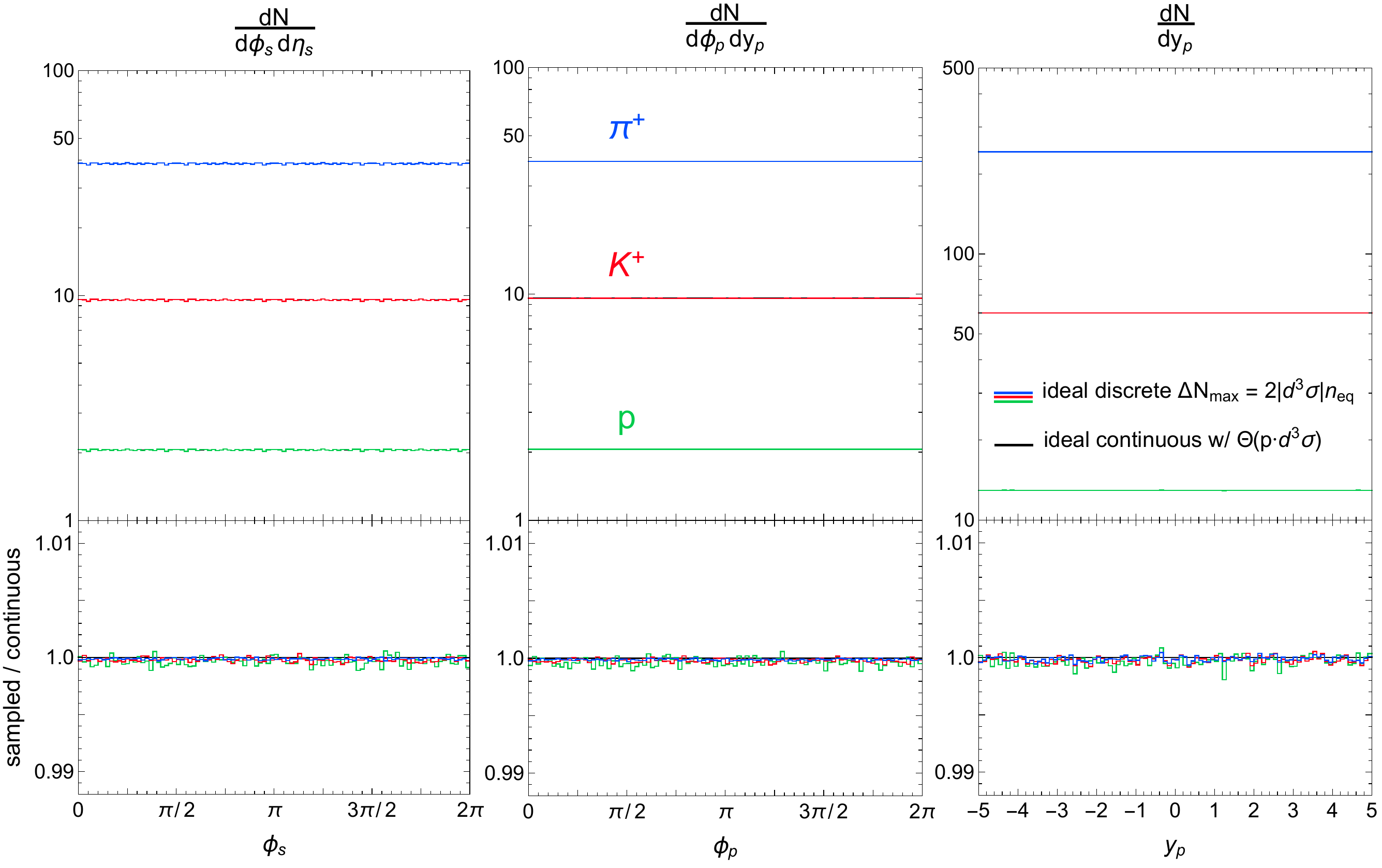}
\centering
\caption{The azimuthal and rapidity distributions of $(\pi^+,K^+,p)$ from the (2+1)--d central Pb+Pb collision with a $\zeta/\mathcal{S}$ peak temperature of $T_p = 180$ MeV. Here the $\delta f_n$ correction is set to zero (labeled \textit{ideal}). The bin widths used are $\Delta \phi_s = \Delta \phi_p = 0.02\, \pi$ and $\Delta y_p = 0.1$. Both the sampled (solid colored) and continuous (solid black) distributions include the outflow correction $\Theta(p\cdot d^3\sigma)$ to the Cooper--Frye formula. The bottom panels show the ratios between the sampled and continuous distributions. 
\label{Fideal_dN_dy}
}
\end{figure}
\subsection{Central Pb+Pb collision}
\label{chap6S6.1}

In the first test, we sample the Cooper--Frye formula for the central Pb+Pb collision with a $(\zeta / \mathcal{S})(T)$ that peaks at a temperature $T_p = 180$\,MeV. We construct the discrete transverse momentum spectra~\eqref{eqch6:central_pT_spectra} by binning the sampled particles in a uniform $p_T$--grid with width $\Delta p_T = 0.03$ GeV, averaging over the events and rapidity. For the sampled temporal and radial distributions (\ref{eqch6:time_formula}) --~(\ref{eqch6:radial_formula}), we bin the particles in the same $(\tau, r)$--grid as the one used in Sec.~\ref{chap6S4}. We have also verified that the event-averaged particle distributions are azimuthally symmetric in position and momentum space and are longitudinally boost-invariant, as shown in Figure~\ref{Fideal_dN_dy}.  
\begin{figure}[htbp]
\includegraphics[width=0.95\textwidth]{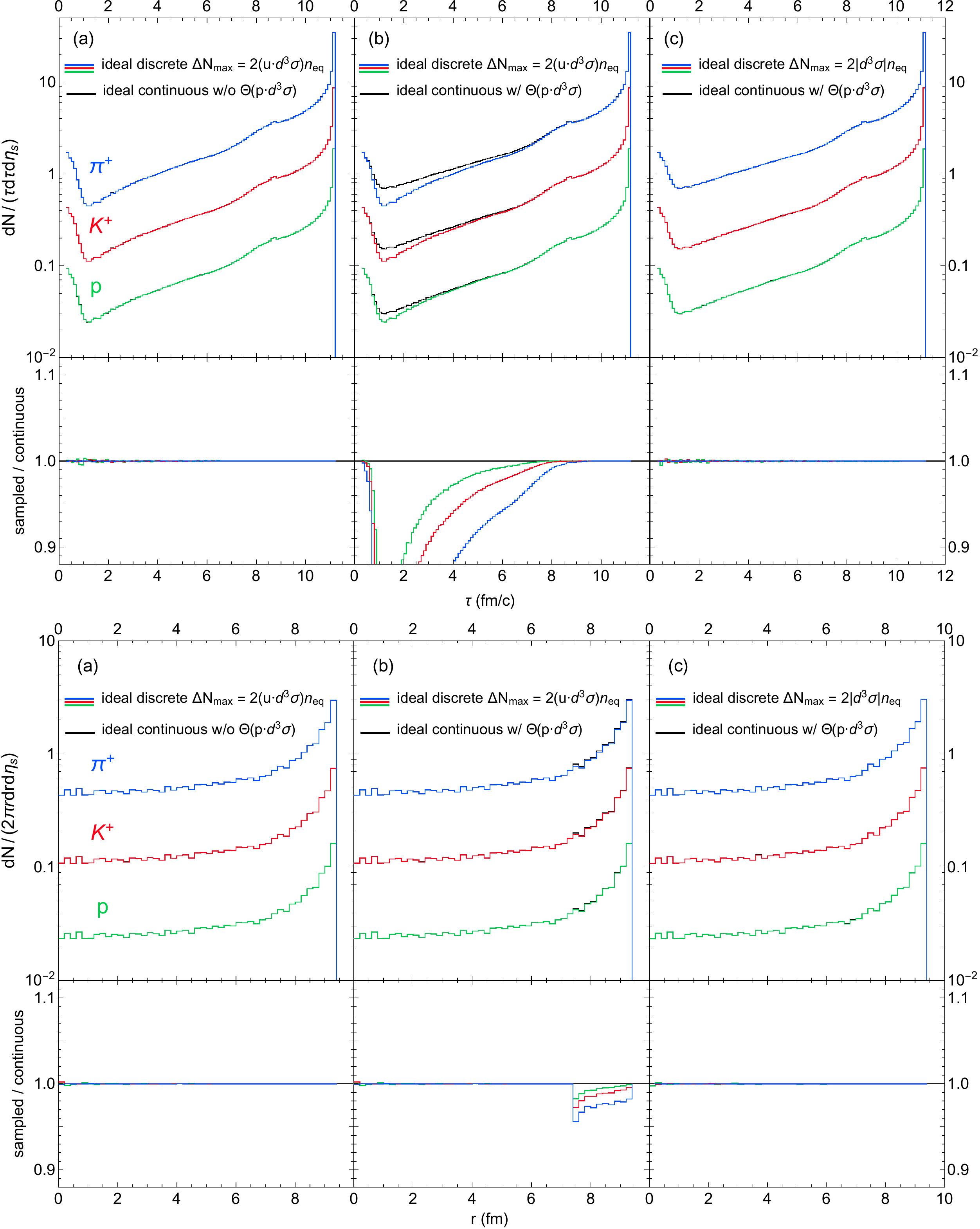}
\centering
\caption{The temporal (top panels) and radial distributions (bottom panels) of $(\pi^+,K^+,p)$ for the (2+1)--d central Pb+Pb collision with a $\zeta/\mathcal{S}$ peak temperature of $T_p = 180$\,MeV. Here the $\delta f_n$ correction is set to zero (\textit{ideal}). The sampled distributions (solid colored) are generated without (a,b) or with (c) the outflow correction to the mean hadron number. The continuous distributions (solid black) are computed without (a) or with (b,c) the $\Theta(p \cdot d^3\sigma)$ function. The bottom subpanels show the ratios between the sampled and continuous distributions for each of the comparisons. 
\label{Fidealspacetime}
}
\end{figure}
\begin{figure}[t]
\includegraphics[width=\textwidth]{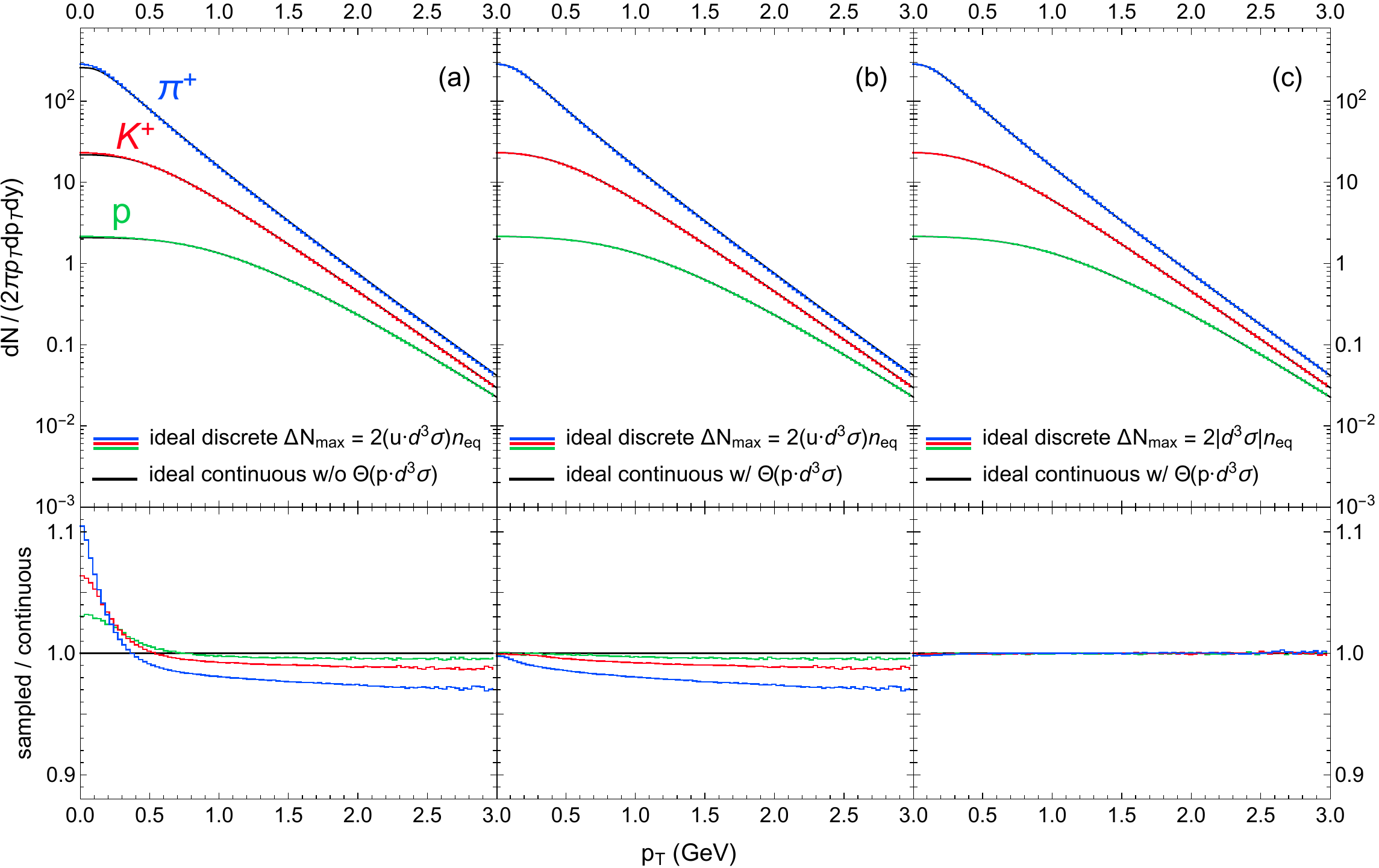}
\centering
\caption{The same comparisons as in Fig.~\ref{Fidealspacetime} but for the azimuthally-averaged transverse momentum spectra \eqref{eqch6:central_pT_spectra}.
\label{Fidealspectra}
}
\end{figure}

\subsubsection{The effect of particle outflow}
\label{chap6S6.1.1}
In this subsection we study the effects of the outflow correction, implemented by the function $\Theta(p \cdot d^3\sigma)$, on the spacetime distributions and momentum spectra. For simplicity, and without loss of insight, we set $\delta f_n = 0$ in this comparison.\footnote{%
    We label spacetime distributions and momentum spectra without any $\delta f_n$ correction as \textit{ideal}, but this does not imply $\eta/\mathcal{S}$ and $\zeta/\mathcal{S}$ are set to zero during the viscous hydrodynamic simulation. We only turn off the $\delta f_n$ correction on the switching surface.}
Figure~\ref{Fidealspacetime} shows the temporal and radial distributions of ($\pi^+, K^+, p$). In Fig.~\ref{Fidealspacetime}a,b we sample the particles without the outflow correction to the mean hadron number (i.e. $\Delta N_n = (u \cdot d^3\sigma)\, n_{\text{eq},n}$).\footnote{%
    If the outflow effect on the particle yield is not included, we replace the freezeout cell volume $|d^3\sigma|$ with $(u \cdot d^3\sigma)$ in Eqs.~\eqref{eqch6:meanHadronsMax} and \eqref{eqch6:meanHadronsMax_mod}. In addition, we move the flux weight $w_{d\sigma}$ from $w_\text{keep}(p)$ to the thermal weight $w_n(p)$.}
We then compare the resulting sampled spacetime distributions to the continuous ones computed from the Cooper--Frye formula with and without the $\Theta(p \cdot d^3\sigma)$ function. Clearly, the continuous method without the $\Theta(p \cdot d^3\sigma)$ function also yields $\Delta N_n = (u \cdot d^3\sigma)\, n_{\text{eq},n}$ for each freezeout cell. Thus, the sampled and continuous distributions are in very good agreement for the first case (Fig.~\ref{Fidealspacetime}a). In the second case (Fig.~\ref{Fidealspacetime}b), where this time we compute the Cooper--Frye formula with the $\Theta(p \cdot d^3\sigma)$ function, the continuous distribution is greater than the sampled one at early times and at large radii. This is because in these spacetime regions the hypersurface elements are space-like. The sampled and continuous distributions are only in agreement for the upper part of the hypersurface in Fig.~\ref{FHydro} ($\tau \gtrsim 9\,\text{fm}/c,\, r \lesssim 7\,\text{fm})$, where the hypersurface elements are time-like and $\Theta(p \cdot d^3\sigma) = 1$ has no effect. For the third case (Fig. \ref{Fidealspacetime}c) we sample particles from freezeout cells of maximum volume $|d^3\sigma|$ defined in Eq.~(\ref{maxvol}), which reproduces the outflow correction to the particle yield after keeping particles with probability $w_\text{keep}$. Correspondingly, the resulting sampled distribution is in excellent agreement with the continuous distribution computed with the $\Theta(p \cdot d^3\sigma)$ function.

Figure~\ref{Fidealspectra} plots the same comparisons for the azimuthally-averaged transverse momentum spectra. In Fig.~\ref{Fidealspectra}a both the sampled and continuous spectra ignore the outflow effect on the mean number of hadrons emitted from the hypersurface; thus, they have the same particle yields. However, sampling the momentum must still be done with the $\Theta(p \cdot d^3\sigma)$ function; this gives preference to the emission of softer particles over harder ones from the space-like regions of the hypersurface. As a result, the sampled spectra are softer than the continuous spectra computed without the $\Theta(p \cdot d^3\sigma)$ function; the discrepancy in the low $p_T$ region is as high as 10\% for pions. This implies that in this mode the particle sampler is not able to conserve energy and momentum, even if the sampled particle yield matches the one given by the original Cooper--Frye Formula. For the second case (Fig.~\ref{Fidealspectra}b) we compare the sampled spectra without the outflow effect on the particle yield to the continuous spectra computed with the $\Theta(p \cdot d^3\sigma)$ function. The sampled and continuous spectra now show better agreement in shape, but the sampled spectra still underestimate the continuous particle yield by a few percent. In the third case (Fig.~\ref{Fidealspectra}c) both the sampled and continuous spectra include the outflow correction to the particle yield. Now that the sampled and continuous methods are consistent with each other, we obtain nearly perfect agreement between the spectra. We conclude that in order to conduct high precision tests on the particle sampler one must include the outflow effect on the mean hadron number in Eq.~\eqref{eqch6:meanHadrons}.

\subsubsection{Small regulated viscous corrections}
\label{central_180}

\begin{figure}[htbp]
\includegraphics[width=\textwidth]{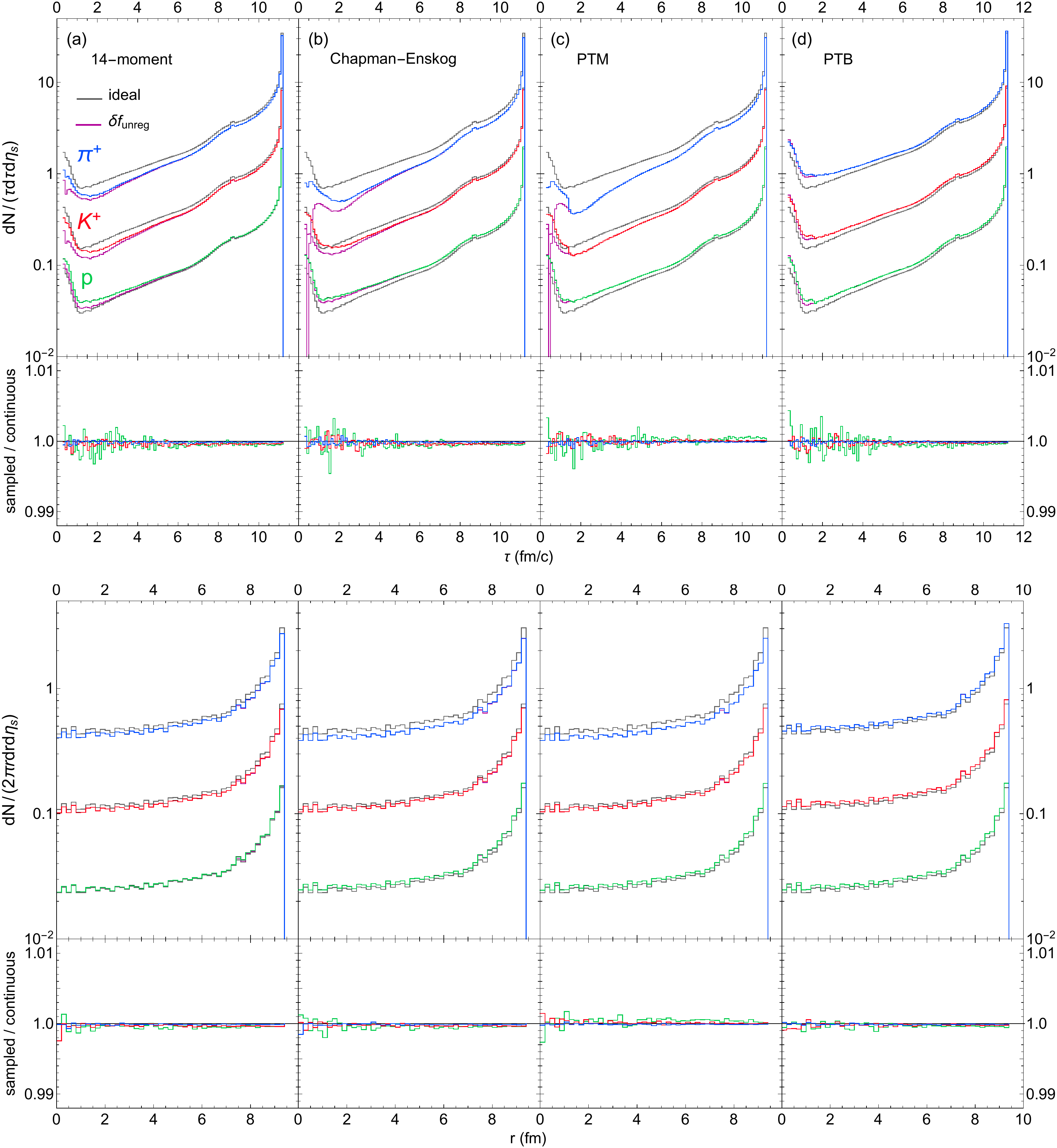}
\centering
\caption{The sampled (solid blue ($\pi^+$), red ($K^+$) and green ($p$)) and continuous (solid black) temporal and radial distributions with both regulated shear and bulk $\delta f_n$ corrections to the \textit{ideal} (solid gray) distributions for a (2+1)--d central Pb+Pb collision with a $\zeta/\mathcal{S}$ peak temperature of $T_p = 180$ MeV. The spacetime distributions with an unregulated $\delta f_n$ correction (solid purple) are also shown. In each panel the lower subpanel shows the ratio between the sampled and continuous distributions with regulated viscous corrections. 
\label{F_df_spacetime_180}
}
\end{figure}

We now add both the regulated shear and bulk viscous corrections, along with the outflow correction, to the sampled spacetime and momentum distributions. Figure~\ref{F_df_spacetime_180} shows the temporal and radial distributions of $(\pi^+, K^+, p)$ computed with each of the first four $\delta f_n$ corrections described in Chapter~\ref{chap5label}. Compared to the \textit{ideal} spacetime distributions with the outflow correction ($\delta f_n = 0$, gray lines), the particle production computed with the 14--moment approximation, RTA Chapman--Enskog expansion and PTM distribution decreases for pions and kaons while it increases for protons; this is mainly due to the form of their bulk viscous corrections. For the PTB distribution, the normalization factor $z_\Pi$ grows with negative bulk pressure, increasing the particle production of all species by the same factor.

We also compare the regulated spacetime distributions to the continuous ones computed with an unregulated $\delta f_n$ correction (purple lines) (both include the particle outflow effect). One sees that the regulated temporal distributions of the 14--moment approximation and RTA Chapman--Enskog expansion are greater than the unregulated ones at early times as a result of enforcing positivity. This is primarily due to the large hydrodynamic gradients of the fireball at early times, leading to the regulation of large bulk viscous corrections.\footnote{%
    The shear viscous correction $\delta f_{\pi,n}$ is also prone to regulation if the isotropic part of distribution function $f_{\eq,n} + \delta f_{\Pi,n}$ is close to the regulation bounds.} 
At later times, the bulk viscous pressure quickly dies down because in this example it peaks far away from the switching temperature $T_\text{sw} = 150$\,MeV (see Fig.~\ref{FHydro}a). As a result, the regulated temporal distributions start to converge to the unregulated ones. The regulated radial distributions are also slightly higher at around $r \sim 7.5 - 8$ fm; this is correlated to the regulation at early times. The modified equilibrium distributions are also regulated at early times ($\tau \lesssim$\,1.5 fm/$c$) since the viscous corrections are so large that we need to switch to a linearized $\delta f_n$ correction, which is regulated. Once we are able to transition to a modified equilibrium distribution, the regulation effects vanish. Compared to the linearized $\delta f_n$ corrections, the regulation has very little effect on the PTM and PTB spacetime distributions.

\begin{figure*}[t]
\includegraphics[width=\textwidth]{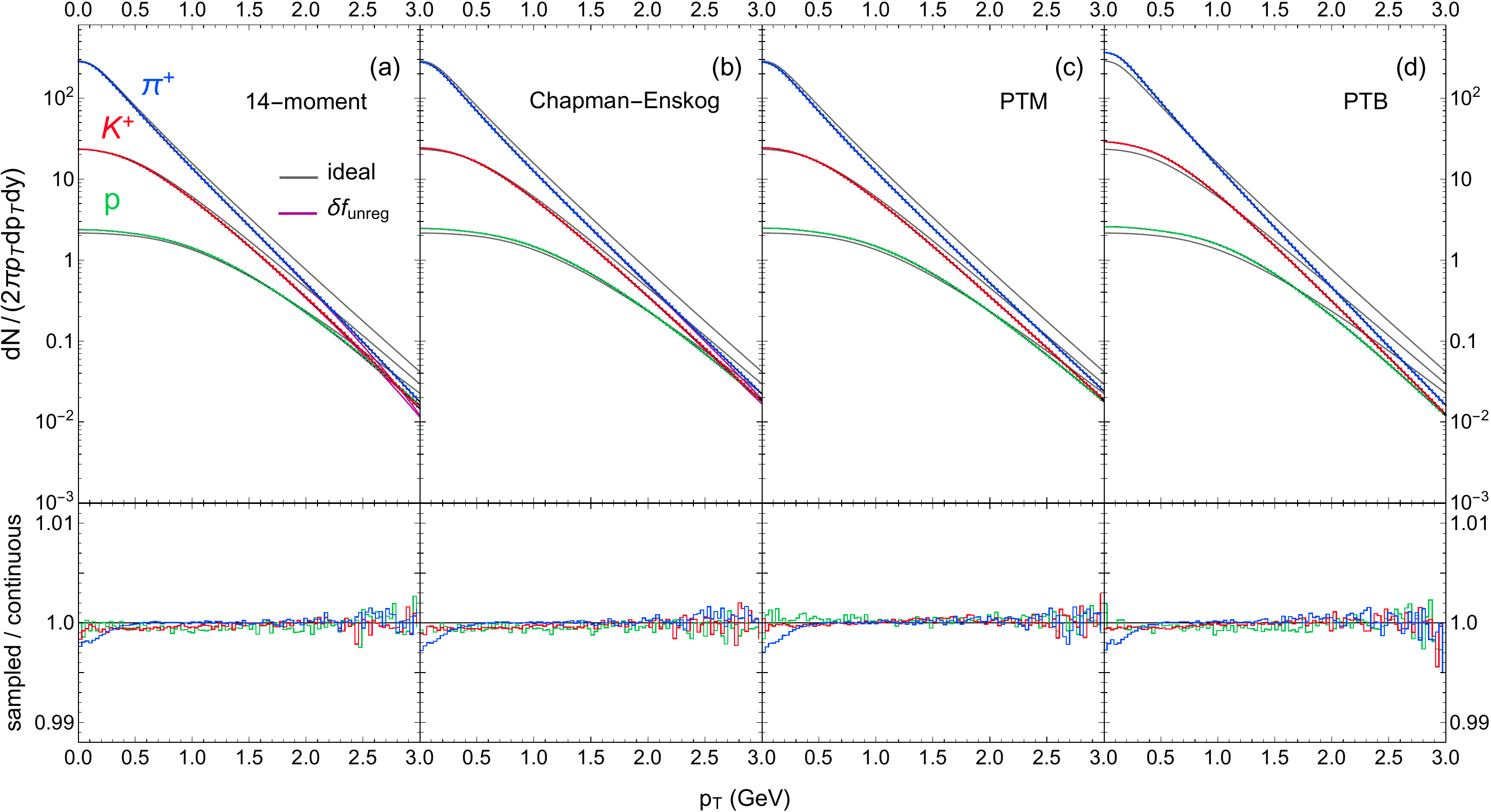}
\centering
\caption{The same comparisons as in Fig.~\ref{F_df_spacetime_180} but for the azimuthally-averaged transverse momentum spectra.
\label{F_df_spectra_180}
}
\end{figure*}

Finally, we compare the sampled spacetime distributions to the regulated distributions. One sees that there is excellent convergence to the continuous distributions (less than 0.5\% error for all values of $(\tau, r)$). One also notices that the sampled-to-continuous ratios fluctuate slightly above (or below) unity at late times and large radii. The reason for this slight discrepancy is unclear; nevertheless a 0.05 - 0.1\% error is more than satisfactory.

Figure~\ref{F_df_spectra_180} shows the corresponding transverse momentum spectra for each $\delta f_n$ method. Similar to the study from the previous chapter (see Fig.~\ref{dNdpT_small_bulk}), the bulk viscous pressure softens the slope of the ideal spectra while the shear stress stress counteracts this by flattening it. Here, the bulk viscous correction exceeds the shear correction, resulting in an overall softening of the $p_T$ spectra. One notices that the sampled spectra of the 14--moment approximation and RTA Chapman--Enskog expansion are regulated at large momentum ($p_T \gtrsim 2$\,GeV for pions and kaons); this is because the 
regulation limits the strength of the bulk viscous correction at high $p_T$. The regulation has virtually no effect on the modified spectra since the changes to the particle production in Fig.~(\ref{F_df_spacetime_180}c-d) were negligible. Once again, we find nearly perfect agreement between the sampled particle spectra and the regulated continuous spectra. The sampled pion spectra, however, slightly dips below the continuous spectra at low values of $p_T$. This effect is caused by finite bin widths. Since the pion spectra is strongly concave at low $p_T$, the average spectra over each of these $p_T$ bins is lower than the midpoint value from Jensen's inequality. One can eliminate this effect simply by using narrower $p_T$ bins in this region.

\subsubsection{Large regulated viscous corrections}
\label{chap6S6.1.3}

In this test we sample the hypersurface from the central Pb+Pb collision with $(\zeta / \mathcal{S})(T)$ peaking at a temperature of $T_p$ = 155 MeV (see Fig.~(\ref{FHydro}b)). In this scenario, the bulk viscous pressure peaks close to the switching temperature $T_\text{sw} = 150$ MeV, resulting in a much larger bulk viscous correction than in the previous test. The nonlinear shear-bulk coupling terms in the standard viscous hydrodynamic equations then also increases the magnitude of the shear stress correction. 
%
\begin{figure}[!t]
\includegraphics[width=\textwidth]{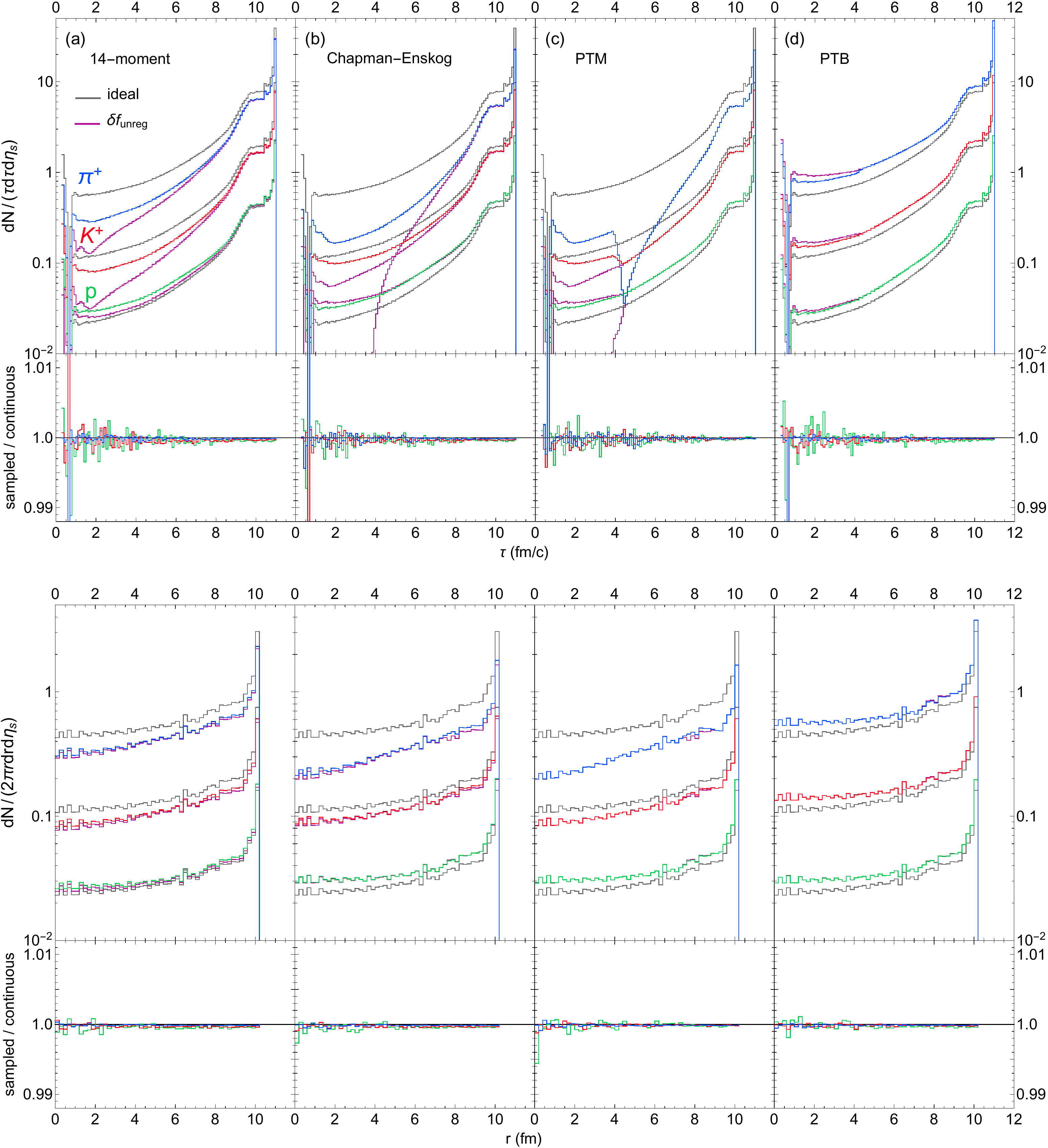}
\centering
\caption{The same as Fig.~\ref{F_df_spacetime_180} but for the central Pb+Pb collision with a peak temperature of $T_p = 155$ MeV for $\zeta/\mathcal{S}$.
\label{F_df_space_155}
}
\end{figure}
%
Figure~\ref{F_df_space_155} shows the resulting spacetime distributions. The effects of the linearized $\delta f_n$ regulation on the particle production rates are much more significant and they have a longer duration, especially for the pion and kaon production rates. For the unregulated RTA Chapman--Enskog $\delta f_n$ correction, the pion production rate is even negative until about $\tau \approx 4$ fm/$c$. Unlike the previous test, the regulation here increases the radial distributions of the linearized $\delta f_n$ corrections for all radii since the bulk viscous corrections are large throughout the entire hypersurface. These results are not so surprising since we expect the linearized $\delta f_n$ approaches to break down when the viscous pressures are large.

For the PTM and PTB distributions, the transition from a linearized $\delta f_n$ correction to a modified equilibrium distribution is delayed until about $\tau \approx 4.25 - 4.5$\,fm/$c$. As a consequence, their particle production rates are more strongly regulated than in the previous scenario; the radial distributions are also modified but only for $r \sim 7.5 - 9$\,fm. In the PTB case, the regulated spacetime distributions turn out to be lower than unregulated distributions.\footnote{%
    Although small, this effect is also observed for protons at $\tau < 4 - 5$\,fm/$c$ and $r \sim 7.5$\,fm for the RTA Chapman--Enskog and PTM distributions.} 
This is because the linearized $\delta f_n$ correction \eqref{eqch5:linear_Jonah} strongly softens the particle distributions at low momentum. Furthermore, the linearized normalization factor $1 + \delta z_\Pi$ increases the overall magnitude of the distribution function. As a result, the linearized PTB correction is regulated mostly by the upper bound $\delta f_n \leq f_{\eq,n}$ in Eq.~\eqref{eqch6:df_reg}. We recall that this choice for the upper bound, which was used to compute the maximum hadron number, is somewhat arbitrary.\footnote{%
    It is partially motivated by the assumption that hydrodynamics is valid on the hypersurface, which implies that $\delta f_n$ should be smaller in magnitude than $f_{\eq,n}$.} 
One could either raise this upper bound by hand or replace it with a more explicit calculation of $\delta f_{n,\text{max}}$ \cite{Shen:2014vra}. However, we will not pursue this here as the effects are only modest. As far as the sampled spacetime distributions are concerned, they are in excellent agreement with the regulated spacetime distributions -- the lower subpanels in Fig.~\ref{F_df_space_155} demonstrate that the particle sampler is sampling the particle yields correctly. The unphysical features in the temporal distributions caused by the sudden transition from modified equilibrium to linearized distributions for the PTM and PTB options wherever the former break down illustrate the conceptual issues arising from particlizing fluids with very large dissipative flows. Whenever such features are observed the user should assess the level of trust to be placed in the results based on physics considerations rather than the sampler algorithm itself.

\begin{figure}[t]
\includegraphics[width=\textwidth]{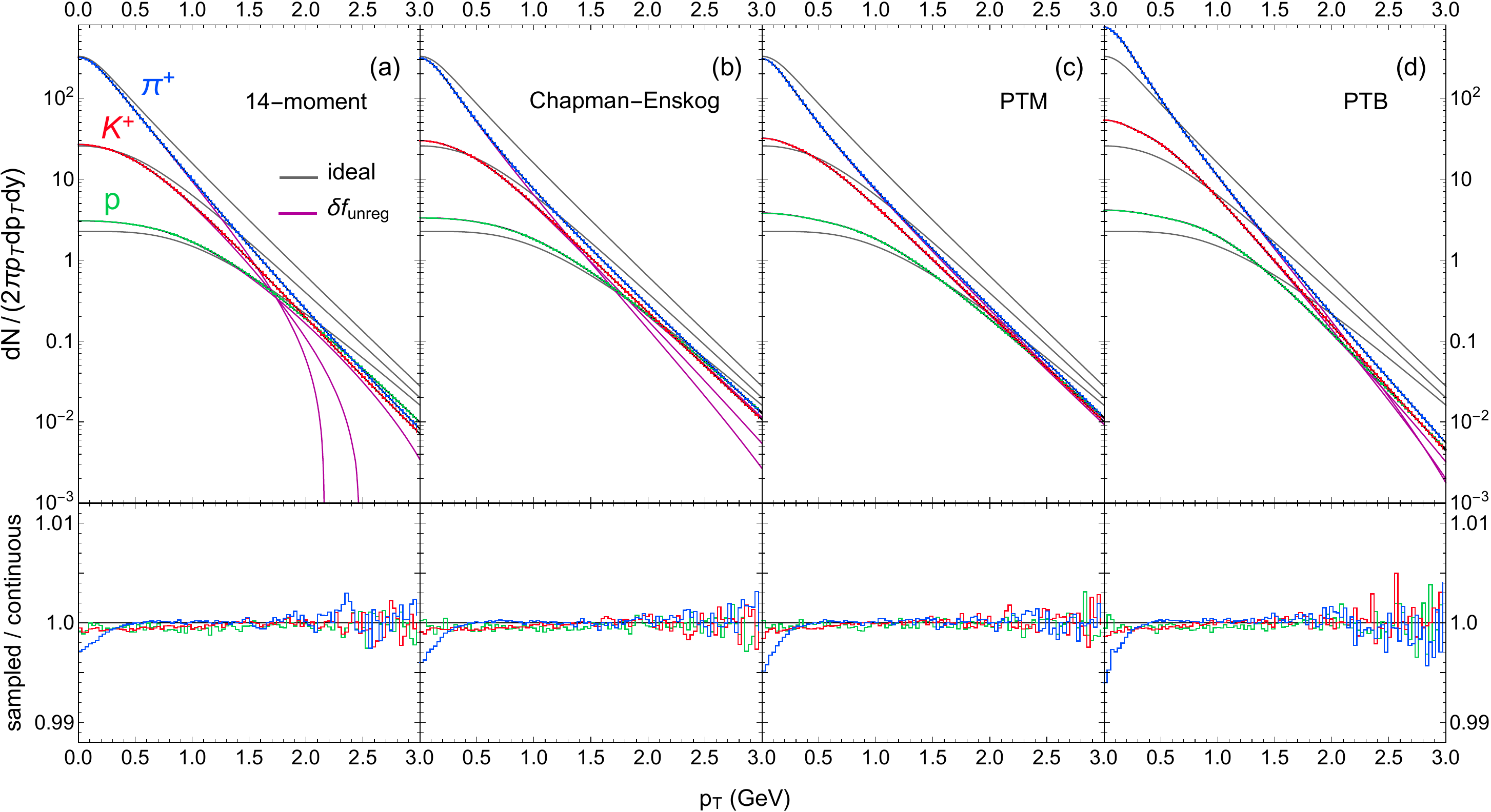}
\centering
\caption{The same comparisons made in Figure~\ref{F_df_space_155} but for the azimuthally-averaged transverse momentum spectra.
\label{F_df_spectra_155}
}
\end{figure}

Figure~\ref{F_df_spectra_155} shows the momentum spectra for this hydrodynamic event. One can see that the spectra of the 14--moment approximation are much more strongly regulated at high $p_T$ than for the event studied in the preceding subsection. The unregulated pion and kaon spectra quickly become negative due to the quadratic momentum dependence of the bulk viscous correction. The RTA Chapman--Enskog spectra are also strongly regulated, although to a somewhat lesser degree since the strength of the bulk viscous correction relative to the thermal distribution grows only linearly with momentum. Out of the four $\delta f_n$ corrections the PTM spectra are the least affected by regulation although there are some modest regulation effects at high $p_T$ resulting from a more frequent breakdown of the modified equilibrium distribution. The regulation has a greater effect on the PTB spectra than the PTM spectra at high $p_T$ since the linearized $\delta f_n$ correction in this case strongly softens the spectra.

Again, we see that the sampled particle spectra agree very well with the regulated continuous spectra. Compared to Fig.~\ref{F_df_spectra_180} in the preceding subsection, the sampling fluctuations are somewhat larger at high $p_T$ since the large viscous corrections lead to fewer particles produced in this region. 

\subsection{Anisotropic flow in non-central Pb+Pb collision}
\label{chap6S6.2}

Turning to non-central collisions, we test in this section the performance of the {\sc iS3D} sampler for the anisotropic flow coefficients $v_2(p_T)$ and $v_4(p_T)$. Again, we first consider the case of small viscous corrections on the particlization hypersurface (by tuning the peak of the specific bulk viscosity to a safe distance from this surface), followed by a case of large bulk viscous effects on the particlization surface. 

\subsubsection{Small regulated viscous corrections}
\label{chap6S6.2.1}

\begin{figure}[htbp]
\includegraphics[width=\textwidth]{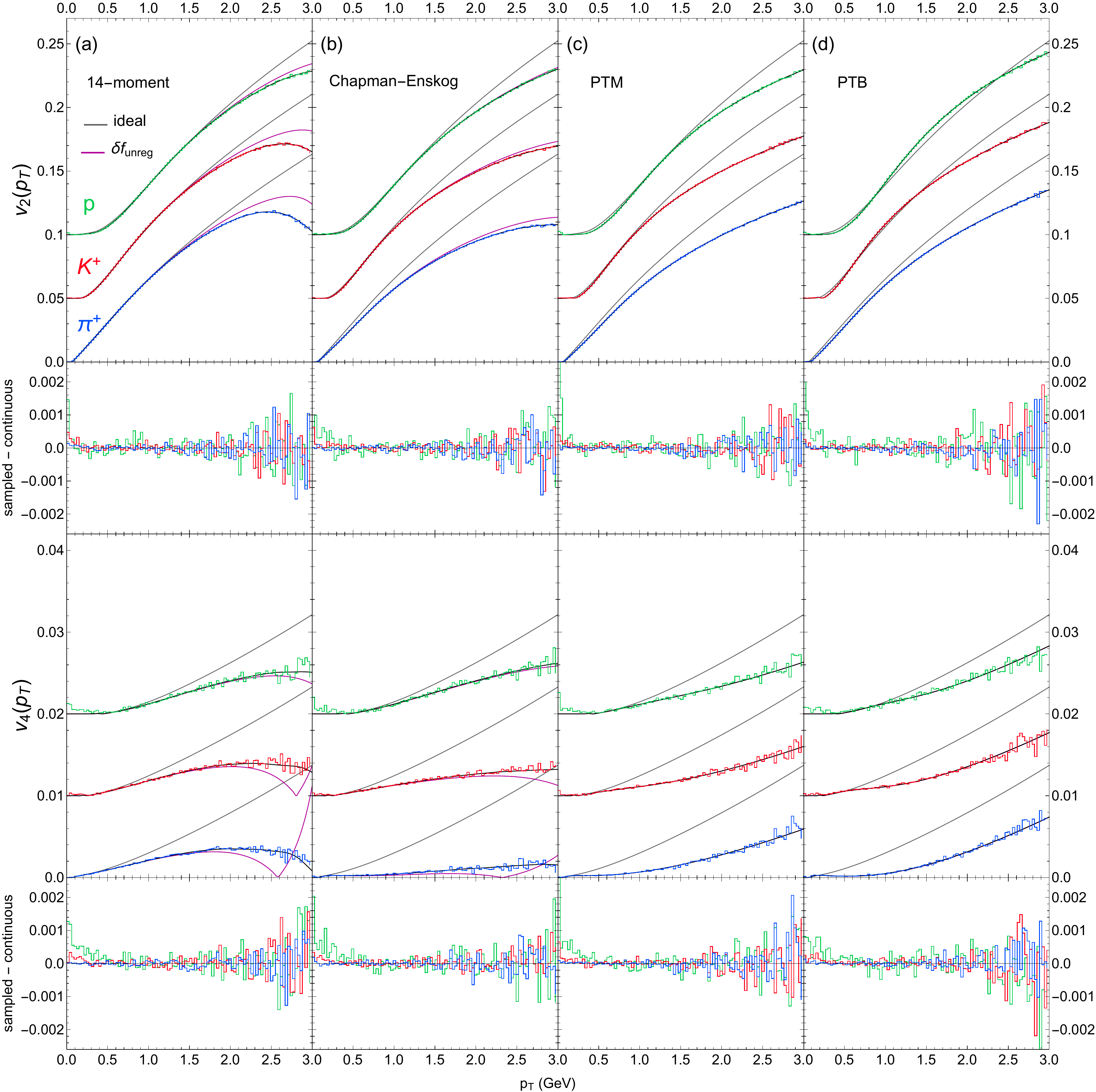}
\centering
\caption{The sampled (solid blue ($\pi^+$), red ($K^+$), and green ($p$)) histograms and continuous (solid black lines) differential $v_{2,4}(p_T)$ with regulated $\delta f_n$ corrections to the \textit{ideal} $v_{2,4}(p_T)$ (solid gray lines), for the (2+1)--d non-central Pb+Pb collision with a peak temperature of $T_p = 180$\,MeV for $\zeta/\mathcal{S}$. For better visibility the results for $v_2$ and $v_4$ for the three particle species are separated vertically by multiples of 0.05 and 0.01, respectively. The $v_{2,4}(p_T)$ with unregulated $\delta f_n$ corrections (solid purple lines) are also shown. The corresponding subpanels show the difference between the sampled and continuous $v_{2,4}(p_T)$ with regulated $\delta f_n$ corrections.
\label{F_vn_180}
}
\end{figure}

Sampling particles from the non-central Pb+Pb collision with a  peak temperature of $T_p = 180$\,MeV for $\zeta / \mathcal{S}$, we compare in Figure~\ref{F_vn_180} the sampled $v_2(p_T)$ and $v_4(p_T)$ to those obtained by integrating the Cooper--Frye formula with a positive-definite integrand, Eq.~(\ref{eqch6:sampledCFF}). For a given hadron species $n$ the event-averaged discrete anisotropic flow coefficients are computed using the formula
\be
  v_{k,n}(p_{T,j}) = 
   \frac{1}{\Delta N_n(p_{T,j})}
  \left|\sum\limits_{m=1}^{\Delta N_n(p_{T,j})} \exp\left(ik\phi_{p,m}\right)\right|\,,
\ee
where $\Delta N_n(p_{T,j})$ is the total number of particles of type $n$ in the transverse momentum bin $[p_{T,j} - \frac{1}{2}\Delta p_T, p_{T,j} + \frac{1}{2}\Delta p_T]$.\footnote{%
    For the purpose of this test we simply sum over all particles sampled from {\em all} events, with $\phi_p$ measured relative to the reaction plane defined by the beam and impact parameter directions, rather than following the experimental procedure of measuring for each event $\phi_p$ relative to the event plane, estimated from the $p_T$--integrated directed or elliptic flow of all charged hadrons.}
The colored histograms in Fig.~\ref{F_vn_180} show the resulting sampled $v_{2}(p_T)$ and $v_{4}(p_T)$ for pions, kaons and protons. The black solid lines show the exact numerical result from the continuous Cooper--Frye integral~\eqref{eqch6:vn} with both regulated viscous and particle outflow corrections whereas the gray solid lines (labeled \textit{ideal}) use the local-equilibrium distribution $f_{\eq,n}$ (i.e. $\delta f_n = 0$) and outflow correction.

From the previous chapter, we know that the suppression of $v_{2,4}(p_T)$ relative to the \textit{ideal} curves (i.e. $\delta f_n$ = 0) is mainly due to the shear stress correction $\delta f_{\pi,n}$~\cite{Schenke:2010rr, Shen:2014lye}. However, the bulk viscous correction $\delta f_{\Pi,n}$ is also important since it tends to counteract this effect by somewhat increasing $v_{2,4}(p_T)$ at higher $p_T$ \cite{Noronha-Hostler:2013gga, Noronha-Hostler:2015qmd}. In the 14--moment approximation and RTA Chapman--Enskog expansion, the regulation limits the bulk viscous corrections, causing the regulated elliptic flow $v_2(p_T)$ to fall at high $p_T$, but the regulation of the shear stress correction partially cancels this effect. In contrast, the regulation causes the quadrangular flow $v_4(p_T)$ to increase slightly at high $p_T$ since it is more sensitive to the regulation of the shear stress correction than the bulk viscous correction. The PTM and PTB predictions for $v_{2,4}(p_T)$ are not affected at all by the regulation since the transition to a modified equilibrium distribution occurs very early and the few particles produced before this time have not yet developed any anisotropic flow.

\begin{figure}[!t]
\includegraphics[width=\textwidth]{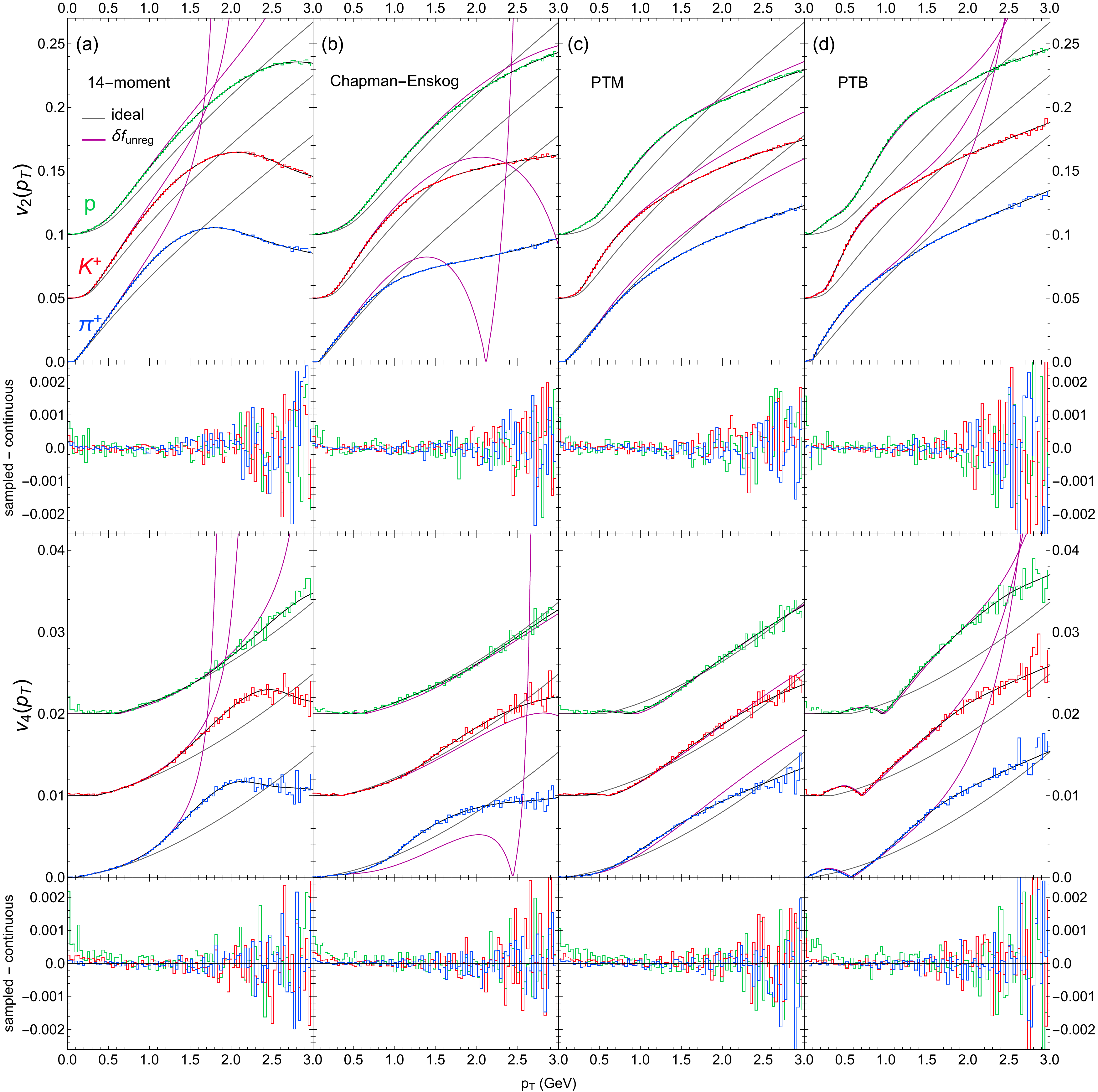}
\centering
\caption{Same as Fig.~\ref{F_vn_180} but for the non-central Pb+Pb collision with a peak temperature of $T_p = 155$\,MeV for $\zeta/\mathcal{S}$.
\label{F_vn_155}
}
\end{figure}
%

Overall, the sampled $v_{2,4}(p_T)$ are in very good agreement with the regulated continuous $v_{2,4}(p_T)$. The larger fluctuations of the sampled $v_{2,4}(p_T)$ in the low and high $p_T$ regions are of statistical nature since there are fewer particles in these $p_T$ bins. In particular, with the numbers of particles sampled for this test, the statistical fluctuations of the sampled quadrangular flow $v_4(p_T)$ are of similar magnitude as its mean value which here is about an order of magnitude smaller than the elliptic flow $v_2(p_T)$, due to the smooth geometric hydrodynamic profile studied in this chapter. This illustrates the increasing statistical demands associated with measurements of higher-order anisotropic flow coefficients from sampled particle distributions.
%
\subsubsection{Large regulated viscous corrections}
\label{chap6S6.2.2}
Figure~\ref{F_vn_155} shows what happens when the bulk viscous correction effects on the particlization surface are increased by moving the peak of $\zeta / \mathcal{S}$ to $T_p = 155$ MeV, i.e. closer to the particlization temperature of 150 MeV. Clearly the bulk viscous corrections now have a considerably stronger impact on the $v_{2,4}(p_T)$ than in the preceding subsection. Nevertheless, we again find excellent agreement between the sampled and continuous regulated elliptic and quadrangalar flows $v_{2,4}(p_T)$. At high $p_T$ the statistical fluctuations are considerably larger than in the preceding test, as a result of the stronger softening of the spectra by the larger bulk viscous pressure effects. Obviously, this could be addressed by sampling more events, at a numerical cost.
\section{Summary}
\label{S7}

In this chapter we have evaluated the performance of a module that can sample hadrons from a particlization hypersurface using one of several dissipative corrections $\delta f_n$ to the hadronic distribution function. The {\sc iS3D} code provides the user with extended capabilities for exploring the shear and bulk viscous effects on experimental observables in a realistic setting, where finite numbers of particles are Monte Carlo sampled from fluctuating heavy-ion collision events. In particular, we can better access the theoretical uncertainties associated with the particlization stage of heavy-ion collisions by considering multiple $\delta f_n$ candidates. A recent model-to-data comparison analysis that utilized our code to reduce model selection bias in the phenomenological constraints on the quark-gluon plasma's transport properties was reported in Refs.~\cite{Everett:2020xug,Everett:2020yty}.

The quantitative precision and reliability of our particle sampler has been demonstrated with a number of tests using two typical particlization hypersurfaces from (2+1)--dimensional viscous hydrodynamic simulations, one for central and the other for non-central collisions. Specifically, our sampling algorithm is able to capture the effects of particle outflow and viscous regulations on the particle spectra with a high degree of accuracy. The simplifying assumption of longitudinal boost-invariance used in this chapter is not a prerequisite for using the {\sc iS3D} sampler. The code can also sample (3+1)--dimensional surfaces, which take only a slightly longer time to generate the desired event or particle statistics. Although this last feature has not yet been rigorously tested, we will not need it for the model-to-data analysis in the next chapter. 

This chapter completes our discussion about the particlization phase. At this point, we have developed two new codes \cpuvah{} and {\sc iS3D}, which together model the pre-equilibrium, fluid dynamic and particlization stages of heavy-ion collisions. We will now deploy our modules to predict experimental observables in event-by-event simulations.

\chapter{Phenomenological applications of anisotropic fluid dynamics}
\label{chapter7label}
With the anisotropic fluid dynamic and particlization modules developed and tested, we are now ready to apply them to heavy-ion phenomenology. In this chapter, we present preliminary results on model-to-data comparisons of mid-rapidity hadronic observables in Pb+Pb collisions at LHC energies ($\sqrt{s_\text{NN}} = 2.76$ TeV). We run the codes {\sc VAH} and {\sc iS3D} within the JETSCAPE modular framework~\cite{Putschke:2019yrg} and run (2+1)--dimensional event-by-event simulations using the Maximum a Posteriori (MAP) model parameters from Refs.~\cite{Everett:2020xug,Everett:2020yty}. First we compare our new framework to the JETSCAPE SIMS hybrid model~\cite{Everett:2020xug}, which uses the same {\sc iS3D} module with the 14--moment approximation $\delta f_n = \delta f^{14}_{n}$ but simulates the pre-equilibrium and fluid dynamic stages with conformal free-streaming and standard viscous hydrodynamics instead of anisotropic hydrodynamics. Since the MAP parameters were optimized for the latter model via Bayesian inference, the initial predictions of our framework underestimate many of the experimental observables, such as the particle multiplicities and anisotropic flow coefficients. As a temporary resolution to this issue, we manually adjust two of the Bayesian model parameters that control the initial conditions. Afterwards, we find very good agreement with most of the experimental data. While this ad-hoc retuning cannot replace a full Bayesian recalibration of the framework, it provides valuable insights about the differences between the \cpuvah{} and SIMS models. Finally, we study the changes to the hadronic observables caused by switching out the 14--moment approximation with the PTMA modified anisotropic distribution in the {\sc iS3D} sampler. 

The material in this chapter has not yet been published. 

\section{Event-by-event simulations with {\sc VAH}}
\label{chap7S1}
In this section, we provide an overview of the components that make up our hybrid model. One difference between our model and the JETSCAPE SIMS package~\cite{Everett:2020xug} is the replacement of standard second-order viscous hydrodynamics by anisotropic hydrodynamics. Since we start anisotropic hydrodynamics at a very early longitudinal proper time, we also do not include the conformal free-streaming module, which models the initial stage with a longitudinally free-streaming gas of massless quarks and gluons prior to the fluid dynamic stage. Compared to fluid dynamics, longitudinal free-streaming initially causes the fireball to cool down at a much slower rate and build up more transverse flow. As a result, our predictions for the experimental observables are significantly altered if we use the same model parameters as those used by the SIMS model.
\subsection{Initial conditions}
We use the code \trento{}~\cite{Moreland:2014oya} to generate 8000 minimum bias Pb+Pb collision events with (2+1)--d fluctuating initial conditions at $\eta_s = 0$. We initialize the energy density profile at $\tau_0 = 0.05$ fm/$c$ as 
\be
\ene(\tau_0,x,y)  = \frac{1}{\tau_0} \times \frac{dE_T}{dxdyd\eta_s}_{\big|\eta_s=0} \,,
\ee
where $dE_T/dxdyd\eta_s$ is the transverse energy distribution~\eqref{eq:trento_perp} provided by the \trento{} model. For more details, we refer the reader to App.~\ref{app4d} and Ref.~\cite{Moreland:2014oya}. 

At first, we set the normalization factor to $N = 14.2$ GeV, the nucleon width to $w = 1.12$ fm, the geometric parameter to $p = 0.063$, the minimum nucleon-nucleon distance to $d_\text{min} = 1.44$ fm and the multiplicity fluctuation parameter to $\sigma_k = 1.05$ (see App.~\ref{app4d} for a physical description of these model parameters). Later, we tune the normalization factor to $N = 20$ GeV and the nucleon width to $w = 0.98$ fm in our model; an explanation for these adjustments will be given in Sec.~\ref{chap7S3}.
\subsection{Anisotropic hydrodynamics}
We start anisotropic hydrodynamics at $\tau_0 = 0.05$ fm/$c$ with an initial $\PL/\Pperp$ ratio of $R = (\PL/\Pperp)_0 = 0.3$ (we initialize the bulk viscous pressure component in $\mathcal{P}_{L,\perp}$ as $\Pi_0 = 0$). For the remaining energy-momentum tensor components, we use the initialization scheme discussed in Sec.~\ref{chap4S3.7}. Then we use \cpuvah{}~\cite{McNelis:2021zji} to evolve the quark-gluon plasma on an Eulerian grid with dimensions $L_x = L_y = 30$ fm and spatial resolution $\Delta x = \Delta y = 0.15 w$ to resolve the initial spatial density fluctuations controlled by the nucleon width parameter $w$. During the simulation, we construct a particlization hypersurface of constant temperature $T_\text{sw} = 0.136$ GeV; we stop the program when all fluid cells are below this switching temperature. 

We use the same viscosity parametrizations~\eqref{eqchap4:etas} --~\eqref{eqchap4:zetas} as the SIMS model. The viscosity model parameters are set to $(\etas)_\text{kink} = 0.096$, $T_\eta = 0.223$ GeV, $a_\text{low} = -0.776$ GeV$^{-1}$, $a_\text{high} = 0.37$ GeV$^{-1}$, $(\zetas)_\text{max} = 0.133$, $T_\zeta = 0.12$ GeV, $w_\zeta = 0.072$ GeV and $\lambda_\zeta = -0.122$ (see Table II in Ref.~\cite{Everett:2020xug}). For now, we will not be changing any of these parameters in our model-to-data comparison.
\subsection{Particlization and hadronic transport}
Next we pass the particlization hypersurface to {\sc iS3D} and sample the hadrons' positions and  momenta using the 14--moment $\delta f_n$ correction~\cite{McNelis:2019auj}. For sufficient particle statistics, we generate from each hypersurface a maximum of $N^{(\text{max})}_\text{ens} = 1000$ particlization events (or fewer events if sufficient to generate more than $N^{(\text{min})}_\text{h} = 10^5$ hadrons in the ensemble). The momentum rapidity $y_p$ of each particle is sampled uniformly in the interval $y_p \in [-2, 2]$ (see Sec.~\ref{chap6S5.3}).

After sampling the particles from the Cooper--Frye formula~\cite{Cooper:1974mv}, we feed them to the Boltzmann transport code {\sc SMASH}~\cite{Weil:2016zrk}. In this simulation, the sampled hadrons and resonances are allowed to decay (via strong interaction only) and scatter stochastically until they completely decouple from each other~\cite{Bass:1998ca,Weil:2016zrk}. The final-state particle yields and their momentum spectra are taken from the output of this module and compared with those experimentally measured by the ALICE detector~\cite{Adam:2016thv,Aamodt:2010cz, Abelev:2013vea, Abelev:2014ckr, ALICE:2011ab} (see Sec.~\ref{chap7S2}).

In our first comparison between SIMS and the new \cpuvah{} framework,  we fix $\delta f_n$ to the 14--moment approximation. Towards the end of the analysis, we will look at how the PTMA distribution affects our initial model. 
\subsection{Differences from the SIMS model}
As we mentioned earlier, the SIMS model uses conformal free-streaming to evolve the pre-hydrodynamic stage during the interval $0 < \tau < \tau_\text{fs}$, where $\tau_\text{fs}$ marks the start of the subsequent fluid dynamic stage~\cite{Liu:2015nwa,Bernhard:2018hnz,Everett:2020xug}. SIMS uses the same initial energy density profile, while the remaining initial conditions are similar to \cpuvah{}: the fluid velocity and residual shear stress profiles are initially $u^\mu_0 = (1,0,0,0)$ and $\pi_{\perp,0}^\munu = 0$. The initial pressure ratio is $R = 0$, which is much smaller than the value used in \cpuvah{} ($R = 0.3$). The absence of particle interactions prevents the dimensionless longitudinal pressure $\PL / \Peq$ from increasing until SIMS makes a transition to standard viscous hydrodynamics at the switching time $\tau_\text{fs} \approx 1.5$ fm/$c$ (i.e. $\PL/\Peq \equiv 0$ for $\tau \leq \tau_\text{fs}$). This primarily affects the energy density normalization in the transverse plane (its impact on longitudinal flow is not considered here). The same effect also maximizes the transverse pressure $\Pperp = \frac{1}{2}\ene$, thus generating in SIMS stronger transverse flow during the pre-equilibrium stage. Furthermore, the transition at $\tau = \tau_\text{fs}$ creates an artificial positive bulk viscous pressure at the beginning of the fluid dynamic stage due to inconsistencies between the conformal and QCD equations of state:
\be
\Pi_\text{fs} = \frac{\ene}{3} - \Peq(\ene) > 0\,.
\ee
This enables the fluid to continue expanding strongly into the transverse plane until $\Pi$ has relaxed to its Navier--Stokes solution. Even though standard viscous hydrodynamics generally has a larger negative bulk viscous pressure than anisotropic hydrodynamics for the remainder of the fluid dynamic stage (see Chapter~\ref{chapter4label}), the SIMS model still produces a much stronger transverse flow than \cpuvah{} given identical initial conditions.

\section{Experimental observables}
For each collision event, we compute the following hadronic observables (all except one are $p_T$--integrated observables):
\label{chap7S2}
\begin{enumerate}
    \item 
    The transverse energy per unit pseudorapidity $\eta = {\tanh^{-1}}{\left(p^z/|\boldsymbol{p}|\right)}$ \cite{Adam:2016thv}
    \be
    \label{eqch7:ET}
    \frac{dE_T}{d\eta} = \frac{1}{2\Delta\eta^{(E)}N_\text{ens}} \,\sum_{i = 1}^{N_\text{h}}\sqrt{m_i^2 + p_{T,i}^2}\,,
    \ee
    where $N_\text{h}$ is the total number of final-state hadrons in the {\sc SMASH} ensemble of $N_\text{ens}$ events after applying the rapidity cut $|\eta_i| \leq \Delta\eta^{(E)} = 0.6$. 
    \item 
    The charged particle multiplicity~\cite{Aamodt:2010cz}
    \be
    \label{eqch7:Nch}
    \frac{dN_\text{ch}}{d\eta} = \frac{N_\text{ch}}{2\Delta\eta^{(\text{ch})}N_\text{ens}} \,,
    \ee
    where $N_\text{ch}$ is the number of charged hadrons in the ensemble after applying the rapidity cut $|\eta_i| \leq \Delta\eta^{(\text{ch})} = 0.5$.
    \item 
    The identified particle yields of pions, kaons and protons~\cite{Abelev:2013vea}
    \be
    \label{eqch7:Nid}
    \frac{dN_\text{id}}{dy_p} = \frac{N_\text{id}}{2\Delta y_p^{(\text{id})}N_\text{ens}}  \,,
    \ee
    where $\text{id} \in (\pi^\pm, K^\pm, p\bar{p})$ and $N_\text{id}$ is the number of hadrons of type ``$\text{id}$" in the ensemble after applying the rapidity cut $|y_{p,i}| \leq \Delta y_p^{(\text{id})} = 0.5$.
    \item 
    The mean transverse momentum of pions, kaons and protons~\cite{Abelev:2013vea}
    \be
    \label{eqch7:mean_pT}
    \langle p_T \rangle_\text{id} = \frac{1}{N_\text{id}} \sum\limits_{i = 1}^{N_\text{id}} p_{T,i} \,.
    \ee
    \item
    The transverse momentum fluctuation of charged hadrons~\cite{Abelev:2014ckr} 
    \be
    \label{eqch7:dpT}
    \frac{\delta p_{T,\text{ch}}}{\langle p_T \rangle_\text{ch}} = \frac{1}{\langle p_T \rangle_\text{ch}}\sqrt{\frac{\sum_{i=1}^{N_\text{ch}^{(\delta)}}\left(p_{T,i} - \langle p_T \rangle_\text{ch}\right)^2}{N_\text{ch}^{(\delta)}-1}} \,,
    \ee
    where
    \be
    \langle p_T \rangle_\text{ch} = \frac{1}{N_\text{ch}^{(\delta)}} \sum\limits_{i = 1}^{N_\text{ch}^{(\delta)}} p_{T,i}
    \ee
    and $N_\text{ch}^{(\delta)}$ is the number of charged hadrons in the ensemble within the cuts $|\eta_i| \leq 0.8$ and $0.15$ GeV $\leq p_{T,i} \leq 2$ GeV.
    \item
    The two-particle cumulant of the $p_T$--integrated anisotropic flow coefficients of charged hadrons~\cite{ALICE:2011ab}
    \be
    \label{eqch7:vn_int}
    \left(v^{(\text{ch})}_k\{2\}\right)^2 = \frac{\big|Q^{(\text{ch})}_k\big|^2 - N^{(Q)}_\text{ch}}{N^{(Q)}_\text{ch}\big(N^{(Q)}_\text{ch} - 1\big)}\,,
    \ee
    where $k \in (2,3,4)$,
    \be
    \label{eqch7:Q_int}
    Q^{(\text{ch})}_k = \sum_{i = 1}^{N^{(Q)}_\text{ch}} e^{ik\phi_{p,i}}
    \ee
    is the $p_T$--integrated anisotropic flow vector, and $N^{(Q)}_\text{ch}$ are the number of charged hadrons within the cuts $|\eta_i| < 0.8$ and $0.2$ GeV $\leq p_{T,i} \leq 5$ GeV. 
    \item
    The two-particle cumulant of the $p_T$--differential elliptic flow of charged hadrons~\cite{ALICE:2011ab}
    \be
    \label{eqch7:v2_diff}
    \left(v^{(\text{ch})}_2\{2\}\big(p_{T,\,j+\frac{1}{2}}\big)\right)^2 = \frac{\big|Q^{(\text{ch})}_2\big(p_{T,\,j+\frac{1}{2}}\big)\big|^2 - N^{(p_T)}_\text{ch}}{N^{(p_T)}_\text{ch}\big(N^{(p_T)}_\text{ch} - 1\big)}\,,
    \ee
    where
    \be
    \label{eqch7:Q_diff}
    Q^{(\text{ch})}_2\big(p_{T,\,j+\frac{1}{2}}\big) =
    \sum_{i = 1}^{N^{(p_T)}_\text{ch}} e^{2i\phi_{p,i}}
    \ee
    is the $p_T$--differential elliptic flow vector and $N^{(p_T)}_\text{ch}$ are the number of charged hadrons in the transverse momentum bin $[p_{T,\,j}, p_{T,\,j+1}]$ after applying the rapidity cut $|y_{p,i}| \leq 1$. 
\end{enumerate}
The collision events are sorted in centrality bins according to their charged particle multiplicity~\eqref{eqch7:Nch}. Then for each centrality bin, we compute the statistical mean and standard deviation of the observables~\eqref{eqch7:ET} --~\eqref{eqch7:dpT}, ~\eqref{eqch7:vn_int} and~\eqref{eqch7:v2_diff} by repeating the above calculations with all the collision events in that centrality bin. We will compare these theoretically computed observables to the experimental measurements made by the ALICE collaboration. The experimental data sets can be found in Refs.~\cite{Adam:2016thv,Aamodt:2010cz, Abelev:2013vea, Abelev:2014ckr, ALICE:2011ab}.

\section{Results}
\label{chap7S3}
\begin{figure}[htbp]
\includegraphics[width=0.95\textwidth]{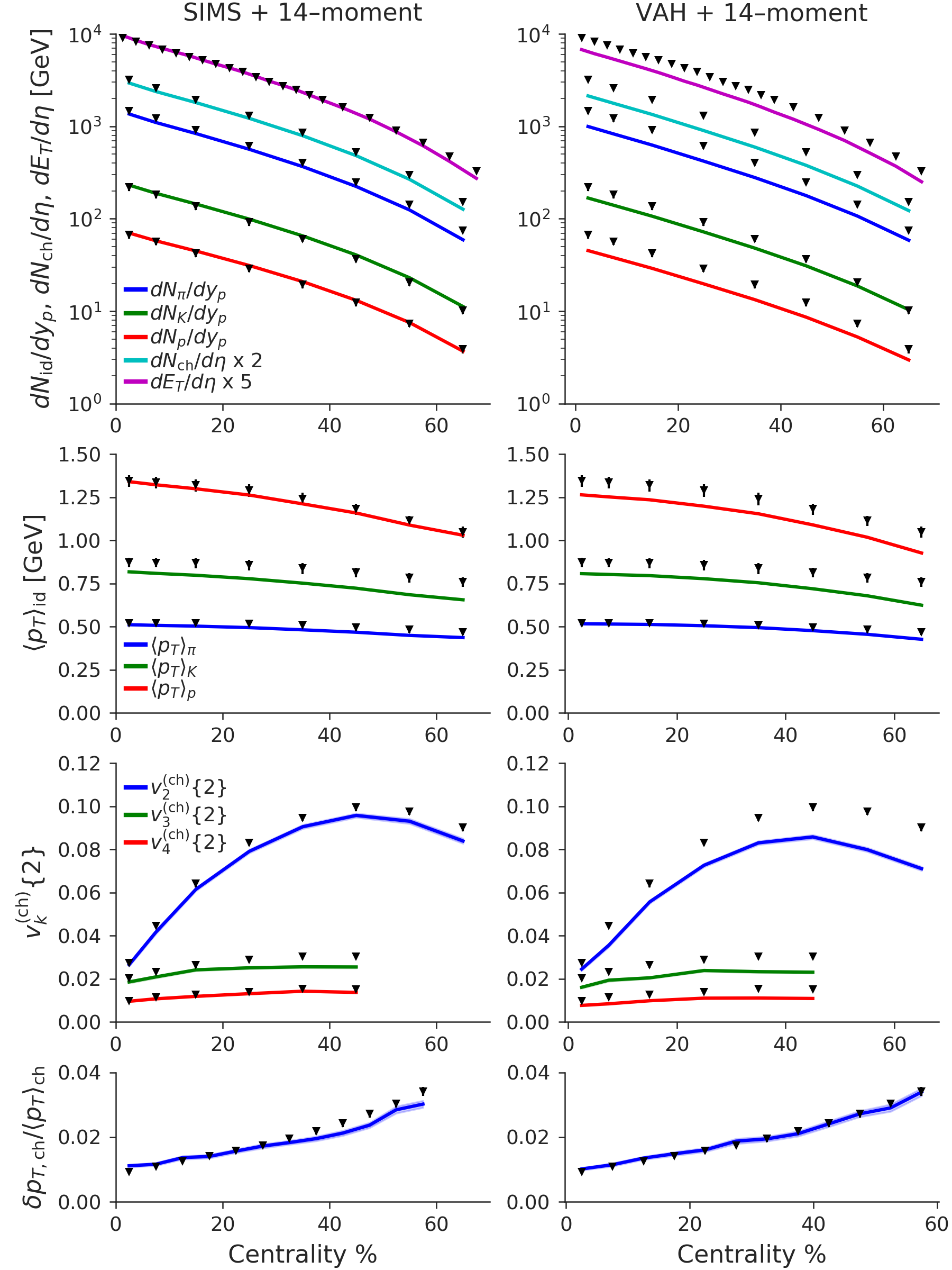}
\centering
\caption{The final-state hadronic observables in Pb+Pb collisions ($\sqrt{s_\text{NN}} = 2.76$ TeV) computed from the SIMS (left) and \cpuvah{} (right) event-by-event simulations (solid color). These are compared to the experimental measurements by the ALICE collaboration (black). Both hybrid models use the Bayesian model parameters from Ref.~\cite{Everett:2020xug} and the 14--moment approximation in the {\sc iS3D} particle sampler. 
\label{SIMS_VAH}
}
\end{figure}
Figure~\ref{SIMS_VAH} shows the predictions of the $p_T$--integrated observables from the SIMS and \cpuvah{} hybrid models, where both use identical \trento{} energy density profiles, the same Bayesian model parameters and the 14--moment approximation for $\delta f_n$. The calibrated SIMS model fits the experimental data quite well~\cite{Everett:2020xug}. In contrast, the \cpuvah{} model grossly underestimates the particle yields, mean transverse momenta and $p_T$--integrated anisotropic flow coefficients when using the same model parameters as SIMS. This should not be surprising since the pre-equilibrium stage is substantially different from the SIMS module. During the early stages of the collision, the build-up of longitudinal pressure in \cpuvah{} causes the system to cool down faster and thus shortens the fireball lifetime relative to SIMS. Thus, the particlization surface shrinks in volume and the particle yields and transverse energy decrease. In addition, the weaker transverse pressure in \cpuvah{} during the stage where SIMS uses free-streaming results in smaller mean transverse momenta (except for pions) and anisotropic flow. We note that \cpuvah{} accurately reproduces the dimensionless $p_T$ fluctuation measure $\delta p_{T,\text{ch}} / \langle p_T\rangle_\text{ch}$ since it is not sensitive to the changes in the build-up of transverse flow during the pre-hydrodynamic stage.

A full Bayesian recalibration of the \cpuvah{} model would provide the best fit for data, but limited time and resources force us to postpone such a project. Instead, we tune two of the \trento{} model parameters by hand after examining the mechanisms behind the model failure shown in the right column of Fig.~\ref{SIMS_VAH}. First, we increase the normalization of the transverse energy deposition to $N = 20$ GeV so that the evolved energy density profile in \cpuvah{} overlaps with the one in SIMS at the switching time $\tau_\text{fs}$. This helps increase the fireball lifetime and particle yields in the \cpuvah{} model. Secondly, we decrease the nucleon width parameter to $w = 0.98$ fm to increase the transverse gradients of the fluctuating hot spots. Larger transverse gradients will induce stronger transverse expansion, which increases the mean transverse momentum and anisotropic flow coefficients.

\begin{figure}[htbp]
\includegraphics[width=0.95\textwidth]{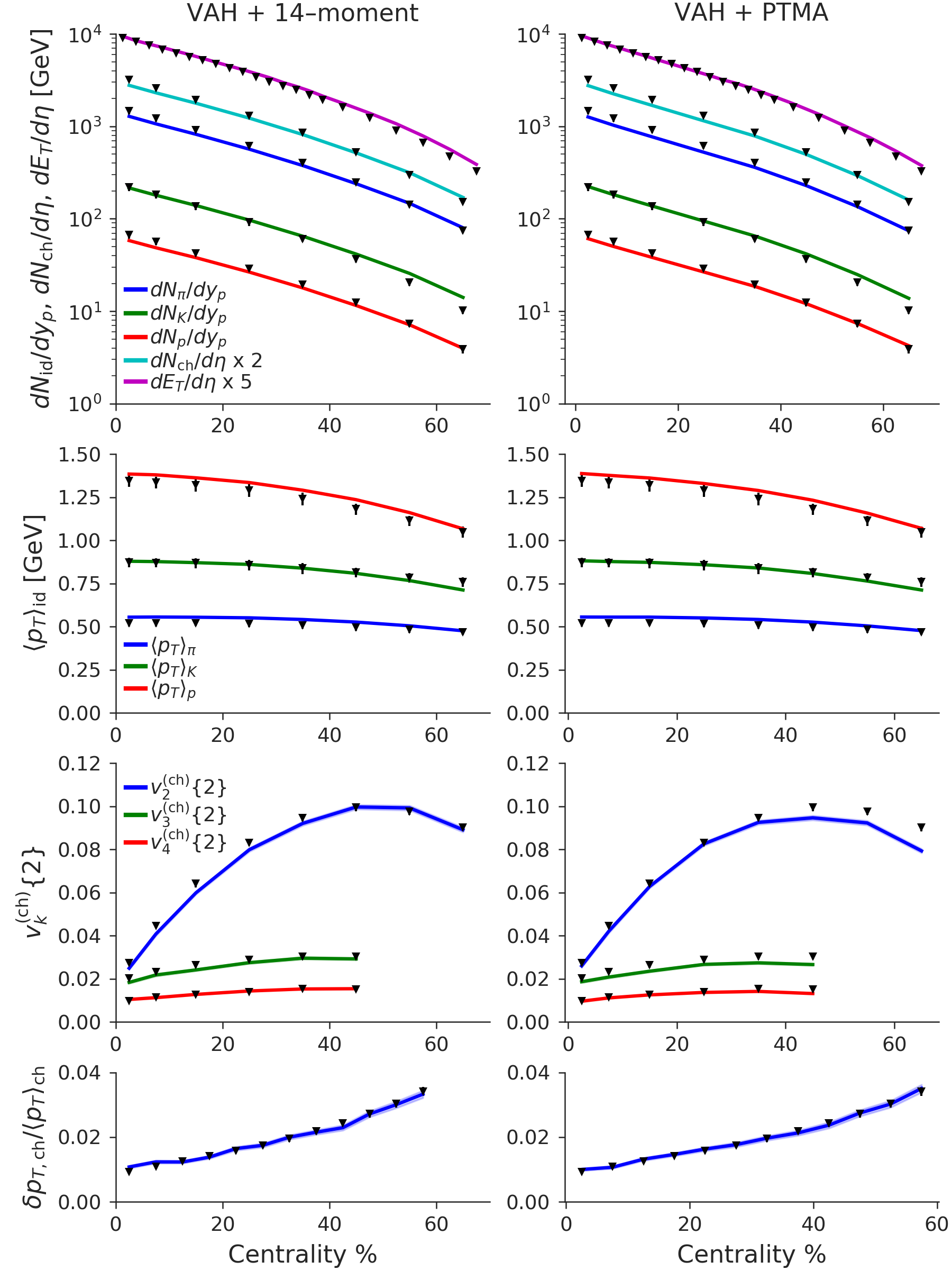}
\centering
\caption{The hadronic observables computed from the \cpuvah{} simulation with the 14--moment approximation (left) or PTMA distribution (right) after adjusting the \trento{} model parameters to $N = 20$ GeV and $w = 0.98$ fm. 
\label{VAH_Grad_PTMA}
}
\end{figure}

The left panel of Figure~\ref{VAH_Grad_PTMA} shows the predictions of the \cpuvah{} model with the tuned model parameters $N = 20$ GeV and $w = 0.98$ fm (the 14--moment approximation is still being used). We see that these simple changes dramatically improve the fit to the experimental data. In particular, \cpuvah{} reproduces the observables $\langle p_T\rangle_K$, $v^{(2)}_\text{ch}\{2\}$, $v^{(3)}_\text{ch}\{2\}$ and $\delta p_{T,\text{ch}} / \langle p_T\rangle_\text{ch}$ better than the SIMS model. However, the fit for the multiplicities is not perfect, especially the transverse energy $dE_T/d\eta$ and kaon yield $dN_K/dy_p$ at high centralities and the pion yield $dN_\pi/dy_p$ and proton yield $dN_p/dy_p$ at low centralities. The model also overestimates $\langle p_T\rangle_p$ for almost all centrality bins. These results are subject to change once a full analysis of the \cpuvah{} framework is completed in the future.

In the right column of Fig.~\ref{VAH_Grad_PTMA} we plot the output of the \cpuvah{} model if we replace the 14--moment $\delta f_n$ correction in {\sc iS3D} by the PTMA distribution. Overall, the agreement between theory and experiment is similar in both panels, with the exception of the charged hadron elliptic flow which, in peripheral collisions, is better described by the 14--moment approximation. The PTMA distribution also slightly underestimates $v_3^{(\text{ch})}\{2\}$ at all centralities. From the SIMS analysis~\cite{Everett:2020xug,Everett:2020yty} we already know that the model parameters are somewhat sensitive to the choice for $\delta f_n$. Since we used the model parameters optimized for the 14--moment approximation, we should expect \cpuvah{} with $\delta f_n = \delta f_n^{14}$ to perform better. 
To optimize the \cpuvah{} model with the PTMA distribution, the Bayesian parameters would therefore require additional recalibration.

\begin{figure}[!t]
\includegraphics[width=0.9\textwidth]{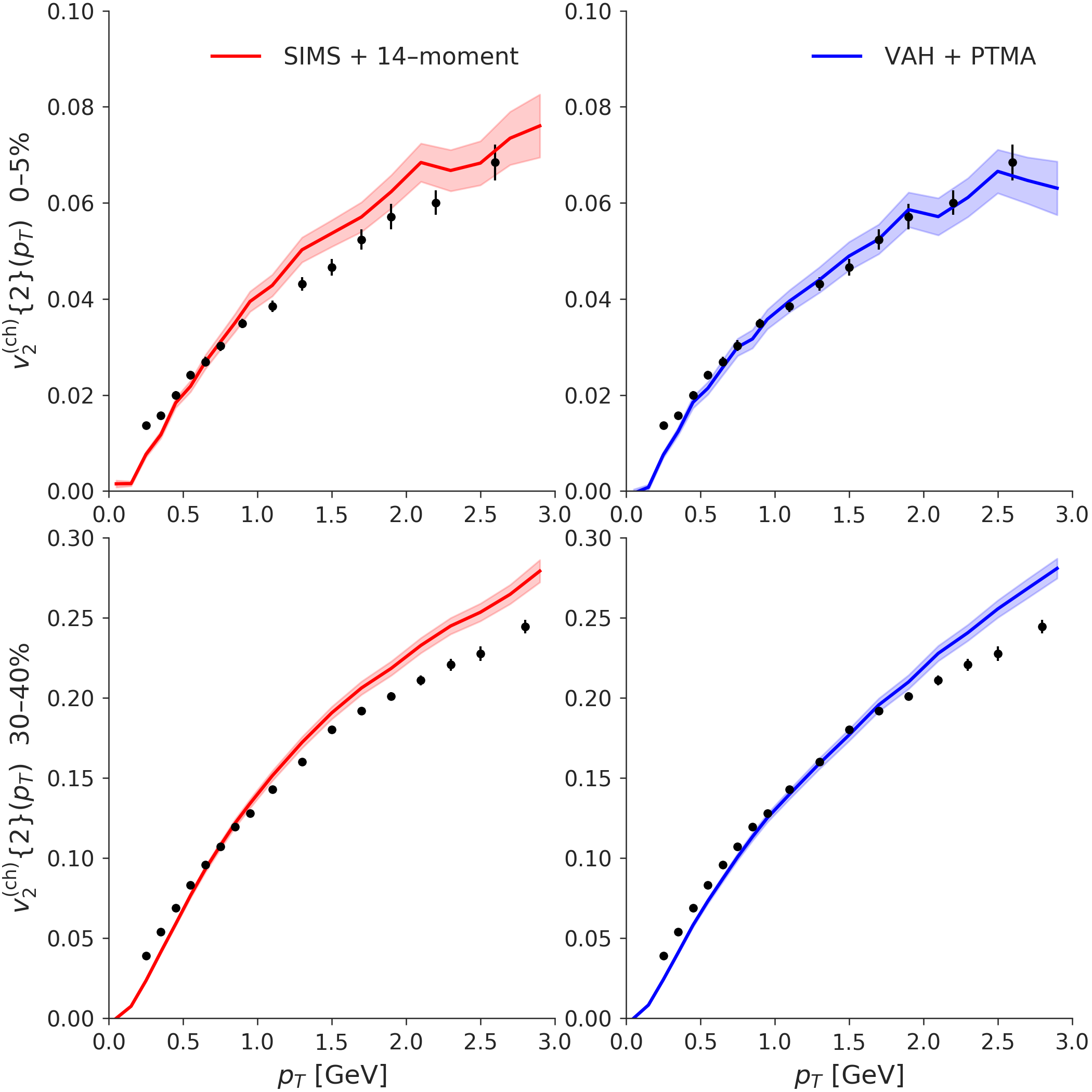}
\centering
\caption{The $p_T$--differential elliptic flow coefficient of charged hadrons $v^{(\text{ch})}_2\{2\}(p_T)$ in Pb+Pb collisions ($\sqrt{s_\text{NN}} = 2.76$ TeV) for the 0--5\% (top row) and 30--40\% (bottom row) centrality bins. Here we compare the model calculations from SIMS (red) and \cpuvah{} (blue) to the experimental data (black), which applied a pseudorapidity cut $\Delta \eta = 1$. The \cpuvah{} model uses the adjusted \trento{} parameters and the PTMA distribution for $\delta f_n$.
\label{v2_diff_JETSCAPE}
}
\end{figure}

Finally, we plot in Figure~\ref{v2_diff_JETSCAPE} the $p_T$--differential charged hadron elliptic flow $v_2^{(\text{ch})}\{2\}(p_T)$ for the $0{-}5\%$ and $30{-}40\%$ centrality bins. Our model calculations use $p_T$--bins that are similar in width to the ones used by ALICE. We see that at $0{-}5\%$ centrality the SIMS model slightly overestimates $v_2^{(\text{ch})}\{2\}$ for intermediate transverse momenta $1$  GeV $\leq p_T \leq 2$ GeV while \cpuvah{} agrees better with the data. The statistical uncertainties at high momentum $p_T \sim 3$ GeV are too large to draw any additional conclusions. At $30{-}40\%$ centrality, both models yield essentially the same curves for $v_2^{(\text{ch})}\{2\}(p_T)$, which underestimate and overestimate the data at low and high $p_T$, respectively.\footnote{%
    We note that the experimental data for $v_2^{(\text{ch})}\{2\}(p_T)$ do not appear to extrapolate to zero at $p_T = 0$ with the theoretically expected quadratic behavior $\sim p_T^2$ at low $p_T$~\cite{Danielewicz:1994nb}.}
We also checked that \cpuvah{} with the 14--moment approximation produces results that are qualitatively similar to \cpuvah{} with the PTMA distribution.

\section{Summary}
In this chapter we have constructed a full event-by-event simulation for heavy-ion collisions that uses \cpuvah{} for both the pre-hydrodynamic and fluid dynamic stages. At very early times this framework is a significant departure from conventional hybrid models, which typically simulate the pre-equilibrium stage separately using either a microscopic kinetic model or the IP--Glasma model~\cite{Schenke:2012wb, Gale:2012rq, Bernhard:2018hnz, Everett:2020xug}. To demonstrate its potential, we ran (2+1)--d event-by-event simulations of Pb+Pb collisions at LHC energies and predicted the final-state hadronic observables. After adjusting two of the model parameters our preliminary results agree very well with the experimental data.

This chapter concludes the main objective of this thesis, which is to fully develop an anisotropic hydrodynamic simulation for heavy-ion phenomenology. In the next chapter, we will discuss our latest contributions to the emerging field of far-from-equilibrium hydrodynamics.

\chapter{Far-from-equilibrium hydrodynamics from Green's functions}
\label{chap8label}
We close this thesis with new developments on far-from-equilibrium hydrodynamics, whose techniques could potentially be generalized to improve the hydrodynamic modeling of non-equilibrium fluids. In the past decade, we have found empirically that viscous hydrodynamics is a quantitatively robust model for the quark-gluon plasma produced in heavy-ion collisions, even though it is subject to large gradients (or Knudsen numbers $\text{Kn} \sim 1$)~\cite{Gale:2012rq, Shen:2014lye, Bernhard:2016tnd, Bernhard:2018hnz,Everett:2020xug,Everett:2020yty}. This has generated a lot of interest in understanding why viscous hydrodynamics has a wider range of applicability than what is traditionally assumed (limited to $\text{Kn} \ll 1$)~\cite{Heller:2016gbp, Romatschke:2016hle, Florkowski:2017olj, Romatschke:2017vte, Romatschke:2017ejr}. Currently, it is widely believed that the success of viscous hydrodynamics is due to the existence of a \textit{hydrodynamic attractor}~\cite{Heller:2015dha}; this attractor reduces to the first-order Navier--Stokes solution in the small-gradient limit but also extends to large gradients (without diverging). Many (0+1)--dimensional studies on Bjorken expansion have shown that dimensionless hydrodynamic quantities (e.g. the temperature $\tau \partial_\tau \ln T$ and shear stress $\pi / (\ene{+}\Peq)$) approach their respective attractor solution within a timescale on the order of the relaxation time $\tau_r$, which is typically much shorter than the thermalization time~\cite{Romatschke:2017vte, Romatschke:2017acs, Strickland:2017kux, Strickland:2018ayk, Jaiswal:2019cju, Chattopadhyay:2019jqj}. Heller and Spalinski first constructed the hydrodynamic attractor by performing a Borel resummation of the original hydrodynamic gradient expansion~\cite{Heller:2015dha}; one cannot always take such a direct approach, however, if the medium's microscopic scattering rate $\tau_r^{-1}$ is too small~\cite{Heller:2018qvh}. Nevertheless, a recent study~\cite{Chattopadhyay:2019jqj} has demonstrated that attractor solutions can exist (across the entire gradient spectrum $\Kn \in [0,\infty)$) even as one approaches this free-streaming limit, suggesting that the hydrodynamic attractor has a more fundamental origin. 

In many-body physics, a Green's function or correlation function between particles (or quantum fields) is the fundamental object that characterizes the physical properties of a non-equilibrium system (e.g. dissipation). It is often very difficult to do a full calculation of the Green's function, especially if the medium's constituents have strong correlations. In this chapter we study a much simpler system of weakly-interacting particles that can be described by relativistic kinetic theory. After specifying the collision kernel as the relaxation time approximation (RTA)~\cite{Anderson_Witting_1974}, the Boltzmann equation for a single-particle distribution\footnote{%
    The particles are massive and on-shell unless stated otherwise.} $f(x,p)$ without external forces in Minkowski spacetime $x^\mu = (t,x,y,z)$ reads
\be
\label{eqch8:RTAmink}
s^\mu(x,p) \partial_\mu f(x,p) = \feq(x,p) - f(x,p)\,,
\ee
where $s^\mu(x,p) = \tau_r(x)\, p^\mu / \big(p {\,\cdot\,} u(x)\big)$ and $\feq(x,p) = \exp\left[- p{\,\cdot\,} u(x) / T(x)\right]$ is the local-equilibrium Boltzmann distribution.\footnote{%
    Here we neglect quantum statistics and conserved charges. The degeneracy factor is set to $g = 1$.}$^,$\footnote{%
    For simplicity, we take the relaxation time $\trel(x)$ to be momentum independent.} 
The RTA Boltzmann equation assumes that the system possesses only one microscopic time scale $\tau_r$ (realistically there is an infinite hierarchy of time scales~\cite{Denicol:2012cn}) and is close to local equilibrium (i.e. $|\delta f| = |f - \feq| \ll \feq$). Although it is an oversimplified, linearized version of the full nonlinear Boltzmann equation, its advantage is that one can obtain a simple expression for the retarded Green's function:
\be
\label{eqch8:Green3D}
G(x,x^\prime, p) = D(x,x^\prime,p)\,\Theta(x - x^\prime)\,.
\ee
Here
\be
\label{eqch8:damping}
D(x,x^\prime,p) = \exp\left[-\int^x_{x'} dx^{\prime\prime}{\cdot\,}s^{-1}(x^{\prime\prime},p)\right]
\ee
is known as the damping function, $s^{-1}_\mu = p_\mu / (s {\,\cdot\,} p)$ is the reciprocal vector of $s^\mu$ (i.e. $s^{-1}_\mu s^\mu = 1$), and $\Theta(x{-}x^\prime)$ is the Heaviside step function. The exponential argument in Eq.~\eqref{eqch8:damping} is a straight path integral between two points $x^{\prime\mu}$ and $x^\mu$; the starting point $x^{\prime\mu}$ is constrained along a past world line parallel to $p^\mu$, with the current position $x^\mu$ being the end point. 

The Green's function~\eqref{eqch8:Green3D} describes the dissipation of particle sources propagating from $x^{\prime\mu}$ to $x^{\mu}$ with momentum $p^\mu$ in a local-equilibrium background. Specifically, it encodes how the medium responds to non-equilibrium deviations from this background, whether they are present initially or caused by global expansion driving the system away from local equilibrium. Thus, the Green's function fully captures the non-equilibrium dynamics of the system. Not only can we use the Green's function to construct a solution $f(x,p)$ to the RTA Boltzmann equation (or an approximate solution to the full Boltzmann equation) but we can also generate a resummed hydrodynamic gradient series from it. In particular, we show that a series expansion of this \textit{hydrodynamic generator} corresponds to the Borel resummed Chapman--Enskog series, as the associated non-hydrodynamic modes decay at late times.

This chapter is mostly based on published material in Ref.~\cite{McNelis:2020jrn}, but it also contains new content regarding applications of the hydrodynamic generator expansion. A new way of expanding the hydrodynamic generator to all orders in the Knudsen number is given in Appendix~\ref{appch8proof}.

\section{Hydrodynamic generator in (0+1)--d Bjorken flow}
\label{chapman_enskog}
We first study a relativistic gas undergoing Bjorken expansion~\cite{Bjorken:1982qr}, which is transversely homogeneous and longitudinally boost-invariant. In Milne coordinates $x^\mu = (\tau, x, y, \eta_s)$, the fluid velocity $u^\mu = (1,0,0,0)$ is static. Thus, the RTA Boltzmann equation~\eqref{eqch8:RTAmink} simplifies to
\be
\label{eqch8:RTA_Bjorken}
  \partial_\tau f(\tau, p) = \frac{\feq(\tau, p) - f(\tau,p)}{\trel(\tau)} \,,
\ee
where the local-equilibrium distribution is
\be
\label{eqch8:feq}
  \feq(\tau,p) = \exp\left[-\frac{p^\tau(\tau)}{T(\tau)}\right]\,,
\ee
with $p^\tau = \sqrt{p_\perp^2 + \tau^2(p^\eta)^2 + m^2}$. The Green's function of this kinetic equation is
\be
\label{eqch8:Green1D}
G(\tau,\tau^\prime) = D(\tau, \tau^\prime)\, \Theta(\tau - \tau^\prime)\,,
\ee
where the damping function simplifies to
\be
  D(\tau,\tau^\prime) = \exp\left[-\int^{\tau}_{\tau^\prime} \frac{d\tau^{\prime\prime}}{\trel(\tau^{\prime\prime})}\right]\,.
\ee
From this, the analytic solution to the RTA Boltzmann equation is \cite{Baym:1984np,Florkowski:2013lya}:
\be
\label{eqch8:exact}
  f(\tau,p) = D(\tau,\tau_0)f_0(\tau_0,p) + {\int^\tau_{\tau_0}}{\frac{d\tp D(\tau,\tp)\feq(\tp, p)}{\trel(\tp)}} \,,
\ee
where $f_0(\tau_0,p)$ is some arbitrary initial distribution. One sees that the first term of the exact solution \eqref{eqch8:exact}, which is sensitive to the initial state, dominates the early-time dynamics. For times $\tau{\,-\,}\tau_0 \gg \trel$, however, the initial-state term decays exponentially. Hence, the second term in Eq.~\eqref{eqch8:exact} describes the long-time behavior of the system. 

The full Boltzmann equation often cannot be solved analytically; instead one usually makes a hydrodynamic approximation using the Chapman--Enskog method~\cite{chapman1990mathematical}. Here we apply this technique to the RTA Boltzmann equation so that we can compare the gradient expansion to the exact RTA solution~\eqref{eqch8:exact}. For a (0+1)--dimensional system with Bjorken symmetry, the Chapman--Enskog expansion of the RTA Boltzmann equation \eqref{eqch8:RTA_Bjorken} takes the form
\be
\label{eqch8:gradient_series}
  f_{\text{CE}}(\tau,p) = \sum_{n=0}^\infty {\left[-\trel(\tau) \partial_\tau\right]^n} \feq(\tau, p) \,.
\ee
In this gradient series, each linear operator $-\trel(\tau) \partial_\tau$ acts on all of the terms to its right. Generally, the series will contain derivatives of not only $\feq(\tau,p)$ but also $\trel(\tau)$. This causes the number of terms to grow like $n!$, which means the gradient series is divergent, even for small Knudsen numbers $\Kn \sim \trel \dt \ll 1$.\footnote{%
    Although the number of distinct gradient terms $\propto (\Kn)^n$ does not grow like $n!$, their prefactors give them the combined appearance of exhibiting $n!$ growth, assuming they have the same magnitude and sign (see for example  Eq.~\eqref{eqch8:fCE3}).}
One can try to resum the divergent series using Borel resummation:
\be
\label{eqch8:borel_sum_general}
  f^{\text B}_{\text{CE}}(\tau,p) = {\int^\infty_0} dz\, e^{-z} \sum_{n=0}^\infty \frac{{z^n}{\left[{-}\trel(\tau) \partial_\tau\right]^n}\feq(\tau, p)}{n!} \,.
\ee
However, it is far from obvious how to sum the entire series to obtain a finite integral representation of the Chapman--Enskog expansion. Instead of computing the Borel sum directly, we analyze the exact solution \eqref{eqch8:exact} to look for a representation of the series. For the simplest case where the relaxation time is constant, the exact distribution function simplifies to
\be
\label{eqch8:exact_const_trel}
  f(\tau,p) =
   \exp\left[-\frac{(\tau - \tau_0)}{\trel}\right] f_0(\tau_0,p)  
   \,+\, \frac{1}{\trel}\int^\tau_{\tau_0} d\tp \exp\left[-\frac{(\tau - \tp)}{\trel}\right]\feq(\tp, p) \,.
\ee
We introduce the dimensionless coordinate $z = (\tau - \tau^\prime)\,/\,\trel$ to rewrite Eq.~\eqref{eqch8:exact_const_trel} as
\be
\label{eqch8:exact_const_trel_z}
  f(\tau,p) = e^{-z_0} f_0(\tau_0,p)\, 
  + \int^{z_0}_{0} dz \, e^{-z}\feq(\tau - \trel z, p) \,,
\ee
where $z_0 = (\tau - \tau_0)\,/\,\trel$, and Taylor expand the second term:
\be
\label{eqch8:exact_expand}
  f(\tau,p) =
  e^{-z_0} f_0(\tau_0,p) + \int^{z_0}_{0} dz \, e^{-z} \sum_{n=0}^\infty \frac{(-z \trel)^n \feq^{(n)}(\tau, p)}{n!} \,,
\ee
with $\feq^{(n)}(\tau,p) \equiv \partial_\tau^n \feq(\tau,p)$. Sure enough, one sees that the expansion of the exact solution \eqref{eqch8:exact_expand} reduces to the Borel resummation \eqref{eqch8:borel_sum_general} in the limit $z_0 \to \infty$ (i.e. $\tau \to \infty$) when all non-hydrodynamic modes have decayed. With this insight, we conjecture that even for non-constant $\tau_r(\tau)$ the \textit{hydrodynamic generator}\footnote{%
    We call Eq.~\eqref{eqch8:general_generator} the hydrodynamic generator since it generates the hydrodynamic gradient series (7) in the limit $z_0 \to \infty$.}
\be
\label{eqch8:general_generator}
  f_\text{G}(\tau,p) = \int^\tau_{\tau_0}\frac{d\tp D(\tau,\tp)\feq(\tp, p)}{\trel(\tp)}
\ee
is, in the limit of vanishing non-hydrodynamic modes, an integral representation of the gradient series \eqref{eqch8:gradient_series}.\footnote{%
    This does not imply that the hydrodynamic generator and RTA Chapman--Enskog series are equivalent in the late time limit. The expansion of the hydrodynamic generator \eqref{eqch8:general_generator} may not have a finite radius of convergence.} 
If the conjecture holds it should be possible to manipulate this expression, as we did for $\tau_r{\,=\,}$const, to obtain a hydrodynamic gradient series. We use the coordinate transformation
\be
\label{eqch8:z_to_tau} 
  z = h(\tp, \tau) = \int^\tau_{\tp} \frac{d\tau^{\prime\prime}}{\tau_r(\tau^{\prime\prime})}
\ee
to rewrite Eq.~\eqref{eqch8:general_generator} as
\be
\label{eqch8:generator_z}
  f_\text{G}(\tau,p) = {\int^{z_0}_0} dz \, e^{-z} \feq\big(h^{-1}(z,\tau), p\big) \,,
\ee
where 
\be
  z_0 = \int^\tau_{\tau_0} \frac{d\tau^{\prime\prime}}{\tau_r(\tau^{\prime\prime})} \,.
\ee
Next, we compute the inverse function $\tp = h^{-1}(z,\tau)$. Physically, the relaxation time is positive and finite, which means that $z$ is a non-negative monotonic function of $\tp \in [\tau_0,\tau]$. Therefore, the function $h(\tp,\tau)$ has an inverse which we expand as a power series:\footnote{\label{chap8fn7}%
    For a given time $\tau$, $z = h(\tp,\tau)$ is a smooth function of $\tp$ when evaluated with the exact solution~\eqref{eqch8:exact}; hence it can be Taylor expanded around $\tp=\tau$, which corresponds to a Taylor expansion of $\tau^\prime = h^{-1}(z,\tau)$ around $z=0$.}
\be
\label{eqch8:power_series_tau}
  \tau^\prime = h^{-1}(z,\tau) = \sum_{n=0}^\infty c_n(\tau) \, z^n \,.
\ee
The coefficients $c_n(\tau)$ can be computed by Taylor expanding Eq.~\eqref{eqch8:z_to_tau} around $\tp = \tau$:
\be
\label{eqch8:z_to_tau_series}
  z = \int^\tau_{\tp} d\tau^{\prime\prime} \sum_{n=0}^\infty \frac{(\tau^{\prime\prime} - \tau)^n}{n!} \partial_\tau^n \big[\trel^{-1}(\tau)\big] \\
  = -\sum_{n=0}^\infty \frac{(\tp - \tau)^{n+1}}{(n+1)!} \partial_\tau^n \big[\trel^{-1}(\tau)\big] \,.
\ee
Inserting the power series \eqref{eqch8:power_series_tau} into Eq.~\eqref{eqch8:z_to_tau_series} we can solve for the coefficients order by order. The first coefficients are
\bs
\allowdisplaybreaks
\label{eqch8:coefficients}
\begin{align}
    c_0 &= \tau \,,\\
    c_1 &= - \tau_r \,,\\
    c_2 &= \frac{\tau_r}{2!}\tau^{(1)}_r  \,,\\
    c_3 &= -\frac{\trel}{3!}\left((\tau^{(1)}_r\big)^2 + \tau_r \tau^{(2)}_r\right) \,,
\end{align}
\es
where $\tau^{(n)}_r \equiv \partial^n_\tau \tau_r(\tau)$; they satisfy the recurrence relation
\bs
\allowdisplaybreaks
\label{eqch8:recursion}
\begin{align}
    c_0 &= \tau \,, \\
    c_n &= - \frac{\tau_r \partial_\tau c_{n{-}1}}{n} \indent \forall\,n\geq 1 \,.
\end{align}
\es
With these coefficients, we can now evaluate the integral \eqref{eqch8:generator_z} after Taylor expanding the integrand:
\be
\label{eqch8:fA_tau_resum}
  f_\text{G}(\tau,p) = \int^{z_0}_0 dz \, e^{-z} \sum_{n=0}^\infty \frac{\big(h^{-1}(z,\tau) - \tau\big)^n \feq^{(n)}(\tau , p)}{n!} \,.
\ee
As a demonstration, we compute the series up to $n = 3$ and truncate the expression at third order in derivatives:
\be
\label{eqch8:generator_truncated}
\begin{split}
  f_\text{G} \approx& \, (1 - e^{-z_0})\feq \,+\, \big(1 - \Gamma(2,z_0)\big)\delta f^{(1)} \,+\, \\
  &\Big(1 - \frac{\Gamma(3,z_0)}{2!}\Big) \delta f^{(2)} \,+\, \Big(1 - \frac{\Gamma(4,z_0)}{3!}\Big)\delta f^{(3)} \,,
\end{split}
\ee
where $\Gamma(n{+}1, z_0) = \int^\infty_{z_0} dz \, e^{-z} z^{n}$ are the upper incomplete gamma functions. After taking the limit $z_0 \to \infty$, Eq.~\eqref{eqch8:generator_truncated} reduces to
\be
\label{eqch8:generator_truncated_limit}
  f_\text{G} \approx \feq + \delta f^{(1)} + \delta f^{(2)} + \delta f^{(3)} \,,
\ee
where 
\bs
\allowdisplaybreaks
\label{eqch8:fCE3}
\begin{align}
    \delta f^{(1)} =& \,- \trel \feq^{(1)} \,, \\
    \delta f^{(2)} =& \ \trel \trel^{(1)} \feq^{(1)} \,+\, \tau_r^2 \feq^{(2)} \,, \\
    \delta f^{(3)} =& -\trel \big(\trel^{(1)}\big)^2 \feq^{(1)} \,-\, \trel^2 \trel^{(2)} \feq^{(1)} 
    -\, 3 \trel^2 \trel^{(1)} \feq^{(2)} \,-\, \trel^3 \feq^{(3)} \,.
\end{align}
\es
These are precisely the non-equilibrium corrections in the Chapman-Enskog series \eqref{eqch8:gradient_series}. Using a computer-generated code\footnote{\label{github}%
    The codes used in this chapter can be downloaded at \url{https://github.com/mjmcnelis/rta_resum}.} 
we verified that the series \eqref{eqch8:generator_truncated} works up to order $O(\Kn^{40}$):
\be
\begin{split}
  f_\G(\tau,p) &\approx {\int_0^{z_0}} dz \, e^{-z} \sum_{n=0}^{40} \frac{z^n{\left[-\trel(\tau) \partial_\tau\right]^n}\feq(\tau, p)}{n!} 
  \\
  &=\sum_{n=0}^{40} \left(1 - \frac{\Gamma(n{+}1,z_0)}{n!}\right){\left[-\trel(\tau) \partial_\tau\right]^n}\feq(\tau, p) \,.
\end{split}
\ee
A proof that the (0+1)--d hydrodynamic generator~\eqref{eqch8:general_generator} maps to the Borel resummed RTA Chapman--Enskog series~\eqref{eqch8:borel_sum_general} at all orders in the Knudsen number as $z_0\to\infty$ has been generously provided by Chandrodoy Chattopadhyay in Appendix~\ref{appch8proof}.

\section{Series expansion of the hydrodynamic generator}
\label{expansion}
In the limit of vanishing non-hydrodynamic modes, the hydrodynamic generator~\eqref{eqch8:general_generator} provides a shortcut to the Borel resummation~\eqref{eqch8:borel_sum_general} without needing to compute it directly. It also satisfies the RTA Boltzmann equation
\be
\begin{split}
  \dt f_\text{G}(\tau,p) 
  =&\,\frac{\feq(\tau,p)}{\trel(\tau)} - \int^\tau_{\tau_0} \frac{d\tp D(\tau,\tp)\feq(\tp,p)}{\trel(\tau)\trel(\tp)}
  =\,\frac{\feq(\tau,p) - f_\text{G}(\tau,p)}{\tau_r(\tau)}\,,
\end{split}
\ee
where we used the identities $\partial_\tau D(\tau,\tp){\,=\,}-D(\tau,\tp)/\trel(\tau)$ and $D(\tau,\tau){\,=\,}1$. However, this alone does not tell us how much hydrodynamics contributes to the dynamics of the system at finite times, before the initial state $f_0(\tau_0,p)$ has completely decayed. As long as the non-hydrodynamic modes contribute, the expansion of the exact distribution function
\be
  f(\tau,p) = e^{-z_0} f_0(\tau_0,p) + f_\text{G}(\tau,p)
\ee
around local equilibrium looks like\footnote{This expansion retains the same transseries-like structure for both early and late times. Transasymptotic solutions for the moments of the distribution function have been studied for Bjorken expansion and have been found to accurately reproduce the numerical solution of the moments equations even when continued back to earlier times~\cite{Behtash:2018moe,Behtash:2019txb}.}
\be
\label{eqch8:anomaly}
  f = \feq + \delta f_\G^{(0)} + \delta f_\G^{(1)} + \delta f_\G^{(2)} + \delta f_\G^{(3)} + O(\Kn^4) \,,
\ee
where
\bs
\allowdisplaybreaks
\begin{align}
  \delta f_\G^{(0)} &= e^{-z_0} \left(f_0 - \feq\right) \,, \\ 
  \delta f_\G^{(n)} &= \left(1 - \frac{\Gamma(n{+}1, z_0)}{n!}\right) \delta f^{(n)} \indent \forall \, n \geq 1 \,.
\end{align}
\es
The zeroth-order correction $\delta f_\G^{(0)}$, which combines the initial-state term with the first term in Eq.~\eqref{eqch8:generator_truncated}, is a purely non-hydrodynamic mode and is only present for a short period of time $\sim\trel$. The other $\delta f_\G^{(n)}$ corrections are the usual hydrodynamic gradient corrections, except they are initially suppressed by their associated non-hydrodynamic mode. These non-perturbative factors control the emerging strengths of the gradient corrections to the distribution function as the particle interactions drive the system towards hydrodynamics over time (i.e. as $z_0$ increases). In particular, as will be discussed below, higher-order gradient corrections are suppressed more strongly and for a longer duration than the lower-order terms.

To study these new non-perturbative effects on the hydrodynamic gradient expansion, we evolve a conformal fluid undergoing Bjorken expansion with the exact solution of the RTA Boltzmann equation \cite{Florkowski:2013lya, Florkowski:2013lza, Tinti:2018qfb}. We initialize the system at $\tau_0 = 0.25$ fm/$c$ with initial temperature $T(\tau_0)=0.6$\,GeV and shear stress $\pi_s(\tau_0)=0$, where $\pi_s\equiv\frac{2}{3}(\mathcal{P}_{\perp}{-}\mathcal{P}_{L})$ (by definition $\pi_{s,\mathrm{eq}}=0$). For the relaxation time we take $\trel=\tpi$ with $\tpi T = 5(\etas)$ and set the specific shear viscosity to $\etas = 3/(4\pi)$. Using these initial conditions we construct the temperature $T(\tau)$ by fixing the exact solution~\eqref{eqch8:exact} to the Landau matching condition $\ene(\tau) = 3T^4(\tau)/\pi^2$ or \cite{Florkowski:2013lya}
\be
\label{eqch8:T_exact}
\begin{split}
T^4(\tau) =&\, D(\tau,\tau_0) \,T^4(\tau_0)\, \mathcal{H}\left(\frac{\tau_0}{\tau}\right) 
+ \int_{\tau_0}^\tau \frac{d\tp}{\tau_\pi({\tp})} D(\tau,\tp) \,T^4(\tp)\, \mathcal{H}\left(\dfrac{\tp}{\tau}\right)\,,
\end{split}
\ee
where
\be
\mathcal{H}(x) = \frac{1}{2}\left(x^2 + \frac{{\tan^{-1}}\sqrt{x^{-2}-1}}{\sqrt{x^{-2}-1}}\right)\,.
\ee
The most straightforward way to solve this integral equation is by using fixed-point iteration. After computing the temperature, we evaluate the normalized shear stress\footnote{%
    This differs from the traditional definition $\bar\pi\equiv\pi/(\ene{+}\Peq)$ which reduces to $\bar\pi = \pi/(4\Peq)$ in the conformal limit.} 
$\bar\pi_s(\tau) = \pi_s(\tau) / \mathcal{P}_\text{eq}(\tau)$, where $\mathcal{P}_\text{eq}(\tau) = \ene(\tau)/3$ is the equilibrium pressure~\cite{Florkowski:2013lya}:
\be
\label{eqch8:pi_exact}
\begin{split}
\bar\pi_s(\tau) =\,& D(\tau,\tau_0) \frac{T^4(\tau_0)}{T^4(\tau)}\left[\frac{1}{2}\mathcal{H}_\perp\Big(\frac{\tau_0}{\tau}\Big) -  \mathcal{H}_L\Big(\frac{\tau_0}{\tau}\Big)\right] \\
& + \int_{\tau_0}^\tau \frac{d\tp}{\tau_\pi({\tp})} D(\tau,\tp)\frac{T^4(\tp)}{T^4(\tau)}\left[\frac{1}{2}\mathcal{H}_\perp\Big(\frac{\tp}{\tau}\Big) -  \mathcal{H}_L\Big(\frac{\tp}{\tau}\Big)\right] \,.
\end{split}
\ee
\begin{figure}[t]
\includegraphics[width=0.65\linewidth]{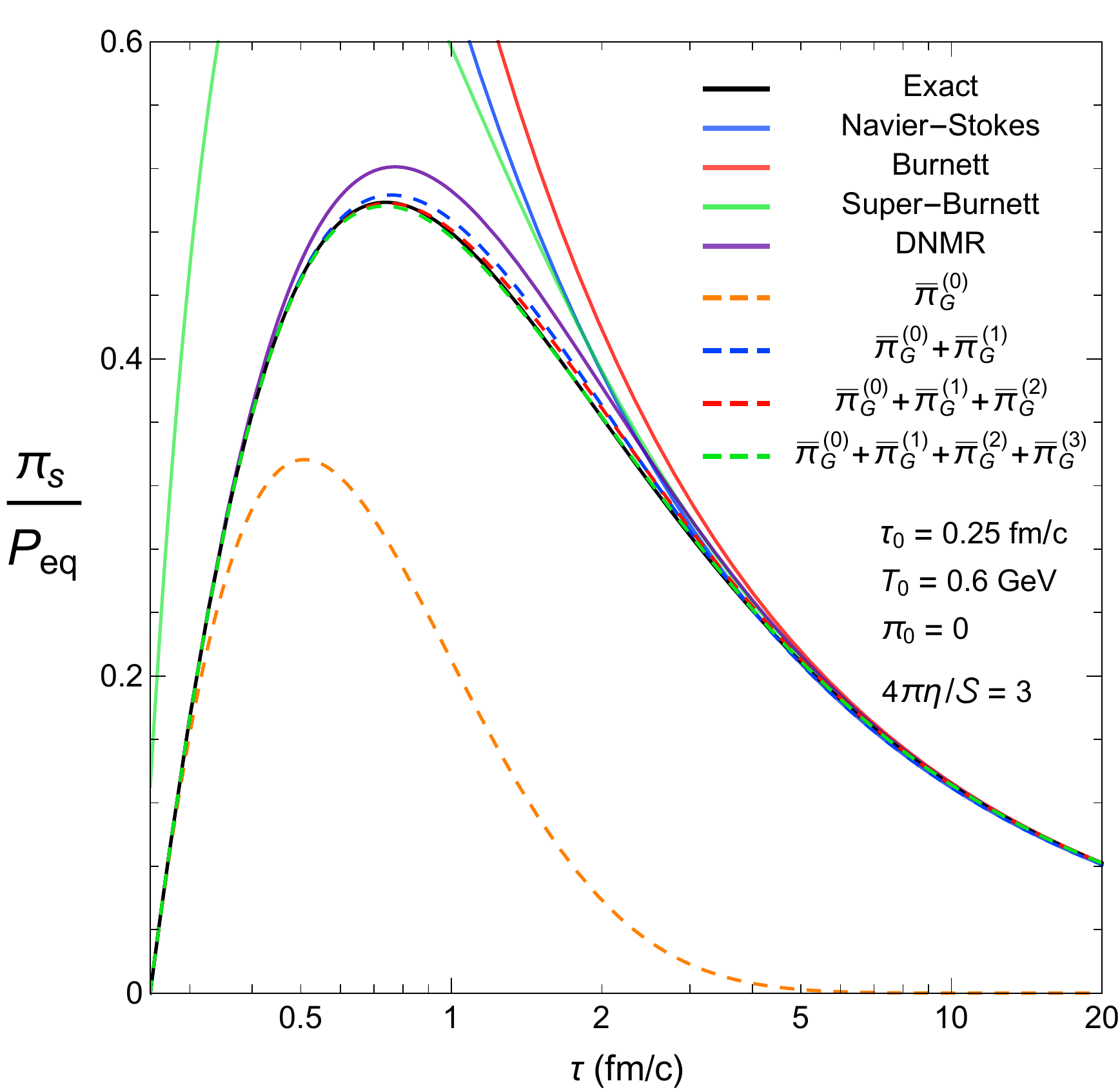}
\centering
\caption{\label{hydro_anomaly}
    The evolution of the exact pressure anisotropy $\bar\pi_s(\tau)$ (solid black) for conformal Bjorken expansion. We plot the contributions of the $\delta f_\G$ corrections (dashed color) to the exact solution and compare them to Navier--Stokes (solid blue), Burnett (solid red), super-Burnett (solid green) and DNMR (solid purple) viscous hydrodynamics.}
\end{figure}
The functions $\mathcal{H}_\perp$ and $\mathcal{H}_L$ are defined in Appendix~\ref{appch8b}. 

The resulting exact evolution of the normalized shear stress is shown as the solid black line in Figure~\ref{hydro_anomaly}. This exact solution is compared with various approximations discussed below. With the exact temperature and shear stress at hand, we evaluate and plot the contributions to the pressure anisotropy from the $\delta f_\G$ corrections up to third order (see Appendices \ref{appch8a} and \ref{appch8b}):
\bs
\allowdisplaybreaks
\label{eqch8:dfg}
\begin{align}
  \bar\pi_{s,\G}^{(0)} =&\, e^{-z_0} \frac{T^4_0}{T^4}\left[\frac{1}{2}\mathcal{H}_\perp\Big(\frac{\tau_0}{\tau}\Big) -  \mathcal{H}_L\Big(\frac{\tau_0}{\tau}\Big)\right] \,, \\
  \bar\pi_{s,\G}^{(1)} =&\, \left(1 - \Gamma(2,z_0)\right) \frac{16\tau_\pi}{15\tau} \,, \\
  \bar\pi_{s,\G}^{(2)} =&\, \left(1 - \frac{\Gamma(3, z_0)}{2!}\right)
  \frac{-16 \tau_\pi^2}{105\tau^2}\,
  \bigl(15 + 49 \tau \partial_\tau{\ln T} \bigr),\\
  \bar\pi_{s,\G}^{(3)} =&\,\left(1 - \frac{\Gamma(4, z_0)}{3!}\right) \frac{16 \tau_\pi^3}{105\tau^3} \bigl(\tau\partial_\tau{\ln T}(135{+}182 \tau\partial_\tau{\ln T}) 
  + 77\tau^2\partial^2_\tau{\ln T} \bigr) \,.
\end{align}
\es
Here the energy conservation law and its time derivative
\bs
\allowdisplaybreaks
\label{eqch8:conservation_law}
\begin{align}
  \tau\partial_\tau{\ln T} &= \frac{\bar\pi_s - 4}{12} \,, \\
  \tau^2\partial^2_\tau{\ln T} &= \frac{4 - \bar\pi_s + \tau \partial_\tau \bar\pi_s}{12} 
\end{align}
\es
are evaluated numerically using the exact solution. We further compare these $\delta f_\G$ corrections to $\bar\pi_s$ to the first-order Navier--Stokes, second-order Burnett and third-order super-Burnett solutions,
\bs
\allowdisplaybreaks
\label{eqch8:Navier_Burnett}
\begin{align}
\bar\pi_s^{(\text{NS})} &= \frac{16 \tau_\pi}{15 \tau} \,, \\
\bar\pi_s^{(\text{B})} &= \frac{16 \tau_\pi}{15 \tau} + \frac{64 \tau_\pi^2}{315 \tau^2} \,, \\
\bar\pi_s^{(\text{SB})} &= \frac{16 \tau_\pi}{15 \tau} + \frac{64 \tau_\pi^2}{315 \tau^2} - \frac{832 \tau_\pi^3}{1575 \tau^3} \,,
\end{align}
\es
as well as to the numerical solution of the standard viscous hydrodynamic equations (DNMR) \cite{Denicol:2012cn,Denicol:2014mca}:
\bs
\allowdisplaybreaks
\label{eqch8:DNMR}
\begin{align}
  \tau\partial_\tau{\ln T} &= \frac{\bar\pi_s - 4}{12} \,, \\
  \partial_\tau \bar\pi_s &= -\frac{\bar\pi_s}{\tau_\pi} + \frac{16}{15\tau} - \frac{10\bar\pi_s}{21\tau} - \frac{\bar\pi_s^2}{3\tau} \,.
\end{align}
\es
At early times, the non-hydrodynamic mode $\delta f_\G^{(0)}$ dominates the evolution of the pressure anisotropy and is responsible for the initial rise away from the local equilibrium initial condition $\bar\pi_{s}(\tau_0)=0$ (see Figure~\ref{hydro_anomaly}). As the system hydrodynamizes,\footnote{%
    \label{fn_hydrodynamization} We define hydrodynamization as the time when the leading non-hydrodynamic mode $\bar\pi_\G^{(0)}$ decays to 10\% of its maximum value. In Fig.~\ref{hydro_anomaly} this occurs at $\tau = 2.47$ fm/$c$ (or $z_0 = 3.6$).} 
the initial-state function decays and the first-order gradient correction $\delta f^{(1)}_\G$ emerges as the leading correction to the local-equilibrium distribution $\feq$ in Eq.~\eqref{eqch8:anomaly}. Already, we see that the addition of $\delta f^{(1)}_\G$ nearly captures the exact pressure anisotropy. This is in stark contrast to the Navier--Stokes solution, which misses both $\delta f_\G^{(0)}$ in (\ref{eqch8:dfg}a) and the prefactor $1-\Gamma(2,z_0)$ in (\ref{eqch8:dfg}b) and hence fails to reproduce the shear stress for $\tau \lesssim 2$ fm/$c$. The reader should also take note of the similarity between the blue-dashed curve and DNMR viscous hydrodynamics, where the $\delta f$ correction used to compute the transport coefficients of the relaxation equation (\ref{eqch8:DNMR}b) is first-order in the shear stress. 

Compared to the second-order correction accounted for in the Burnett solution (\ref{eqch8:Navier_Burnett}b), the full $\delta f_\G^{(2)}$ gradient correction to the shear stress is much weaker at early times since it is strongly suppressed by the corresponding non-hydrodynamic mode. By the time the non-perturbative factor $1-\frac{1}{2!}\Gamma(3,z_0)$ has decayed by 90\% (at around $\tau = 3.9$ fm/$c$), the gradients characterized by the Knudsen number $\Kn = \tau_\pi/\tau \approx 0.2$ have already greatly diminished. As a result, the $\delta f_\G^{(2)}$ correction ends up having little overall impact on the evolution of the system. A similar observation holds for the third-order correction $\delta f_\G^{(3)}$. The combined low-order $\delta f_\G$ corrections to the local-equilibrium distribution are seen to provide excellent agreement with the exact solution. While we caution the reader that this does not necessarily mean the rest of the series \eqref{eqch8:anomaly} will converge, take this observation as justification to truncate the new expansion scheme \eqref{eqch8:anomaly} at a low order: Figure~\ref{hydro_anomaly} makes it clear that, at least for Bjorken flow, gradient corrections beyond first order have almost negligible influence on the fluid's dynamics during the early stages of evolution even though there the expansion rate is large. We conclude that the non-perturbative corrections in Eq.~\eqref{eqch8:anomaly} are essential ingredients in constructing an accurate solution for the RTA Boltzmann equation, along with the usual Chapman--Enskog expansion. 

\section{Applications for conformal Bjorken flow}
\subsection{Far-from-equilibrium hydrodynamics}

The study from the previous section showed that only the initial-state term $\delta f_\G^{(0)}$ and first-order generator correction $\delta f_\G^{(1)}$ are needed to capture most of the shear stress (at least for Bjorken expansion). Therefore, we can apply our new expansion scheme~\eqref{eqch8:anomaly} to derive hydrodynamic equations for a conformal gas subject to Bjorken flow that are just as accurate as anisotropic hydrodynamics with $\PL$--matching.

In conformal Bjorken expansion, there are only two hydrodynamic degrees of freedom: the energy density $\ene$ and the shear stress $\pi_s$. Their kinetic definitions are
\bs
\allowdisplaybreaks
\label{eqch8:kinetic_def}
\begin{align}
    \ene(\tau) &= \int_p \frac{v^2}{\tau^2}\, f(\tau,p_\perp,w) \,,\\
    \pi_s(\tau) &= \frac{2}{3} \int_p \left[\frac{1}{2}p_\perp^2 - \frac{w^2}{\tau^2}\right] \delta f(\tau,p_\perp,w)\,,
\end{align}
\es
where $w = \tau^2 p^\eta$, $v = \sqrt{\tau^2 p_\perp^2 + w^2}$, $\int_p = \int \dfrac{d^2 p_\perp dw}{v(2\pi)^3}$ is the momentum integration measure and $\delta f = f - \feq$ (we excluded the term $\feq$ in the integrand of Eq.~(\ref{eqch8:kinetic_def}b) since it vanishes by symmetry). The evolution equation for $\ene$ is given by the energy conservation law~\eqref{eqch8:conservation_law}, rewritten as
\be
\label{eqch8:edot}
\partial_\tau \ene = \frac{3\pi_s - 4\ene}{3\tau}\,,
\ee
where we used the conformal relation $\ene = 3T^4/\pi^2$. For the shear relaxation equation, we follow Denicol's method~\cite{Denicol:2010xn,Denicol:2012cn} by taking the time derivative of Eq.~(\ref{eqch8:kinetic_def}b). After expanding the r.h.s. of Eq.~(\ref{eqch8:kinetic_def}b) and substituting $\partial_\tau\delta f$ from the RTA Boltzmann equation~\eqref{eqch8:RTA_Bjorken}
\be
\partial_\tau \delta f = -\frac{\delta f}{\tau_\pi} - \partial_\tau \feq \,,
\ee
the relaxation equation for the shear stress is\footnote{%
    In Eq.~\eqref{eqch8:shear_denicol}, there is an additional term $\propto \delta \ene$ (see Eq.~\eqref{eqch8:landau_energy}), which vanishes after Landau matching.}
\be
\label{eqch8:shear_denicol}
\partial_\tau\pi_s = - \frac{\pi_s}{\tau_\pi} + \frac{16\ene}{45\tau} + \frac{1}{\tau}\int_p\left[\frac{8w^2}{3\tau^2} - \frac{w^4}{v^2\tau^2}\right]\delta f\,.
\ee
To close the system of equations~\eqref{eqch8:edot} and~\eqref{eqch8:shear_denicol}, we need some approximation for $\delta f$, for which we take the first two terms of the expansion~\eqref{eqch8:anomaly}:
\be
\label{eqch8:df_FFE}
\delta f = c_0 (f_0 {-} \feq) + c_1 \delta f^{(1)}\,.
\ee
Here we replaced the non-perturbative factors with two coefficients $c_0$ and $c_1$, which are adjusted to match the energy density and shear stress. We solve for the coefficients by inserting Eq.~\eqref{eqch8:df_FFE} into the Landau matching condition for the energy density,
\be
\label{eqch8:landau_energy}
\delta \ene = \int_p \frac{v^2}{\tau^2} \,\delta f = 0\,,
\ee
and into the kinetic shear stress~(\ref{eqch8:kinetic_def}b). After decoupling the algebraic equations one obtains
\be
\label{eqch8:c01}
    c_0 = \frac{\pi^2_s}{\pi_s^{(f_0)} \pi_s + \frac{16}{45}(\ene^{(f_0)} {-} \ene) \ene} \,,\qquad
    c_1 = \frac{\tau}{\tau_\pi} \times \frac{\pi_s (\ene^{(f_0)} {-} \ene)}{\pi_s^{(f_0)} \pi_s + \frac{16}{45}(\ene^{(f_0)} {-} \ene) \ene}\,,
\ee
where
\be
    \ene^{(f_0)} = \int_p \frac{v^2}{\tau^2} \, f_0(\tau_0,p_\perp,w) \,,\qquad
    \pi_s^{(f_0)} = \frac{2}{3} \int_p \left[\frac{1}{2}p_\perp^2 - \frac{w^2}{\tau^2}\right] f_0(\tau_0,p_\perp,w) \,,
\ee
Initially, the coefficients are $c_0(\tau_0) = 1$ and $c_1(\tau_0) = 0$, while at late times we find that they qualitatively follow the evolution of the non-perturbative factors $e^{-z_0}$ and $1-\Gamma(2,z_0)$ (i.e. $c_0 \to 0$ and $c_1 \to 1$).

With the expansion coefficients computed, we can finally write down the final relaxation equation for the shear stress:
\be
\label{eqch8:shear_FFE}
\partial_\tau{\pi_s}_{|\text{FFE}} = - \frac{\pi_s}{\tau_\pi} + \frac{16\ene}{45\tau} + \frac{\zeta_\pi}{\tau}\,,
\ee
which we label as ``far-from-equilibrium" (FFE) hydrodynamics. The shear transport coefficient $\zeta_\pi$ is
\be
\label{eqch8:zeta_pi}
\zeta_\pi = c_0\Big(\frac{8}{3} \mathcal{P}_L^{(f_0)} - \I_{240}^{(f_0)} - \frac{31}{45}\ene\Big) \,-\, \frac{c_1 \tau_\pi}{945\tau} \left(608\ene + 651\pi_s\right)\,,
\ee
where
\be
    \mathcal{P}_L^{(f_0)} =  \int_p \frac{w^2}{\tau^2}\, f_0(\tau_0,p_\perp,w) \,,\qquad
    \I_{240}^{(f_0)} = \int_p \frac{w^4}{v^2\tau^2}\, f_0(\tau_0,p_\perp,w)\,.
\ee
Although the FFE hydrodynamic equations~\eqref{eqch8:edot} and~\eqref{eqch8:shear_FFE} do not make any reference to the exact solution~\eqref{eqch8:exact}, we still need to specify the initial distribution $f_0(\tau_0,p_\perp,w)$ in Eq.~\eqref{eqch8:df_FFE}. This means that the macroscopic evolution equations require some microscopic input, which is unavoidable since we are taking directly into account the evolution of non-hydrodynamic modes. For the following example, we simply use the leading-order anisotropic distribution
\be
\label{eqch8:f0_as_fa}
f_0(\tau_0,p_\perp,w) = f_a(\tau_0, p_\perp,w) = \exp\left[-\frac{\sqrt{\tau_0^2 p_\perp^2 + (1{+}\xi_0)w^2}}{\tau_0\Lambda_0}\right]\,,
\ee
where the initial effective temperature and anisotropy parameter ($\Lambda_0$, $\xi_0$) can vary depending on the initial conditions. However, it is often the case that one only has information about the macroscopic energy-momentum tensor $T^\munu_0$ (and charged currents $J^\mu_0$ if any) on an initial-state hypersurface. For these situations, one could set the initial $\delta f$ correction to the 14--moment approximation while initializing higher moments of the distribution function as zero.

We will compare Eq.~\eqref{eqch8:shear_FFE} to DNMR second-order viscous hydrodynamics~\cite{Denicol:2012cn,Denicol:2014mca}, Jaiswal's third-order viscous hydrodynamics~\cite{Jaiswal:2013vta} and anisotropic hydrodynamics with $\PL$--matching~\cite{Molnar:2016gwq}:
\bs
\allowdisplaybreaks
\label{eqch8:compare_hydro}
\begin{align}
    \partial_\tau {\pi_s}_{|\text{DNMR}} &= - \frac{\pi_s}{\tau_\pi} + \frac{16\ene}{45\tau} - \frac{38\pi_s}{21\tau} \,,\\
    \partial_\tau {\pi_s}_{|\text{Jaiswal}} &= - \frac{\pi_s}{\tau_\pi} + \frac{16\ene}{45\tau} - \frac{38\pi_s}{21\tau} - \frac{54\pi_s^2}{49\ene\tau}\,,\\
    \partial_\tau {\pi_s}_{|\text{AH}} &= - \frac{\pi_s}{\tau_\pi} + \frac{16\ene}{45\tau} + \frac{\frac{8}{3}\PL - \I_{240}^{(a)} - \frac{31}{45}\ene}{\tau}\,,
\end{align}
\es
where
\be
\I_{240}^{(a)} = \int_p \frac{w^4}{v^2\tau^2} \,f_a(\tau,p_\perp,w)
\ee
and we have rewritten Eq.~(\ref{eqchap4:semi_Bjorken}b) in terms of $\pi_s$ using the conformal relation $\PL = \frac{1}{3}\ene - \pi_s$. We note that the transport coefficient $\zeta_\pi$ in Eq.~\eqref{eqch8:zeta_pi} has essentially the same structure as the $\PL$--matching coefficient at the initial time $\tau_0$. In the near-equilibrium limit at late times $\pi_s/\ene \approx 16\tau_\pi / (45\tau) \ll 1$, $\zeta_\pi \to - 38\pi_s / 21$, which is the DNMR coefficient in Eq.~(\ref{eqch8:compare_hydro}a).

\begin{figure}[t]
\includegraphics[width=\linewidth]{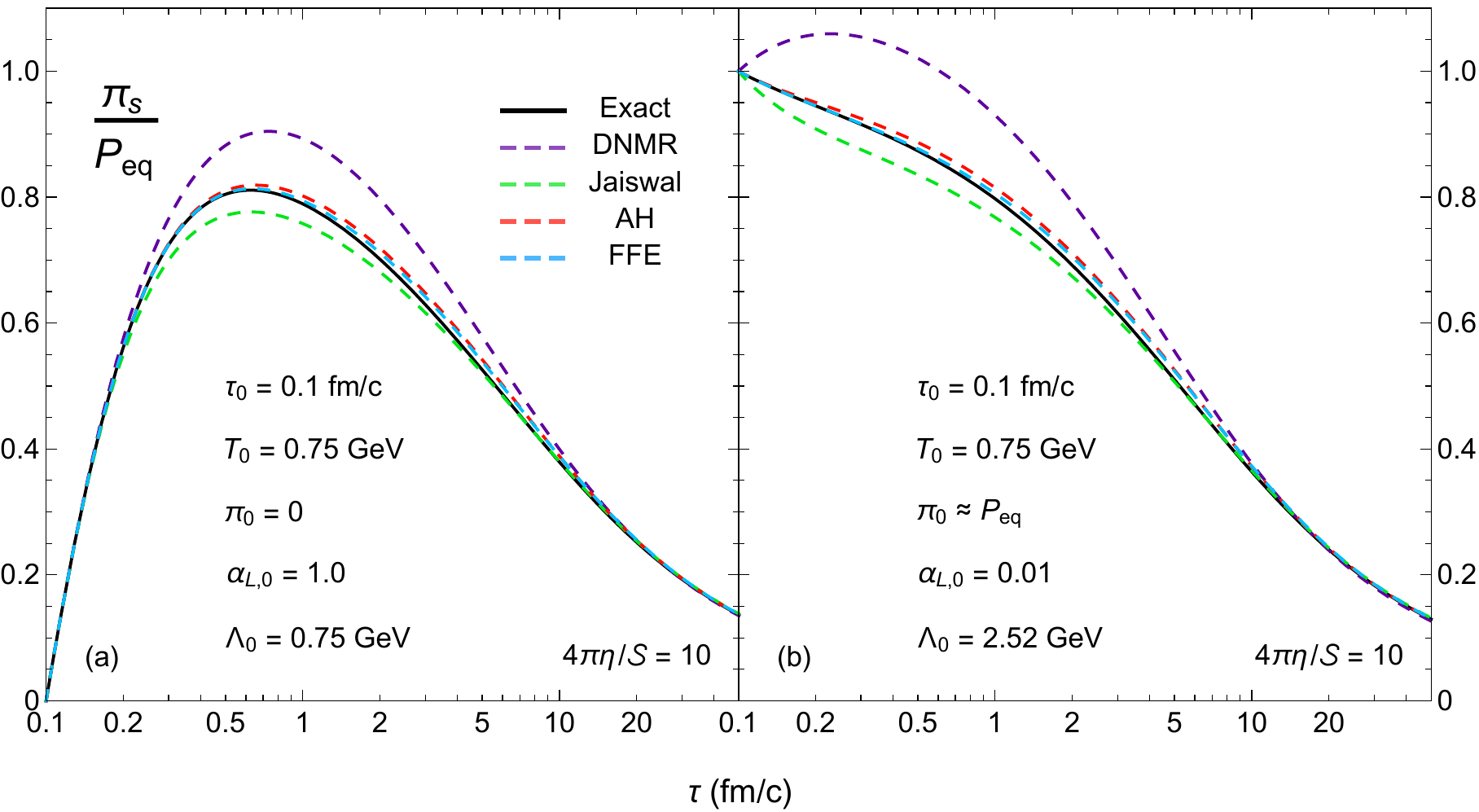}
\centering
\caption{\label{FFE_plot}
    Conformal Bjorken evolution of the normalized shear stress with equilibrium (a) and longitudinally free-streaming (b) initial conditions. We compare the results from DNMR second-order viscous hydrodynamics (dashed purple), Jaiswal's third-order viscous hydrodynamics (dashed green), anisotropic hydrodynamics with $\PL$--matching (dashed red) and FFE hydrodynamics (dashed blue) to the exact RTA solution (solid black).}
\end{figure}

In Figure~\ref{FFE_plot}a we plot the normalized shear stress evolution given by the four hydrodynamic models against the exact conformal RTA solution~\eqref{eqch8:pi_exact}. Compared to Fig.~\ref{hydro_anomaly}, we start at equilibrium ($\bar\pi_s(\tau_0) = 0$) but use more extreme initial conditions: we set the initial time to $\tau_0 = 0.1$ fm/$c$, the initial temperature to $T_0 = 0.75$ GeV and use a larger shear viscosity $\etas = 10/(4\pi)$. By driving the system farther from equilibrium, we can better distinguish the hydrodynamic models. We see that DNMR is qualitatively correct but it overshoots the peak of the exact solution at $\tau \sim 1$ fm/$c$. Jaiswal's third-order method performs better than DNMR hydrodynamics but is still about 5\% below the peak. In contrast, we see that FFE hydrodynamics and the $\PL$--matching scheme almost exactly reproduce the RTA solution. The two approaches also yield nearly identical results in Fig.~\ref{FFE_plot}b, where we start far from equilibrium ($\bar\pi_s(\tau_0) \approx 1$). This indicates a strong connection between anisotropic hydrodynamics and the resummed expansion technique~\eqref{eqch8:anomaly}. 

The study here shows that for a hydrodynamic model to make precise predictions, including the initial state's non-equilibrium dynamics directly in the transport equations is essential. By considering microscopic input in the $\delta f$ expansion~\eqref{eqch8:df_FFE}, the FFE approach reliably captures the exact hydrodynamic evolution. We note that FFE hydrodynamics is similar to another expansion scheme recently proposed by Alwawi and Strickland~\cite{Alalawi:2020zbx}; their ansatz for the $\delta f$ correction is
\be
\label{eqch8:alwawi}
\delta f = e^{-z_0}(f_0 - \feq) \,+ \,(1 - e^{-z_0})(f_a - \feq)\,.
\ee
Compared to Eq.~\eqref{eqch8:df_FFE}, they use strictly exponential coefficients and replace the first-order Chapman--Enskog correction $\delta f^{(1)}$ by the non-equilibrium part of the anisotropic distribution. Their approach~\cite{Alalawi:2020zbx} turns out to be more effective than ordinary anisotropic hydrodynamics. Thus, optimizing the expansion schemes~\eqref{eqch8:df_FFE} and \eqref{eqch8:alwawi} and generalizing them to non-conformal fluids would be worth exploring in future research.
\begin{figure}[t]
\includegraphics[width=0.7\linewidth]{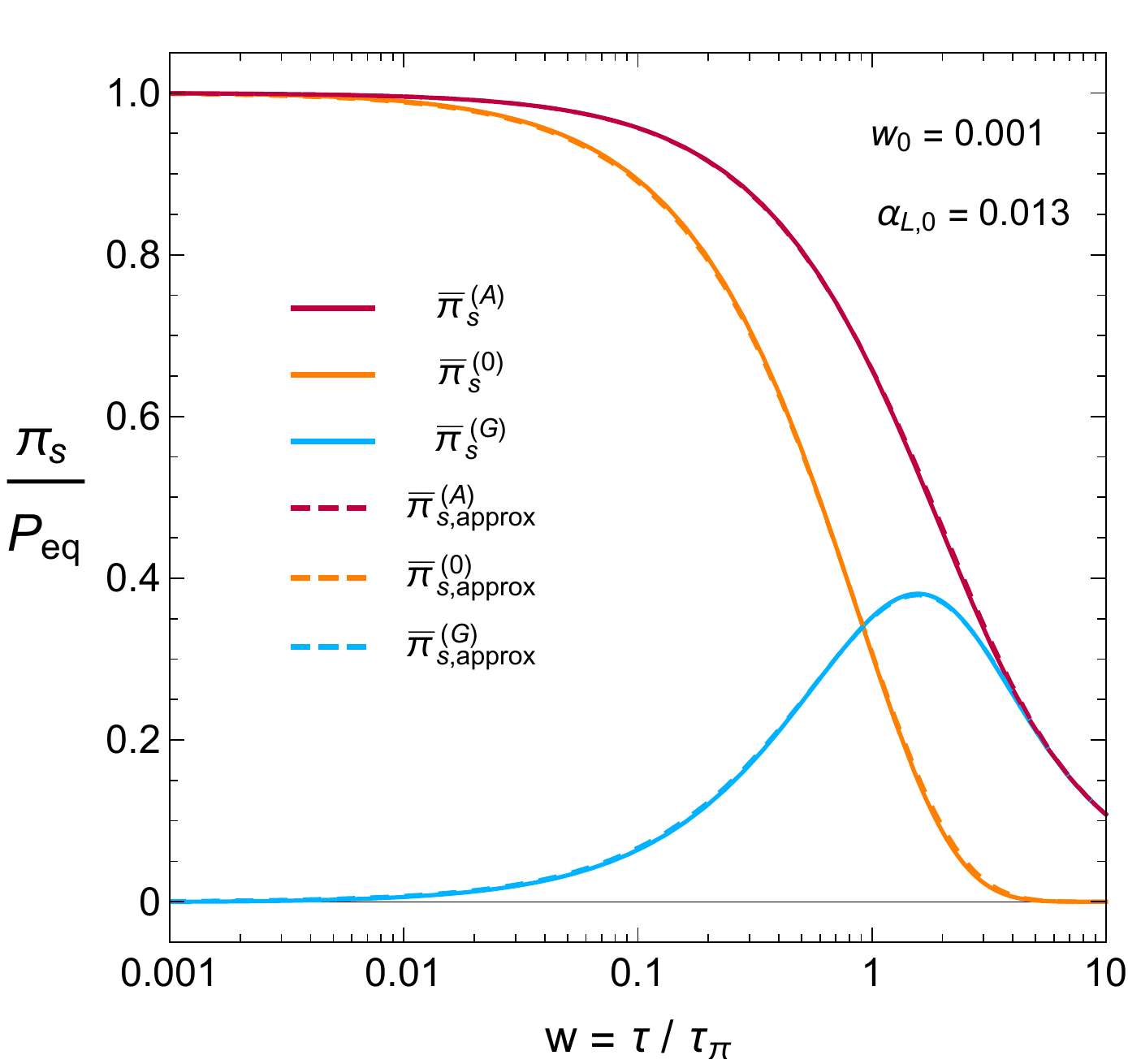}
\centering
\caption{\label{attractor}
    Conformal RTA attractor of the shear inverse Reynolds number as a function of the inverse Knudsen number $\text{Kn}^{-1} = w$ (solid purple). We also plot the contributions from the initial-state term (solid orange) and hydrodynamic generator (solid blue). Our approximations for the total attractor and its constituents are given by the dashed-color curves.}
\end{figure}
\subsection{Hydrodynamic attractor}
We can also use the hydrodynamic generator expansion to construct the hydrodynamic attractor~\cite{Heller:2015dha}, specifically the conformal RTA attractor of the normalized shear stress $\bar\pi_s$. Numerically, the RTA attractor can be extracted by evolving the exact solution~\eqref{eqch8:exact} whose initial conditions satisfy the slow-roll approximation~\cite{Romatschke:2017vte}
\be
\frac{d\bar\pi_s}{dw}_{\big|w = w_0} = 0\,,
\ee
where $w = \tau / \tau_\pi$ and we set the initial value to $w_0 = 10^{-3}$. For an initial anisotropic distribution~\eqref{eqch8:f0_as_fa}, the anisotropy parameter is tuned to $\alpha_{L,0} = (1\,{+}\,\xi_0)^{-1/2} = 0.013$, which corresponds to the longitudinal free-streaming limit $\bar\pi_s(w_0) \approx 1$ or $\PL(w_0) \approx 0$. Figure~\ref{attractor} shows the exact RTA attractor as a function of $w$ (solid purple). We see that the attractor remains finite as the Knusden number $\text{Kn} = 1/w$ increases (i.e. $w\to 0$) instead of diverging like the Navier--Stokes solution in Eq.~(\ref{eqch8:Navier_Burnett}a) (rewritten as $\bar\pi_{s}^{(\text{NS})} = 16/(15w)$).

We approximate the hydrodynamic attractor by first decomposing it as
\be
\bar\pi_{s}^{(\text{A})}(w) = \bar\pi_{s}^{(0)}(w) + \bar\pi_{s}^{(\G)}(w) \,,
\ee
where $\bar\pi_{s}^{(0)}$ and $\bar\pi_{s}^{(\G)}$ are the first and second terms of the exact shear stress~\eqref{eqch8:pi_exact}, respectively. The two contributions are plotted in Fig.~\ref{attractor}; as expected, the initial-state term dominates the attractor for small values of $w$ while the generator term takes over at large $w$. Our strategy is to model these two curves as a function of $w$ by applying our expansion scheme~\eqref{eqapp8:pi0_aniso} and~(\ref{eqch8:dfg}b-d) for a system initialized with $\bar\pi_s(w_0) = 1$ at $w_0\to 0$.

We start by taking the initial-state and first-order generator corrections from Eqs.~\eqref{eqapp8:pi0_aniso} and (\ref{eqch8:dfg}b):
\bs
\allowdisplaybreaks
\label{eqch8:attractor_start}
\begin{align}
    \bar\pi_{s}^{(0)} &= e^{-z_0}\,\left(\frac{T_0}{T}\right)^{4}\times\frac{\mathcal{H}_\perp{\left(\dfrac{w_0 \alpha_{L,0} T}{w T_0}\right)} -  2\mathcal{H}_L{\left(\dfrac{w_0 \alpha_{L,0} T}{w T_0}\right)}}{2\mathcal{H}(\alpha_{L,0})} \,,\\
    \bar\pi_{s}^{(\G)} &= \left(1- \Gamma\left[2,z_0\right]\right) \frac{16}{15w}\,.
\end{align}
\es
In the first line, we replaced $\Lambda_0$ by Eq.~\eqref{eqapp8:lambda_0} and used the relation $\tau_0/\tau = w_0T/(wT_0)$. After taking the limit $\alpha_{L,0} \to 0$, Eq.~(\ref{eqch8:attractor_start}a) reduces to
\be
\label{eqch8:trial_aL_0}
\bar\pi_{s}^{(0)} = e^{-z_0}\,\frac{w_0}{w} \left(\frac{T_0}{T}\right)^3\,.
\ee
The temperature's evolution is given by the energy conservation law~\eqref{eqch8:conservation_law}, rewritten as
\be
\label{eqch8:conservation_w}
w \frac{d\ln T}{dw} = \frac{\bar\pi_s - 4}{8 + \bar\pi_s}\,,
\ee
where we used the chain rule identity $12\tau \frac{d}{d\tau} = w(8 + \bar\pi_s)\frac{d}{dw}$. About $w = 0$, the longitudinal pressure is approximately zero (i.e. $\bar\pi_s \approx 1$); thus, the temperature initially follows the free-streaming solution
\be
\label{eqch8:T_fs}
T \approx T^{(0)} = T_0\left(\frac{w_0}{w}\right)^{1/3}\,.
\ee
As a result, the factors in Eq.~\eqref{eqch8:trial_aL_0} cancel at zeroth order and we can safely take the limit $w_0 \to 0$:
\be
\label{eqch8:zero_order}
\lim_{w_0\to\, 0}\,\frac{w_0}{w} \left(\frac{T_0}{T}\right)^3 = 1 + O(w)\,.
\ee
The first-order expression for the particle interaction measure $z_0$ is 
\be
z_0 = \int_{\tau_0}^\tau \frac{d\tau^\prime}{\tau_\pi(\tau^\prime)} = \int_{w_0}^w \frac{12\,dw^\prime}{8+\bar\pi_s(w^\prime)} \approx z_0^{(1)} = \frac{4w}{3}\,.
\ee
Our first trial solution (not shown in Fig.~\ref{attractor}) then simplifies to
\bs
\allowdisplaybreaks
\label{eqch8:trial_attractor}
\begin{align}
    \bar\pi_{s}^{(0)} &= e^{-z_0^{(1)}} \,,\\
    \bar\pi_{s}^{(\G)} &= \left(1- \Gamma\big[2,z_0^{(1)}\big]\right) \frac{16}{15w}\,.
\end{align}
\es
The combined expression works well for small and large values of $w$ but not for intermediate values. Before considering the higher-order generator corrections, we make a first-order correction to the temperature evolution in Eq.~\eqref{eqch8:zero_order} via perturbation. After inserting the trial solution $\bar\pi_{s} = \bar\pi_{s}^{(0)} + \bar\pi_{s}^{(\G)}$ from Eq.~\eqref{eqch8:trial_attractor} into Eq.~\eqref{eqch8:conservation_w} and solving the differential equation to first order in $w$, one obtains
\be
T \approx T^{(1)} = T_0\left(\frac{w_0}{w}\right)^{1/3}\left(1 - \frac{208w}{3645}\right)\,.
\ee
Thus, our approximation for the initial-state term becomes
\be
\label{eqch8:trial_refined_0}
\bar\pi_{s}^{(0)} = e^{-z_0^{(1)}}\left(1 + \frac{208w}{1215}\right)\,,
\ee
which is given by the orange-dashed curve in Fig.~\ref{attractor}. One sees that it captures the exact $\bar\pi_{s}^{(0)}$ term very well even for intermediate values of $w \sim 0.1 - 1$. Although this solution breaks down for large $w$, the errors relative to the total attractor in that region are exponentially small. 

Now we can finally turn our attention to the higher-order generator corrections~(\ref{eqch8:dfg}c-d), which will further close the gap at intermediate values of $w$. We solve for the second-order generator correction perturbatively by inserting our refined trial solution~\eqref{eqch8:trial_refined_0} and~(\ref{eqch8:trial_attractor}b) into Eq.~(\ref{eqch8:dfg}c):
\be
\bar\pi_{s,\G}^{(2)} = \Bigg(1 - \frac{\Gamma\big[3,z_0^{(1)}\big]}{2!}\Bigg) \frac{16}{105w^2}\,
  \left[\frac{4}{3} - \frac{49\big(\bar\pi_{s}^{(0)} {+} \bar\pi_{s}^{(\G)}\big)}{12}\right]\,.
\ee
Then we add this correction to our current approximation for $\bar\pi_s^{(\G)}$ in Eq.~(\ref{eqch8:trial_attractor}b). We repeat this procedure for the third-order generator correction $\bar\pi_{s,\G}^{(3)}$ in Eq.~(\ref{eqch8:dfg}d), but the expression is too cumbersome to show here. 

Figure~\ref{attractor} shows the reconstruction of the hydrodynamic generator contribution $\bar\pi_s^{(\G)}$ (dashed blue) and the total RTA attractor $\bar\pi_s^{(\text{A})}$ (dashed purple), with errors staying below $2\%$. This is accomplished by using significantly fewer perturbative corrections than what is typically needed to compute the attractor via Borel resummation~\cite{Heller:2015dha,Heller:2016rtz}. Another potential advantage of this technique is that it could be applied to Bjorken expanding systems with different relaxation times. For example, it is known that a direct Borel resummation fails when the relaxation time $\tau_r \propto T^{-\Delta}$ and $\Delta > 3$~\cite{Heller:2018qvh}. Additional work would be needed to verify whether or not this method is capable of constructing the attractor in these situations.  


\section{Hydrodynamic generator in (3+1)--dimensions}
\label{hydro_generator_Minkowski}

\begin{figure}[t]
\includegraphics[width=0.7\linewidth]{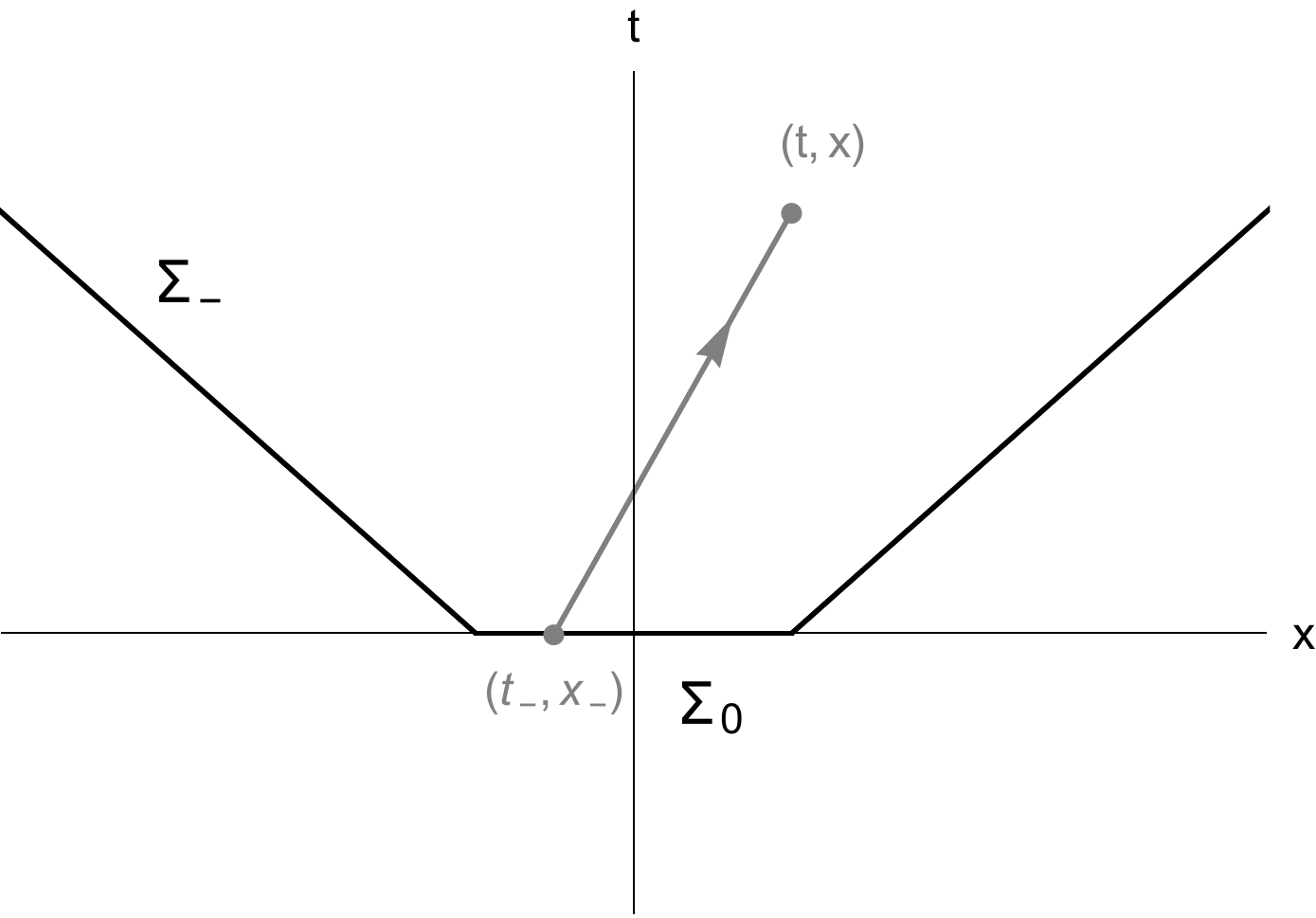}
\centering
\caption{
    \label{minkowski_diagram}
    An illustration of the path integral \eqref{eqch8:generator_minkowski} (solid gray) running from a point $(t_-,\bm{x}_-)$ on the hypersurface $\Sigma_-$ (solid black) to the current position $(t,\bm{x})$. The path is parallel to the particle momentum $p^\mu$ at point $(t,\bm{x})$. (Note that $x$, $x_-$ in the figure stand for 3-dimensional spatial vectors.)
}
\end{figure}

In (3+1)--dimensional Minkowski spacetime $x^\mu = (t,x,y,z)$ without Bjorken symmetry, the hydrodynamic generator in the relaxation time approximation can be further generalized as a path integral along free-streaming characteristics:
\be
\label{eqch8:generator_minkowski}
  f_\text{G}(x,p) = \int^x_{x_-}{dx^\prime {\cdot\,}s^{-1}(x^\prime, p) D(x,x^\prime,p)\feq(x^\prime, p)} \,.
\ee
Here the starting point $x^\mu_- = x^\mu - (t {-} t_-) p^\mu / E$ lies on a hypersurface $t_- = \Sigma_-(x,y,z)$ consisting of the initial-state boundary $\Sigma_0$ and the future light cone enclosing it (see Figure \ref{minkowski_diagram}).
Note that, in contrast to Eq.~(\ref{eqch8:general_generator}), the (3+1)--dimensional generator is not constrained by any symmetries and (after Landau matching) can accommodate any flow velocity profile $u^\mu(x)$, including ones with non-vanishing vorticity (which for Bjorken flow is forbidden by symmetry). The path integral (\ref{eqch8:generator_minkowski}) runs over a straight time-like characteristic line parallel to the particle momentum $p^\mu$, with a measure $dx'\cdot s^{-1}(x',p)$ that (unlike the one in Eq.~(\ref{eqch8:general_generator})) depends on momentum. The fraction of particles with momentum $p^\mu$ emitted from the thermal source $\feq(x^\prime,p)$ that travel freely through the medium and arrive at the current position $x^\mu$ unscathed is given by the damping function~\eqref{eqch8:damping}. For short relaxation times this damping function decays very rapidly, limiting the range of influence on the fluid's dynamics at position $x$ to points $x'$ in the past light cone of $x$ with small spacetime separations $x{-}x'$.  

Let us verify that the expansion of Eq.~\eqref{eqch8:generator_minkowski} reduces to the more general Borel resummed RTA Chapman--Enskog expansion (up to some finite order):
\be
\label{eqch8:Chapman_series_Minkowski}
  f^{\text{B}}_\text{CE}(x,p) = \int_0^\infty dz\,e^{-z} \sum_{n=0}^\infty \frac{z^n\left[- s^\mu(x,p)\partial_\mu\right]^n \feq(x,p)}{n!} \,.
\ee
Following the same steps outlined in the previous section, we use the coordinate transformation
\be
\label{eqch8:z_minkowski}
  z = h(x^\prime, x, p) = \int^x_{x^\prime} dx^{\prime\prime}{\cdot\,}s^{-1}(x^{\prime\prime}, p)
\ee
to rewrite Eq.~\eqref{eqch8:generator_minkowski} as
\be
\label{eqch8:generator_z_minkowski}
  f_\text{G}(x,p) = \int_0^{z_-} dz\, e^{-z} \feq(h^{-1}(z, x, p), p) \,,
\ee
where
\be
  z_- = \int^x_{x_-} dx^{\prime\prime}{\cdot\,}s^{-1}(x^{\prime\prime}, p) \,.
\ee
The inverse function $h^{-1}_\mu(z, x, p)$ is now promoted to a four-vector that can be expanded as a power series:
\be
  x^\prime_\mu = h^{-1}_\mu(z, x, p) = \sum_{n=0}^\infty c_{n,\mu}(x,p)\,z^n \,,
\ee
which can be inserted in the Taylor expansion of Eq.~\eqref{eqch8:z_minkowski}:
\be
\begin{split}
  z =& {\int^{\lambda^\prime}_0} d\lambda^{\prime\prime}\, g(x^{\prime\prime}(\lambda^{\prime\prime}),p) 
  = {\int^{\lambda^\prime}_0} d\lambda^{\prime\prime}\, \sum_{n=0}^\infty  \frac{(-\lambda^{\prime\prime})^n p^n {\cdot}\, \partial^n g(x,p)}{n!} \\
  =& -\sum_{n=0}^\infty  \frac{(-\lambda^\prime)^{n+1} p^n {\cdot}\, \partial^n g(x,p)}{(n+1)!} \,,
\end{split}
\ee
where we used the parameterization
\be
  x^{\prime\prime\mu}(\lambda^{\prime\prime}) = x^\mu - \lambda^{\prime\prime} p^\mu\,, \indent\indent (0 \leq \lambda^{\prime\prime} \leq \lambda^\prime),
\ee
with $\lambda^\prime = (x{-}x^\prime)\cdot p / p^2$ and $g(x,p) = p \cdot u(x) / \tau_r(x)$. Using the product rule identities
\bs
\allowdisplaybreaks
\begin{align}
  s^\mu\partial_\nu s^{-1}_\mu =& - s^{-1}_\mu\partial_\nu s^\mu\,, \\
  s^\mu\partial_\alpha \partial_\nu s^{-1}_\mu =& - s^{-1}_\mu\partial_\alpha\partial_\nu s^\mu - (\partial_\alpha s^{-1}_\mu) (\partial_\nu s^\mu) - (\partial_\nu s^{-1}_\mu) (\partial_\alpha s^\mu) \,,
\end{align}
\es
one obtains, after some algebra, the following first coefficients of the series:
\bs
\allowdisplaybreaks
\begin{align}
    c_0^{\,\mu} &= x^\mu \,, \\
    c_1^{\,\mu} &= - s^\mu \,, \\
    c_2^{\,\mu} &= \frac{1}{2!} s^\nu\partial_\nu s^\mu \,, \\
    c_3^{\,\mu} &= -\frac{1}{3!}\left((s^\alpha\partial_\alpha s^\nu)\partial_\nu s^\mu + s^\alpha s^\nu \partial_\alpha \partial_\nu s^\mu \right) \,,
\end{align}
\es
analogous to the coefficients \eqref{eqch8:coefficients}. They obey the recurrence relation
\bs
\label{eqch8:recursion_general}
\allowdisplaybreaks
\begin{align}
     c_0^{\,\mu} &= x^\mu \,, \\
     c_n^{\,\mu} &= - \frac{s^\nu \partial_\nu c_{n{-}1}^{\,\mu}}{n} \indent \forall \, n \geq 1 \,,
\end{align}
\es
the proof of which is given in Appendix~\ref{appch8proof}. The integral \eqref{eqch8:generator_z_minkowski} can then be Taylor expanded as
\be
\label{eqch8:generator_expand}
  f_\text{G}(x,p) {=} \int_0^{z_-}{dz\, e^{-z} \sum_{n=0}^\infty \frac{(h^{-1}(z,x,p) - x)^n \cdot \feq^{(n)}(x, p)}{n!}} \,,
\ee
where $\feq^{(n)}(x,p) = \partial^n \feq(x,p)$. The series expansion of the hydrodynamic generator up to third order in the Knudsen number is
\be
\begin{split}
  f_\text{G} \approx& \, (1 - e^{-z_-})\feq \,+\, \big(1 - \Gamma(2,z_-)\big)\delta f^{(1)} \,+\, \\
  &\Big(1 - \frac{\Gamma(3,z_-)}{2!}\Big) \delta f^{(2)} \,+\, \Big(1 - \frac{\Gamma(4,z_-)}{3!}\Big)\delta f^{(3)} \,,
\end{split}
\ee
which has the same structure as Eq.~(\ref{eqch8:generator_truncated}). For spacetime regions far in the future from the hypersurface $\Sigma_-$ in Fig.~\ref{minkowski_diagram} we assume we can take the limit $z_- \to \infty$:
\be
\label{eqch8:minkowski_generator_truncated}
  f_\text{G} \approx \feq + \delta f^{(1)} + \delta f^{(2)} + \delta f^{(3)} \,,
\ee
where
\bs
\allowdisplaybreaks
\begin{align}
    \delta f^{(1)} =& \,- s^\mu\partial_\mu \feq \,, \\
    \delta f^{(2)} =& \, (s^\nu\partial_\nu s^\mu) \partial_\mu \feq \,+\, s^\nu s^\mu\partial_\nu \partial_\mu \feq \,, \\
    \delta f^{(3)} =& \,- \left((s^\alpha\partial_\alpha s^\nu)\partial_\nu s^\mu\right)\partial_\mu \feq \,-\, (s^\alpha s^\nu \partial_\alpha \partial_\nu s^\mu) \partial_\mu \feq \\\nonumber &\,-\, 3 s^\mu (s^\alpha \partial_\alpha s^\nu) \partial_\nu \partial_\mu \feq \,-\, s^\alpha s^\nu s^\mu \partial_\alpha \partial_\nu \partial_\mu \feq \,.
\end{align}
\es
As expected, these agree with the corresponding gradient corrections from the RTA Chapman--Enskog expansion when worked out to third order. A proof that the (3+1)--d hydrodynamic generator expansion reduces to the Borel resummation~\eqref{eqch8:Chapman_series_Minkowski} at all orders in the Knudsen number is given in Appendix~\ref{appch8proof}.

The distribution function $f_\text{G}(x,p)$ given in Eq.~\eqref{eqch8:generator_minkowski} is also a particular solution of the RTA Boltzmann equation in Minkowski spacetime~\eqref{eqch8:RTAmink}:
\be
\label{eqch8:verify_fG}
\begin{split}
  s^\mu(x,p) \partial_\mu f_\text{G}(x,p) &= \feq(x,p) - {\int^x_{x_-}}dx^\prime {\cdot\, }s^{-1}(x^\prime, p) D(x,x^\prime,p)\feq(x^\prime, p)
  \\
  &= \feq(x,p) - f_\text{G}(x,p) \,.
\end{split}
\ee
Here we used the identities $s^\mu(x,p)\partial_\mu D(x,x^\prime,p) = - D(x,x^\prime,p)$
and $D(x,x,p)=1$.\footnote{%
    Note that the directional derivative $s^\mu \partial_\mu$ does not act on the lower limit of the path integral \eqref{eqch8:generator_minkowski} since the current position $x^\mu$ varies infinitesimally only along the direction of $s^\mu$, which means that the starting point $x^\mu_-$ remains fixed.}
Unlike Eq.~\eqref{eqch8:exact}, the distribution function \eqref{eqch8:generator_minkowski} approaches zero on the entire initial-state surface $\Sigma_0$ since it does not include any initial-state information. One can make use of the diagram in Fig.~\ref{minkowski_diagram} to construct and add such an initial-state term. We know that the system is initialized at time $t_0$ as $f_0(x_0,p)$, with $x^\mu_0 \in \Sigma_0$. In addition, we assume that the medium on $\Sigma_0$ is surrounded by vacuum so that $f(x,p)$ vanishes on its entire future light cone. Therefore, only characteristic lines that are connected to the initial-state surface $\Sigma_0$ as shown in Fig.~\ref{minkowski_diagram} will pick up an initial source that decays over time:
\be
  f_\text{I}(x,p) = D(x,x_-,p) f_0(x_-,p) \Theta(t_0 - t_-)\,.
\ee
Here the Heaviside step function $\Theta(t_0 - t_-)$ excludes those characteristic lines that end on the light cones in Fig.~\ref{minkowski_diagram} and thereby enforces $f_\text{I}(x,p) = 0$ if $x_- \notin \Sigma_0$. The full (3+1)--dimensional solution of the RTA Boltzmann equation \eqref{eqch8:RTAmink} is then
\be
\label{eqch8:rta_solution}
  f(x,p) = f_\text{I}(x,p) + f_\text{G}(x,p)\,.
\ee
One can check that
\be
  s^\mu(x,p) \partial_\mu f(x,p) = -f_\text{I}(x,p) + \feq(x,p) - f_\G(x,p) = \feq(x,p) - f(x,p)\,,
\ee
where we used the relation $s^\mu(x,p) \partial_\mu D(x,x_-,p) = - D(x,x_-,p)$. A more formal derivation of this solution can be found in Appendix~\ref{appch8c}.\footnote{%
    Equation~\eqref{eqch8:rta_solution} generalizes the RTA Bjorken solution~\eqref{eqch8:exact} to (3+1)--dimensional systems, by replacing the integration over the fluid's history in $\tau^\prime$ with one over a path parameter $\lambda^\prime$ along a set of free-streaming past world lines. Each world line's direction depends on the momentum of the incoming particle, emitted by either an initial source $f_0(x_0,p)$ or a thermal source $\feq(x^\prime,p)$. Macroscopic observables at a given spacetime coordinate $x^\mu$ are influenced by the fluid's history encoded in these world lines.} In the free-streaming limit $\tau_r \to \infty$ ($z_- \to 0)$, the distribution function takes on the free-streaming solution $f_0(x_-,p) \Theta(t_0 {-} t_-)$. In the ideal hydrodynamic limit $\tau_r \to 0$ ($z_- \to \infty)$, $f(x,p) \to f_\G(x,p)$, which reduces to $\feq(x,p)$ since $s^\mu\partial_\mu f_\G(x,p) \to 0$ in Eq.~\eqref{eqch8:verify_fG}.\footnote{In the ideal hydrodynamic limit, the local equilibrium density operator can only accommodate flow profiles that are irrotational~\cite{Becattini:2019dxo}.}

This completes our formal argument for the RTA Boltzmann equation in the (3+1)--dimensional case. We leave its numerical implementation to future work but close this section with some thoughts about how such an implementation might look. Just like the Bjorken solution \eqref{eqch8:exact}, the distribution function \eqref{eqch8:rta_solution} for (3+1)--dimensional expansion is an {\it implicit} solution of the RTA Boltzmann equation since it depends on the temperature $T(x)$ and fluid velocity $u^\mu(x)$. In principle, the hydrodynamic fields can be reconstructed by matching the solution to the Landau frame:
\bs
\label{eqch8:Landau}
\begin{align}
  \mathcal{E}(x) &= \int_p (p\cdot u(x))^2 f(x,p)\,, \\
  u^\mu(x) &= \dfrac{\int_p (p\cdot u(x)) \,p^\mu f(x,p)}{\int_p (p\cdot u(x))^2 f(x,p)} \,,
\end{align}
\es
where $\int_p = \int \dfrac{d^3p}{E}$. Similar to Eq.~\eqref{eqch8:T_exact}, these integral equations can then be solved numerically with a root-finding algorithm such as fixed-point iteration. Starting with an approximate solution for $T(x)$ and $u^\mu(x)$, which can be provided e.g. by a viscous hydrodynamic simulation, one would repeatedly update the solution by evaluating the right-hand-side of Eq.~\eqref{eqch8:Landau}. Since the initial guess and exact solution share the same initial condition but may differ greatly for later times, this numerical scheme is likely to converge faster at times near $t_0$ than at later times. Instead of computing a single iteration across the entire evolution, as is commonly done \cite{Florkowski:2013lya}, it would here be more efficient to perform these iterations at a given time step until the solution is within the desired error tolerance, before proceeding to the next time step. Faster rates of convergence might be achievable if a more accurate hydrodynamic model is used to evolve the initial guess for the fluid's energy density and flow profiles. 

Due to the momentum dependence of the characteristic lines and their associated damping functions, solving the integral equations \eqref{eqch8:Landau} is much more involved than for the Bjorken case. Unfortunately, there does not seem to be a way of reducing the momentum-space integral without invoking additional symmetries like in Eq.~\eqref{eqch8:T_exact}. This leaves us with the computationally intensive task of numerically evaluating a four-dimensional integral for each spacetime point: three for the momentum and one for the path parameter along the associated characteristic line. One possible way to reduce the computing time is to parallelize at each time step the computation over the spatial grid points. Doing this on a GPU, however, still faces memory limitations: for a uniform spacetime grid, the memory required to do a full calculation of the distribution function grows rapidly with the volume of the future light cone, $V \propto t^4$. For short relaxation times, rapid damping will reduce the need for RAM to only a fraction of the fluid's evolution history. Still, the task looks formidable and will likely require a highly advanced algorithm and significant computing resources.

Based on the structure of our formal (3+1)--d solution we anticipate that effects qualitatively similar to those described in Sec.~\ref{expansion} will also be found for RTA kinetic fluids without Bjorken symmetry: at early times, the dynamics of the fluid is dominated by the non-hydrodynamic mode associated with the initial state $f_0(x_0,p)$. As time moves away from the initial-state surface $\Sigma_0$ the local-equilibrium distribution $\feq(x,p)$ and first-order gradient correction $\delta f^{(1)}(x,p)$ quickly take over, with the higher-order corrections emerging more slowly. A quantitative analysis of the contributions from the non-equilibrium corrections $\delta f_\G$ to macroscopic observables will need to wait until the corresponding codes have been developed. Intermediate studies of systems with reduced symmetry (for example undergoing spherical expansion) may be useful for developing intuition and computational tools.
    
\section{Summary}

In this chapter we established a correspondence between the hydrodynamic generator and the Borel resummed Chapman--Enskog series for the (3+1)--d RTA Boltzmann equation. Instead of computing the Borel resummation directly, we used the Green's function method to construct a map to the integral representation of the hydrodynamic gradient series in the late-time limit. Relaxing this limit, we find at earlier times a set of non-hydrodynamic modes that are coupled to the RTA Chapman--Enskog expansion. For Bjorken flow, we showed that these non-hydrodynamic modes control the emergence of hydrodynamics as an effective field theory description of non-equilibrium fluids. As the initial-state memory decays, the local-equilibrium distribution and its first-order gradient correction emerge as the leading contributors to the fluid's dynamics. Higher-order gradient corrections to the particle distribution function are suppressed during the hydrodynamization process, especially at early times. This means that even in far-off-equilibrium situations, where the Knudsen number is initially large, these higher-order corrections are not as severe as traditionally thought. 

Our formula \eqref{eqch8:rta_solution} for the (3+1)--dimensional solution of the RTA Boltzmann equation in Minkowski spacetime has the nice features of being positive definite and finite for both small and large values of the Knudsen number. It is also causal since it only depends on the present and past hydrodynamic fields. While it is not immediately obvious how to numerically implement this solution, it can potentially serve as a reference to test the validity of known viscous hydrodynamic approximations, as well as the new expansion scheme described in this chapter, in the relaxation time approximation without the need for Bjorken symmetry.

\chapter{Conclusions and outlook}

In this thesis, we have developed a set of tools to complete an event-by-event anisotropic hydrodynamic simulation of ultrarelativistic heavy-ion collisions. This was accomplished by deriving a new macroscopic set of equations for anisotropic hydrodynamics and numerically implementing them in \cpuvah{} to model realistic nuclear collisions. We showed that our model is applicable at much earlier times when standard second-order viscous hydrodynamics breaks down. Thus, we were able to use anisotropic hydrodynamics to evolve most of the far-off-equilibrium stage $\tau \in (0,1]$ fm/$c$ without needing to use a separate pre-hydrodynamic module. We then merged our hydrodynamic code with the particlization module {\sc iS3D} to study the effects our PTMA modified anisotropic distribution had on the sampled particle spectra (e.g. elliptic flow at high centralities decreased relative to the 14--moment approximation in Fig.~\ref{VAH_Grad_PTMA}). In our preliminary analysis, we launched these two modules together in an end-to-end simulation to model Pb+Pb collisions at the LHC ($\sqrt{s_\text{NN}} = 2.76$ TeV) and accurately reproduced the experimental hadronic observables. This marks a significant advancement in the field of anisotropic hydrodynamics, which only recently has started being used in heavy-ion phenomenology~\cite{Alqahtani:2017jwl, Alqahtani:2017tnq, Almaalol:2018gjh}. 

A more comprehensive model-to-data analysis that would involve both a re-optimization of the Bayesian model parameters and the inclusion of experimental data from other collision systems (e.g. Au+Au $\sqrt{s_\text{NN}} = 200$ GeV) will be left to future studies. Once completed, this will allow us to investigate how different selections among a discrete set of hydrodynamic models affect both the theoretical description of heavy-ion experimental observables and the shear and bulk viscosity constraints inferred from such data--theory comparisons. In particular, it would be interesting to test the predictions from anisotropic hydrodynamics against a variety of initializations of second-order viscous hydrodynamics, using different pre-hydrodynamic evolution models (or no pre-hydrodynamic stage at all), similar in spirit to the recent study done in Ref.~\cite{NunesdaSilva:2020bfs}.

The current version of our code only evolves the quark-gluon plasma's energy-momentum tensor components and ignores the effects of net baryon density and diffusion. Second-order viscous hydrodynamic codes with nonzero $n_B$ and $V_B^\mu$ have already been practically implemented \cite{Monnai:2012jc, Denicol:2018wdp}, and others that initialize viscous hydrodynamics with dynamical sources \cite{Hirano:2012kj, Shen:2017bsr, Shen:2018pty,Du:2018mpf, Du:2019obx} are currently under development. The \beshydro{} code \cite{Du:2018mpf, Du:2019obx}, in particular, shares a common ancestry \cite{Bazow:2017ewq} and therefore a number of similar features with \cpuvah{}. However, we have not yet considered the effects of nonzero chemical potentials for net baryon number, electric charge and strangeness in our formulation of anisotropic hydrodynamics. An upgrade of our code package that includes them is left to future work. 

We note that anisotropic hydrodynamics is a sub-field of far-from-equilibrium hydrodynamics, which has only been around for about a decade. There are still many things left to learn about the remarkable applicability of hydrodynamics in systems with large gradients. In Chapter~\ref{chap8label}, we introduced a new technique based on Green's functions to extract the resummed gradient series of the linear RTA Boltzmann equation. It would be much more difficult to derive a hydrodynamic generator for the full nonlinear Boltzmann equation, which generally has a much more complex collision kernel. For strongly coupled fluids whose multi-particle correlations cannot be neglected, we would also have to consider stochastic fluctuations and the microscopic correlations that they generate \cite{Akamatsu:2016llw, Schlichting:2019abc, Bluhm:2020mpc}. However, if we perturb the system around local equilibrium (or generally, around some known background~\cite{Jeon:2015dfa, Kamata:2020mka}), it might be possible to at least derive an approximation for the corresponding Green's function from either the Boltzmann equation or a Kadanoff--Baym type equation~\cite{PhysRev.124.287}. It remains an open question whether or not the concept of a hydrodynamic generator can be extended to other systems, but if it turns out that constructing them is possible, we could potentially use them to develop better hydrodynamic models for far-from-equilibrium fluids.
\begin{appendices}
\chapter{}
\section{Anisotropic integrals in conformal Gubser flow}
\label{appch2:anisint}
Here we calculate the conformal anisotropic thermal integrals $\Ih_{nlq}$ and $\hint_{nlr}$ that appear Chapter~\ref{chap2label}. First we define
\be
\label{eq:ch2anisint}
   \Ih_{nlq}\big(\hat{\Lambda},\xi\big)
   = \int_{\pp} (\pp^\rho)^{n-l-2q}\,\pp_\eta^{\,l} \bigg(\frac{\po}{\cosh^2\!\rho}\bigg)^q f_a\,,
\ee
where the anisotropic distribution $f_a$ is given by Eq.~\eqref{eq:ch2RSansatz}. The change of variables 
\bs
\allowdisplaybreaks
\label{eq:ch2changecoor}
\beal
\frac{\pth}{\cosh\rho}&=\lambda\,\sin\alpha\,\cos\beta\,,\\
\frac{\pph}{\cosh\rho\,\sin\theta}&=\lambda\,\sin\alpha\,\sin\beta\,,\\
\pp_\eta &=\lambda\,(1{+}\xi)^{-1/2}\cos\alpha
\end{align}
\es
leads to the following factorization of the integral~\eqref{eq:ch2anisint}:
\be
\label{eq:ch2anisint2}
\Ih_{nlq}\big(\hat{\Lambda},\xi \big)
   =\,\Jhat_{n}\big(\hat{\Lambda}\big)\,\Rhat_{nlq}\left(\xi\right)\,,
\ee
where
\bea
\allowdisplaybreaks
\label{eq:ch2Jfunc}
\Jhat_{n}\big(\hat{\Lambda}\big)
  &=& \int_0^\infty \frac{d\lambda}{2\pi^2}\,\lambda^{n+1}\, e^{-\lambda/\hat{\Lambda}} 
    = \frac{(n{+}1)!}{2\pi^2}\,\hat{\Lambda}^{n+2},
\\
\label{eq:ch2Rfunc}
\nonumber
\Rhat_{nlq}\left(\xi\right)
  &=&\frac{1}{2(1{+}\xi)^{\left(n-2q\right)/2}} \int_{-1}^1 d\cos\alpha\left(\sin\alpha\right)^{2q}\left(\cos\alpha\right)^l
\\
  &&\times\left[(1{+}\xi)\sin^2\!\alpha+\cos^2\!\alpha\right]^{(n-l-2q-1)/2}\,.
\eea
The $\Rhat_{nlq}$ functions that appear in Chapter~\ref{chap2label} are
\bs
\allowdisplaybreaks
\label{eq:ch2Rfunctions}
\beal
\label{eq:ch2R}
\Rhat_{200}(\xi)&=\frac{1}{2}\left(
\frac{1}{1{+}\xi}+t(\xi)
\right) &
\partial_\xi \Rhat_{200}(\xi)&=\frac{1}{4\xi}\left(
\frac{1{-}\xi}{(1{+}\xi)^2}-t(\xi)
\right) \\
\Rhat_{201}(\xi)&=\frac{1}{2\xi}\left( 1+ (\xi{-}1) t(\xi) \right) &
\Rhat_{220}(\xi)&=\frac{1}{2\xi}\left(
-\frac{1}{1{+}\xi}+t(\xi)
\right) \\
\Rhat_{240}(\xi)&=\frac{1}{2\xi^2}\left(
\frac{3{+}2\xi}{1{+}\xi}-3t(\xi)
\right) &
\Rhat_{301}(\xi)&=\frac{2}{3(1{+}\xi)^{3/2}} \\
\Rhat_{320}(\xi)&=\frac{1}{3(1{+}\xi)^{1/2}} &
 \Rhat_{340}(\xi)&=\frac{1}{\xi^2(1{+}\xi)^{1/2}}\left(
-\frac{(3{+}4\xi)}{3(1{+}\xi)}+h(\xi)
\right) \\
\Rhat_{400}(\xi)&=\frac{1}{8}\left(
\frac{5{+}3\xi}{(1{+}\xi)^2}+3t(\xi)
\right) &
\Rhat_{420}(\xi)&=\frac{1}{8\xi}\left(
\frac{\xi {-} 1}{(1{+}\xi)^2}+t(\xi)
\right)\\
\Rhat_{440}(\xi)&=\frac{1}{8\xi^2}\left(
-\frac{3{+}5\xi}{(1{+}\xi)^2}+3t(\xi)
\right)  &
\Rhat_{460}(\xi)&=\frac{1}{8\xi^3}\left(
\frac{8\xi^2{+}25\xi{+}15}{(1{+}\xi)^2}-15t(\xi)
\right) \,,
\end{align}
\es
where $t(\xi) = \arctan\sqrt{\xi} / \sqrt{\xi}$ and $h(\xi) = \text{arctanh}\sqrt{\frac{\xi}{1{+}\xi}} / \sqrt{\frac{\xi}{1{+}\xi}}$. In Section~\ref{sec:fluid} we needed the moments $\Ih_{nlq}^\mathrm{eq}$ associated with the equilibrium distribution function. They are obtained as the isotropic limit of Eq.~\eqref{eq:ch2anisint2}:
\be
\label{eq:ch2equilmom}
\Ih_{nlq}^\mathrm{eq}(\temhat)\equiv \lim_{\xi\to 0} \Ih_{nlq}\big(\hat{\Lambda},\xi \big)= \Ih_{nlq}\big(\temhat,0)\,.
\ee
Finally, we define the anisotropic integrals
\be
\label{eq:ch2Hfunc}
  \hint_{nlr}\big(\hat{\Lambda},\xi\big) 
  = \int_{\pp}\! \hat{E}_{\pp}^{\,n{-}l}\,\pp_\eta^{\,l}\,E_\mathrm{RS}^r\,f_a\,.
\ee
With the change of variables~\eqref{eq:ch2changecoor} one can show that
\be
\label{eq:ch2Hfunc2}
\hint_{nlr}\big(\hat{\Lambda},\xi\big) = \Jhat_{n+r}\big(\hat{\Lambda}\big)\,\Rhat_{nl0}\left(\xi\right).
\ee
%
\section{14--moment approximation in the NLO NRS scheme}
\label{app:14mom}

In this section we present the calculation of the four coefficients which enter into the 14--moment approximation \eqref{eq:ch214gradgub} for $\tdf$. We begin with the most general 14--moment ansatz for anisotropic fluids
\cite{Bazow:2013ifa} and decompose it in LRF coordinates for Gubser flow, using the notation $\hat{E}_{\pp} \equiv \pp^\rho=-\hat{u}\cdot\pp$:
\be
\label{eq:ch214grad}
\begin{split}
\tdf  \approx \tdf_{14} 
&= \left[\hat{\alpha} + \hat{\beta}_\mu \hat{p}^\mu + \hat{\omega}_{\mu\nu} \hat{p}^\mu \hat{p}^\nu \right] f_a \\
&= \left[ \hat{\alpha} - \hat{\beta} \hat{E}_{\pp} 
  + \hat{\omega}\Big(\hat{E}_{\pp}^2{+}\textstyle{\frac{1}{3}}\hat{\mathbf{p}}^2\Big)
  + \Big(\hat{\beta}_{\langle\mu\rangle}{-}2\hat{E}_{\pp}\,\hat{\omega}_{\langle\mu\rangle}\Big)
             \hat{p}^{\langle\mu\rangle}
  + \hat{\omega}_{\langle\mu\nu\rangle}\hat{p}^{\langle\mu}\hat{p}^{\nu\rangle} \right] f_a \,.
\end{split}
\ee
Here we introduced $\hat{\beta}\equiv-\hat{\beta}_\mu\hat{u}^\mu$, $\hat{\mathbf{p}}^2\equiv\po/\cosh^2\rho+\pp_\eta^2$, $\hat{\omega}\equiv\hat{\omega}^{\mu\nu}\hat{u}_\mu\hat{u}_\nu$, $\hat{\omega}_{\langle\mu\rangle}= - \hat{\Delta}_{\mu\nu}\hat{\omega}^{\nu\lambda}\hat{u}_\lambda$ $\hat{\omega}_{\langle\mu\nu\rangle} = \hat{\Delta}^{\alpha\beta}_\munu \hat{\omega}_{\alpha\beta}$. This expression can be simplified by using the constraints imposed by Gubser symmetry and Landau matching. In the Landau frame and in the absence of chemical potentials, the spatial vectors $\hat{\beta}_{\langle\mu\rangle}$ and $\hat{\omega}_{\langle\mu\rangle}$ vanish, respectively. Due to the $\text{SO}(3)_q$ symmetry, the traceless symmetric spatial tensor $\hat{\omega}_{\langle\mu\nu\rangle}$ has only one independent component, namely $\hat{\omega}_{\langle\eta\eta\rangle}$:
\be
\hat{\omega}_{\langle\mu\nu\rangle}=\hat{\omega}_{\langle\eta\eta\rangle}\,\text{diag}\left(0,-\cosh^2\rho,-\cosh^2\rho\sin^2\theta,1\right)\,.
\ee
Using the mass-shell condition to eliminate $\po$, Eq.~\eqref{eq:ch214grad} thus reduces to
\be
\label{eq:ch214momentC}
\tdf_{14} = \left[\hat{\alpha} - \hat{\beta}\hat{E}_{\pp} 
              + \hat{\omega}\left(\hat{E}_{\pp}^2{+}\textstyle{\frac{1}{3}}\hat{\mathbf{p}}^2\right)  
              + \textstyle{\frac{1}{2}}\hat{\omega}_{\langle\eta\eta\rangle}\left(3\pp_\eta^2{-}\hat{\mathbf{p}}^2 \right) \right] f_a\,.
\ee
The four coefficients $\hat{\alpha}$, $\hat{\beta}$, $\hat{\omega}$, and $\hat{\omega}_{\langle\eta\eta\rangle}$ must be matched to the contributions of the residual deviation $\tdf$ to the particle and energy densities as well as the bulk and shear viscous pressures \cite{Bazow:2013ifa}:
\bs
\allowdisplaybreaks
\label{eq:ch2match4}
\begin{align}
  \delta\hat{\tilde{n}} &\equiv -\langle (\hat{u}\cdot\pp) \rangle_{\tilde{\delta}} = 0\,,
\\
  \delta\hat{\tilde{\epsilon}} &\equiv \langle(\hat{u}\cdot\pp)^2 \rangle_{\tilde{\delta}} = 0 \,,
\\
  \hat{\tilde{\Pi}} &\equiv \frac{1}{3} \langle \hat{\Delta}_{\mu\nu} \pp^\mu \pp^\nu \rangle_{\tilde{\delta}} 
     = \frac{1}{3} \langle \hat{\mathbf{p}}^2 \rangle_{\tilde{\delta}} \,,
\\
  \hat{\tilde{\pi}} 
  &\equiv \langle\pp^{\langle\eta}\pp^{\eta\rangle}\rangle_{\tilde{\delta}} 
     = \langle \hat{p}^2_\eta - \frac{1}{3} \hat{\mathbf{p}}^2 \rangle_{\tilde{\delta}}\,.
\end{align}
\es
For conformal systems the bulk viscous pressure vanishes; for technical reasons we introduce an infinitesimal fictitious mass $m$ for the particles, approaching the conformal limit at the end by setting $m\to0$. This is the reason why in Eq.~\eqref{eq:ch214momentC} we did not replace $\hat{\mathbf{p}}^2$ by $\hat{E}_{\pp}^2$.

Inserting the ansatz \eqref{eq:ch214momentC} into Eq.~\eqref{eq:ch2match4} and using $\hat{\mathbf{p}}^2 = \hat{E}_{\pp}^2{-}m^2$ one obtains the matrix equation $\mathcal{A}\,b = c$, where 
\be
b = \left(
\begin{array}{c c c c}
\hat{\alpha} & \hat{\beta} & \hat{\omega} & \hat{\omega}_{\langle\eta\eta\rangle}
\end{array} \right)^T,
\qquad
c = \big(
\begin{array}{c c c c}
0 & 0 & 3\hat{\tilde{\Pi}} & \hat{\tilde{\pi}}
\end{array} \big)^T,
\ee
\be
\mathcal{A} =
\left[
\begin{array}{c c c c}
\Ih_{100}  & -\Ih_{200} & \frac{4}{3} \Ih_{300} & \frac{3}{2}\Ih_{320} - \frac{1}{2}\Ih_{300}  \\ \\
\Ih_{200}  & -\Ih_{300} & \frac{4}{3} \Ih_{400} {-} \frac{m^2}{3}\Ih_{200} & \frac{3}{2}\Ih_{420} {-} \frac{1}{2}\Ih_{400} {+} \frac{m^2}{2}\Ih_{200}\\ \\
\Ih_{200} {-} m^2\Ih_{000} & m^2\Ih_{100} {-}\Ih_{300} &  \frac{4}{3} \Ih_{400} {-} \frac{5m^2}{3}\Ih_{200} & \frac{3}{2}\Ih_{420} {-} \frac{1}{2}\Ih_{400} {+} m^2(\Ih_{200}{-}\frac{3}{2}\Ih_{220}) \\ \\
\Ih_{220} {-} \frac{1}{3} \Ih_{200} & \frac{1}{3} \Ih_{300} {-} \Ih_{320} & \frac{4}{3} \Ih_{420} {-} \frac{4}{9}\Ih_{400} & \frac{3}{2}\Ih_{440} {-} \Ih_{420} {+} \frac{1}{6}\Ih_{400} \\
\end{array} \right]
\ee
Here we expanded the $2^{\text{nd}}$ and $3^{\text{rd}}$ rows in $m$, keeping only terms up to $\mathcal{O}(m^2)$. After subtracting the $2^{\text{nd}}$ from the $3^{\text{rd}}$ row and rescaling it by $-1/m^2$, the matrix equation becomes $\mathcal{A}^\prime \,b = c^\prime$, where
\be
c^\prime = \big(
\begin{array}{c c c c}
0 & 0 & \hat{\tilde{\rho}}_0 & \hat{\tilde{\pi}}
\end{array} \big)^T,
\ee
\be
\label{eq:ch2matrixsimple}
\mathcal{A}^\prime = \left[
\begin{array}{c c c c}
\Ih_{100}  & -\Ih_{200} & \frac{4}{3} \Ih_{300} & \frac{3}{2}\Ih_{320} {-} \frac{1}{2}\Ih_{300}  \\ \\
\Ih_{200} & -\Ih_{300} & \frac{4}{3} \Ih_{400} {-} \frac{m^2}{3}\Ih_{200} &  \frac{3}{2}\Ih_{420} {-} \frac{1}{2}\Ih_{400} {+} \frac{m^2}{2}\Ih_{200}\\ \\
\Ih_{000}  & - \Ih_{100} & \frac{4}{3}\Ih_{200} & \frac{3}{2}\Ih_{220} {-} \frac{1}{2}\Ih_{200} \\ \\
\Ih_{220} {-} \frac{1}{3} \Ih_{200} & \frac{1}{3} \Ih_{300} {-} \Ih_{320} & \frac{4}{3} \Ih_{420} {-} \frac{4}{9}\Ih_{400} & \frac{3}{2}\Ih_{440} {-} \Ih_{420} {+} \frac{1}{6}\Ih_{400} \\
\end{array} \right]\,,
\ee
with $\hat{\tilde{\rho}}_0 \equiv -3\hat{\tilde{\Pi}}/m^2$ being the irreducible $\tdf$-moment associated with the residual bulk viscous pressure. In the conformal limit we set in Eq.~\eqref{eq:ch2matrixsimple} the mass $m$ and the scalar moment $\hat{\tilde{\rho}}_0$ to zero. The matrix equation can now be easily inverted. One finds
\bea
\label{eq:ch214gradgubcoeff}
\hat{\alpha} = \frac{\hat{\tilde{\pi}} \, \alpha_{\tilde{\pi}}(\xi)}{\J_{2}(\hat{\Lambda})}\,,\qquad
\hat{\beta} = \frac{\hat{\tilde{\pi}} \, \beta_{\tilde{\pi}}(\xi)}{\J_{3}(\hat{\Lambda})} \,,\qquad
\hat{\omega} = \frac{\hat{\tilde{\pi}} \, \gamma_{\tilde{\pi}}(\xi)}{\J_{4}(\hat{\Lambda})} \,,\qquad
\hat{\omega}_{\langle\eta\eta\rangle} = \frac{\hat{\tilde{\pi}} \, \kappa_{\tilde{\pi}}(\xi)}{\J_{4}(\hat{\Lambda})},
\eea
where the functions $\alpha_{\tilde{\pi}}$, $\beta_{\tilde{\pi}}$, $\gamma_{\tilde{\pi}}$ and $\kappa_{\tilde{\pi}}$ are given by
\be 
\label{eq:ch2alphapi}
\begin{gathered}
\textstyle
\alpha_{\tilde{\pi}} =
 \frac{-3\xi\left(2(39{+}76\xi){-}(213 + 343 \xi + 104 \xi^2) t(\xi)%
                {+}90(1{+}\xi)t^2(\xi){+}45(1{+}\xi)^3 t^3(\xi)\right)}
        {192 + 6(131{+}212\xi)t(\xi){-}(1932{+}2817\xi{+}712 \xi^2)t^2(\xi)%
             {+}630(1{+}\xi)t^3(\xi){+}315(1{+}\xi)^3t^4(\xi)},
\end{gathered}
\ee
\be 
\label{eq:ch2betapi}
\begin{gathered}
\textstyle
\beta_{\tilde{\pi}} =
 \frac{24\xi\sqrt{1{+}\xi}\left(12{+}(15{+}13\xi)t(\xi){-}(27{+}26\xi{-}\xi^2)t^2(\xi)\right)}
        {192 + 6(131{+}212\xi)t(\xi){-}(1932{+}2817\xi{+}712 \xi^2)t^2(\xi)%
             {+}630(1{+}\xi)t^3(\xi){+}315(1{+}\xi)^3t^4(\xi)},
\end{gathered}
\ee
\be 
\label{eq:ch2gammapi}
\begin{gathered}
\textstyle
\gamma_{\tilde{\pi}} = \frac{5\xi}{2}
             \frac{3(57{+}95\xi{+}4\xi^2){-}2(99{+}234\xi{+}231\xi^2{+}122\xi^3)t(\xi)%
                   {+}9(9{+}19\xi{+}31\xi^2{+}21\xi^3)t^2(\xi){+}54(1{+}\xi)^3(2\xi{-}1)t^3(\xi)}
       {192{+}6(212\xi{+}131) t(\xi){-}(1923{+}2817\xi{+}712\xi^2) t^2(\xi)%
               {+}630(1{+}\xi)t^3(\xi){+}315(1{+}\xi)^3 t^4(\xi)},
\end{gathered}
\ee
\be
\label{eq:ch2kappapi}
\begin{gathered}
\textstyle
\kappa_{\tilde{\pi}} = \frac{20\xi^2}{3}
            \frac{3(4\xi{-}13){-}(162{+}297\xi{+}122\xi^2)t(\xi)%
                   {+}9(31{+}52\xi{+}21\xi^2)t^2(\xi){+}108(1{+}\xi)^3 t^3(\xi)}
        {192{+}6(212\xi{+}131) t(\xi){-}(1923{+}2817\xi{+}712\xi^2) t^2(\xi)%
               {+}630(1{+}\xi)t^3(\xi){+}315(1{+}\xi)^3 t^4(\xi)}.
\end{gathered}
\ee
In the massless limit Eq.~\eqref{eq:ch214momentC} coincides with Eq.~\eqref{eq:ch214gradgub} in the text. As a cross-check we consider the isotropic limit $\xi \rightarrow 0$ of the coefficients \eqref{eq:ch214gradgubcoeff}: 
\bea
\hat{\alpha} \to 0,\qquad 
\hat{\beta}\to 0,\qquad
\hat{\omega} \to 0,\qquad
\hat{\omega}_{\langle\eta\eta\rangle} 
     \to \frac{\hat{\pi}}{2(\enehat{+}\hat{\mathcal{P}}_\eq)\hat{T}^2}.
\eea
With this we see that Eq.~\eqref{eq:ch214gradgub} reduces in the isotropic limit to the well-known Israel-Stewart result \cite{Israel:1979wp} for conformal systems \cite{Baier:2007ix}:
\be
\label{eq:ch2ISlimit}
\delta \tilde{f}_{14} \to 
\left[\frac{\hat{\pi}}{4(\enehat{+}\hat{\mathcal{P}}_\eq)\hat{T}^2}\left(3\pvv^2 - \hat{E}_{\pp}^2 \right) \right] \feq.
\ee

\chapter{}
\section{Nonconformal anisotropic integrals}
\label{appch3a}
In this section, we define the nonconformal anisotropic integrals that appear in Chapters~\ref{ch3label} --~\ref{chap5label} and show how to integrate them numerically. The anisotropic integrals $\I_{nrqs}$ and $\J_{nrqs}$ are defined as
\be
\label{eqapp3:A1}
   \I_{nrqs} =  \int_p \frac{(\up)^{n-r-2q}}{(2q)!!} \mzp^r \pxp^q (p \cdot \Omega \cdot p)^{s/2} f_a \,,
\ee
\be
\label{eqapp3:A2}
   \J_{nrqs} = \int_p \frac{(\up)^{n-r-2q}}{(2q)!!} \mzp^r \pxp^q (p \cdot \Omega \cdot p)^{s/2} f_a \bar{f}_a\,,
\ee
where $f_a$ is given by Eq.~\eqref{eqch3:15}. For particles with Boltzmann statistics $\I_{nrqs}=\J_{nrqs}$. Although there is no known analytical solution for these integrals for massive particles, their dimensionality can be reduced to one. After substituting the spherical coordinates
\be
\label{eqapp3:A4}
\begin{aligned}
   & p_{x,\mathrm{LRF}} = \alpha_\perp \Lambda \, \bar{p} \sin\theta \cos\phi\,, \\
   & p_{y,\mathrm{LRF}} = \alpha_\perp \Lambda \, \bar{p} \sin\theta \sin\phi\,, \\ 
   & p_{z,\mathrm{LRF}} = \alpha_L \Lambda \, \bar{p} \cos\theta\,,
\end{aligned}
\ee
with $\bar{p} = p / \Lambda$, the angular integrals in Eqs.~(\ref{eqapp3:A1}) --~(\ref{eqapp3:A2}) can be evaluated analytically, yielding 
\be
\label{eqapp3:A5}
\begin{split}
   \I_{nrqs} =\,& \frac{\alpha_\perp^{2q+2}\alpha_L^{r+1}\Lambda^{n+s+2}}{4\pi^2(2q)!!}
   \int_0^\infty d\bar p \, \bar p^{n+1}\, (\bar{p}^2{+}\bar{m}^2)^{s/2} 
\R_{nrq}(\alpha_\perp, \alpha_L; \bar m / \bar p)\, f_{\eq}(\bar{p},\bar{m})\,,
\end{split}
\ee
\be
\label{eqapp3:A6}
\begin{split}
   \J_{nrqs} =\,& \frac{\alpha_\perp^{2q+2}\alpha_L^{r+1}\Lambda^{n+s+2}}{4\pi^2(2q)!!}
   \int_0^\infty d\bar p \, \bar p^{n+1}\, (\bar{p}^2{+}\bar{m}^2)^{s/2} 
\R_{nrq}(\alpha_\perp, \alpha_L; \bar m / \bar p)\, f_{\eq}(\bar{p},\bar{m})
    \bar{f}_{\eq}(\bar{p},\bar{m})\,,
\end{split}
\ee
where $\bar m = m / \Lambda$, $\feq(\bar{p},\bar{m})= \bigg(\exp\left[\sqrt{\bar{p}^2+\bar{m}^2}\right]+\Theta\bigg)^{-1}$, $\bar{f}_\eq(\bar{p},\bar{m}) = 1 - \Theta\feq(\bar{p},\bar{m})$ and the functions $\R_{nrq}$ are defined as
\be
\label{eqapp3:A7}
   \R_{nrq}(\alpha_\perp, \alpha_L; \bar m / \bar p) =  w^{\,n-r-2q-1}\int_{-1}^1 {d}{\cos\theta} (\cos\theta)^r (\sin\theta)^{2q} (1{+}z \sin^2\theta)^{(n-r-2q-1)/2}\,,
\ee
with $w = \sqrt{\alpha_L^2 + (\bar m / \bar p)^2}$ and $z = (\alpha_\perp^2-\alpha_L^2) / w^2$. The radial momentum integral can be computed numerically with generalized Gauss-Laguerre quadrature. For reference, we list the functions $\R_{nrq}$ that appear in the anisotropic hydrodynamic equations:
\bs
\allowdisplaybreaks
\label{eqapp3:A8}
\beal
\R_{200} & = w \big(1+(1{+}z)t(z)\big) \,,\\
\R_{220} & = \frac{-1+(1{+}z)t(z)}{z w} \,,\\
\R_{201} & = \frac{1+(z{-}1)t(z)}{z w} \,,\\
\R_{240} & = \frac{3+2z-3(1{+}z)t(z)}{z^2w^3} \,,\\
\R_{202} & = \frac{3+z+(1{+}z)(z{-}3)t(z)}{z^2(1{+}z)w^3} \,,\\
\R_{221} & = \frac{-3+(3{+}z)t(z)}{z^2w^3} \,,\\
\R_{441} & = \frac{-15+13z+3(1{+}z)(5{+}z)t(z)}{4z^3w^3} \,,\\
\R_{402} & = \frac{3(z{-}1)+(z(3z{-}2)+3)t(z)}{4z^2w} \,,\\
\R_{421} & = \frac{3+z+(1{+}z)(z{-}3)t(z)}{4z^2w} \,,\\
\R_{422} & = \frac{15+z+(z(z{-}6)-15)t(z)}{4z^3w^3} \,,\\
\R_{403} & = \frac{(z{-}3)(5{+}3z)+3(1{+}z)(z(z{-}2)+5)t(z)}{4z^3(1{+}z)w^3}\,,
\end{align}
\es 
where $t(z) = \arctan\!{\sqrt{z}} / \sqrt{z}$.
%

\section{Anisotropic integral identities}
\label{appch3b}

Here we show how to derive the identities (\ref{eqchap4:jacobian_ids}). First, we express the anisotropic integrals (\ref{eqapp3:A1}) --~(\ref{eqapp3:A2}) as
\be
\label{eqapp3:Iint}
   \I_{nrqs} = \frac{1}{(2q)!!} \int_p E_\mathrm{LRF}^{n-r-2q} \, p_{z,\mathrm{LRF}}^r \, p_{\perp,\mathrm{LRF}}^{2q} \, E_a^s \, f_a\,,
\ee
\be
\label{eqapp3:Jint}
   \J_{nrqs} = \frac{1}{(2q)!!} \int_p E_\mathrm{LRF}^{n-r-2q} \, p_{z,\mathrm{LRF}}^r \, p_{\perp,\mathrm{LRF}}^{2q} \, E_a^s \, f_a \bar{f}_a\,,
\ee
where 
\be
E_a = \sqrt{m^2 + \dfrac{p_{\perp,\mathrm{LRF}}^2}{\alpha_\perp^2}+\dfrac{p_{z,\mathrm{LRF}}^2}{\alpha_L^2}}\,.
\ee
From here on out, we will suppress the LRF subscripts. To obtain the first identity (\ref{eqchap4:jacobian_ids}a), we introduce rapidity coordinates
\be
\label{eqapp3:rapidity}
\begin{aligned}
   E= m_\perp \cosh{y}\,, \qquad
   p_z = m_\perp \sinh{y} \,,
\end{aligned}
\ee
where $m_\perp = \sqrt{m^2 + p_\perp^2}$, to rewrite Eq.~\eqref{eqapp3:Jint} as
\be
\label{eqapp3:Jint1}
	\J_{nrqs} = g \int \frac{dy \,d^2 p_\perp}{(2\pi)^3(2q)!!} \, m_\perp^{n-2q} (\cosh{y})^{n-r-2q}  \,(\sinh{y})^r \, p_\perp^{2q} \, E_a^s \, f_a \bar{f}_a \,.
\ee
Next, one can write the distribution term $f_a \bar{f}_a$ as
\be
\label{eqapp3:ffbar_rapidity}
	f_a \bar{f}_a = - \frac{\Lambda \, \alpha_L^2 \, E_a}{m_\perp^2 \cosh{y} \, \sinh{y}} \frac{\partial f_a}{\partial y} \,.
\ee
After integrating by parts with respect to the variable $y$ (where the boundary term vanishes for $r \geq 2$) one obtains the relation 
\be
\label{eqapp3:id1} 
    \J_{nrqs} =\Lambda \alpha_L^2 (n{-}r{-}2q{-}1)\I_{n-2,r,q,s+1} + \Lambda \alpha_L^2 (r{-}1)\I_{n-2,r-2,q,s+1} + \Lambda(s{+}1)\I_{n,r,q,s-1} \,,
\ee
which for $(n,r,q,s) = (4,2,0,-1)$ yields Eq.~(\ref{eqchap4:jacobian_ids}a).

For the second identity (\ref{eqchap4:jacobian_ids}b) one uses cylindrical coordinates
\be
\label{eqapp3:cylindrical}
\begin{aligned}
    E = \sqrt{m^2+p_\perp^2+p_z^2}\,,  \qquad
    p_x = p_\perp \cos\phi\,, \qquad
    p_y = p_\perp \sin\phi 
\end{aligned}
\ee
to express the integral~\eqref{eqapp3:Jint} as
\be
\label{eqapp3:Jint2}
\begin{split}
	\J_{nrqs} = \, & g \int \frac{dp_\perp dp_z}{(2\pi)^2(2q)!!} \, E^{n-r-2q-1} \,
	p_z^{r} \, p_\perp^{2q+1} \, E_a^s \, f_a \bar{f}_a \,.
\nonumber
\end{split}
\ee
The term $f_a \bar{f}_a$ can be written as
\be
\label{eqapp3:ffbar_cylinder}
	f_a \bar{f}_a = - \frac{\Lambda \, \alpha_\perp^2 \, E_a}{p_\perp} \frac{\partial f_a}{\partial p_\perp} \,.
\ee
Integration by parts with respect to $p_\perp$, with the boundary term vanishing for $q \geq 1$, gives
\be
\label{eqapp3:id2} 
\J_{nrqs} = \Lambda \alpha_\perp^2(n{-}r{-}2q{-}1)\I_{n-2,r,q,s+1} +\Lambda \alpha_\perp^2 \I_{n-2,r,q-1,s+1} + \Lambda(s{+}1)\I_{n,r,q,s-1} \,,
\ee
which for $(n,r,q,s) = (4,0,1,-1)$ yields Eq.~(\ref{eqchap4:jacobian_ids}b). 

Finally, to get the third identity (\ref{eqchap4:jacobian_ids}c) we use spherical coordinates
\be
\label{eqapp3:spherical}
\begin{aligned}
   &E = \sqrt{m^2 + p^2}\,,  \qquad p_z = p \cos\theta\,, \qquad p_x = p \sin\theta \cos\phi\,, \qquad  p_y = p \sin\theta \sin\phi 
\end{aligned}
\ee
in the integral~\eqref{eqapp3:Jint}
\be
\label{eqapp3:Jint3}
	\J_{nrqs} = g \int \frac{dp \, d\cos\theta}{(2\pi)^2(2q)!!} \, E^{n-r-2q-1} \, p^{r+2q+2} \, (\cos\theta)^r \, (\sin\theta)^{2q} \, E_a^s \, f_a \bar{f}_a 
\ee
and write the term $f_a \bar{f}_a$ as
\be
\label{eqapp3:ffbar_spherical}
    f_a \bar{f}_a = - \frac{\Lambda \, \alpha_\perp^2 \alpha_L^2}{\alpha_\perp^2{-}\alpha_L^2} \, 
                              \frac{E_a}{p^2\cos\theta}\, \frac{\partial f_a}{\partial \cos\theta} \,.
\ee
Integrating by parts with respect to $\cos\theta$, where the boundary term vanishes for $q \geq 1$, gives the following relation:
\be
\label{eqapp3:id3} 
   \J_{nrqs} = \frac{\Lambda \, \alpha_\perp^2 \alpha_L^2}{\alpha_\perp^2{-}\alpha_L^2} 
                          \Bigl((r{-}1)\,\I_{n-2,r-2,q,s+1}{-}\I_{n-2,r,q-1,s+1}\Bigr)
	            + \Lambda\,(s{+}1)\,\I_{n,r,q,s-1} \,,
\ee
which for $(n,r,q,s) = (4,2,1,-1)$ yields Eq.~(\ref{eqchap4:jacobian_ids}c).
 
\section{Anisotropic transport coefficients (kinetic part)}
\label{appch3c}

Here we list the kinetic contribution to the transport coefficients (labeled as $(k)$) appearing in the relaxation equations (\ref{eqchap3:relax_2}a-d) for the kinetic longitudinal pressure $\PL^{(k)}$:
\bs
\allowdisplaybreaks
\label{eqapp3:B1}
\beal
\bar{\zeta}^{L(k)}_z & = \I_{2400} - 3 \PL^{(k)} + m\frac{dm}{d\ene}(\ene{+}\PL)\I_{0200}\,,  \\
\bar{\zeta}^{L(k)}_\perp & = \I_{2210} - \PL^{(k)} + m\frac{dm}{d\ene}(\ene{+}\Pperp)\I_{0200}\,, \\
\bar{\lambda}^{L(k)}_{Wu} & = \frac{\J_{4410}}{\J_{4210}} + m\frac{dm}{d\ene}\I_{0200}\,,\\
\bar{\lambda}^{L(k)}_{W\perp} & =  1- \bar{\lambda}^{L(k)}_{Wu}\,,\\
\bar{\lambda}^{L(k)}_\pi & = \frac{\J_{4220}}{\J_{4020}} + m\frac{dm}{d\ene}\I_{0200}
\end{align}
\es
and kinetic transverse pressure $\Pperp^{(k)}$:
\bs
\allowdisplaybreaks
\label{eqapp3:B2}
\beal
\bar{\zeta}^{\perp(k)}_z & = \I_{2210} - \Pperp^{(k)} + m\frac{dm}{d\ene}(\ene{+}\PL)\I_{0010}\,, \\
\bar{\zeta}^{\perp(k)}_\perp & = 2\bigl(\I_{2020} - \Pperp^{(k)} \bigr) 
                                         + m\frac{dm}{d\ene}(\ene{+}\Pperp)\I_{0010}\,, \\
\bar{\lambda}^{\perp(k)}_{W\perp} & = \frac{2 \,\J_{4220}}{\J_{4210}} + m\frac{dm}{d\ene}\I_{0010}\,,\\
\bar{\lambda}^{\perp(k)}_{Wu} & = \bar{\lambda}^{\perp(k)}_{W\perp} - 1\,, \\
\bar{\lambda}^{\perp(k)}_\pi & = 1 - \frac{3\,\J_{4030}}{\J_{4020}} -  m\frac{dm}{d\ene}\I_{0010}\,.
\end{align}
\es
%
\section{Second-order viscous hydrodynamics}
\label{appch3e}

Here we derive the second-order viscous hydrodynamic equations and their transport coefficients. We start with the quasiparticle case (\vh{}) and derive the relaxation equation for $\delta B$ and its second-order solution \eqref{eqch3:dB2nd}. The general evolution equation for $\delta B$ is given by \cite{Tinti:2016bav}
\be
\delta\dot B = - \frac{\delta B}{\tau_\Pi} + \frac{\dot m}{m} \left(3 \Pi + 4 \,\delta B\right)\,,
\ee
where the $\dot m / m$ term can be written as
\be
\frac{\dot m}{m} = \frac{1}{m} \frac{dm}{d\ene} \dot\ene\,.
\ee
We replace the time derivative $\dot\ene$ with the energy conservation law in viscous hydrodynamics
\be
\label{eqapp3:energyvhydro}
\dot\ene + (\ene {+} \Peq {+} \Pi)\theta - \pi^\munu \sigma_\munu = 0 \,,
\ee
where $\theta = \partial_\mu u^\mu$ is the scalar expansion rate and $\sigma_\munu = \Delta^{\alpha\beta}_\munu \partial_\beta u_\alpha$ is the velocity shear tensor, where $\Delta^{\alpha\beta}_\munu = \frac{1}{2}(\Delta^{\alpha}_\mu \Delta^{\beta}_\nu  + \Delta^{\beta}_\mu \Delta^{\alpha}_\nu  - \frac{2}{3}\Delta^{\alpha\beta} \Delta_\munu)$ is the traceless double spatial projector and $\Delta^{\munu} = g^\munu - u^\mu u^\nu$ the spatial projector. For the second-order relaxation equation, we only need the first-order approximation $\dot\ene \approx - (\ene {+} \Peq)\theta$, thus
\be
\frac{\dot m}{m} \approx - \frac{1}{m} \frac{dm}{d\ene} (\ene {+} \Peq)\theta \,.
\ee
The equation of motion for $\delta B$ then reduces to
\be
\label{eqapp3:dBdot1}
\delta\dot B = - \frac{\delta B}{\tau_\Pi} - \frac{\ene{+}\Peq}{m} \frac{dm}{d\ene} \left(3 \Pi {+} 4\delta B\right) \theta\,.
\ee
To first order in deviations from equilibrium $\delta B$ = 0. The second-order solution is given by Eq.~\eqref{eqch3:dB2nd} \cite{Tinti:2016bav}. In Eq.~\eqref{eqapp3:dBdot1}, we truncate the third-order term $\propto \delta B \, \theta$ to arrive at the second-order relaxation equation
\be
\label{eqapp3:dBdot2}
\delta\dot B = - \frac{\delta B}{\tau_\Pi} - \frac{3(\ene{+}\Peq)}{m} \frac{dm}{d\ene} \Pi\, \theta\,.
\ee

Next, we derive the relaxation equations for the viscous components $\Pi$ and $\pi^\munu$ in the same manner as in Sec. \ref{ch3sec3}. We start by taking the co-moving time derivative of their quasi-kinetic definitions \cite{Tinti:2016bav}
\bs
\allowdisplaybreaks
\label{eqapp3:vhydrodot1}
\beal
&\dot{\Pi} =  \frac{1}{3} \, D \int_p (- p \cdot \Delta \cdot p)  \delta f - \delta \dot B\,, \\
&\dot{\pi}^{\langle\munu\rangle} =  \frac{1}{3} \Delta^\munu_{\alpha\beta} \, D \int_p  p^{\langle\alpha} \, p^{\beta\rangle}  \delta f \,,
\end{align}
\es
where $\delta f$ is the non-equilibrium correction to the distribution function
\be
\label{eqapp3:fvhydro}
f = f_\eq + \delta f \,,
\ee
with $f_\eq = \exp\big({-}u{\,\cdot\,}p/T\big)$ being the local-equilibrium Boltzmann distribution. The time derivative $\delta \dot B$ is given by Eq.~\eqref{eqapp3:dBdot2}. We substitute the terms containing $\delta\dot{f}$ using the Boltzmann equation~\eqref{eqch3:13} and Eq.~\eqref{eqapp3:fvhydro}:
\be
\delta\dot{f} = - \dot{f}_\eq + \frac{C[f] - m \, \partial^\mu m \, \partial_\mu^{(p)} f}{\up} - \frac{p^{\langle\mu\rangle} \nabla_\mu f_\eq}{\up} - \frac{p^{\langle\mu\rangle} \nabla_\mu \delta f}{\up} \,,
\ee
where $p^{\langle\mu\rangle} = \Delta^\mu_\nu p^\nu$ and $\nabla_\mu = \Delta^\nu_\mu \partial_\nu$ is the spatial gradient. To close the system of equations we use the 14--moment approximation for $\delta f$ \cite{Denicol:2014vaa}:
\be
\label{eqapp3:14ansatz}
   \frac{\delta f}{f_\eq} = c_\ene (\up)^2 + \frac{1}{3}c_\Pi \,(- p \cdot \Delta \cdot p) 
   + c_{\pi}^{\langle\munu\rangle} p_{\langle\mu} \, p_{\nu\rangle}\,,
\ee
where $p_{\langle\mu} \, p_{\nu\rangle}$ = $\Delta^{\alpha\beta}_\munu p_\alpha p_\beta$. To solve for the coefficients we insert the expression~\eqref{eqapp3:14ansatz} into the energy matching condition and the definitions of $\Pi$ and $\pi^\munu$:  
\bs
\allowdisplaybreaks
\label{eqapp3:14momenteq}
\beal
\delta\ene &= \int_p (\up)^2 \delta f = 0\,, \\
\Pi &=  \frac{1}{3} \int_p (-p \cdot \Delta \cdot p)\, \delta f\,, \\
\pi^\munu &= \int_p p^{\langle\mu} \, p^{\nu\rangle} \delta f\,.
\end{align}
\es
Since the 14--moment approximation is first-order in the dissipative flows, we neglect the second-order contribution ${\sim\,}\delta B$ to the energy density and bulk viscous pressure in Eq.~\eqref{eqapp3:14momenteq}. After some algebra, the coefficients are
\bs
\allowdisplaybreaks
\beal
c_\ene &= - \frac{\I_{41} \, \Pi}{\frac{5}{3}\I_{40} I_{42}- \I_{41}^2}\,, \\
c_\Pi &= \frac{\I_{40} \, \Pi}{\frac{5}{3}\I_{40} I_{42}- \I_{41}^2}\,, \\
c_\pi^{\langle\munu\rangle} &= \frac{\pi^\munu}{2\,\I_{42}}\,,
\end{align}
\es
where we defined the thermodynamic integrals
\be
\label{eqapp3:intI}
  \I_{nq} =  \int_p \frac{(\up)^{n-2q}}{(2q{+}1)!!}(- p \cdot \Delta \cdot p)^q f_\eq\,.
\ee
The final expression for $\delta f$ is
\be
\label{eqapp3:14momenteq2}
\frac{\delta f}{f_\eq} = \Big(\bar{c}_\ene (\up)^2 + \frac{1}{3}\bar{c}_\Pi \,(- p \cdot \Delta \cdot p)\Big)\Pi \,+\, \frac{1}{2}\bar{c}_{\pi} p_{\langle\mu} \, p_{\nu\rangle} \pi^\munu \,,
\ee
where $\bar{c}_\ene = c_\ene / \Pi$, $\bar{c}_\Pi = c_\Pi / \Pi$ and $\bar{c}_\pi = 1 / \I_{42}$. After integration by parts and inserting for $\delta f$ the 14--moment approximation~\eqref{eqapp3:14momenteq2}, the relaxation equations for $\Pi$ and $\pi^\munu$ reduce to
\bs
\allowdisplaybreaks
\label{eqapp3:relaxvhydro}
\beal
\dot{\Pi} &= -\frac{\Pi}{\tau_\Pi} - \beta_\Pi \theta - \delta_{\Pi\Pi} \Pi \theta + \lambda_{\Pi\pi}\pi^\munu \sigma_\munu\,,
\\
\dot{\pi}^{\langle\munu\rangle} &= -\frac{\pi^\munu}{\tau_\pi} - \beta_\pi \sigma^\munu + 2\pi^{\lambda\langle\mu} \omega^{\nu\rangle}_\lambda - \tau_{\pi\pi} \pi^{\lambda\langle\mu} \sigma^{\nu\rangle}_\lambda - \delta_{\pi\pi} \pi^\munu \theta + \lambda_{\pi\Pi} \Pi \sigma^\munu\,.
\end{align}
\es
Here $\omega^\munu = \Delta^\mu_\alpha \Delta^\nu_\beta \partial_{[\beta} u_{\alpha ]}$ is the vorticity tensor. The relaxation times $\tau_\pi$ and $\tau_\Pi$ are given by Eq.~\eqref{eqch3:81}. The remaining transport coefficients in quasiparticle viscous hydrodynamics (\vh{}) are
\bs
\allowdisplaybreaks
\beal
\beta_\pi &= \frac{\I_{32}}{T} \,,\\
\beta_\Pi &= \frac{5}{3}\beta_\pi - c_s^2(\ene{+}\Peq) + c_s^2 m \frac{dm}{dT} \, \I_{11}\,, \\
\delta_{\Pi\Pi} &= 1 - c_s^2 - \frac{m^4}{9}(\bar{c}_\ene \, \I_{00} {+} \bar{c}_\Pi \, \I_{01}) - m \frac{dm}{d\ene}(\ene{+}\Peq)\Big(\bar{c}_\ene \I_{21} {+} \frac{5}{3} \bar{c}_\Pi \I_{22} {+} \frac{3}{m^2} \Big)\,, \\
\lambda_{\Pi\pi} &= \frac{1}{3} - c_s^2 + \frac{\bar{c}_\pi m^2 \I_{22}}{3}\,, \\
\tau_{\pi\pi} &= \frac{10}{7} + \frac{4 \,\bar{c}_\pi m^2 \I_{22}}{7}\,, \\
\delta_{\pi\pi} &= \frac{4}{3} + \frac{\bar{c}_\pi m^2\I_{22}}{3} - \bar{c}_\pi m \frac{dm}{d\ene}(\ene{+}\Peq) \I_{22} \,,\\
\lambda_{\pi\Pi} &= \frac{6}{5} - \frac{2 m^4}{15} (\bar{c}_\ene \, \I_{00} {+} \bar{c}_\Pi \, \I_{01})\,.
\end{align}
\es
In standard viscous hydrodynamics (\vh{}2), one performs an expansion in powers of $z = m/T \ll 1$ and takes the fixed-mass limit $dm/dT \to 0$. The leading terms in the second-order transport coefficients are \cite{Denicol:2014vaa}
\bs
\allowdisplaybreaks
\beal
&\beta_\pi = \frac{\ene{+}\Peq}{5} + \order(z^2)\,, \\
&\beta_\Pi = 15\Big(\frac{1}{3} - c^2_s\Big)^2\big(\ene{+}\Peq\big) + \order(z^5)\,,\\
&\delta_{\Pi\Pi} = \frac{2}{3} + \order(z^2 \ln z)\,, \\
&\lambda_{\Pi\pi} = \frac{8}{5}\Big(\frac{1}{3} -  c^2_s\Big) + \order(z^4)\,, \\
&\tau_{\pi\pi} = \frac{10}{7} + \order(z^2) \,,\\
&\delta_{\pi\pi} = \frac{4}{3} + \order(z^2)\,, \\
&\lambda_{\pi\Pi} = \frac{6}{5} + \order(z^2 \ln z)\,.
\end{align}
\es
Note that both viscous hydrodynamic models use the same shear and bulk viscosities as anisotropic hydrodynamics (e.g. Figs.~\ref{FV} and~\ref{viscosities}).

For Bjorken flow, the energy conservation law~\eqref{eqapp3:energyvhydro} and relaxation equations~\eqref{eqapp3:relaxvhydro} simplify greatly. The gradient terms are $\sigma^\munu = \mathrm{diag}(0,\frac{1}{3\tau}, \frac{1}{3\tau},-\frac{2}{3\tau^3})$, $\theta = 1 / \tau$, and $\omega^\munu = 0$. The shear stress components are $\pi^\munu = \mathrm{diag}(0, - \frac{1}{2} \tau^2 \pi^{\eta\eta},- \frac{1}{2} \tau^2 \pi^{\eta\eta},\pi^{\eta\eta})$. As a result, the viscous hydrodynamic equations reduce to \eqref{eqch3:vhydroeqs}.

\chapter{}
\section{Gradient source terms}
\label{app4a}
In this appendix, we list the gradients that appear in the source terms of the anisotropic hydrodynamic equations. The derivatives of the fluid velocity component $u^\tau$ are
\bs
\allowdisplaybreaks
\beal
\partial_\tau u^\tau &= v^x \partial_\tau u^x + v^y \partial_\tau u^y + \tau^2 v^\eta \partial_\tau u^\eta + \tau v^\eta u^\eta \,,\\
\partial_x u^\tau &= v^x \partial_x u^x + v^y \partial_x u^y + \tau^2 v^\eta \partial_x u^\eta \,,\\
\partial_y u^\tau &= v^x \partial_y u^x + v^y \partial_y u^y + \tau^2 v^\eta \partial_y u^\eta \,,\\
\partial_\eta u^\tau &= v^x \partial_\eta u^x + v^y \partial_\eta u^y + \tau^2 v^\eta \partial_\eta u^\eta \,.
\end{align}
\es
The derivatives of the longitudinal basis vector $z^\mu$ are
\bs
\allowdisplaybreaks
\beal
\partial_\tau z^\tau &= \frac{\tau}{\sqrt{1{+}u_\perp^2}}\left(\partial_\tau u^\eta - \frac{u^\eta\left(u^x \partial_\tau u^x {+} u^y \partial_\tau u^y \right)}{1{+}u_\perp^2}\right) + \frac{z^\tau}{\tau} \,,\\  
\partial_x z^\tau &= \frac{\tau}{\sqrt{1{+}u_\perp^2}}\left(\partial_x u^\eta - \frac{u^\eta\left(u^x \partial_x u^x {+} u^y \partial_x u^y \right)}{1{+}u_\perp^2}\right) \,,\\ 
\partial_y z^\tau &= \frac{\tau}{\sqrt{1{+}u_\perp^2}}\left(\partial_y u^\eta - \frac{u^\eta\left(u^x \partial_y u^x {+} u^y \partial_y u^y \right)}{1{+}u_\perp^2}\right) \,,\\ 
\partial_\eta z^\tau &= \frac{\tau}{\sqrt{1{+}u_\perp^2}}\left(\partial_\eta u^\eta - \frac{u^\eta\left(u^x \partial_\eta u^x {+} u^y \partial_\eta u^y \right)}{1{+}u_\perp^2}\right) \,,\\
\partial_\tau z^\eta &= \frac{1}{\tau\sqrt{1{+}u_\perp^2}}\left(\partial_\tau u^\tau - \frac{u^\tau\left(u^x \partial_\tau u^x {+} u^y \partial_\tau u^y \right)}{1{+}u_\perp^2}\right) - \frac{z^\eta}{\tau} \,,\\
\partial_x z^\eta &= \frac{1}{\tau\sqrt{1{+}u_\perp^2}}\left(\partial_x u^\tau - \frac{u^\tau\left(u^x \partial_x u^x {+} u^y \partial_x u^y \right)}{1{+}u_\perp^2}\right) \,,\\ 
\partial_y z^\eta &= \frac{1}{\tau\sqrt{1{+}u_\perp^2}}\left(\partial_y u^\tau - \frac{u^\tau\left(u^x \partial_y u^x {+} u^y \partial_y u^y \right)}{1{+}u_\perp^2}\right) \,,\\ 
\partial_\eta z^\eta &= \frac{1}{\tau\sqrt{1{+}u_\perp^2}}\left(\partial_\eta u^\tau - \frac{u^\tau\left(u^x \partial_\eta u^x {+} u^y \partial_\eta u^y \right)}{1{+}u_\perp^2}\right) \,.
\end{align}
\es
The divergence of the spatial fluid velocity $v^i = u^i/u^\tau$ is
\be
\partial_i v^i = \frac{\partial_x u^x - v^x \partial_x u^\tau + \partial_y u^y - v^y \partial_y u^\tau + \partial_\eta u^\eta - v^\eta \partial_\eta u^\tau}{u^\tau} \,.
\ee
The longitudinal expansion rate is
\be
\begin{split}
\theta_L &=  z_\mu D_z u^\mu = - z_\mu z^\nu \partial_\nu u^\mu - z_\mu z^\nu \Gamma^\mu_{\nu\lambda} u^\lambda \\
&= - (z^\tau)^2 \partial_\tau u^\tau + z^\tau z^\eta (\tau^2 \partial_\tau u^\eta - \partial_\eta u^\tau) + (\tau z^\eta)^2 \partial_\eta u^\eta + \tau (z^\eta)^2 u^\tau\,.
\end{split}
\ee
The transverse expansion rate is
\be
\theta_\perp = \nabla_{\perp\mu}u^\mu = \theta - \theta_L\,,
\ee
where
\be
\label{eq:theta}
\begin{split}
\theta &= D_\mu u^\mu = \partial_\mu u^\mu + \Gamma^\mu_{\mu\nu} u^\nu = \partial_\tau u^\tau + \partial_x u^x + \partial_y u^y + \partial_\eta u^\eta + \frac{u^\tau}{\tau}
\end{split}
\ee
is the scalar expansion rate.

The components of the fluid acceleration
\be
a^\mu = Du^\mu = u^\nu\partial_\nu u^\mu + u^\nu \Gamma^\mu_{\nu\lambda} u^\lambda
\ee
are
\bs
\allowdisplaybreaks
\beal
a^\tau &= u^\tau \partial_\tau u^\tau + u^x \partial_x u^\tau + u^y \partial_y u^\tau + u^\eta \partial_\eta u^\tau + \tau(u^\eta)^2 \,,\\
a^x &= u^\tau \partial_\tau u^x + u^x \partial_x u^x + u^y \partial_y u^x + u^\eta \partial_\eta u^x \,,\\
a^y &= u^\tau \partial_\tau u^y + u^x \partial_x u^y + u^y \partial_y u^y + u^\eta \partial_\eta u^y \,,\\
a^\eta &= u^\tau \partial_\tau u^\eta + u^x \partial_x u^\eta + u^y \partial_y u^\eta + u^\eta \partial_\eta u^\eta + \frac{2u^\tau u^\eta}{\tau}\,.
\end{align}
\es
The components of the longitudinal vector's co-moving time derivative
\be
\dot{z}^\mu = D z^\mu = u^\nu\partial_\nu z^\mu + u^\nu \Gamma^\mu_{\nu\lambda} z^\lambda
\ee
are
\bs
\allowdisplaybreaks
\beal
\dot{z}^\tau &= u^\tau \partial_\tau z^\tau + u^x \partial_x z^\tau + u^y \partial_y z^\tau + u^\eta \partial_\eta z^\tau + \tau u^\eta z^\eta \,,\\
\dot{z}^\eta &= u^\tau \partial_\tau z^\eta + u^x \partial_x z^\eta + u^y \partial_y z^\eta + u^\eta \partial_\eta z^\eta + \frac{u^\tau z^\eta {+} u^\eta z^\tau}{\tau}\,.
\end{align}
\es
The components of the fluid velocity's longitudinal derivative
\be
D_z u^\mu = - z^\nu\partial_\nu u^\mu - z^\nu \Gamma^\mu_{\nu\lambda} u^\lambda
\ee
are
\bs
\allowdisplaybreaks
\beal
D_z u^\tau &= -z^\tau \partial_\tau u^\tau - z^\eta \partial_\eta u^\tau - \tau u^\eta z^\eta \,,\\
D_z u^x &= -z^\tau \partial_\tau u^x - z^\eta \partial_\eta u^x \,,\\
D_z u^y &= -z^\tau \partial_\tau u^y - z^\eta \partial_\eta u^y \,,\\
D_z u^\eta &= -z^\tau \partial_\tau u^\eta - z^\eta \partial_\eta u^\eta - \frac{u^\tau z^\eta{+} u^\eta z^\tau}{\tau} \,.
\end{align}
\es
The transverse gradient of the fluid velocity projected along the longitudinal direction is
\be
z_\nu \nabla_\perp^\mu u^\nu = \Xi^\mu_\alpha z_\nu D^\alpha u^\nu \,,
\ee
where the components of
\be
z_\nu D^\alpha u^\nu = g^{\alpha\beta} z_\nu \partial_\beta u^\nu + g^{\alpha\beta}z_\nu \Gamma^\nu_{\beta\lambda} u^\lambda
\ee
are
\bs
\allowdisplaybreaks
\beal
z_\nu D^\tau u^\nu &= z^\tau \partial_\tau u^\tau - \tau^2 z^\eta \partial_\tau u^\eta - \tau u^\eta z^\eta \,,\\
z_\nu D^x u^\nu &= - z^\tau \partial_x u^\tau + \tau^2 z^\eta \partial_x u^\eta \,,\\
z_\nu D^y u^\nu &= - z^\tau \partial_y u^\tau + \tau^2 z^\eta \partial_y u^\eta \,,\\
z_\nu D^\eta u^\nu &= -\frac{1}{\tau^2}\left(z^\tau \partial_\eta u^\tau - \tau^2 z^\eta \partial_\eta u^\eta + \tau \left(u^\eta z^\tau {-} u^\tau z^\eta\right) \right)
\end{align}
\es
The transverse velocity-shear tensor is\footnote{%
    In the code we obtain $\sigma_\perp^\munu$ by applying the projector $\Xi^\munu_{\alpha\beta}$ onto $D^{(\alpha} u^{\beta)}$ directly, rather than simplifying the expression \eqref{eq:sigmaT}.}
\be
\label{eq:sigmaT}
\sigma_\perp^\munu = \Xi^\munu_{\alpha\beta} D^{(\alpha} u^{\beta)} \,,
\ee
where the components of
\be
D^{(\alpha} u^{\beta)} = \frac{1}{2}\left(g^{\alpha\rho}\partial_\rho u^\beta + g^{\beta\rho}\partial_\rho u^\alpha + g^{\alpha\rho} \Gamma^\beta_{\rho\lambda}u^\lambda + g^{\beta\rho} \Gamma^\alpha_{\rho\lambda}u^\lambda\right)
\ee
are
\bs
\allowdisplaybreaks
\beal
D^{(\tau} u^{\tau)} &= \partial_\tau u^\tau  \,,\\
D^{(\tau} u^{x)} &= \frac{1}{2}\left(\partial_\tau u^x - \partial_x u^\tau \right) \,,\\
D^{(\tau} u^{y)} &= \frac{1}{2}\left(\partial_\tau u^y - \partial_y u^\tau \right) \,,\\
D^{(\tau} u^{\eta)} &= \frac{1}{2}\left(\partial_\tau u^\eta - \frac{\partial_\eta u^\tau}{\tau^2} \right) \,,\\
D^{(x} u^{x)} &= - \partial_x u^x \,,\\
D^{(x} u^{y)} &= -\frac{1}{2}\left(\partial_x u^y + \partial_y u^x \right) \,,\\
D^{(x} u^{\eta)} &= -\frac{1}{2}\left(\partial_x u^\eta + \frac{\partial_\eta u^x}{\tau^2} \right) \,,\\
D^{(y} u^{y)} &= -\partial_y u^y \,,\\
D^{(y} u^{\eta)} &= -\frac{1}{2}\left(\partial_y u^\eta + \frac{\partial_\eta u^y}{\tau^2} \right) \,,\\
D^{(\eta} u^{\eta)} &= -\frac{1}{\tau^2}\left(\partial_\eta u^\eta + \frac{u^\tau}{\tau} \right) \,.
\end{align}
\es
The transverse vorticity tensor is
\be
\begin{split}
\omega_\perp^\munu &= \Xi^\mu_\alpha \Xi^\nu_\beta D^{[\alpha} u^{\beta]} \\
&= D^{[\mu} u^{\nu]} - \frac{u^\mu a^{\nu} {-} u^\nu a^{\mu} {+} z^\mu (D_zu^\nu {+} z_\alpha D^\nu u^\alpha) {-} z^\nu (D_z u^\mu {+} z_\alpha D^\mu  u^\alpha)}{2}
\end{split}
\ee
where the components of
\be
\label{eq:anti_Du}
D^{[\alpha} u^{\beta]} = \frac{1}{2}\left(g^{\alpha\rho}\partial_\rho u^\beta - g^{\beta\rho}\partial_\rho u^\alpha + g^{\alpha\rho} \Gamma^\beta_{\rho\lambda}u^\lambda - g^{\beta\rho} \Gamma^\alpha_{\rho\lambda}u^\lambda\right)
\ee
are
\bs
\allowdisplaybreaks
\beal
D^{[\tau} u^{\tau]} &= 0  \,,\\
D^{[\tau} u^{x]} &= \frac{1}{2}\left(\partial_\tau u^x + \partial_x u^\tau \right) \,,\\
D^{[\tau} u^{y]} &= \frac{1}{2}\left(\partial_\tau u^y + \partial_y u^\tau \right) \,,\\
D^{[\tau} u^{\eta]} &= \frac{1}{2}\left(\partial_\tau u^\eta + \frac{\partial_\eta u^\tau}{\tau^2} \right) + \frac{u^\eta}{\tau}\,,\\
D^{[x} u^{x]} &= 0 \,,\\
D^{[x} u^{y]} &= -\frac{1}{2}\left(\partial_x u^y - \partial_y u^x \right) \,,\\
D^{[x} u^{\eta]} &= -\frac{1}{2}\left(\partial_x u^\eta - \frac{\partial_\eta u^x}{\tau^2} \right) \,,\\
D^{[y} u^{y]} &= 0 \,,\\
D^{[y} u^{\eta]} &= -\frac{1}{2}\left(\partial_y u^\eta + \frac{\partial_\eta u^y}{\tau^2} \right) \,,\\
D^{[\eta} u^{\eta]} &= 0 \,.
\end{align}
\es
\section{Geometric source terms}
\label{app4b}
Here we list the components of the geometric source terms $\mathcal{G}_W^\mu$ and $\mathcal{G}_\pi^\munu$ (Eq.~\eqref{eqchap4:geometric_source}) that appear in the relaxation equations~\eqref{eqchap4:relax_residual} for $\Wperp$ and $\piperp$, respectively:
\bs
\allowdisplaybreaks
\beal
\mathcal{G}_W^\tau &= \tau u^\eta W_{\perp z}^\eta \,, \\
\mathcal{G}_W^x &= 0 \,,\\
\mathcal{G}_W^y &= 0 \,,\\
\mathcal{G}_W^\eta &= \frac{u^\tau W_{\perp z}^\eta + u^\eta W_{\perp z}^\tau}{\tau} \,.
\end{align}
\es
\bs
\allowdisplaybreaks
\label{eq:geometric_piperp_comps}
\beal
\mathcal{G}_\pi^{\tau\tau} &= 2 \tau u^\eta \pi_\perp^{\tau\eta} \,, \\
\mathcal{G}_\pi^{\tau x} &= \tau u^\eta \pi_\perp^{x\eta} \,,\\
\mathcal{G}_\pi^{\tau y} &= \tau u^\eta \pi_\perp^{y\eta} \,,\\
\mathcal{G}_\pi^{\tau\eta} &= \tau u^\eta \pi_\perp^{\eta\eta}  +  \frac{u^\tau \pi_\perp^{\tau\eta} {+} u^\eta \pi_\perp^{\tau\tau}}{\tau} \,,\\
\mathcal{G}_\pi^{xx} &= 0 \,,\\
\mathcal{G}_\pi^{xy} &= 0 \,,\\
\mathcal{G}_\pi^{x\eta} &= \frac{u^\tau \pi_\perp^{x\eta} + u^\eta \pi_\perp^{\tau x}}{\tau} \,, \\
\mathcal{G}_\pi^{yy} &= 0 \,,\\
\mathcal{G}_\pi^{y\eta} &= \frac{u^\tau \pi_\perp^{y\eta} + u^\eta \pi_\perp^{\tau y}}{\tau} \,, \\
\mathcal{G}_\pi^{\eta\eta} &= \frac{2\left(u^\tau \pi_\perp^{\eta\eta} + u^\eta \pi_\perp^{\tau\eta}\right)}{\tau} \,.
\end{align}
\es
\section{Conformal anisotropic transport coefficients}
\label{app4c}

In the conformal limit $m = B = 0$, the anisotropic transport coefficients~\eqref{eqchap4:pl_coeff} -- \eqref{eqB4} only depend on the functions $\I_{nrqs}$ given in Eq.~\eqref{eqapp3:A1}, which reduce to 
\be
    \I_{nrqs} = \frac{g(n{+}s{+}1)!\,\alpha_L^{r+1}\Lambda^{n+s+2}\mathcal{R}_{nrq}}{4\pi^2(2q)!!}\,,
\ee
where $g$, $\alpha_L$ and $\Lambda$ are given by Eqs.~\eqref{eqch3:46},~\eqref{eqchap4:aL_conformal} and~\eqref{eqchap4:Lambda_conformal}, respectively. We list the functions $\mathcal{R}_{nrq}$ used in Chapter~\ref{chapter4label}~\cite{McNelis:2018jho}: 
\bs
\allowdisplaybreaks
\beal
\mathcal{R}_{200} & = \alpha_L \big(1+(1+\xi_L)t_L\big) \,,\\
\mathcal{R}_{220} & = \frac{-1+(1+\xi_L)t_L}{\xi_L \alpha_L} \,,\\
\mathcal{R}_{201} & = \frac{1+(\xi_L-1)t_L}{\xi_L \alpha_L} \,,\\
\mathcal{R}_{240} & = \frac{3+2\xi_L-3(1+\xi_L)t_L)}{\xi_L^2\alpha_L^3} \,,\\
\mathcal{R}_{202} & = \frac{3+\xi_L+(1+\xi_L)(\xi_L-3)t_L}{\xi_L^2(1+\xi_L)\alpha_L^3} \,,\\
\mathcal{R}_{221} & = \frac{-3+(3+\xi_L)t_L)}{\xi_L^2\alpha_L^3} \,,\\
\mathcal{R}_{441} & = \frac{-15+13\xi_L+3(1+\xi_L)(5+\xi_L)t_L}{4\xi_L^3\alpha_L^3} \,,\\
\mathcal{R}_{402} & = \frac{3(\xi_L-1)+(\xi_L(3\xi_L-2)+3)t_L}{4\xi_L^2\alpha_L} \,,\\
\mathcal{R}_{421} & = \frac{3+\xi_L+(1+\xi_L)(\xi_L-3)t_L}{4\xi_L^2\alpha_L} \,,\\
\mathcal{R}_{422} & = \frac{15+\xi_L+(\xi_L(\xi_L-6)-15)t_L}{4\xi_L^3\alpha_L^3} \,,\\
\mathcal{R}_{403} & = \frac{(\xi_L-3)(5+3\xi_L)+3(1+\xi_L)(\xi_L(\xi_L-2)+5)t_L}{4\xi_L^3(1+\xi_L)\alpha_L^3}\,,
\end{align}
\es 
where $\xi_L = \alpha_L^{-2} - 1$ and $t_L = \mathrm{arctan}\sqrt{\xi_L}/\sqrt{\xi_L}$. These are the same functions as Eq.~\eqref{eqapp3:A7} except here we set $\alpha_\perp = 1$; this translates to $w = \alpha_L$ and $z = \xi_L$.

\section{\trento{} energy deposition model}
\label{app4d}
In the \trento{} model, the transverse energy deposition (GeV/fm$^2$) of a single fluctuating nuclear collision event in the mid-rapidity region is~\cite{Moreland:2014oya}
\be
\label{eq:trento_perp}
\frac{dE_T}{dxdyd\eta_s}_{\big|{\eta_s=0}} = N \times T_R(x,y)\,,
\ee
where $N$ is the normalization parameter and 
\be
T_R(x,y) = \left(\frac{T^p_A(x,y) + T^p_B(x,y)}{2}\right)^{1/p}
\ee
is the reduced nuclear thickness function, with $p$ being the geometric parameter. The nuclear thickness function of nucleus $A$ ($B$) is\footnote{In Chapters~\ref{chapter4label} and~\ref{chapter7label}, we only consider Pb+Pb collisions ($A=B=208$) at LHC energies $\sqrt{s_\text{NN}} = 2.76$ TeV; the inelastic nucleon--nucleon cross section is set to $\sigma_\text{NN} = 6.4$ fm$^{-2}$.}
\be
T_{A,B}(x,y) = \sum_{n=1}^{N_{\text{part},A,B}} \gamma_n \,T_p(x-x_n,y-y_n)\,,
\ee
where $N_{\text{part},A,B}$ are the number of participant nucleons from nucleus $A$ ($B$); the nucleon positions are sampled from a Woods--Saxon distribution under the constraint that each nucleon--nucleon pair in nucleus $A$ ($B$) maintains a minimum separation $d_\text{min}$~\cite{Moreland:2014oya}. The participant nucleon's thickness function is centered around its sampled transverse position ($x_n$, $y_n$):
\be
\label{app4:Tpp}
    T_p(x-x_n, y - y_n) = \frac{1}{2\pi w^2} \times \exp\left[-\frac{(x{-}x_n)^2+(y{-}y_n)^2}{2w^2}\right]\,,
\ee
where $w$ is the nucleon width parameter.\footnote{This parameter does not control the nucleon's charge radius but rather the spread of thermal energy it deposits in the collision zone along the transverse directions.} Furthermore, the multiplicity factor $\gamma_n$ of each participant nucleon is sampled from the gamma distribution
\be
P(\gamma) = \frac{k^k \gamma^{k-1}\exp\left[-k\gamma\right]}{\Gamma[k]}\,,
\ee
where $k = \sigma_k^{-2}$ and $\sigma_k$ is the standard deviation. 

For a very brief period $\tau_0$ after the collision, we assume the system is longitudinally free-streaming and static in Milne coordinates (i.e. $\PL/\Peq \sim 0$ and $\boldsymbol{u} \sim \boldsymbol{0}$) so that $\ene(x) \propto 1/\tau_0$. Therefore, we initialize the energy density profile of the hydrodynamic simulation at the starting time $\tau_0$ as
\be
\label{eq:trento_3d}
\ene(\tau_0,x,y,\eta_s) = \frac{1}{\tau_0} \times \frac{dE_T}{dxdyd\eta_s}_{\big|\eta_s=0} \times f_L(\eta_s)\,.
\ee
In Chapter~\ref{chapter4label} we extend the transverse energy density profile along the spacetime rapidity direction with a smooth plateau distribution (unitless)~\cite{Pang:2018zzo}:
\be
\label{app4:plateau}
f_L(\eta_s) = \exp\left[-\frac{\big(|\eta_s| - \half\eta_\text{flat}\big)^2\, \Theta\big(|\eta_s| - \half\eta_\text{flat}\big)}{2\sigma_\eta^2}\right]\,,
\ee
where $\eta_\text{flat}$ is the plateau length, $\sigma_\eta$ is the standard deviation of the half-Gaussian tails and $\Theta$ is the Heaviside step function. The initial condition parameter values used in Chapter~\ref{chapter4label} are $N = 14.19$ GeV, $p = 0.06$, $w = 1.11$ fm, $d_\text{min} = 1.45$ fm, $\sigma_k = 1.03$, $\eta_\text{flat} = 4.0$ and $\sigma_\eta = 1.8$~\cite{Pang:2018zzo,Everett:2020yty,Everett:2020xug}.  

In Chapter~\ref{chapter4label} we set the lattice spacings to $\Delta x = \Delta y = \frac{1}{5}w$ and $\Delta \eta_s = \frac{1}{5}\sigma_\eta$ to resolve the fluctuating energy density profile~\eqref{eq:trento_3d} (or event-averaged profile). For the grid size, we set the longitudinal length to $L_\eta = \eta_\text{flat} + 10\sigma_\eta$ to fit the rapidity plateau~\eqref{app4:plateau}. The transverse lengths $L_x$ and $L_y$ are automatically configured by the algorithm described in Sec. 5.2 of Ref.~\cite{McNelis:2021zji}.
\section{Numerical implementation of second-order viscous hydrodynamics}
\label{app4e}

In this appendix, we summarize how second-order viscous hydrodynamics is implemented in the \cpuvah{} code. We evolve viscous hydrodynamics with the same numerical algorithm discussed in Sec.~\ref{chap4S3} except the energy-momentum tensor is decomposed as
\be
\label{eq:Tmunu_VH_VH2}
T^\munu = \ene u^\mu u^\nu - (\Peq {+} \Pi) \Delta^\munu + \pi^\munu \,,
\ee
where $\pi^\munu = \Delta^\munu_\ab T^{\alpha\beta}$ is the shear stress tensor and $\Pi = -\frac{1}{3}\Delta_\munu T^\munu - \Peq$ is the bulk viscous pressure. We also define the spatial projector $\Delta^\munu = g^\munu - u^\mu u^\nu$ and traceless double spatial projector $\Delta^\munu_\ab = \half(\Delta^\mu_\alpha\Delta^\nu_\beta +\Delta^\nu_\beta\Delta^\mu_\alpha - \frac{2}{3}\Delta^\munu\Delta_{\alpha\beta})$. The corresponding dynamical variables
\be
\label{eq:q_viscous}
    \boldsymbol{q} = (T^{\tau\mu}, \pi^\munu, \Pi)
\ee
are propagated along with $\ene$ and $\boldsymbol{u}$ (the mean-field $B$ and anisotropic variables ($\Lambda$, $\alpha_\perp$, $\alpha_L$) are not propagated). Although $\pi^\munu$ has only five independent components, we propagate all ten components in the simulation~\cite{Bazow:2016yra}.\footnote{%
    For longitudinally boost-invariant systems, we do not propagate the shear stress components $\pi^{\tau\eta}$, $\pi^{x \eta}$ and $\pi^{y \eta}$.}
%
%
\subsection{Hydrodynamic equations}

Here we list the evolution equations for the dynamical variables~\eqref{eq:q_viscous} (we refer the reader to Refs.~\cite{Bazow:2016yra, Denicol:2012cn} and App.~\ref{appch3e} for details on their derivation):
\bs
\allowdisplaybreaks
\label{eq:viscous_hydro_equations}
\beal
\partial_\tau T^{\tau\tau} + \partial_i (v^i T^{\tau\tau}) =& - \frac{T^{\tau\tau} {+} \tau^2 T^{\eta\eta}}{\tau} + (\pi^{\tau\tau} {-} \Peq {-} \Pi) \partial_i v^i \\ \nonumber 
& + v^i \partial_i (\pi^{\tau\tau} {-} \Peq {-} \Pi) - \partial_i \pi^{\tau i} \,, \\\nonumber\\
\partial_\tau T^{\tau x} + \partial_i (v^i T^{\tau x}) =& - \frac{T^{\tau x}}{\tau} - \partial_x (\Peq{+}\Pi) + \pi^{\tau x} \partial_i v^i + v^i \partial_i \pi^{\tau x} - \partial_i \pi^{xi} \,,\\\nonumber\\
\partial_\tau T^{\tau y} + \partial_i (v^i T^{\tau y}) =& - \frac{T^{\tau y}}{\tau} - \partial_y(\Peq{+}\Pi) + \pi^{\tau y} \partial_i v^i + v^i \partial_i \pi^{\tau y} - \partial_i \pi^{yi} \,,\\\nonumber\\
\partial_\tau T^{\tau\eta} + \partial_i (v^i T^{\tau\eta}) =& - \frac{3T^{\tau\eta}}{\tau} - \frac{\partial_\eta (\Peq{+}\Pi)}{\tau^2} + \pi^{\tau\eta} \partial_i v^i + v^i \partial_i \pi^{\tau\eta} - \partial_i \pi^{\eta i}\,,\\\nonumber\\
\partial_\tau \pi^\munu + \partial_i (v^i \pi^\munu) =& \,\pi^\munu \partial_i v^i + \frac{1}{u^\tau}\left[-\frac{\pi^\munu}{\tau_\pi} + \I_{\pi^\prime}^\munu - \mathcal{P}_{\pi^\prime}^\munu - \mathcal{G}_{\pi^\prime}^\munu \right] \,,\\\nonumber\\
\partial_\tau \Pi + \partial_i (v^i \Pi) =& \,\Pi \partial_i v^i + \frac{1}{u^\tau}\left[-\frac{\Pi}{\tau_\Pi} + \I_\Pi\right]  \,,
\end{align}
\es
where $T^{\eta\eta} = (\ene{+}\Peq{+}\Pi)(u^\eta)^2 + (\Peq{+}\Pi)/\tau^2+\pi^{\eta\eta}$,
\bs
\beal
\mathcal{I}^\munu_{\pi^\prime} =&\, 2 \beta_\pi \sigma^\munu  +\Delta^\munu_{\alpha\beta}\big(2 \pi^{\lambda (\alpha} \omega^{\beta)}_{\,\,\,\lambda} - \bar{\tau}_{\pi\pi} \pi^{\lambda(\alpha} \sigma^{\beta)}_{\,\,\,\lambda}\big) - \bar{\delta}_{\pi\pi} \pi^\munu \theta  + \bar{\lambda}_{\pi\Pi} \Pi \sigma^\munu \,,\\\nonumber\\  
\mathcal{I}_\Pi =&\, -\beta_\Pi \theta - \bar{\delta}_{\Pi\Pi}\Pi\theta + \bar{\lambda}_{\Pi\pi} \pi^\munu \sigma_\munu\,,
\end{align}
\es
are the gradient source terms for $\pi^\munu$ and $\Pi$ and
\bs
\label{eq:geometric_shear_prime}
\beal
\mathcal{P}^\munu_{\pi^\prime} =&\, \left(\pi^{\mu\alpha}u^\nu + \pi^{\nu\alpha}u^\mu\right)a_\alpha \,,\\
\mathcal{G}^\munu_{\pi^\prime} =&\, u^\gamma \Gamma^\mu_{\gamma\lambda} \pi^{\nu\lambda} 
+ u^\gamma \Gamma^\nu_{\gamma\lambda} \pi^{\mu\lambda}\,,
\end{align}
\es
are the spatial projection and geometric source terms for $\pi^\munu$. We also define the velocity-shear tensor $\sigma^\munu = \Delta^\munu_\ab D^{(\alpha}u^{\beta)}$ and vorticity tensor $\omega^\munu = \Delta^\mu_\alpha \Delta^\nu_\beta D^{[\alpha}u^{\beta]}$. The transport coefficients are listed in App.~\ref{appch3e}.

The components of $\sigma^\munu$ and $\mathcal{G}^\munu_{\pi^\prime}$ are the same as $\sigma_\perp^\munu$ and $\mathcal{G}^\munu_{\pi}$ after replacing $\Xi^\munu_\ab \to \Delta^\munu_\ab$ and $\piperp \to \pi^\munu$ in Eqs.~\eqref{eq:sigmaT} and~\eqref{eq:geometric_piperp_comps}, respectively. The components of $\omega^\munu$ are
\be
\omega^\munu =  D^{[\mu}u^{\nu]} - \frac{u^\mu a^\nu {-} u^\nu a^\mu}{2}\,,
\ee
where $D^{[\mu}u^{\nu]}$ is given by Eq.~\eqref{eq:anti_Du}.
\subsection{Reconstructing the energy density and fluid velocity}
\label{app4:energy}
We reconstruct the energy density by solving the following nonlinear equation via Newton's method~\cite{Shen:2014vra, Bazow:2016yra}:
\be
\label{eq:newton_energy}
f(\ene) = 0 \,,
\ee
where
\be
f(\ene) = \left(\bar{M}^\tau {-} \ene\right)\left(\bar{M}^\tau {+} \Peq {+} \Pi\right) - (\bar{M}^x)^2 - (\bar{M}^y)^2 - (\tau \bar{M}^\eta)^2\,,
\ee
with $\bar{M}^\mu = T^{\tau\mu} - \pi^{\tau\mu}$ and $\Peq = \Peq(\ene)$ being the QCD equation of state. Using the previous energy density for the initial guess, we iterate the solution to Eq.~\eqref{eq:newton_energy} as
\be
\label{eq:energy_iteration}
\ene \leftarrow \ene - \frac{f(\ene)}{df /d\ene}\,,
\ee
where
\be
\frac{df}{d\ene} = c_s^2 (\bar{M}^\tau {-} \ene) - \bar{M}^\tau - \Peq - \Pi\,,
\ee
with $c_s^2 = d\Peq/d\ene$ being the QCD speed of sound squared. We repeat the iteration~\eqref{eq:energy_iteration} until we achieve sufficient convergence or the energy density falls below $\ene_\text{min}$. If the bulk viscous pressure $\Pi < - \Peq$, we regulate it so that $\Peq + \Pi = 0$; this allows us to solve for $\ene$ explicitly~\cite{Shen:2014vra}:
\be
\ene = \bar{M}^\tau - \frac{(\bar{M}^x)^2 {+} (\bar{M}^y)^2 {+} (\tau \bar{M}^\eta)^2}{{\bar M}^\tau}\,.
\ee
Afterwards, we regulate the energy density via Eq.~\eqref{eqchap4:energy_reg} and evaluate the fluid velocity components as
\bs
\allowdisplaybreaks
\beal
u^x &= \frac{\bar{M}^x}{\sqrt{\left(\ene{+}\Peq{+}\Pi\right)\left(\bar{M}^\tau{+}\Peq{+}\Pi\right)}} \,,\\
u^y &= \frac{\bar{M}^y}{\sqrt{\left(\ene{+}\Peq{+}\Pi\right)\left(\bar{M}^\tau{+}\Peq{+}\Pi\right)}} \,,\\
u^\eta &= \frac{\bar{M}^\eta}{\sqrt{\left(\ene{+}\Peq{+}\Pi\right)\left(\bar{M}^\tau{+}\Peq{+}\Pi\right)}}  \,.
\end{align}
\es
\subsection{Regulating the shear stress and bulk viscous pressure}
In this regulation scheme, we first adjust the shear stress components
\bs
\allowdisplaybreaks
\beal
\pi^{\eta\eta} \leftarrow&\, \frac{1}{\tau^2\left(1 {+} u_\perp^2\right)}\Big[\pi^{xx}\big((u^x)^2 {-} (u^\tau)^2\big) + \pi^{yy}\big((u^y)^2 {-} (u^\tau)^2\big) \nonumber\\
&+ 2\big(\pi^{xy} u^x u^y + \tau^2 (\pi^{x\eta}u^x {+} \pi^{y\eta}u^y)u^\eta\big)\Big] \,,\\
\pi^{\tau x} \leftarrow&\, \frac{\pi^{xx} u^x + \pi^{xy}u^y + \tau^2 \pi^{x\eta} u^\eta}{u^\tau}\,,\\
\pi^{\tau y} \leftarrow&\, \frac{\pi^{xy} u^x + \pi^{yy}u^y + \tau^2 \pi^{y\eta} u^\eta}{u^\tau}\,,\\
\pi^{\tau \eta} \leftarrow&\, \frac{\pi^{x\eta} u^x + \pi^{y\eta}u^y + \tau^2 \pi^{\eta\eta} u^\eta}{u^\tau}\,,\\
\pi^{\tau\tau} \leftarrow&\, \frac{\pi^{\tau x} u^x + \pi^{\tau y}u^y + \tau^2 \pi^{\tau\eta} u^\eta}{u^\tau} \,,
\end{align}
\es
so that $\pi^\munu$ satisfies the orthogonality and tracelessness conditions
\bs
\beal
\pi^\munu u_\nu &= 0 \,,\\
\pi^\mu_\mu &= 0 \,.
\end{align}
\es
Then we regulate the shear stress and bulk viscous pressure as
\bs
\beal
\pi^\munu &\leftarrow \gamma_\text{reg} \pi^\munu \,,\\
\Pi &\leftarrow \gamma_\text{reg} \Pi\,,
\end{align}
\es
where
\be
\label{eq:rescale_viscous}
\gamma_\text{reg} = \min\Bigg(1, \sqrt{\frac{3\mathcal{P}_\text{eq}^2}{\pi {\,\cdot\,} \pi + 3 \Pi^2}}\Bigg) \,,
\ee
with $\pi {\,\cdot\,} \pi = \pi_{\munu} \pi^\munu$. The regulation factor $\gamma_\text{reg}$ usually suppresses $\pi^\munu$ and $\Pi$ around the edges of the fireball at early times $\tau < 1$ fm/$c$, especially in standard viscous hydrodynamics (e.g. see Figs.~(\ref{freezeout_x}c) and~(\ref{freezeout_z}c)).
\chapter{}
\section{Moments of the distribution function}
\label{app5:integrals}

We define the thermal, isotropic and anisotropic integrals that appear in Chapter~\ref{chap5label}. For each species $n$ let
\be
  J_{kq,n} \equiv \int_p \frac{(\up)^{k-2q} ({-}p{\,\cdot\,}\Delta{\,\cdot\,}p)^q}{(2q{+}1)!!} f_{\eq,n} \bar{f}_{\eq,n}\,.
\ee
The {\it thermal integrals} over the local-equilibrium distribution are then given by
\bs
\allowdisplaybreaks
\begin{align}
  \mathcal{J}_{kq} &= \sum_n J_{kq,n}\,,
\\
  \mathcal{N}_{kq} &= \sum_n b_n J_{kq,n}\,,
\\
  \mathcal{M}_{kq} &= \sum_n b^2_n J_{kq,n}\,,
\\
  \mathcal{A}_{kq} &= \sum_n m_n^2 J_{kq,n}\,,
\\
  \mathcal{B}_{kq} &= \sum_n b_n m_n^2 J_{kq,n}\,.
\end{align}
\es
The {\it isotropic integrals} over the PTB distribution \eqref{eqch5:Jonah} without shear stress modifications (i.e. for $\pi_{ij} = 0$) are
\be
  \mathcal{L}_{kq} = \sum_n \int_p \frac{(\up)^{k-2q} ({-}p{\,\cdot\,}\Delta{\,\cdot\,}p)^q}{(2q{+}1)!!} f_{\lambda,n}\,,
\ee
where 
\be
  f_{\lambda,n} = \frac{g_n(1{+}\lambda_\Pi)^{-3}}{\exp\left[\dfrac{1}{T}\sqrt{m_n^2 - \dfrac{p {\,\cdot\,} \Delta {\,\cdot\,} p}{(1{+}\lambda_\Pi)^2}}\right]+\Theta_n}\,.
\ee
In particular, the modified energy density and isotropic pressure are given by $\ene^\prime(\lambda_\Pi,z_\Pi,T) = z_\Pi \mathcal{L}_{20}(\lambda_\Pi,T)$ and $\mathcal{P}^\prime(\lambda_\Pi,z_\Pi,T) = z_\Pi \mathcal{L}_{21}(\lambda_\Pi,T)$. Finally, the {\it anisotropic integrals} are given by 
\be
  \J^{(R)}_{krqs} =  \sum_n  \frac{1}{(2q)!!} \int_p (\up)^{k{-}r{-}2q} (- Z {\,\cdot\,} p)^r ({-}p {\,\cdot\,} \Xi {\,\cdot\,} p)^q \pOp^{s/2} f_{a,n} \bar{f}_{a,n}\,.
\ee
\section{Conservation laws in first-order approximation}
\label{app5:conservation}

Here we derive the expression for the time derivatives~\eqref{eq15} by making a first-order approximation to the conversation laws~\eqref{eqch5:conservation_laws}. The conservation equations for the net baryon number and energy up to first order in gradients are
\be
  \dot{n}_B = - n_B \theta,\qquad
  \dot \ene = - (\ene{+}\Peq)\theta\,.
\ee
Taking the time derivative of the kinetic definitions for the net baryon and energy densities,
\be
\!\!
  n_B = \sum_n b_n \int_p (\up) f_{\eq,n}, \quad 
  \ene = \sum_n \int_p (\up)^2 f_{\eq,n},\!\!
\ee
one obtains
\be
  \dot\alpha_B = \mathcal{G}(T,\alpha_B) \theta\,, 
  \qquad
  \dot T = \mathcal{F}(T,\alpha_B) \theta \,.
\ee
The coefficients $\mathcal{G}$ and $\mathcal{F}$ appear in the RTA Chapman--Enskog expansion \eqref{eqch5:Chapman_Enskog}:
\bs
\allowdisplaybreaks
\begin{align}
  \mathcal G &= T\left(\frac{(\ene{+}\Peq)\N_{20} - n_B \J_{30}}{\J_{30} \mathcal{M}_{10} - \N_{20}^2}\right) \,,
\\
  \mathcal F &= T^2\left(\frac{n_B\N_{20} - (\ene{+}\Peq)\mathcal{M}_{10}}{\J_{30} \mathcal{M}_{10} - \N_{20}^2}\right) \,.
\end{align}
\es
The evolution equation for the fluid velocity up to first-order in gradients is given by
\be
\label{eqapp5:fluid_velocity}
  \dot{u}^\mu = \frac{\nabla^\mu \Peq}{\ene{+}\Peq}\,.
\ee
After computing the spatial gradient of the equilibrium pressure 
\be
\Peq = \frac{1}{3} \sum_n \int_p (-p{\,\cdot\,} \Delta{\,\cdot\,}p) f_{\eq,n}\,,
\ee
Eq.~\eqref{eqapp5:fluid_velocity} can be rewritten as
\be
\dot u^\mu = \nabla^\mu {\ln}T + \frac{n_B T}{\ene {+} \Peq} \nabla^\mu \alpha_B\,,
\ee
where we made use of the identities~\cite{Jaiswal:2014isa} $\mathcal{J}_{31} = (\ene{+}\Peq)T$ and $\mathcal{N}_{21} = n_B T$.
\chapter{}
\section{Expansion of the hydrodynamic generator}
\label{appch8proof}
In this section we show that the correspondence between the hydrodynamic generator and Borel resummed RTA Chapman--Enskog series holds for all orders in the Knudsen number. We thank Chandrodoy Chattopadhyay for first working out the proof in the Bjorken case; we further generalize it to (3+1)--dimensions.
\subsection{(0+1)--d hydrodynamic generator}
For Bjorken flow, we want to prove the map
\be
f_\G(\tau,p) = \int^\tau_{\tau_0}\frac{d\tp D(\tau,\tp)\feq(\tp, p)}{\trel(\tp)} \to {\int^\infty_0} dz\, e^{-z} \sum_{n=0}^\infty \frac{{z^n}{\left[{-}\trel(\tau) \partial_\tau\right]^n}\feq(\tau, p)}{n!} \,,
\ee
given that the particle interaction measure $z_0 \to \infty$ at late times. We will work with the expression given by the coordinate transformation in Eq.~\eqref{eqch8:generator_z}:
\be
\label{eqapp8:generator_transform}
  f_\text{G}(\tau,p) = {\int^{z_0}_0} dz \, e^{-z} \feq\big(h^{-1}(z,\tau), p\big) \,,
\ee
where
\be
\label{eqapp8:z_transform} 
  z = h(\tp, \tau) = \int^\tau_{\tp} \frac{d\tau^{\prime\prime}}{\tau_r(\tau^{\prime\prime})}
\ee
and
\be
\label{eqapp8:tau_prime}
  \tau^\prime = h^{-1}(z,\tau) = \sum_{n=0}^\infty c_n(\tau) \, z^n \,.
\ee
First we show the recurrence relation~\eqref{eqch8:recursion} between the expansion coefficients $c_n$:
\bs
\allowdisplaybreaks
\label{eqapp8:recurrence}
\begin{align}
    c_0 &= \tau \,, \\
    c_n &= - \frac{\tau_r(\tau) \partial_\tau c_{n{-}1}}{n} \indent \forall\,n\geq 1 \,.
\end{align}
\es
We apply the partial derivative $\partial_\tau$ on both sides of Eq.~\eqref{eqapp8:tau_prime} (while keeping $\tau^\prime =$ const):
\be
\label{eqapp8:partial_on_tau_prime}
  0 = \sum_{n=0}^\infty\left( \partial_\tau c_n \, z^n + \frac{n c_n \,z^{n-1}}{\tau_r(\tau)} \right) = \sum_{n=0}^\infty z^n \left( \partial_\tau c_n  + \frac{(n{+}1) c_{n+1}}{\tau_r(\tau)} \right) \,,
\ee
where we used the relation $\partial_\tau z = 1/\tau_r(\tau)$ and incremented the index $n$ in the second term. Setting the expression in parentheses to zero gives us Eq.~(\ref{eqapp8:recurrence}b). We compute the coefficient $c_0$ by setting $z = 0$ in Eq.~\eqref{eqapp8:tau_prime} so that $c_0 = \tau^\prime$. From Eq.~\eqref{eqapp8:z_transform}, $\tau^\prime = \tau$ when $z = 0$, thus establishing the recurrence relation~\eqref{eqapp8:recurrence}.

Next we expand the function $\feq\big(h^{-1}(z,\tau),p\big)$ in the integrand~\eqref{eqapp8:generator_transform} as a power series:
\be
\label{eqapp8:feqh_series}
\feq\big(h^{-1}(z,\tau),p\big) = \sum_{n=0}^\infty b_n(\tau,p) z^n
\ee
and compute the coefficients $b_n$. With the recurrence relation~\eqref{eqapp8:recurrence}, we can construct the identity
\be
\left(\frac{\partial h^{-1}}{\partial z}\right)_\tau \,+\,  \tau_r(\tau)\left(\frac{\partial h^{-1}}{\partial \tau}\right)_{z} = 0\,.
\ee
Thus, after the applying the chain rule on $\feq\big(h^{-1}(z,\tau),p\big)$, we have
\be
\label{eqapp8:feqh_id}
\left(\frac{\partial \feq(h^{-1},p)}{\partial z}\right)_\tau \,+\, \tau_r(\tau)\left(\frac{\partial \feq(h^{-1},p)}{\partial\tau}\right)_z = 0\,.
\ee
Inserting the power series~\eqref{eqapp8:feqh_series} into Eq.~\eqref{eqapp8:feqh_id} yields the recurrence relation
\be
b_n = - \frac{\tau_r(\tau)\partial_\tau b_{n-1}}{n} \indent \forall\,n\geq 1 \,,
\ee
which is identical to Eq.~(\ref{eqapp8:recurrence}b). By setting $z = 0$ in Eq.~\eqref{eqapp8:feqh_series}, we get the first coefficient
\be
b_0 = \feq(\tau,p)\,.
\ee
As a result, the series expansion of the hydrodynamic generator~\eqref{eqapp8:generator_transform} becomes
\be
f_\G(\tau,p) = {\int^{z_0}_0} dz \, e^{-z} \sum_{n=0}^\infty \frac{{z^n}{\left[{-}\trel(\tau) \partial_\tau\right]^n}\feq(\tau, p)}{n!} \,.
\ee
Finally, we take the limit $z_0 \to \infty$ to recover the Borel resummed RTA Chapman--Enskog series~\eqref{eqch8:borel_sum_general}.

\subsection{(3+1)--d hydrodynamic generator}
The generalization of the correspondence to (3+1)--dimensions 
\be
\begin{split}
f_\G(x,p) &= \int^x_{x_-}{dx^\prime {\cdot\,}s^{-1}(x^\prime, p) D(x,x^\prime,p)\feq(x^\prime, p)} \\
&\to \int_0^\infty dz\,e^{-z} \sum_{n=0}^\infty \frac{z^n\left[- s^\mu(x,p)\partial_\mu\right]^n \feq(x,p)}{n!} 
\end{split}
\ee
is completely analogous. We start from Eq.~\eqref{eqch8:generator_z_minkowski}
\be
\label{eqapp8:generator_z_minkowski}
  f_\text{G}(x,p) = \int_0^{z_-} dz\, e^{-z} \feq\big(h^{-1}(z, x, p), p\big) \,,
\ee
where
\be
\label{eqapp8:z_minkowski}
  z = h(x^\prime, x, p) = \int^x_{x^\prime} dx^{\prime\prime}{\cdot\,}s^{-1}(x^{\prime\prime}, p)
\ee
and
\be
\label{eqapp8:h_inverse}
  x^\prime_\mu = h^{-1}_\mu(z, x, p) = \sum_{n=0}^\infty c_{n,\mu}(x,p)\,z^n \,.
\ee
After taking the directional derivative $s^\nu(x,p)\partial_\nu$ of Eq.~\eqref{eqapp8:h_inverse} while keeping the coordinate $x^\prime_\mu$ fixed, we get
\be
0 = \sum_{n=0}^\infty z^n\left( s^\nu(x,p) \partial_\nu c_{n,\mu} + (n{+}1) c_{n+1,\mu} \right)\,,
\ee
where we used the identity $s^\nu(x,p)\partial_\nu z = 1$. This gives us the recurrence relation~\eqref{eqch8:recursion_general} (we get $c_{0,\mu} = x_\mu$ by setting $z = 0$ in Eqs.~\eqref{eqapp8:z_minkowski} --~\eqref{eqapp8:h_inverse})). 

Then we expand the function in Eq.~\eqref{eqapp8:generator_z_minkowski} as a power series:
\be
\label{eqapp8:feqh_series_gen}
\feq\big(h^{-1}(z, x, p), p\big) = \sum_{n=0}^\infty b_n(x,p) z^n\,.
\ee
From the recurrence relation~\eqref{eqch8:recursion_general}, we have the identity
\be
\left(\frac{\partial h^{-1}_\mu}{\partial z}\right)_x \,+\,  s^\nu(x,p)\left(\frac{\partial h^{-1}_\mu}{\partial x^\nu}\right)_{z} = 0\,,
\ee
or equivalently
\be
\label{eqapp8:feqh_3d}
\left(\frac{\partial \feq(h^{-1},p)}{\partial z}\right)_x \,+\, s^\nu(x,p)\left(\frac{\partial \feq(h^{-1},p)}{\partial x^\nu}\right)_z = 0\,.
\ee
After inserting the power series~\eqref{eqapp8:feqh_series_gen} into Eq.~\eqref{eqapp8:feqh_3d}, we find the recurrence relation
\be
b_n = - \frac{s^\nu(x,p) \partial_\nu b_{n-1}}{n} \indent \forall\,n\geq 1 \,,
\ee
while the first coefficient is (after setting $z = 0$ in Eq.~\eqref{eqapp8:feqh_series_gen})
\be
b_0 = \feq(x,p)\,.
\ee
Therefore, the expansion of the (3+1)--d hydrodynamic generator~\eqref{eqapp8:generator_z_minkowski} simplifies to 
\be
f_\G(x,p) = \int_0^{z_-} dz\,e^{-z} \sum_{n=0}^\infty \frac{z^n\left[- s^\mu(x,p)\partial_\mu\right]^n \feq(x,p)}{n!} 
\ee
After taking the limit $z_- \to \infty$, we arrive at Eq.~\eqref{eqch8:Chapman_series_Minkowski}.
\section{Standard gradient corrections}
\label{appch8a}
Here we compute the standard gradient corrections to the normalized shear stress for a conformal system undergoing Bjorken expansion:
\be
  \bar\pi_s = \bar\pi_s^{(1)} + \bar\pi_s^{(2)} + \bar\pi_s^{(3)} + O(\Kn^4) \,.
\ee
Before proceeding, we make a change in variables $w = \tau^2 p^\eta$ to rewrite the local-equilibrium distribution~\eqref{eqch8:feq} as
\be
\feq(\tau,p_T,w) = \exp\Bigg[-\frac{\sqrt{\tau^2p_\perp^2 + w^2}}{\tau T(\tau)}\Bigg] \,.
\ee
The first-order gradient correction to the distribution function is
\be
\label{eqapp8:df1}
\delta f^{(1)} = - \trel \partial_\tau \feq = - \frac{\tpi \feq}{\tau^2 T} \bigg(\frac{w^2}{v} \,+\, v \tau \partial_\tau \ln T\bigg) \,,
\ee
where we set $\trel = \tau_\pi = 5(\etas)/T$ (taking $\etas$ as a constant) and $v = \sqrt{\tau^2p_\perp^2 + w^2}$. The first-order correction to the shear stress is then
\be
\label{eqapp8:dPL1}
\bar\pi^{(1)}_s = \frac{2\pi^2}{3T^4} \int_p \left(\frac{p_\perp^2}{2} - \frac{w^2}{\tau^2}\right) \delta f^{(1)} \,,
\ee
where $\int_p = \int \dfrac{d^2 p_\perp dw}{v(2\pi)^3}$. After inserting $\delta f^{(1)}$ in Eq.~\eqref{eqapp8:dPL1} and substituting the spherical coordinates
\bs
\allowdisplaybreaks
\begin{align}
\tau p^x &= v \sin\theta \cos\phi \,, \\
\tau p^y &= v \sin\theta \sin\phi \,, \\
w &= v \cos\theta \,,
\end{align}
\es
one obtains
\be
\bar\pi^{(1)}_s = \frac{16\tpi}{15\tau} \,.
\ee
The $\delta f^{(2)}$ and $\delta f^{(3)}$ corrections are too cumbersome to list here. We simply state the results for the second and third-order shear corrections (for the derivation see the auxiliary materials available in the link referenced in footnote~\ref{github}):
\bs
\allowdisplaybreaks
\label{eqapp8:pi23}
\begin{align}
\bar\pi^{(2)}_s &= - \frac{16 \tau_\pi^2}{105\tau^2} \left(15 + 49 \tau \partial_\tau{\ln T} \right) \,, \\
\bar\pi^{(3)}_s &= \frac{16 \tau_\pi^3}{105\tau^3} \big(\tau\partial_\tau{\ln T}(135 + 182\tau\partial_\tau{\ln T}) + 77\tau^2\partial^2_\tau{\ln T} \big) \,.
\end{align}
\es
Here we used the relation $\partial_\tau \tau_\pi = - \tau_\pi \partial_\tau{\ln T}$ to eliminate time derivatives of the shear relaxation time. At late times, the gradients $\Kn = \tau_\pi / \tau  \sim \tau^{-2/3}$ become small. Hence, the asymptotic solutions for the energy conservation law and its time derivative \eqref{eqch8:conservation_law} are
\bs
\allowdisplaybreaks
\begin{align}
  \tau \, \partial_\tau{\ln T} &= -\frac{1}{3} + \frac{4\tpi}{45\tau} + O\big(\Kn^2\big) \,, \\
  \tau^2 \partial^2_\tau{\ln T} &= \frac{1}{3} + O\left(\Kn\right) \,.
\end{align}
\es
The shear corrections \eqref{eqapp8:pi23} then reduce to
\bs
\allowdisplaybreaks
\label{eqapp8:shear_asymptotic}
\begin{align}
  \bar\pi^{(2)}_s &\approx \frac{64 \tau_\pi^2}{315\tau^2} - \frac{448 \tau_\pi^3}{675\tau^3} \,,\\
  \bar\pi^{(3)}_s &\approx \frac{128 \tau_\pi^3}{945\tau^3} \,.
\end{align}
\es
Finally, the non-hydrodynamic modes in the exact solution \eqref{eqch8:anomaly} decay at late times since $z_0 \sim \tau^{2/3}$. Using this, the second and third-order corrections in Eq.~\eqref{eqapp8:shear_asymptotic} can be regrouped as
\bs
\begin{align}
  \bar\pi^{(2)}_s &\to \frac{64 \tau_\pi^2}{315\tau^2} \,, \\
  \bar\pi^{(3)}_s &\to -\frac{832 \tau_\pi^3}{1575\tau^3} \,,
\end{align}
\es
which appear in the Burnett and super-Burnett solutions (\ref{eqch8:Navier_Burnett}b,c).

\section{Leading non-hydrodynamic mode correction}
\label{appch8b}

Here we compute the shear stress correction from the leading non-hydrodynamic mode $\delta f_\G^{(0)} = e^{-z_0}(f_0 - \feq)$:
\be
\label{eqapp8:shear_nonhydro}
  \bar\pi^{(0)}_{s,\G} = \frac{2\pi^2 e^{-z_0}}{3T^4} \int_p \left(\frac{p_\perp^2}{2} - \frac{w^2}{\tau^2}\right) f_0(\tau_0,p_T,w) \,,
\ee
where the second term $\propto \feq(\tau,p_T,w)$ vanishes by symmetry. The code \cite{Florkowski:2013lya, Florkowski:2013lza,Tinti:2018qfb} that evolves the RTA Bjorken solution \eqref{eqch8:exact} gives the user the option to initialize the distribution as
\be
  f_0(\tau_0, p_T, w) = \exp\left[-\frac{\sqrt{\tau_0^2p_\perp^2 + (1{+}\xi_0)w^2}}{\tau_0 \Lambda_0}\right] \,,
\ee
where
\be
\label{eqapp8:lambda_0}
\Lambda_0 = T_0 \, \mathcal{H}\big((1{+}\xi_0)^{-1/2}\big)^{-1/4}
\ee
is the effective temperature and $\xi_0$ is the momentum anisotropy parameter. After substituting the spherical coordinates
\bs
\begin{align}
\tau_0 p^x &= v_0 \sin\theta \cos\phi \,, \\
\tau_0 p^y &= v_0 \sin\theta \sin\phi \,, \\
(1+\xi_0)^{1/2}w &= v_0 \cos\theta \,,
\end{align}
\es
where $v_0 = \sqrt{\tau_0^2p_\perp^2 + (1{+}\xi_0)w^2}$, Eq.~\eqref{eqapp8:shear_nonhydro} can be rewritten as
\be
\label{eqapp8:pi0_aniso}
\bar\pi^{(0)}_{s,\G} = e^{-z_0}\frac{\Lambda^4_0}{T^4}\left[\frac{1}{2}\mathcal{H}_\perp\Big(\frac{\tau_0 \alpha_{L,0}}{\tau}\Big) {-}  \mathcal{H}_L\Big(\frac{\tau_0\alpha_{L,0}}{\tau}\Big)\right] \,,
\ee
where $\alpha_{L,0} = (1\,{+}\,\xi_0)^{-1/2}$ and the hypergeometric functions \cite{Florkowski:2013lya} 
\bs
\begin{align}
\mathcal{H}_\perp(x) &=\, x \int_{{-}1}^1 \frac{d{\cos\theta} \, (1 - \cos^2\theta)}{\sqrt{1 + (x^2{-}1)\cos^2\theta}} \,, \\
\mathcal{H}_L(x) &=\, x^3 \int_{{-}1}^1 \frac{d{\cos\theta} \, \cos^2\theta}{\sqrt{1 + (x^2{-}1)\cos^2\theta}}
\end{align}
\es
are 
\bs
\begin{align}
  \mathcal{H}_\perp(x) &=\, \frac{1}{1{-}x^2}\big(x^2 + (1{-}2x^2)\, t(x^{-2}{-}1)\big) \,, \\ 
  \mathcal{H}_L(x) &=\,\frac{x^2}{1{-}x^2}\big({-}x^2 + t(x^{-2}{-}1)\big) \,,
\end{align}
\es
with $t(y) = \tan^{-1}{\sqrt{y}} / \sqrt{y}$. In Sec.~\ref{expansion}, we had initialized the shear stress to $\pi_s(\tau_0) = 0$ so that $\xi_0 = 0$ and $\Lambda_0 = T_0$. Then Eq.~\eqref{eqapp8:pi0_aniso} reduces to Eq. (\ref{eqch8:dfg}a).

\section{Solution of the RTA Boltzmann equation in (3+1)--dimensions}
\label{appch8c}

Here we derive the (3+1)--dimensional solution of the RTA Boltzmann equation in Minkowski spacetime. First, we multiply both sides of Eq.~\eqref{eqch8:RTAmink} by the function
\be
\label{eqapp8:C2}
  q(x,p) = \exp\left[{\int_{x_\star}^x} dx^{\prime\prime} {\cdot\, } s^{-1}(x^{\prime\prime},p)\right] \,,
\ee
where the path integral runs over a straight line that is parallel to $p^\mu$; the coordinate $x_\star^\mu$ is a fixed point on the characteristic line (see Figure~\ref{minkowski_diagram})
\be
\label{eqapp8:characteristic}
{x^\prime}^\mu(\lambda^\prime) = x^\mu - \lambda^\prime p^\mu  \indent \indent 0 \leq \lambda^\prime \leq \lambda_- \,,
\ee
with $\lambda_- = (t - t_-) / E$. Equation~\eqref{eqch8:RTAmink} can be rewritten as
\be
s^\mu(x,p)\partial_\mu \left[q(x,p) f(x,p)\right] = q(x,p) \feq(x,p)\,. 
\ee
Now we integrate this equation along the characteristic line~\eqref{eqapp8:characteristic}:
\be
\label{eqapp8:integrate_character}
\int_{x_-}^x dx^\prime {\cdot\, } s^{-1}(x^\prime,p)\, s^\mu(x^\prime,p)\frac{\partial\left[q(x^\prime,p) f(x^\prime,p)\right]}{\partial {x^\prime}^\mu} = \int_{x_-}^x dx^\prime {\cdot\, } s^{-1}(x^\prime,p)\, q(x^\prime,p) \feq(x^\prime,p) \,.
\ee
The l.h.s. of Eq.~\eqref{eqapp8:integrate_character} can be parameterized in terms of $\lambda^\prime$:
\be
\int_{\lambda_-}^0 d\lambda^\prime \frac{d\left[q(x^\prime(\lambda^\prime),p) f(x^\prime(\lambda^\prime),p)\right]}{d\lambda^\prime} = q(x,p) f(x,p) - q(x_-,p) f(x_-,p) \,,
\ee
where we used the relations $d{x^\prime}^\nu = -p^\nu d\lambda^\prime$ and $\dfrac{\partial}{\partial {x^\prime}^\mu} = -\dfrac{p_\mu}{p^2} \dfrac{d}{d\lambda^\prime}$. For the distribution function on the hypersurface $\Sigma_-$ (see Figure~\ref{minkowski_diagram}) we take
\be
f(x_-,p) = f_0(x_-,p) \Theta(t_0 - t_-) \,.
\ee
The solution of the RTA Boltzmann equation is then
\be
\label{eqapp8:C8}
  f(x,p) = \,\frac{q(x_-,p)}{q(x,p)} f_0(x_-,p) \Theta(t_0 - t_-) \,+\, \int_{x_-}^x dx^\prime {\cdot\, } s^{-1}(x^\prime,p)\, \frac{q(x^\prime,p)}{q(x,p)} \feq(x^\prime,p) \,.
\ee
After using the identity $D(x_2,x_1,p) = q(x_1,p) / q(x_2,p)$, one arrives at Eq.~\eqref{eqch8:rta_solution}.
\end{appendices}

\backmatter
\bibliography{templatebib}
\bibliographystyle{elsarticle-num}

\end{document}